\documentclass[twoside,a4paper,11pt]{book}

\usepackage{etex}

\usepackage{amssymb,amsfonts,amsthm}
\usepackage{extarrows}

\usepackage[british,spanish,es-nosectiondot]{babel} 
\usepackage[T1]{fontenc}
\usepackage[utf8]{inputenc} 

\usepackage{fancyhdr} 
\usepackage[nottoc]{tocbibind} 
\usepackage{graphicx}
\usepackage{graphbox}
\usepackage{epsfig}\parskip 5pt plus 1pt
\usepackage{multirow}
\usepackage{booktabs}  
\usepackage{epsfig}
\usepackage{epstopdf}
\usepackage{tikz-cd}
\usepackage{commath}
\usepackage{xcolor}
\usepackage{cancel}
\usepackage{tikz-feynman}

\usepackage[utf8]{inputenc} 

\usepackage{enumerate}
\usepackage{mathtools}
\usepackage{amsmath}
\usepackage{amssymb}
\usepackage{amsfonts}
\usepackage{bbm}
\usepackage{mathrsfs}
\usepackage{ytableau}
\usepackage{subfigure} 
\usepackage[font=small,labelfont=bf]{caption} 
\usepackage{comment}
\usepackage{fancybox}
\usepackage{braket}
\usepackage[skins,theorems]{tcolorbox}
\tcbset{highlight math style={enhanced,
		colframe=black,colback=white,arc=0pt,boxrule=1pt}}

\newtheorem{theorem}{Claim}

\usepackage{amsbsy}
\usepackage{color}

\definecolor{dark-gray}{gray}{0.20}
\definecolor{gray}{gray}{0.30}
\definecolor{light-gray}{gray}{0.80}
\definecolor{dark-red}{rgb}{0.7,0,0}
\definecolor{dark-green}{rgb}{0.1,0.4,0}
\definecolor{dark-blue}{rgb}{0.3,0.3,0.7}
\definecolor{light-blue}{rgb}{0.8,0.8,1}
\definecolor{swamp}{RGB}{240, 199, 197}

\usepackage{slashed}

\usepackage[hmargin=3cm,vmargin=3.4cm]{geometry}

\usepackage{tabularx} 

\newcolumntype{M}[1]{>{\centering\arraybackslash}m{#1}}
\newcolumntype{N}{@{}m{0pt}@{}}

\definecolor{MyGrey}{rgb}{0,0,0} 
\definecolor{MyDarkBlue}{rgb}{0.23,0.21,0.69} 
\definecolor{MyLightBlue}{rgb}{0.22,0.51,0.86}
\definecolor{MyDarkRed}{rgb}{0.69,0.21,0.23}

\usepackage[colorlinks=true,linkcolor=MyDarkRed,citecolor=MyDarkRed,urlcolor=MyLightBlue,bookmarksnumbered=true,bookmarksopen]{hyperref}

\usepackage{cite} 
\usepackage{cleveref}
\usepackage{float}

\usetikzlibrary{mindmap,trees}

\usepackage[times]{quotchap}

\def\beq{\begin{equation}}
\def\eeq{\end{equation}}
\def\beqa{\begin{eqnarray}}
\def\eeqa{\end{eqnarray}}

\def\-{\hphantom{-}}

\def\s2{\frac{1}{\sqrt2}}

\def\IC{\mathbb{C}}

\def\IF{\relax{\rm I\kern-.18em F}}
\def\II{\relax{\rm I\kern-.18em I}}
\def\IP{\relax{\rm I\kern-.18em P}}
\def\IC{\relax\hbox{\kern.25em$\inbar\kern-.3em{\rm C}$}}
\def\IR{\relax{\rm I\kern-.18em R}}

\newcommand{\bmat}{\left(\begin{array}}
\newcommand{\emat}{\end{array}\right)}

\def\yzero{\smash{\hbox{$y\kern-4pt\raise1pt\hbox{${}^\circ$}$}}}

\def\b{\beta}
\def\g{\gamma}

\def\beq{\begin{equation}}
\def\eeq{\end{equation}}
\def\beqa{\begin{eqnarray}}
\def\eeqa{\end{eqnarray}}

\def\-{\hphantom{-}}

\def\s2{\frac{1}{\sqrt2}}

\def\beq{\begin{equation}}
\def\eeq{\end{equation}}
\def\beqa{\begin{eqnarray}}
\def\eeqa{\end{eqnarray}}

\def\IF{\relax{\rm I\kern-.18em F}}
\def\II{\relax{\rm I\kern-.18em I}}
\def\IP{\relax{\rm I\kern-.18em P}}
\def\IC{\relax\hbox{\kern.25em$\inbar\kern-.3em{\rm C}$}}
\def\IR{\relax{\rm I\kern-.18em R}}

\def\Dsl{\,\raise.15ex\hbox{/}\mkern-13.5mu D} 

\def\LQG{\Lambda_{\text{QG}}}
\def\LSP{\Lambda_{\text{sp}}}

\def\M6{\mathcal{M}_6}
\def\i{\text{i}}

\def\Mpd{M_{\text{Pl;}\, d}}
\def\Mpf{M_{\text{Pl;}\, 4}}
\def\Mpt{M_{\text{Pl;}\,  10}}

\def\MKK{m_{\text{KK}}}

\def\LQG{\Lambda_{\text{QG}}}
\def\LSP{\Lambda_{\text{sp}}}
\def\Ns{\mathsf{N}_s}
\def\Ms{m_s}
\def\Ntot{N_{\text{tot}}}
\def\MBH{M_{\text{BH}}}

\def\Mt{m_{\text{tow}}}
\def\Mti{m_{\text{tow,}\, i}}
\def\peff{p_{\text {eff }}}
\def\Mteff{m_{\text {tow, eff }}}
\def\fdim{2^{\lfloor d/2 \rfloor}}

\def\b{\beta}

\def\g{\gamma}

\def\s{\sigma}

\def\beq{\begin{equation}}
\def\enq{\end{equation}}

\newcommand{\cK}{{\cal K}}  
\newcommand{\cM}{{\cal M}}

\newcommand{\be}{\begin{equation}} \newcommand{\ee}{\end{equation}}
\newcommand{\bea}{\begin{eqnarray}} \newcommand{\eea}{\end{eqnarray}}
\newcommand{\beann}{\begin{eqnarray*}}  \newcommand{\eeann}{\end{eqnarray*}}
\newcommand{\bfig}{\begin{figure}} \newcommand{\efig}{\end{figure}}
\newcommand{\ba}{\begin{array}} \newcommand{\ea}{\end{array}}
\newcommand{\bcen}{\begin{center}} \newcommand{\ecen}{\end{center}}
\newcommand{\btab}{\begin{tabular}} \newcommand{\etab}{\end{tabular}}

\newcommand{\matt}{\left ( \begin{array}{ccc}}
\newcommand{\ematt}{\end{array} \right )} \newcommand{\matf}{\left
(\begin{array}{cccc}}
    \newcommand{\ematf}{\end{array} \right )} \newcommand{\vect}{\left (
\begin{array}{c}}
    \newcommand{\evect}{\end{array} \right )} 
     \def\beqn{\begin{eqnarray}}
 \def\eeqn{\end{eqnarray}}  

\def\IC{{\bf C}}

\def\IR{{\bf R}}
\def\IZ{{\mathbb{Z}}}

\def\IP{\bf P}

   \def\dim{\operatorname{dim}}

\newcommand{\dd}{{\rm d}}
\newcommand{\e}{{\rm e}}

\newtheorem{Theorem}{Theorem}[section]
\newtheorem{Lemma}{Lemma}[section]
\newtheorem{Corrolary}{Corrolary}[section]

\newcommand{\bt}{\begin{Theorem}}   \newcommand{\et}{\end{Theorem}}
\newcommand{\bl}{\begin{Lemma}}     \newcommand{\el}{\end{Lemma}}
\newcommand{\bc}{\begin{Corrolary}} \newcommand{\ec}{\end{Corrolary}}

\newcommand{\AC}[1]{{\color{red}{\textbf{AC: }#1}}}

\graphicspath{{./Introduction/}{./1-Axion/}{./2-Instanton/}{./3-BFT/}{./4-AdS/}}

\usepackage{dsfont}

\renewcommand{\thesection}{\arabic{chapter}.\arabic{section}{}}

\title{Titulo}
\author{Alberto Castellano Mora}
\makeindex

\begin{document}

\pagenumbering{roman} 
\pagestyle{headings}
\pagestyle{empty}

\newcommand{\HRule}{\rule{\linewidth}{1mm}}
\setlength{\parindent}{1cm}
\setlength{\parskip}{1mm}
\noindent




\pagestyle{fancy}
\fancyhead{}
\fancyfoot{}
\fancyhead[LE] {\itshape\nouppercase\leftmark}
\fancyhead[RO] {\itshape\nouppercase\rightmark}
\fancyfoot[C]{\thepage}
\renewcommand{\headrulewidth}{0.3pt}
\thispagestyle{empty}

\fancypagestyle{simple}{
  \fancyhead{}
  \fancyfoot[C]{\thepage}
 \renewcommand{\headrulewidth}{0.0pt} 
}

\noindent
\HRule
\begin{center}
\huge{\textbf{  The Quantum Gravity Scale\\
and the Swampland}}
 \vspace{0.2cm}
\end{center}
\HRule

\vspace{1.0cm}

\begin{center}

\large{	   
Memoria de Tesis Doctoral realizada por \\[3mm]
\textbf{\large{Alberto Castellano Mora}} \\[3mm]
presentada ante el Departamento de F\'isica Te\'orica \\[1mm]                  
de la Universidad Aut\'onoma de Madrid \\[1mm]
para optar al T\'itulo de Doctor en F\'isica Te\'orica \\[1mm]
}

\vspace{1.3cm}
Tesis Doctoral dirigida por \textbf{\large{Luis E. Ib\'a\~nez Santiago}}, \\[2mm]
Catedr\'atico del Departamento de F\'isica Te\'orica\\[1mm] de la Universidad Aut\'onoma de Madrid \\

\end{center}

\vspace{0.7cm}

\begin{center}
{\Large {Departamento de F\'isica Te\'orica\\[1mm] Universidad Aut\'onoma de Madrid }}\\
\vspace{0.5cm}
{\Large {Instituto de F\'isica Te\'orica UAM-CSIC}}\\
\end{center}
\vspace{1.2cm}

\begin{figure}[ht]
\centering
\begin{tabular}{cc}
\includegraphics[height=2.7cm]{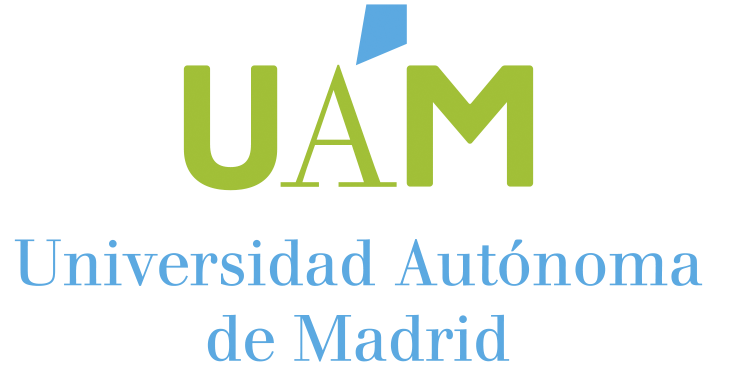}\qquad\qquad
\includegraphics[height=2.5cm]{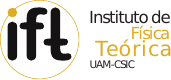}
%
%
%
\end{tabular}
\end{figure}

\vspace{0.5cm}

\begin{center}
{\large Mayo de 2024}
\end{center}

\newpage

\newpage
\thispagestyle{empty}
\phantom{lala}

\cleardoublepage




\thispagestyle{empty}
\clearpage
\begin{center}
     \thispagestyle{empty}
    \vspace*{30mm}
  \textit{A mis padres, Mari Cruz y Miguel Ángel; a mi hermano, Miguel; y a mi todo, Teresa.}
    \vspace*{\fill}   
\end{center}


\clearpage
\thispagestyle{empty}



\newpage
\vspace*{120pt}
\thispagestyle{empty}
\newpage
\thispagestyle{empty}

Esta tesis doctoral está basada en los siguientes artículos:
\begin{enumerate}
\item[{\hypersetup{hidelinks}\cite{Castellano:2022bvr}}]{\emph{The Emergence Proposal in Quantum Gravity and the Species Scale},\\
\textbf{A. Castellano}, A. Herr\'aez, L. Ib\'añez\\
\href{https://link.springer.com/article/10.1007/JHEP06(2023)047}{\emph{JHEP} \textbf{06} (2023) 047} 
[\href{https://arxiv.org/abs/2212.03908}{arXiv:2212.03908}]}

\item[{\hypersetup{hidelinks}\cite{Calderon-Infante:2023ler}}]{\emph{Entropy Bounds and the Species Scale Distance Conjecture}, \\
J. Calder\'on-Infante, \textbf{A. Castellano}, A. Herr\'aez, L. Ib\'añez\\
\href{https://link.springer.com/article/10.1007/JHEP01(2024)039}{\emph{JHEP} \textbf{01} (2024) 039 } 
[\href{https://arxiv.org/abs/2306.16450}{arXiv:2306.16450}]}

\item[{\hypersetup{hidelinks}\cite{Castellano:2023aum}}]{\emph{On the Species Scale, Modular Invariance and the Gravitational EFT expansion}, \\
\textbf{A. Castellano}, A. Herr\'aez, L. Ib\'añez\\ 
\href{https://arxiv.org/abs/2310.07708}{arXiv:2310.07708}}

\item[{\hypersetup{hidelinks}\cite{PhysRevLett.132.181601}}]{\emph{Universal Pattern in Quantum Gravity at Infinite Distance}, \\
\textbf{A. Castellano}, I. Ruiz, I. Valenzuela \\ 
\href{https://journals.aps.org/prl/abstract/10.1103/PhysRevLett.132.181601}{\emph{Phys. Rev. Lett.} \textbf{132} (2024) 181601 } 
[\href{https://arxiv.org/abs/2311.01501}{arXiv:2311.01501}]}

\item[{\hypersetup{hidelinks}\cite{Castellano:2023jjt}}]{\emph{Stringy Evidence for a Universal Pattern at Infinite Distance}, \\
\textbf{A. Castellano}, I. Ruiz, I. Valenzuela \\ 
\href{https://link.springer.com/article/10.1007/JHEP06(2024)037?utm_source=rct_congratemailt&utm_medium=email&utm_campaign=oa_20240608&utm_content=10.1007/JHEP06(2024)037}{\emph{JHEP} \textbf{06} (2024) 037 } 
[\href{https://arxiv.org/abs/2311.01536}{arXiv:2311.01536}]}

\end{enumerate}

El autor también contribuyó a los siguientes artículos no incluidos en la tesis:

\begin{enumerate}
\item[{\hypersetup{hidelinks}\cite{Castellano:2021yye}}]{\emph{A Gravitino Distance Conjecture},\\
\textbf{A. Castellano}, A. Font, A. Herr\'aez, L. Ib\'añez\\
\href{https://link.springer.com/article/10.1007/JHEP08(2021)092}{\emph{JHEP} \textbf{08} (2021) 092} 
[\href{https://arxiv.org/abs/2104.10181}{arXiv:2104.10181}]}

\item[{\hypersetup{hidelinks}\cite{Castellano:2021mmx}}]{\emph{IR/UV Mixing, Towers of Species and Swampland Conjectures},\\
\textbf{A. Castellano}, A. Herr\'aez, L. Ib\'añez\\
\href{https://link.springer.com/article/10.1007/JHEP08(2022)217}{\emph{JHEP} \textbf{08} (2022) 217} 
[\href{https://arxiv.org/abs/2112.10796}{arXiv:2112.10796}]}

\item[{\hypersetup{hidelinks}\cite{Castellano:2023qhp}}]{\emph{Towers and Hierarchies in the Standard Model from Emergence in Quantum Gravity},\\
\textbf{A. Castellano}, A. Herr\'aez, L. Ib\'añez\\
\href{https://link.springer.com/article/10.1007/JHEP10(2023)172}{\emph{JHEP} \textbf{10} (2023) 172} 
[\href{https://arxiv.org/abs/2302.00017}{arXiv:2302.00017}]}

\end{enumerate}

\selectlanguage{spanish}
\thispagestyle{empty}
\chapter*{Agradecimientos}

Esta tesis no podría comenzar de otra manera que no fuera dando las gracias. En primer lugar, gracias Luis por haberme acogido desde el minuto uno, ya fuera para realizar una beca de colaboración, visitar el ift durante los veranos de grado, o para hacer y descubrir nueva física juntos. De ti he aprendido qué preguntas hacer, cómo abordar los problemas, superando cualquier obstáculo que pudiera surgir por el camino, y sobre todo disfrutar haciendo investigación. Gracias por haberme hecho sentir partícipe de todos y cada uno de los proyectos, ideas y discusiones, por animarme a desarrollar mis propias inquietudes, y por compartir conmigo esa intuición y emoción mágicas que posees y desprendes. Me siento realmente un afortunado de haber podido trabajar y aprender de ti durante los últimos 4 años, y espero poder seguir haciéndolo en el futuro.

Asimismo, quisiera dar mi más sincero agradecimiento a quien considero ha sido mi otro mentor desde que un día cualquiera decidiera acercarme al final de una clase de problemas a preguntar y pedir alguna referencia. Gracias Álvaro por haber dedicado una infinidad de tiempo a explicarme los entresijos de las cuerdas, las compactificaciones y el Swampland. Por haberme invitado a visitarte en París, por tener siempre dos minutos para discutir cualquier cosilla, por tu enorme honestidad y también amistad. En resumen, por ser sin duda uno de los mejores físicos y amigos que he tenido el placer de descubrir. Poder conocer y colaborar con gente como tú hace que la experiencia (a veces dura) en la Academia merezca mucho más la pena. También te estoy agradecido por tus comentarios sobre la versión preliminar de la tesis, que no han hecho sino mejorarla.

Por supuesto, una mención especial al resto de geniales colaboradores de los que he podido disfrutar durante estos años: Anamaría, Fernando, Irene, José, Lorenzo, Luca y Nacho. En particular, quisiera agradecer a Irene por ofrecer apoyo continuo y por haberme dado la inmensa oportunidad de visitar y trabajar en el CERN durante tres fantásticos meses; ha sido realmente un sueño cumplido. También a José, por compartir muchas de sus ideas conmigo, por tomarse en serio las mías, y por sus agudas críticas y comentarios sobre la tesis. De todos vosotros me llevo bonitos recuerdos y muchísimo aprendizaje en cada proyecto. ¡Gracias!

Asimismo, quisiera darle las gracias a todos los miembros del grupo de cuerdas del ift. A Fernando por organizar los mejores workshops, por ayudarnos a navegar la bucrocracia y por toda la física que nos enseñas. A Ángel, por compartir toda tu sabiduría durante los seminarios, por tu curso de teoría de cuerdas y por liderar el mejor canal de física de YouTube que pueda existir. Y gracias por supuesto a Miguel, por animarnos a ser más críticos, a participar y hacer preguntas, por instaurar la cosmo-pizza, por tu buen rollo y por haberme hecho sentir siempre arropado, sobre todo en los momentos más difíciles.

Igualmente agradecido estoy por los grandes amigos y compañeros que hacen que el día a día en el instituto sea enriqucedor y mucho más divertido. En particular, quisiera agradecerle a Max por su amistad y por haber sido mi otro gran apoyo durante el primer año de doctorado, cuando el Covid y la falta de presencialidad hacían aún más difícil la entrada al mundo de la investigación. Gracias a Matilda, por ser mi `partner in crime' en este negocio, por su alegría contagiosa todos los días del año y por enseñarme tantas cosas, dentro y fuera de la física. Y por supuesto a Luca, a quien admiro profundamente como científico pero sobre todo como amigo. (Pd: aún tenemos unos cuantos memes pendientes.) Mención especial también para David, con quien compartí apartamento y aventuras durante los tres meses de estancia en Ginebra. También quisiera acordarme aquí de mis compis del 413: Luca, Manu y Pablo. Echaré muchísimo de menos los ``bueeenoooos díiiiiaas'' de Pablo, las comidas a horas intempestivas de Manu y los ``koffings'' con Luca cada tarde. No quisiera olvidarme por supuesto del resto de la horda del ift con los que he compartido muchos momentos inolvidables durante estos cuatro años: Alexa, Alexander, Andriana, Bernardo, Bruno, Camilo, Christian, David Alonso, David Pereñíguez, Edu, Fer, Florent, Gonzalisto, Gonchi, Jesús, Joan, Jonathan, Matteo, Michelangelo, Mikel, Naredo, Pau, Roberta, Sergio, Víctor y Thibaut. Gracias también a Lars Aalsma, por toda la ayuda, esfuerzo y dedicación prestados durante el tercer y cuarto años de doctorado.  

Quisiera darme el lujo en este momento de mencionar y agradecer como se merece a Pablo García Abia, ya que de tu mano tuve mi primer contacto directo con la investigación. Fue un enorme placer poder aprender de ti, de observar en vivo y en directo cómo se resuelven los problemas reales en física, tanto teóricos como experimentales. Y sobre todo, gracias por haber mantenido el contacto conmigo todos estos años, por poder ser ahora no sólo colegas de profesión sino también amigos. Has sido y sigues siendo mi modelo a seguir en este mundillo. ¡Gracias por tanto!

Fuera del ámbito de la física, son muchas las personas cuyo cariño y apoyo han resultado imprescindibles para poder estar hoy aquí escribiendo estos agradecimientos. En primer lugar, gracias a mis colegas de toda la vida: Arturo, Carlota, Dani, Mayte y Piña. Estoy tremendamente orgulloso de todos y cada uno de vosotros, de haber crecido a vuestro lado, de disfrutar y celebrarnos cada finde con una cervecita, de sentir vuestro aliento y apoyo en cada paso que doy; en resumen, de poder considerarme vuestro amigo. Gracias también a los grandes amigos que la carrera me trajo: Ángela, David, Diego, Pablo y Sami. Sois todos una fuente de inspiración para mi, y aunque no nos veamos tanto como quisiera, os guardo siempre en un rinconcito importante de mi cora. Quisiera dedicar una mención especial a la Residencia de Estudiantes por el fantástico año en que pude disfrutar de vivir en un edificio histórico rodeado de gente brillante y a la que admiro sobremanera. En especial a los becarios: Adri, Alba (historia), Alba (mates), Alberto (mi tocayo), Alicia (bio), Alicia (bailarina), Ana, Guille, Indira, Juan, Laura, Lucía, Mónica y Rocío. ¡Sois un equipazo!

Por supuesto, un lugar especial dentro de estos agradecimientos queda reservado a mi familia. Gracias a mis padres, por quererme y educarme desde siempre, por animarme a perseguir mis sueños, ya fuera en España o en el extranjero. Gran parte de la persona que soy hoy refleja todo el trabajo, cariño y esfuerzo que emanan de vosotros. En especial quisiera agradecerle a mi madre por haber compartido conmigo las alegrías y tristezas de este trabajo, por haberme acompañado desde que era niño a charlas de divulgación, e incluso haber conseguido ponerme en contacto con investigadores de la talla de Pablo. Gracias a mi hermano, por ser como es, cariñoso, inteligente, trabajador y un cachondo. Estoy muy orgulloso de ti y de todos tus logros. Y gracias al resto de mi familia, a mis tíos, abuelos, primos (y primitas :), por siempre estar ahí e incluso interesarse por mi trabajo, aunque suene a lengua extraterrestre. A todos vosotros os quiero y os adoro muchísimo.

Finalmente, y precisamente por ello aún más importante si cabe, gracias por absolutamente todo Teresa. Has sido mi gran descubrimiento, mi gran apoyo durante estos cuatro años, tanto en los momentos de celebración como en aquellos en que el mundo se venía abajo. Eres la alegría de mis días y la ilusión de mi vida. Gracias por ser mi mejor amiga, mi compi de aventuras, mi pinche de cocina y mi otaku favorita. Gracias por animarme y empujarme en cada decisión, por acompañarme a vivir la aventura americana, por quererme y hacerme crecer. Te quiero como los patos, Teresa, ¡pato' la vida!

\thispagestyle{empty}
\chapter*{Abstract}
\thispagestyle{empty}

In this thesis we investigate the role of the quantum gravity cut-off in effective descriptions of gravity at low energies, also in connection with the Swampland program. The focus is placed on understanding in a model-independent way what is the maximum regime of validity of generic effective field theories weakly-coupled to Einstein gravity, as well as characterizing any possible universal behaviour exhibited by the aforementioned cut-off. 

After reviewing some background material on string theory and the Swampland program, we then discuss in great generality the energy scale that supposedly captures the point where quantum-gravitational effects cannot be neglected. Based on various kinds of arguments, which are perturbative and non-perturbative in nature, we arrive at the species scale as the most natural candidate for the latter. This should be understood as the energy cut-off controlling generically the EFT expansion in gravity, therefore signalling the maximum energies/curvatures that can be reliably accommodated by the semi-classical effective description.

Later on, we proceed to check systematically the above picture in consistent theories arising from string compactifications, finding a non-trivial agreement with the former. In addition, we study various formal applications of the concept of the species scale in quantum gravity, including the conjectured phenomenon of Emergence, which posits that all kinematics in the low energy field theory arise from integrating out the massive dual degrees of freedom up to the quantum gravity scale. Indeed, we find that a naive field-theoretic analysis requires from the identification of the quantum gravity cut-off with the species scale, so as to be able to recover the singular behaviour exhibited by the different kinetic functions in the theory, when approaching various kinds of infinite distance limits in field space.

Finally, we perform a thorough analysis and characterization of the species cut-off close to infinite distance boundaries in moduli space, where certain universal properties emerge. In particular, we are able to motivate and provide non-trivial evidence for a lower bound on the exponential decay rate of the species scale, which forces the quantum gravity cut-off to fall off at infinity at least exponentially with the canonical distance defined therein. Relatedly, we are able to uncover some intriguing pattern relating the variation of the species cut-off and the characteristic mass of the lightest tower in the theory for any infinite distance limit. This is moreover satisfied in all up to now explored string theory constructions with at least eight supercharges, even though a purely bottom-up argument for the latter is still missing.

\newpage
\thispagestyle{empty}

\chapter*{Resumen}
\selectlanguage{spanish}\thispagestyle{empty}

En esta tesis se investiga el papel que la escala de gravedad cuántica juega en las descripciones efectivas de gravedad a bajas energías, así como su conexión con el programa de la Ciénaga. El enfoque principal se centra en comprender de manera independiente de cualquier modelo particular de gravedad cuántica, cuál es el régimen máximo de validez de las teorías de campo efectivo acopladas débilmente a la gravedad de Einstein, así como en caracterizar cualquier posible comportamiento universal que esta pueda exhibir.

Una vez sentadas las bases y explicado el material introductorio necesario para la comprensión de esta tesis, procedemos a discutir de forma general la escala de energías precisa que capturaría el punto donde los efectos cuántico-gravitacionales no pueden ser ignorados. Así, basándonos en varios argumentos teóricos (perturbativos y no perturbativos), concluimos que dicha escala se corresponde con la conocida como escala de especies. De esta manera, sería la escala de especies la que controlaría de manera genérica la expansión de toda teoría efectiva de campos incluyendo la gravedad, señalando así las energías/curvaturas máximas que pueden ser descritas por la misma.

A continuación, tratamos de verificar sistemáticamente las ideas presentadas anteriormente usando teorías consistentes que surgen de compactificaciones de la teoría de cuerdas como laboratorio, confirmando así nuestras expectativas. Asimismo, estudiamos varias aplicaciones formales del concepto de la escala de especies dentro de gravedad cuántica, incluyendo el conjeturado fenómeno de Emergencia, que postula que toda la cinemática en la teoría de campos efectiva (incluyendo la propia interacción gravitacional) surgiría de integrar los grados de libertad duales masivos hasta la escala de gravedad cuántica. De hecho, encontramos que un análisis a primer orden usando el formalismo de teoría de campos e identificando dicha escala con la propia de especies, nos permite recuperar de forma no trivial el comportamiento singular exhibido por las diferentes funciones cinéticas que la teoría presenta cuando probamos límites a distancia infinita en el espacio de módulos.

Finalmente, realizamos un análisis exhaustivo así como una caracterización de la escala de especies cerca de los límites a distancia infinita en el espacio de módulos, donde pudieran emerger ciertas propiedades universales. En particular, somos capaces de motivar y proporcionar evidencia significativa acerca de la existencia de un límite inferior en la tasa de decaimiento exponencial de la escala de especies, que obligaría a la misma a decrecer al menos exponencialmente a lo largo de dichos límites. Asimismo, se discute un patrón interesante que relaciona la variación de la escala de especies y la masa característica de la torre más ligera en la teoría, para cualquier límite de distancia infinita. Esto, además, parece cumplirse en todas las construcciones consistentes de teoría de cuerdas exploradas hasta la fecha con al menos ocho supercargas. No obstante, no somos capaces de proporcionar un argumento puramente desde la perspectiva infrarroja. 

\newpage

\selectlanguage{british}

\newpage

\selectlanguage{british}
\thispagestyle{empty}
{\hypersetup{hidelinks}
\tableofcontents}
\clearpage
\thispagestyle{empty}
\cleardoublepage
\pagenumbering{arabic}

\part{\scshape \textcolor{MyDarkRed}{Preliminaries}} 
\label{partI}


\chapter{Introduction}	\thispagestyle{simple}

The quest for a unified theory of fundamental interactions remains one of the most ambitious endeavors in theoretical physics. At the heart of this pursuit lies the challenge of finding a fully-fledged theory of quantum gravitational interactions, which of course reduces to General Relativity \cite{EinsteinGR} for low enough energies --- equivalently curvatures, but also crucially resolves the pathologies found therein, such as the endemic presence of spacetime singularities (e.g., black holes). Over the last fifty years or so, remarkable progress has been made in understanding the quantum nature of spacetime, thanks in part to the (still ongoing) development of string theory, here understood as the richer set of ideas which incorporate the quantum physics of extended objects (e.g, strings, D-branes, etc.), as well as make manifest certain non-trivial features of quantum gravity such as the concept of holography \cite{Maldacena:1997re,Witten:1998qj}. However, despite the enormous successes of string theory in uncovering new phenomena both in gravity as well as in (supersymmetric) field theory, the challenge remains to reproduce the physics observed in particle accelerators, where gravity plays no important role. This has led to various puzzles over the years, since it seems that one can get a priori as many as $10^{272000}$ inequivalent 4d vacua directly from string theory constructions \cite{Taylor:2015xtz}, thus suggesting a poor predictive power of the theory. In this regard, a key role has been played by the realization that in fact, the amount of consistent vacua in string theory (more generally in quantum gravity) comprises a set of measure zero within the complete set of possibilities that are allowed by field theory arguments such as gauge anomaly cancellation, etc. This has led to the idea of the Swampland program \cite{Vafa:2005ui}, which aims to find what are the quantum gravity consistency conditions that arise purely in gravitational theories and must be satisfied by any effective field theory weakly coupled to Einstein gravity. The present thesis aims to explore and extend these recent developments, focusing on a particular quantity that seems to play an starring role within this story: The quantum gravity cut-off.

\subsubsection*{Current status of high energy physics}

High energy physics has undergone significant advancements over the last decades, fueled by both experimental discoveries and theoretical developments. The discovery of the Higgs boson \cite{ATLAS:2012yve, CMS:2012qbp} at the Large Hadron Collider (LHC) in 2012 provided the final piece of the Standard Model, confirming our understanding of particle physics up to energies of approximately 14 TeV. Despite this triumph, the Standard Model leaves several profound questions unanswered, such as the nature of dark matter/energy \cite{Planck:2015fie}, the origin of neutrino masses \cite{Super-Kamiokande:1998uiq,SNO:2002tuh,KamLAND:2002uet}, and the hierarchy/cosmological constant problems \cite{Weinberg:1988cp}.

In parallel, the detection of gravitational waves by the LIGO and Virgo collaborations \cite{LIGO} has opened a new observational window into the cosmos. These ripples in spacetime, predicted by the classical theory of gravity \cite{GW1,GW2}, have provided crucial insights and precision tests concerning the dynamics of strongly-curved spacetimes, such as black holes and neutron stars. The first direct observation of a binary black hole merger in 2015 and the subsequent detection of numerous gravitational wave events  provide us with new experimental tools that can potentially revolutionize our understanding of strong-field gravity and compact astrophysical objects. Furthermore, the observation of gravitational waves has not only confirmed general relativity in extreme curvature regimes but has also spurred a deeper investigation into the nature of black holes. The Event Horizon Telescope's image of the M87 black hole's shadow in 2019 \cite{EventHorizonTelescope:2019dse} provided a direct visual confirmation of black hole horizons. These observations challenge us to understand at a deeper level how the classical descriptions of these objects connect with the quantum theory, especially in relation with outstanding questions such as the origin of black hole entropy \cite{Bekenstein:1972tm,Hawking:1975vcx} or the information paradox \cite{hawking2015information}.

String theory, positing that fundamental particles are not point-like but rather one-dimensional strings, remains the leading candidate for a quantum theory of gravity. It naturally incorporates a spin-2 massless particle mediating long-range gravitational interactions, and moreover unifies it with other fundamental forces within a consistent quantum framework. Hence, it predicts a rich spectrum of particles and suggests the existence of extra dimensions, which could have profound implications for our understanding of the Universe at the smallest distance scales in case they are experimentally confirmed.


\subsubsection*{A golden era for quantum gravity}

Importantly, it seems that we are living now in a particularly auspicious time for studying the (hard) problem of quantum gravity. The convergence of theoretical advancements and experimental discoveries mentioned before has created a fertile ground for new insights. The ability to observe and measure gravitational waves and black hole phenomena provides empirical data that can inform and constrain theoretical models. On the other hand, the rapidly growing set of knowledge gathered from the string theory point of view offers a robust framework for understanding the quantum aspects of gravity.

This thesis aims to delve into the role of the maximum energy cut-off that effective field theory descriptions of gravity at low energies can have, with a particular emphasis on its connections to the Swampland program. The former must be understood as the energy scale beyond which local effective field theory breaks down due to the appearance of purely quantum gravitational effects, such as the presence of strings or extra dimensions. Therefore, by examining various recents developments in the theoretical front, we seek to understand the limits of effective field theories weakly coulpled to gravity and explore the universal properties that this quantum gravity cut-off can exhibit. Through this investigation, we hope that we can contribute to the ongoing efforts to uncover the fundamental nature of spacetime and gravity.

In summary, the current era in high energy physics is marked by a synergy between theory and experiment that is driving forward our understanding of the Universe at its most fundamental level. This thesis is positioned within this vibrant landscape, aiming to humbly contribute to some of the most pressing questions in the field.

\subsubsection*{Plan of the Thesis}

To be concrete, in this thesis we will consider different string theory compactifications preserving 32, 16 or 8 supercharges in dimensions ranging from ten to four. In particular, the rest of Part \ref{partI} includes a detailed review on the basic aspects and ingredients that the aforementioned string theory constructions present, placing special emphasis on the two-derivative dynamics as well as the massive (non-perturbative) content of the theories. We also draw some deep connections between the latter that are captured by the phenomenon of (string) dualities, which will be used at many instances in the thesis. To finish, we briefly introduce the Swampland program, focusing on the conjectures that will play a major role on the rest of this work, namely the Distance  \cite{Ooguri:2006in} and Weak Gravity conjectures \cite{Heidenreich:2015nta, Heidenreich:2016aqi, Montero:2016tif, Andriolo:2018lvp}. 

The bulk of the results reported in this thesis are contained in Parts \ref{part:QGscale}-\ref{part:pattern}. Hence, in Chapter \ref{ch:SpeciesIntro} we introduce and discuss in detail the concept of the quantum gravity cut-off. We first explain what are the basic expectations coming from the non-renormalizable character of General Relativity, so as to later confront this intuition with several ideas that are believed to play a fundamental role in quantum gravity, such as the holographic principle. This leads us instead to propose this quantity to be given by a seemingly different energy scale, usually denoted as the species cut-off \cite{Han:2004wt, Dvali:2007hz, Dvali:2007wp}. Interestingly, we review and extend several perturbative and non-perturbative arguments pointing toward the species scale as encapsulating the minimum length-scale describable by any effective field theory weakly coupled to gravity. Moreover, this turns out to be in agreement with our familiar intuition based on theories of extra dimensions and string theory itself, where the UV cut-off is given either by the higher-dimensional Planck mass or the fundamental string scale. Many of the discussions presented in this part of the thesis build on material already existing in the literature, whilst the new contributions are based on the publications \cite{Castellano:2022bvr,Castellano:2023aum}.

Part \ref{part:StringTheoryTests} is devoted to a careful study and application of the ideas introduced in Part \ref{part:QGscale} within the context of string theory. In particular, in Chapter \ref{ch:Higherdimops} and using a large set of the string compactifications described in Chapter \ref{ch:reviewstringtheory}, we test whether the species scale indeed arises as the ultra-violet cut-off in gravity. This is signalled by the appearance of the latter as the energy scale controlling the EFT expansion of higher-dimensional and higher-curvature operators in the low energy EFT. Crucially, we find perfect agreement with the expectations based on our discussion from Part \ref{part:QGscale}. In addition, we also investigate in Chapter \ref{ch:Emergence} the precise role of the species cut-off within certain conjectural criteria proposed in the Swampland program. More precisely, we analyze how the Emergence mechanism \cite{Harlow:2015lma, Grimm:2018ohb, Heidenreich:2018kpg, Palti:2019pca} is realized in string theory, which hinges on the precise identification of the gravitational cut-off in the EFT with the species scale. The material presented in this part of the thesis is based on earlier publications by the author \cite{Castellano:2022bvr,Castellano:2023aum}.

In Part \ref{part:pattern} of the thesis we turn our attention to finding universal constraints and patterns concerning the species cut-off close to infinite distance boundaries in field space. In particular, in Chapter \ref{ch:bounds} we introduce and discuss certain lower bound on the exponential decay rate that the species scale seems to satisfy along any infinite distance trajectory in moduli space. This non-trivial constraint can be reformulated as a convex hull condition and indeed exhibits lots of geometric structure which is deeply rooted in the duality properties of the theories under consideration. Later on, in Chapter \ref{ch:pattern} we follow up on these ideas and present another seemingly universal pattern that relates the aforementioned decay rates of the species and the (lightest) tower mass scales. This latter property is seen to be satisfied in all up to now explored supersymmetric vacua in string theory, being moreover intimately related (although not completely equivalent) to the Emergent String Conjecture \cite{Lee:2019wij}. The material of these chapters builds on earlier results by the author contained in references \cite{Calderon-Infante:2023ler, Castellano:2023jjt, PhysRevLett.132.181601}.

Finally, in Part \ref{part:Conclussions} we draw some general conclusions that follow naturally from the work presented here, whereas in Part \ref{part:appendices} several technical details which are relevant for the analysis in the bulk of the thesis are presented.

\selectlanguage{spanish}
\chapter*{Introducción}	\thispagestyle{simple}

La búsqueda de una teoría unificada de las interacciones fundamentales sigue siendo uno de los esfuerzos más ambiciosos dentro del marco de la física teórica. En el corazón de esta búsqueda se hallaría el desafío de encontrar una teoría completa de las interacciones gravitacionales a nivel cuántico, que por supuesto se reduzca a la ya conocida Relatividad General \cite{EinsteinGR} para energías suficientemente bajas, pero que también resuelva crucialmente toda patología presente en la misma, como la inevitable presencia de singularidades en el espacio-tiempo (por ejemplo, agujeros negros). Durante los últimos cincuenta años, se ha logrado un progreso notable en la comprensión de la naturaleza cuántica del espacio-tiempo, gracias en parte al desarrollo (aún en curso) de la teoría de cuerdas, entendida globalmente como el conjunto de ideas que incorporan la física de objetos extendidos (por ejemplo cuerdas, D-branas, etc.); así como manifestar ciertas características no triviales de la gravedad cuántica, como el concepto de holografía \cite{Maldacena:1997re, Witten:1998qj}. Sin embargo, a pesar del enorme éxito de la teoría de cuerdas al descubrir nuevos fenómenos tanto en gravedad como en teoría de campos, todavía perdura el importante desafío de reproducir de forma teórica la física observada en los aceleradores de partículas, donde la gravedad no juega ningún papel importante. Esto habría conducido a varios enigmas importantes a lo largo de los años, ya que pareciera que uno puediera obtener a priori hasta $10^{272000}$ vacíos cuatridimensionales inequivalentes directamente de teoría de cuerdas \cite{Taylor:2015xtz}, lo que sugiere un poder predictivo casi nulo de la teoría. En este sentido, un papel crucial lo ha jugado la realización de que, de hecho, la cantidad de vacíos consistentes que teoría de cuerdas puede proporcionar comprende en realidad un conjunto de medida cero dentro de la completitud de posibilidades que serían permitidas por argumentos puramente de teoría de campos, como la cancelación de anomalías gauge, etc. Esto ha conducido a proponer la interesante idea del programa de la Ciénaga \cite{Vafa:2005ui}, que tendría como objetivo encontrar cuáles son las condiciones de consistencia que surgen de tener gravedad acoplada a nuestras teorías, y que por tanto deben ser satisfechas por cualquier teoría de campos efectiva débilmente acoplada a la misma. La presente tesis tiene como objetivo principal explorar y extender estos desarrollos, centrándose en una cantidad particular que parece jugar un papel protagonista dentro de esta historia: la escala de gravedad cuántica.

\subsubsection*{Estado actual de la física de altas energías}

La física de altas energías ha experimentado avances significativos en las últimas décadas, impulsados a la par por descubrimientos experimentales así como desarrollos teóricos. El descubrimiento del bosón de Higgs \cite{ATLAS:2012yve, CMS:2012qbp} en el Gran Colisionador de Hadrones en 2012 proporcionó la pieza final que confirmaba el Modelo Estándar, completando así nuestra comprensión de la física de partículas hasta energías de aproximadamente 14 TeV. A pesar de este triunfo, el Modelo Estándar dejaría varias preguntas profundas sin responder, como la naturaleza de la materia/energía oscura \cite{Planck:2015fie}, el origen de las masa de los neutrinos \cite{Super-Kamiokande:1998uiq, SNO:2002tuh, KamLAND:2002uet} o el problema de la jerarquía/constante cosmológica \cite{Weinberg:1988cp}, entre otros.

En paralelo, la detección de ondas gravitacionales por las colaboraciones LIGO y Virgo \cite{LIGO} habría abierto una nueva ventana de observación hacia el cosmos. Estas ondulaciones en el espacio-tiempo, predichas por la teoría clásica de la gravedad de Einstein \cite{GW1, GW2}, han proporcionado asimismo conocimientos cruciales y pruebas de alta precisión sobre la dinámica de espacio-tiempos fuertemente curvados, por ejemplo en presencia de agujeros negros o estrellas de neutrones. De hecho, la observación directa de fusiones de agujeros negros y la detección subsiguiente de numerosos eventos de ondas gravitacionales nos proporcionan nuevas herramientas experimentales que pueden potencialmente revolucionar nuestra comprensión de la gravedad en presencia campos fuertes. Además, la observación de ondas gravitacionales no solo ha confirmado la relatividad general en regímenes de curvatura extrema, sino que también ha impulsado una investigación más profunda sobre la naturaleza de los agujeros negros. La imagen del horizonte del agujero negro situado en el centro de la galaxia M87, que fue tomada por el Telescopio de Horizonte de Sucesos en 2019 \cite{EventHorizonTelescope:2019dse}, proporcionó una confirmación visual directa de la existencia de los horizontes predichos por la teoría. Estas observaciones nos desafían a comprender a un nivel más profundo las descripciones clásicas de estos objetos en relación con la teoría cuántica, especialmente teniendo en cuenta ciertas preguntas abiertas como el origen de la entropía de los agujeros negros \cite{Bekenstein:1972tm, Hawking:1975vcx} o la paradoja de la información \cite{hawking2015information}.

La teoría de cuerdas, la cual postula que las partículas fundamentales no serían objetos puntuales sino filamentos vibrantes con estructura unidimensional, sigue siendo además la principal candidata para proporcionar una teoría unificada de la gravedad con el resto de interacciones. Esta teoría incorpora por tanto de forma natural una partícula sin masa y de espín 2 que mediaría las interacciones gravitatorias de largo alcance, proporcionando un entendimiento cuántico de la dinámica gravitacional. Además, la teoría de cuerdas predice un espectro rico de partículas y sugiere la existencia de dimensiones extra, lo que podría tener, en caso de confirmarse de forma experimental, consecuencias de enorme impacto para nuestra comprensión del universo.

\subsubsection*{Una era dorada para la gravedad cuántica}

En efecto, vivimos ahora en un momento particularmente excitante para abordar el difícil problema de la gravedad cuántica. La convergencia tanto de avances teóricos como de descubrimientos experimentales habría creado un terreno fértil para el desarrollo de nuevo conocimiento. La capacidad de observar y medir ondas gravitacionales así como fenómenos de agujeros negros proporciona datos empíricos directos que pueden informar y restringir los modelos teóricos que podamos construir. Por otro lado, el conocimiento teórico rápidamente creciente proporcionado por la teoría de cuerdas ofrece un marco robusto para entender los aspectos cuánticos más sutiles de la gravedad.

Esta tesis tiene como objetivo profundizar en el papel de la escala de gravedad cuántica dentro de las descripciones de teorías de campos efectivas que incorporan gravedad a bajas energías, con énfasis en sus conexiones con el programa de la Ciénaga. Esta escala debe entenderse además como la escala de energía más allá de la cual toda teoría de campos efectiva local quedaría invalidada debido a la aparición de efectos puramente de gravedad cuántica, como la presencia de cuerdas o dimensiones adicionales. Por lo tanto, al examinar varios desarrollos recientes en el frente teórico, buscamos comprender los límites de las teorías de campos efectivas débilmente acopladas a la gravedad y explorar las propiedades universales que dicha escala pueda exhibir. A través de esta investigación, esperamos contribuir a los esfuerzos aún en curso para descubrir la naturaleza fundamental del espacio-tiempo y la gravedad.

En resumen, la era actual en que se ve inmersa la física de altas energías está marcada por una sinergia entre teoría y experimento, lo que impulsaría de forma considerable nuestra comprensión del universo a su nivel más fundamental. Esta tesis se posiciona dentro de este vibrante paisaje, con el objetivo de contribuir humildemente a algunas de las preguntas más apremiantes en el campo.

\subsubsection*{Organización de la tesis}

Concretamente, en esta tesis consideraremos diferentes compactificaciones de la teoría de cuerdas que preservan 32, 16 u 8 supercargas en dimensiones que van desde diez hasta cuatro. En particular, el resto de la Parte \ref{partI} incluye una revisión detallada sobre los aspectos básicos y los ingredientes que presentan las construcciones de teoría de cuerdas antes mencionadas, poniendo un énfasis especial en la dinámica así como en el contenido masivo (no perturbativo) de estas teorías. También establecemos algunas conexiones entre estas construcciones que son capturadas por el fascinante fenómeno de dualidad, las cuales se utilizarán en varias ocasiones a lo largo de la tesis. Para finalizar, introducimos brevemente el programa de la Ciénaga, centrándonos en las conjeturas que juegan un papel principal en el resto de este trabajo, a saber, las conjeturas de la Distancia \cite{Ooguri:2006in} y de la Gravedad Débil \cite{Heidenreich:2015nta, Heidenreich:2016aqi, Montero:2016tif, Andriolo:2018lvp}.

La mayor parte de los resultados reportados en esta tesis se encuentran en las Partes \ref{part:QGscale}-\ref{part:pattern}. Así, en el Capítulo \ref{ch:SpeciesIntro} introducimos y discutimos en detalle el concepto de escala de gravedad cuántica. Primero explicamos cuáles son las expectativas básicas provenientes del carácter no renormalizable de la Relatividad General, para luego confrontar esta intuición con varias ideas que se cree juegan un papel fundamental en gravedad cuántica, como el principio holográfico. Esto nos lleva a proponer que esta cantidad esté dada por una escala de energía aparentemente diferente, usualmente denominada como escala de especies \cite{Han:2004wt, Dvali:2007hz, Dvali:2007wp}. En consecuencia, revisamos y extendemos varios argumentos perturbativos y no perturbativos que apuntan a la escala de especies como aquella que encapsularía la longitud mínima describible por cualquier teoría de campos efectiva débilmente acoplada a la gravedad. Además, esto resulta estar en concordancia con nuestra intuición física basada en teorías de dimensiones extra y en la propia teoría de cuerdas, donde el corte ultravioleta estaría dado por la masa de Planck de dimensiones superiores o por la escala de cuerdas fundamental. Muchas de las discusiones presentadas en esta parte se apoyan en material ya existente en la literatura, mientras que las nuevas contribuciones se basan en las publicaciones \cite{Castellano:2022bvr, Castellano:2023aum}.

La Parte \ref{part:StringTheoryTests} está dedicada a un estudio y aplicación cuidadosos de las ideas introducidas en la Parte \ref{part:QGscale} dentro del contexto de la teoría de cuerdas. En particular, utilizando un gran conjunto de compactificaciones de teoría de cuerdas descritas en el Capítulo \ref{ch:reviewstringtheory}, probamos en el Capítulo \ref{ch:Higherdimops} si la escala de especies realmente surge como el corte ultravioleta en gravedad. Esto quedaría patente con la aparición de esta última como la escala de energía que controla la expansión efectiva de operadores de dimensiones superiores dentro de la descripcióne efectiva a bajas energías. Asimismo, encontramos un acuerdo perfecto con las expectativas basadas en nuestra discusión de la Parte \ref{part:QGscale}. Además, investigamos en el Capítulo \ref{ch:Emergence} el papel preciso de la escala de especies dentro de ciertos criterios conjeturales propuestos en el programa de la Ciénaga. Más concretamente, analizamos cómo funcionaría el mecanismo de Emergencia \cite{Harlow:2015lma, Grimm:2018ohb, Heidenreich:2018kpg, Palti:2019pca} en construcciones de teoría de cuerdas (al menos en su versión más débil), que depende de la identificación precisa del corte ultravioleta en la teoría efectiva con la escala de especies. El material presentado en esta parte de la tesis se basa en publicaciones anteriores del autor \cite{Castellano:2022bvr, Castellano:2023aum}.

En la Parte \ref{part:pattern} de la tesis, dirigimos nuestra atención a encontrar restricciones y patrones universales relacionados con la escala de especies cerca de las fronteras a distancia infinita en el espacio de campos. En particular, en el Capítulo \ref{ch:bounds} introducimos y discutimos un límite inferior en la tasa de decaimiento exponencial que parece satisfacer la escala de especies a lo largo de cualquier trayectoria de distancia infinita en el espacio de módulos. Esta restricción no trivial puede reformularse como una condición de envoltura convexa y, de hecho, exhibe mucha estructura geométrica que estaría profundamente enraizada en las propiedades de dualidad de las teorías bajo consideración. Posteriormente, en el Capítulo \ref{ch:pattern} proseguimos con estas ideas y presentamos otro patrón aparentemente universal que relaciona las tasas de decaimiento antes mencionadas de la escala de especies y la propia asociada a la torre más ligera en la teoría. Esta última propiedad se ha observado en todos los vacíos supersimétricos explorados hasta la fecha en teoría de cuerdas, estando además íntimamente relacionada con la Conjetura de la Cuerda Emergente \cite{Lee:2019wij}. El material de estos dos capítulos se basa en resultados anteriores del autor contenidos en las referencias \cite{Calderon-Infante:2023ler, Castellano:2023jjt, PhysRevLett.132.181601}.

Finalmente, en la Parte \ref{part:Conclussions} extraemos algunas conclusiones generales que surgen naturalmente del trabajo aquí presentado, mientras que en la Parte \ref{part:appendices} se presentan varios detalles técnicos que serían relevantes para el análisis en la mayor parte de la tesis.

\selectlanguage{british}



\chapter{From String Theory to the Swampland Program} \label{ch:reviewstringtheory} In the present chapter we provide the essential background material for both understanding the context of this thesis as well as the main contributions discussed in the upcoming parts.

To begin with, in Section \ref{s:basicconcepts} we briefly introduce string theory and string compactifications, focusing mostly on conceptual issues. Later on, in Sections \ref{s:maxsugraintro} and \ref{s:CYcompact} we describe the main compactification backgrounds that will be heavily employed to illustrate and study the physics presented in this thesis. In particular, we provide detailed discussions of the low energy effective actions describing the dynamics of maximally supersymmetric theories in eleven, ten, nine and eight spacetime dimensions; as well as those preserving 8 unbroken supercharges obtained from e.g., compactifying Type II string theory on Calabi--Yau manifolds. We also briefly comment in Section \ref{s:dualities} on the rich interconnections between the aforementioned theories, which arise in the form of (non-)perturbative string dualities. For a more in-depth treatment of the material presented here we refer the interested reader to the references \cite{Green:2012oqa,Green:2012pqa,Aspinwall:1996mn,Polchinski:1998rq,Polchinski:1998rr,Hori:2003ic,Becker:2006dvp,Ibanez:2012zz,Blumenhagen:2013fgp}.

Finally, to end the chapter, we introduce in Section \ref{s:SwamplandProgram} the Swampland program as well as the most important Swampland criteria that will play a starring role in this work. More concretely, we discuss in detail the Distance (Section \ref{s:SDC}) and the Weak Gravity (Section \ref{s:WGC}) conjectures. Let us mention that by now there already exists a good amount of reviews of the Swampland approach to quantum gravity \cite{Brennan:2017rbf,Palti:2019pca,vanBeest:2021lhn,Grana:2021zvf,Harlow:2022gzl,Agmon:2022thq,VanRiet:2023pnx}, which we recommend for further details on the ideas surrounding this interesting program.

\section{Basics of string theory} \label{s:basicconcepts}

The foundational premise of string theory is the conceptual replacement of point-like particles, which usually comprise the irreducible constituents in more familiar field-theoretic approaches to fundamental physics, with an entity that is additionally extended along one spatial dimension: a \emph{vibrating string}. This shift has lead to profound consequences in our understanding of the very high energy realm of Nature, uncovering fascinating phenomena that has impacted even quantum field theory itself. In this section we provide a brief introduction to the subject, highlighting the key concepts that will pay a major role in this thesis.

Crucially, unlike particles that are topologically trivial --- i.e. points without any further structure, one-dimensional objects can exhibit two inequivalent topologies: the loop (circle) and the interval (segment). Consequently, this allows for the existence of two different types of strings that can be considered within the theory: closed and open strings, respectively.

In order to describe the dynamics of relativistic strings on a $d$-dimensional spacetime $\mathcal{M}$, one usually starts from the extension of the worldline action appropriate for point-like objects, therefore adapting it to accommodate the extended nature of strings. Such a natural generalization associates an action functional to the 2d surface $\Sigma$ (i.e. the worldsheet) swept out by the string as it propagates through $\mathcal{M}$,\footnote{Note that the type of string considered, namely whether it is open or closed, is encapsulated in the topology of the worldhseet $\Sigma$ via the presence or absence of a boundary $\partial \Sigma$.} which can be locally parametrized by a set of embedding functions $X^\mu : \Sigma \to \mathcal{M}$, with $\mu=0,\dots, d-1$ and where $\sigma^a=(\tau, \sigma)$ are local coordinates on the worldsheet. It is known as the Nambu-Goto action \cite{Nambu:1986ze, Goto:1971ce}, and reads as follows
\begin{align}\label{eq:NambuGoto}
 S_{\text{NG}} = -\frac{1}{2\pi \alpha'} \int_\Sigma \dd^2 \sigma \sqrt{-h}\, , 
\end{align}
where $h$ is the (determinant of the) metric on the worldsheet induced by the one associated to the target space --- namely $g_{\mu \nu} (X)$, and $T_s$ denotes the tension of the string. The latter can be related to the more familiar string length $\ell_s=2\pi \sqrt{\alpha'}$ by $T_s=2\pi/\ell_s^2$, and it sets the energy scale beyond which non-local effects associated to the extended nature of the string must become apparent.

The action \eqref{eq:NambuGoto} captures in the most elementary way the physics of a relativistic string. Nevertheless, the explicit appearance of a square root within the 2d lagrangian formulation complicates the quantization process considerably. To address this challenge, one can benefit from the classical conformal invariance of the worldsheet theory and rewrite the Nambu-Goto action in its Polyakov version \cite{Polyakov:1981rd}, which takes the form
\begin{align}\label{eq:Polyakov}
 S_{\text{Poly}} = -\frac{T_s}{2} \int_\Sigma \dd^2 \sigma \sqrt{-h} h^{ab} \partial_a X^{\mu} \partial_b X^{\nu} g_{\mu \nu}\, , 
\end{align}
where the worldhseet metric $h_{ab}$ is now regarded as an independent field. Interestingly, eq. \eqref{eq:Polyakov} defines the perturbative string in terms of a local 2d quantum field theory living on the worldsheet, which can in principle incorporate further ingredients beyond the bosonic fields $\{X^{\mu}, h_{ab} \}$, such as fermionic superpartners $\{\psi^{\mu}, \chi_a\}$. In that case, the Polyakov action is found to be
\begin{equation}\label{eq:Polyakovsuperstring}
	\begin{split}
		S = & - \frac{T_s}{2} \int_\Sigma \dd^2 \sigma \sqrt{-h} \bigg(h^{ab} \partial_a X^\mu \partial_b X^\nu g_{\mu\nu} + \i \bar{\psi}^\mu \slashed{D} \psi_\mu \\ & \hspace{1cm}- 2 \i \bar{\chi}_a\rho^b\rho^a \psi^\mu \partial_b X_\mu  + \frac{1}{2}(\bar{\chi}_a\rho^b \rho^a \chi_b)(\bar{\psi}^\mu \psi_\mu)\bigg) \,,
	\end{split}
\end{equation}
where $\rho^a$ are two-dimensional gamma matrices satisfying the Clifford algebra (see Appendix \ref{ap:conventions} for conventions). The above action in turn defines a \emph{superstring} theory, which has many advantages over its bosonic counterpart, one of those being the incorporation of fermionic fields (from the spacetime point of view) in their quantized spectrum, which is arguably a crucial ingredient in Nature.

In fact, at the classical level, the equations of motion set the field $h_{ab}$ to be equal to the pulled-back metric from spacetime, hence recovering the Nambu-Goto action. Quantum-mechanically, however, the conformal symmetry (which is gauged in the 2d theory) can suffer from anomalies. Ensuring that this is not the case --- i.e. that the quantum string theory is consistent --- restricts the matter content as seen from the two-dimensional perspective in a non-trivial way. For instance, since the local fields in the 2d CFT are simply bosons and fermions, requiring the conformal anomaly to cancel in a flat Minkowski background $g_{\mu \nu} = \eta_{\mu \nu}$ imposes constraints on e.g., the dimension of the target spacetime, which is fixed to be $d=26$ for the bosonic string and $d=10$ for superstrings.

More importantly, when quantizing the string theory one discovers an infinite hierarchy of massive string oscillators, which may be interpreted as elementary particle excitations. In fact, separating between left-moving and right-moving modes (in the closed string case), one finds
\begin{subequations}\label{eq:modeexpansionclosedstring}
	\begin{align}
		X^\mu (\tau-\sigma)& = \frac12 x^{\mu} + \frac12 \alpha' p^{\mu} (\tau-\sigma) + \i\, \sqrt{\frac{\alpha'}{2}} \sum_{n\in \mathbb{Z}} \frac{1}{n} \alpha^\mu_n \, e^{-2\pi \i n (\tau-\sigma)} \,, \\[1mm]
		X^\mu (\tau+\sigma) & = \frac12 x^{\mu} + \frac12 \alpha' p^{\mu} (\tau+\sigma) + \i\, \sqrt{\frac{\alpha'}{2}} \sum_{n\in \mathbb{Z}} \frac{1}{n} \tilde{\alpha}^\mu_n \, e^{-2\pi \i n (\tau+\sigma)} \,, 
	\end{align}
\end{subequations}
for the bosonic fields, where $\{x^\mu, p^{\mu} \}$ are the position and momentum of the centre of mass; as well as 
\begin{subequations}\label{eq:fermionmodeexpansionstrings}
	\begin{align}
		\psi^\mu_L (\tau+\sigma) & = \sqrt{\alpha'} \sum_{k \in \mathbb{Z} + s} \tilde{b}^\mu_k\, e^{-2\pi \i n (\tau+\sigma)} \,, \\[2mm]
		\psi^\mu_R (\tau-\sigma) & = \sqrt{\alpha'} 	\sum_{k \in \mathbb{Z} + s} {b}^\mu_k\, e^{-2\pi \i n (\tau-\sigma)} \,,
	\end{align}
\end{subequations}	
for the fermionic superpartners, where $s = 0, \frac12$, accounts for Ramond (R) and Neveu--Schwarz (NS) boundary conditions, respectively.

Similarly to what happens in the worldline formalism for quantum fields, the worldsheet theory can also describe quantum interactions between string states by joining and splitting 2d surfaces, which manifest themselves via the non-trivial topology of $\Sigma$. Hence, when performing the path integral $\mathcal{Z}$ so as to derive any scattering amplitude in the theory, one is instructed to sum not only over worldsheet geometries (associated to the fluctuations of the metric $h_{ab}$) but also taking into account different topologies
\begin{align}\label{eq:pathintegralstring}
    \mathcal{Z}= \sum_{\rm w.s.\ topology} \int \mathcal{D}X \mathcal{D}h\, e^{ \i S_{\rm Poly}}\, .
\end{align}
Luckily, two-dimensional oriented real surfaces (i.e. Riemann surfaces) without boundary are completely classified at the topological level by the number of handles, namely the \emph{genus}. Hence, the formal sum over topologies in \eqref{eq:pathintegralstring} can be equivalently phrased --- in the closed string case --- as a sum over genera, and the expansion parameter controlling the series defines the string coupling constant $g_s = e^{-\phi}$, which is itself the vacuum expectation value for a dynamical field of the theory: The dilaton, see discussion around eq. \eqref{eq:RR4formfieldstrength} below.

Let us finally mention that the internal consistency of the worldsheet theory, even at the interacting level, significantly restricts the variety of viable superstring theories in ten dimensions. In fact, it has been shown that there exist precisely five distinct superstring theories in 10d, typically identified as Type I, Heterotic $\mathsf{SO(32)}$,\footnote{More correctly, such superstring theory is denoted as Heterotic $\mathsf{Spin}(32)/\mathbb{Z}_2$ \cite{Gross:1985fr}.} Heterotic $\mathsf{E8}\times \mathsf{E8}$, Type IIA, and Type IIB. These theories are distinguished by their constituents (i.e. the type of strings included and the 2d matter content) as well as the number of off-shell supersymmetries they possess: the first three have 16 unbroken supercharges (in a Minkowski background), while Type IIA and Type IIB are endowed with 32 supercharges. At low energies (compared to the string scale $m_s = \ell_s^{-1}$), the effective dynamics of the massless excitations in the aforementioned theories is described by the five existing supergravity theories in ten dimensions, see Section \ref{s:maxsugraintro} below.

\subsubsection*{String theory compactifications}

As already argued, all five known supersymmetric string theories live naturally in ten spacetime dimensions. This does not mean, however, that there cannot exist string constructions in lower dimensions, but it suggests instead that the backgrounds we must consider should be in fact more complicated than just flat Minkowski space. This observation opens up various interesting possibilities, ranging from non-geometric constructions, where one assumes that (part of) the worldsheet theory is comprised by some intricate conformal field theory which has no direct geometric interpretation as a target Riemannian space; to geometric compactifications, where some portion of the spacetime is assumed to be compact and possibly curved.

In this thesis we follow the second route, oftentimes scanning over different effective theories that can arise from string theory when expanded over backgrounds of this sort. In what follows, we present some brief introduction to the concept of \emph{string compactifications}, so as to set up both the notation and terminology that will be heavily used in later parts of this work (see \cite{Grana:2005jc,Douglas:2023yof,Douglas:2006es} for reviews). Let us mention in passing, though, that the idea of compactification and dimensional reduction does not pertain solely to string theory, and in fact it was originally envisaged using a purely field-theoretic approach. The latter set of ideas are usually referred to as Kaluza-Klein (KK) theories, see e.g., \cite{Bailin:1987jd,Duff:1986hr} for comprehensive reviews on the topic.

Let us illustrate this point using very simple and general considerations. Thus, our aim is to find some vacuum configuration of string theory which exhibits $d$ large non-compact \emph{external} dimensions, that we assume to be flat for simplicity. In order to obtain an effective theory expanded around this background --- which is assumed to solve the equations of motion of the theory, we propose our spacetime to have (at least locally) the following product form
\begin{align}\label{prodansatz}
    \mathbb{R}^{1,d-1} \times X_{10-d}\, ,
\end{align}
where $X_{10-d}$ denotes some compact Riemannian space. To describe the physics occurring at low energies, what one can do is expand the massless spectrum of the original 10d theory using a basis of eigenfunctions of the appropriate Laplace operator $\Delta_{\rm 10}$ acting on tensor fields defined over $X_{10-d}$
\begin{align}
 \Delta_{10} = \Delta_{X_{10-d}} + \Delta_{d} \, .
\end{align}
For instance, consider the case where $X_{10-d} \cong \mathbf{S}^{10-d}$, and introduce the familiar scalar laplacian on the sphere, namely
\begin{align}\label{eq:scalarlaplaciansphere}
    \Delta_{\mathbf{S}^{10-d}} = \frac{1}{\sqrt{g}} \frac{\partial}{\partial \xi^i} \left( \sqrt{g} g^{ij} \frac{\partial}{\partial \xi^j}\right)\, ,
\end{align}
where $\{\xi^i\}$ are local coordinates on $\mathbf{S}^{10-d}$ and $g_{ij}$ defines the line element
\begin{align}\label{eq:lineelementsphere}
    ds^2_{10-d}= g_{ij} d\xi^i d\xi^j = \mathsf{R}^2\, d\Omega^2_{10-d}\, ,
\end{align}
with $\mathsf{R}$ being the radius of the sphere. Thus, given this set-up we find the corresponding eigenmodes of $\Delta$ to be given precisely by the (scalar) spherical harmonics $\mathsf{Y}_{(10-d)}^{I_1 \ldots I_\ell}$, which arise for any $\ell=0,1,\ldots$, they satisfy the condition
\begin{align}\label{eq:harmonics}
    -\Delta_{\mathbf{S}^{10-d}} \mathsf{Y}_{(10-d)}^{I_1 \ldots I_\ell} = \frac{\ell (\ell+9-d)}{\mathsf{R}^2} \mathsf{Y}_{(10-d)}^{I_1 \ldots I_\ell}\, ,
\end{align}
and are in general degenerate, with the total degeneracy of modes given by
\begin{align}\label{eq:deg10-dsphere}
    \frac{(9-d+2\ell)(8-d+\ell)!}{\ell! (9-d)!}\, .
\end{align}
The latter is precisely the number of independent components of a traceless symmetric tensor of rank $\ell$. Notice that this means, in particular, that the effective mass --- as seen from the $d$-dimensional point of view --- of any non-zero mode is controlled by the quantity $m_{\rm KK} = \frac{1}{\mathsf{R}}$, which we refer to as the Kaluza-Klein scale in here. Therefore, if we only care about energies well below $m_{\rm KK}$, it is actually more convenient to integrate out (in the path integral sense) all the massive modes and define an effective field theory (EFT) for the massless --- i.e. $\ell=0$ --- states. The process just described is what we usually understand as compactification.

A crucial question at this point is what are the type of compactification manifolds $X_{10-d}$ we can actually place our string theory on. As already mentioned, a minimal requirement would be to ask for the total ten-dimensional background \eqref{prodansatz} to solve the equations of motion in the low energy effective theory.\footnote{More appropriately, one should ask for the 2d worldsheet theory defined over such background to provide for a reliable string background, namely to exactly preserve conformal invariance.} Hence, the easiest possibility involves flat compact backgrounds --- namely tori, leading to lower-dimensional effective  theories that preserve all the original supersymmetries, since they present trivial holonomy. This possibility will be further explored in Section \ref{s:maxsugraintro} below. Another, more interesting route, would be to select non-trivial Ricci-flat manifolds which break certain amount of the supercharges but still preserve some others. This can lead ultimately to theories manifestly exhibiting 16, 8, 4 or even 0 supercharges, see Section \ref{s:CYcompact} for more on this. As an example, let us briefly discuss how to obtain four-dimensional theories preserving 8 on-shell supercharges. Starting from e.g., Type II string theory, this can be accomplished upon breaking the original holonomy group as follows $\mathsf{SO(10)} \rightarrow \mathsf{SU(2)}\times \mathsf{SU(2)} \times  \mathsf{SU(3)}$, where the last piece corresponds to the (reduced) holonomy associated to the internal dimensions. Consequently, the compact six-dimensional manifold $X_6$ we choose in \eqref{prodansatz} allows for a globally well-defined spinor $\eta(y^m)$ that is moreover covariantly constant along $X_6$, namely it satisfies
\begin{align}\label{eq:covariantspinor}
	\nabla_m \eta(y^m) = 0\, , 
\end{align}
such that the original supercharges split into the group-theoretic spinorial representation $\mathbf{16} \rightarrow (\boldsymbol{1},\boldsymbol{2},\boldsymbol{1})+ (\boldsymbol{1},\boldsymbol{1},\boldsymbol{2} )$. Therefore, every 10d Majorana-Weyl spinor $\epsilon$ gives rise to one conserved supersymmetry as follows
\begin{equation}\label{eq:TypeIIspinorsonCY}
\begin{split}
&\epsilon^1=\xi^1_{+} \otimes \eta_{+} + \, \mathrm{h.c.} \, ,\\
&\epsilon^2=\xi^2_{+} \otimes \eta_{-,+} +\,  \mathrm{h.c.}\, ,
\end{split}
\end{equation}
where $\xi^A$ are 4d Weyl spinors and the lower indices indicate their respective chiralities (depending on whether we consider the Type IIA or Type IIB string). The compact six-dimensional spaces featuring the above necessary conditions are known as Calabi--Yau manifolds, see Section \ref{ss:8supercharges} for details.

\section{Maximally supersymmetric theories}\label{s:maxsugraintro}

A very concrete arena in which many of our discussions will take place arises from string theory compactifications preserving the maximal possible amount of supersymmetry, namely 32 supercharges. These theories can be systematically obtained from the Type IIA (or Type IIB) string on trivial compact backgrounds, i.e. tori. From a phenomenological perspective, such constructions are not very interesting, either because they live in higher dimensions or rather because they fail to exhibit the chiral spectrum that is needed to match with observations, as described by the Standard Model of Particle Physics. However, they provide a very interesting set of UV consistent quantum gravity vacua which are both computationally tractable as well as under control with regard to quantum corrections \cite{Cecotti:2015wqa}. In fact, given that the target spacetime is Riemann flat, one can quantize the Type II string \emph{exactly}, thus having access to purely stringy (or quantum gravitational) effects --- i.e. $\alpha'$ and $g_s$ corrections. 

We will review here their construction as well as the features that are most relevant for the upcoming chapters. To do so, we follow an alternative route, namely we first introduce 11d supergravity, understood as the low energy limit of M-theory \cite{Witten:1995ex}, and subsequently we consider toroidal compactifications thereof. The reason is that maximal supergravity in $d\leq 9$ is unique, and can be equivalently obtained upon compactifying M-theory instead of Type II string theory, the two constructions being related by (non-perturbative) dualities, as reviewed in Section \ref{s:dualities}. We present the relevant piece of the Type IIA(B) supergravity actions in 10d, as well as their dimensionally reduced relatives in nine and eight spacetime dimensions, since they will be extensively used in Parts \ref{part:StringTheoryTests} and \ref{part:pattern} of the thesis.

\subsection{11d M-theory}\label{ss:Mthy11d}

Even though there is no fully-fledged microscopic description of M-theory as of today (see e.g., \cite{Townsend:1995kk,Banks:1996vh, Seiberg:1997ad,Nicolai:1998ic,Dasgupta:2002iy} for old attempts), one can study its dynamics at low energies --- compared with the 11d Planck scale, which is given in terms of $\mathcal{N}=1$ supergravity in eleven dimensions. The field content is completely fixed by supersymmetry \cite{Nahm:1977tg}: it contains the gravitational field $g_{\mu \nu}$, a rank-3 antisymmetric tensor $C_3$ and a Majorana gravitino $\Psi_{\mu}$, thus including a total number of 256 degrees of freedom (d.o.f.s), 128 bosonic and 128 fermionic. This theory admits a lagrangian description first obtained by Cremmer, Julia and Scherk \cite{Cremmer:1978km}, which at the two-derivative order reads as follows
\begin{equation}\label{eq:11daction}
    \begin{aligned}
       S_\text{M-th}^{\text{11d}} &= \frac{1}{2\kappa_{11}^2} \int \mathcal{R} \star 1 - \frac{1}{2} G_4 \wedge \star G_4 -\frac{1}{3} C_3 \wedge G_4 \wedge G_4\\
       &+\frac{1}{2\kappa_{11}^2} \int \dd^{11}x\, \sqrt{-g} \left( -2i \bar\Psi_{\mu} \Gamma^{\mu \nu \sigma} D_{\nu} \left( \frac{\omega + \hat \omega}{2}\right) \Psi_{\sigma}\right)\\
       &+ \frac{1}{2\kappa_{11}^2} \int \dd^{11}x\, \sqrt{-g}\, \frac{\i}{96} \left(\bar\Psi_{\mu_1} \Gamma^{\mu_1 \mu_2 \mu_3 \mu_4 \mu_5 \mu_6} \Psi_{\mu_2} + 12 \bar\Psi^{\mu_3} \Gamma^{\mu_4 \mu_5} \Psi^{\mu_6}\right) \left( G_4 + \hat G_4\right)_{\mu_3 \mu_4 \mu_5 \mu_6}\, ,
    \end{aligned}
\end{equation}
see Appendix \ref{ap:conventions} for conventions. Here $G_4=dC_3$ denotes the 4-form field strength associated to the 3-form gauge potential and $\hat G_4$ is its supercovariant counterpart, with components
\begin{equation}\label{eq:supercovariant4form}
       (\hat{G}_4)_{\mu_1 \mu_2 \mu_3 \mu_4} =(G_4)_{\mu_1 \mu_2 \mu_3 \mu_4} -3 \bar\Psi_{[\mu_1} \Gamma_{\mu_2 \mu_3} \Psi_{\mu_4]}\, .
\end{equation}
On the other hand, $D_{\nu}$ denotes the supercovariant derivative acting on the gravitino field
\begin{equation}\label{eq:supercovariantderivative}
       D_{\nu} \left(\omega \right) \Psi_{\sigma} = \partial_{\nu} \Psi_{\sigma} + \frac{1}{4} \omega_{\nu}^{a b} \Gamma_{a b} \Psi_{\sigma}\, ,
\end{equation}
where $\omega$ and $\hat \omega$ are the spin and supercovariant connections, respectively. The former is determined by the equations of motion from the action \eqref{eq:11daction} upon treating it as an independent field, whereas the latter reads as
\begin{equation}\label{eq:supercovariantconnection}
       \hat\omega_{\nu a b} = \omega_{\nu a b} + \frac{1}{8} \bar\Psi^{\mu} \Gamma_{\mu \nu a b \sigma} \Psi^{\sigma}\, .
\end{equation}
Notice that the normalization of fields other than the metric has been chosen so that the supersymmetry transformations --- under which eq. \eqref{eq:11daction} is manifestly invariant --- do not include any additional factor of the gravitational constant $\kappa_{11}$, namely
\begin{equation}\label{eq:11dsusyvariations}
    \begin{aligned}
       &\delta e^{a}_{\mu} = \i \bar \epsilon \Gamma^{^a}\Psi_{\mu}\, ,\\
       &\delta \Psi_{\mu} = D_{\mu} (\hat \omega) \epsilon - \frac{1}{12 \cdot 4!} \left( \Gamma^{\nu_1 \nu_2 \nu_3 \nu_4}_{\mu} + 8 \Gamma^{\nu_1 \nu_2 \nu_3} \delta ^{\nu_4}_{\mu}\right) (\hat{G}_4)_{\nu_1 \nu_2 \nu_3 \nu_4} \epsilon \, ,\\
       &\delta (C_3)_{\mu \nu \sigma} = 3 \i \bar \epsilon \Gamma_{[\mu \nu}\Psi_{\sigma]}\, ,
    \end{aligned}
\end{equation}
where $e^{a}_{\mu}$ is the 11d vielbein and $\epsilon(x)$ denotes any spacetime-dependent 11d Majorana spinor.

For most of our purposes here, it will be enough to focus on the bosonic part of the theory. Hence, in what follows we only display the latter when writing down any supergravity action, keeping in mind that the fermionic terms can be obtained upon completing the corresponding supermultiplets and imposing local supersymmetry, see e.g., \cite{Cecotti:2015wqa} for details on this point.

Let us also briefly comment on the massive spectrum of the theory, since we will make use of it at several instances in this work. Indeed, it is possible to argue --- either via explicit black brane solutions in 11d supergravity \cite{Gueven:1992hh} or through dualities, see Section \ref{s:dualities} below --- that the relevant BPS states admitted by the 11d supersymmetry algebra arise as electrically and magnetically charged objects under the 3-form gauge field $C_3$. These are usually referred to as M2 and M5-branes, respectively, and constitute the fundamental objects of M-theory. Their low energy dynamics is controlled by the tension and charge density of the corresponding object, and it is determined by the appropriately supersymmetrized version of the Nambu-Goto plus Chern-Simons action (see e.g., \cite{Becker:2006dvp})
\begin{equation}\label{eq:DBICSactionMpbranes}
				S_\text{M-brane} = -\frac{2 \pi}{\ell_{11}^{p+1}} \int_{\Sigma_{p+1}} \dd^{p+1}x\, \sqrt{- g} + \frac{2 \pi}{\ell_{11}^{p+1}} \int_{\Sigma_{p+1}}  C_{p+1}\, ,\qquad p=2, 5\, ,
\end{equation}
where $\Sigma_{p+1}$ denotes the $(p+1)$-dimensional worldvolume of the associated M2 or M5-brane and $C_6$ is the magnetic dual of $C_3$.

\subsection{Type II supergravity in 10d}\label{ss:IIAB10d}

As already mentioned in Section \ref{s:basicconcepts}, conformal invariance of the 2d worldsheet theory requires that the superstring lives in ten spacetime dimensions,\footnote{Barring possible non-geometric constructions, see e.g., \cite{Gepner:1989gr,Flournoy:2004vn,Hull:2004in,Font:2017cya,Plauschinn:2018wbo} and references therein for some background on this topic.} where the minimal representation of $\mathsf{Spin(1,9)}$ is a 16-dimensional Majorana-Weyl spinor. This means that for theories with 32 supercharges, the supersymmetry generators can be arranged into two such independent spinors, with equal or different chirality (c.f. \eqref{eq:TypeIIspinorsonCY}). This partially accounts for the distinction between the two Type II string theories, which at low energies reduce to either $\mathcal{N}=(1,1)$ or $\mathcal{N}=(2,0)$ 10d supergravity, as we briefly review in the following.

\subsubsection*{Type IIA supergravity}

Let us start with non-chiral maximal supergravity in ten dimensions. This theory describes the dynamics of the massless spectrum of Type IIA string theory at energies well below the string scale. The relevant bosonic d.o.f.s consist of the metric $g_{\mu \nu}$, the Kalb-Ramond 2-form $B_2$ and the dilaton $\phi$ from the Neveu-Schwarz/Neveu-Schwarz (NSNS) sector, as well as the Ramond/Ramond (RR) $p$-forms $C_p$ with $p=1,3$. The bosonic part of the action reads in the string frame as \cite{Polchinski:1998rr}\footnote{For a more democratic formulation of the Type II string theories, whose bosonic content is extended so as to include also the magnetic $p$-form potentials along with some additional duality relations at the level of the e.o.m, see \cite{Townsend:1995gp,Bergshoeff:2001pv}.}
\begin{equation}\label{eq:IIA10dstringframeaction}
\begin{aligned}
S_\text{IIA, s}^{\text{10d}}\, =\, &\frac{2\pi}{\ell_s^8} \int \dd^{10}x\sqrt{-g}\ e^{-2\phi} \left(\mathcal{R}+4(\partial \phi)^2\right)-\frac{\pi}{\ell_s^8}\int e^{-2\phi} H_3\wedge \star H_3 \\
&-\frac{2\pi}{\ell_s^8}\int \left[F_2 \wedge \star F_2 + \tilde F_4 \wedge \star \tilde F_4 + B_2\wedge F_4 \wedge F_4\right]\, . 
\end{aligned}
\end{equation}
As is customary, $H_3=dB_2$ denotes the 3-form field strength associated to the $B_2$-field whilst $F_{p+1}$ are the field strengths of RR $p$-forms, i.e. $F_{p+1}=dC_p$. In addition, the kinetic term for the RR 3-form involves the following antisymmetric tensor 
\begin{align}\label{eq:RR4formfieldstrength}
 \tilde F_4 = dC_3 - C_1\wedge  H_3\, , 
\end{align}
which mixes with the RR 1-form and the Kalb-Ramond field due to the $\mathsf{U(1)}$ gauge invariance associated to the former. Notice that, contrary to the 11d case above, the theory has a non-trivial moduli space of vacua parametrized by the string coupling constant $g_s= e^{\phi_0}$, where $\phi_0 \equiv \braket{\phi}$ denotes the vacuum expectation value (v.e.v.) the dilaton field.

As seen from \eqref{eq:IIA10dstringframeaction}, the string frame action includes an exponential coupling to the dilaton in front of the Ricci scalar. On the other hand, whenever we want to make any (quantum) gravity statement, it is always convenient to switch to the more `canonical' Einstein frame, where the Einstein-Hilbert term is accompanied just by a constant prefactor. In the present case, this can be achieved by performing a Weyl rescaling of the form $g_{\mu \nu} \to e^{\phi/2}\, g_{\mu \nu}$, thus leading to
\begin{equation}
			\begin{aligned}\label{eq:IIA10dEinsteinframeaction}
				S_\text{IIA, E}^{\text{10d}} \, =\, &\frac{1}{2\kappa_{10}^2} \int \text{d}^{10}x\sqrt{-g} \left(\mathcal{R}-\frac{1}{2}(\partial \phi)^2\right)-\frac{1}{4\kappa_{10}^2}\int e^{-\phi}\, H_3\wedge \star H_3 \\
				&-\frac{1}{4\kappa_{10}^2}\int \left[e^{\frac{3}{2}\phi}\, F_2 \wedge \star F_2 + e^{\frac{1}{2}\phi}\, \tilde F_4 \wedge \star \tilde F_4 + B_2\wedge F_4 \wedge F_4\right]\, . 
			\end{aligned}
\end{equation}
Notice that the gravitational strength, encapsulated by the coupling constant $2\kappa_{10}^2= 2 \Mpt^{-8}= (2 \pi)^{7} \alpha'^4\ e^{2 \phi_0}$, is hence controlled by the dilaton v.e.v., which thus determines the Planck-to-string scale ratio. 

Regarding the non-perturbative massive spectrum on the theory, let us discuss now D$p$-branes. Contrary to what happened in 11d M-theory, where our only guide to deduce any possible massive excitation was supersymmetry, here there are various different ways to motivate the existence of such non-perturbative extended objects. Hence, beyond the low energy supergravity analysis, where it is possible to construct BPS black brane solutions for $p$ even \cite{Horowitz:1991cd}, one can also consider the possibility of allowing for the presence of open strings in the theory, which can end in principle on certain submanifolds $\Sigma_{p+1} \subset \mathbb{R}^{1, 9}$ \cite{Dai:1989ua,Horava:1989ga,Polchinski:1995mt} --- that supersymmetry constrain to be precisely such that $p=2k$, with $k=0,1,2,3, 4$. The inclusion of these objects breaks explicitly the 10d Lorentz symmetry, and in fact they can be viewed as $(p+1)$-dimensional states in the theory that we call D$p$-branes. Crucially, the supersymmetry algebra requires these objects to be charged under the RR $p$-forms, similarly to what happened in M-theory for the M2 and M5-branes.\footnote{Let us note that the exact same argument requires from the presence of a six-dimensional object charged magnetically under the Kalb-Ramond $B_2$-field, which is usually referred to as a NS5-brane \cite{Strominger:1990et}.}
 
A remarkable characteristic of these objects is their capacity to manifest gauge symmetries within their worldvolume theories. This ultimately arises from open strings states stretching between (possibly different) $p$-branes. In fact, it has been shown that the quantization of the open strings leads to non-trivial gauge theories living on the brane, which are described by suitable extensions of Yang-Mills theory and are frequently accompanied by additional \emph{charged} matter content. Their dynamics at low energies can be effectively described by the (supersymmetrized version of the) Dirac-Born-Infeld (DBI) action \cite{Born:1934gh, Dirac:1962iy}, which reads
\begin{align}\label{eq:DBITypeIIA}
S_\text{DBI} = -\frac{2\pi}{\ell_s^{p+1}} \int_{\Sigma_{p+1}} \dd^{p+1} x\ e^{-\phi} \sqrt{-\text{det} \left(g+B_2 -2\pi \alpha' \mathcal{F} \right)}\, ,
\end{align}
where $\mathcal{F}$ describes a field strength restricted to the worldvolume $\Sigma_{p+1}$. This action represents an extension of the Nambu-Goto action (c.f. eq. \eqref{eq:NambuGoto}), as a applied to higher-dimensional objects, and incorporates additional complexities suitable for describing the properties of D$p$-branes. Thus, upon expanding the square root of the determinant to leading order in $\mathcal{F}$ one indeed recovers (super-)Yang--Mills theory for the latter, plus an infinite number of higher-dimensional and higehr-derivative corrections in $\alpha'$. Interestingly, notice that quantizing the worldvolume theory of the D$p$-branes in a similar manner to what we did for the fundamental string in Section \ref{s:basicconcepts}, is strictly speaking not possible, since the theory \eqref{eq:DBITypeIIA} is not conformally invariant, such that it cannot be recasted in the Polyakov form. Let us also note that the DBI action includes an additional coupling to the dilaton, which implies that the physical tension of the branes grow like $g_s^{-1}$, thus suggesting that they are actually non-perturbative in nature, at least from the string theory point of view. 

On the other hand, the interaction between the D-branes and the RR $p$-forms is determined by the Chern--Simons (or Wess--Zumino) action 
\begin{align}\label{eq:CSWZactionTypeIIA}
S_\text{CS} = \frac{2\pi}{\ell_s^{p+1}} \int_{\Sigma_{p+1}} \left(\sum_q C_q \right) \wedge e^{2\pi \alpha' \mathcal{F} -B_2} \wedge \sqrt{\frac{\hat A (\ell_s^2 \mathcal{R}_T)}{\hat A (\ell_s^2 \mathcal{R}_N)}}\, ,
\end{align}
where the ${\text{A}}$-roof genus depends on the curvature 2-forms $\mathcal{R}_T$ and $\mathcal{R}_N$ of the tangent and normal bundles (respectively) of $\Sigma_{p+1}$ as follows 
\begin{align}
\hat A \left(\ell_s^2 \mathcal{R}_{N(T)} \right) = \frac{1}{(2\pi)^{\frac{p+1}{2}}} \sqrt{\det \left( \frac{\ell_s^2 \mathcal{R}_{N(T)}/2}{\sinh \left( \ell_s^2 \mathcal{R}_{N(T)}/2\right)}\right)}\, .
\end{align}
Therefore, depending on the topology and geometry of $\Sigma_{p+1}$, there can appear induced lower $q$-form charges on the D$p$-brane, see e.g., \cite{Witten:1998cd}. 

\subsubsection*{Type IIB supergravity}

Type IIB String Theory presents a chiral spectrum, such that at energies well below the string sale it reduces to 10d $\mathcal{N}=(2,0)$ supergravity in ten dimensions. The bosonic fields in the NSNS sector agree with those of Type IIA string theory, whilst the RR one provides for a different set of $p$-form gauge fields, namely $C_p$ with $p=0,2,4$. The 4-form $C_4$ has moreover a self-dual field strength satisfying $\tilde F_5= \star \tilde F_5$, and the dynamics (i.e. equations of motion) of the theory can be encapsulated (at the two derivative level) by the following pseudo-action \cite{Polchinski:1998rr} 
\begin{equation}\label{eq:IIB10d}
			\begin{aligned}
				S_\text{IIB}^{\text{10d}}\, =\, & \frac{1}{2\kappa_{10}^2} \int \dd^{10}x\sqrt{-g} \left(\mathcal{R}-\frac{1}{2}(\partial \phi)^2\right) -\frac{1}{4\kappa_{10}^2}\int e^{-\phi} H_3\wedge \star H_3 \\
				&-\frac{1}{4\kappa_{10}^2}\int \left[e^{2 \phi}F_1 \wedge \star F_1 + e^{\phi} \tilde F_3 \wedge \star \tilde F_3 + \frac{1}{2} \tilde F_5 \wedge \star \tilde F_5  +C_4\wedge H_3 \wedge F_3\right]\,,
			\end{aligned}
\end{equation}
where we have already switched to the Einstein frame upon performing the same Weyl rescaling as in Type IIA. The different field strengths are defined as follows
\begin{equation}
\begin{aligned}
&H_3=dB_2\, , \qquad  \tilde{F}_3=dC_2-C_0 H_3\, ,\\
&F_1=d C_0\, , \qquad \tilde{F}_5= dC_4-\frac{1}{2} C_2 \wedge F_3 + \frac{1}{2} B_2 \wedge H_3\, . 
\end{aligned}
\end{equation}
Analogously to the Type IIA case, the existence of exactly massless scalar fields in the theory --- i.e. the dilaton and the axion $C_0$, implies that there is a non-trivial moduli space (of complex dimension one) parametrized by the former. Several features of such moduli space will be discussed later on in Section \ref{s:dualities} and exploited in the rest of this thesis.

Finally, let us turn to the D-brane spectrum of the theory. Notice that, since the RR field content is different for Type IIB string theory, the possible extended objects saturating the BPS conditions can become quite distinct. Indeed, following the same logic as outlined before, one finds a D$(-1)$-brane --- i.e. an instanton --- coupling electrically to $C_0$, a D1-string coupling electrically to $C_2$, and a D3-brane coupled to the self-dual 4-form, together with their magnetic duals: the D5 and the D7-branes. Again, their stability is ensured by supersymmetry and the low energy dynamics is described in terms of the DBI and Chern--Simons actions, c.f. eqs. \eqref{eq:DBITypeIIA} and \eqref{eq:CSWZactionTypeIIA}. 

\subsection{Maximal supergravity in 9d}\label{ss:9dmaxsugra}

In nine dimensions there is a unique theory of gravity with 32 supercharges, namely 9d $\mathcal{N}=2$ supergravity \cite{Gates:1984kr}. It can be easily obtained from any of the previous 10d Type II supergravities after dimensionally reducing on a circle $\mathbf{S}^1$, which washes away the original chirality distinction in ten dimensions. Alternatively, if we only care about the low energy lagrangian description, it is equivalent (up to field redefinitions) to start from 11d $\mathcal{N}=1$ supergravity and compactify two spatial directions on a torus $\mathbf{T}^2$. Here we choose this second route for convenience. Of course, ultimately the fact that the 9d theory can be retrieved from any of these corners of the string theory landscape is simply a manifestation of the intricate dualities relating the aforementioned descriptions (see Section \ref{s:dualities} below). 

Hence, we consider M-theory at low energies, described by the action \eqref{eq:11daction} and we impose the following ansatz for the 11d metric 
\begin{equation}\label{eq:11dmetric}
	ds^2_{11} = \mathcal{V}_2^{-2/7} ds_9^2 + \mathcal{V}_2 \left( \tau_2^{-1}\left( dy^1- A_1^{(1)}+\tau_1 dy^2\right)^2+ \tau_2\left( dy^2 - A_1^{(2)}\right)^2\right)\, ,
\end{equation}
where the fields $A^{(i)}_1$ are Kaluza-Klein 1-forms in the non-compact 9d space. Notice that the purely internal piece of the above line element can be written as
\begin{equation}
	ds^2_{\mathbf{T}^2}= g_{mn} dy^m dy^n\, , \qquad \text{with}\quad g_{mn}= \frac{\mathcal{V}_2}{\tau_2} \left(
	\begin{array}{cc}
		1 & \tau_1  \\
		\tau_1 & |\tau|^2  \\
	\end{array}
	\right) \, ,
\end{equation}
with $\tau=\tau_1+{\rm i}\tau_2$ being the complex structure of the torus and $\mathcal{V}_2$ denoting its overall volume (in M-theory units). After doing so, one obtains
\begin{equation}\label{eq:9daction}
\begin{aligned}
    S_\text{M-th}^{\text{9d}} &= \frac{1}{2\kappa_9^2} \int \dd^{9}x\, \sqrt{-g}\,  \left( \mathcal{R} - \frac{9}{14} \left( \partial \log \mathcal{V}_2 \right)^2 -\frac{\partial \tau \partial \bar \tau}{2 \tau_2^2} \right) - \frac{1}{4\kappa_9^2} \int \frac{\mathcal{V}_2^{9/7}}{\text{Im}\, \tau}F_2 \wedge \star \bar{F}_2\\
    & - \frac{1}{4\kappa_9^2} \int \mathcal{V}_2^{-12/7} dC_1 \wedge \star dC_1+\frac{\mathcal{V}_2^{4/7}}{\text{Im}\, \tau}F_3 \wedge \star \bar{F}_3 + \mathcal{V}_2^{3/7}dC_3 \wedge \star dC_3\\
    & - \frac{1}{2\kappa_9^2} \int C_1 \wedge dC_3 \wedge dC_3\, ,
\end{aligned}
\end{equation}
where we have absorbed an overall volume factor into the 9d gravitational coupling constant.\footnote{\label{fnote:11d-9dPlancklengths}More precisely, the 11d and 9d Planck lengths are related by $\ell_{11}^7=\ell_9^7\, \mathcal{V}_2$, see Appendix \ref{ap:conventions} for conventions.} The nine-dimensional bosonic fields other than the metric arrange into a scalar sector $\{ \mathcal{V}_2, \tau\}$ parametrizing the vacuum manifold; the two KK 1-forms with a complex field strength defined by $F_2= dA^{(1)}_1- \tau dA^{(2)}_1$, as well as an additional 1-form $C_1$ obtained from the reduction of the 11d 3-form on the torus; a complex 2-form with field strength $F_3= dC^{(1)}_2- \tau dC^{(2)}_2$ arising from the 11d 3-form with one leg along any internal direction; and a single 3-form gauge field $C_3$.

Before moving on, let us mention that the moduli space of the 9d theory can be seen to be isomorphic to the group coset $\mathcal{M}_{\text{9d}}=\mathbb{R}_+ \times \mathsf{SL(2, \mathbb{R})}/\mathsf{SO(2)}$, where the first factor corresponds to the torus volume and the second one is identified with its complex structure. To see this, one first defines a moduli-dependent upper triangular matrix as follows
\begin{equation}\label{eq:Pmatrix}
	\mathcal{P} = \left( \begin{array}{cc}
		\tau_2^{-1/2} & \tau_1 \tau_2^{-1/2} \\
		0 & \tau_2^{1/2}  \\
	\end{array}
	\right) \, ,
\end{equation}
as well as the symmetric $2\times 2$ matrices $\mathcal{Q} = \mathcal{P}^{\text{T}} \mathcal{P}$ and its inverse $\mathcal{Q}^{-1}$
\begin{equation}
	\mathcal{Q} = \left( \begin{array}{cc}
		\tau_2^{-1} & \tau_1 \tau_2^{-1} \\
		\tau_1 \tau_2^{-1} & \tau_2 + \tau_2^{-1} \tau_1^2  \\
	\end{array}
	\right) \, , \qquad 
   \mathcal{Q}^{-1} = \left( \begin{array}{cc}
		 \tau_2 + \tau_2^{-1} \tau_1^2 & -\tau_1 \tau_2^{-1} \\
		-\tau_1 \tau_2^{-1} & \tau_2^{-1}  \\
	\end{array} \right)\, ,
\end{equation}
which both have unit determinant. Using these objects, the complex structure piece of the 9d lagrangian can be written in matrix form as
\begin{equation}\label{eq:sl2matrixform}
	-\frac{1}{2 \tau_2^2} \partial \tau \cdot \partial \bar \tau= \frac{1}{4} \text{tr}\, \left( \partial\mathcal{Q}^{-1} \cdot \partial \mathcal{Q} \right)\, .
\end{equation}
Therefore, one may alternatively describe the 9d moduli space in terms of unimodular matrices of the upper triangular form displayed in \eqref{eq:Pmatrix}. In addition, notice that under generic $\mathsf{SL(2, \mathbb{R})}$ transformations acting from the right on $\mathcal{P}$, one finds that $\mathcal{Q} \to \mathcal{A}^{\text{T}} \mathcal{Q} \mathcal{A}$, with $\mathcal{A} \in \mathsf{SL(2, \mathbb{R})}$. These transformations thus leave invariant the kinetic term \eqref{eq:sl2matrixform}, preserving moreover the unimodular condition. However, they do not, in general, maintain the upper triangular form of $\mathcal{P}$. Upon further accounting for such `compensating' $\mathsf{SO(2)}$ rotations one concludes that the complex modulus $\tau$ can be equivalently described by the group of $\mathsf{SL(2, \mathbb{R})}$ matrices up to 2d rotations, as claimed. In fact, at the classical level, one can easily get convinced that the action \eqref{eq:9daction} above is symmetric under the group $\mathsf{SL(2, \mathbb{R})}$, with scalars transforming as indicated above whilst the different $p$-forms transform linearly, either as a doublet (the complex 1- and 2-forms $F_2$, $F_3$) or as a singlet ($C_1$ and $C_3$).

Regarding the massive spectrum of the theory, it can be deduced directly from that of the parent 11d description, as will be explained in more detail in Section \ref{ss:MthyT2SSDC}, so that we refrain from repeating it here and refer the interested reader to Chapter \ref{ch:bounds} of the thesis.

\subsection{Maximal supergravity in 8d}\label{ss:8dmaxsugra}

Following the analysis of the previous section, a good strategy to obtain the maximal supergravity action for a given spacetime dimension $d\leq 9$ is to consider M-theory compactified on a $k=11-d$ dimensional torus, with a metric ansatz of the form 
\begin{equation}\label{eq:torusmetric}
	ds^2_{11} = \left(\text{det}\, g_{mn}\right)^{-\frac{1}{d-2}} ds_d^2 + g_{mn} dy^m dy^n\, .
\end{equation}
where $g_{mn}$ parametrizes the metric on the internal torus. As usual, the prefactor in the $d$-dimensional line element has been chosen so as to obtain the resulting lower dimensional theory in the Einstein frame. Therefore, in order to find the correct description of maximal supergravity in eight dimensions one simply needs to compactify M-theory on $\mathbf{T}^3$.

Notice that we did not include in \eqref{eq:torusmetric} the fluctuations in the metric associated to the Kaluza-Klein photons (c.f. eq. \eqref{eq:11dmetric}). The reason for this is because we will focus in this section solely on the gravitational-scalar sector of the theory, thus forgetting about any other tensorial field, since it will play no role in the analysis performed in the upcoming chapters (see \cite{Salam:1984ft} for the full supersymmetric action). After these manipulations, one finally arrives at
\begin{align}\label{eq:8dsl3}
	S_\text{M-th}^{\text{8d}}\, \supset\, &\frac{1}{2\kappa_8^2} \int \dd^{8}x\, \sqrt{-g}\,  \left( \mathcal{R} + \frac{1}{4} \text{tr} \left( \partial \tilde{g} \cdot \partial \tilde{g}^{-1} \right) -\frac{\partial \mathcal{T} \cdot \partial \bar{\mathcal{T}}}{2 \mathcal{T}_2^2} \right)\, ,
\end{align}
where the $3\times3$ matrix $\tilde{g}$ is obtained from the internal metric of the $\mathbf{T}^3$ with the overall volume extracted, i.e. $\tilde{g}_{m n}= \mathcal{V}_3^{-2/3} g_{mn}$. More precisely it reads as
\begin{align}\label{eq:trace8d}
	-\text{tr} \left( \partial \tilde{g} \cdot \partial \tilde{g}^{-1} \right) = \left(\tilde{g}^{m p} \tilde{g}^{n q}+ \frac{1}{6} \tilde{g}^{mn} \tilde{g}^{pq} \right) \partial \tilde{g}_{mn} \cdot \partial \tilde{g}_{pq}\, .
\end{align}
In addition, we have defined the complex field $\mathcal{T}=C_{123}^{(3)} + \text{i} \mathcal{V}_3$, which contains the compact scalar $C_{123}^{(3)}$ arising from the reduction of the 11d 3-form $C_3$ along the torus as well as the volume modulus. Note that the parametrization in \eqref{eq:8dsl3} is rather useful since it already makes manifest certain symmetries of the classical action. In fact, to detect those it is enough to realize that the moduli $\{ \tilde{g}_{m n}, \mathcal{T}\}$, parametrize the coset space\footnote{One quick way to understand the $\mathsf{SL(3, \mathbb{R})}$ symmetry of the action \eqref{eq:8dsl3} is by realizing that the fields $\tilde g_{mn}$ transform in the adjoint representation, namely $\tilde g \to \mathcal{A}^{\text{T}} \tilde g \mathcal{A}$, with $\mathcal{A} \in \mathsf{SL(3, \mathbb{R})}$. Such transformations thus leave the trace \eqref{eq:trace8d} invariant.}
\begin{align}
	\mathcal{M}_{\text{8d}} = \frac{\mathsf{SL(3, \mathbb{R})}}{\mathsf{SO(3)}} \times \frac{\mathsf{SL(2, \mathbb{R})}}{\mathsf{SO(2)}}\, ,
\end{align}
which exhibits some nice structure that will play an important role in Parts \ref{part:StringTheoryTests} and \ref{part:pattern} of the thesis.

For future reference, it is also convenient to rewrite \eqref{eq:8dsl3} using an alternative set of fields that arise more naturally when directly reducing the 9d action introduced in Section \ref{ss:9dmaxsugra} above, on $\mathbf{S}^1$. Therefore, starting from eq. \eqref{eq:9daction} and upon further compactification on a circle of radius $R_3$ (measured in 9d Planck units), we find
\begin{equation}\label{eq:8dalternativeaction}
\begin{aligned}
	S_\text{M-th}^{\text{8d}}\, &\supset\, \frac{1}{2\kappa_{8}^2} \int \dd^{8}x\sqrt{-g} \Bigg[\mathcal{R}-\frac{9}{14} \left( \partial \log \mathcal{V}_2\right)^2 - \frac{7}{6} \left( \partial \log R_3\right)^2 -\frac{\partial \tau \cdot \partial \bar \tau}{2 \tau_2^2}\\
      &- \frac{\mathcal{V}_2^{-12/7} R_3^{-2}}{2} \left( \partial C_{123}^{(3)}\right)^2 -\frac{\mathcal{V}_2^{9/7} R_3^{-2}}{2 \tau_2} \left| \partial A^{(1)}_0-\tau \partial A^{(1)}_0 \right|^2\Bigg]\, ,
\end{aligned}
\end{equation}
where $C_{123}^{(3)}$ is defined after eq. \eqref{eq:trace8d} and the compact scalars $\{ A^{(1)}_0, A^{(1)}_0 \}$ parametrize the orientation of the two-dimensional torus within the $\mathbf{T}^3$. From the parent nine-dimensional theory, these arise from the reduction of the singlet and doublet of 1-forms along the extra circle, see discussion around after eq. \eqref{eq:9daction}. Moreover, the overall $\mathbf{T}^3$ volume can be expressed in terms of those of the submanifolds $\mathbf{T}^2$ and $\mathbf{S}^1$ as follows
\begin{equation} \label{eq:T3volume}
     \mathcal{V}_3 = \mathcal{V}_2\, R_3\, \frac{\ell_9}{\ell_{11}} = \mathcal{V}_2^{6/7}\, R_3\, ,
\end{equation}
where the relation between the 9d and 11d Planck lengths can be found in footnote \ref{fnote:11d-9dPlancklengths}.

Let us conclude by mentioning that the (non-)perturbative massive spectrum of the present 8d theory will be discussed in more detail in Section \ref{ss:MthyT3SSDC} (see in particular Table \ref{tab:BPSstates}), to which we refer in here.

\section{Theories with reduced supersymmetry}\label{s:CYcompact}

In this section we proceed by introducing and discussing effective field theories in diverse spacetime dimensions which preserve less amount of supersymmetry. In particular, we focus on EFTs that exhibit either 16 (Section \ref{ss:16supercharges}) or 8 unbroken supercharges (Section \ref{ss:8supercharges}). They are less constrained than their maximally supersymmetric counterparts, which already implies that there can be significant quantum corrections (both perturbative and non-perturbative), even at the two-derivative level. On the other hand, they exhibit rich dynamics accommodating for further ingredients that are moreover phenomenologically attractive. We have chosen certain representative examples in each case, based on the analysis carried in later parts of the thesis.

\subsection{Compactifications preserving 16 supercharges} \label{ss:16supercharges}

There are in fact many ways to obtain low energy effective theories preserving 16 supercharges. One economic way to do so is to start from (any of) the Heterotic string(s) --- or Type I string theory --- in ten spacetime dimensions, which precisely exhibit that amount of supersymmetries, and subsequently reduce the theory on flat tori. This already provides for various interesting phenomena that are absent in maximally supersymmetric constructions, such as gauge enhancements at special points in moduli space (see \cite{Font:2020rsk} for a recent analysis of this family of string theory compactifications). Even though we will utilize these theories to illustrate some of the physics described in this work (see in particular Section \ref{ss:het s1} below), we refrain from presenting here a detailed analysis of their low energy dynamics. Instead, we have chosen to describe another set-up preserving 16 supercharges which arises from M-theory compactifications on $K3$ surfaces, since it also appears at several instances in the thesis, playing a major role in our discussions.



\subsubsection{M-theory on a $K3$ surface}\label{sss:MtheoryonK3}

Let us consider M-theory compactified on a $K3$ manifold down to seven (non-compact) spacetime dimensions. The resulting theory displays (minimal) $\mathcal{N}=1$ supersymmetry in 7d, namely it preserves 16 of the original 32 supercharges of 11d supergravity. In fact, a useful way to think about $K3$ is in terms of a $\mathbf{T}^4/\mathbb{Z}_2$ orbifold, where the discrete group acts as $z^i \to -z^i$ on the two complex coordinates parametrizing the torus. The $\mathbb{Z}_2$ group acts non-freely, such that there are 16 fixed points in total which locally look like $\mathbf{R}^4/\mathbb{Z}_2$, that are responsible for breaking half of the original supersymmetries. One can moreover associate 3 parameters to each of these points, which can be used to resolve the singularity into some $\mathbb{P}^1$ blow-up, thus leaving some smooth space behind of the Eguchi-Hanson type \cite{Eguchi:1978gw}
\beq\label{eq:EguchiHanson}
	ds^2_{\rm{EH}}= f(r)^{-1}dr^2+r^2 f(r) \left( d\psi + \cos \theta d\phi \right)^2 + r^2 \left( d\theta^2 + \sin^2 \theta d\phi^2\right)\, ,
\eeq
with $f(r)=1-\left(\frac{t}{r}\right)^4$ and where $(\theta, \phi, \psi)$ parametrize a 3-sphere (modded out by $\mathbb{Z}_2$ due to the identification $\psi \sim \psi + 2\pi$). The free parameters correspond to $t$, which provides the size of the 2-cycle, together with two additional angles that control the orientation of the $\mathbb{P}^1$ inside $\mathbb{R}^4$. 

Here we will only need to know a few topological and geometric facts that are specific to $K3$ manifolds (see e.g., \cite{Aspinwall:1996mn} for more details).\footnote{In the mathematical literature, $K3$ surfaces are defined as compact K\"ahler manifolds of complex dimension two with vanishing first Chern class. Therefore, they are nothing but Calabi--Yau two-folds, see Section \ref{ss:8supercharges} below.} First, for a given complex structure, the Hodge decomposition of $H^2(K3, \mathbb{R})$ is such that the Hodge numbers read
\beq\label{eq:hodgeK3}
	h^{2,0}(K3) = h^{0,2}(K3) = 1\,,\qquad h^{1,1}(K3) = 20\, .
\eeq
To these we can associate a basis of harmonic 2-forms, i.e. $\lbrace \omega_A \rbrace= \lbrace \Omega_2, \overline \Omega_2, \omega_a \rbrace$, where $\Omega_2 \in H^{2,0}(K3)$ is the unique holomorphic (2,0)-form and $\omega_a \in H^{1,1}(K3)$, $a=1, \ldots, h^{1,1}(K3)$. In addition, the Hodge star operator maps $H^2(K3, \mathbb{R})$ to itself, and moreover satisfies $\star^2=1$. Therefore, one may divide the second cohomology group of $K3$ into self-dual and anti self-dual  forms, namely $H^2(K3, \mathbb{R}) = H_+^2(K3, \mathbb{R}) \oplus H_-^2(K3, \mathbb{R})$, where $\dim H_+^2(K3, \mathbb{R})=3$ and $H_-^2(K3, \mathbb{R})=19$. The former admits a natural basis spanned by the real and imaginary parts of $\Omega_2$, together with the K\"ahler 2-form $J= t^a \omega_a$, which controls the volume of the different holomorphic cycles within the $K3$. 
		
With the above mathematical background, we are now ready to discuss both the field content and the low energy dynamics of the resulting theory. In a supersymmetric language, one is left with $d=7$, $\mathcal{N}=1$ supergravity coupled to 19 abelian vector multiplets (at generic points in moduli space). The gravity multiplet contains the metric field, a 2-form (which can be dualized to a 3-form gauge field in 7d), an $\mathsf{SO(3)}$ triplet of 1-forms and a scalar field. On the other hand, each vector multiplet consists of a 1-form potential as well as an $\mathsf{SO(3)}$ triplet of scalar fields. The scalar sector arises from the deformations of the $K3$ metric, which comprise 58 parameters in total, one of which corresponding to the overall volume --- which belongs to the gravity multiplet.\footnote{In the orbifold description, the scalar fields arise from the internal metric of the $\mathbf{T}^4$ (10 in total) together with each of the 3 parameters associated to the 16 fixed points (48 scalars).} Finally, the 3-form and 1-forms both descend from the 11d 3-form field $C_3$:
\beq\label{eq:7dreductionC3}
		C_3=A^a \wedge \omega_a + A_3\, .
\eeq
Notice that the latter arise upon reducing $C_3$ on a basis of harmonic 2-forms in $K3$. One thus obtains $h^2(K3)=22$ 1-form gauge potentials in total, three of which belong to the gravity multiplet.

All in all, the relevant piece of the 7d (bosonic) action reads as follows
\begin{equation}\label{eq:7dMthyK3}
	\begin{aligned}
		S_\text{M-th}^{\text{7d}}\, =\, & \frac{1}{2\kappa^2_7} \int \dd^{7}x\, \sqrt{-g}\,  \left( \mathcal{R} - \frac{9}{20} \left( \partial \log \mathcal{V}_{K3} \right)^2 - G_{i j} \partial \phi^i \cdot \partial \phi^j\right)\\
        &  -\frac{1}{4\kappa_{7}^2} \int \mathcal{V}_{K3}^{6/5}\, dA_3 \wedge \star dA_3 + \mathcal{V}_{K3}^{-3/5} \mathsf{G}_{a b} F^a \wedge F^b + 2\eta_{a b} A_3 \wedge F^a \wedge F^b\, ,
	\end{aligned}
\end{equation}
where $\mathcal{V}_{K3} = \frac{1}{2} \eta_{ab} t^a t^b$ is the overall volume of the internal space and $\eta_{ab}= \omega_a \cdot \omega_b \equiv \int_{K3} \omega_a \wedge \omega_b$ denotes the intersection form of the $K3$ surface. The fields $\phi^i$, $i=1, \ldots, 57$, parametrize the vector multiplet moduli space, i.e. the space of deformations of Ricci-flat K\"ahler metrics on $K3$ with fixed overall volume. This is given by the group coset \cite{Aspinwall:1996mn}
\begin{align}\label{eq:cosetspace7d}
	\mathcal{M}_{\text{VM}} = \mathsf{O(\Gamma_{3,19})}\backslash \mathsf{O(3,19)} / (\mathsf{O(3)} \times \mathsf{O(19))}\, ,
\end{align}
where $\Gamma_{3,19}$ denotes the lattice with signature $(3,19)$ that is isomorphic to the integer (non-trivial) cohomology on K3. The tensor $G_{ij}$ is the canonical metric on $\mathcal{M}_{\text{VM}}$, whose explicit form can be found in the original references, see \cite{Duff:1983vj}. On the other hand, the gauge kinetic matrix for the vectors depends solely on the K\"ahler deformations through the functions $\mathsf{G}_{a b}$. Its explicit form can be readily computed to be \cite{Lee:2019xtm}
\begin{equation}\label{eq:7dmodspacemetric}
	\begin{aligned}
		\mathsf{G}_{a b} = \int_{K3} \omega_a \wedge \star \omega_b = \frac{t_a t_b}{\mathcal{V}_{K3}}- \eta_{a b} = \tilde{t}_a \tilde{t}_b -\eta_{a b}\, ,
	\end{aligned}
\end{equation}
where the indices are lowered with the intersection form $\eta_{a b}$. In addition, one can argue that the action \eqref{eq:7dMthyK3} is classically exact \cite{Witten:1995ex, Cadavid:1995bk} and hence does not receive any further quantum corrections. This is in contrast to e.g., Type IIA string theory on the same two-fold, which probes the quantum $H^2(K3)$-cohomology due to the extra modes arising from the $B_2$-field.

An important simplification occurs when the $K3$ surface is \emph{attractive} \cite{Moore:1998pn}, namely when the rank of its Picard group is maximal.\footnote{The Picard group is defined as $\text{Pic}(K3)= H^{1,1}(K3) \cap H^2(K3,\mathbb{Z})$, such that it corresponds to (dual) curve classes which have some holomorphic representative \cite{Aspinwall:1996mn}. For attractive $K3$ two-folds, $\text{rk}(\text{Pic}(K3))=20$.} For such manifolds, the complex structure is completely fixed (see e.g., \cite{Moore:1998zu} for details on this), so that both the 7d lagrangian as well as the mass of the different (non-)perturbative states depend solely on the K\"ahler moduli. In those cases, the scalar lagrangian in \eqref{eq:7dMthyK3} reduces to
\begin{equation}\label{eq:7dMthyattractive}
	\begin{aligned}
		\mathcal{L}_\text{M-th}^{\text{7d}}\, \supset\, - \frac{1}{2\kappa^2_7} \sqrt{-g}\,  \left[ \frac{9}{20} \left(\frac{\partial \mathcal{V}_{K3}}{\mathcal{V}_{K3}} \right)^2 + \mathsf{G}_{a b}\, \partial \tilde{t}^a \cdot \partial \tilde{t}^b \right]\, ,
	\end{aligned}
\end{equation}
where $\tilde{t}^a= t^a/\mathcal{V}_{K3}^{1/2}$ are rescaled moduli subject to the constraint $ \frac{1}{2} \eta_{ab} \tilde{t}^a \tilde{t}^b \stackrel{!}{=} 1$ and $\mathsf{G}_{a b}$ is given in \eqref{eq:7dmodspacemetric}.

\subsection{Compactifications preserving 8 supercharges} \label{ss:8supercharges}

We finally turn to theories preserving $1/4$ or $1/2$ of the original supersymmetries, depending on whether they arise from Type II or Heterotic string compactifications. Their significance within the context of the present thesis lies in the fact that whilst they are still under computational control, the dynamics is far richer than their higher supersymmetric cousins --- even at the two-derivative level. In fact, the geometry of their moduli spaces can receive a plethora of quantum and stringy corrections, which can be then used to learn non-trivial aspects of (non-)perturbative quantum gravity. 

In this subsection we will particularize to two sets of theories preserving eight unbroken supercharges, which live in four and five non-compact spacetime dimensions. These set-ups can be obtained upon compactifying e.g., Type II string theory or M-theory on a Calabi--Yau threefold, as explained in Section \ref{s:basicconcepts}. In the following, we will first review the main topological and geometrical data that characterizes this kind of complex manifolds. Subsequently, we consider Type IIA string theory (Section \ref{sss:4dN=2basics}) and M-theory (Section \ref{sss:5dMtheory}) on these geometric backgrounds, providing all necessary details which will be needed in later parts of the thesis.

\subsubsection*{Calabi--Yau Manifolds}

Mathematically, a Calabi--Yau (CY) space $X_n$ is defined as a compact K\"ahler manifold of complex dimension $n$ with trivial canonical bundle $\mathscr{K}_{X_n}$. The latter comprises the bundle of $(n,0)$-forms, and its triviality implies that it can be identified with $X_n \times \mathbb{C}$. Therefore, the manifold $X_n$ possesses a \emph{unique} globally defined holomorphic $n$-form, denoted $\Omega_n$, corresponding to the unit section in $\mathscr{K}_{X_n}$. Equivalently, the Calabi--Yau condition may be stated as the triviality of the first Chern class associated to the tangent bundle, i.e. $c_1 (TX_n)=0$. Hence, since $c_1$ is defined in terms of the trace of the curvature connection, one concludes that Calabi--Yau $n$-folds have $\mathsf{SU(n)}$ holonomy, which according to our discussion in Section \ref{s:basicconcepts} is required to preserve some amount of unbroken supersymmetry in the resulting $(d-2n)$-dimensional compactified theory. However, from the Einstein field equations, one would like to find such a restricted manifold which moreover satisfies the condition $\text{tr}_{\mathbb{C}}\, \mathcal{R}=0$ \emph{pointwise} and not just in cohomology. The fact that this is actually possible for any given K\"ahler class was first conjectured by Calabi \cite{calabi} and later proved by Yau \cite{yau}, and it tells us that the space of deformations of Ricci-flat K\"ahler metrics is isomorphic to the space of allowed complex structures and, once the latter is fixed, the K\"ahler deformations. This is what we study next.

Despite the existence of a Ricci-flat metric $\mathcal{R}_{i \bar j}=0$ in a CY manifold for any given K\"ahler and complex structure --- which is ensured by Yau's theorem, the explicit form of the associated K\"ahler metric is usually not known except for very simple examples like tori or orbifolds thereof. However, as stated above, the usefulness of the theorem lies in the fact that the information about its deformation space, which determines part of the light degrees of freedom of the resulting lower dimensional theory, is completely encoded in certain topological data. In particular, the dimension of the aforementioned spaces depends on the Hodge numbers $h^{p,q} (X_n)$ associated to the Calabi--Yau manifold. These are defined as the dimension of the \u Cech-Dolbeault cohomology groups $H^{p,q} (X_n)$, which are related to the more familiar de Rahm cohomology as follows
\begin{align}
  H^r_{\rm{dR}}(X_n) = \sum_{k=0}^r H^{k,r-k}(X_n)\, ,
\end{align}
such that $b_r(X_n)=\sum_{p+q=r} h^{p,q} (X_n)$, where $b_r(X_n)$ denote the Betti numbers. Henceforth, we will stick to the case of three-dimensional CY spaces, being these the most relevant ones for the upcoming sections. We will additionally restrict ourselves to the irreducible Calabi--Yau case, namely those three-folds which are simply connected and thus have $b_1=h^{1,0}=h^{0,1}=0$. By Poincaré and Serre duality, one also concludes that $\text{dim}\ H^{2,0} (X_3)=0$ as well $\text{dim}\ H^{2,2} (X_3) = \text{dim}\ H^{1,1} (X_3)$, thus leaving us with the following simple Hodge diamond
 \begin{align}\label{Hodgediamond}
\begin{matrix}&&&h^{0,0} & & &\\ &&h^{1,0} & &h^{0,1} && \\ &h^{2,0}&&h^{1,1}&&h^{0,2}&\\ h^{3,0}&&h^{2,1} &&h^{1,2}&&h^{0,3}\\ &h^{3,1}&&h^{2,2}&&h^{1,3}&\\ &&h^{3,2}&&h^{2,3}&& \\ &&&h^{3,3} &&&\end{matrix}\; \stackrel{ \text{}}{=}\;\begin{matrix}&&&1& & &\\ &&0 & &0&& \\ &0&&h^{1,1}&&0&\\ 1&&h^{2,1} &&h^{2,1}&&1\\ &0&&h^{1,1}&&0&\\ &&0&&0&& \\ &&&1 &&&\end{matrix}\, ,
\end{align}
which depends only on the numbers $\{ h^{2,1}, h^{1,1}\}$, parametrizing the third and second cohomologies of the three-fold $X_3$. Each of these quantities will be in fact associated to certain subspaces of the Calabi--Yau moduli space (see \cite{Candelas:1990pi} for details).

Let us start with the set of K\"ahler deformations. These are encoded into the K\"ahler $(1,1)$-form $J = (\i/2) g_{i \bar j} dz^i \wedge d\bar{z}^j$, which must be cohomologically non-trivial. The latter statement follows from the K\"ahler condition, namely the fact that $dJ=0$, together with the definition of the overall volume $\mathcal{V}_{X_3}$ of the CY
\begin{align} \label{eq:CYvolume}
   \mathcal{V}_{X_3} = \frac{1}{3!}\int_{X_3} J\wedge J \wedge J \, .
\end{align}
Hence, exactness of the K\"ahler 2-form would imply a vanishing three-fold volume, which is inadmissible since $\mathcal{V}_{X_3}$ controls the effective lower-dimensional Planck scale.

Therefore, it becomes useful to introduce a basis of harmonic 2-forms and 4-forms
\begin{equation}
\label{eq:h11forms}
\left\{\omega_{a}\right\} \in H^{1,1}(X_3, \mathbb{R}), \quad\left\{\tilde{\omega}^{a}\right\} \in H^{1,1}(X_3, \mathbb{R}), \quad a=1, \ldots, h^{1,1}\, .
\end{equation}
These two bases can be chosen to satisfy the relation
\begin{equation}
\label{eq:2formsproductN=2}
\int_{X_3} \omega_{a} \wedge \tilde{\omega}^{b}=\delta_{a}^{b}\, ,
\end{equation}
and they can be used to expand the different relevant forms appearing in any of our starting 10d action. In particular, the 2-dorm $J$ can be written as
\begin{align}\label{Kahlerexpand}
J= t^a \omega_a \, ,
\end{align}
where the (real) coefficients $t^a$ are usually referred to as K\"ahler moduli. Their domain of definition must be such that the K\"ahler metric associated to $J$ is positive definite,\footnote{As discussed in the main text, the K\"ahler form takes values inside a strongly convex polyhedron, whose interior we denote by $\mathcal{K}(X_3)$. If the number of generators $N$ of such cone is equal to the dimension $h^{1,1}$ of the K\"ahler cone itself (as in \eqref{Kahlerexpand}), then we say that it is \emph{simplicial}. If $N > h^{1,1}$, the cone is \emph{non-simplicial} \cite{Rudelius:2014wla,Heidenreich:2020ptx}.} which is ensured if the volumes of all complex curves $\mathcal{C}$, surfaces $\mathcal{S}$ and $X_3$ itself are positive, i.e.
\begin{align}
\int_{\mathcal{C}} J>0 \, ,\qquad \int_{\mathcal{S}} J\wedge J>0\, ,\qquad \int_{X_3} J\wedge J\wedge J>0\, . 
\end{align}
Notice that if for some choice of K\"ahler moduli $\{ t^a\}$, $J$ satisfies this condition then $\lambda J$ with $\lambda\in\mathbb{R}_+$ also satisfies the latter. Hence, the values of the K\"ahler moduli for which the K\"ahler metric is positive definite form a cone, the so-called K\"ahler cone, $\mathcal{K}(X_3)$. Moreover, any $(1,1)$-form belonging to the closure of $\mathcal{K}(X_3)$ is said to be \emph{nef}. These forms enjoy the property of having non-negative (triple) intersection product. Additionally, the K\"ahler form $J$ is oftentimes able to cross certain codimension-1 boundaries in $\mathcal{K}(X_3)$. When this happens, one enters into the K\"ahler cone $\mathcal{K}(\tilde{X}_3)$ of a birationally equivalent\footnote{Two spaces are said to be birationally equivalent if upon removing suitable codimension one subsets from each they indeed become isomorphic \cite{Aspinwall:1993xz}.} three-fold $\tilde{X}_3$, which is related to the original one by a so-called flop transition (see e.g., \cite{Greene:1996cy}). The union of all these birationally equivalent CY three-folds is commonly referred to as the \emph{extended K\"ahler cone} of $X_3$, $\mathcal{K}_{\cup}(X_3)$.

In order to make the link with the CY moduli space more explicit, let us mention that, for a given complex structure $\mathcal{I}^j_i$, the deformations of the K\"ahler form parametrized by the $t^a$ in \eqref{Kahlerexpand} translate into deformations of the Ricci-flat K\"ahler metric as follows
\begin{align}\label{Kahlerdeform}
 \delta g_{i\bar \jmath} = -\i t^a (\omega_a)_{i\bar \jmath}\, . 
\end{align}
Apart from those, one can similarly consider variations of $\mathcal{I}^j_i$ itself, which translate into a deformation of the holomorphic $(3,0)$-form, whose components read as \cite{Koerber:2010bx}
\begin{align}\label{eq:variation3form}
   \delta\Omega_{ij \bar{k}}= -\frac{3\i}{2} \Omega_{[ij|l}\delta g^l_{\bar{k}]}\, ,
 \end{align}
where $\delta  g^i_{\bar{k}}= g^{i \bar{l}} \delta  g_{\bar{l} \bar{k}}$ controls the variation of the metric when expressed in the original coordinates. Notice that $\delta \Omega_3$, whose components are shown in eq. \eqref{eq:variation3form} above, defines some $(2,1)$-form. Furthermore, the Calabi--Yau condition $d\left(\Omega_3 + \delta \Omega_3\right)=0$ together with integrability of the deformed complex structure $\mathcal{I} + \delta \mathcal{I}$, requires that $\delta \Omega_3 \in H^{2,1}(X_3, \mathbb{R})$. Hence, one can parametrize the complex structure deformations as follows
\begin{align}\label{complexstructuredef}
 \i \Omega_{ijk}\delta g^k_{\bar l} = z^K (\chi_{K})_{ij\bar l}\, , \qquad K= 1, \ldots, h^{2,1} (X_3)\, ,
\end{align}
where $\{\chi_{K}\}$ denotes some basis of harmonic $(2,1)$-forms.

For future reference, it is also convenient to introduce a set of 3-forms $\{\alpha_K\}$ that together with their duals $\{\beta^K\}$ form a symplectic basis of $H^3(X_3, \mathbb{Z})$. They thus satisfy 
\begin{align}\label{eq:symplecticbasis}
	\int_{X_3} \alpha_K \wedge \beta^L =\delta_K^L\, . 
\end{align}

\subsubsection{4d $\mathcal{N}=2$ theories} \label{sss:4dN=2basics}

Let us start with theories preserving 8 supercharges in four spacetime dimensions, since they will be extensively used to check various ideas throughout this thesis. As already mentioned, they arise upon compactifying e.g., Type II string theory on a Calabi--Yau three-fold $X_3$. For future reference, we will specifically consider in what follows Type IIA compactifications.

As discussed in Section \ref{s:basicconcepts}, the low energy dynamics after the process of compactification is dominated by the massless fields, which are associated to the zero modes of the appropriate Laplace operator defined over the internal space. Therefore, to organize the 4d effective action we need to expand the different 10d Type IIA fields using some basis of harmonic forms, which yields \cite{Grimm:2005fa}
\begin{equation}\label{eq:IIAansatz}
\begin{aligned}
\phi(x,y)& =\phi(x)\, ,\quad B_2(x,y) = B_2(x) + b^a(x) \omega_a\, ,\\
C_1(x,y)&= \hat{A}^0(x)\, ,\qquad C_3(x,y)= C_3(x) + \hat{A}^a(x) \omega_a + \xi^I \alpha_I- \tilde{\xi}_J \beta^J \, .
\end{aligned}
\end{equation}
These should also be supplemented with the metric deformations of the Calabi--Yau manifold, c.f. eqs. \eqref{Kahlerdeform} and \eqref{complexstructuredef}. In 4d $\mathcal{N}=2$ language, the above fields arrange into several supermultiplets: a gravity multiplet, whose bosonic components correspond to the 4d metric as well as the graviphoton;\footnote{To be precise, the additional vector boson belonging to the gravity multiplet (usually referred to as the graviphoton) is not $\hat{A}^0(x)$ itself but rather an appropriate linear combination of all vectors in the vector multiplet sector\cite{Ceresole:1995ca}.} $h^{1,1}$ vector multiplets, each of them containing one vector from the set $\{ \hat{A}^a \}$ in \eqref{eq:IIAansatz} together with the complex scalars $z^a=b^a+{\rm i}t^a$ --- which describe the (complexified) K\"ahler sector of the theory; and finally $h^{2,1} + 1$ hypermultiplets, whose scalar components read 
\begin{center}
\renewcommand{\arraystretch}{2.00}
\begin{tabular}{r | c l}
universal hypermultiplet & $\phi\, ,\;\; \varrho\, ,\;\; \xi^0\, ,\;\; \tilde{\xi}_0$\\ 
\hline  
$h^{2,1}$ hypermultiplets & $z^i\, ,\;\; \xi^i\, ,\;\; \tilde{\xi}_i$\\
\end{tabular}\label{tab:4dhypers}
\end{center}  
%
where $z^i$ are local complex coordinates parametrizing the complex structure deformations and $\varrho$ is a compact scalar field dual to the NSNS 2-form $B_2$.

Therefore, upon inserting the ansatz \eqref{eq:IIAansatz} into \eqref{eq:IIA10dEinsteinframeaction}, one finds for the bosonic part of the 4d action \cite{Bodner:1990zm}
\begin{equation}\label{eq:IIAaction4d}
	\begin{aligned}
		\ S^{\text{4d}}_{\rm IIA}\, =\, & \frac{1}{2\kappa^2_4} \int \mathcal{R} \star 1 + \frac{1}{2} \text{Re}\, \mathcal{N}_{AB} F^A \wedge F^B + \frac{1}{2} \text{Im}\, \mathcal{N}_{AB} F^A \wedge \star F^B \\
		& - \frac{1}{\kappa^2_4} \int G_{a\bar b}\, d z^a\wedge \star d\bar z^b + h_{pq}\, d q^p \wedge \star d q^q\, ,
	\end{aligned}
\end{equation}
with $A=0,1, \ldots, h^{1,1}$. We collectively denote the scalars belonging to the various hypermultiplets by $q^p$, while the field strengths $F^B=dA^B$ correspond to $\mathsf{U(1)}$ gauge bosons having integrally-quantized charges, which are defined as
\begin{align}\label{eq:rotatedbasisvectorfields4d}
 A^0=\hat{A}^0\, , \qquad  A^a= \hat{A}^a+b^a \hat{A}^0\, .
\end{align} 
Furthermore, consistency with 4d $\mathcal{N}=2$ supersymmetry requires that the total moduli space $\mathcal{M}_{\text{mod}}$ splits --- at the two derivative level --- as follows \cite{Candelas:1990pi}\footnote{\label{fnote:modspaceprod}To get a product structure like \eqref{productmoduli} one might need to consider a multiple cover of ${\cal M}_{\rm VM}$ or ${\cal M}_{\rm HM}$ \cite{Seiberg:1996ns}.}
\begin{align}\label{productmoduli}
 \mathcal{M}_{\text{mod}} = \mathcal{M}_\text{VM} \times \mathcal{M}_\text{HM}\, ,
\end{align} 
where the first factor corresponds to the vector multiplets whilst the second denotes the hypermultiplet sector. In the following, we will discuss each of them in turn.
 
Regarding the vector multiplets, it is convenient to choose a basis \eqref{eq:h11forms} of integral 2-forms which is dual to a basis of Mori cone generators in $H_2(X_3,\mathbb{Z})$ \cite{Lanza:2021udy}. With this choice, one finds a metric for the scalars within the vector multiplets of the form \cite{Candelas:1990pi,Strominger:1985ks}:
\beq\label{eq:kahlersectormetric}
	G_{a\bar b}=\partial_a \partial_{\bar{b}} K_{\text{ks}}=\partial_a \partial_{\bar{b}} \left(- \log \frac{4}{3} \mathcal{K} \right) = \frac{3}{2} \left(\frac{3}{2}\frac{\cK_a\cK_b}{\cK^2}-\frac{\cK_{ab}}{\cK}\right)\, ,
\eeq
where $\frac{\mathcal{K}}{6}=\frac{1}{6} \mathcal{K}_{a b c}t^a t^b t^c = \mathcal{V}_{X_3}$ denotes the volume of the classical three-fold in string units, $K_{\text{ks}}$ is the K\"ahler potential and $\mathcal{K}_{a b c}$ are the triple intersection numbers of the Calabi--Yau $X_3$, given by
\begin{equation}
	\mathcal{K}_{a b c}=\int_{X_3}\omega_a\wedge\omega_b\wedge \omega_c\;.
\end{equation}
We have also defined the following useful contractions
\begin{align}
\cK_{ab}=\mathcal{K}_{a b c}t^c \, ,\qquad \cK_{a}=\mathcal{K}_{a b c}t^b t^c \, ,\qquad \cK=\mathcal{K}_{a b c}t^a t^b t^c\, ,
\end{align}
which are related to certain volumes of even-dimensional cycles within $X_3$. In addition, the vector multiplet moduli space describes some special K\"ahler manifold \cite{deWit:1984wbb}, such that one can introduce a set of local projective coordinates $X^A$ in terms of which the K\"ahler moduli read as
\begin{equation}\label{eq:projcoords}
	z^A = \frac{X^A}{X^0}\, .
\end{equation}
This also means, in practice, that its geometry is completely encoded into some holomorphic quantity $\mathcal{F} (X^A)$, dubbed the prepotential, which is moreover a homogeneous function of degree two, namely $\mathcal{F}= \frac{1}{2} X^A \mathcal{F}_A$, with $\mathcal{F}_A = \partial_{X^A} \mathcal{F}$. Classically, it has a simple cubic form
\begin{equation}\label{eq:classicalprepotential}
	\mathcal{F}_{\rm cl} = - \frac{1}{6} \frac{\mathcal{K}_{a b c} X^a X^b X^c}{X^0}\, .
\end{equation}

As a consequence of $\mathcal{N}=2$ supersymmetry, the gauge kinetic function $\mathcal{N}_{AB}$ is determined by the K\"ahler structure deformations via the relation 
\begin{equation}\label{eq:holomorphicgauge kineticfunction4d}
\begin{aligned}
\mathcal{N}_{AB}= \overline{\mathcal{F}}_{AB} + 2i \frac{(\text{Im}\, \mathcal{F})_{AC}X^C (\text{Im}\, \mathcal{F})_{BD} X^D}{X^C (\text{Im}\, \mathcal{F})_{CD} X^D}\, , 
\end{aligned}
\end{equation}
where $\mathcal{F}_{KL}=\partial_{X^K}\partial_{X^L} \mathcal{F}$. Substituting \eqref{eq:classicalprepotential} into eq. \eqref{eq:holomorphicgauge kineticfunction4d} one finds
\begin{equation}\label{eq:gaugetopologicalterm}
	\text{Re}\, \mathcal{N}\, = \, \left(
		\begin{array}{cc}
			-\frac{1}{3} \mathcal{K}_{abc}b^a b^b b^c & \frac{1}{2} \mathcal{K}_{abc}b^b b^c  \\
			\frac{1}{2}\mathcal{K}_{abc}b^b b^c & -\mathcal{K}_{abc} b^c  \\
		\end{array}
		\right) \, ,
\end{equation}
for the topological theta-like term and
\begin{equation}\label{eq:gaugekineticmatrix}
	\text{Im}\, \mathcal{N}\, = -\frac{\mathcal{K}}{6} \, \left(
		\begin{array}{cc}
			1+4G_{a\bar b}b^a b^b & -4G_{a\bar b} b^b  \\
			-4G_{a\bar b} b^b & 4G_{a\bar b}  \\
		\end{array}
		\right) \, ,\qquad 
		-(\text{Im}\, \mathcal{N})^{-1}\, = \frac{6}{\mathcal{K}} \, \left(
		\begin{array}{cc}
			1 &  b^a  \\
			b^a & \frac{1}{4} G^{a\bar b} + b^a b^b  \\
		\end{array}
		\right) \, .
\end{equation}
for the kinetic matrix (as well as its inverse).

Let us now turn to the second factor in \eqref{productmoduli}, namely the hypermultiplet moduli space. Geometrically, $\mathcal{M}_{\rm HM}$ describes a quaternionic-K\"ahler space of real dimension $4(h^{2,1} (X_3)+1)$ \cite{Bagger:1983tt}. These fields contain, in particular, the complex structure deformations, which are parametrized by complex coordinates, $z^I= \left(1, z^i \right)$, arising from the periods of the holomorphic $(3,0)$-form $\Omega_3$
\begin{equation}\label{eq:CSmoduli}
	Z^I (z^i) = \int_{X_3} \Omega_3 \wedge \beta^I\, , \quad \mathcal{F}_J (z^i) = \int_{X_3} \Omega_3 \wedge \alpha_J\, .
\end{equation}
as $z^I = \frac{Z^I}{Z^0}$. Classically, the sigma-model metric for this set of fields reads \cite{Ferrara:1989ik,Cecotti:1988zx}
\begin{align}\label{eq:classicalhypermetric}
	h_{p q}\, d q^p d q^q &= \left( \dd \varphi_4\right)^2 + G_{i \bar j} \dd z^i \dd z^{\bar j} + \frac{e^{4\varphi_4}}{4} \left( d \varrho - \left( \tilde{\xi}_J d\xi^J-\xi^J d\tilde{\xi}_J \right)\right)^2 \notag\\
	& -\frac{e^{2\varphi_4}}{2} \left( \text{Im}\, \mathcal{U}\right)^{-1\ IJ} \left( d\tilde{\xi}_I -\mathcal{U}_{IK} d\xi^K\right) \left( d\tilde{\xi}_J -\bar{\mathcal{U}}_{JL} d\xi^L\right)\, ,
\end{align}
where $e^{-2\varphi_4}= e^{-2\phi}\mathcal{V}_{X_3}$ is the 4d dilaton, $G_{i \bar j}$ denotes the metric on the space of complex structures \cite{Candelas:1990pi}
\begin{equation}\label{eq:CSmetric}
	G_{i\bar j} = \partial_{z^i}\partial_{\bar z^j} K_{\text{cs}} \, ,\qquad \text{with}\qquad K_{\text{cs}}=-\log \left[{\rm i} \int_{X_3}\Omega_3 \wedge \bar \Omega_3 \right]\, ,
\end{equation}
and $\mathcal{U}_{IJ} (z^i)$ is a complex matrix which is implicitly defined as follows
\begin{equation}\label{eq:hypermatrix}
\begin{aligned}
\int_{X_3} \alpha_K\wedge \star \alpha_L &= -\left(\text{Im}\, \mathcal{U} +(\text{Re}\, \mathcal{U})(\text{Im}\, \mathcal{U})^{-1}(\text{Re}\, \mathcal{U})\right)_{KL}\, ,\\
\int_{X_3} \beta^K\wedge \star \beta^L &=- \left(\text{Im}\, \mathcal{U}\right)^{-1\;KL}\, ,\\
\int_{X_3} \alpha_K\wedge \star \beta^L &=- \left((\text{Re}\, \mathcal{U})(\text{Im}\, \mathcal{U})^{-1}\right)^K_L\, . 
\end{aligned}
\end{equation}

Before closing this subsection, let us mention that the geometry of the 4d $\mathcal{N}=2$ moduli space so far described does receive both $\alpha'$ and $g_s$ corrections, which can further modify the classical expressions here presented. In particular, due to the splitting between the vector and hypermultiplet moduli spaces and the fact that the 4d dilaton belongs to the latter, one can argue that these two sectors receive quantum modifications which are very different in origin: $\alpha'$-corrections for the case of $\mathcal{M}_{\rm VM}$ and string $g_s$ corrections --- both perturbative and non-perturbative --- for $\mathcal{M}_{\rm HM}$. (See \cite{Cortes:2021vdm} for a recent description of the Type IIA quantum moduli space metric in a regime with mutually local instanton corrections).

\subsubsection{5d $\mathcal{N}=1$ theories}
\label{sss:5dMtheory}
	
Another interesting set of theories preserving 8 supercharges can be constructed in five dimensions upon compactifying M-theory on a Calabi--Yau three-fold instead of Type II string theory. In fact, they share with the latter many of the geometrical and topological data characterizing the effective theory, so that we will use the same notation as in the previous section.

The low energy EFT is described by 5d $\mathcal{N}=1$ supergravity, whose bosonic action reads as \cite{Cadavid:1995bk,Ferrara:1996hh,Ferrara:1996wv}
\begin{equation}\label{eq:Mthyaction5d}
	\begin{aligned}
		S^{\text{5d}}_{\text{M-th}}= \dfrac{1}{2\kappa^2_5} \int &  \mathcal{R} \star 1 -  G_{a b} (\tilde t)\left( d \tilde{t}^a\wedge \star d \tilde{t}^b  + F^a \wedge \star F^b \right) \\ 
		&- \frac{1}{6} \mathcal{K}_{a b c} A^a \wedge F^b \wedge F^c- 2 h_{pq}d  q^p \wedge \star d  q^q \, , 
	\end{aligned}
\end{equation}
where $ \tilde{t}^a=t^a/\mathcal{V}_5^{1/3}$, $a=1, \ldots, h^{1,1}$, are real scalars within the vector multiplets --- parametrizing the K\"ahler deformations, c.f. eq. \eqref{Kahlerexpand}; $F^a=dA^a$ denote the corresponding field strengths of the $\mathsf{U(1)}$ gauge bosons plus the graviphoton; and the scalars in the various hypermultiplets are represented by $q^p$. In our conventions $(2\kappa^2_5)^{-1}= 2 \pi/ \ell_{5}^3$, where $\ell_{5}$ denotes the five-dimensional Planck length.

The scalar fields belonging to the vector multiplets $\{ \tilde{t}^a\}$ are subject to the following non-linear constraint
\beq\label{eq:veryspecialgeometry}
	\mathcal{F}=\frac{1}{3!} \mathcal{K}_{a b c} \tilde{t}^a \tilde{t}^b \tilde{t}^c \stackrel{!}{=} 1\, ,
\eeq
where $\mathcal{K}_{a b c}$ are the triple intersection numbers of the three-fold, which also appear within the Chern-Simons coupling in \eqref{eq:Mthyaction5d}. In fact, the moduli $\{ \tilde{t}^a\}$ locally parametrize a manifold with \emph{very special geometry}, such that it is endowed with a metric (which also controls the gauge kinetic functions of the abelian gauge fields) given explicitly by
\beq\label{eq:5dVMmetric}
	G_{a b} (\tilde t)=\frac{1}{2 \mathcal{V}_5^{1/3}} \int_{X_3} \omega_a \wedge \star \omega_b=- \frac{1}{2} \left( \partial_{\tilde{t}^a} \partial_{\tilde{t}^a} \log \mathcal{F}\right)\bigg\rvert_{\mathcal{F}=1} \, .
\eeq
Interestingly, the vector multiplet piece of the 5d action \eqref{eq:Mthyaction5d} does not receive any quantum corrections whatsoever, in contrast to the Type IIA set-up discussed before. This is similar to the case of $K3$ compactifications of M-theory, and precisely the same argument ensures this classical exactness, see discussion after eq. \eqref{eq:7dmodspacemetric}.
		
On the other hand, in the hypermultiplet sector one obtains the following classical kinetic terms
\begin{align}\label{eq:classicalhypermetric5d}
		\nonumber h_{pq} dq^p \wedge \star dq^q =\, &\frac{1}{4\mathcal{V}_5^2}d\mathcal{V}_5\wedge \star d\mathcal{V}_5 + G_{l \bar{k}} dz^l \wedge \star d\bar z^k + \frac{1}{4} \mathcal{V}_5^2 dA_3\wedge \star dA_3 \\
		\nonumber&+ \frac{1}{4}\left(\xi^K d \tilde{\xi}_K- \tilde{\xi}_K d \xi^K\right) \wedge dA_3 \\
		&-\frac{1}{4\mathcal{V}_5} (\text{Im}\ \mathcal{U}^{-1})^{KL} \left(d \tilde{\xi}_K-\mathcal{U}_{KN}d \xi^N  \right)\wedge \star \left(d \tilde{\xi}_L-\mathcal{U}_{LM}d \xi^M  \right)\, ,
\end{align}
where $\mathcal{V}_5$ denotes the volume of the three-fold measured in 11d Planck units, the $\{z^l\}$, with $l=1, \ldots, h^{2,1}$ refer to the complex structure moduli and the rest of the fields arise from the remaining pieces in the reduction of the 11d 3-form that are not part of the $\mathsf{U(1)}$ gauge fields, namely
\beq \label{eq:3-formexpansion}
	C_3-A^a \wedge \omega_a= \xi^K \alpha_K - \tilde{\xi}_L \beta^L + A_3\, .
\eeq
Moreover, the matrix $\mathcal{U}^{KL}$ appearing in \eqref{eq:classicalhypermetric5d} depends explicitly on the complex structure of the $X_3$, and can be computed via eq. \eqref{eq:hypermatrix}. Notice that in order to have an actual quaternionic-K\"ahler metric in the hypermultiplet sector above it is necessary to first dualize the 3-form $A_3$ into a scalar, $\varrho$, which belongs to the so-called universal hypermultiplet, as so does $\mathcal{V}_5$ \cite{Cadavid:1995bk}, giving a total of $(h^{2,1}+1)$ hypermultiplets.

All in all, we deduce that the 5d theory comprises $(h^{1,1}-1)$ vector multiplets, $(h^{2,1}+1)$ hypermultiplets as well as one gravity multiplet. In fact, the similarity of the massless spectrum and the supergravity action between Type IIA and M-theory compactified on $X_3$ can be easily understood in terms of the (non-perturbative) duality explained in Section \ref{s:dualities} below.

\section{String dualities} \label{s:dualities}

One of the main discoveries during the second string revolution in the mid 1990's was the realization that all different superstring theories in ten dimensions, which had been defined perturbatively in an obviously independent manner, were in fact related to one another upon exploring different corners of their moduli spaces. This hinted towards the fact that the underlying physics was the same, which could then be in principle encapsulated into some unified description. In practice, however, these relations involve oftentimes non-perturbative stringy physics, such as the inclusion of D-branes or strong coupling effects, which prevents us from having a rigorous proof of the precise duality statements. However, the evidence gathered in favour of string dualities have been growing since their conception, giving us strong confidence that the main ideas behind are in fact correct. Furthermore, these duality symmetries offer not only a beautiful picture of the string theory landscape, but also serve as very powerful tools in order to analyze certain highly non-trivial aspects of the physics associated to a given theory. This becomes particularly useful when exploring corners of the moduli space which are way out of the regime of validity for the initial (perturbative) definition.

In this section we will briefly review some of the most important string dualities that have been proposed in the literature, placing more emphasis on those which will be used at several instances in the thesis.

\subsection{Duality symmetries with higher supersymmetry}\label{ss:dualitieswithhighersusy}

\subsubsection*{Type IIA/M-theory duality}

Let us start by considering the eleven-dimensional M-theory introduced in Section \ref{ss:Mthy11d} and compactify it on a circle. We will concentrate on the action functional that arises for the zero modes, corresponding to 10d massless fields. Hence, we impose the familiar ansatz for the 11d metric
\begin{align}\label{eq:MthyS1ansatz}
	ds^2_{11} = e^{-\rho/4} ds^2_{10}+ e^{2\rho} (dy-C_1)^2\, ,    
\end{align}
where $y \in [0, 2 \pi R_{11})$ parametrizes the circular direction, $\rho$ is the radion and $C_1$ denotes the KK 1-form. The reason for the choice of notation in \eqref{eq:MthyS1ansatz} will become clear in the following. Thus, upon dimensionally reducing the 11d bosonic action \eqref{eq:11daction} on $\mathbf{S}^1$, one finds
\begin{equation}
	\label{eq:3formddimgravity}
	\begin{aligned}
		S^{\text{10d}}_{\text{M-th}} = &-\frac{2 \pi R_{11}}{4\kappa_{11}^2} \int \dd^{10}x\sqrt{-g} \left(\mathcal{R}-\frac{9}{8}(\partial \rho)^2\right) - \frac{2 \pi R_{11}}{4\kappa_{11}^2} \int e^{\frac{9}{4}\rho}\, dC_1 \wedge \star dC_1\\
        &- \frac{2 \pi R_{11}}{4\kappa_{11}^2} \int e^{\frac{3}{4}\rho} \left( dC_3-C_1 \wedge dB_2 \right) \wedge \star \left( dC_3-C_1 \wedge dB_2 \right) + e^{-\frac{3}{2}\rho}\, dB_2 \wedge \star dB_2\\
		& - \frac{2 \pi R_{11}}{4\kappa_{11}^2} \int  B_2 \wedge dC_3 \wedge dC_3\, ,
	\end{aligned}
\end{equation}
where $B_2$, $C_3$ are 2-form and 3-form fields, respectively, that arise from the zero-mode components of the 11d 3-form potential with/without one leg in the 11-th direction. It is easy to see now that if we perform the following identifications
\begin{align}\label{eq:IIAMthyident}
    \rho=2 \phi/3\, ,\qquad \frac{2 \pi R_{11}}{2\kappa^2_{11}}\,=\,\frac{1}{2\kappa^2_{10}}\,=\, \frac{2 \pi}{g_s^2 \ell_s^8}\, , 
\end{align}
we immediately recover the 10d Type IIA supergravity written in the Einstein frame, c.f. eq. \eqref{eq:IIA10dEinsteinframeaction}. This formal similarity was conjectured in \cite{Witten:1995ex} to be a full-fledged duality statement
\begin{align} \label{IIAMduality}
\text{10d Type IIA} \qquad \stackrel{\text{dual}}{\longleftrightarrow} \qquad \text{M-theory on}\; \mathbf{S}^1\, . 
\end{align}
In particular, notice that the moduli dictionary in \eqref{eq:IIAMthyident} suggests that the strong coupling limit of Type IIA string theory can be understood as a decompactification limit in which the eleventh dimension grows large.  

In addition, the duality \eqref{IIAMduality} can also account for the matching between different BPS states in both theories. Thus, at this level, one is prompted to identify the D0-brane states with Kaluza-Klein replicas of the 11d gravity multiplet, since both couple to the massless 1-form $C_1$. Analogously, the fundamental Type IIA string arises from wrapping one direction of the M2-brane along the circle, whereas the D2-brane is simply the unwrapped version of the latter. Similar considerations lead us to conclude that the D4 and NS5-branes arise from the reduction of the 11d M5-brane with/without one leg along the circle, whereas the D6-brane becomes a KK-monopole \cite{Gross:1983hb,Sorkin:1983ns}.  

\subsubsection*{Type IIB S-duality}

Another interesting non-perturbative duality arises in the case of the Type IIB string and relates, among other things, the weak and strong coupling regimes of the theory. In fact, the 10d $\mathcal{N}=(2,0)$ supergravity is famously invariant under the non-compact group $\mathsf{SL (2,\mathbb{R})}$ \cite{Schwarz:1995dk}. This can be made manifest upon rewriting the 10d action \eqref{eq:IIB10d} in a slightly different way
\begin{equation}\label{eq:IIB10dSL2intro}
	\begin{aligned}
			S_\text{IIB}^{\text{10d}}\, =\, & \frac{1}{2\kappa_{10}^2} \int \dd^{10}x\sqrt{-g} \left(\mathcal{R}-\frac{\partial \tau \cdot \partial \bar \tau}{2 (\text{Im}\, \tau)^2}\right) -\frac{1}{4\kappa_{10}^2}\int \frac{1}{\text{Im}\, \tau} G_3\wedge \star \bar{G}_3 + \frac{1}{2} \tilde F_5 \wedge \star \tilde F_5\\
			&+\frac{1}{8 \text{i} \kappa_{10}^2}\int \frac{1}{\text{Im}\, \tau} C_4\wedge G_3 \wedge \bar{G}_3\, ,
	\end{aligned}
\end{equation}
where $\tau=C_0 + \text{i} e^{-\phi}$ is the axio-dilaton and $G_3=dC_2 - \tau dB_2$ defines a complex combination of the NSNS and RR 2-form gauge fields. The $\mathsf{SL (2,\mathbb{R})}$ group acts on the Type IIB variables as follows
\begin{align}\label{eq:Sdualityrules10dIIB}
	&\tau \rightarrow \frac{a\, \tau + b}{c\, \tau+d}\,,\qquad \begin{pmatrix}
		C_2\\ B_2
	\end{pmatrix}
	\rightarrow \mathcal{A} \begin{pmatrix}
		C_2\\ B_2
	\end{pmatrix}\, , \qquad \mathcal{A}= \begin{pmatrix}
		a \quad  b\\c \quad  d
	\end{pmatrix} \in \mathsf{SL (2,\mathbb{R})}\, ,
\end{align}
whilst both the (Einstein frame) metric and the RR 4-form are acted on trivially.  On the other hand, even though the above set of transformations leave the action \eqref{eq:IIB10dSL2intro} invariant, the full \emph{quantum} theory is only expected to preserve a discrete $\mathsf{SL (2,\mathbb{Z})}$ duality,\footnote{More appropriately, the duality group is $\mathsf{GL^+(2, \mathbb{Z})}$, i.e. the $\mathsf{Pin}^+$ double cover of $\mathsf{GL(2, \mathbb{Z})}$ \cite{Tachikawa:2018njr}. In fact, it has been seriously proposed by studying potential global anomalies associated to the latter, that there might be alternative UV completions that agree at the supergravity level with Type IIB string theory but could differ in its massive spectrum \cite{Debray:2021vob}.} since otherwise there would be some incompatibility with the Dirac quantization of fundamental and D1-string charges. In fact, the aforementioned symmetry reduction can be seen to explicitly arise in the effective action due to D$(-1)$-instanton effects, which couple to $C_0$ and thus require $\tau$ and $\tau + k$ to be identified, with $k \in \mathbb{Z}$ (see Section \ref{ss:10dIIB} for more on this).

At the level of the spectrum, S-duality exchanges fundamental and D1-strings, which couple to $B_2$ and $C_2$, respectively. More importantly, it requires the existence of an infinite number of $(p, q)$ bound states thereof \cite{Schwarz:1995dk}, where $p$ and $q$ are co-prime integers \cite{Witten:1995im} and which can be related to either fundamental or D1-branes via some $\mathsf{SL (2,\mathbb{Z})}$ transformation. Similarly, one may construct $(p, q)$ 7-branes, defined as hypersurfaces where $(p, q)$-strings can end, therefore inducing some $\mathsf{SL (2,\mathbb{Z})}$ monodromy of the form\cite{Weigand:2010wm}
\begin{align}
	\mathcal{A}_{p,q}=\begin{pmatrix}
		1-pq \quad  p^2\\-q^2 \quad  1-pq
	\end{pmatrix}\, .
\end{align}
Finally, D5-branes are left invariant under the duality group, since $C_4$ transforms as a singlet.

\subsubsection*{T-duality}

The previous dualities were non-perturtative in nature, since they manifested themselves in the strong coupling regime of Type II string theory. Still, they could be detected --- up to a certain point --- just by looking at the relevant low energy descriptions, i.e. the supergravity theories. In what follows, we will discuss a third string duality, which is usually referred to as T-duality, that involves perturbative physics of the string.\footnote{Of course, this does not mean that non-perturbative objects, such as D-branes, do not transform under the symmetry. In fact, T-duality both acts non-trivially on and is preserved by the non-perturbative setor as well, but it can be exhibited already at the perturbative level.} Nonetheless, as we will see, this symmetry is difficult to appreciate at the supergravity level, since it requires from stringy ingredients.

The simplest instance where T-duality effects can be made manifest arise when considering e.g., bosonic closed string theory on a $\mathbf{S}^1$ of radius $R$, measured in units of $\ell_s$. Denoting by $X^y = X^9$ the embedding coordinate of the string parametrizing the compact direction, we obtain for its associated oscillation modes the following expansion (c.f. eq. \eqref{eq:modeexpansionclosedstring})
\begin{align}
	X^y (\tau, \sigma)& = x^{y} + \alpha' \frac{k}{R} \tau + \omega R \sigma\ +\ \text{oscillators}\, ,
\end{align}
where the momentum $p^y$ is now quantized in units of the inverse radius and there is an additional quantum number $\omega \in \mathbb{Z}$ that accounts for the possible topologically non-trivial wrappings of the string along the $\mathbf{S}^1$, thus ensuring that $X^y (\tau, \sigma + 2\pi) = X^y (\tau, \sigma)$. From this perspective it is already evident that the spectrum of the theory remains invariant under the map
\begin{align}\label{eq:Tdualtransf}
R\rightarrow \frac{\alpha'}{R} \, ,\qquad \text{and}\quad k\leftrightarrow w\, . 
\end{align} 
In fact, the above transformation can be argued to be a full symmetry of the worldsheet theory as well, even at the interacting level \cite{Ginsparg:1986wr}. For instance, the energy-momentum tensor and other basic properties of the 2d SCFT are left invariant under such duality. In physical terms, this has the striking consequence that the closed string theory compactified on a circle with small radius is completely equivalent to the same theory at large radius provided that we exchange winding and momentum modes.  

When applied to the superstring case, a similar analysis leads to the conclusion that the two ten-dimensional Type II string theories are T-dual of one another. This becomes readily apparent already at the massless level, since both theories yield, when compactified on $\mathbf{S}^1$, the same supergravity description, which is unique in dimensions smaller than or equal to 9. Similarly, one can show that the $\mathsf{E8}\times \mathsf{E8}$ Heterotic string is T-dual to its $\mathsf{SO(32)}$ counterpart, thereby completing the web of dualities for the 10d superstrings. 

For completeness, let us show here explicitly how the T-duality map \eqref{eq:Tdualtransf}, when applied to the massless modes of the theory, looks like in simple circle reductions. The exact set of transformations go under the name of Buscher's rules \cite{Buscher:1987sk,Buscher:1987qj}, and read
\begin{equation}\label{eq:BuscherrulesNS}
			\begin{aligned}
				\tilde{g}_{yy} &= \frac{1}{g_{yy}}\, , \qquad e^{2\tilde{\phi}} = \frac{e^{2\phi}}{g_{yy}}\, , \qquad \tilde{g}_{\mu y} = \frac{B_{\mu y}}{g_{yy}}\, , \qquad \tilde{B}_{\mu y} = \frac{g_{\mu y}}{g_{yy}}\, ,\\
              \tilde{g}_{\mu \nu} &= g_{\mu \nu} - \frac{g_{\mu y} g_{\nu y}-B_{\mu y} B_{\nu y}}{g_{yy}}\, ,\\
               \tilde{B}_{\mu \nu} &= B_{\mu \nu} - \frac{B_{\mu y} g_{\nu y}-g_{\mu y} B_{\nu y}}{g_{yy}}\, ,
			\end{aligned}
\end{equation}
for the NSNS sector, whereas for the RR fields one finds instead \cite{Meessen:1998qm}
\begin{equation}\label{eq:BuscherrulesRR}
			\begin{aligned}
				\tilde{C}^{(p)}_{\mu_1 \ldots \mu_{p-1} y} &= C^{(p-1)}_{\mu_1 \ldots \mu_{p-1}} - (p-1) \frac{C^{(p-1)}_{[\mu_1 \ldots \mu_{p-2} | y} g_{|\mu_{p-1}] y}}{g_{yy}}\, ,\\
               \tilde{C}^{(p)}_{\mu_1 \ldots \mu_{p}} &= C^{(p+1)}_{\mu_1 \ldots \mu_{p} y} + p C^{(p-1)}_{[\mu_1 \ldots \mu_{p-1}} B_{|\mu_{p}] y}+p(p+1)\frac{C^{(p-1)}_{[\mu_1 \ldots \mu_{p-2}] y} B_{[\mu_{p-1} | y} g_{|\mu_{p}] y}}{g_{yy}}\, .
			\end{aligned}
\end{equation}

\subsubsection*{U-duality}

Interestingly, the T- and S-dualities previously discussed can be sometimes combined together so as to enlarge the duality group of the theory, also incorporating M-theory into the game \cite{Hull:1994ys}. In fact, it was precisely in the M-theoretic framework where this enhancement of the global symmetry group in maximal supergravity actions for $d \leq 8$ was first identified, see \cite{Cremmer:1978ds, Cremmer:1979up,deWit:1985np,deWit:1986mz} for the original references. 

To illustrate this point in a specific example, let us consider the map between M-theory compactified on $\mathbf{T}^3$ and Type IIB string theory on $\mathbf{T}^2$. The main reason to choose this particular set-up is because we will need the precise U-duality relations between the different equivalent frames later on in Chapter \ref{ch:Higherdimops}, when studying certain higher-curvature corrections in eight spacetime dimensions. Hence, we start from the 8d action as derived from M-theory, namely eq. \eqref{eq:8dalternativeaction}, which we repeat here for convenience
\begin{equation}\label{eq:8dap}
			\begin{aligned}
				S_\text{M-th}^{\text{8d}}\, &=\, \frac{1}{2\kappa_{8}^2} \int \dd^{8}x\sqrt{-g} \Bigg[\mathcal{R}-\frac{9}{14} \left( \partial \log \mathcal{V}_2\right)^2 - \frac{7}{6} \left( \partial \log R_3\right)^2 -\frac{\partial \tau \cdot \partial \bar \tau}{2 \tau_2^2}\\
                &- \frac{\mathcal{V}_2^{-12/7} R_3^{-2}}{2} \left( \partial C_{123}^{(3)}\right)^2 -\frac{\mathcal{V}_2^{9/7} R_3^{-2}}{2 \tau_2} \left| \partial \left( \frac{\text{Im}\, (\tau \bar{\xi}_{\text{M}})}{\tau_2}\right) + \tau\, \partial \left( \frac{\text{Im}\, (\xi_{\text{M}})}{\tau_2}\right)\right|^2\Bigg]\, ,
			\end{aligned}
\end{equation}
where we have defined $\xi_{\text{M}}=-A_0^{1}+\text{i} A_0^{2} \tau_2$, with $\lbrace A_0^{1}, A_0^{2} \rbrace$ being compact scalar fields parametrizing the orientation of the two-dimensional torus within the overall $\mathbf{T}^3$. To relate this with Type IIA string theory we first use the M-theory/Type IIA duality by taking the circle with radius $R_3$ to provide for the M-theory circle. Upon carefully doing so (c.f. eq. \eqref{eq:IIAMthyident}), one finds the following moduli identifications
\begin{center}
\renewcommand{\arraystretch}{2.00}
\begin{tabular}{r c l}
M-theory on $\mathbf{T}^3$ & $\longleftrightarrow $ & Type IIA on $\mathbf{T}^2$\\ 
\hline
\hline  
$\mathcal{T}$ & $ \longleftrightarrow$ & $T=b+\text{i} T_2$ \\
$\tau$ & $\longleftrightarrow $&  $U=U_1+\text{i} U_2$\\
$\mathcal{V}_2^{9/7}\, R_3^{-2}$ & $\longleftrightarrow $ & $e^{-2\varphi_8}=e^{-2\phi}\, T_2$
\end{tabular}\label{tab:8dMthy/IIA}
\end{center}  
The variables shown on the right hand side correspond to the Type IIA moduli, which consist on the usual (complexified) K\"ahler modulus $T$ --- whose imaginary part controls the volume of the Type IIA torus in string units, the complex structure modulus $U$ as well as $\varphi_8$, which is the 8d dilaton. Hence, after performing the above field transformation one arrives at an action functional of the form
\begin{equation}\label{eq:8dIIA}
			\begin{aligned}
				S_\text{IIA}^{\text{8d}}\, &=\, \frac{1}{2\kappa_{8}^2} \int \dd^{8}x\sqrt{-g} \Bigg[\mathcal{R} - \frac{2}{3} \left( \partial \varphi_8\right)^2-\frac{\partial T \cdot \partial \bar T}{2 T_2^2}  -\frac{\partial U \cdot \partial \bar U}{2 U_2^2}\\
                & -\frac{e^{-2\varphi_8}}{2 U_2} \left| \partial \left( \frac{\text{Im}\, (U \bar{\xi}_{\text{A}})}{U_2}\right) + U\, \partial \left( \frac{\text{Im}\, (\xi_{\text{A}})}{U_2}\right)\right|^2\Bigg]\, ,
			\end{aligned}
\end{equation}
with $\xi_{\text{A}}=-C_0^{1}+\text{i} C_0^{2} U_2$. The scalars $\lbrace C_0^{1},C_0^{2} \rbrace$ now arise from the reduction of the RR 1-form $C_1$ on any of the two 1-cycles within the (dual) $\mathbf{T}^2$.

On a next step, we can perform a T-duality along any 1-cycle within the torus, thus relating the two Type II string theories in eight dimensions. Therefore, as may be easily checked, the Buscher's rules given in eqs. \eqref{eq:BuscherrulesNS} and \eqref{eq:BuscherrulesRR} translate into the exchange of K\"ahler and complex structure moduli --- i.e. $T \leftrightarrow U$, whilst the 8d dilaton is left unchanged. Additionally, one finds for the complex $\xi_{\text{B}}$ field the following new expression
\begin{equation}
    \xi_{\text{B}}= -b+\text{i} \tau_1\, T_2\, ,
\end{equation}
where $\tau_1=C_0$ corresponds to the Type IIB RR 0-form. This yields an action for the scalar-tensor sector of the theory that reads as follows
\begin{equation}\label{eq:8dIIB}
			\begin{aligned}
				S_\text{IIB}^{\text{8d}}\, &=\, \frac{1}{2\kappa_{8}^2} \int \dd^{8}x\sqrt{-g} \Bigg[\mathcal{R} - \frac{2}{3} \left( \partial \varphi_8\right)^2 -\frac{\partial U \cdot \partial \bar U}{2 U_2^2}-\frac{\partial T \cdot \partial \bar T}{2 T_2^2}\\
                & -\frac{e^{-2\varphi_8}}{2 T_2} \left| \partial \left( \frac{\text{Im}\, (T \bar{\xi}_{\text{B}})}{T_2}\right) + T\, \partial \left( \frac{\text{Im}\, (\xi_{\text{B}})}{T_2}\right)\right|^2\Bigg]\, .
			\end{aligned}
\end{equation}

At the level of the spectrum, one can easily deduce how the different states get exchanged upon performing the chain of dualities and combining our knowledge gained from previous discussions. One of the most interesting facts about U-duality is precisely that it reshuffles both non-perturbative and perturbative massive states among each other, since it mixes both kind of string dualities. We will have more to say about these issues in Part \ref{part:pattern} of this thesis.

\subsubsection*{Completing the duality web in ten dimensions}

\begin{figure}[htb]
\begin{center}
\includegraphics[width=0.6\textwidth]{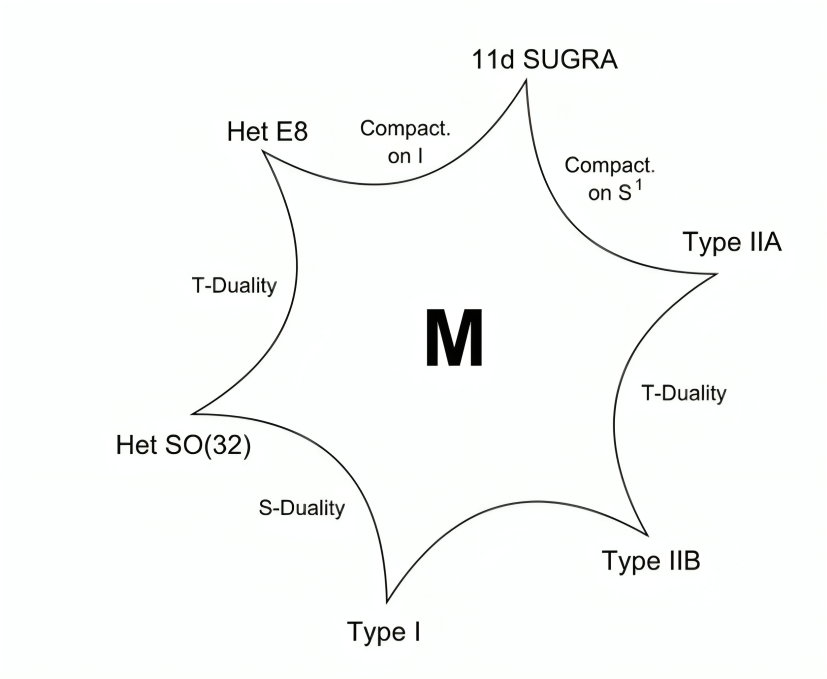}
\caption{\small Schematic depiction of the web of dualities relating the different superstring theories in ten dimensions among each other, as well as to 11d M-theory.}
\label{fig:dualitystar}
\end{center}
\end{figure}

Let us finally mention that as beautifully explained in \cite{Witten:1995ex}, it is indeed possible to retrieve all the existing 10d superstring theories upon considering 11d M-theory reduced on different compact backgrounds, after taking suitable limits in moduli space. For instance, by compactifying M-theory on an interval $\mathbf{I} \cong \mathbf{S}^1/\mathbb{Z}_2$ we break half of the original supersymmetries, such that accounting for gauge and gravitational anomalies in the resulting 10d $\mathcal{N}=1$ theory as well, ones arrives at the $\mathsf{E8}\times \mathsf{E8}$ Heterotic string \cite{Horava:1995qa,Horava:1996ma}. Interestingly, from the 11d perspective, each of the $\mathsf{E8}$ groups is realized at one boundary of spacetime (i.e. the fix points of the $\mathbb{Z}_2$ orbifold), and the fundamental Heterotic string arises from M2-branes stretched between the two Ho\v rava-Witten walls. The full duality map in ten dimensions is shown in Figure \ref{fig:dualitystar} for completeness.

\subsection{Duality symmetries with lower supersymmetry}\label{ss:dualitieswithlowersusy}

As it was early realized, the existence of string dualities is actually a very generic phenomenon which is not tied to the high level of supersymmetry of our starting superstring theories in ten dimensions. In fact, the duality web described in the previous section rapidly grows both in broadness and complexity once we start considering non-trivial compactifications to lower spacetime dimensions. 

Here we will only comment on those symmetries which persist (or arise) when the compactification process breaks some of the original supersymmetries of the theory. In particular, and with an eye to future applications, we will pay special attention to string dualities that appear in 4d $\mathcal{N}=2$ settings.
 
\subsubsection*{Type IIA/Heterotic duality}

The first duality in lower dimensions that we want to discuss here involves the Type II and the $\mathsf{E8}\times \mathsf{E8}$ Heterotic strings. However, it is useful to argue for this using an intermediate relation between M-theory and Heterotic string theory in one dimension higher, since this will also appear at several instances of the thesis. 

Let us thus consider M-theory compactified on a $K3$ surface. As already discussed in Section \ref{sss:MtheoryonK3}, this leads to a seven-dimensional theory preserving 16 supercharges, whose moduli space is classically exact and is moreover described, in general, by the space of Ricci-flat metrics on $K3$. The latter exhibits a group coset structure of the form
\begin{align}\label{eq:modspaceK3surfaces}
	\mathcal{M}_{\text{K3}} = \mathsf{O(\Gamma_{3,19})}\backslash \mathsf{O(3,19)} / (\mathsf{O(3)} \times \mathsf{O(19))} \times \mathbb{R}_+\, ,
\end{align}
where the precise meaning of the different mathematical objects involved is discussed around eq. \eqref{eq:cosetspace7d}, and the $\mathbb{R}_+$ factor accounts for the overall $K3$ volume --- which belongs to the gravity multiplet. Crucially, the exact same moduli space arises when compactifying the Heterotic string on a flat 3-torus. This can be easily seen by recalling that the (Narain) moduli space of string theory compactifications on $k$-dimensional tori $\mathbf{T}^k$, when accounting for the full set of T-duality transformations, is indeed isomorphic to \cite{Narain:1985jj,Narain:1986am}
\begin{align}
	\mathcal{M}_{\mathbf{T}^k} = \mathsf{O(\Gamma_{k,k})}\backslash \mathsf{O(k,k)} / (\mathsf{O(k)} \times \mathsf{O(k))} \times \mathbb{R}_+\, ,
\end{align}
where the extra factor $\mathbb{R}_+$ is now associated to the 7d string coupling constant $g_7$. Hence, since the Heterotic string (in its bosonized description) includes 16 additional directions in e.g., the right-moving sector --- parametrizing a 16d compact torus with fixed radius of the order of the string scale, the resulting moduli space for $k=3$ indeed matches the one shown in \eqref{eq:modspaceK3surfaces}. Furthermore, taking into account that the graviton-dilaton piece of the Heterotic string action reads as
\begin{equation}
			\begin{aligned}
				S_\text{Het}^{\text{7d}} \, \supset\, &\frac{1}{2\kappa_{7}^2} \int \text{d}^{10}x\sqrt{-g} \left(\mathcal{R}-\frac{4}{5}(\partial \varphi_7)^2\right)\, , 
			\end{aligned}
\end{equation}
and upon comparing with \eqref{eq:7dMthyK3}, we deduce the following moduli identification
\begin{align}
\label{eq:mthyHetmoduli}
 g_7 = \mathcal{V}_{K3}^{3/4}\, , 
\end{align}
which relates the Heterotic string coupling constant with the overall internal volume in the M-theory frame. This means, in turn, that the small $K3$ limit, as seen from M-theory, would correspond to a weak coupling regime for a dual Heterotic string, which arises from a solitonic M5-brane wrapping the entire $K3$ surface,\footnote{Note that the tension of both objects agree exactly upon using the identification \eqref{eq:mthyHetmoduli}, namely
\begin{align}
 \notag \frac{T_{\text{M5, str}}}{M_{\text{Pl};\, 7}^2} = (4\pi)^{-2/5}\, \mathcal{V}_{K3}^{3/5} = (4\pi)^{-2/5}\,  g_7^{4/5} = \frac{T_{\text{het}}}{M_{\text{Pl};\, 7}^2}\, . 
\end{align}} whereas the strong $g_7$ limit (from the Heterotic point of view), induces some full decompactification to 11d M-theory instead. All this suggests that in fact both theories might be S-dual to each other, and indeed there exist by now multiple non-trivial checks of the proposed duality
\begin{align}\label{eq:duality7dMthHet}
\text{M-theory on}\; K3 \qquad \stackrel{\text{dual}}{\longleftrightarrow} \qquad \text{Heterotic string on}\; \mathbf{T}^3\, , 
\end{align}
see e.g., \cite{Witten:1995ex,Cherkis:1997bx,Park_2009} for an incomplete list of references.

Interestingly, one can propagate the above relation to theories living in lower dimensions and with less amount of supersymmetry as well. Doing so requires from considering M-theory on e.g., some $K3$-fibered Calabi--Yau thereefold, where the $K3$ is adiabatically fibered over a rational $\mathbb{P}^1$ curve, which should be much larger than the fibre itself. This allows us to perform the duality \eqref{eq:duality7dMthHet} \emph{fiberwise}, thus yielding the following relation between 5d $\mathcal{N}=1$ theories
\begin{align}\label{eq:duality5dMthHet}
\text{M-theory on}\; X_3 \cong K3\rightarrow \mathbb{P}^1 \qquad \stackrel{\text{dual}}{\longleftrightarrow} \qquad \text{Heterotic string on}\; K3 \times \mathbf{S}^1\, , 
\end{align}
where the internal space in the Heterotic side of the duality arises by fibering a $\mathbf{T}^2$ inside the $\mathbf{T}^3$ over the $\mathbb{P}^1$ base, which famously gives rise to a $K3$ surface.\footnote{Indeed, the K3 two-fold is known to have a locus on its moduli space where it arises as an elliptic fibration over a one-fold base $\mathbb{P}^1$ \cite{Weigand:2010wm}.}

Finally, let us mention that upon further compactification on $\mathbf{S}^1$, and by employing M-theory/Type IIA duality (c.f. eq. \eqref{IIAMduality}), one can argue for the existence of the following third S-dual relation \cite{Hull:1994ys,Ferrara:1995yx,Kachru:1995wm} (see \cite{Aspinwall:1996mn} for a comprehensive review)
\begin{align}\label{eq:IIA/HETduality4d}
\text{Type IIA on}\; X_3 \cong K3\rightarrow \mathbb{P}^1 \qquad \stackrel{\text{dual}}{\longleftrightarrow} \qquad \text{Heterotic string on}\; K3\times \mathbf{T}^2\, , 
\end{align}
where now it is the limit of large $\mathbb{P}^1$ volume (on the Type IIA side) the one corresponding to a weak coupling point for the dual Heterotic string, i.e. $g_4\rightarrow 0$. We will heavily make use of the above chain of dualities when discussing several important aspects studied in this thesis, see Chapters \ref{ch:Higherdimops}, \ref{ch:Emergence} and \ref{ch:pattern}.

\subsubsection*{M-/F-theory duality}

Let us discuss now one of the most interesting string dualities that have been derived so far. It relates in a highly non-trivial way the physics of M-theory with a non-perturbative description of Type IIB string theory (denoted F-theory) --- i.e. including D-branes, and in fact it is precisely via this duality how many phenomenologically viable scenarios have been constructed up to date within string theory, which are close to the Standard Model of Particle Physics (see e.g., \cite{Cvetic:2022fnv,Marchesano:2022qbx,Marchesano:2024gul} and references therein).

The crucial realization that motivated the development of F-theory was the intimate relation between the $\mathsf{SL(2,\mathbb{Z})}$ self-duality of Type IIB string theory (see discussion around eq. \eqref{eq:Sdualityrules10dIIB}) and the group of large diffeomorphisms of the torus, i.e. the modular transformations. The latter act solely on the complex structure of $\mathbf{T}^2$, such that one useful and very geometrical way to think about Type IIB string theory is to `embed' the description into some twelve-dimensional theory where two compact dimensions define some elliptic curve $\mathcal{C}$, with frozen volume and varying complex structure, as determined by the axio-dilaton profile $\tau (x)$ \cite{Vafa:1996xn}. However, the crucial insight is that this a priori auxiliary identification, does indeed have physical meaning via its connection to 11d M-theory, as we explain next.

Therefore, in order to establish the correspondence between F-theory and M-theory (or rather its 11d low-energy supergravity limit), we start by recalling that M-theory compactified on a circle $\mathbf{S}^1_a$ reduces to Type IIA string theory in the limit where the radius $R_a$ of the circle vanishes, namely when $R_a\rightarrow 0$. If we further reduce on an additional circle $\mathbf{S}^1_b$, one effectively arrives at a $\mathbf{T}^2$ compactification of M-theory, which after taking $R_a\rightarrow 0$, becomes Type IIA on $\mathbf{S}^1_b$. This is entirely equivalent to Type IIB on the T-dual circle $\overline{\mathbf{S}}^1_b$, with radius $\overline R_b =\alpha'/R_b$, as explained in Section \ref{ss:dualitieswithhighersusy}. Hence, by additionally approaching the small $R_b$ regime, we essentially decompactify $\overline R_b$ in Type IIB dual frame. Thus, we conclude that M-theory on $\mathbf{T}^2$ reduces to Type IIB string theory in the limit of vanishing torus volume, $\mathcal{V}_{\mathbf{T}^2}$, whereas the complex structure $\tau$ is left untouched. Additionally, one can see that the Kaluza-Klein tower associated to the decompactifying circle in the Type IIB frame corresponds, from the M-theory perspective, to M2-branes wrapping the whole torus-fibre $n \in \mathbb{Z}$ times, whose mass
\begin{align}\label{eq:massM2branes}
			m_{\text{M2}}\,  (n \mathbf{T}^2) = \left | \int_{n \mathbf{T}^2} J \right |= |n\, \mathcal{V}_{\mathbf{T}^2}|= \left |\frac{n}{\overline R_b} \right |\, ,    
\end{align}
goes then to zero in the limit $\overline R_b \to \infty$.
		
More generally, let us consider M-theory compactified on an elliptically-fibered $n$-fold $X_{n}$, over some $(n-1)$-fold base $C_{n-1}$, which we denote as follows
\begin{equation}\label{eq:fibration}
			\begin{aligned}
				\pi: \qquad \mathbf{T}^2 \hookrightarrow &\;X_{n} \\
				&\;\; \downarrow \qquad . \\ &\;C_{n-1}
			\end{aligned}
\end{equation}
One can then easily see that in order to preserve some unbroken supersymmetry in the low energy EFT, the $n$-fold needs to have at least vanishing first Chern class, i.e. it has to be Calabi--Yau  \cite{Denef:2008wq, Weigand:2010wm, Weigand:2018rez, Cvetic:2018bni}. This results in a supersymmetric EFT living in $\mathbb{R}^{1,10-2n}$. Performing now adiabatically the duality discussed above, namely by taking the limit of zero fibre volume from the M-theory perspective, one obtains Type IIB string theory compactified on $C_{n-1}$, possibly with some non-perturbative defects (i.e. 7-branes) wrapping certain cycles of the internal geometry (given by the singular loci of the torus fibration) \cite{Morrison:1996na,Morrison:1996pp}.

\subsubsection*{Mirror symmetry}

Finally, we turn to a fascinating property exhibited by certain 4d $\mathcal{N}=2$ theories, which is inherited from the simple T-duality relation described in Section \ref{ss:dualitieswithhighersusy} above. The proposal is to identify Type IIA and Type IIB string theories compactified on mirror (dual) Calabi--Yau manifolds. Let us briefly elaborate on this point. 

In fact, when discussing the low energy effective field descriptions arising from the aforementioned compactified theories (c.f. Section \ref{s:4dN=2}) it becomes readily apparent that they are essentially the same up to an exchange of the (extended) K\"ahler and complex structures (see e.g., \cite{Grimm:2005fa} for a review). This led originally \cite{Candelas:1989hd, Greene:1989cf,Greene:1990ud,Aspinwall:1990xe,Candelas:1990rm,Aspinwall_1994,kontsevich1994homological} to propose that (fully-fledged) Type IIA compactified on a CY three-fold $X_3$ is equivalent to (fully-fledged) Type IIB reduced on a different three-fold $Y_3$, satisfying 
\begin{align}
 h^{1,1}(X_3) = h^{2,1}(Y_3)\, ,\qquad h^{2,1} (X_3) = h^{1,1}(Y_3)\, ,
\end{align}
where the manifold $Y_3$ is usually referred to as the mirror of $X_3$. This implies, in turn, a
map between even-dimensional and middle-dimensional cohomologies on $X_3$ and $Y_3$, $H^{2p}(X_3) \leftrightarrow H^3(Y_3)$, with $p=0, \ldots,3$, and consequently, an analogous relation between integral, symplectic bases of the aforemetioned spaces
\begin{align}
\mathcal{C} \in H_{2p}(X_3) \longleftrightarrow \gamma \in H_{3}(Y_3)\, . 
\end{align}
The duality exchanging Type IIA on $X_3$ and Type IIB on $Y_3$ is referred to as \textit{(quantum) mirror symmetry},\footnote{The word quantum refers to the assertion that the full quantum string theories should be taken to be equivalent, which goes beyond the usual statement of the equivalence between the associated perturbative CFTs.} see \cite{Hori:2003ic,Hosono:1994av} for a detailed review on the topic. Therefore, applied to Type II string theories, this duality exchanges
\begin{align}\label{eq:mapmirrorsymmetry}
\mathcal{M}_{\text{VM}}^\text{IIA} \longleftrightarrow \mathcal{M}_{\text{VM}}^\text{IIB} \qquad \text{and} \qquad \mathcal{M}_{\text{HM}}^\text{IIA} \longleftrightarrow \mathcal{M}_{\text{HM}}^\text{IIB}\, . 
\end{align}
Notice that this implies that we can identify the \emph{quantum-corrected} K\"ahler moduli space $\mathcal{M}_{\text{QK}}$ pertaining to some three-fold $X_3$, with the \emph{classical} moduli space, $\mathcal{M}_{\text{CS}}$, of complex structures of the mirror manifold $Y_3$. We can hence identify e.g., the complexified K\"ahler moduli $\{z^a = b^a + \i t^a\}$ of Type IIA compactified on $X_3$ with the periods $\{Z^A\}$, used to describe the complex structure moduli space of Type IIB on $Y_3$, as follows
\begin{align}\label{mirrormap}
 z^a = \frac{Z^a}{Z^0}\, .
\end{align}
This is known as the mirror map. Notice, however, that Mirror Symmetry is \textit{a priori} defined in the large volume limit of $X_3$, i.e. $z^a \rightarrow \i \infty$, corresponding to the large complex structure limit of $Y_3$, i.e. $Z^a \rightarrow \i \infty$. This is due to the fact that the vector multiplet moduli space of Type IIA string theory receives $\alpha'$-corrections which are suppressed in this limit. On the other hand, the vector multiplet moduli space of Type IIB does not receive corrections of any sort and thus can be described in purely classical terms. Since the above duality states that the two moduli spaces need to be the same, the calculation of the periods $\{Z^A, \mathcal{F}_A\}$ of $Y_3$ can be actually used so as to infer $\alpha'$-corrections to the IIA moduli space, in particular worldsheet instanton contributions, see below. 

Close to the large complex structure point (LCS) of $Y_3$, one can find a basis of periods such that they split into a \textit{unique} power series plus single logs
\begin{align}\label{largecomplexperiods}
 Z^a = \frac{1}{2\pi \i } \log y^a + \mathcal{O}(y^a) \, ,\qquad Z^0 = 1 + \mathcal{O}(y^a)\, ,
\end{align}
with the large complex structure point corresponding to $y^a=0$. In practice it is a non-trivial task to find the correct basis of periods with this leading logarithmic behaviour, which is commonly referred to as integral basis. --- due to its properties under the (maximal unipotent) monodromies induced transforming the $y^a$-hyperplane as $y^a\rightarrow e^{2\pi \i }y^a$.

With this (rather special) basis one can now define inhomogeneous, flat coordinates on $\mathcal{M}_{\text{CS}}$ as follows 
\begin{align}\label{eq:coordinates}
 \tau^a = \frac{Z^a}{Z^0}=\frac{1}{2\pi \i } \log y^a + \mathcal{O}(y^a)\, ,
\end{align}
which coincide near $y=0$ with the classical complexified K\"ahler parameters $\tau^a (y \to 0) = b^a + \i t^a$ near the large volume limit of $X_3$. Away from $y=0$, the $\tau^a$ then define, via analytic continuation, the (multi-valued) coordinates over the full quantum K\"ahler moduli space, $\mathcal{M}_{\text{QK}}$.

From the periods \eqref{largecomplexperiods} one can then infer the prepotential $\mathcal{F}= (Z^{0})^2 F (z^a)$ that underlies the special K\"ahler geometry of the vector multiplet moduli space of Type IIA compactified on $X_3$ which takes the form
\begin{align}\label{prepotentialIIA}
F =-\frac{1}{6} \cK_{abc} z^a z^b z^c + K_{ab}^{(1)} z^a z^b + K_{a}^{(2)} z^a + K^{(3)} - \frac{1}{(2 \pi \i)^3}\sum_{\boldsymbol{k}\geq \boldsymbol{0}} n_{\boldsymbol{k}}^{(0)} \sum_{m \geq 1} \frac{1}{m^3} e^{2\pi \i m k_a z^a}\, , 
\end{align} 
where in terms of a dual basis $\lbrace \omega_a \rbrace$ of $H^2(X_3, \mathbb{Z})$ we have
\begin{align}
 \cK_{abc}= \omega_a \cdot \omega_b \cdot \omega_c \, ,\qquad K_{a}^{(2)}=\frac{1}{24} c_2(X_3) \cdot \omega_a\, ,\qquad K^{(3)}= \frac{\i \zeta(3)}{2(2\pi)^3} \chi_E(X_3)\, .
\end{align}
The $K_{ab}^{(1)}$ are in general not specified geometrically but can be determined (up to monodromy \cite{deWit:1992wf,Harvey:1995fq}) by requiring good symplectic transformation properties of the period vector
\begin{align}
    \Pi(z) &= \begin{pmatrix}
           Z^{0} \\
           Z^{a} \\
           \partial_a \mathcal{F} \\
           \partial_0 \mathcal{F}
         \end{pmatrix}=Z^{0} \begin{pmatrix}
           1 \\
           z^{a} \\
           \partial_a F \\
           2F- z^a \partial_a F
         \end{pmatrix}\, .
  \end{align}
Note that already the perturbative $\alpha'$-corrections in \eqref{prepotentialIIA} break, in general, the no-scale structure of the classical K\"ahler potential (c.f. eq. \eqref{eq:kahlersectormetric}). Thus, the term given by $K^{(3)}$ encodes contributions which descend from $\alpha'^3 \mathcal{R}^4$ curvature corrections already present in the 10d supergravity action (see e.g., \cite{Green:1999pv, Palti:2008mg,Grimm:2017okk}), and it turns out to be the only effective perturbative contribution to the K\"ahler potential $K_{\rm ks}$. As opposed to this, the terms $K_{ab}^{(1)}$ and $K_{a}^{(2)}$ correspond respectively to one-loop and two-loop corrections in $\alpha'$, yet do not have a ten-dimensional counterpart due to the lack of a ten-dimensional curvature polynomial with the appropriate features. Their presence is however physically irrelevant at the level of the K\"ahler metrics\cite{Escobar:2018rna},\footnote{However, both $K_{ab}^{(1)}$ and $K_{a}^{(2)}$ do appear in the 4d effective action \eqref{eq:IIAaction4d} through the kinetic and topological terms for the $\mathsf{U(1)}$ gauge fields determined by eq. \eqref{eq:holomorphicgauge kineticfunction4d} (see e.g., Appendix B of \cite{Marchesano:2022axe}).} as confirmed by their absence in the K\"ahler potential that results from \eqref{prepotentialIIA}
\begin{align}\label{eq:alpha'correctedKahlerpot}
 K_{\rm ks}=- \log \left(\frac{4}{3}\mathcal{K} +4 \i K^{(3)}\right)\, .
\end{align}
On the other hand, the non-perturbative terms in \eqref{prepotentialIIA} correspond to worldsheet instanton contributions to the prepotential, and can be deduced from the corrections to the periods that are polynomial in the $y^a$ (c.f. eq. \eqref{eq:coordinates}). There, $\boldsymbol{k}$ denotes a vector of length $h^{1,1}(X_3)$ that scans over positive homology classes $k_a\gamma^a \in H_2^+(X_3,\mathbb{Z})$. The coefficients $n_{\boldsymbol{k}}^{(0)}$ are known as genus-zero Gopakumar--Vafa invariants and count the BPS degeneracy of (bound states of) D2-branes wrapped on cycles in the homology class $k_a \gamma^a$\cite{Gopakumar:1998ii, Gopakumar:1998jq}.

Finally, as promised at the beginning of this subsection, let us stress that a useful way to think about Mirror Symmetry is via some particular chain of T-dualities \cite{Strominger:1996it}. In fact, this can be shown very explicitly in toroidal orbifold models, where upon explicitly performing three T-dualities via the familiar Bushcer rules (c.f. eqs. \eqref{eq:BuscherrulesNS}-\eqref{eq:BuscherrulesRR}) one arrives precisely at the map \eqref{eq:mapmirrorsymmetry}, see e.g., \cite{Ibanez:2012zz} for details. 

\section{The Swampland program} \label{s:SwamplandProgram}

In previous sections we have discussed various possibilities for low energy effective field theories that string or M-theory allows us to construct. These may differ in the number of non-compact spacetime dimensions --- depending on the internal manifold we place our theory on, the type and amount of gauge symmetries exhibited by the effective description (i.e. supersymmetry, non-Abelian interactions, etc.), or even the matter content. This diversity creates an enormous set of physically distinct vacua, which are collectively known as the \emph{String Landscape} \cite{Susskind:2003kw}. Crucially, all these theories share the common feature that can be embedded in string theory, and thus are compatible (by construction) with an ultra-violet completion of gravity. 

Historically though, the mere existence of this rich structure of vacua was considered to be detrimental for the theory, since it suggested that essentially every semi-classically consistent EFT that one may think of could be found within some corner of the String Landscape. Hence, trying to understand any physics beyond the Standard Model of Particle Physics using a top-down approach was regarded to be meaningless, since the theory did not seem to have any predictive power at all. However, as pointed out in \cite{Vafa:2005ui}, it turns out that coupling generic effective field theories to gravity can introduce certain inconsistencies --- which go beyond the usual gravitational anomaly analysis \cite{Alvarez-Gaume:1983ihn} --- that are not visible form the field theory prism. In fact, it is the aim of the \emph{Swampland Program} \cite{Vafa:2005ui} to delineate the boundary between the set of EFTs that can be derived from quantum gravity/string theory (thereby belonging to the Landscape) and those that are not consistent with gravitational interactions at a deeper level (referred to as the Swampland). 

The pursuit of distinguishing these two sets of effective field theories involves establishing definite criteria that any consistent EFT must satisfy in order to belong to the Landscape. In this regard, even though string theory plays a prominent role for uncovering universal constraints in quantum gravity, the idea of the Swampland can be formulated regardless of the former. In fact, heuristic arguments based on black hole physics can also contribute significantly to this discourse, providing insights which should be a priori independent of any underlying microscopic theory of quantum gravity. For instance, the \emph{no global symmetries conjecture} states that in a consistent theory of quantum gravity there cannot be exact global (continuous or discrete) symmetries \cite{Israel:1967za,Zeldovich:1976vq,Kallosh:1995hi,Susskind:1995da,Banks:1988yz, Banks:2010zn,Yonekura:2020ino}.\footnote{See e.g., \cite{Polchinski:1998rr} for a proof in string perturbation theory as well as \cite{Harlow:2018jwu, Harlow:2018tng} for an analogous statement AdS holographic spacetimes.} (See also \cite{McNamara:2019rup} for a more refined version of the conjecture including higher-form topological charges as well.) This can be motivated by studying black holes, which are singular solutions in general relativity that are independent of any global charge present in the theory \cite{Israel:1967za}, therefore leading to various puzzles in relation to charge conservation and the finiteness of the black hole entropy \cite{Bekenstein:1972tm,Hawking:1975vcx}. Remarkably, all these problems can be easily resolved once we preclude any conserved global charge from existing in the first place. Despite this, string theory remains a crucial testing ground for verifying the applicability of any proposed Swampland condition, as well as for exploring the intricate relationships and structures that underpin all these criteria. 

One of the main objectives of this thesis will be to understand the precise role that the quantum gravity cut-off $\LQG$, here understood as the energy scale where semi-classical Einstein gravity --- coupled to any sort of matter --- breaks down completely, plays within the Swampland program. In particular, we would like to study how it behaves in generic effective theories of gravity, so as to ensure the fulfillment of these quantum gravity (or Swampland) constraints, and hopefully extract some general and useful lessons. To do so, we will restrict ourselves to string theory constructions mostly in flat spacetime backgrounds, although many of the considerations applied in this work could be in principle extended to other approaches of quantum gravity such as holography, via the AdS/CFT correspondence  \cite{Maldacena:1997re,Witten:1998qj}. For reasons that will become more apparent as we progress in the thesis, we will be particularly interested in two conjectures: the Distance Conjecture and the Weak Gravity Conjecture. They are both introduced and discussed in detail in Sections \ref{s:SDC} and \ref{s:WGC}, respectively. For an in-depth review of these and other Swampland conjectures, we refer the reader to the comprehensive discussions that can be found in \cite{Brennan:2017rbf,Palti:2019pca,vanBeest:2021lhn,Grana:2021zvf,Harlow:2022gzl,Agmon:2022thq,VanRiet:2023pnx}.

\subsection{The Weak Gravity Conjecture}\label{s:WGC}

The Weak Gravity Conjecture (WGC) was originally proposed in \cite{Arkani-Hamed:2006emk} and subsequently studied in many follow up works, resulting in several generalizations and refinements thereof (see \cite{Palti:2020mwc, Harlow:2022gzl} for some recent reviews on this conjecture and related ideas). Here we will focus on a formulation of the conjecture for $\mathsf{U(1)}$ gauge theories, see however \cite{Heidenreich:2017sim} for a discussion involving non-abelian gauge groups and product groups as well. There are in fact two versions of the original conjecture:\footnote{We focus here in the four-dimensional formulation. Note that the factors and the powers of the Planck mass would change in different dimensions \cite{Heidenreich:2015nta}, and we will include them explicitly when dealing with cases where $d \neq 4$, see e.g., Chapter \ref{ch:Emergence}.} 
\begin{itemize}
\item[] \textbf{Electric Weak Gravity Conjecture:} 
In a $\mathsf{U(1)}$ gauge theory coupled to gravity there must exist a charged state of mass and charge given by $\{m, q\}$, respectively, and whose charge-to-mass ratio is \emph{superextremal}, i.e. 
\begin{align}\label{WGCel}
 \frac{q^2 g^2}{m^2} \geq \left. \frac{Q_{\rm BH}^2 g^2}{M_{\rm BH}^2}\right|_{\text{ext}}\, ,
\end{align}
where the right hand side of the inequality refers to the charge-to-mass ratio of a extremal black hole in the theory, and $g$ denotes the gauge coupling constant. 
\item[] \textbf{Magnetic Weak Gravity Conjecture:} 
For a $\mathsf{U(1)}$ gauge theory coupled to gravity, there exists an upper bound for the UV cut-off $\Lambda_{\rm EFT}$ of the effective field theory, which is given by 
\begin{align}\label{WGCmag}
\Lambda_{\rm EFT} \leq g \Mpf\, . 
\end{align}
\end{itemize}
The importance of this conjecture lies in the fact that it attempts to make sharp statements about the charged spectrum of a $\mathsf{U(1)}$ gauge theory coupled to Einstein gravity. For instance, one direct implication of the condition \eqref{WGCel} is that pure Einstein-Maxwell gravity without matter (valid all the way up to $\Mpf$) must belong to the Swampland. As for the magnetic version, notice that it can be motivated in two different ways. One can either apply the electric version \eqref{WGCel} to the Hodge dual field strength $\tilde{F} = \star F$, thus forcing the theory to have a \emph{magnetic monopole} whose mass $m_{\rm mon}$ is smaller than its (physical) charge, or rather it can be recovered by directly imposing that the theory has some monopole that is not yet a black hole state. The EFT would then break down at energies close to $\Lambda_{\rm EFT} \approx m_{\rm mon}$, because it is there where it becomes sensitive to the monopole degrees of freedom, which can no longer be treated as solitonic objects.

For concreteness, let us consider Einstein-Maxwell theory coupled to a massive charged scalar $\phi$ in 4d, with an action that reads (at leading order)
\begin{equation}
\label{eq:EinsteinMaxwell}
S_{\mathrm{EM}}=\int \dd^{4} x \sqrt{-g}\left(\frac{1}{2 \kappa_4^2}\, \mathcal{R}-\frac{1}{4 g^{2}} F^{2} - \overline{D_\mu \phi} D^\mu \phi - m^2 |\phi|^2 \right)\, ,
\end{equation}
where the covariant derivative is defined as follows
\begin{equation}
D_\mu \phi= \left(\partial_\mu + \i q A_\mu \right) \phi\, , \qquad q \in \IZ\, .
\end{equation}
To get a grasp on what the conjecture is telling us, let us try to understand what goes wrong from the gravitational physics perspective when e.g., eq. \eqref{WGCmag} is not satisfied. Notice that the global\footnote{The global part of a gauge symmetry group $G$ (as well as the so-called large gauge transformations) is strictly speaking not included in the definition of a global symmetry (when $g$ is finite), since there is no gauge invariant charged local operator $\mathcal{O}(x)$ that is acted non-trivially by $G$ (see \cite{vanBeest:2021lhn} for more details).} part of this gauge symmetry acts on the field as
\begin{equation}\label{eq:globalpartsymm}
\phi \rightarrow e^{2\pi \i q \alpha} \phi\, ,
\end{equation}
with $\alpha$ being any constant parameter. Hence, if we insist on taking the limit $g\rightarrow 0$ we can actually recover an exact $\mathsf{U(1)}$ global symmetry, since the gauge boson $A_{\mu}$ decouples, such that \eqref{eq:globalpartsymm} survives as a remnant. To prevent this pathological behaviour from happening, the regime of validity of the EFT must shrink to zero size (in energies), which is precisely what eq. \eqref{WGCmag} imposes. Let us remark, though, that the above arguments are merely heuristic considerations. Nevertheless, as already commented at the beginning of this section, they serve the purpose of illustrating some of the universal ideas behind quantum gravity, which ultimately lead to non-trivial constraints which can be motivated independently of string theory. On the other hand, the Weak Gravity Conjecture is actually supported by a rich amount of top-down constructions within string theory, and its connection to the no-global symmetries proposal can be made much more rigorous, see e.g, \cite{Harlow:2018jwu, Harlow:2018tng,Montero:2018fns}.

Let us also comment here that an alternative, and perhaps more insightful, viewpoint on the Weak Gravity Conjecture is to interpret the latter as imposing that there should not be stable charged black hole remnants, which requires in turn to ask that the self-interaction of the state with highest charge-to-mass ratio is repulsive. This condition is usually referred to as the Repulsive Force Conjecture \cite{Palti:2017elp,Heidenreich:2019zkl, Lee:2018spm}, and roughly speaking it requires gravity to provide for the \emph{weakest} force (which is precisely the behaviour observed in Nature), namely
\begin{align}\label{CoulombGrav}
 |\vec{F}_\text{Coulomb}|  \geq |\vec{F}_\text{grav}| \Longleftrightarrow \frac{q^2 g^2}{m^2} \geq \frac{1}{\Mpd^{d-2}} \frac{d-3}{d-2}\, ,
\end{align}
precisely reducing to \eqref{WGCel} when $d=4$.\footnote{See also \cite{Palti_2017,Lee:2018spm,Gonzalo:2019gjp,Gonzalo:2020kke} for further extensions so as to include attractive Yukawa-like interactions as well.}

Let us finally introduce certain classes of refinements of the WGC which will play an important role in later parts of this thesis, also in connection with the Distance Conjecture discussed in Section \ref{s:SDC} below. Up to now, we have only asked for the theory to contain a \emph{single} (possibly very massive) state whose charge-to-mass ratio is superextremal. However, there exist various versions of the conjecture which indeed ask for the presence of an \emph{infinite} number of states --- with increasing mass --- satisfying \eqref{WGCel} instead \cite{Cheung:2014vva, Heidenreich:2015nta, Heidenreich:2016aqi,Montero:2016tif}. The rationale for this would be the requirement to have a stable statement under dimensional reduction (thereby including additional gauge fields that can appear in the compactification process, such as the Kaluza-Klein photon). This motivates tower \cite{Andriolo:2018lvp} and/or sub-Lattice versions of the WGC \cite{Heidenreich:2015nta, Heidenreich:2016aqi}, which have been tested up to date with an impressive level of accuracy in all known string theory compactifications, see e.g., \cite{Lee:2018urn,Lee:2019tst,Lee:2019skh,Lee:2019xtm,Heidenreich:2019bjd,Klaewer:2020lfg,Alim:2021vhs,Heidenreich:2021yda,Gendler:2022ztv,Cota:2022yjw,Cota:2022maf,Heidenreich:2024dmr} for an incomplete list of references.\footnote{See however \cite{Cota:2023uir} for a recent alternative statement that does not always impose the necessity of having an infinite number of superextremal particle states.}

\subsection{The Distance Conjecture}\label{s:SDC}

The Distance Conjecture \cite{Ooguri:2006in}, on the other hand, articulates some conjectural behaviour concerning effective gravitational theories characterized by some moduli space $\mathcal{M}$. (See however \cite{Calderon-Infante:2020dhm} for extensions and checks in theories lacking a quantum exact moduli space.) As already mentioned, this space is parameterized by the massless scalar fields within the theory, and it possesses a natural intrinsic metric defined by the kinetic terms associated to those.\footnote{Strictly speaking, the kinetic terms of the scalar fields are given by the pullback $f^*$ of the map $f: \mathbb{R}^{1, d-1} \to \mathcal{M}$ applied to the non-linear $\sigma$-model metric parameterizing the scalar manifold geometry $\mathcal{M}$, to the physical spacetime $\mathbb{R}^{1, d-1}$, whose coordinatization is given precisely by the scalar fields themselves.} The conjecture posits that, first of all, the diameter of the moduli space should be infinite, meaning that there should exist \emph{at least} one geodesic exploring infinite distance. Therefore, strictly compact moduli spaces, which are perfectly fine from the field theory point of view (for instance, one could simply consider Einstein gravity coupled to a massless and shift-symmetric scalar field $\theta \sim \theta + 2\pi$), should not be allowed in quantum gravity. Secondly, it states that, upon starting at some point $P \in \mathcal{M}$ and after traversing an infinite distance toward a second point $Q$ along some geodesic direction, one should encounter an infinite tower of states becoming light in Planck units. More precisely, this fall-off in the mass scale associated to the tower must be such that 
\begin{equation}
m_{\rm tow}(Q)\ \sim \ m_{\rm tow}(P) \ \, e^{-\lambda \,  d(P,Q)}\, , \qquad \text{as}\ \ d(P,Q)\rightarrow \infty\, ,
\label{eq:masslesstower}
\end{equation}
is satisfied, with $\lambda$ being some $\mathcal{O}(1)$ constant (in Planck units) that has been moreover conjectured to be greater than or equal to $\frac{1}{\sqrt{d-2}}$ \cite{Etheredge:2022opl}, where $d$ is the number of spacetime dimensions. Thus, at infinite distance, the effective field theory --- with dynamical gravity --- must break down due to the presence of an (infinite tower of) nearly massless states, which were not accounted for in the original EFT. This can be stated formally by saying that the naive effective field theory cut-off decreases exponentially with the moduli space distance, as follows
\begin{equation}
\Lambda_{\text{EFT}}(Q)\ \sim \ \Lambda_{\text{EFT}}(P) \ \, e^{-\lambda\,  d(P,Q)}\, ,
\end{equation}
where we have identified $\Lambda_{\text{EFT}} = m_{\rm tow}$ in the above equation. Notice that, strictly speaking, this does \emph{not} necessarily mean that a purely field-theoretic approach cannot be employed for energies well above $\Lambda_{\text{EFT}}$ (or $m_{\rm tow}$), since it may be possible to define some different local field theory description when e.g., the theory decompactifies and the tower verifying \eqref{eq:masslesstower} is comprised by Kaluza-Klein replica; for instance a higher-dimensional field theory. What it \emph{does} imply is the necessity of abandoning our original EFT construction.

A very simple realization of the Distance conjecture can be found when compactifying $d$-dimensional gravity on a circle of radius $R$. For concreteness, and in order to make contact with our discussion in Section \ref{s:maxsugraintro}, let us consider 11d supergravity on $\mathbf{S}^1$. Once we sit in the 10d Einstein frame, the scalar-gravitational sector of the theory reads as (c.f. eq. \eqref{eq:3formddimgravity})
\begin{equation}
	\label{eq:scalartensorMthonS1}
	\begin{aligned}
		S^{\text{10d}}_{\text{M-th}}\, \supset\ -\frac{1}{2\kappa_{10}^2} \int \dd^{10}x\sqrt{-g} \left(\mathcal{R}-\frac{9}{8}(\partial \rho)^2\right)\, ,
	\end{aligned}
\end{equation}
so that the moduli space of the theory corresponds to the possible v.e.v.s of the radius field $R= e^{\rho}$, whose metric has the form
\begin{align}
  G_{R R} = \frac{9}{8}\, \frac{1}{R^2}\, .
\end{align}
Therefore, the associated geodesic distance between any two values for the radius modulus, $R_i$ and $R_f$, is thus computed to be
\begin{equation}
d(R_i, R_f)=  \frac{3}{\sqrt{8}}\, \int_{R_i}^{R_f} \dfrac{\dd R}{R} =  \frac{3}{\sqrt{8}}\, \log \left( \frac{R_f}{R_i} \right)\, ,
\end{equation}
which presents two infinite distance points, namely at $R \to 0, \infty$. In the latter case, the limit corresponds to a decompactification back to eleven dimensions, where as already mentioned, it is the Kaluza-Klein states on the $\mathbf{S}^1$ the ones that become light at a rate
\begin{equation}
\frac{m_{\rm KK}}{M_{\text{Pl};\, 10}}=  (4\pi)^{-1/8}\, R^{-9/8} = (4\pi)^{-1/8}\, e^{-\frac{3}{\sqrt{8}}\, d(1,\, R)}\, ,
\end{equation}
hence exhibiting some $\lambda = \frac{3}{\sqrt{8}} \geq \frac{1}{\sqrt{8}}$. From this perspective, the Distance conjecture does not seem to impose any non-trivial requirement. Note, however, that this is not true anymore if probing the remaining infinite distance limit, i.e. $R \to 0$, where there seems to be a priori no field-theoretic degrees of freedom that can make \eqref{eq:masslesstower} hold. In fact, we already know what is the tower fulfilling the conjecture, since the small radius limit actually implements the duality with Type IIA string theory, whose fundamental object becomes weakly coupled precisely along this regime, such that we find
\begin{equation}
\frac{T_{\rm M2,\, str}}{M_{\text{Pl};\, 10}^2}=  (4\pi)^{-1/4}\, R^{3/4} \Longrightarrow \frac{m_{\rm tow}}{M_{\text{Pl};\, 10}} = \frac{m_s}{M_{\text{Pl};\, 10}} = (4\pi)^{-1/8}\, e^{-\frac{1}{\sqrt{8}}\, d(1,\, R)}\, ,
\end{equation}
where now $\lambda = \frac{1}{\sqrt{8}}$. This simple example gives us various important insights. First, the conjecture seems to require from additional ingredients beyond the original EFT description, namely extended objects, such as the M2-brane. Second, it also becomes intimately related with the concept of dualities in quantum gravity, as explained in Section \ref{s:dualities} above.

Let us stress here the fact that, despite the good amount of evidence gathered in favour of the conjecture (see e.g., \cite{Montero:2015ofa,Baume:2016psm,Klaewer:2016kiy,Valenzuela:2016yny,Blumenhagen:2017cxt,Palti:2017elp,Hebecker:2017lxm,Grimm:2018ohb,Heidenreich:2018kpg,Blumenhagen:2018nts,Landete:2018kqf,Lee:2018urn,Reece:2018zvv,Lee:2018spm,Ooguri:2018wrx,Grimm:2018cpv,Buratti:2018xjt,Hebecker:2018fln,Gonzalo:2018guu,Corvilain:2018lgw,Lee:2019tst,Blumenhagen:2019qcg,Joshi:2019nzi,Font:2019cxq,Lee:2019xtm,Grimm:2019wtx,Lee:2019wij, Erkinger:2019umg, EnriquezRojo:2020hzi,Gendler:2020dfp,Grimm:2020cda,Bastian:2020egp,Baume:2020dqd, Perlmutter:2020buo, Calderon-Infante:2020dhm,Grimm:2020ouv, Ooguri:2024ofs,Aoufia:2024awo,Lanza:2020qmt, Lanza:2021udy, Lanza:2022zyg} for an incomplete set of references), there is still no completely satisfactory microscopic argument explaining why it should hold in general, even though currently there are some interesting proposals trying to address this point (see Chapter \ref{ch:Emergence} for more on this).\footnote{See \cite{Rudelius:2024vmc} for recent arguments based on generalized symmetries.}

\subsubsection*{Emergent String Conjecture}

One simple yet important question that the previous analysis raises concerns the type of (lightest) towers that can appear upon exploring infinite distance limits in quantum gravitational field spaces. Indeed, already in the example mentioned before, we saw two different kind of behaviours arising in our theory: either the former decompactifies and it is the Kaluza-Klein tower of states who furnish the Distance Conjecture, or rather we end up probing a tensionless limit for a (possibly dual) fundamental string (an emergent string limit). Hence, it is natural to wonder at this point whether there could be any other behaviour (i.e. different from the aforementioned ones), that could arise in quantum gravity at infinite distance; for instance, the appearance of some asymptotically tensionless membrane dominating the light spectrum of the theory. Indeed, this would be both striking and very interesting at the same time, since it could hint toward yet unknown UV completions of gravity not involving strings at all. 

However, in \cite{Lee:2019wij} the authors proposed that this observed pattern is in fact all that can actually occur in quantum gravity, such that there could not be any other possible UV complete theory arising at infinite distance. This conjecture is referred to as the Emergent String Conjecture, and states that any equi-dimensional infinite distance limit\footnote{\label{fnote:equidimensional}Here equi-dimensional means that the asymptotic physics is governed by a theory defined in the same number of space-time dimensions as the original theory in the interior of the moduli space.} needs to correspond to some infinite distance point in moduli space where a critical string becomes tensionless and weakly coupled. Therefore, along these degenerations the tower of states predicted by \eqref{eq:masslesstower} comprise the excitations of the light, critical string (c.f. eq. \eqref{eq:modeexpansionclosedstring}-\eqref{eq:fermionmodeexpansionstrings}). Interestingly, as argued in \cite{Alvarez-Garcia:2021pxo}, consistency of the Emergent String Conjecture under dimensional reduction seems to forbid e.g., membrane limits where it is a fundamental 2-brane which dominates the light spectrum of the theory.

Crucially, there seems be always a \textit{unique} critical string becoming tensionless at the fastest rate, such that in principle no two different massless gravitons can arise in the spectrum of light states.




\part{\scshape \textcolor{MyDarkRed}{The Quantum Gravity Scale}}\label{part:QGscale}
\chapter{The Species Scale as the Quantum Gravity cut-off}\label{ch:SpeciesIntro} 
In the remainder of this thesis we will be most interested in certain non-trivial aspects that gravitational interactions exhibit, which are quantum in origin and moreover single out gravity from the rest of the interactions observed in Nature. However, in order to properly understand what makes gravity so special, it is crucial to have in mind which properties are actually shared by other fundamental forces. In this regard, the most interesting statement concerns the fact that both kind of interactions can be accommodated --- at low enough energies --- within the same theoretical framework, namely that of \emph{effective field theory} (EFT).

Indeed, we are by now familiar with the idea of \emph{separation of scales}. This concept turns out to be crucial in our modern understanding of physics, acting moreover as a useful organizing principle. It posits that physics itself can be essentially understood in an independent manner at each energy scale (or more accurately regime), therefore avoiding the important but subtle question of what is happening at much higher energies (equivalently very short distances), which cannot be resolved by our experimental apparatus. A clear example of this is the theory of hydrodynamics, which correctly describes how certain fluids (e.g., water) flow in streams or rivers, without caring too much about the details of the processes that occur at the molecular or even (sub-)atomic level. In a sense, what we do is isolate the relevant degrees of freedom that are needed in order to describe some physical phenomena and then use an effective `coarse-grain' description. Of course, this does not mean that the physics at higher energies does not matter, since after all the low energy theory may be, in principle, retrieved from this more fundamental description. 

Furthermore, one can argue that the conclusion drawn in the previous paragraph is in fact unavoidable. More precisely, if we want our effective description to be compatible with \emph{(i)} Lorentz invariance, \emph{(ii)} quantum mechanics (causality, unitarity, crossing symmetry) and \emph{(iii)} locality (in the form of cluster decomposition), we must resort to quantum field theory \cite{Weinberg:2016kyd, Weinberg:2021exr}. Therefore, the EFTs we usually have to deal with are defined as a path integral
\beq
\mathcal{Z}_{\Lambda} = \int \mathcal{D} \Phi \mathcal{D} \Psi \mathcal{D} A \mathcal{D} g\ e^{\text{i} S_{\Lambda} \left[ \Phi,\, \Psi,\, A,\, g\right]/\hbar}\, ,
\label{eq:EFTpathintegral}
\eeq
where the action functional includes terms of the form (assuming the existence of some lagrangian description) 
\beq
\begin{aligned}
    S_{\Lambda} \left[ \Phi,\, \Psi,\, A,\, g\right] = \int \dd^4x \sqrt{-g} \Bigg( & \frac{1}{2 \kappa_4^2} \left(\mathcal{R} - 2 \Lambda_{\rm c.c.} \right) - \frac{1}{4} F^2 - \frac{1}{2} \left(\partial \Phi \right)^2 + \text{i} \bar{\Psi}  \slashed{\partial} \Psi\\
    &- V(\Phi) + \mathcal{Y}_{\rm yuk} \Phi \bar{\Psi} \Psi + \ldots\Bigg)\, ,
\end{aligned}
\label{eq:EFTaction}
\eeq
with $\Phi(x)$ describing scalar degrees of freedom, $\Psi^{\alpha}(x)$ is Grassmann-valued, $A_{\mu}(x)$ accounts for gauge interactions, $g_{\mu \nu} (x)$ denotes the gravitational field and the ellipsis indicates any further local higher-dimensional operator that may appear in the action. The subscript `$\Lambda$' both in \eqref{eq:EFTpathintegral} and \eqref{eq:EFTaction} signify that the path integral is meant to be restricted to field variations which probe energy scales no greater than the ultra-violet (UV) cut-off $\Lambda$, beyond which new physics (in the form of e.g., very massive degrees of freedom) may arise. From this perspective, it is clear that gravity is not much different than other interactions appearing in the effective action $S_{\Lambda}$, and in fact one must include gravitational degrees of freedom in the path integral so as to account for classical dynamical phenomena such as gravitational waves \cite{GW1,GW2}.

The actual difference between gravity and say (non-)Abelian gauge theories becomes manifest when studying how the amplitudes of physical processes behave as we vary the energy of the external particles. Indeed, the gravitational coupling $\kappa_4$ can be related to the more familiar Newton's constant $G_N$ as follows
\beq
 \kappa_4^2 = 8 \pi G_N\, ,
\label{eq:gravconstant4d}
\eeq
which has units of $[E]^{-2}$ in four dimensions and is extremely small. As a consequence, we deduce that General Relativy (GR) is \emph{non-renormalizable}, thus implying that it is a priori more sensitive to the relevant UV physics than its renormalizable counterparts. However, when viewed from the prism of effective field theory, this simply means that one can organize the theory into an energy expansion which \emph{must} include higher-curvature as well as higher-dimensional operators, whose form is moreover highly constrained by the symmetries of the theory (i.e. general covariance). Furthermore, given the smallness of $G_N$ at ordinary energies, one deduces that gravity is very well-suited for perturbation theory, as long as we remain within its range of applicability.

The aim of this chapter is to revisit these issues and try to pinpoint, from the perspective of a low energy observer, what would be the precise regime of validity of any EFT weakly coupled to Einstein gravity. In other words, the main question that will occupy us in the following is what is the maximum energy cut-off $\Lambda$ of any such effective description. The discussion is thus organized as follows. In Section \ref{s:genrelEFT} we will propose a first potential candidate for the quantum gravity cut-off $\LQG$, based on our experience with other non-remormalizable theories, such as the Fermi theory of weak interactions or the chiral lagrangian in quantum chromo-dynamics (QCD). There we also elaborate on the EFT-like expansion that one would expect to arise within these gravitational theories, since it will play an important role in Chapter \ref{ch:Higherdimops} of this thesis. Later on, in Section \ref{s:speciesmotivation} we reconsider the validity of this answer in the presence of a large amount of light degrees of freedom, both from a perturbative and non-perturbative perspective. In particular, this introduces certain puzzles related to (the minimal) black hole entropy, and leads to a seemingly different proposal for $\LQG$. Finally, in Section \ref{s:speciesscale}, we elaborate on the possibility that the energy scale introduced in Section \ref{s:speciesmotivation} be precisely identified with the quantum gravity cut-off. We also argue that this observation may have profound impact when embedded into the Swampland program, as we investigate more thoroughly in Parts \ref{part:StringTheoryTests} and \ref{part:pattern} of the thesis.

Most of the material included in this chapter is based on results existing in the literature, whilst the rest has been taken from the publications \cite{Castellano:2022bvr, Castellano:2023aum}. We will include the appropriate references accordingly, so as to distinguish the original contributions in this work from already known results.   

\section{General relativity as an effective field theory}
\label{s:genrelEFT}	

In this section we review what is the typical structure of non-renormalizable EFTs in order to hopefully extract some general lessons regarding their predictive power. To do so, we briefly consider in Section \ref{ss:nonrenormalizableEFTs} a simple and familiar example, namely 4-Fermi theory. Subsequently, in Section \ref{ss:basics} we apply these ideas in the context of gravitational interactions. The discussion here is based on refs. \cite{Burgess:2020tbq,Donoghue:1994dn,Donoghue:2022eay} (see also references therein), to which we refer the reader interested in more details. 

\subsection{Non-renormalizable field theories}
\label{ss:nonrenormalizableEFTs}

For any given effective field theory, what is (non-)renormalizable strongly depends on the dimension of the target spacetime. Hence, in order to be as general as possible (and with an eye to future applications in the context of string theory), let us consider here some $d$-dimensional EFT with an action functional of the form   
\beq
\begin{aligned}
    S_{\rm EFT} = \int \dd^dx \left( \mathcal{L}_{\rm ren} + \mathcal{L}_{\cancel{\rm ren}}\right)\, ,
\end{aligned}
\label{eq:EFTactionII}
\eeq
where $\mathcal{L}_{\rm ren}$ ($\mathcal{L}_{\cancel{\rm ren}}$) denotes the (non-)renormalizable lagrangian density. On the one hand, $\mathcal{L}_{\rm ren}$ typically includes local operators whose classical mass dimension is smaller than or equal to $d$, such as kinetic or mass terms as well as the first few low-lying interactions, i.e. scalar potentials, Yukawa or gauge interactions, etc. (c.f. eq. \eqref{eq:EFTaction}). Their significance lies in the fact that they capture the most \emph{relevant} physics (in the Wilsonian sense) at low energies. Moreover, even if they do not provide for the full UV complete theory, their physical predictions are shielded from the very high energy degrees of freedom, which may enter when computing quantum (e.g., loop) corrections in \eqref{eq:EFTpathintegral}. This follows from the Appelquist-Carazzone theorem \cite{Appelquist:1974tg}, which ensures that all such effects can be either effectively encoded into the `running' of a finite number of physical parameters that enter into the original lagrangian and can be measured in experiments (i.e. the masses and couplings of the theory), or are rather highly suppressed at low energies. Hence, such renormalizable theories are very appealing from the theoretical point of view, since one can in principle make an infinite number of predictions once a few coupling constants have been determined experimentally. A canonical example of this class of theories is quantum electro-dynamics (QED) in four dimensions \cite{Feynman:1949zx, Feynman:1950ir}
\beq
\begin{aligned}
    S_{\rm QED} \left[ A, \psi\right] = \int \dd^4x \left( -\frac{1}{4 e^2} F^{\mu \nu} F_{\mu \nu} + \bar{\psi} \left( \text{i} \slashed{D} -m\right) \psi\right)\, ,
\end{aligned}
\label{eq:QED}
\eeq
where $F_{\mu \nu} = \partial_{\mu} A_{\nu}- \partial_{\nu} A_{\mu}$ denotes the field strength of the photon, $D_{\mu} = \partial_{\mu} - \text{i} A_{\mu}$ is the $\mathsf{U(1)}$ covariant derivative and $e$ determines the gauge coupling constant. This theory is interacting, describes the electromagnetic force observed in our world and offers many interesting predictions. Among them one finds effective photon interactions, which appear to be forbidden at the classical level but are instead generated by quantum corrections due to the electron-positron field $\psi (x)$. In fact, some of these can be calculated exactly at one-loop, yielding the following local lagrangian due to Heisenberg (see e.g., \cite{Schwartz:2014sze})
\beq
\begin{aligned}
    \mathcal{L}_{\rm EH}\, =\, \frac{e^2}{360 \pi m^4} \left[ \left( F^{\mu \nu} F_{\mu \nu} \right)^2+ \frac74 \left( F^{\mu \nu} \tilde{F}_{\mu \nu}\right)^2\right]\, ,
\end{aligned}
\label{eq:Euler-Heisenberg}
\eeq
where $\tilde{F}_{\mu \nu} = \frac12 \epsilon_{\mu \nu \rho \sigma} F^{\rho \sigma}$ is the Hodge dual field strength.

On the other hand, the non-renormalizable piece frequently consists in an infinite number of local operators of mass dimension $n>d$, as follows
\beq
\begin{aligned}
    \mathcal{L}_{\cancel{\rm ren}}\, \supset\, \sum_{n>d}^{\infty} c_n \frac{\mathsf{O}_n \left( \Phi,\, \Psi,\, \ldots\right)}{\Lambda_{\rm UV}^{n-d}}\, ,
\end{aligned}
\label{eq:nonrenlagrangian}
\eeq
where we have absorbed the dimensions into some energy scale $\Lambda_{\rm UV}$, thus leaving us with a set of \emph{Wilson coefficients} $\{c_n\}$, which are dimensionless parameters generically of order 1 (or at least not parametrically large/small except possibly for a few of them). Notice that the introduction of the same dimensionful quantity $\Lambda_{\rm UV}$ for each operator in \eqref{eq:nonrenlagrangian} can be done without any loss of generality. The non-trivial statement concerns the order of magnitude of the expansion coefficients, and the fact that they become roughly of the same order implies that they agree on the maximum energy scale at which the theory can be trusted, namely the UV cut-off discussed around eq. \eqref{eq:EFTaction}. That this indeed happens for any existing non-renormalizable EFT has not been proven rigorously, but rather may be regarded as a general expectation coming from our field theory experience and is moreover supported by examples, see below. Let us also note that the terms appearing in $\mathcal{L}_{\cancel{\rm ren}}$ are usually referred to as \emph{irrelevant} operators, since their contribution to physical processes at low enough energies (i.e. well below $\Lambda_{\rm UV}$) is negligible,\footnote{Of course, this is true as long as the renormalizable piece contributes to the process under consideration, since otherwise it is the non-renormalizable operator the one providing for the leading order term in the energy expansion. A well-known example being the case of dimension-6 operators allowing for the proton decay in supersymmetric grand unified theories (see e.g., \cite{Raby:2002wc}).} whereas they become important in the ultra-violet regime --- being moreover sensitive to the UV completion of the theory.

From our previous discussion, one could be tempted to conclude that indeed non-renormalizable field theories seem to be less appealing than their renormalizable counterparts. However, the modern Wilsonian understanding of effective field theories actually flips this logic around and suggests that it is indeed the non-renormalizable EFTs the ones that are most interesting as well as predictive. The reason for this is twofold: first, from the `top-down' perspective, it may be sometimes useful to not bother about all the details and complications associated to the UV complete theory, and just construct --- using e.g., symmetry principles --- a simpler EFT for the relevant light degrees of freedom, which typically is of the non-renormalizable type and yields the same answers. Second, from the `bottom-up' point of view, these theories oftentimes predict their own failure, in the sense that they tend to organize into some energy expansion in terms of the UV cut-off $\Lambda_{\rm UV}$ as in \eqref{eq:nonrenlagrangian}, which signals the appearance of new physics around that scale which cannot be accounted for by the original theory. In the following, we will illustrate these ideas using perhaps the most simple example of a non-renormalizable EFT (at least in the context of particle physics), namely the 4-Fermi effective theory of weak interactions.

\subsubsection*{Example: Fermi theory of weak interactions}

The effective theory of weak interactions was originally introduced by Fermi \cite{Fermi:1934hr} as an extension of QED, in order to account for certain phenomena (like $\beta$-decay) which could not be explained within the theoretical framework existing at that time. It was moreover constructed purely from a `bottom-up' perspective, since the non-Abelian character of the underlying gauge theory was not known yet, and indeed it is very practical for all purposes as long as we want to describe physical processes involving weak interactions (e.g., muon decay) at energies well below the mass of the $W$ boson. However, in order to better connect with our previous discussion, we will derive this effective theory from the `top-down', i.e. starting with the Standard Model (SM) of particle physics.

As we now know, the weak interactions are mediated by the $W^{\pm}$ and $Z^0$ bosons, which in the SM describe an $\mathsf{SU(2)}$ gauge theory. The superscripts indicate the associated $\mathsf{U(1)}$ charge of the corresponding vector boson under the electromagnetic (EM) field. Hence, they couple to quark and lepton fields, which can be arranged in $\mathsf{SU(2)}$ doublets as follows (using a flavour eigenbasis)
\beq
\begin{aligned}
   \text{Leptons}: \qquad \begin{pmatrix}
		\nu_e\\ e
	\end{pmatrix}\, ,& \qquad
     \begin{pmatrix}
		\nu_{\mu}\\ \mu
	\end{pmatrix}\, , \qquad
      \begin{pmatrix}
		\nu_{\tau}\\ \tau
	\end{pmatrix}\, ,\\
     \\
    \text{Quarks}: \qquad \begin{pmatrix}
		u\\ d
	\end{pmatrix}\, ,& \qquad
     \begin{pmatrix}
		c\\ s
	\end{pmatrix}\, , \qquad
      \begin{pmatrix}
		t\\ b
	\end{pmatrix}\, ,\\
\end{aligned}
\label{eq:quarks&leptons}
\eeq
via some EM and left-handed $\mathsf{SU(2)}$ 1-form currents, which read
\beq
\begin{aligned}
    j^{\mu}_{\rm EM} &= \sum_{i} \frac23 \bar{u}_i \gamma^{\mu} u_i - \frac13 \bar{d}_i \gamma^{\mu} d_i - \bar{e}_i \gamma^{\mu} e_i\, ,\\
    j^{\mu}_{\rm a} &= \sum_{\rm fermions} \bar{\Psi} \gamma^{\mu} \left( \frac{1- \gamma_5}{2}\right) \frac{\sigma_{\text{a}}}{2} \Psi\, , \qquad \text{a}=1,2, 3\, ,\\ 
\end{aligned}
\label{eq:EM&SU(2)currents}
\eeq
where $i=1, 2, 3$ runs over the three generations of fermions, $\Psi$ denotes any doublet from \eqref{eq:quarks&leptons}, $\sigma_{\text{a}}$ are the Pauli matrices and the operator $P_L = \left( \frac{1- \gamma_5}{2}\right)$ projects on the left-handed pieces of the associated Dirac fields. More precisely, the vector bosons couple to the aforementioned currents through the following lagrangian
\beq
\begin{aligned}
    \mathcal{L}_{\rm SM} \supset \frac{e}{\sin \theta_w} \left( W_{\mu}^+ J^{\mu}_- + W_{\mu}^- J^{\mu}_+ \right) + \frac{e}{\sin \theta_w\, \cos \theta_w} Z_{\mu} \left( j^{\mu}_{3} -\sin^2 \theta_w j^{\mu}_{\rm EM}\right)\, ,\\ 
\end{aligned}
\label{eq:WZbosonscoupling}
\eeq
where $\cos \theta_w = \frac{m_W}{m_Z}$ defines the Weinberg (or mixing) angle in terms of the masses of the $W$ and $Z$ bosons, and we have introduced the currents $J^{\mu}_{\pm} = \frac{j^{\mu}_1 \mp \text{i} j^{\mu}_2}{\sqrt{2}}$ in the previous expression.

Therefore, consider some four-point amplitude involving the exchange of a $W$ boson at tree-level. For instance, one may want to describe a $\beta^-$--\,process, where a neutron decays into a proton, an electron and an antineutrino, i.e. $\rm n \to p + e^- + \bar{\nu}_e$. At the level of the SM fields, such decay can be described by a four-point function schematically of the form $\braket{\bar{d} u \bar{\nu}_e e}$, and is precisely accounted for by the interaction lagrangian \eqref{eq:WZbosonscoupling}, yielding an amplitude which reads 
\beq
   \mathcal{A}= \left( \frac{e}{\sin \theta_w}\right)^2 \frac{J_{\mu\, +} J^{\mu}_-}{p^2 - m_W^2}\, ,
\label{eq:4pointamplitude}
\eeq
where $p^2$ denotes the momentum exchange and one should understand the quantity $J_{\mu\, +} J^{\mu}_-$ as the corresponding operator evaluated on the external spinor states. Following the logic explained before, if the energies involved in the process are much lower than the mass scale of the $W$ boson (that is of order $80.4$ GeV), one can expand the denominator in \eqref{eq:4pointamplitude} into a power series, with the leading term being
\beq
   \mathcal{A}= -\left( \frac{e}{\sin \theta_w m_W}\right)^2 J_{\mu\, +} J^{\mu}_-\, +\, \mathcal{O} \left(\frac{p^2}{m_W^2}\right)\, ,
\label{eq:4pointamplitudeleadingterm}
\eeq
which can be reproduced by the following effective four-fermion contact interaction 
\beq
   \mathcal{L}_{\rm Fermi} = - \left( \frac{e}{\sin \theta_w m_W}\right)^2 J_{\mu\, +} J^{\mu}_- = \frac{8}{\sqrt{2}} G_F J_{\mu\, +} J^{\mu}_-\, .
\label{eq:fermilagrangian}
\eeq
Here we have introduced Fermi coupling constant $G_F = \frac{\sqrt{2}}{8} \frac{g^2}{m_W^2} \approx 10^{-5}\ \text{GeV}^{-2}$, which depends on the $\mathsf{SU(2)}$ gauge coupling $g$ and the mass of the $W$ boson. This essentially defines 4-Fermi theory (written in a perhaps more modern language \cite{Schwartz:2014sze}), consisting of a dimension-6 operator that is irrelevant, in the Wilsonian sense. Consequently, it can be seen as a non-renormalizable theory including, apart from the leading term \eqref{eq:fermilagrangian}, an infinite number of higher-dimensional and higher-derivative operators (which can be deduced from the series expansion \eqref{eq:4pointamplitudeleadingterm}) of the schematic form
\beq
   \mathcal{L}_{\rm EFT} \supset c_6\, G_F\, \bar{\psi}\psi \bar{\psi} \psi + c_8\, G_F^2\, \bar{\psi} \psi \partial^2 \bar{\psi} \psi + c_{10}\, G_F^3\, \bar{\psi} \slashed{\partial} \psi \partial^2 \bar{\psi} \slashed{\partial} \psi + \ldots\, ,
\label{eq:fermiEFT}
\eeq
where $\{c_n\}$ are Wilson coefficients that can be computed exactly (as in eq. \eqref{eq:fermilagrangian}) and are given by $\mathcal{O}(1)$ numbers. Note that if we define some energy scale associated to the dimensionful coupling constant $G_F$, as follows
\beq
   \Lambda_{\rm UV} := G_F^{-1/2} \approx 300\ \text{GeV}\, ,
\label{eq:UVcutofffermi}
\eeq
then the EFT expansion in \eqref{eq:fermiEFT} can be precisely recast in the form advocated in eq. \eqref{eq:nonrenlagrangian}. Moreover, such $\Lambda_{\rm UV}$ correctly captures the order of magnitude of the maximum regime of validity in energy of the effective theory, and thus can be regarded as the true cut-off of the theory.

This simple yet illustrative example provides us with an important and fairly general lesson. Indeed, whenever we are given some low energy non-renormalizable EFT whose coupling constant is \emph{dimensionful}, and we face the question of what is the regime of validity of the theory, a good estimate for the relevant UV cut-off can be obtained by looking at the energy scale associated to the coupling constant itself --- which can be measured in experiments, i.e. at low energies. This is precisely the case of the Fermi theory here described, where $G_F^{-1/2}$ provides the right order of magnitude of the energy scale where new physics arises, namely the electro-weak scale of the underlying $\mathsf{SU(2)}$ gauge theory. Let us note in passing that, in fact, the same conclusion extends to other more sophisticated EFTs, such as the chiral lagrangian of QCD. This moreover gives us a feeling of the great predictive power that non-renomalizable EFTs typically have.

\subsection{The gravitational EFT}
\label{ss:basics}

Let us now turn to gravitational interactions. Indeed, as already discussed in the introduction of this chapter, GR must be regarded as a low energy effective field theory describing the dynamics of a massless spin-2 field which couples universally (per the Equivalence Principle) to energy and momentum \cite{Donoghue:1994dn,Donoghue:2022eay}. In $d$ spacetime dimensions, the two-derivative action reads
\beq
\begin{aligned}
    S_{\rm EH} \left[ g_{\mu \nu} \right]= \frac{1}{2 \kappa_d^2} \int \dd^dx \sqrt{-g} \left( \mathcal{R} - 2 \Lambda_{\rm c.c.} \right)\, ,
\end{aligned}
\label{eq:EinsteinHilbertaction}
\eeq
where $\kappa_d^{2} = 8 \pi G_N$ is again related to Newton's gravitational constant, $\mathcal{R}$ is the Ricci scalar (see Appendix \ref{ap:conventions} for conventions) and we have allowed for the presence of some cosmological constant $\Lambda_{\rm c.c.}$. This theory has a coupling constant, i.e. $G_N$, with dimensions of $[E]^{2-d}$, and therefore it is non-renormalizable whenever $d >2$ (as in the cases of interest here). One can readily see this either by coupling \eqref{eq:EinsteinHilbertaction} to some other matter action $S_{\rm matter}$, giving rise to the following equations of motion
\beq
\begin{aligned}
  \mathcal{R}_{\mu \nu}- \frac12 g_{\mu \nu} \mathcal{R} + \Lambda_{\rm c.c.}\, g_{\mu \nu} = 8 \pi G_N T_{\mu \nu}\, ,
\end{aligned}
\label{eq:Einsteineqs}
\eeq
where the energy-momentum tensor is defined as usual, namely
\beq
\begin{aligned}
  T_{\mu \nu} = - \frac{2}{\sqrt{-g}} \frac{\delta S_{\rm matter}}{\delta g^{\mu \nu}}\, ,
\end{aligned}
\label{eq:Energymomentum}
\eeq
or rather by directly expanding the Einstein-Hilbert action around some classical background $\bar{g}_{\mu \nu}$ and looking at the $\kappa_d\,$--\,dependence of the graviton self-interactions. Indeed, upon doing so one finds
\beq
\begin{aligned}
    g_{\mu \nu} &= \bar{g}_{\mu \nu} + 2\kappa_d h_{\mu \nu}\, ,\\
    g^{\mu \nu} &= \bar{g}^{\mu \nu} - 2\kappa_d h^{\mu \nu} + \mathcal{O}(\kappa_d^2)\, ,
\end{aligned}
\label{eq:metricfluctuations}
\eeq
where the indices are raised with the background metric, and which yields the following action \cite{Fierz:1939ix} after plugging it back to \eqref{eq:EinsteinHilbertaction}
\beq
\begin{aligned}
    S_{\rm EH} \left[ h_{\mu \nu} \right] &= S_0 + \int \dd^dx \sqrt{-\bar{g}}\, \Bigg( -\frac12 \bar{\nabla}_{\rho} h_{\mu \nu } \bar{\nabla}^{\rho} h^{\mu \nu } + \frac12 \bar{\nabla}_{\mu} h \bar{\nabla}^{\mu} h - \bar{\nabla}_{\mu} h \bar{\nabla}_{\nu} h^{\mu \nu}\\
    &+ \bar{\nabla}_{\rho} h_{\mu \nu} \bar{\nabla}^{\nu} h^{\mu \rho} - \bar{\mathcal{R}} \left( \frac14 h^2 - \frac12 h_{\mu \nu} h^{\mu \nu} \right) - \left( 2h^{\rho}_{\mu} h_{\nu \rho} - h h_{\mu \nu}\right) \bar{\mathcal{R}}^{\mu \nu} + \mathcal{O} (\kappa_d)\Bigg)\, .
\end{aligned}
\label{eq:Pauliaction}
\eeq
Here we have defined $S_0$ as being the classical on-shell contribution to the action $S_{\rm EH}$, $\bar{\nabla}_{\mu}$ is the covariant derivative with respect to the background field metric, and $h= h^{\mu}_{\mu}$ is the trace of the metric fluctuations. Crucially, at higher orders there are further local contact terms involving at least three gravitons, thus accounting for an infinite series of self-interactions, which carry positive powers of the coupling constant $\kappa_d$.

From the EFT point of view, it is clear then that the action \eqref{eq:EinsteinHilbertaction} captures just the first few terms in a derivative/energy expansion which should include, a priori, any term that respects the underlying symmetries of the theory --- general covariance. In particular, one would expect to find the following effective field theory expansion
\beq
\begin{aligned}
    S_{\rm EFT} \left[ g_{\mu \nu} \right]= \frac{1}{2 \kappa_d^2} \int \dd^dx \sqrt{-g} \left( \mathcal{R} - 2 \Lambda_{\rm c.c.} + \sum_{n\geq 2} \frac{\mathsf{O}_{n} (\mathcal{R})}{\Lambda_{\rm QG}^{n-2}}\right)\, ,
\end{aligned}
\label{eq:gravEFT}
\eeq
where $\mathsf{O}_n (\mathcal{R})$ represents any dimension-$n$ higher-curvature operator --- of mass dimension greater than or equal to 2, such as $\mathcal{R}^2$ or $\mathcal{R}_{\mu \nu} \mathcal{R}^{\mu \nu}$ for $n=4$ \cite{tHooft:1974toh}, and $\Lambda_{\text{QG}}$ gives precisely the energy scale controlling the suppression of such quantum corrections with respect to the Einstein-Hilbert term. Furthermore, according to our discussion in the previous subsection and taking into account the general lessons extracted from the Fermi example above, the natural conclusion at this point would be to claim that the order of magnitude of the UV cut-off of the theory --- i.e. the quantum gravity cut-off --- is the one associated to the gravitational coupling constant $\kappa_d$. In other words,
\beq
   \Lambda_{\rm QG} := \kappa_d^{-\frac{1}{d-2}} = \Mpd\, ,
\label{eq:UVcutoffgravity}
\eeq
which is nothing but the (reduced) $d$-dimensional Planck scale. Note that this naive answer actually makes perfect sense. Indeed, we are by now familiar with the existence of spacetime singularities where the semi-classical description of gravity fails drastically. In fact, as soon as the curvatures involved, say in a small region of some background spacetime (for instance inside black holes), reach the Planck scale, one can no longer trust the original description. Relatedly, whenever the energies probed by some scattering process in particle physics get close to $\Mpd$, the induced gravitational back-reaction becomes significantly strong, and can even lead to gravitational collapse. Hence, we expect that precisely around that energy scale a more fundamental quantum gravity description may be needed that supersedes the low energy effective field theory \eqref{eq:EinsteinHilbertaction}, and thus provides for a UV completion of the gravitational interactions.

\section{Gravity in the presence of a large number of species}
\label{s:speciesmotivation}

In this section we want to reconsider these issues in order to pinpoint precisely which quantity should capture the UV cut-off associated to quantum gravity, based on what we have learned so far about the latter. The upshot will be that, in the presence of a large amount of light degrees of freedom in the theory, $\LQG$ will significantly depart from the expected value $\Mpd$. To argue for this, we employ two seemingly different class of arguments that nonetheless lead to the same answer. In particular, in Section \ref{ss:nonperturbative} we analyze this problem from the perspective of black hole physics and holography, whereas in Section \ref{ss:perturbative} we follow a more perturbative approach and we seek for the energy scale where the quantum perturbative series of the graviton naively breaks down.

\subsection{Non-perturbative analysis}
\label{ss:nonperturbative}

\begin{figure}[htb]
\begin{center}
\includegraphics[width=0.25\textwidth]{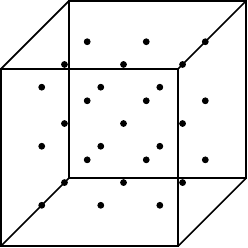}
\caption{\small Schematic depiction of a box of size $\mathsf{R}$ including $N$ particle species.}
\label{fig:boxparticles}
\end{center}
\end{figure}

Let us consider a gravitational theory weakly coupled to some matter fields in $d \geq 4$ spacetime dimensions. Within this theory, one may choose to study some closed spacelike hypersurface of e.g., $\mathbf{S}^{d-2}$ topology, and analyze all possible field and metric configurations inside the latter. In what follows, we will require the allowed states to be observable from the outside region, namely we ask the system to have a total energy $E$ bounded from above by the size of its surrounding surface
\beq
   E \leq \frac{(d-2) A}{16 \pi G_N \mathsf{R}}\, , \qquad A= \frac{2 \pi^{\frac{d-1}{2}}\mathsf{R}^{d-2}}{\Gamma(\frac{d-1}{2})}\, ,
\label{eq:boundEnergyBH}
\eeq
where $\mathsf{R}$ denotes the radius of the $\mathbf{S}^{d-2}$ hypersurface and $A$ its area. This restriction imposes that the system has not yet collapsed into a (Schwarzschild) black hole, or it is at the verge of doing so, see Figure \ref{fig:boxparticles}.
Given these state of affairs, one can easily estimate the number of possible configurations compatible with the above requirements. Indeed, if there are $N$ fundamental particle species, a simple thermodynamic analysis reveals that the macroscopic properties of the system would read as
\beq
\begin{aligned}
    E &= \mathfrak{c}_1\, N\, \mathsf{R}^{d-1}\, T^d\, ,\\
    S &= \mathfrak{c}_2\, N\, \left(\mathsf{R}\, T \right)^{d-1}\, ,
\end{aligned}
\label{eq:gasthermodynamics}
\eeq
where $T$ is the temperature of the system --- which we take to be much bigger than the masses of the particles involved, $S$ denotes its entropy and $\{ \mathfrak{c}_1, \mathfrak{c}_2\}$ are certain numerical constants that will not be important for our considerations here. Furthermore, upon solving for the temperature and imposing the restriction \eqref{eq:boundEnergyBH}, one obtains the following upper bound for the entropy inside the spherical region (in units where $G_N =1$)
\beq
\begin{aligned}
    S\, \leq\, \mathfrak{c}_3\, \left(N\, A^{d-1} \right)^{\frac{1}{d}}\, ,
\end{aligned}
\label{eq:entropyvsA}
\eeq
which only depends on the area of the $\mathbf{S}^{d-2}$ surface and the number of species. It is important to stress the fact that the bound \eqref{eq:entropyvsA} can be saturated precisely when the system is close to collapse into a black hole. 

Crucially, it has been argued that the maximum information content of any given spacetime region with dynamical gravity must be bounded from above by its surrounding area, namely
\beq
\begin{aligned}
    S\, \leq\, \frac{A}{4}\, .
\end{aligned}
\label{eq:entropybound}
\eeq
This has been conjectured to be a fundamental property of quantum-gravitational systems and is moreover formulated in terms of the \emph{holographic principle} \cite{tHooft:1993dmi,Susskind:1994vu,Bousso:1999xy, Bousso:1999cb}, which posits that the entropy bound \eqref{eq:entropybound} hinges on the total number of independent degrees of freedom in an underlying theory of quantum gravity. Notice that an area-law behaviour rather than the common volume growth in field theory is familiar from our experience with certain gravitational systems such as black holes, which indeed saturate  \eqref{eq:entropybound}. Moreover, if we assume the number of species $N$ to be small (say of $\mathcal{O}(1)$), then for spacetime regions which are large in Planck units, the entropy of the system satisfies the gravitational bound with room to spare, namely
\beq
\begin{aligned}
    S\, \lesssim\, A^{\frac{d-1}{d}}\, \ll\, \frac{A}{4}\, .
\end{aligned}
\eeq
On the other hand, precisely when the size of the $\mathbf{S}^{d-2}$ hypersurface becomes of order $\ell_d$, one finds a naive violation of the bound \eqref{eq:entropybound}, which tells us that this is the minimum length where semi-classical gravity makes sense, in accordance with our discussion in Section \ref{ss:basics}. Interestingly, if $N$ becomes instead very large, then a comparison between \eqref{eq:entropyvsA} and the maximum holographic entropy leads to the conclusion that the minimum length-scale that Einstein gravity can reliably describe grows with the number of species as follows (see also \cite{Vafa:2024fti})\footnote{\label{fnote:minentropyBH}Notice that this simple argument implies that the minimum holographic entropy in the presence of a large number of species grows like $S_{\rm min} \gtrsim N$.}
\beq
\begin{aligned}\label{eq:specieslength}
    \ell_{\rm sp} = \ell_d\, \left(\frac{N}{4\pi} \right)^{\frac{1}{d-2}}\, ,
\end{aligned}
\eeq
which in terms of energy cut-offs would read as
\beq
\begin{aligned}\label{species}
    \LSP = \frac{\Mpd}{N^{\frac{1}{d-2}}}\, .
\end{aligned}
\eeq
This energy scale is usually referred to as the \emph{species scale}, and was introduced (and further discussed) in the context of gravitational interactions in \cite{Han:2004wt,Arkani-Hamed:2005zuc, Distler:2005hi, Dimopoulos:2005ac, Dvali:2007hz, Dvali:2007wp, Dvali:2010vm}. Note that when the number of light species becomes parametrically large --- which happens e.g., when probing infinite distance limits in field space --- the separation between the Planck scale and the actual quantum gravity cut-off may be rendered parametric as well. 

Alternatively, one may argue for the existence of a species scale using black hole physics, by defining the latter as determining the smallest possible black hole in the theory, which we review now. The argument relies on the fact that black holes of size given by $\ell_{\rm sp}=\LSP^{-1}$ (thus much larger than the Planck length, $\ell_d$) have a lifetime --- due to Hawking radiation --- roughly of $\mathcal{O}(\ell_{\rm sp})$, and hence they should already probe the microscopic theory of gravity \cite{Dvali:2007hz, Dvali:2007wp}. That is, the smallest size for a semi-classical black hole is given by $\ell_{\rm sp}$ instead of $\ell_d$. To see this, let us consider the decay rate of a $d$-dimensional black hole of mass $\MBH$ into $N$ light species. This is computed to be \cite{Dvali:2007wp} 
\beq\label{eq:BHmassloss}
	\frac{d\MBH}{dt}\, \sim\, -N T_{\text{BH}}^2\, ,
\eeq
where $T_{\text{BH}}=R_{\text{BH}}^{-1} \sim \left( \Mpd^{d-2}\, \MBH^{-1}\right)^{\frac{1}{d-3}}$ denotes its Hawking temperature, which is assumed to be much larger than the masses of the individual particle states (see Section \ref{ss:Planck&string} for a more refined computation). From this, one can estimate the lifetime of the black hole by integrating \eqref{eq:BHmassloss} up to a temperature of order of the species cut-off, as follows
\beq
	\tau\ \sim\ \frac{1}{N} \int_0^{\Mpd ^{d-2}/\LSP^{d-3}} \text{d}\mu\ \left( \frac{\mu}{\Mpd^{d-2}}\right)^{\frac{2}{d-3}}\ \sim\  \ell_{\rm sp}\, ,
\eeq
where the upper limit corresponds to the mass associated to $T_{\text{BH}} = \LSP$, and in the last step we have used \eqref{species}. This strongly supports the idea that semi-classical, neutral, non-rotating black holes must be larger than $\LSP^{-1}$, which should therefore be identified as the quantum gravity cut-off. 

\subsection{Perturbative analysis}
\label{ss:perturbative}

\begin{figure}[htb]
\begin{center}
\includegraphics[width=0.5\textwidth]{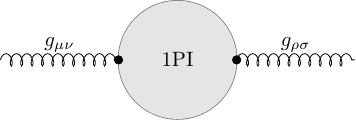}
\caption{\small 1PI graviton self-energy. At one-loop order, matter fields contribute to the diagram as explained in the text, c.f. eqs. \eqref{eq:gravitonpropagator} and \eqref{eq:selfenergy}.}
\label{fig:1PIgraviton}
\end{center}
\end{figure}

The existence of an energy cut-off in gravity which is smaller than the Planck scale (in the presence of a large number of light degrees of freedom) can be motivated as well using a different class of arguments that are perturbative in nature. The idea hinges on studying the contribution of quantum corrections to the Einstein-Hilbert term in the Wilsonian effective action due to the interaction of gravity with $N$ particle species (see also \cite{Caron-Huot:2022ugt} for a related discussion in the context of S-matrix bootstrap). Strictly speaking, since gravity is non-renormalizable (c.f. Section \ref{ss:basics}), there is no sensible way to absorb the momentum dependence of loop corrections into a running coupling constant \cite{Anber:2011ut, Heidenreich:2017sim}. However, one can still estimate the scale at which amplitudes including loops become significant, thus breaking down the perturbative series. Including such loop corrections from the $N$ species minimally coupled to the gravitational field then results in a weakening of gravity by a factor of $1/N$ \cite{Donoghue:1994dn,Aydemir:2012nz,Anber:2011ut,Calmet:2017omb,Han:2004wt}. To see this, one can take the resummed one-loop propagator of the graviton in Lorentzian signature (for concreteness we consider a 4d flat background here although the computation can be easily extended to higher dimensions)
\beq \label{eq:gravitonpropagator}
	\i \Pi^{\mu \nu \rho \sigma}= \i \left(P^{\mu \rho} P^{\nu \sigma} + P^{\mu \sigma} P^{\nu \rho}-P^{\mu \nu} P^{\rho \sigma} \right)\, \pi(p^2)\ ,
\eeq
with $P^{\mu \nu}= \eta^{\mu \nu}-\frac{p^{\mu} p^{\nu}}{p^2}$ a projection operator onto polarization states transverse to $p^{\sigma}$ (i.e. it satisfies $P^{\rho}_{\sigma} P^{\sigma}_{\kappa}=P^{\rho}_{\kappa}$) and\footnote{Here $N$ is a weighted sum of light degrees of freedom. In four dimensions one has $N=N_s/3+N_f+4N_V$, with $N_s$ the number of real scalars, $N_f$ that of Weyl spinors and $N_V$ denotes the number of vectors \cite{Han:2004wt,Aydemir:2012nz}.}
\beq \label{eq:selfenergy}
	\pi^{-1}(p^2)=2p^2 \left( 1-\frac{N p^2}{120 \pi \Mpf^2}\, \log (-p^2/\mu^2) \right )\, .
\eeq
Therefore, one can now ask what is the momentum $p^2$ of the external graviton states for which the perturbative expansion breaks down, which happens precisely when the second term inside the parenthesis of \eqref{eq:selfenergy} becomes comparable to the first one, i.e. the tree-level contribution. Notice that in the one-loop propagator above there is an additional energy scale, $\mu$, which is related to the renormalization of certain quadratic terms in the curvature that typically appear at next order in the derivative expansion of the gravitational EFT \cite{Aydemir:2012nz}. Ignoring the logarithmic factor, one indeed recovers \eqref{species} up to $\mathcal{O}(1)$ coefficients from the above expression, signalling the energy scale where gravity presumably becomes strongly coupled.\footnote{See also Appendix A.2 in \cite{Heidenreich:2017sim} for a simple argument extending these perturbative considerations to higher-point functions derived from the on-shell graviton S-matrix.}

It is important to stress here the fact that there is actually nothing special about the previous perturbative argument, which clearly resembles other similar calculations of the wave-function renormalization of massless scalar fields or in (non-Abelian) gauge theories. There, depending on the number of colours and/or matter content, one can also encounter similar enhancement factors. The crucial difference in the context of gravitational theories is that \emph{(i)} gravity couples to everything that carries energy and momentum, in contrast to the scalar/gauge scenario where only `charged' particles contribute, and \emph{(ii)} the interaction itself is universal (at least in its minimal version) --- per the Equivalence Principle \cite{Wald:1984rg}, which is captured in the linearized lagrangian by a term of the form
\begin{equation}
\mathcal{L}_{\rm int} \supset \kappa_d \left( T^{\mu \nu} h_{\mu \nu}\right)\, ,
\end{equation}
where $h_{\mu \nu}(x)$ is the perturbation around the background metric $\bar{g}_{\mu \nu}$, c.f. eq. \eqref{eq:metricfluctuations}.

Oftentimes, when trying to compute explicitly what is the species cut-off associated to a certain spectrum of relatively light fields (or towers), one resorts to simple counting techniques that are related to the above perturbative calculation. In fact, what one does in practice is to compute both the number of species and the cut-off $\LSP$ at the same time in a self-consistent way, namely upon imposing $N$ to account for those degrees of freedom which lie below the species scale itself. Later on in Section \ref{ss:Planck&string}, we will provide explicit examples of this procedure. However, it is important to remark that the counting technique should only be taken as a book-keeping device yielding just an approximate answer which can actually differ from the exact result, especially when dealing with infinite towers of massive states.

\section{The species scale in quantum gravity}
\label{s:speciesscale} 

The main goal of this thesis is to fully grasp the role and provide a proper definition of the relevant energy cut-off in gravity, understood as the scale at which quantum gravitational effects can no longer be neglected and local effective field theory breaks down. We have already introduced in Section \ref{s:speciesmotivation} some particular energy scale which depends on the number of light degrees of freedom existing in our theory, which we dub the species scale $\LSP$ (c.f. eq. \eqref{species}). Such cut-off was motivated using various different arguments, ranging from the perturbative physics of the massless graviton in the presence of a large amount of degrees of freedom, to a non-perturbative black hole/holography analysis which imposes some minimal length-scale in gravity. In this section, we want to provide a unifying picture from which all these considerations would naturally follow, as well as discuss the value of $\LSP$ in realistic string theory/quantum gravity vacua.

Therefore, the precise statement would be, following the logic of Section \ref{ss:basics}, that the lagrangian density of any gravitational effective field theory is organized according to the energy expansion
\beq
\mathcal{L}_{\mathrm{EFT}}\, =\, \sqrt{-g}\, \dfrac{1}{2\kappa_d^2}\left(\mathcal{R} + \sum_{n \geq 2} \frac{\mathsf{O}_n (\mathcal{R})}{\LSP^{n-2}}\right)\, +\, \ldots\, ,
\label{eq:gravEFTexpansion}
\eeq
where we have substituted $\LQG = \LSP$ in eq. \eqref{eq:gravEFT}, such that the species scale controls/suppresses \emph{generic} local gravitational corrections. The ellipsis in \eqref{eq:gravEFTexpansion} indicates any other couplings and matter fields present in the effective field theory. From this perspective it becomes clear how the previous (non-)perturbative arguments arise. In the former case, when computing any scattering amplitude involving graviton states in the external legs, as soon as we consider energies close to $\LSP$ we find that an infinite number of higher-curvature and higher-dimensional corrections become relevant, spoiling any naive analysis involving just the two-derivative term. On the other hand, for black holes which are small enough so as to probe the cut-off scale, one typically finds curvatures of order $\ell_{\rm sp}^{-2}$, thus inducing important corrections to e.g., the Bekenstein-Hawking entropy obtained solely from the Einstein-Hilbert action \cite{Sen:2005wa}.

Let us also mention that, in principle, different kinds of higher-dimensional operators can arise in a given gravitational theory, and not all of them need to be a priori suppressed by a single ultra-violet scale \cite{Donoghue:1995cz}. For instance, one may get threshold contributions in the Wilsonian effective action once we integrate out some massive particle(s), and which can ultimately dominate over the suppression given by the quantum gravity cut-off in \eqref{eq:gravEFTexpansion}. Still, this should not be regarded as a drawback in the identification of the species scale from higher-curvature operators, since these couplings naturally depend on the energy scale at which they are measured. Hence, a more accurate statement would be that the suppression of generic higher-curvature operators in the gravitational EFT is controlled by $\LSP$ when measured at energies close to the cut-off itself, see Chapter \ref{ch:Higherdimops} for details on this point.

\subsection{Why you get old}
\label{ss:Planck&string}

The fact that the Planck scale does not actually provide for the ultra-violet cut-off in effective field theories consistent with quantum gravity is strongly supported by the string theory landscape. There, the lower dimensional Planck mass usually depends on the details of the string theory embedding, such as the fundamental vibrating string that we start with as well as the size of the compact internal dimensions. In this section we analyze the behaviour of the species cut-off \eqref{species} in realistic string theory constructions. Along the way, we will realize that it matches the expectations both in Kaluza-Klein theories of gravity and string theory, therefore providing for a unifying concept that defines the maximum regime of validity of any given EFT weakly coupled to Einstein gravity, regardless of its UV completion. Our strategy here will consist in focusing on asymptotic regions within field space, where the species counting proceeds in an easier way thanks to the weak coupling behaviour exhibited by the theory, which helps in determining both its massive spectrum as well as in organizing the EFT expansion according to eq. \eqref{eq:gravEFTexpansion}. Indeed, in these extreme regimes the Distance Conjecture \cite{Ooguri:2006in} predicts the appearance of infinite towers of exponentially light particles, which become asymptotically stable resonances of the theory. Moreover, per the Emergent String Conjecture \cite{Lee:2019wij}, one only expects two such different scenarios to emerge: either the theory decompactifies to a higher-dimensional one, or one reaches a weak coupling limit for an emergent critical string (not necessarily the original one). The relevant towers of nearly massless states thus become Kaluza-Klein replica --- with spin $s \leq 2$, or rather the excitation modes of a fundamental string, which include higher-spin particles as well. Hence, in what follows we will consider both scenarios in turn, analyzing the behaviour of the species scale in the presence of the aforementioned light towers.
	
\subsubsection{Species scale from Kaluza-Klein towers}\label{sss:KKtowersspecies}
	
Consider a $d$-dimensional EFT describing the physics of $N_0$ massless/light modes, weakly coupled to Einstein gravity. In particular, let us assume this theory as coming from the dimensional reduction of some $(d+k)$-dimensional gravitational EFT. For simplicity, we take the higher-dimensional theory to be compactified on an isotropic and rectangular torus $\mathbf{T}^k$, of radius $R$ measured in $(d+k)$-dimensional Planck units. That is, in our conventions $R$ is dimensionless and the physical size of the $k$-torus is given by $\left( 2\pi R\, \ell_{d+k}\right)^k$, where $\ell_{d+k}$ denotes the higher-dimensional Planck length. For this set-up, the Planck scales of the lower and higher dimensional theories are related as follows
\beq
	\Mpd^{d-2}\, =\, M_{\text{Pl};\, d+k}^{d-2}\, (2\pi R)^k\, ,
	\label{eq:Planckscales}
\eeq
whereas the mass scale of the $k$ corresponding  KK towers reads\footnote{Each of these towers is indeed associated to internal momentum of the higher dimensional fields along the $k$ different directions in the compactification space. Notice that particles with multiple KK charges do exist, and thus the degeneracy of states grows roughly as the product of the maximum excitation numbers, see Section \ref{ss:MultipleTowers}.}
\beq\label{eq:KKmass}
	\MKK\, =\, \dfrac{M_{\text{Pl};\, d+k}}{R}\, .
\eeq
If we approach now the large volume point, namely the limit $R\to \infty$, the Kaluza-Klein towers become light in an exponential fashion with the proper field distance, as predicted by the Distance Conjecture (see Section \ref{s:SDC}). With this, we can then determine the species scale associated to such a dense tower of light states. We will use perturbative and non-perturbative arguments to compute $\LSP$ in the present case, leading both ultimately to the same quantitative answer.

\subsubsection*{Determining $\LSP$ via species counting}

Let us assume that every light mode in the $d$-dimensional EFT has its own KK replicas. Equivalently, we think of the lower dimensional theory to be propagating as well in the higher-dimensional bulk, in contrast to when e.g., it is localized on a brane. As a consequence, one can estimate the total number of states $N$ below the UV cut-off as follows
\beq
	N\, \simeq\, N_0 \left( \frac{\LSP}{\MKK} \right)^{k}\, .
	\label{eq:numberKKmodes}
\eeq
Combining now eqs. \eqref{species} and \eqref{eq:Planckscales}, we arrive at the following species cut-off
\beq\label{eq:speciesscaleKK}
	\LSP^{d+k-2}\, \simeq\, \frac{\Mpd^{d-2}\, \MKK^k}{N_0}\, \simeq\, \frac{M_{\text{Pl};\, d+k}^{d+k-2}}{N_0}\, .
\eeq
Notice that what we obtain is precisely the species scale associated to the higher dimensional theory, including the $N_0$ fields already present there. That is, the species scale of the $d$-dimensional EFT provides the right cut-off that one would expect to find in the UV theory, namely the $(d+k)$-dimensional $\LSP$.\footnote{Notice that this simple argument makes manifest the fact that the definition/concept of species scale is consistent (or preserved) under dimensional reduction.} 
Hence, the intuitive result that the higher-dimensional Planck mass captures the quantum gravity cut-off is thus recovered when $N_0=\mathcal{O}(1)$.

\subsubsection*{Determining $\LSP$ via black holes}

One can arrive at essentially the same result for $\LSP$ upon studying black holes of minimal size instead (c.f. Section \ref{ss:nonperturbative}). This allows us to give more evidence for the picture advocated in this chapter. In fact, we can argue for the relation \eqref{eq:speciesscaleKK} in two different ways: \emph{(i)} by determining the size of the black hole whose microstates saturate the minimum holographic entropy $S_{\rm min} \gtrsim N$ (see \cite{Blumenhagen:2023yws, Calderon-Infante:2023uhz} for similar considerations), and \emph{(ii)} upon finding those black holes whose typical evaporation time is already of the order of their size \cite{BHdecaynotes}. 

Let us consider first the microscopic entropy associated to the smallest possible neutral black hole that can be constructed in the theory, namely one with radius $\mathsf{R}= \ell_{\rm sp}$. Its associated mass would read as
\beq\label{eq:massminimalBH}
	M_{\rm BH,\, min} = \frac{(d-2) \pi^{\frac{d-3}{2}}}{8\, G_N\, \Gamma (\frac{d-1}{2})}\, \ell_{\rm sp}^{d-3}\, .
\eeq
We can now estimate the contribution of \emph{uncharged} Kaluza-Klein modes to the Bekenstein-Hawking entropy of such minimal black holes by finding the total number of microstates compatible with the constraint \eqref{eq:massminimalBH}. For simplicity, we work in the following with a KK spectrum corresponding to a single decompactifying dimension of topology $\mathbf{S}^1/\mathbb{Z}_2$. In that case, the maximum occupation number in the tower that can contribute to the BH entropy is
\beq
	\mathsf{N}= \frac{M_{\rm BH,\, min}}{\MKK}\, ,
\eeq
whilst the total number of microstates comprised by combinations of Kaluza-Klein modes is given by the number of partitions of $\mathsf{N}$. In the decompactification limit, where this quantity becomes large, we can approximate the total amount of microscopic configurations as follows
\beq
	\Omega = p(\mathsf{N}) \sim \frac{1}{4\, \sqrt{3}\, \mathsf{N}} \exp \left( \sqrt{\frac{2\, \pi^2\, \mathsf{N}}{3}} \right)\, , \qquad \text{for}\ \ \mathsf{N} \gg 1\, ,
\eeq
where we used the Hardy-Ramanujan formula above. Therefore, the entropy associated to a black hole of mass $M_{\rm BH,\, min}$ would be given by 
\begin{align}\label{eq:statentropyKK}
	S_{\rm BH,\, min} := \log \Omega = \sqrt{\frac{2\, \pi^2\, \mathsf{N}}{3}}\, +\, \mathcal{O} \left(\log \mathsf{N}\right)\, \sim\, \left( \frac{\ell_{\rm sp}^{d-3} \Mpd^{d-2}}{\MKK}\right)^{1/2}\, \sim\, N\, ,
\end{align}
where in the last step we made use of the definition of $\LSP$ as well as $N= \LSP/\MKK$. This is in precise agreement (up to numerical factors) with the minimum holographic entropy in the presence of $N$ Kaluza-Klein species, such that we conclude that $\ell_{\rm sp}$ --- as defined in \eqref{eq:specieslength} --- correctly determines the minimum black hole size. Equivalently, one may ask for the point where the statistical entropy \eqref{eq:statentropyKK} equals that of a black hole of mass $M_{\rm BH}$, which happens when 
\begin{align}\label{eq:transitionpointBHtoKK}
	M_{\rm BH}^{d-1}\, \sim\, \frac{\Mpd^{2(d-2)}}{m_{\rm KK}^{d-3}} \Longrightarrow R_{\rm BH}\, \sim\, m_{\rm KK}^{\frac{-1}{d-1}}\, \Mpd^{\frac{2-d}{d-1}}\, ,
\end{align}
i.e. precisely when the black hole has a size of order of the species cut-off $\ell_{\rm sp}$, c.f. eq. \eqref{eq:speciesscaleKK}.

On the other hand, it is possible to argue for \eqref{eq:massminimalBH} as capturing the minimal semi-classical black hole mass by studying its typical evaporation time (via Hawking radiation) in the presence of the KK tower. Indeed, consider the generic case where the internal space is given by some Ricci-flat $k$-dimensional compact manifold $\mathcal{X}_k$, with a mass spectrum of the form\footnote{The relation \eqref{eq:spectrumLaplaciannmfd} follows essentially from the WKB approximation, which can be understood here as simply saying that for highly excited modes, the spectrum of the laplacian behaves roughly as in e.g., a toroidal manifold with the same number of dimensions.}
\begin{align}\label{eq:spectrumLaplaciannmfd}
	m_n\, \sim\, n\, \MKK\, , \qquad \text{for}\ \ n \gg 1\, .
\end{align}
Thus, to compute the evaporation time one needs to solve the following differential equation
\begin{equation}\label{eq: deg t}
    \frac{d M_{\rm BH}}{d t}\, \sim\, - T_{\rm BH}^2\sum_{n=0}^{\infty}d_{n}(\mathcal{X}_k)\left(e^{-\frac{\MKK}{T_{\rm BH}}}\right)^{n}\,,
\end{equation}
where $d_n(\mathcal{X}_k)$ denotes the degeneracy of each mass eigenmode $m_n$ in the tower (i.e. the number different states with $m=m_n$) and we have included an additional Boltzmann suppression factor $e^{-\frac{m_n}{T_{\rm BH}}}$ with respect to \eqref{eq:BHmassloss}. Moreover, using Weyl's asymptotic formula
\begin{equation}
    N(\lambda)\, \sim\, \lambda^{k/2}\, \text{vol} \left( \mathcal{X}_k\right),\qquad \text{for}\ \ \lambda\to\infty\, .
\end{equation}
which accounts for the number of accumulated eigenvalues $\lambda$ of the Laplace-Beltrami operator acting on functions defined over $\mathcal{X}_k$, we can readily estimate the degeneracy to be\footnote{Even though \eqref{eq:asymptoticdegeneracyKK} gives only an approximation for large KK masses, in certain cases it may be possible to determine $d_n(\mathcal{X}_k)$ exactly. For instance, if $\mathcal{X}_k$ is a $k$-sphere, one finds (c.f. eq. \eqref{eq:deg10-dsphere})
\begin{equation}\label{eq:asymptoticdegeneracyKK}
    \notag d_n(\mathbf{S}^k)=\frac{(n+k-2)!(2n+k-1)}{n!(k-1)!}\, \sim\, \frac{2}{(k-1)!}n^{k-1}\, .
\end{equation}
}
\begin{equation}
    d_n(\mathcal{X}_k) = N(m_n)-N(m_{n-1})\, \sim\, n^{k-1}\, .
\end{equation}
With this in mind, we can then go back to \eqref{eq: deg t} and rewrite it as follows
\begin{align}
     \frac{d M_{\rm BH}}{d t}\, &\sim\, -T_{\rm BH}^2\sum_{n=0}^{\infty}n^{k-1}\, e^{-n\, \frac{\MKK}{T_{\rm BH}}}\, =\, -T_{\rm BH}^2\, \left( -\frac{d}{dx}\right)^{k-1} \frac{1}{1-e^{-x}}\bigg\rvert_{x=\frac{\MKK}{T_{\rm BH}}} \notag\\
     &=-T_{\rm BH}^2\, (k-1)!\,  \left(\frac{T_{\rm BH}}{\MKK}\right)^k\, +\, \mathcal{O} \left( \left(\frac{\MKK}{T_{\rm BH}}\right)^0 \right)\, ,
\end{align}
where we have first resummed the series in terms of an auxiliary variable $x=\frac{\MKK}{T_{\rm BH}}$ and subsequently expanded the result assuming $T_{\rm BH} \gg \MKK$ all along the evaporation process. Hence, integrating the previous differential equation yields 
\begin{align}\label{eq: tau int}
    \tau&\sim \MKK^k\int_0^{M_{\rm BH,\, min}}T_{\rm BH}(\mu)^{-k-2}\dd \mu\, \sim\, \MKK^k\, M_{{\rm Pl;}\, d}^{-\frac{(k+2)(d-2)}{d-3}}\int_0^{M_{\rm BH,\, min}}\mu^{\frac{k+2}{d-3}}\dd \mu\notag\\
    &\simeq\, \MKK^k\, M_{{\rm Pl;}\, d}^{-\frac{(k+2)(d-2)}{d-3}} M_{\rm BH,\, min}^{\frac{d+k-1}{d-3}}\, \sim\, \MKK^k\, M_{{\rm Pl;}\, d}^{d-2}\, \LSP^{-d-k+1}\,,
\end{align}
where in the last step we have substituted \eqref{eq:massminimalBH}. Finally, upon inserting \eqref{eq:numberKKmodes} as well as the definition of the species scale we find that
\begin{equation}
    \tau\, \sim\, \ell_{\rm sp}\, ,
\end{equation}
in agreement with our previous arguments.
	
	
\subsubsection{Species scale from string towers}\label{sss:stringtowersspecies}
	
We turn now to the case in which the tower of states is provided by the excitation modes of a fundamental critical string that becomes asymptotically tensionless. Here the situation is dramatically different, since as it is well-known the associated higher-spin towers present a much denser spectrum than their Kaluza-Klein counterparts. The reason for this is twofold: first, in flat backgrounds, the string oscillators are characterized by having a mass given roughly by the Regge excitation pattern 
\beq
	m_n^2\,=\, 8\pi T_s \left(n-1\right)\, ,
\label{eq:stringmasses}
\eeq
where $T_s=2 \pi m_s^2$ denotes the string tension and $n \in \mathbb{N}$ refers to the excitation/oscillator level of the critical string. Second, one has to take into account the high level density of modes, $d_n$, associated to the very massive excited states. In fact, to leading order, for large $n$  this degeneracy behaves in an exponential fashion \cite{Green:2012oqa}
\begin{align}\label{eq:exactleveldensitystrings}
     d_n & \sim\, n^{-11/2}\, \e^{4\pi \sqrt{2n}}\, , \qquad \qquad \ \text{for Type II strings}\, , \notag\\
     d_n\, &\sim\, n^{-11/2}\, e^{2 \pi (2+\sqrt{2}) \sqrt{n}}\, , \qquad \text{for Heterotic strings}\, .
\end{align}
With these ingredients we can then study both perturbatively and non-perturbatively the behaviour of the species scale in the presence of string oscillator towers. As we will explicitly demonstrate, both approaches lead to the same \emph{qualitative} answer, even though they might differ quantitatively due to the validity of the corresponding computational methods.

\subsubsection*{Determining $\LSP$ via species counting}
	
We proceed first via the usual state counting algorithm, which is based on the perturbative argument provided in Section \ref{ss:perturbative}. Let us stress that since that discussion was phrased using purely field-theoretic considerations, a direct application of these ideas to an infinite set of particles containing higher-spin states is strictly speaking not suitable anymore. However, this crude prescription will provide us with some potential candidate for $\LSP$ that is qualitatively similar to the exact one, see below. Moreover, the route followed in this section will also be amenable to certain string theory applications discussed in Part \ref{part:StringTheoryTests} of this thesis. Hence, it is instructive to perform the exercise at this point.

In order to be as general as possible, let us consider some $d$-dimensional EFT where we take the weak coupling limit corresponding to some fundamental string. Consequently, the associated oscillator modes become asymptotically massless when measured in Planck units, since their masses are proportional to 
\beq\label{eq:ddimdilaton}
	\frac{m_s}{\Mpd} = (4 \pi)^{2-d}\, g_d^{2/d-2} \to 0\, ,\qquad \text{as}\ \ g_d \to 0\, ,
\eeq
where $g_d$ denotes the $d$-dimensional string coupling. To compute $\LSP$ we first need to know what is the maximum excitation level $\Ns$ whose mass lies at or below the species cut-off. Subsequently, we need to count the number of accumulated string modes up to such $\Ns$, which we denote by $N$ in the following. The species scale thus fulfills
\beq
\label{eq:speciescale}
	\LSP^{d-2}\, \simeq\, \frac{\Mpd^{d-2}}{N}\, \simeq\, \Ns^{\frac{d-2}{2}}\, \Ms^{d-2}\, .
\eeq
Rearranging the above equation we arrive at
\beq
\label{eq:Ns}
	\Ns^{\frac{d-2}{2}} \sum_{n=1}^{\Ns} d_n \, \simeq\, \left ( \frac{\Mpd}{\Ms}\right )^{d-2}\, ,
\eeq
where we have substituted $N =\sum_{n=1}^{\Ns} d_n$ in terms of the sum over the density levels of physical states in the string tower. Next, one needs to substitute the explicit form \eqref{eq:exactleveldensitystrings} of the degeneracy of the oscillators, and then solve for $\Ns$. In what follows, we will approximate $d_n$ by the simplified quantity
\beq
	d_n\, \sim\, e^{ \sqrt{n}}\, ,
	\label{eq:leveldensity}
\eeq
since this is enough for our purposes here and already captures the key asymptotic behaviour of $\LSP$.\footnote{See Appendix A of \cite{Castellano:2022bvr} for a more accurate computation in all the relevant string theories considered in this thesis.} Upon doing so, one finds 
\beq
	\left ( \dfrac{ \Mpd}{\Ms}\right )^{d-2}\, \simeq\, \Ns^{\frac{d-2}{2}} \sum_{n=1}^{\Ns} e^{\sqrt{n}} \, \sim\, 2\, \Ns^{\frac{d-1}{2}} e^{\sqrt{\Ns}}\, ,
\label{eq:maxstringlevel}
\eeq
where we have retained just the leading order term in the sum, which is justified in the limit where $\Ns \to \infty$. From this, we can obtain an explicit expression for $\Ns$ that reads
\beq
	\sqrt{\Ns}\, \sim\, (d-1)\, W_0 \left( \dfrac{1}{(d-1)\, 2^{\frac{1}{d-1}} } \left[ \dfrac{\Mpd}{\Ms} \right]^{\frac{d-2}{d-1}} \right)\, ,
\eeq
where $W_0$ refers to the principal branch of Lambert $W$ function.\footnote{\label{fn:Lambert} The Lambert $W$ function is defined as a solution to the equation
\beq
	\notag y\ e^y = x \iff y= W (x)\, .
\eeq
It has two real branches, namely  the principal branch, denoted  $W_0 (x)$ (defined for $x\geq 0$),  and the (-1)-branch, denoted $W_{-1}(x)$ (defined for $ -\frac{1}{e} \leq x < 0$). The asymptotic expansions that are relevant for this work take the form \cite{Lambert}
\beq
\label{eq:Wexpansion}
		\begin{split}
			&W_{0}(x\to \infty)\, =\, \log (x)-\log (\log(x))+\ldots\, ,\\
			&W_{-1}(x\to 0^-)\, = \, \log (-x)-\log (-\log(-x))+\ldots\, .
		\end{split}
\eeq
}
Crucially, the above equation reveals that the maximum excitation level of the string that falls below the cut-off scale diverges when $\Ms/\Mpd \to 0$, thereby confirming the approximations taken so far. Indeed, upon using the relevant expansion of the $W$-function from eq. \eqref{eq:Wexpansion}, such divergence can be seen to behave essentially in a logarithmic fashion, i.e.
\beq
	\sqrt{\Ns}\, \sim\, (d-2)\, \log \left(\dfrac{\Mpd}{\Ms}\right) + \mathcal{O}\left(\log \left( \log (\Mpd/\Ms) \right) \right)\, .
\eeq
This means that, according to this prescription, the species scale for a critical emergent string in $d$ spacetime dimensions would behave as  
\beq\label{eq:speciesforstringspert}
	\frac{\LSP}{\Mpd}\, \simeq\, \sqrt{\Ns}\, \frac{\Ms}{\Mpd}\, \sim\,  (d-2)\,  \dfrac{\Ms}{\Mpd}\  \log \left(\dfrac{\Mpd}{\Ms}\right)\, ,
\eeq
thus confirming our expectations that the quantum gravity cut-off should be given in this case by the string scale itself. Nonetheless, there seems to be important logarithmic corrections in \eqref{eq:speciesforstringspert} which are crucial in order to make sense of the state counting between $\LSP$ and $\Ms$, since they encode the asymptotic behaviour of $\Ns$. As we argue in what follows, however, we believe these additional factors to be unphysical as well as an artifact of using the field theoretic approach for a higher-spin tower.

\subsubsection*{Determining $\LSP$ via black holes}

We study now the same question from the perspective of black hole physics, namely we want to know what is the minimum size for a semi-classical black hole in the presence of a nearly tensionless fundamental string. Following the same logic as in the Kaluza-Klein scenario, we can solve the problem using two different approaches.

First, we want to find the transition point where the minimum holographic entropy is attained. Taking advantage of what we learned in the previous case, we can easily 
estimate this upon determining the mass (equivalently the radius) of the black hole where its entropy can be completely accounted for by the string oscillator modes \cite{Susskind:1993ws}. Let us thus confront both entropy functions. On the one hand, the entropy of a free string is proportional to its length \cite{Green:2012oqa} 
\beq
	S_{\rm string}\, \sim\, \frac{L}{\ell_s}\, \sim\, \frac{M_{\rm BH,\, min}}{m_s}\, ,
\eeq
whilst that of a (minimal) black hole reads as follows
\beq
	S_{\rm BH}\, \sim\, \left( R_{\rm BH,\, min}\, \Mpd \right)^{d-2}\, \sim\, \left(\frac{M_{\rm BH,\, min}}{\Mpd} \right)^{d-2}\, .
\eeq
Hence, by comparison between the two, we find
\begin{align}
	M_{\rm BH,\, min}\, \sim\, \frac{\Mpd^{d-2}}{m_s^{d-3}} \Longrightarrow R_{\rm BH,\, min}\, \sim\, \ell_s\, ,
\end{align}
namely the minimum black hole size is precisely of order of the string length. Stated differently, we reach the conclusion that $\LSP \simeq m_s$. Note that the previous analysis is actually rather crude, essentially because we ignored the self-gravitating effects of the string when comparing both behaviours.\footnote{In a sense, we were entitled to do so since by assumption we consider a regime of weak string coupling.} It turns out, however, that one can do slightly better and properly take into account these effects. This was done in \cite{Horowitz:1996nw,Horowitz:1997jc}, which ultimately led to the black hole-string correspondence, where it is claimed that in fact a typical uncharged black hole undergoes a transition to a highly excited and long (but compact) string precisely when it reaches a size of the order of $\ell_s$. (See also \cite{Chen:2021dsw} for a recent analysis of the black hole-string transition in Heterotic and Type II string theories.)

On the other hand, one may also argue for $\LSP \simeq m_s$ by studying the typical decay time of a black hole of stringy size. Indeed, the Hawking evaporation process in the presence of the string oscillator modes would read as
\begin{equation}\label{eq:BHtostringdecay}
    \frac{d M_{\rm BH}}{d t}\, \sim\, - T_{\rm BH}^2\sum_{n=1}^{\infty}n^{-\frac{11}{2}}\, \e^{\left(\beta_{\rm H}-\beta_{\rm BH}\right)\, m_n}\, ,
\end{equation}
where $\beta_{\rm BH} = 1/T_{\rm BH}$ and $\beta_{\rm H}$ denotes the inverse Hagedorn temperature of the corresponding critical string (which is thus proportional to the string length $\ell_s$, c.f. eq. \eqref{eq:exactleveldensitystrings}). Therefore, it is easy to check that when $T_{\rm BH} \ll T_{\rm H}$ (equivalently $\beta_{\rm BH} \gg \beta_{\rm H}$) one obtains
\begin{equation}
    \tau\, \sim\, \frac{e^{\frac{T_{\rm BH}}{T_{\rm H}}}}{T_{\rm BH}}\, \left( \frac{\Mpd}{T_{\rm H}}\right)^{d-2} \gg T_{\rm BH}^{-1}\, \sim\, R_{\rm BH}\, ,
\end{equation}
whereas if $T_{\rm BH} > T_{\rm H}$ the calculation \eqref{eq:BHtostringdecay} breaks down completely since at the Hagedorn point the thermal ensemble should stop being well-defined, even at large distances from the black hole. In fact, precisely when $\beta_{\rm BH} \gtrsim \beta_{\rm H}$ (with $\beta_{\rm BH}-\beta_{\rm H} \ll \ell_s$) it is the Horowitz-Polchinski saddle the one that dominates the dynamics of the system \cite{Horowitz:1997jc}.

\subsection{An algorithmic procedure in the presence of multiple towers}
\label{ss:MultipleTowers}

To end this section, let us address the more general scenario in which several towers with different charges and in principle distinct spectra are present. As we will discuss in later parts of this thesis, the following analysis is especially relevant in the context of string theory, since it is typically the case that not just one but actually several towers become light when probing asymptotic regions in field space. In particular, one could imagine a situation where as we approach an infinite distance point in moduli space, multiple towers get asymptotically massless and lie below the species cut-off, such that each one of them should contribute a priori to its computation --- since they all couple to the gravitational field. Given this state of affairs, one can think of two qualitatively different situations that may arise, depending on whether or not the towers form bound states between each other. We discuss each of them in turn in what follows.
	
\subsubsection*{Case I: Additive species}

 Let us first consider a set-up in which there is no mixing between the towers. This happens e.g., when we have two (infinite) sets of particles that couple to different fields and such that no states with both charges are present in the spectrum (of quasi-stable modes). In this case, the total number of species below the quantum gravity cut-off is given by $N_{\mathrm{tot}}= \sum_i N_i$, where $N_i$ is the number of states associated to the $i$-th tower. The species scale, as defined by eq. \eqref{species}, should then be computed by including all light particles arising from every tower. However, since the number of species --- below some fixed energy scale $\Lambda_{\rm UV}$ --- of any given set of light states diverges in the asymptotic limit, $N_{\mathrm{tot}}$ will be dominated essentially by just one of the towers (unless all $N_i$ scale in the same way with respect to the relevant moduli, in which case the following calculation would simply be modified by $\mathcal{O}(1)$ factors). In practice, we can calculate the would-be cut-off associated to each tower separately. Thus, consider a set of states with mass spectrum given by
\begin{equation}
\label{eq:Mni}
		m_{n,\, i}\, =\, n^{1/p_i}\ \Mti\, .
\end{equation}
Here, the parameter $p \in \mathbb{R}$ characterizes the spacing between different steps within the tower (of constant degeneracy), or equivalently it can be regarded as counting the number of towers with identical mass gap. (For instance, a standard Kaluza-Klein tower associated to a circular extra dimension has $p=1$.) In practical terms, however, it should be seen just as a book-keeping device which is useful when solving counting exercises as in the present case.\footnote{See \cite{Casas:2024ttx} for a recent example of a truly $p=2$ tower of particles.} We can then compute the species scale as follows
\begin{equation}\label{eq:speciesscaleithtower}
		\LSP{}_{,\, i}\, \simeq\, N_i^{1/p_i}\Mti\, \simeq\, \dfrac{\Mpd}{N_i^{\frac{1}{d-2}}}\, .
\end{equation}
The physical value for $\LSP$ would be given by the minimum out of the set $\{ \LSP{}_{,\, i} \}$, since it is dominated by the states corresponding to the tower with the lightest mass scale. For concreteness, let us label the leading tower by the index 1, characterized by the density parameter $p_1$ and mass gap $\frac{m_{\text{tow,}\, 1}}{\Mpd} \sim t^{-a_1}$, which goes to zero as we take some modulus $t$ to infinity. For all the remaining towers, we consider $\frac{\Mti}{\Mpd} \sim t^{-a_i}$. Since the set of light states is dominated by the first tower alone, we have $N_{\text{tot}}=N_1+\ldots\, \sim\, N_1$, and
\begin{equation}\label{eq:speciesadditive}
		N_1\, \sim\, t^{\frac{a_1 p_1(d-2)}{ d-2+p_1}}\, , \qquad  \LSP\, \sim\, \Mpd\, t^{-\frac{ a_1 p_1}{d-2+ p_1}}\, .
\end{equation}
To check the consistency of this picture, we can recalculate the number of states associated to the subleading towers --- i.e. the ones with $i\geq 2$ --- that lie below the $\LSP$ just determined. These read
\begin{equation}
		\tilde{N}_i\, \sim\, t^{\frac{(d-2) a_i p_i+ p_1 p_i (a_i - a_1)}{d-2+p_1}}\, ,
\end{equation}
and we find $\tilde{N}_i \ll N_i$, as expected. Note that this expression contains negative contributions in the exponent, corresponding to the case where a tower is still too heavy to provide any mode below $\LSP$.
	
Let us remark also that a stringy tower would correspond in this language to having $p \to \infty$.\footnote{To be precise, one would need to include the exponential degeneracy together with $p=2$, as we did in the previous section. Nonetheless, for the case at hand most of the calculations yield the correct result if we model such high degeneracy by just taking the limit $p \to \infty$. } It is thus clear that such a spectrum would completely dominate any set-up in which the string oscillators have to be taken into account, thus recovering the expected result $\LSP \sim m_s$ (see Section \ref{sss:stringtowersspecies} for details).

\subsubsection*{Case II: Multiplicative species}
Second, we consider the scenario in which the different towers are such that states with mixed charges can be present. This is the case that is actually relevant in most set-ups. It is realized by e.g., several KK towers since we can have states with non-vanishing momentum along various internal directions at the same time; or even in the presence of a fundamental string and Kaluza-Klein modes. In this case, the total number of species is not just additive, but rather multiplicative, i.e. $N_{\text{tot}}\simeq \prod_i N_i$. Thus, let us assume that we have $k \in \mathbb{N}$ generating towers of (asymptotically) light particles behaving again like \eqref{eq:Mni}, such that states with more than one non-vanishing occupation number appear in the spectrum, with a mass formula of the general form
\begin{equation}\label{eq:massmixedsprectra}
		m_{n_1\,  \ldots  n_k}^2\, =\, \sum_{i=1}^k n_i^{2 / p_i} \Mti^2\, .
\end{equation}
Then, we can compute an effective mass and density parameters which read as follows
\begin{equation}
		\label{eq:Meffpeff}
		\Mteff =\left(m_1^{p_1} m_2^{p_2} \ldots m_k^{p_k}\right)^{1 / \sum_i p_i}\, , \qquad \peff =\sum_i p_i\, ,
\end{equation}
with the species number and cut-off scale given in terms of these by
\begin{equation}
		\label{eq:NtotLQGeff}
		\Ntot\, \simeq\, \left( \dfrac{\Mpd}{ \Mteff} \right) ^{\frac{(d-2)\peff}{d-2+\peff}}\, , \qquad   \LSP\, \simeq\, \Mpd \, \left( \dfrac{\Mpd}{ \Mteff} \right)^{-\frac{\peff}{d-2+\peff}} \, .
\end{equation}
The maximum occupation number for each tower can also be determined by combining this last equation with $\LSP \simeq N_i^{1/p_i}\Mti$. Additionally, one can again parameterize the mass scale of the towers by $\frac{\Mti}{\Mpd} \sim t^{-a_i}$ and obtain the expression for the species cut-off as a function of the modulus $t$ when it becomes large.
	
Let us emphasize that, as opposed to the previous case, it is not true anymore that the leading tower always determines \emph{alone} the value of the species scale. In fact, even in the case of a parametrically smaller mass for one tower, additional ones can still significantly contribute with a divergent number of states below $\LSP$. Hence, it is in general not enough to know which tower has the lightest mass scale, but rather all the towers that lie below the cut-off. To study systematically which towers actually lie below the species scale and contribute to eqs. \eqref{eq:Meffpeff}-\eqref{eq:NtotLQGeff}, one can perform an iterative process by starting with the lightest one first, calculating its associated species scale, and checking whether it lies above the first step of the second to lightest tower. If so, the latter must be included, the species scale should be properly recalculated and then the third tower must be subsequently checked. The algorithm must be carried on until we find a tower that lies above the species scale and thus need not be included. If $i$ towers contribute to the species scale (and have associated effective mass scale and density parameter denoted by $m_{\text{tow,}\, (i)}$ and $p_{\text{eff,}\, (i)}$), the condition to check whether the $(i+1)$-th tower also lies below the such scale can be easily stated as follows
\begin{equation}\label{eq:algorithmstops}
		m_{\text {tow,}\, (i)}\, \leq\, m_{\text {tow, }\,  i+1}^{\frac{d-2+p_{\text{eff,}\, (i)}}{p_{\text{eff,}\, (i)}}}\, .
\end{equation}
If the inequality is fulfilled, the $(i+1)$-th tower lies above the species scale and can be safely ignored for the purposes of the present computation, otherwise it must be included and the process continues until we find some other tower satisfying \eqref{eq:algorithmstops}. To get an intuition of what this algorithm does in physical terms, let us consider the case of two single Kaluza-Klein towers of $p_1=p_2=1$, and assume that they become light at different rates. In that case, the species cut-off computed via the lightest KK tower is roughly given by the Planck scale of the higher dimensional theory where we decompactify the corresponding internal cycle. Hence, the fact that the second KK would still be needed to be accounted for means that it becomes actually asymptotically massless when measured in higher dimensional Planck units. This would then signal the necessity of a further decompactification, such that the resulting species scale would correspond to the highest Planck mass, which can be easily checked upon substituting \eqref{eq:KKmass} into eqs. \eqref{eq:Meffpeff} and \eqref{eq:NtotLQGeff} (see Part \ref{part:pattern} for explicit examples of this).

Before concluding, let us mention that in the presence of some critical string becoming light, the inequality is automatically fulfilled for heavier towers (since $p_{\text{eff,}\, (i)} \to \infty\, $), and the species scale is saturated by the string oscillator modes, as expected. Notice, however, that this still allows for the possibility of having a limit in which the lightest tower is of KK-type but we still find $\LSP \sim m_s$ if the former does not saturate the species scale before the tensionless string kicks in. 

\section{Summary}

In this chapter of the thesis, we have investigated the regime of validity of generic effective field theories in the presence of gravitational interactions. The main focus have been placed on elucidating which energy scale signals the breakdown of semi-classical Einstein gravity, which we dubbed the \emph{quantum gravity cut-off}, $\LQG$. To do so, we first took advantage of our experience with other non-renormalizable quantum field theories so as to propose some potential candidate for the latter, which turned out to be given by the Planck scale $\Mpd$. 

Later on, in Section \ref{s:speciesmotivation} we revisited this question in the presence of a large amount of light degrees of freedom, dubbed \emph{species}. Thereby, using both perturbative arguments (based on the behaviour of the quantum series of the graviton), as well as non-perturbative considerations (rooted in black hole physics and the holographic principle), we arrived at an alternative energy cut-off that takes into account the matter content in the theory, which is usually referred to as the \emph{species scale}, $\LSP$ \cite{Dvali:2007hz, Dvali:2007wp}. Moreover, this quantity appears to be bounded from above by our previous estimation, namely $\Mpd$, but crucially can become parametrically lower than the former when the number of species grows in an unbounded fashion (as has been conjectured to happen every time we probe an infinite distance limit in quantum gravity \cite{Ooguri:2006in}).

In order to properly understand what kind of behaviour is exhibited by this species cut-off in realistic quantum gravity constructions, we analyzed in Section \ref{s:speciesscale} the problem using our intuition gained from string theory. Therefore, we considered two different scenarios where the number of particles in the theory may grow exponentially, thus corresponding to either decompactification or emergent string limits \cite{Lee:2019wij}. There, a careful treatment of these matters reveals that it matches with our naive intuition. In particular, we showed that it seems to agree with the higher-dimensional Planck scale or the fundamental string scale probed by the limit, respectively. We further proposed a bottom-up algorithm in Section \ref{ss:MultipleTowers} so as to determine $\LSP$ upon knowing the kind of towers becoming light in a given set-up, which will be important in later parts of this work.

The rest of this thesis will be devoted to test this idea further in string theory so as to give more evidence for the picture advocated in this chapter. Our aim will also be to refine our understanding of the species scale, with an eye to future applications both in string phenomenology and the Swampland program.


\part{\scshape  \textcolor{MyDarkRed}{String Theory Tests and Applications}}
\label{part:StringTheoryTests}

\chapter{Higher-curvature Corrections in String Theory}\label{ch:Higherdimops} 
In Part \ref{part:QGscale} of this thesis we have described in detail some particular energy scale, namely the species scale $\LSP$, which is gravitational in origin and moreover depends crucially on the light spectrum of the theory under consideration. Furthermore, the existence of such energy cut-off introduces important conceptual differences which strongly modify both semi-classical black hole physics as well as low energy effective field theory (EFT) considerations. In particular, since $\LSP$ can become arbitrarily smaller than the Planck scale in the presence of a large number of light species, it offers interesting resolutions to old theoretical problems (like the species problem, see Section \ref{s:speciesmotivation}), and even potential phenomenological explanations for the appearance of certain hierarchies in the Standard Model of particle physics \cite{Dvali:2007hz,Dvali:2007iv, Castellano:2023qhp}. On the other hand, from a more modern perspective, this energy scale becomes also interesting in relation with the Swampland program, since it links two key ingredients together in a rather direct way: The quantum gravity cut-off and the number of light degrees of freedom in our theories. Indeed, many of the most interesting quantum gravity conjectures that have been proposed and thoroughly tested in the literature --- using mostly string theory constructions and holography, such as the Distance conjecture \cite{Ooguri:2006in} or (the tower versions of) the Weak Gravity conjecture \cite{Arkani-Hamed:2006emk,Heidenreich:2016aqi,Andriolo:2018lvp,Montero:2016tif}, require from the existence of infinite towers of particle states that become light in Planck units when approaching certain limiting regimes within the gravitational EFT. Therefore, it is important to pinpoint what is the precise role that this energy cut-off plays within quantum gravity in general and, more specifically, in connection with the Swampland conjectures.

Regarding this important question, in Chapter \ref{ch:SpeciesIntro} we elaborated on how the identification of the maximum cut-off energy for any effective field theory weakly coupled to Einstein gravity with the species scale allows us to understand all these matters within the same theoretical framework. This means, in particular, that when writing any EFT expansion for gravity, the scale suppressing generic higher-curvature corrections should be given precisely by $\LSP$, such that one would expect to find the following expression in $d$ spacetime dimensions
\beq
\mathcal{L}_{\mathrm{EFT}}\, \supset\, \sqrt{-g} \left[\frac{1}{2 \kappa_d^2} \left(\mathcal{R} + \sum_{n >2} \frac{\mathsf{O}_n (\mathcal{R})}{\LSP^{n-2}}\right) -\frac{1}{2} G_{ij} (\phi) \partial_{\mu} \phi^i \partial^{\mu} \phi^j \right]\, ,
\label{eq:scalargravDlag}
\eeq
where as usual $\kappa_d^{2} =\Mpd^{2-d}$ controls the gravitational coupling constant, $\mathsf{O}_n (\mathcal{R})$ denotes any dimension-$n$ local operator involving higher powers of curvature invariants, and we have also included explicitly the kinetic terms for potentially massless/light scalar fields that may be present in the theory. Notice that this implies that the energy suppression of the aforementioned higher-dimensional operators can be smaller than naively expected, since depending on the vacuum we expand our theory around, $\LSP$ might be well below the Planck scale, thus providing for some gravitational `enhancement' of the Wilson coefficients associated to the set $\{ \mathsf{O}_n (\mathcal{R})\}$. In addition, one can argue that such UV cut-off \emph{must} depend non-trivially on the parameters defining the low energy EFT, which are typically controlled by the v.e.v.s of the light scalar fields $\phi^i$ in eq. \eqref{eq:scalargravDlag} above. This follows from the no-global symmetry conjecture (see Section \ref{s:SwamplandProgram} for details), since the existence of an absolute energy scale $\LSP$ in Planck units would imply the presence of some constant and physical parameter that cannot change dynamically in the theory, to which we can therefore associate an exact $(-1)$-form global symmetry. Hence, a more accurate statement would be that there exists some QG cut-off that generically varies over the moduli space of the theory, i.e. $\LSP= \LSP (\phi^i)$, which is indeed in agreement with general Swampland expectations, given that it is precisely when we probe some infinite distance limit in field space that we see a significant decrease in the quantum gravity scale.

Accordingly, our aim in this chapter we will to test this idea further using specific string theory constructions as a quantum gravity laboratory. Furthermore, given that this requires us to know the exact moduli dependence of the higher-curvature corrections under consideration, the strategy that we adopt will consist in focusing on those operators which are somehow protected, so that we can be sure that we are not missing any important information. This means, in practice, that we will restrict ourselves to analyze BPS operators in highly supersymmetric theories, for which the exact moduli dependence is known. Note that, strictly speaking, by doing so one cannot be entirely sure that the resulting behaviour indeed captures the scale we seek for, since there could be strong cancellations depending on the model we consider that would prevent us from extracting the relevant physics. Consequently, one should be careful to claim as general any quantitative information obtained from just observing a few low-lying higher-dimensional operators, but rather consider them to provide at least for some \emph{upper bound} on $\LSP$. 

The chapter is hence organized as follows. In Section \ref{s:Exampleshighdim} we systematically analyze the moduli dependence of the first non-trivial higher-curvature corrections that have been already computed in the literature for all string theory constructions preserving 32 supercharges. More specifically, we consider ten-dimensional Type II string theories and toroidal compactifications thereof, with the relevant operators involving certain contractions of four Riemann tensors. In Section \ref{s:4dN=2} we consider instead theories preserving less amount of supersymmetry. Therefore, we focus mostly on Type II compactifications on Calabi--Yau three-folds, leading to 4d $\mathcal{N}=2$ supergravity EFTs, and study the moduli dependence of BPS operators involving fields only within the gravity multiplet. Finally, in Section \ref{s:summaryhigherops} we summarize our findings, putting special emphasis on the general lessons obtained and discussing certain important questions that our analysis raises as well as potentially interesting directions for future research. 

The material presented in this chapter is based on the publication \cite{Castellano:2023aum} adapted to better fit in the broader context of this thesis. (See also \cite{vandeHeisteeg:2023dlw} for a complementary viewpoint on these matters.)

\section{String theory examples in higher dimensions}
\label{s:Exampleshighdim}

In this section we restrict ourselves to maximally supersymmetric set-ups describing the low energy dynamics of certain string theory constructions in ten, nine and eight spacetime dimensions. The reason for doing so is that these theories are highly constrained (c.f. Section \ref{s:maxsugraintro}), which allows us to determine the moduli dependence of the Wilson coefficients associated to some higher-dimensional operators in an exact way. 

\subsection{ 10d Type IIB string theory}
\label{ss:10dIIB}

As our first example, we consider Type IIB string theory in ten dimensions, whose bosonic (two-derivative) pseudo-action in the Einstein frame is displayed by eq. \eqref{eq:IIB10d}. Moreover, as discussed in Section \ref{s:dualities}, this theory enjoys some non-perturbative $\mathsf{SL(2,\mathbb{Z})}$ duality symmetry, allowing us to arrange the supergravity fields into different representations of the duality group. In particular, the gravitational and scalar sectors of the theory can be written as follows 
\begin{equation}\label{eq:IIB10dSL2}
	S_\text{IIB}^{\text{10d}}\, \supset\, \frac{1}{2\kappa_{10}^2} \int \dd^{10}x\sqrt{-g} \left(\mathcal{R}-\frac{\partial \tau \cdot \partial \bar \tau}{2 (\text{Im}\, \tau)^2}\right)\, ,
\end{equation}
with $\tau=C_0 + \text{i} e^{-\phi}$ denoting the axio-dilaton. This pair of fields transform under $\mathsf{SL(2,\mathbb{Z})}$ as \cite{Schwarz:1995dk}
\begin{align}\label{eq:SdualitytransIIBdilaton}
	&\tau \rightarrow \frac{a\, \tau + b}{c\, \tau+d}\,,\qquad g_{\mu \nu} \rightarrow g_{\mu \nu}\, ,
\end{align}
where the constants $\{ a, b, c, d\}$ are integers satisfying $ad-bc=1$.

Crucially, since this theory has only one dimensionful parameter entering the supergravity action at the two-derivative level --- i.e. the Planck mass, we cannot obtain directly from it any useful information about the quantum gravity cut-off. To do this, what we should try instead is to look at higher-curvature operators which may be present in the theory, since those are expected to be suppressed by the quantum gravity scale.

Before doing so, let us take advantage from our knowledge acquired in precious chapters and try to guess how this function should look like. In particular, if the QG cut-off actually coincides with the species scale, one would expect it to be given by some sort of automorphic function\footnote{See Appendix \ref{ss:mathdefs} for the precise mathematical definition of an automorphic form.} that respects the duality symmetries of the theory. Furthermore, since at infinite distance the fundamental Type IIB string becomes weakly coupled, this function should behave asymptotically like the string scale. Our aim in the following will be to see whether or not these expectations are borne out in the present set-up.

\subsubsection*{The $\mathcal{R}^4$--\,operator}

Let us start by looking at the first non-trivial higher-curvature operator appearing in the 10d effective action. Such correction is $\frac12$-BPS protected, involves four powers of the Riemann tensor and has the following functional form \cite{Green:1999pv,Green:1997di,Pioline:1998mn}
\beq
S_\text{IIB}^{\text{10d}}\, \supset\, \frac{1}{\ell_{10}^2} \int \dd^{10}x \sqrt{-g}\, E_{3/2}^{sl_2} (\tau, \bar \tau)\, t_8 t_8 \mathcal{R}^4\, ,
\label{eq:10dR^4IIB}
\eeq
where $t_8 t_8 \mathcal{R}^4 \equiv t^{\mu_1 \ldots \mu_8} t_{\nu_1 \ldots \nu_8} \mathcal{R}^{\nu_1 \nu_2}_{\mu_1 \mu_2} \ldots \mathcal{R}^{\nu_7 \nu_8}_{\mu_7 \mu_8}$,\footnote{In fact, the actual term arising in the Type IIB effective action involves the $\mathcal{N}=(2,0)$ superinvariant $\mathcal{J}_0 = t_8 t_8 \mathcal{R}^4+ \frac{1}{8} \epsilon_{10}\epsilon_{10} \mathcal{R}^4$ \cite{Kiritsis:1997em}. Here $\epsilon_{10}$ denotes the Levi-civita tensor in ten dimensions and we have defined $\epsilon_{10}\epsilon_{10} \mathcal{R}^4 \equiv \epsilon^{\nu_1 \nu_2 \mu_1 \ldots \mu_8} \epsilon_{\nu_1 \nu_2 \rho_1 \ldots \rho_8} \mathcal{R}^{\rho_1 \rho_2}_{\mu_1 \mu_2} \ldots \mathcal{R}^{\rho_7 \rho_8}_{\mu_7 \mu_8}$.} and the tensor $t^{\mu_1 \ldots \mu_8}$ reads \cite{Green:1981ya}
\begin{equation}\label{eq:t8tensor}
\begin{aligned}
    t^{\mu_1 \ldots \mu_8} &= \frac{1}{5} \Big[ -2 \left( g^{\mu_1 \mu_3} g^{\mu_2 \mu_4} g^{\mu_5 \mu_7} g^{\mu_6 \mu_8} + g^{\mu_1 \mu_5} g^{\mu_2 \mu_6} g^{\mu_3 \mu_7} g^{\mu_4 \mu_8} + g^{\mu_1 \mu_7} g^{\mu_2 \mu_8} g^{\mu_3 \mu_5} g^{\mu_4 \mu_6}\right)\\
    & + 8 \left( g^{\mu_2 \mu_3} g^{\mu_4 \mu_5} g^{\mu_6 \mu_7} g^{\mu_1 \mu_8} + g^{\mu_2 \mu_5} g^{\mu_3 \mu_6} g^{\mu_4 \mu_7} g^{\mu_1 \mu_8} + g^{\mu_2 \mu_5} g^{\mu_6 \mu_7} g^{\mu_3 \mu_8} g^{\mu_1 \mu_4}\right)\\
    & - \left(\mu_1 \leftrightarrow \mu_2 \right) - \left(\mu_3 \leftrightarrow \mu_4 \right) - \left(\mu_5 \leftrightarrow \mu_6 \right) - \left(\mu_7 \leftrightarrow \mu_8 \right) \Big]\, .
\end{aligned}
\end{equation}
On the other hand, the quantity $E_{3/2}^{sl_2} (\tau, \bar \tau)$ appearing in \eqref{eq:10dR^4IIB} denotes the order--$\frac32$ non-holomorphic Eisenstein series of $\mathsf{SL(2,\mathbb{Z})}$, which is an automorphic form that can be defined as a series expansion in the complex valued field $\tau$ as follows (see Appendix \ref{ap:Massform} for details)
\begin{align}\label{eq:nonpertexpansionE3/2}
	E_{3/2}^{sl_2} =\, 2\zeta(3) \tau_2^{3/2} + 4\zeta(2) \tau_2^{-1/2} + \mathcal{O} \left( e^{-2\pi \tau_2}\right)\, .
\end{align}
Due to automorphicity, i.e. the fact that it remains invariant under the transformations \eqref{eq:SdualitytransIIBdilaton}, it is actually enough to restrict ourselves to the fundamental domain $\mathscr{F}$ of $\mathsf{SL(2,\mathbb{Z})}$ when studying e.g., the asymptotic behaviour of the function \eqref{eq:nonpertexpansionE3/2}. This leaves us with only one possible infinite distance limit to analyze, namely the weak coupling point $\tau_2 \to \infty$. Hence, at leading order, the $\mathcal{R}^4$--\,correction behaves like $\tau_2^{3/2}$ for large $\tau_2$. Now, since this operator has mass dimension $n=8$, eq. \eqref{eq:scalargravDlag} implies that its associated (generalized) Wilson coefficient should grow like $\left(\frac{\Mpt}{\LSP}\right)^{6}$. Therefore, given that the species scale coincides in the present case with the string scale asymptotically
\beq\label{eq:stringscale}
m_s=\frac{\Mpt}{\left(4\pi \tau_2^{2}\right)^{1/8}} \, ,
\eeq
we conclude that $\LSP^{-6} \sim \Mpt^{-6}\, \tau_2^{3/2}$, in agreement with the leading-order term in \eqref{eq:nonpertexpansionE3/2}. 

More generally, one can rewrite --- after Poisson resummation --- the series expansion \eqref{eq:nonpertexpansionE3/2}  in the following compact form (c.f. eq. \eqref{eq:nonholoEisenstein})
\beq\label{eq:E3/2Poisson}
E_{3/2}^{sl_2}(\tau, \bar \tau) = \sideset{}{'}\sum_{(p, q) \in \mathbb{Z}^2 }\frac{\tau_2^{3/2}}{\left| p+q\tau\right|^{3}}\, ,
\eeq
where the prime in the sum indicates that we should exclude the point $(0,0)$. Crucially, we can recognize the above expression as a  formal sum over all relevant $(p,q)$-string tensions 
\beq
E_{3/2}^{sl_2}(\tau, \bar \tau) = \left(4 \pi^{\frac34}\right)^3 \sideset{}{'}\sum_{(p, q) \in \mathbb{Z}^2 } \left( \frac{\Mpt}{\sqrt{T_{p,q}}}\right)^6\, , \qquad \text{with}\ \ T_{p,q}= \frac{2\pi}{\ell_{10}^2} \frac{\left| p+q\tau\right|}{\sqrt{\tau_2}}\, ,
\eeq
which indeed provide for the leading tower of states upon exploring other infinite distance points outside the fundamental domain $\mathscr{F}$, and hence determine the species scale asymptotically.

\subsubsection*{Further quantitative tests}

Additionally, one can try to extend the previous analysis by looking at the next few contributions to the four-(super)graviton effective action in 10d Type IIB string theory. These terms --- which still preserve some reduced amount of supersymmetry --- involve respectively four and six derivatives of $\mathcal{R}^4$ and they receive both perturbative and non-perturbative corrections. The first one, which corresponds to a gravitational operator of mass dimension $n=12$, can be computed to be \cite{Green:1999pu}
\beq
S_\text{IIB}^{\text{10d}}\, \supset\, \frac{\ell_{10}^2}{2} \int \dd^{10}x \sqrt{-g}\,E_{5/2}^{sl_2} (\tau, \bar \tau)\, \partial^4 \mathcal{R}^4\, ,
\label{eq:10dpartial4R^4IIB}
\eeq
and its moduli dependence is captured this time by the order--$\frac52$ non-holomorphic Eisenstein series. As it was also the case for the $\mathcal{R}^4$--\,term before, in order to check whether the expected expansion \eqref{eq:scalargravDlag} holds for this case as well we only need to study the large $\tau_2$ behaviour. Upon doing so, one finds (c.f. eq. \eqref{eq:nonpertexpansion})
\beq
E_{5/2}^{sl_2} (\tau, \bar \tau) =  2\zeta(5) \tau_2^{5/2} + \frac{4\pi^4}{135} \tau_2^{-3/2} + \mathcal{O}(e^{-4\pi\tau_2})\, ,
\eeq
which to leading order agrees with $\left(\frac{\Mpt}{\LSP}\right)^{10}$, where $\LSP=m_s$.

On the other hand, the second operator involving six derivatives of the Riemann tensor reads as follows\cite{Green:2005ba,Green:2010wi,Green:2010kv}
\beq
S_\text{IIB}^{\text{10d}}\, \supset\, \ell_{10}^4 \int \dd^{10}x \sqrt{-g}\, \mathcal{E} (\tau, \bar \tau)\, \partial^6 \mathcal{R}^4\, .
\label{eq:10dpartial6R^4IIA}
\eeq
where $\mathcal{E} (\tau, \bar \tau)$ is some particular modular function not of the Eisenstein type. It can be nevertheless expanded around $\tau_2\to \infty$, yielding
\beq
\mathcal{E} (\tau, \bar \tau) = \frac{2\zeta(3)^2}{3} \tau_2^{3} + \frac{4}{3}\tau_2 + \frac{8\zeta(2)^2}{5}\tau_2^{-1} + \frac{4\zeta(6)}{27}\tau_2^{-3} + \mathcal{O}(e^{-4\pi\tau_2})\, ,
\eeq
where the first term corresponds to the tree-level contribution, whilst the remaining pieces --- except for the exponentially suppressed corrections --- include up to three-loop contributions in $g_s$ (see \cite{Green:2010wi} and references therein). Notice that, since the above operator has mass dimension $n=14$, one expects according to eq. \eqref{eq:scalargravDlag} a dependence of the form $\LSP^{-12}\, \partial^6 \mathcal{R}^4$ in Planck units, which indeed matches asymptotically with the species scale computation.

Beyond four-point graviton scattering one may also consider higher-dimensional operators mixing the gravitational field with the Ramond-Ramond $p$-forms. In particular, there is an infinite family of such terms in the 10d Type IIB effective action which are of the form $\mathcal{R}^4 |G_3|^{4g-4}$ (c.f. discussion after eq. \eqref{eq:IIB10dSL2intro} for a precise definition of $G_3$). Note that these operators can be alternatively expressed, upon compactification on any hyper-K\"ahler manifold, in terms of $\mathcal{N}=4$ topological string theory \cite{Berkovits:1994vy} (see also \cite{Ooguri:1995cp}).\footnote{A similar phenomenon happens in 4d $\mathcal{N}=2$ theories, where certain higher derivative F-terms can be computed exactly by the $\mathcal{N}=2$ topological string, see Section \ref{s:4dN=2} below.} Moreover, their precise moduli dependence has been conjectured to be \cite{Berkovits:1998ex}
\begin{equation}\label{eq:BerkovitsVafa}
	\begin{aligned}
		S_\text{IIB}^{\text{10d}}\, \supset\, & \int \dd^{10}x\sqrt{-g}\, \mathcal{R}^4 \sum_{g\geq1} \ell_{10}^{4g-6} \alpha_{g} \sum_{k=2-2g}^{2g-2} (-1)^k \tau_2^{-2g+2}G_3^{2g-2+k} \bar{G}_3^{2g-2-k}\\
            &\sum_{(m, n) \in \mathbb{Z}^2 \setminus \lbrace (0,0) \rbrace} \frac{\tau_2^{g+1/2}}{\left( m+n\tau\right)^{g+1/2+k} \left( m+n \bar \tau\right)^{g+1/2-k}}\, ,
	\end{aligned}
\end{equation}
where $\{\alpha_g\}$ denote some unknown normalization coefficients. Notice that for $k=0$, the operators reduce to $\mathcal{R}^4 \left|\tau_2^{-1/2} G_3 \right|^{4g-4}$, which are manifestly modular invariant, have mass dimension $n=4g+4$ and their accompanying coefficients become $E_{g + 1/2}^{sl_2} (\tau, \bar \tau)$. Therefore, at infinite distance, each of these higher-derivative terms behaves like $\tau_2^{g +1/2} \propto \left(m_s/\Mpt \right)^{-4g-2}$, in perfect agreement with eq. \eqref{eq:scalargravDlag}.

\subsubsection*{A closer look into the EFT expansion}

There are a couple of important lessons that one can extract already from this simple example. On the one hand, if we just restrict ourselves to the first non-trivial gravitational correction described in \eqref{eq:10dpartial4R^4IIB}, it is natural to propose the modular form
\beq\label{eq:speciesIIB}
\LSP = \left(E_{3/2}^{sl_2} (\tau, \bar \tau)\right)^{-1/6}\, ,
\eeq
as being the perfect candidate for a species scale function globally defined over the entire 10d Type IIB moduli space. Furthermore, the function \eqref{eq:speciesIIB} satisfies the two minimal requirements that any bona-fide species scale must fulfill in the present case:
\begin{itemize}
    \item[$\circ$] $\LSP$ is bounded from above (since it cannot exceed $\Mpd$) and it vanishes asymptotically at infinite distance, namely $\LSP (\tau, \bar \tau) \to 0$ as $\tau \to \text{i} \infty$.

    \item[$\circ$] It is given by some \emph{automorphic} form, namely a modular invariant function of $\tau$ satisfying
\beq
 \notag \LSP \left(\frac{a \tau +b}{c\tau +d}\, , \, \frac{a \bar \tau +b}{c \bar \tau +d} \right) = \LSP (\tau, \bar \tau)\, , \qquad ad-cd=1\, ,
\eeq
where $a,b,c,d \in \mathbb{Z}$.
\end{itemize}
However, once we go beyond and consider additional higher-curvature corrections, we readily realize that they do not strictly organize in powers of the \emph{same} cut-off function, as the naive expectation from \eqref{eq:scalargravDlag} would suggest. Indeed, these moduli-dependent coefficients are seen to be given by certain automorphic forms of $\mathsf{SL(2,\mathbb{Z})}$, which nonetheless cannot be written as powers of one another. This precludes a priori from a canonical identification of the species scale function as any one of the aforementioned generalized Wilson coefficients. In any event, what remains undoubtedly true is that the gravitational EFT expansion proposed in eq. \eqref{eq:scalargravDlag} emerges when approaching any infinite distance point in the present 10d example. This is of course in agreement with our quantum field theory experience, since it is there where the weak coupling behaviour is usually restored \cite{Gendler:2020dfp}, and the classical dimensions of the different gravitational operators are actually reliable. On the contrary, when venturing towards the bulk of the moduli space, further perturbative and non-perturbative corrections may become important and thus significantly change the expectations from \eqref{eq:scalargravDlag}, which can be thought of as some sort of `anomalous dimensions'.

On the other hand, it should be mentioned that despite these difficulties for selecting any canonical representative for $\LSP (\tau, \bar \tau)$, certain qualitative features such as the presence and location of a \emph{desert point} \cite{Long:2021jlv, vandeHeisteeg:2022btw} --- i.e. the locus where the quantum gravity cut-off is maximized, do not crucially depend on which operator one chooses to focus on. In particular, all these modular forms present an absolute minimum at the cusp $\tau= e^{\frac{2\pi \text{i}}{3}}$, thus coinciding with the point where the the BPS gap of $(p,q)$-strings is maximized (c.f. eq. \eqref{eq:E3/2Poisson}).

\subsection{ 10d Type IIA string theory}
\label{ss:10dIIA}

We now turn to ten-dimensional Type IIA string theory, whose bosonic action can be found in eq. \eqref{eq:IIA10dEinsteinframeaction}. This theory has simpler moduli space than its Type IIB counterpart, which it is actually isomorphic to the real line when parametrized in terms of the dilaton v.e.v. $\braket{\phi}$. Relatedly, the Type IIA theory does not enjoy any non-trivial U-duality properties, preventing us from inferring a priori which kind of functions should appear in the gravitational EFT expansion. In any event, we can still consider the first few non-zero corrections to the supergravity action and explicitly analyze its moduli dependence. The first one of these happens to be protected again by supersymmetry, and it reads as follows \cite{Grisaru:1986dk,Grisaru:1986kw,Gross:1986iv} 
\beq
S_\text{IIA}^{\text{10d}}\, \supset\, \frac{1}{\ell_{10}^2} \int \dd^{10}x \sqrt{-g}\, \left( 2\zeta(3) e^{-3\phi/2} + \frac{2\pi^2}{3} e^{\phi/2}\right) t_8 t_8 \mathcal{R}^4\, ,
\label{eq:10dR^4IIA}
\eeq
which is nothing but the expression \eqref{eq:10dR^4IIB} with the instanton sum excluded. In fact, the first term corresponds to the tree-level contribution (which arises at fourth-loop order in the 2d sigma-model perturbation theory), whilst the second piece is a one-loop correction in $g_s$. 

Let us now check what are the relevant asymptotics of this dimension-8 operator. At weak coupling, namely when $\phi \to -\infty$ (equivalently $g_s \to 0$), the tree-level term dominates and we obtain
\beq
\mathcal{L}_{\mathcal{R}^4}^{\text{10d}}\, \sim\, \sqrt{-g} \left(\frac{2\zeta(3)}{\ell_{10}^2} e^{-3\phi/2} \right) t_8 t_8 \mathcal{R}^4\ .
\label{eq:R^4weakcoupling}
\eeq
Comparing this with eq. \eqref{eq:scalargravDlag}, we deduce that the coefficient accompanying such coupling in the effective action should behave like $\left(\frac{\Mpt}{\LSP}\right)^{6}$ asymptotically. Therefore, since the species scale coincides with the string scale along the weak coupling limit, we again find that
\beq
\left(\frac{\Mpt}{\LSP}\right)^{6}\, =\, (4\pi)^{3/4}e^{-3\phi/2}\, ,
\eeq
in agreement with eq. \eqref{eq:R^4weakcoupling} above.

On the contrary, at strong coupling, it is the one-loop correction which becomes more important, thus leading to the following asymptotic dilaton dependence
\beq
\mathcal{L}_{\mathcal{R}^4}^{\text{10d}}\, \sim\, \sqrt{-g} \left(\frac{4\zeta(2)}{\ell_{10}^2} e^{\phi/2} \right) t_8 t_8 \mathcal{R}^4 \ .
\label{eq:R^4strongcoupling}
\eeq
Crucially, the species counting is now dominated by the tower of D0-brane bound states instead, since the fundamental string becomes infinitely heavy in 10d Planck units. Following the original definition of $\LSP$ in \eqref{species}, one recovers that the species scale is capturing the 11d M-theory Planck mass, which is given by
\beq \label{eq:QGscaleandN} 
		\LSP\, \sim\, m_{\text{D}0}^{1/9}\ \Mpt^{8/9}\, \sim\,  e^{-\phi/12} \Mpt\, ,
\eeq
such that the quantity $\LSP^{-6}\, \mathcal{R}^4$ precisely reproduces the leading-order piece \eqref{eq:R^4strongcoupling}. Note that the previous conclusion strongly rests on the fact that there are no further loop corrections to $t_8 t_8 \mathcal{R}^4$ in Type IIA string theory, since otherwise the matching with the 11d Planck scale would be automatically spoiled. 

Finally, let us comment that by performing a similar analysis to the one done for the 10d Type IIB theory regarding the corrections of the form $\partial^4 \mathcal{R}^4$ and $\partial^6 \mathcal{R}^4$, it can be readily seen that they are in fact not suppressed by the species scale to the expected power when the M-theory limit is probed. In contrast, it is the mass scale of the lightest tower (i.e. the D0-branes) the one controlling the suppression. For emergent string limits, however, we still find that they always appear with the correct power of $\LSP$, but one must take into account that in such cases the species scale and that of the tower coincide, namely $\LSP = m_s$. We elaborate further on this point in Section \ref{ss:gravEFTexpansion} below.

\subsection{M-theory on $\mathbf{T}^2$}
\label{ss:MthyT2}

Let us now turn to the unique 9d $\mathcal{N}=2$ supergravity theory, which may be obtained by e.g., compactifying M-theory on a $\mathbf{T}^2$ with an internal metric of the form
\begin{equation}\label{eq:T2metric}
	g_{m n}= \frac{\mathcal{V}_2}{\tau_2} \left(
	\begin{array}{cc}
		1 & \tau_1  \\
		\tau_1 & |\tau|^2  \\
	\end{array}
	\right) \, ,
\end{equation}
where $\tau=\tau_1+{\rm i}\tau_2$ denotes the complex structure of the torus and $\mathcal{V}_2$ its overall volume. The scalar and gravitational sectors in the 9d Einstein frame read (see Section \ref{ss:9dmaxsugra} for details)
\begin{equation}\label{eq:9dchap5}
	S_\text{M-th}^{\text{9d}} \supset \frac{1}{2\kappa_9^2} \int \dd^{9}x\, \sqrt{-g}\,  \left[ \mathcal{R} - \frac{9}{14} \frac{\left( \partial \mathcal{V}_2 \right)^2}{\mathcal{V}_2^2} -\frac{\partial \tau \cdot \partial \bar \tau}{2 \tau_2^2} \right]\, ,
\end{equation}
therefore exhibiting some non-trivial moduli space which is moreover classically exact and parametrizes the manifold $\mathcal{M}_{\text{9d}}=\mathsf{SL(2,\mathbb{Z})}\backslash \mathsf{SL(2,\mathbb{R})}/\mathsf{U(1)} \times \mathbb{R}_{+}$ (after quotient out the $\mathsf{SL(2,\mathbb{Z})}$ U-duality symmetry \cite{Schwarz:1995dk,Aspinwall:1995fw}). 

Our strategy here will again consist in studying the asymptotic moduli dependence featured by certain protected quantities in the low energy action, so as to see whether the expectations arising from the species counting are furnished in the present example as well. Hence, we look at the next non-trivial correction to the two-derivative lagrangian \eqref{eq:9dchap5}, which again behaves schematically like $\mathcal{R}^4$, and is still BPS protected. Its dependence with respect to the moduli space parametrized by $\lbrace \mathcal{V}_2, \tau \rbrace$ has been already computed in \cite{Green:1997tv,Green:1997as}, and is captured by the following non-trivial function (see also \cite{Green:2010wi})
\beq
S_\text{M-th}^{\text{9d}} \supset \frac{1}{\ell_{9}} \int \dd^{9}x \sqrt{-g}\, \left( \frac{2\pi^2}{3} \mathcal{V}_2^{6/7} + \mathcal{V}_2^{-9/14} E_{3/2}^{sl_2} (\tau, \bar \tau)\right) t_8 t_8 \mathcal{R}^4\, ,
\label{eq:9dR^4MthT2}
\eeq
where $\mathsf{SL(2,\mathbb{Z})}$ invariance is readily manifest (recall that the volume modulus is left unchanged under a modular transformation). The above operator has mass dimension $n=8$, such that the function in parenthesis is expected to behave like $\left(\frac{M_{\rm Pl;\, 9}}{\LSP}\right)^{6}$ at any infinite distance boundary of $\mathcal{M}_{\text{9d}}$. In the following, we consider each of these limits in turn (see Figure \ref{fig:MthyT2} below), with the important simplification that one can restrict to a subset of the latter for which $\tau_2 \to \infty$, thanks to $\mathsf{SL(2,\mathbb{Z})}$ invariance. Furthermore, we note that depending on whether $\tau_2 > \mathcal{V}_2$ or $\tau_2 < \mathcal{V}_2$, the second/first term in eq. \eqref{eq:9dR^4MthT2} dominates, which allows us to divide the fundamental domain of the phase diagram into two different subregions.
\newline

\begin{figure}[htb]
\begin{center}
\includegraphics[width=0.6\textwidth]{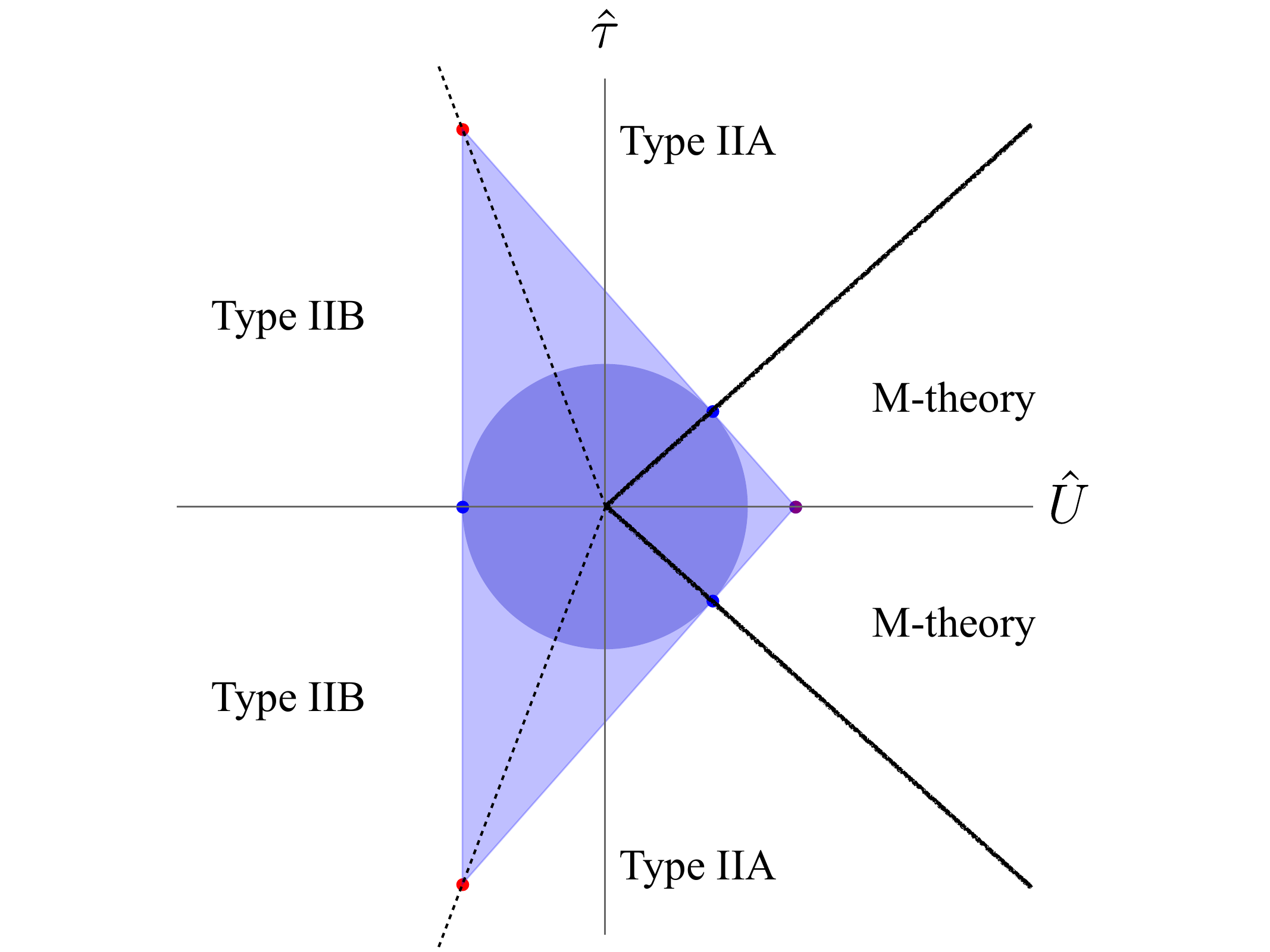}
\caption{\small Phase diagram for the (asymptotic) species scale in M-theory on $\mathbf{T}^2$, parametrized by the canonical variables $\{ \hat U = \frac{3}{\sqrt{14}} \log \mathcal{V}_2\, ,\, \hat \tau= \frac{1}{\sqrt{2}}\log \tau_2\}$. The blue dots are associated to circle decompactifications (possibly to a dual frame), whereas the red and purple ones signal emergent Type II string limits and full decompactification to eleven dimensions, respectively. The self-dual line $\hat{\tau}=0$ is fixed under the U-duality symmetry.}
\label{fig:MthyT2}
\end{center}
\end{figure}

\underline{\textit{The M-theory regime}}
\newline

Let us start with the region $\mathcal{V}_2 > \tau_2$. In this case, the dominant term in the expression for the $\mathcal{R}^4$--\,operator depends solely on the internal volume, so that one can restrict in practice to the large radius limit at fixed (and finite) complex structure. This is nothing but a full decompactification to 11d supergravity, and thus the species scale should coincide (up to order one factors) with the 11d Planck scale, which depends on the moduli fields as follows
\beq
\frac{M_{\rm Pl;\, 11}}{M_{\rm Pl;\, 9}}\, =\, (4\pi)^{-2/9}\, \mathcal{V}_2^{-1/7}\, .
\label{eq:11dPlanckmass}
\eeq
Therefore, according to eq. \eqref{eq:scalargravDlag} we expect the asymptotic behavior $\LSP^{-6} \sim \mathcal{V}_2^{6/7}$ arising in front of the quartic correction to the 9d action, which indeed matches the correct result. 
\newpage

\underline{\textit{The Type II regime}}
\newline

In the opposite regime, namely when $\tau_2 > \mathcal{V}_2$, the species scale should be controlled by the fundamental string mass (defined here as $m_{\text{str}} \equiv \sqrt{T_{\text{str}}}$), which corresponds to the red dot in the upper half-plane in Figure \ref{fig:MthyT2}. This can be readily computed, yielding
\beq
\frac{m_{\text{str}}}{M_{\rm Pl;\, 9}}\, =\, \frac{(4\pi)^{5/14}}{\sqrt{2}}\, \mathcal{V}_2^{3/28} \, \tau_2^{-1/4}\, .
\label{eq:fundstringmass9d}
\eeq
Hence, focusing in the second term in \eqref{eq:9dR^4MthT2} and using the asymptotic behaviour exhibited by the order--$\frac32$ non-holomorphic Eisenstein series (c.f. eq. \eqref{eq:nonpertexpansionE3/2}), we conclude that the coefficient of the $\mathcal{R}^4$--\,operator behaves asymptotically as $\mathcal{V}_2^{-9/14}\, \tau_2^{3/2} \sim m_{\text{str}}^{-6}$, in agreement with \eqref{eq:scalargravDlag}.
\newline

\underline{\textit{Decompactification to 10d}}
\newline

For completeness, let us also discuss the two boundaries between the different asymptotic regions in moduli space, as seen from the diagram in Figure \ref{fig:MthyT2}. These are moreover associated to certain asymptotic directions signalling towards partial decompactification to either 10d Type IIA or Type IIB string theory. On the one hand, precisely when $\tau_2 = \mathcal{V}_2 \to \infty$, a subset of KK modes become light and the theory decompactifies to 10d Type IIA supergravity. The 10d Planck scale presents the following moduli dependence
\beq
\frac{M^{\rm IIA}_{\rm Pl;\, 10}}{M_{\rm Pl;\, 9}}\, =\, (4\pi)^{1/56}\, \mathcal{V}_2^{-9/112} \tau_2^{-1/16} \sim \mathcal{V}_2^{-1/7}\, ,
\label{eq:10dPlanckmassIIA}
\eeq
which agrees asymptotically with both $M_{\rm Pl;\, 11}$ and $m_{\text{str}}$ along the aforementioned limit. Therefore, upon inserting $\LSP \sim M^{\rm IIA}_{\rm Pl;\, 10}$ into eq. \eqref{eq:scalargravDlag}, one reproduces the behaviour exhibited by \eqref{eq:9dR^4MthT2}.

On the other hand, for the limit $\mathcal{V}_2 \to 0$ the species counting is dominated by M2-branes wrapping the internal space. These states correspond to the KK tower implementing the M-theory/F-theory duality (see Section \ref{ss:dualitieswithlowersusy} for details), such that the quantum gravity scale becomes identical to the 10d Type IIB Planck mass, which reads
\beq
\frac{M^{\rm IIB}_{\rm Pl;\, 10}}{M_{\rm Pl;\, 9}}\, =\, (4\pi)^{1/56}\, \mathcal{V}_2^{3/28}\, .
\label{eq:10dPlanckmass}
\eeq
Hence, along such limit the $\mathcal{R}^4$--\,operator should be controlled by $\LSP^{-6} \sim \mathcal{V}_2^{-9/14}$, thus matching the behaviour observed in the second term of \eqref{eq:9dR^4MthT2}.

Before turning to higher-dimensional operators other than $t_8 t_8 \mathcal{R}^4$, let us make one more comment. Indeed, from the discussion above one would be tempted to propose $\LSP$ to be defined in the present nine-dimensional set-up precisely by
\beq
\LSP = \left( \frac{2\pi^2}{3} \mathcal{V}_2^{6/7} + \mathcal{V}_2^{-9/14} E_{3/2}^{sl_2} (\tau)\right)^{-1/6}\, ,
\label{eq:9dfullspeciesfn}
\eeq
which is automorphic and moreover reproduces the correct asymptotic behaviour in every infinite distance corner of $\mathcal{M}_{\text{9d}}$, as we just demonstrated. However, let us stress that this would imply, via eq. \eqref{eq:scalargravDlag}, that all subsequent higher order gravitational operators in the 9d effective action should also be accompanied by appropriate powers of the function \eqref{eq:9dfullspeciesfn}, which is known to be \emph{not} the case, see below. What seems to be true, though, is the fact that the species scale, as one typically defines it close to the infinite distance boundaries in moduli space (c.f. Section \ref{ss:Planck&string}), is the energy scale organizing the asymptotic EFT expansion of our quantum-gravitational theories, and thus should be taken as the true QG cut-off.  

%
%
\subsubsection*{Further quantitative tests}

Proceeding as in the previous ten-dimensional examples, let us now look at the next few contributions to the four-(super)graviton effective action in 9d M-theory. For the four-derivative quartic term, the exact moduli dependence has already been obtained in the literature, leading to \cite{Green:2010wi}
\beq
S_\text{M-th}^{\text{9d}} \supset \ell_{9}^3 \int \dd^{9}x \sqrt{-g}\, \left( \frac{1}{2} \mathcal{V}_2^{-15/14} E_{5/2}^{sl_2} (\tau) + \frac{2\zeta(2)}{15} \mathcal{V}_2^{27/14} E_{3/2}^{sl_2} (\tau) + \frac{4\zeta(2)\zeta(3)}{15} \mathcal{V}_2^{-18/7}\right) \partial^4 \mathcal{R}^4\, .
\label{eq:9dpartial4R^4}
\eeq
Notice that such term is again compatible with $\mathsf{SL(2,\mathbb{Z})}$ invariance, so that we can restrict to the fundamental domain $\{ \tau \in \mathscr{F}\, ,\, \mathcal{V}_2 \geq 0\}$ in what follows. Moreover, the mass dimension of the $\partial^4 \mathcal{R}^4$--\,operator is $n=12$, such that according to \eqref{eq:scalargravDlag}, we expect an asymptotic dependence for the generalized Wilson coefficient of the form $\left(\frac{M_{\rm Pl;\, 9}}{\LSP}\right)^{10}$. Interestingly, this prediction is fulfilled in all the cases in which we probe an emergent string limit, namely when $\LSP = m_{\rm str}$, whereas in decompactification limits (either to ten or eleven dimensions) one cannot directly identify $\LSP^{-10}$ as the coefficient appearing in front of this operator. Still, this might be understood in terms of the dimensionality of the aforementioned operators, as we comment in Section \ref{ss:gravEFTexpansion} with more detail. 

\subsection{M-theory on $\mathbf{T}^3$}
\label{ss:MthyT3}

As our final example, we consider maximal supergravity in eight spacetime dimensions. This theory arises upon compactifying e.g., Type IIB supergravity on a two-dimensional torus, leading to the following 8d action in the scalar-tensor sector
\begin{equation}\label{eq:IIB8d}
		\begin{aligned}
			S_\text{IIB}^{\text{8d}}\, \supset\, & \frac{1}{2\kappa_{8}^2} \int \dd^{8}x\sqrt{-g} \left[\mathcal{R}-\frac{1}{6}\frac{(\partial \nu)^2}{\nu^2} -\frac{\partial \tau \cdot \partial \bar \tau}{2 \tau_2^2} -\frac{\partial U \cdot \partial \bar U}{2 U_2^2} - \nu \frac{\left| \tau \partial b + \partial c\right|^2}{2\tau_2}\right]\, ,
		\end{aligned}
\end{equation}
where $U$ denotes the complex structure of the torus, $\nu= \left( \tau_2 V_2^2\right)^{-1}$ is an $\mathsf{SL(2,\mathbb{Z})}_{\tau}\,$--\, invariant volume, and $\{ b,c\}$ are compact scalar fields arising from the reduction of the NS and RR 2-form fields of 10d $\mathcal{N}=(2,0)$ supergravity along the internal 2-cycle. Note that there are two modular symmetries visible from the action \eqref{eq:IIB8d} above: that associated to the axio-dilaton --- which is inherited from ten dimensions, as well as an additional one which transforms the complex $U$ field in a fractional linear fashion. There is, however, an extra `hidden' $\mathsf{SL(2,\mathbb{Z})}_{T}$ symmetry associated to the K\"ahler modulus $T=b+ \i V_2$, 
which can be made manifest upon changing variables from $\lbrace \nu, \tau\rbrace \leftrightarrow \lbrace \varphi_8, T\rbrace$ (c.f. eq. \eqref{eq:8dIIB}), where $\varphi_8$ denotes the 8d dilaton
\begin{equation}\label{eq:8ddilaton}
     e^{-2\varphi_8} = e^{-2\phi} V_2\, .
\end{equation}
What is important for us here is that the \emph{full} U-duality symmetry of the theory is actually larger, namely it consists of $\mathsf{SL(2,\mathbb{Z})} \times \mathsf{SL(3,\mathbb{Z})}$, where the modular factor acts solely on the complex structure modulus. In fact, upon introducing the following symmetric  $3\times3$ matrix with unit determinant \cite{Liu:1997mb}
\beq\label{eq:SL3matrix}
 \mathcal{B}= \nu^{1/3} \begin{pmatrix}
		\frac{1}{\tau_2} \quad  \frac{\tau_1}{\tau_2} \quad \frac{c+\tau_1 b}{\tau_2}\\ \frac{\tau_1}{\tau_2} \quad  \frac{|\tau|^2}{\tau_2} \quad \frac{\tau_1 c+|\tau|^2 b}{\tau_2}\\ \frac{c+\tau_1 b}{\tau_2} \quad  \frac{\tau_1 c+|\tau|^2 b}{\tau_2} \quad \frac{1}{\nu} + \frac{|c+\tau b|^2}{\tau_2}
	\end{pmatrix}\, ,
\eeq
which transforms in the adjoint representation of $\mathsf{SL(3,\mathbb{Z})}$, one can rewrite the above action in a manifestly $\mathsf{SL(2,\mathbb{Z})} \times \mathsf{SL(3,\mathbb{Z})}$ invariant way\footnote{Both $\mathsf{SL(2,\mathbb{Z})}_{\tau}$ and $\mathsf{SL(2,\mathbb{Z})}_{T}$ transformations are embedded within $\mathsf{SL(3,\mathbb{Z})}$ as upper and lower block-diagonal subgroups \cite{Kiritsis:1997em}.} 
\begin{equation}\label{eq:IIB8dSL3}
			\begin{aligned}
				S_\text{IIB}^{\text{8d}}\, \supset\, & \frac{1}{2\kappa_{8}^2} \int \dd^{8}x\sqrt{-g} \left[\mathcal{R} -\frac{\partial U \cdot \partial \bar U}{2 U_2^2} + \frac{1}{4} \text{tr} \left( \partial \mathcal{B} \cdot \partial \mathcal{B}^{-1} \right) \right]\, .
			\end{aligned}
\end{equation}
Therefore, we conclude that the moduli space of the theory is described by a group coset of the form $\mathcal{M}_{\text{8d}}=\mathsf{SL(2,\mathbb{Z})}\backslash \mathsf{SL(2,\mathbb{R})}/\mathsf{U(1)} \times \mathsf{SL(3,\mathbb{Z})}\backslash \mathsf{SL(3,\mathbb{R})}/\mathsf{SO(3)}$, where we have modded out by the U-duality group of the 8d theory.
	
In the following, it will be useful to phrase all our discussion using a dual description in terms of M-theory compactified on $\mathbf{T}^3$, whose bosonic action was already introduced in Section \ref{ss:8dmaxsugra}. Recall that the relevant scalar degrees of freedom, as seen from this dual perspective, are associated to the internal metric $g_{mn}$, the overall volume modulus $\mathcal{V}_3$ and a compact field $C^{(3)}_{123}$, which arises by reducing the antisymmetric 3-form field along the torus. In particular, one can make direct contact with the previous Type IIB description using a chain of dualities (see discussion around eqs. \eqref{eq:8dap} and \eqref{eq:8dIIA}), thus relating the complex structure modulus $U$ with some complex-valued field $\mathcal{T}$, namely
\beq\label{eq:T3complexvolume}
\mathcal{T}= C^{(3)}_{123}+ \text{i} \mathcal{V}_3\, ,
\eeq
as well as the matrix $\mathcal{B}$ in \eqref{eq:SL3matrix} with the \emph{unimodular} metric components of the compact space, i.e. $\tilde{g}_{m n}= \mathcal{V}_3^{-2/3} g_{mn}$. This recovers the action \eqref{eq:8dsl3} written in a manifestly $\mathsf{SL(3,\mathbb{Z})}$ invariant form. Furthermore, we will choose a parametrization of the scalar manifold that breaks the underlying duality symmetry of the theory, since it is better adapted for the discussion that is to follow as well as for the rest of the thesis. This amounts to selecting some $\mathbf{T}^2$ within the $\mathbf{T}^3$, compactify the 11d theory on it, and subsequently reduce the 9d supergravity EFT on an extra $\mathbf{S}^1$, thus leading to the action displayed in eq. \eqref{eq:8dalternativeaction}, which we repeat here for the comfort of the reader
\begin{equation}\label{eq:8dalternativeactionII}
	\begin{aligned}
			S_\text{M-th}^{\text{8d}}\, &\supset\, \frac{1}{2\kappa_{8}^2} \int \dd^{8}x\sqrt{-g} \Bigg[\mathcal{R}-\frac{9}{14} \left( \partial \log \mathcal{V}_2\right)^2 - \frac{7}{6} \left( \partial \log R_3\right)^2 -\frac{\partial \tau \cdot \partial \bar \tau}{2 \tau_2^2}\\
            &- \frac{\mathcal{V}_2^{-12/7} R_3^{-2}}{2} \left( \partial C_{123}^{(3)}\right)^2 -\frac{\mathcal{V}_2^{9/7} R_3^{-2}}{2 \tau_2} \left| \partial A^{(1)}_0-\tau \partial A^{(1)}_0 \right|^2\Bigg]\, ,
	\end{aligned}
\end{equation}
where the geometrical meaning of the different fields above can be found in Section \ref{ss:8dmaxsugra}.

%
%

\subsubsection*{Higher-derivative corrections}

\begin{figure}[htb]
		\begin{center}
			\subfigure{
				\includegraphics[width=0.45\textwidth]{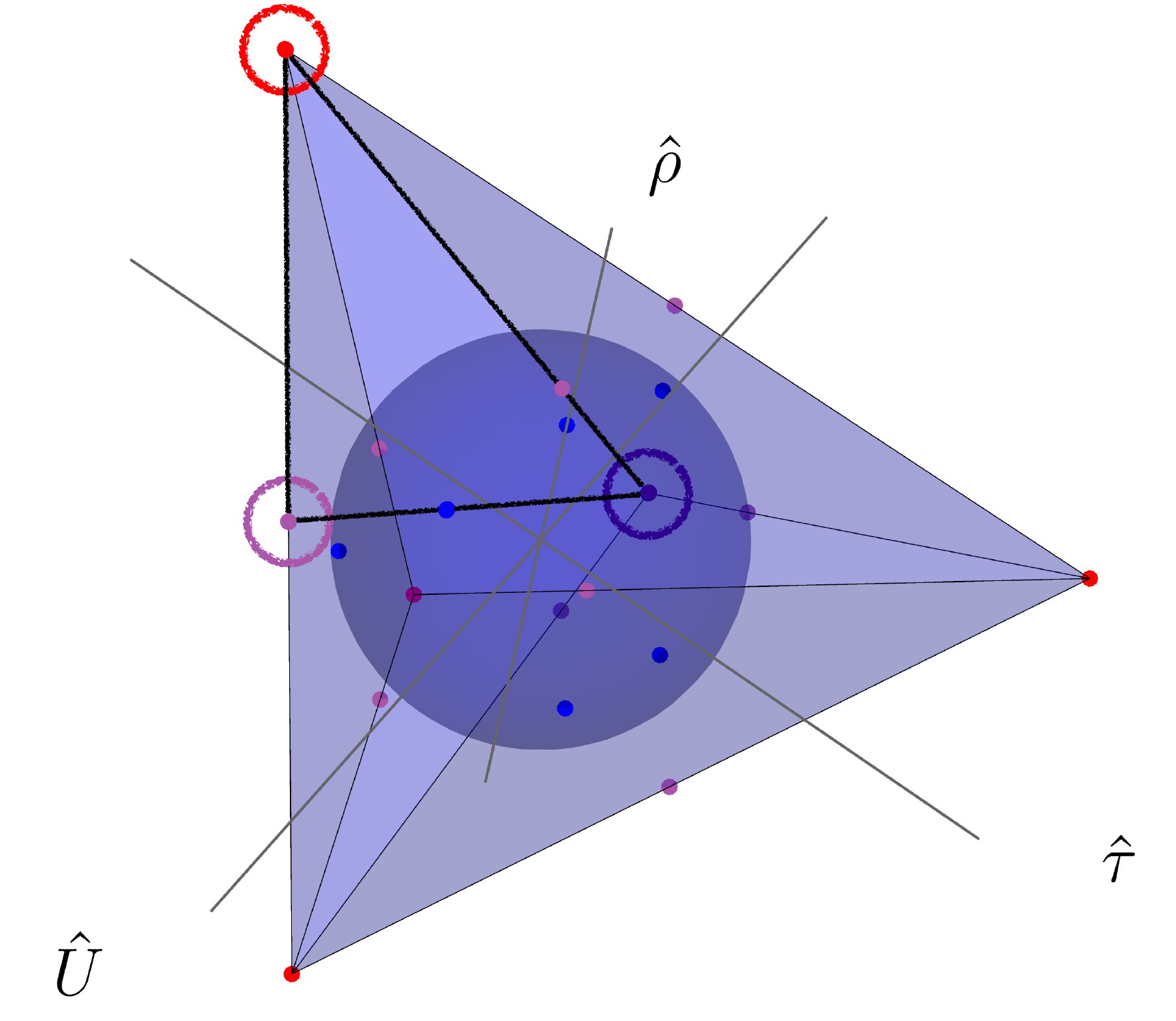}
			}
			\subfigure{
				\includegraphics[width=0.45\textwidth]{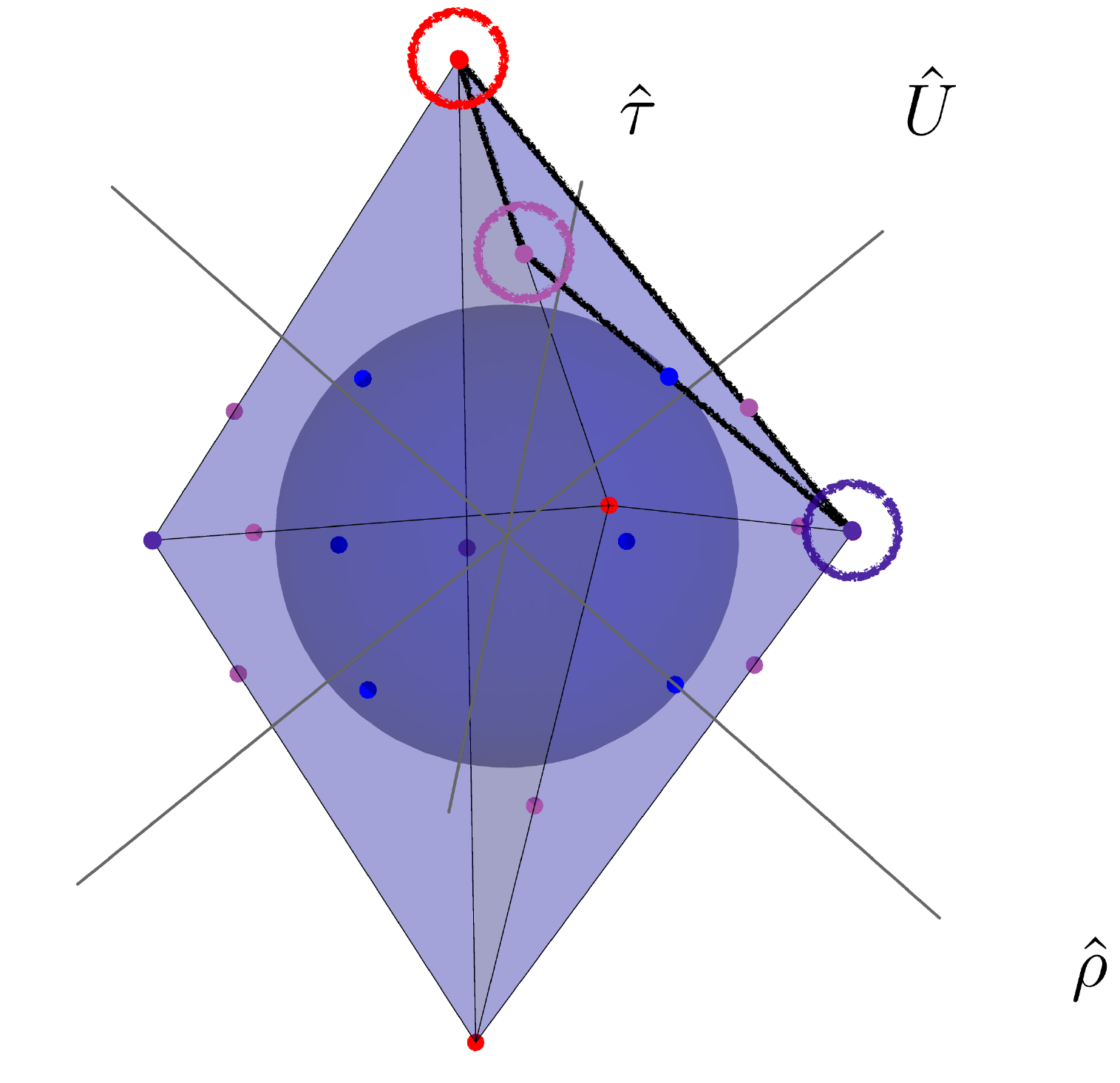}
			}
			\caption{\small Phase diagram for the (asymptotic) species scale in M-theory on $\mathbf{T}^3$, as seen from two different angles. The axes correspond to the canonical variables $\lbrace \hat U = \frac{3}{\sqrt{14}} \log \mathcal{V}_2\, ,\, \hat \tau= \frac{1}{\sqrt{2}}\log \tau_2\, ,\, \hat \rho = \sqrt{\frac76} \log R_3\rbrace$. The blue dots are associated to circle decompactifications, the light purple ones signal some double decompactification to 10d, whilst the purple and red dots correspond to full decompactification to 11d and emergent Type II string limits, respectively. We have selected three particular directions which span a cone determining some fundamental domain $\mathscr{F}_8$.}
			\label{fig:MthyT3}
		\end{center}
\end{figure}

Given this set-up --- and based on our previous analysis, one would expect the species scale $\LSP$ to be captured by some automorphic function of the moduli v.e.v.s, whose asymptotic behaviour should match the usual species counting procedure. This is summarized in Figure \ref{fig:MthyT3}, where the phase diagram of the 8d theory is depicted. In order to check this intuition, we are going to look at the lowest-order gravitational operators not encoded within the two-derivative effective action \eqref{eq:IIB8d}. The first non-trivial such correction includes an operator involving four Riemann tensors contracted in a particular way, and which has been computed in the past using a variety of methods, ranging from one-loop calculations in M-theory to non-perturbative instanton computations. The result, in the Type IIB frame, is the following \cite{Green:1997as,Green:2005ba} (see also \cite{Kiritsis:1997em,Basu:2007ru,Basu:2007ck})
\beq
S_\text{IIB}^{\text{8d}}\, \supset\, \int \dd^{8}x \sqrt{-g}\, \left( \hat{E}_{3/2}^{sl_3} +2\hat{E}_{1}^{sl_2}\right) t_8 t_8 \mathcal{R}^4\, ,
\label{eq:8dR^4IIB}
\eeq
where the explicit form of $t_8 t_8 \mathcal{R}^4$ is discussed around eq. \eqref{eq:t8tensor}. Moreover, the functions $\hat{E}_{3/2}^{sl_3}$ and $\hat{E}_{1}^{sl_2}$ are (appropriately regularized) Eisenstein series of order--$\frac32$ and 1 for the duality groups $\mathsf{SL(3,\mathbb{Z})}$ and $\mathsf{SL(2,\mathbb{Z})}$, respectively. They can be expanded as follows (see Appendix \ref{ap:Massform} for details):
\begin{align}\label{eq:instexpSL3ap}
	\notag \hat{E}_{3/2}^{sl_3} &= 2\zeta(3) \frac{\tau_2^{3/2}}{\nu^{1/2}} + \frac{2 \pi^2}{3} T_2 + \frac{4\pi}{3} \log \nu\\
 &+  4\pi \sqrt{\frac{\tau_2}{\nu}} \sum_{m,n \neq 0} \left| \frac{m}{n}\right| e^{2\pi \text{i} m n\tau_1}\, K_{1} (2\pi |m n| \tau_2)\, +\, \sideset{}{'}\sum_{m, n \in \mathbb{Z}} \mathcal{I}^{3/2}_{m, n}\, ,
\end{align}
with
\begin{align}\label{eq:I3/2mn}
	\mathcal{I}^{3/2}_{m, n} = 2\frac{\pi^{3/2} \nu^{-1/4}}{\Gamma({3/2}) \tau_2^{1/4}} \sum_{k \neq 0} \left| \frac{m+n\tau}{k}\right|^{1/2} e^{2\pi \text{i} k \left[n(c+\tau_1 b)- (m+n\tau_1)b \right]}\, K_{1/2} \left(2\pi |k|\frac{\left| m+n\tau \right|}{\sqrt{\nu \tau_2}}\right)\, ,
\end{align}
for the $\mathsf{SL(3,\mathbb{Z})}$--\,invariant piece, whilst the second factor in \eqref{eq:8dR^4IIB} reads as
\beq
2\hat{E}_{1}^{sl_2}\, =\, -2\pi \text{log} \left(U_2\,|\eta(U)|^4\right)\, .
\eeq
The physical origin of both terms, as seen from the Type IIB perspective, can be easily understood by looking at the relevant quantities entering into each of the two series expansions. Indeed, $\hat{E}_{1}^{sl_2}$ encodes certain KK threshold corrections, which depend on the complex structure of the torus, whilst $\hat{E}_{3/2}^{sl_3}$ is richer: It contains both $\alpha'$ and $g_s$ \emph{perturbative} contributions, together with non-perturbative D$(-1)$- as well as $(p,q)$-string instantons series, whose action is controlled by the same quantity appearing in the modified Bessel function in \eqref{eq:I3/2mn}, namely $\frac{\left| m+n\tau \right|}{\sqrt{\nu \tau_2}}=\left| m+n\tau \right|T_2$.

Before proceeding any further, let us note that if instead of the $\lbrace \nu, \tau\rbrace$--\,parametrization one chooses the alternative $\lbrace \varphi_8, T\rbrace$ coordinates --- whose leading order action is shown in eq. \eqref{eq:8dIIB}, the $\mathsf{SL(3,\mathbb{Z})}$--\,series can be expanded as follows
\begin{align}\label{eq:Eisenstein3/2-2ap}
	\hat{E}_{3/2}^{sl_3} &= 2\zeta(3) e^{-2\varphi_8} + 2 \hat{E}_{1}^{sl_2} (T) + \frac{4 \pi}{3} \varphi_8 + \mathcal{O} \left( \exp(-e^{-\varphi_8})\right)\, ,
\end{align}
which of course agrees with \eqref{eq:instexpSL3ap}.

In order to connect with our discussion in the M-theory picture, we need to rewrite the previous $\mathcal{R}^4$ correction in terms of the variables adapted to the action \eqref{eq:8dalternativeactionII}. Hence, upon using the map between Type IIB and M-theory (c.f. Section \ref{s:dualities})
\begin{center}
\renewcommand{\arraystretch}{2.00}
\begin{tabular}{r c l}
M-theory on $\mathbf{T}^3$ & $\longleftrightarrow $ & Type IIB on $\mathbf{T}^2$\\ 
\hline
\hline  
$\mathcal{T}$ & $ \longleftrightarrow$ & $U$ \\
$\tau$ & $\longleftrightarrow $&  $T$\\
$\mathcal{V}_2^{9/7}\, R_3^{-2}$ & $\longleftrightarrow $ & $e^{-2\varphi_8}$
\end{tabular}\label{tab:8dMthy/IIB}
\end{center}  
as well as the expression for the $\mathbf{T}^3$ volume in terms of $\mathcal{V}_2$ and $R_3$ (see eq. \eqref{eq:T3volume}), one finds
\begin{equation} \label{eq:usefulmap}
      \mathcal{V}_2 = T_2^{\frac{2}{3}}\, e^{-\frac{2\varphi_8}{3}}\, , \qquad R_3 = T_2^{\frac{3}{7}}\, e^{\frac{4\varphi_8}{7}}\, .
\end{equation}
From these we can also deduce how the Type IIB coordinate $\nu$ relates to the M-theory variables, namely $\nu=\mathcal{V}_2^{-18/7}\, R_3^4\, \tau_2^{-3/2}$, thus allowing us to rewrite the first few terms in $\hat{E}_{3/2}^{sl_3}$ as
\begin{align}\label{eq:instexpSL3Mth1}
	\hat{E}_{3/2}^{sl_3} &= 2\zeta(3) \mathcal{V}_2^{9/7}\, R_3^{-2} + \frac{2\pi^2}{3} \tau_2 - 2\pi \text{log} (\tau_2) - \frac{2 \pi}{3} \log \left( \mathcal{V}_2^{9/7}\, R_3^{-2}\right) + \ldots\, ,
\end{align}
where the ellipsis indicates further non-perturbative contributions.

\subsubsection*{Asymptotic checks}

With this information, we are now ready to analyze whether the moduli-dependent Wilson coefficient associated to the  $t_8 t_8 \mathcal{R}^4$ operator in \eqref{eq:8dR^4IIB} exhibits the behaviour predicted by \eqref{eq:scalargravDlag}. To do so, we follow the same procedure as in the 9d example from Section \ref{ss:MthyT2}, therefore studying different representative infinite distance limits within $\mathcal{M}_{\rm 8d}$. Importantly, notice that by U-duality we can actually restrict ourselves to some fundamental domain $\mathscr{F}_8$ containing the minimal non-redundant information captured by the asymptotic phase diagram shown in Figure \ref{fig:MthyT3} above.\footnote{In Part \ref{part:pattern} of the thesis we will explain how to systematically construct such fundamental domains by looking precisely at the tangent bundle of the moduli space, see in particular the discussion in Section \ref{sss:sketch}.} For concreteness, we take a cone within $\mathcal{M}_{\rm 8d}$ spanned by geodesic directions associated to \emph{(i)} some emergent Type II string limit, \emph{(ii)} a pure decompactification to 10d (implementing M/F-theory duality) \emph{(iii)} as well as a third direction involving full decompactification to 11d M-theory. In particular, using an orthonormal frame adapted to the non-compact scalar fields appearing in \eqref{eq:8dalternativeactionII}, the aforementioned region corresponds to the cone generated by the following three directions in Figure \ref{fig:MthyT3}: one red dot at a vertex (emergent string limit), one purple dot at other vertex sharing a common edge with the former (decompactification to 11d), and finally one light purple dot (decompactification to 10d) belonging to the same facet as the other two and sharing an edge with the red dot but not with the purple one. Note that any such domain automatically includes additional asymptotic directions associated to yet another partial decompactification of two extra dimensions (light purple dot), and one partial decompactification of some $\mathbf{S}^1$ within the $\mathbf{T}^3$ (blue dot at the interior of the cone). A specific choice for $\mathscr{F}_8$ is shown in Figure \ref{fig:MthyT3}, which in 8d M-theory variables $\lbrace \tau_2, R_3, \mathcal{V}_2 \rbrace$ is defined by the following inequalities
\begin{equation}
    \label{eq:funddomainT3}
\mathcal{V}_2^{9/7} \, R_3^{-2}\,  \geq\, \tau_2\, \geq 1 \, ,  \qquad \mathcal{V}_2 \geq R_3^{-7/6} \, .
\end{equation}
Moreover, the three limiting directions spanning the cone \eqref{eq:funddomainT3} correspond to the following asymptotic species scales (in 8d Planck units) 
\begin{equation} \label{eq:QGcutoffs8d}
	\begin{split} 
		\frac{m_{\text{str}}}{M_{\rm Pl;\, 8}} &= \frac{(4\pi)^{1/3}}{\sqrt{2}} R_3^{1/3}\, \mathcal{V}_2^{-3/14} \, , \qquad \frac{M_{\rm Pl;\, 10}}{M_{\rm Pl;\, 8}} = (4\pi)^{1/24} R_3^{1/12}\, \mathcal{V}_2^{-3/56}\, \tau_2^{-1/8} \, ,\\
		\frac{M_{\rm Pl;\, 11}}{M_{\rm Pl;\, 8}} &= (4\pi)^{-1/18} R_3^{-1/6}\, \mathcal{V}_2^{-1/7} = (4\pi)^{-1/18} \mathcal{V}_3^{-1/6}\, ,
	\end{split}
\end{equation}
which eq. \eqref{eq:8dR^4IIB} should reproduce upon taking appropriate limits within $\mathscr{F}_8$. We study each of them in turn. 
\newline

\underline{\textit{The M-theory regime}}
\newline

Let us first consider the asymptotic regime of the selected fundamental domain where $M_{\rm Pl;\, 11}$ is the lightest of these three scales, and therefore fixes the species cut-off, i.e. $\LSP=M_{\rm Pl;\, 11}$. Geometrically, this corresponds to the $\mathbf{T}^3$ decompactification limit $\mathcal{V}_3 \to \infty$ with $\mathcal{V}_2 \leq R_3^7$, restricted to \eqref{eq:funddomainT3}. It is moreover associated to the $\mathsf{SL(2,\mathbb{Z})}$ sector of the theory, since the $\hat{E}_{1}^{sl_2}$ term within \eqref{eq:8dR^4IIB} clearly dominates. Hence, for all such limits we find (c.f. eq. \eqref{eq:asymptotic behavior})
\beq
2\hat{E}_{1}^{sl_2}(\mathcal{T}, \bar{\mathcal{T}}) = -2\pi\text{log} \left(\mathcal{T}_2\,|\eta(\mathcal{T})|^4\right)\, \sim\, \frac{\pi^2}{3} \mathcal{V}_3\, \propto\, \left(\frac{M_{\rm Pl;\, 11}}{M_{\rm Pl;\, 8}}\right)^{-6}\, ,
\eeq
where the asymptotic dependence should be understood when taking limit $\mathcal{T}_2\to \infty$. This matches exactly with eq. \eqref{eq:scalargravDlag}, thus recovering the expected suppression of the $\mathcal{R}^4$--\,term with $\LSP^6$.
\newline

\underline{\textit{The Type II regime}}
\newline
 
The asymptotic region in which the Type II fundamental string scale is the lightest within $\mathscr{F}_8$ is given by $\mathcal{V}_2 \geq R_3^7$, and hence corresponds to $\LSP=m_{\text{str}}$. In this regime, the leading contribution to \eqref{eq:8dR^4IIB} comes from the order--$\frac32$ $\mathsf{SL(3,\mathbb{Z})}$ Eisenstein series, such that all encompassing asymptotic boundaries are mapped to the limit $\nu \to \mathcal{V}_2^{-18/7}\, R_3^4\, \tau_2^{-3/2}\to 0$ in the Type IIB dual frame (see discussion around eq. \eqref{eq:usefulmap}). Thus, upon using the expansion \eqref{eq:instexpSL3Mth1}, one obtains the following leading order contribution to the $\mathcal{R}^4$--\,operator
\begin{align}
\label{eq:8dtotypeII}
	\hat{E}_{3/2}^{sl_3}\, \sim\, 2\zeta(3) \mathcal{V}_2^{9/7}\, R_3^{-2}\, \propto\, \left(\frac{m_{\text{str}}}{M_{\rm Pl;\, 8}}\right)^{-6}\, ,
\end{align}
which in turn reproduces the expected suppression with the species scale.
\newline

\underline{\textit{Decompactification to 10d}}
\newline

Note that the two regimes described so far already cover the entire asymptotic region associated to the fundamental domain defined in \eqref{eq:funddomainT3}. However, along certain directions, one may actually find additional partial decompactification limits, as we discuss in what follows. For instance, within the selected domain, the region where the 10d Planck mass sets the species cut-off corresponds to the boundary where it coincides with the string scale, namely $\tau_2=\, \mathcal{V}_2^{9/7}\, R_3^{-2}$ and $\mathcal{V}_2 \geq R_3^7$. In that case, the mass scale of the (double) KK tower represented by the light purple dot in Figure \ref{fig:MthyT3} is actually lighter than the string scale. Thus, what we end up seeing is actually a decompactification to ten dimensions, with the 10d Planck mass being parametrically of the same order as the string scale, i.e. $\LSP=M_{\rm Pl;\, 10}\sim m_{\text{str}}$. The dominant contribution to the $\mathcal{R}^4$ term thus takes the same form as in eq. \eqref{eq:8dtotypeII}, which can be equivalently expressed as
\beq
\hat{E}_{3/2}^{sl_3}\, \sim\, \left(\frac{M_{\rm Pl;\, 10}}{M_{\rm Pl;\, 8}}\right)^{-6} \, ,
\eeq
in agreement with \eqref{eq:scalargravDlag}.
\newline

\underline{\textit{Decompactification to 9d}}
\newline

Finally, we study the special direction corresponding to the center of the facet within $\mathscr{F}_8$, to which we can associate a single KK tower signalling decompactification from 8d to 9d as shown in Figure \ref{fig:MthyT3}. Furthermore, this asymptotic geodesic, which is parametrized by $\tau_2=\mathcal{V}_2=R_3^7 \to \infty$, is such that all potential candidates for the species cut-off in eq. \eqref{eq:QGcutoffs8d} scale in the same way. Hence, along this limit, we have 
\begin{equation}
   \hat{E}_{3/2}^{sl_3}\, \sim\, \hat{E}_{1}^{sl_2}\, \sim\,  \left(\frac{M_{\rm Pl;\, 9}}{M_{\rm Pl;\, 8}}\right)^{-6} \, ,
\end{equation}
yielding once again the correct dependence with the number of species for the $\mathcal{R}^4$--\,term.

All in all, we conclude that the function 
\beq
\LSP = \left( \hat{E}_{3/2}^{sl_3} +2\hat{E}_{1}^{sl_2}\right)^{-1/6}\, ,
\label{eq:8dfullspeciesfn}
\eeq
captures every single relevant asymptotic behaviour of the species scale in 8d maximal supergravity, as arising from e.g., M-theory compactified on $\mathbf{T}^3$. It is moreover invariant under the $\mathsf{SL(2,\mathbb{Z})}\times \mathsf{SL(3,\mathbb{Z})}$ duality group and thus reproduces precisely the phase diagram depicted in Figure \ref{fig:MthyT3}.

Before closing this section, let us remark that one can also consider higher order curvature corrections to the 8d effective action and perform a similar analysis, since some of these terms have been already computed in the literature (see e.g., \cite{Green:2010wi}). Upon doing so, one finds, similarly to the previous examples in ten and nine dimensions, that they are not in general suppressed by the appropriate power of the species scale, at least for certain type of limits (i.e. decompactification limits), but rather by the scale associated to the lightest tower. A potential physical explanation for this discrepancy will be explained later on in Section \ref{ss:gravEFTexpansion}. 

\section{String theory examples in lower dimensions}
\label{s:4dN=2}

In this section we will focus on 4d $\mathcal{N}=2$ settings arising from Type IIA string theory compactified on a Calabi--Yau three-fold $X_3$. Such theories are known to present, beyond the two-derivative lagrangian discussed in Section \ref{sss:4dN=2basics}, interesting higher-dimensional and higher-curvature corrections. In particular, there is an infinite number of F-terms, which are $\frac{1}{2}$-BPS and thus protected by supersymmetry, ensuring that their dependence with respect to the vector multiplet moduli can be computed in an exact manner. They read as follows \cite{Bershadsky:1993ta, Bershadsky:1993cx,Antoniadis:1993ze,Antoniadis:1995zn}:
\beq
\label{eq:lagrangian}
	S_{\text{IIA}}^{\text{4d}} \supset \int \dd^4x\, \sqrt{-g}\, \int \dd^4\theta\, \sum_{g\geq 1} \mathcal{F}_g (\mathcal{X}^A)\, \mathcal{W}^{2g}\ +\ \text{h.c.}\, ,
\eeq
where $\mathcal{F}_g (\mathcal{X}^A)$ is a chiral superfield that is related to the $g$-loop topological free energy of the supersymmetric closed string, $\theta^{\alpha}$ denote the fermionic $\mathcal{N}=2$ superspace coordinates and $\mathcal{W}_{\mu \nu} = F^+_{\mu \nu} - \mathcal{R}^+_{\mu \nu \rho \sigma} \theta \sigma^{\rho \sigma} \theta + \ldots\,$, is the Weyl superfield, which in Euclidean signature depends on the self-dual components of the graviphoton field strength and the Riemann tensor \cite{Antoniadis:1995zn}. Thus, upon performing the integration over the fermionic variables, one finds terms within \eqref{eq:lagrangian} of the form
\beq
\label{eq:GVterms}
	S_{\text{IIA}}^{\text{4d}} \supset \int \dd^4x\, \sqrt{-g}\, \left( \sum_{g\geq 1} \mathcal{F}_g(X^A)\, \mathcal{R}_+^2\, F_+^{2g-2} \right)\ +\ \text{h.c.}\, ,
\eeq
where $X^A$, $A=0,\ldots, h^{1,1}$, denote the bottom (i.e. scalar) components of the chiral superfields $\mathcal{X}^A$, c.f. eq. \eqref{eq:projcoords}.

As originally proposed in \cite{Gopakumar:1998ii,Gopakumar:1998jq}, one can alternatively compute the quantities $\mathcal{F}_g$ for $g\geq 0$ using the duality between Type IIA string theory on $X_3$ and M-theory on $X_3 \times \mathbf{S^1}$. For a single BPS particle of mass $m=|Z|$ --- where $Z$ denotes its central charge, one indeed obtains a generating function via a Schwinger-type one-loop computation in the presence of a constant self-dual graviphoton background, as follows
\begin{align}
\label{eq:generatingseries}
	\notag \sum_{g\geq 0}\mathcal{F}_g\, F_+^{2g-2} &= -\frac{1}{4} \int_{0^+}^{i \infty}\frac{\dd\tau}{\tau} \frac{1}{\sin^2{\frac{\tau F_+ \bar Z}{2}}} e^{-\tau m^2}\\
    &= \frac{1}{4} \int_{0^+}^{\infty}\frac{\dd\tau}{\tau} \sum_{g\geq0} \frac{2^{2g} (2g-1)}{(2g)!} (-1)^{g} B_{2g} \left( \frac{\tau F_+}{2}\right)^{2g-2} e^{-\tau Z}\, +\, \mathcal{O}\left(e^{-\frac{Z}{F_+}}\right)\, ,
\end{align}
where in the second step we rotated the integration contour 
and we have performed a perturbative expansion using the Laurent series for $\csc^2(x)$ around zero, namely
\begin{align}
	\frac{1}{\sin^{2}(x)} = \sum_{n=0}^{\infty} \frac{2^{2n}(2n-1)}{(2n)!} (-1)^{n-1} B_{2n} x^{2n-2}\, ,
\end{align}
which is valid for $0<|x|<\pi$. Notice that the coupling of the BPS particle to the background field crucially involves the anti-holomorphic piece of the mass \cite{Dedushenko:2014nya}. The $B_{2g}$ are referred to as Bernouilli numbers, which are given by
\begin{align}\label{eq:bernouilli}
	B_{2g}= \frac{(-1)^{g+1} 2 (2g)!}{(2\pi)^{2g}} \zeta(2g)\, .
\end{align}
From eq. \eqref{eq:generatingseries} one may already get a feeling of which $\mathcal{F}_g$ are UV sensitive/divergent versus those which actually provide for a convergent contribution. The claim would be that for $g\geq2$, the above integral converges in the UV, whilst for $g=0,1,$ one needs to adopt some regularization scheme. Indeed, one finds
\begin{align}
\label{eq:divergence}
	\mathcal{F}_g \propto \int_{\varepsilon}^{\infty} \dd\tau\, \tau^{2g-3}e^{-\tau Z} = Z^{2-2g} \Gamma(2g-2, \varepsilon Z)\, ,
\end{align}
where $\varepsilon=\Lambda_{\rm UV}^{-2}$ is nothing but the Schwinger implementation of the UV cut-off. Therefore, for $g > 1$, the incomplete gamma function converges to $\Gamma(2g-2) = (2g-3)!\,$, whilst for the remaining cases one finds a UV divergence that needs to be carefully dealt with.

In what follows, we will study the moduli dependence of the coefficients $\mathcal{F}_g(X^A)$ when probing certain representative infinite distance limits within the vector multiplet moduli space \cite{Lee:2019wij}.\footnote{The quantities $\mathcal{F}_g$, as computed by the topological string theory, are holomorphic in the chiral coordinates $\mathcal{X}^A$. There exists, however, a holomorphic anomaly in the quantum effective action associated to the contribution of the massless fields \cite{Bershadsky:1993cx}. For our purposes here, it will be enough to focus just on the holomorphic piece.} We will distinguish between operators that are relevant/marginal (in the Wilsonian sense), from those which are irrelevant (and thus UV convergent). The focus will be placed on understanding whether the general EFT expansion proposed in \eqref{eq:scalargravDlag} is fulfilled or not in the present 4d set-up.

\subsection{The $\mathcal{R}^2$--\,operator}
\label{ss:threshold4d}

Let us start with the only relevant/marginal BPS operator within the 4d lagrangian \eqref{eq:GVterms}, i.e. the one associated to $g=1$. It is proportional (in Euclidean signature) to the self-dual part of the curvature tensor squared, and its Wilson coefficient can be identified with the A-model topological free energy at genus one. In what follows, we review the mathematical definition and basic properties of $\mathcal{F}_1$, as computed from the topological string side.

This quantity can be defined as a supersymmetric index in the $\mathcal{N}=(2,2)$ superconformal field theory (SCFT) living on the worldsheet \cite{Cecotti:1992vy,Cecotti:1993stu}
\begin{align}
\label{eq:defF_1}
	\mathcal{F}_1 = \frac12 \int_{\mathscr{F}} \frac{\dd^2\tau}{\tau_2}\, \text{tr} \left( (-1)^F\, F_L\, F_R\, e^{2\pi \i H_0}\, e^{-2\pi \i \bar{H}_0}\right)\, ,
\end{align}
where $F_{L(R)}$ corresponds to the left-(right-)moving fermion number in the 2d theory, $F= F_L+F_R$ and $H_0$ denotes the associated Hamiltonian. Actually, the above index turns out to be IR divergent due to the contribution of the string massless states, and thus it should be properly regularized. This prescription only determines $\mathcal{F}_1$ up to an overall additive constant. Moreover, despite the apparent holomorphicity on the background flat coordinates $\{t^i\}$ --- i.e. the vector multiplet moduli, there is in fact some holomorphic anomaly which is captured by the following differential equation 
\begin{align}
\label{eq:holoanomalyF1}
	\frac{\partial^2\mathcal{F}_1}{\partial t^i \partial \bar{t}^j} = \text{tr}\, (-1)^F C_i \bar{C}_{\bar{j}} - \frac{1}{12} G_{i \bar{j}}\, \text{tr}\, (-1)^F\, ,
\end{align}
with $C_i (\bar{C}_{\bar{j}})$ being the structure constants of the (anti-)chiral ring of supersymmetric ground states and $G_{i \bar{j}}$ denotes the moduli space metric (see \cite{Bershadsky:1993ta} for details). Importantly, the above equation can be integrated exactly, thus fixing the moduli dependence of $\mathcal{F}_1$ up to some holomorphic function $f(t^i)$ that can be determined by confronting the solution to \eqref{eq:holoanomalyF1} with its expected boundary behaviour \cite{Bershadsky:1993ta,Bershadsky:1993cx}. The resulting expression would read as
\beq\label{eq:F1IIA}
	\mathcal{F}_1 =   \frac {1}{2}\left( 3+h^{1,1}-\frac {\chi_{E} (X_3)}{12}\right)K_{\text{ks}} + \frac {1}{2}\log \det G_{i \bar j} + \log|f|^2\, ,
\eeq
where $\chi_E(X_3)$ is the Euler characteristic of the three-fold $X_3$, whilst $K_{\text{ks}}$ denotes the K\"ahler potential for the moduli fields. Crucially, it turns out that the asymptotic properties exhibited by $\mathcal{F}_1$ indeed match the behaviour predicted by \eqref{eq:scalargravDlag}, where one should take into account that the corresponding $\mathcal{R}^2$--operator has mass dimension $n=4$, such that we expect the above quantity to behave like 
\beq\label{eq:F1QGcutoff}
	\mathcal{F}_1\, \sim\, \left(\frac{\Mpf}{\LSP}\right)^2\, .
\eeq
For illustrative purposes, let us briefly consider Type IIA string theory compactified on the Enriques Calabi--Yau $\left(K3 \times \mathbf{T}^2\right)/\mathbb{Z}_2$ \cite{Klemm:2005pd}, which is known to be dual to a Heterotic compactification on $K3 \times \mathbf{T}^2$. In this case, one finds that the moduli space metric behaves as $G_{T \bar{T}}=\frac{1}{4 T_2^2}\, $ (at large volume), whereas the genus-one topological free energy takes the following simple form \cite{vandeHeisteeg:2023ubh,Grimm:2007tm}
\beq \label{4dtopologicalfreeenergy}
\mathcal{F}_1 =-6 \log \left( T_2 |\eta(T)|^4\right) + \text{const.}\, , 
\eeq
where $T$ denotes the complexified K\"ahler modulus of the internal torus. In the original Type IIA frame, an emergent string limit arises when taking $T_2 \to \infty$ --- corresponding to a large volume limit for the internal $\mathbf{T}^2$, where the critical string is featured by a NS5-brane wrapping the $K3$-fibre, with tension
\beq
	T_{\rm NS5,\, str} = \frac{\Mpf^2}{2T_2}\, .
\label{eq:hettensionEnriques}
\eeq
In the dual Heterotic frame, such infinite distance degeneration is mapped to a perturbative weak coupling point for the fundamental string. Therefore, upon using eq. \eqref{eq:asymptotic behavior} we find
\beq
\mathcal{F}_1 =2\pi T_2 + \mathcal{O}\left( \log T_2\right)\, , 
\eeq
which is in perfect agreement with \eqref{eq:F1QGcutoff} above.

Let us remark here that an analysis along these lines can be analogously performed when probing other kind of infinite distance limits within the vector multiplet sector of 4d $\mathcal{N}=2$ theories, including partial decompactifications to M-/F-theory, see below. This was done in detail in refs. \cite{vandeHeisteeg:2022btw,vandeHeisteeg:2023ubh}, so we refrain from repeating it here and refer the interested reader to the original works.

\subsection{The irrelevant operators}
\label{ss:UVconvergent}

We turn now to the BPS operators in eq. \eqref{eq:GVterms} with $g>1$. Using eq. \eqref{eq:generatingseries}, we find that the contribution to $\mathcal{F}_{g>1}$ due to a particle of mass $m= |Z|$ is
\begin{equation}
\label{eq:Fg>1}
	\mathcal{F}_{g>1}= \frac{(2g-1)}{(2g)!} (-1)^g B_{2g} \int_0^{\infty} \dd\tau\, \tau^{2g-3}e^{-\tau Z}\, ,
\end{equation}
where one should substitute the appropriate Bernouilli numbers from \eqref{eq:bernouilli}. In the following, we will extract the relevant asymptotic behaviour of these higher-dimensional operators depending on the infinite distance limit that we approach. Later on, in Section \ref{ss:gravEFTexpansion} we comment on how these examples fit within the general framework discussed in Chapter \ref{ch:SpeciesIntro}.

\subsubsection*{M-theory limit}

The large volume point (with 4d dilaton fixed and finite) corresponds to a decompactification limit to 5d M-theory, where the M-theory circle grows large. This can be easily understood by looking at the light spectrum of the theory along the aforementioned limit, where it is precisely the tower of D0-branes which become light the fastest (see e.g., \cite{Font:2019cxq}). These states are $\frac12$-BPS, with a mass given by
\begin{equation}
\label{eq:D0mass}
	m_n = 2\pi |n|\, \frac{m_s}{g_s} = |n|\, m_{\rm D0}\, ,
\end{equation}
where $m_s$ is the fundamental Type IIA string scale and $n\in \mathbb{Z} \setminus \{ 0\}$ denotes the D0-brane charge. Notice that we are excluding the contribution of the massless (i.e. $n=0$) fields here, which would be actually part to the 4d EFT. After substituting in \eqref{eq:Fg>1}, one finds
\begin{align}
\label{eq:Fg>1D0}
	\mathcal{F}^{\rm D0}_{g>1}&= \chi_E(X_3) \frac{(2g-1) \zeta(2g)}{(2\pi)^{2g}} \sideset{}{'}\sum_{n \in \mathbb{Z}}\int_0^{\infty}d\tau\, \tau^{2g-3}e^{-\tau\, n\,  m_{\rm D0}} \notag\\
    &=\chi_E(X_3) \frac{(2g-1) \zeta(2g)}{(2\pi)^{2g}} \Gamma(2g-2)  m_{\rm D0}^{2-2g} \sideset{}{'}\sum_{n \in \mathbb{Z}} \frac{1}{n^{2g-2}} \notag\\
    &= \chi_E(X_3)\frac{2(2g-1) \zeta(2g) \Gamma(2g-2)}{(2\pi)^{2g}} \frac{\zeta(2g-2)}{m_{\rm D0}^{2g-2}}\, ,
\end{align}
which is of course convergent and moreover depends solely on the mass scale of the infinite tower of states, namely $m_{\rm D0}$, instead of the UV cut-off given by the species scale.

\subsubsection*{Partial decompactification limits}

Let us next consider the possibility that our Calabi--Yau three-fold presents some elliptic fibration $\pi: X_3 \to B_2$ (for simplicity we assume it to be non-singular). This means, in particular, that there exists an infinite distance limit at large volume within the vector multiplet moduli space where the base of the fibration blows up, whilst the volume of the elliptic fibre remains constant. Such limit corresponds to a partial decompactification limit to 6d F-theory \cite{Lee:2019wij}, where a tower of bound states with arbitrary D2 and D0-brane quantum numbers become asymptotically light (in 4d Planck units). The mass spectrum for such tower reads
\begin{equation}
\label{eq:ellipticmass}
	m_{n,\omega} = \frac{2\pi m_s}{g_s} \left| \omega z+n\right|\, ,
\end{equation}
where $z$ is the K\"ahler modulus associated to the elliptic fibre and $(n,\omega) \in \mathbb{Z}^2$ correspond to D0 and D2-brane charge, respectively. To properly account for the effect of such a tower one needs to sum over the integer set $(\omega, n)$, yielding
\begin{align}
\label{eq:Fg>1elliptic}
	\mathcal{F}^{\rm ell}_{g>1} (z)&= \chi_E(X_3)(-1)^{g-1} \frac{(2g-1)B_{2g} \Gamma(2g-2)}{2(2g)!} m_{\rm D0}^{2-2g} \sideset{}{'}\sum_{(\omega,n) \in \mathbb{Z}^2} \left(\omega z+n \right)^{2-2g}\notag\\
    &=\chi_E(X_3)(-1)^{g-1} \frac{(2g-1)B_{2g} \Gamma(2g-2)}{2(2g)!} \frac{G_{2g-2}(z)}{m_{\rm D0}^{2g-2}}\, ,
\end{align}
where in the last equality we have introduced the holomorphic Eisenstein series 
\begin{align}
\label{eq:holoEisenstein}
	G_{2k}(z) = \sideset{}{'}\sum_{(\omega, n) \in \mathbb{Z}^2} \frac{1}{\left(\omega z +n \right)^{2k}}\, .
\end{align}
Notice that the resulting operator is modular invariant,\footnote{The particular case of $g=2$ is a bit subtle, since it appears to be proportional to $G_2(z)$ which by itself is not a modular form. However, the holomorphic anomaly \cite{Bershadsky:1993cx} crucially solves this problem by promoting $G_2(z)$ in eq. \eqref{eq:Fg>1elliptic} to its non-holomorphic cousin, namely $\tilde{G}_2 (z, \bar z)= G_2(z)- \frac{\pi}{\text{Im}\, z}$, which now has definite modular weight.} as one can see from the fact that the terms in the lagrangian \eqref{eq:GVterms} are homogeneous in the fields $X^A$ of degree $2-2g$, whose K\"ahler transformation is exactly compensated by that of the graviphoton background field strength $F^+_{\mu \nu}$. Indeed, the relevant $\mathsf{SL(2,\mathbb{Z})}$ transformation corresponds to some generalized (double) T-duality which acts on the K\"ahler modulus as follows
\begin{align}\label{eq:Tdualitytrans}
	&z \rightarrow \frac{a\, z + b}{c\, z+d}\,,\qquad \text{with}\ \ \mathcal{A}= \begin{pmatrix}
		a \quad  b\\c \quad  d
	\end{pmatrix} \in \mathsf{SL(2,\mathbb{Z})}\, ,
\end{align}
whilst the vectors get transformed linearly through the matrix $\mathcal{A}$, which is moreover embedded into the symplectic K\"ahler group $\mathsf{Sp(2h^{1,1}+2, \mathbb{Z})}$, thus leaving all the F-terms in \eqref{eq:GVterms} invariant.\footnote{In general, these transformations are more complicated and also take into account the non-trivial fibration structure of the three-fold. This involves promoting the (double) T-duality in \eqref{eq:Tdualitytrans} to a Fourier-Mukai transform \cite{Andreas:2004uf, Cota:2019cjx}.} Note that the scale suppressing the tower of BPS operators becomes again that of the D2-particles (equivalently D0-branes, since both have asymptotically the same mass), and not the quantum gravity cut-off.

\subsubsection*{Emergent string limits}

Finally, we come to analyze infinite distance points in the vector multiplet moduli space corresponding to emergent string limits. These arise when the Calabi--Yau three-fold exhibits some $K3/\mathbf{T}^4$-fibration \cite{Lee:2019wij}, with the leading tower of asymptotically light states being the excitation modes of a dual critical string obtained by wrapping a NS5-brane on the generic fibre. As a concrete example, we consider here Type IIA compactified on $\mathbb{P}^{1,1,2,8,12}[24]$, which has $h^{1,1}=3$, $h^{2,1}=243$ and moreover exhibits a $K3$-fibration over a $\mathbb{P}^1$-base. The triple intersection numbers are \cite{Hosono:1994ax}
\beq
	\mathcal{K}_{111}=8\, , \quad \mathcal{K}_{112}=2\, , \quad \mathcal{K}_{113}=4\, , \quad \mathcal{K}_{133}=2,\, \quad \mathcal{K}_{123}=1\, .
\label{eq:triplenumbers}
\eeq
Upon probing the limit $t_b \to \infty$, where we denote by $t_b:= t^2$ the K\"ahler modulus associated to the $\mathbb{P}^1$-base, one encounters an infinite distance boundary of the emergent string kind, as discussed before. The corresponding tension of the wrapped NS5-brane reads
\beq
	T_{\rm NS5,\, str} = \Mpf^2 \frac{\mathcal{V}_{K3}}{2\mathcal{V}}\, ,
\label{eq:hettension}
\eeq
where $\mathcal{V}_{K3} = \frac{1}{2} \mathcal{K}_{2ij} t^i t^j = (t^1)^2 + t^1 t^3$ denotes the (classical) volume of the generic fibre and $\mathcal{V}$ that of the three-fold. Along the aforementioned limit, the 4d theory admits a dual interpertation in terms of a perturbative (i.e. weak coupling) limit for an $\mathsf{E_8} \times \mathsf{E_8}$ Heterotic compactification on $K3 \times \mathbf{T}^2$, with some $\mathsf{SU(2)}$ bundle (of instanton number 12) embedded in each of the $\mathsf{E_8}$ factors and such that all non-abelian symmetries are higgsed. The remaining $\mathsf{U(1)}$ factors come from the 3 vector multiplets associated to the geometric $\{T,U\}$ moduli of the internal torus and the complex dilaton $S= \frac{1}{2}\left(\varrho + \i e^{-2\varphi_4}\right)$, with $\varrho$ being a compact scalar dual to the Neveu-Schwarz 2-form $B_2$; as well as the graviphoton. Moreover, in the dual frame, the quantities $\mathcal{F}_{g>1}$ arise at one-loop order in string perturbation theory \cite{Antoniadis:1993ze,Antoniadis:1995zn}.

Incidentally, the moduli dependence of all the relevant higher-derivative couplings can be encapsulated at once upon defining the generating function 
 \beq
    F(\lambda, T, U)= \sum_{g=1}^{\infty} \lambda^{2g} \mathcal{F}^{\rm het}_g(T,U)\, ,
 \eeq
for which a formal exact expression is available in the case of interest \cite{Marino:1998pg}
\beq
	F(\lambda, T, U) = \frac{1}{2\pi^2} \int_{\mathscr{F}} \frac{\dd^2\tau}{\tau_2} \left( \frac{G_4 G_6}{\eta^{24}} \sum_{\Gamma^{2,2}} q^{\frac12 |p_L|^2} \bar{q}^{\frac12 |p_R|^2}\right) \left[ \left( \frac{2\pi \i \lambda \eta^3}{\theta_1(\tilde \lambda| \tau)}\right)^2 e^{-\frac{\pi \tilde \lambda^2}{\tau_2}}\right]\, .
\label{eq:Fheteroticoneloop}
\eeq
Here $\mathscr{F}$ denotes the $\mathsf{SL(2,\mathbb{Z})}$ fundamental domain, $G_4(\tau)$ and $G_6(\tau)$ are holomorphic Eisenstein series (c.f. eq. \eqref{eq:holoEisenstein}), $\theta_1$ is the Jacobi theta function with characteristics $(1/2,1/2)$, and we have defined the quantities
\beq
	\tilde \lambda = \frac{p_R \tau_2 \lambda}{\sqrt{2T_2 U_2}}\, , \qquad q=e^{2\pi \i \tau}\, ,
\eeq
where $p_{L,R}$ are the right-/left-moving momenta along the torus: 
\begin{align}
   p_L&= \frac{1}{\sqrt{2T_2U_2}} \left( n_1 +n_2 \bar{T} + m_2 U +m_1 \bar{T}U\right)\, ,\notag \\
   p_R&= \frac{1}{\sqrt{2T_2U_2}} \left( n_1 +n_2 T + m_2 U +m_1 T U\right)\, .
\end{align}
The important point for us is that all Wilsonian couplings captured by $F(\lambda, T, U)$ arise at one-loop and are thus proportional to $(S-\bar S)^0$ when written in the string frame. Therefore, upon switching to the 4d Einstein frame and taking the perturbative string limit, namely when $S \to \i \infty$ --- for generic values of the moduli $\{ T,U\}$, the functions $\mathcal{F}_{g>1}$ behave as follows
\beq
	\mathcal{F}^{\rm het}_{g>1} \sim \Mpf^{2-2g}(\text{Im}\, S)^{g-1} \sim \left(m_{\text{str, het}}\right)^{2-2g}\, ,
\eeq
where $m_{\text{str, het}}$ denotes the fundamental Heterotic string scale. In the original Type IIA description, this means that upon probing the limit $t_b \to \infty$, one finds
\beq \label{eq:F_g>1emergentstring}
	\mathcal{F}^{\rm IIA}_{g>1} \sim \Mpf^{2-2g} t_b^{g-1} \sim (T_{\rm NS5, \, str})^{1-g}\, ,
\eeq
with $T_{\rm NS5,\, str}$ given in eq. \eqref{eq:hettension} above. In this case, the scale suppressing the F-terms \eqref{eq:GVterms} does coincide with the species scale along the limit $t_b \to \infty$, since the latter is given by the (emergent) Heterotic string scale. Still, the particular power exhibited by \eqref{eq:F_g>1emergentstring} does not seem to agree with the expectations coming from eq. \eqref{eq:scalargravDlag}.

\subsection{General lessons}
\label{ss:gravEFTexpansion}

After the previous general considerations, we are now in good position to discuss the asymptotic mass dependence of the irrelevant F-terms appearing in generic 4d $\mathcal{N}=2$ theories arising from quantum gravity. We argue in the following that the relevant UV scale suppressing these terms in the EFT is the characteristic mass of the tower, such that the series expansion reads
\beq
\mathcal{L}_{\mathrm{IIA}}^{\text{4d}} \supset \sqrt{-g} \left(\sum_{n > 4} \frac{\mathsf{O}_n (\mathcal{R})}{m_{\rm tow}^{n-4}} \right) \, ,
\label{eq:gravEFTexpansionIII}
\eeq
where $n$ denotes the mass dimension of the corresponding coupling. To see this, we first note that the dimension of the BPS operators $\mathcal{R}_+^2\, F_+^{2g-2}$ in \eqref{eq:GVterms} is $n=2g+2$. Hence, for $g>1$ we find that $n > 4$, such that they are all \emph{irrelevant} in the Wilsonian sense. Furthermore, the asymptotic moduli dependence arising in the three possible types of infinite distance boundaries within the vector multiplet moduli space follows the behaviour displayed in \eqref{eq:gravEFTexpansionIII}, where $m_{\rm tow}$ becomes either $m_{\rm D0}$ or $T_{\rm NS5,\, str}^{1/2}$. For instance, in the large volume limit we obtained $\mathcal{F}_{g>1} \propto m_{\rm D0}^{2-2g}$, c.f. eq. \eqref{eq:Fg>1D0}, whilst for the emergent Heterotic string the dependence was of the form $\mathcal{F}_{g>1} \propto \left(T_{\rm NS5,\, str}^{1/2} \right)^{2-2g}$, see eq. \eqref{eq:F_g>1emergentstring}.

Our aim in this section will be to take the first steps towards understanding whether this observed suppression by the lightest scale in the ultra-violet is an artifact due to some sort of fine tuning \cite{Heckman:2019bzm}, or it rather captures some general behaviour. We distinguish between decompactification and emergent string limits in what follows, for reasons that will become clear along the way.

\subsubsection*{Decompactification limits}

Let us consider first infinite distance limits signalling towards decompactification of one or more internal dimensions. Along these, the dominant tower of states becomes the Kaluza-Klein replica, whose masses scale inversely with the volume of the growing cycle. In particular, in the supersymmetric examples studied both in Sections \ref{s:Exampleshighdim} and \ref{ss:UVconvergent}, it was found that for such decompactification limits the suppression exhibited by the \emph{irrelevant} operators was of the form
\beq
\mathcal{L}_{\mathrm{EFT}} \supset \sqrt{-g} \left(\sum_{n > d} \frac{\mathsf{O}_n (\mathcal{R})}{\Lambda_{\rm UV}^{n-d}}\right) \, ,
\label{eq:gravEFTexpansionII}
\eeq
which follows the usual rules of EFT expansions with $\Lambda_{\rm UV}=m_{\rm KK} \ll \LSP$, see Section \ref{ss:nonrenormalizableEFTs}. This includes the BPS operators $\partial^4 \mathcal{R}^4$ and $\partial^6 \mathcal{R}^4$ appearing in maximally supersymmetric set-ups in $8\leq d \leq 10$, as well as those of the form $\mathcal{R}_+^2\, F_+^{2g-2}$ (for $g>1$) arising in 4d $\mathcal{N}=2$ theories. What we want to point out here is that this behaviour actually fits with our quantum field theory experience and, if properly interpreted, it does not imply any violation of the expected suppression given by the quantum gravity cut-off, as shown in eq. \eqref{eq:scalargravDlag}.

Let us elaborate more on this important point. The idea would be that in general, when integrating out a (tower of) particle(s) in quantum field theory, one expects to get threshold corrections plus possibly divergent contributions in the UV, depending on the precise dimension of the operator under study. The latter must be properly regularized, and in quantum gravity the natural prescription would be to cut them off at the species scale, according to \eqref{eq:scalargravDlag}. The former, on the other hand, always follow the pattern shown in \eqref{eq:gravEFTexpansionIII}. Therefore, when starting from some definite $d$-dimensional EFT and upon exploring a decompactification limit, the behaviour exhibited by any higher-curvature coupling will crucially depend on which of these two terms dominates. In fact, it is easy to get convinced --- upon using the relation between the higher-dimensional Planck mass and the tower scale, namely eq. \eqref{eq:speciesscaleKK} --- that the suppression controlled by the Kaluza-Klein scale becomes dominant over the `bare' quantum gravity contribution if and only if $n > d+k$, where $k$ is the number of internal dimensions that get decompactified along the limit. Otherwise, the dimensionally reduced term associated to the UV divergence takes over, in accordance with the series expansion \eqref{eq:scalargravDlag}. This of course agrees with the examples studied in previous sections and can be tested even further.\footnote{See also \cite{Aoufia:2024awo} for a recent extension of the argument to non-geometric phases using worldsheet CFT techniques.} For instance, recall that the operator $t_8 t_8 \mathcal{R}^4$ has mass dimension $n=8$, such that it stops being relevant/marginal for $d\leq 7$. In particular, in the case of seven-dimensional maximal supergravity as obtained from Type IIB on $\mathbf{T}^3$, one finds the following generalized Wilson coefficient for the latter \cite{Green:2010wi}
\beq\label{eq:R4in7d}
 E_{3/2}^{sl_5} = \sideset{}{'}\sum_{\mathbf{m} \in \mathbb{Z}^2,\, \mathbf{n} \in \mathbb{Z}^3} \left[ V_3^{2/5} \left( \frac{\left| m_1 + m_2\tau + \mathbf{B}\cdot \mathbf{n}\right|^2}{\tau_2} + \frac{\mathbf{n}^{\rm T} \cdot \tilde{g}^{-1} \cdot \mathbf{n}}{V_3^{2/3}}\right)\right]^{-3/2}\, ,
\eeq
where $\tau$ denotes the axio-dilaton (see discussion after eq. \eqref{eq:IIB10dSL2}), $\tilde{g}_{mn}=(\det g)^{-1/3}\, g_{m n}$ is the unimodular metric defined on $\mathbf{T}^3$, $V_3$ refers to the volume of the torus measured in 10d Planck units, whilst the moduli-dependent vector $\mathbf{B}$ contains three scalar fields
\beq
 B^i= c_2^i + \tau b_2^i\, , \qquad i=1,2,3\, ,
\eeq
which can be obtained by reducing on any 2-cycle the complex combination of 2-forms given by $C_2+\tau B_2$. The function \eqref{eq:R4in7d} is a special case of the Epstein series and is moreover manifestly invariant under the $\mathsf{SL(5,\mathbb{Z})}$ U-duality group of the theory. Let us now take the full decompactification limit to 11d M-theory. Along this limit, the dominant term in the $E_{3/2}^{sl_5}$ series corresponds to the zero mode, which reads \cite{Green:2010wi}
\beq\label{eq:zeromodeR4in7d}
 E_{3/2}^{sl_5} = \mathcal{V}_4^{9/20}\, E_{3/2}^{sl_4} + 4 \zeta(2)\mathcal{V}_4^{6/5} + \ldots\, ,
\eeq
where it is easy to see that the first contribution is controlled by the mass of the Kaluza-Klein tower and the second matches \eqref{eq:scalargravDlag} with $\LSP = M_{\text{Pl}\,; 11}$.\footnote{Recall that the 7d and 11d Planck masses are related as follows $M_{\text{Pl}\,; 11} = M_{\text{Pl}\,; 7}\, \mathcal{V}_4^{-1/5}$, whereas the overall Kaluza-Klein scale associated to the internal $\mathbf{T}^4$ reads as $m_{\rm KK} = M_{\text{Pl}\,; 7}\, \mathcal{V}_4^{-9/20}$.} Here we denote by $\mathcal{V}_4$ the volume of the $\mathbf{T}^4$ upon which we compactify M-theory to arrive at the aforementioned 7d theory. Hence, we see that since the $t_8 t_8 \mathcal{R}^4$ operator is relevant in the decompactified theory, the suppression is controlled by the species cut-off rather than the tower scale.

To provide more evidence in favour of the previous claims, let us consider another simple example: $k$-(super-)graviton scattering in $d$ spacetime dimensions. We focus here on the one-loop contribution induced by the Kaluza-Klein replica of some massless particle to $\frac12$-BPS operators of the schematic form $\mathcal{R}^k$, which can be nicely expressed using the worldline formalism (or Schwinger method), see e.g., \cite{Schwartz:2014sze}. This amounts to computing the following one-loop amplitude 
\beq
\mathcal{A}_{k,\,d}\, =\, \int_0^{\infty} \dd^dp \int_{0}^{\infty} \frac{\dd\tau}{\tau}\, \sum_n e^{-\tau \left( p^2+\frac{n^2}{\mathsf{R}^2}\right)}\, \text{tr}\, \Bigg \langle \prod_{r=1}^k \left( \int_0^{\tau} \dd t_r\, V_{g_{\mu \nu}} (t_r)\right) \Bigg \rangle\, ,
\label{eq:wordlineamplitude}
\eeq
where $\tau$ parametrizes the worldline proper time, whilst $\mathbf{p}$ and $n \in \mathbb{Z}$ denote the momenta along the non-compact and $\mathbf{S}^1_{\mathsf{R}}$ directions, respectively. The quantities appearing in brackets contain the $k$ vertex operators associated to the massless (super-)gravitons, which are inserted at every possible proper time in the worldline, and the trace involves an integral over fermionic zero modes \cite{Green:1999by}. For BPS operators, the vertex insertions in \eqref{eq:wordlineamplitude} already saturate the fermionic trace, yielding a simpler expression of the form
\beq
\mathcal{A}_{k,\,d}\, =\, \tilde K \sum_{n \in \mathbb{Z} \setminus \{ 0\}} \int \dd^dp \int_{0}^{\infty} \frac{\dd\tau}{\tau}\, \tau^k\, e^{-\tau \left( p^2+m_n^2\right)}\, ,
\label{eq:kgravscattering}
\eeq
where $m_n= \frac{|n|}{\mathsf{R}} = |n|\, m_{\rm KK}$ and $\tilde K$ denotes the appropriate kinematic factor that accounts for the Lorentz structure of the amplitude.\footnote{The kinematic factor $\tilde K$ can be easily obtained as a linearized version of the relevant operator whose corrections we are computing.} After performing the integration over the loop momenta one is left with
\beq\label{eq:kgravscatteringKKtower}
\mathcal{A}_{k,\,d}\, =\, \tilde K \sum_{n \in \mathbb{Z} \setminus \{ 0\}} \int_{0}^{\infty} \dd \tau\, \tau^{k-1-d/2}\, e^{-\tau\, m_n^2}\, .
\eeq
Notice from the above expression that the amplitude converges in the UV (i.e. the $\tau \to 0$ region) if and only if $2k-1-d >0$, which is equivalent to ask for the operator $\mathcal{R}^k$ to be irrelevant already in the parent $D=d+1$ dimensional theory. Finally, after performing the summation over the quantized momentum number we find
\beq\label{eq:kgravscatteringKKtowerII}
\mathcal{A}_{k,\,d}\, =\, 2\tilde K\, \Gamma\left(k-\frac{d}{2}\right) \sum_{n >0} \frac{1}{m_n^{2k-d}} = 2\tilde K\, \Gamma\left(k-\frac{d}{2}\right) \frac{\zeta(2k-d)}{m_{\rm KK}^{2k-d}}\, ,
\eeq
in agreement with \eqref{eq:gravEFTexpansionII}. Let us remark that if instead of summing over all Kaluza-Klein modes we instead truncate up to some maximum excitation number $N_{\rm max}$ such that $N_{\rm max} m_{\rm KK} \sim \LSP$, one still recovers the same result minus a correction of the form $\mathcal{A}^{\rm corr}_{k,\,d} \sim \Mpd^{d-2}\,\LSP^{2-2k}$. Nonetheless, despite the similarity with \eqref{eq:scalargravDlag}, thi
s latter contribution is of course subleading with respect to \eqref{eq:kgravscatteringKKtowerII}.

In a similar vein, one could consider operators involving spacetime derivatives of the Riemann tensor, such as $\partial^{2\ell} \mathcal{R}^k$, whose mass dimension is $2(\ell+k)$. Indeed, essentially the same argument as before leads to the following schematic form of the amplitude (after integrating over loop momenta)
\beq
\mathcal{A}_{k + \ell,\,d}\, =\, \tilde K \sum_{n \in \mathbb{Z} \setminus \{ 0\}} \int_{0}^{\infty} \dd\tau\, \tau^{k+ \ell-1-d/2}\, e^{-\tau\, m_n^2} = 2\tilde K\, \Gamma\left(k+\ell-\frac{d}{2}\right) \frac{\zeta(2(k+\ell)-d)}{m_{\rm KK}^{2(k+\ell)-d}}\, ,
\label{eq:kgravscatteringwithlderivatives}
\eeq
whose convergence properties now depend on whether $2(k+\ell)-d-1 >0$ or not, which again is equivalent to the corresponding operator being irrelevant or otherwise in the higher-dimensional theory.

To conclude, we point out that even in those cases where one finds the Kaluza-Klein term \eqref{eq:gravEFTexpansionIII} to provide for the dominant suppression of the corresponding higher-curvature operator, once we decompactify the theory and resum the highly non-local effects induced by the KK modes, one should find agreement with the EFT series \eqref{eq:scalargravDlag}. Therefore, a more refined statement would to be to claim that the quantum gravity cut-off $\LSP$ actually suppresses generic gravitational operators of dimension higher than two --- with respect to the Einstein-Hilbert term --- when measured \emph{at} the species scale itself \cite{Wiesnerprivate}.

\subsubsection*{Emergent string limits}

Next, we turn to emergent string limits, where the dominant tower of asymptotically light states becomes the excitation modes of a (possibly dual) fundamental string. Recall that it was precisely along these limits where the EFT expansion in \eqref{eq:scalargravDlag} was fulfilled in maximal supergravity set-ups, even for irrelevant operators such as $\partial^4 \mathcal{R}^4$ or $\partial^6 \mathcal{R}^4$ (c.f. Section \ref{s:Exampleshighdim}).

A first indication that the situation is dramatically different for string towers compared to the Kaluza-Klein scenario is provided by the mass degeneracy of states. Namely, even if the one-loop contribution to say $\mathcal{A}_{k,\,d}$ converges in the UV for each mode in the tower, the summation over the oscillator number $n$ diverges, a priori
\beq
\mathcal{A}^{\rm osc}_{k,\,d} \propto \tilde K\, \Gamma\left(k-\frac{d}{2}\right) \sum_{n >1} \frac{d_n}{m_n^{2k-d}} \to \infty\, ,
\eeq
where $m_n^2= 16\pi^2 m_s^2 (n-1)$ and $d_n \sim n^{-11/2} e^{4\pi \sqrt{2n}}$ in the case of e.g., the Type II fundamental string (c.f. eq. \eqref{eq:exactleveldensitystrings}). This means, in particular, that a naive analysis using worldline/QFT techniques is strictly speaking not valid in this case, as we already knew. Instead, a worldsheet computation should be more appropriate. In this regard, it is useful to investigate what string perturbation theory teaches us about the suppression of these terms in the low energy regime.  

To address this question, let us briefly come back to the 4d example analyzed in Section \ref{ss:UVconvergent}, where we considered Type IIA string theory on $\mathbb{P}^{1,1,2,8,12}[24]$. From this simple set-up we can extract two important lessons. First, even for emergent string limits, one can in principle find an \emph{infinite} number of gravitational terms in the effective action which do not accommodate the ansatz proposed in \eqref{eq:scalargravDlag}. This follows from a straightforward genus counting, such that whenever the higher-curvature operator in question receives a \emph{tree-level} contribution in string perturbation theory, the behaviour predicted by \eqref{eq:scalargravDlag} will be automatically fulfilled. On the contrary, if the leading term --- whenever $g_s^{-1} \gg 1$ --- comes at \emph{one-loop} order, one rather finds agreement with eq. \eqref{eq:gravEFTexpansionIII}, thus giving some a priori smaller suppression. 
However, this does not cause any real problem though, since the scale controlling these operators relative to the Einstein-Hilbert term behaves like $m_s^{\frac{n-d}{n-2}}$, such that for large enough classical dimension $n$ it asymptotes to $m_s$, which precisely provides for the species cut-off $\LSP$.


\section{Summary}\label{s:summaryhigherops}

In the present chapter we have tested the idea of identifying the species scale $\LSP$ as the energy cut-off controlling the EFT expansion of any effective description weakly coupled to Einstein gravity. To do so, we systematically analyzed the first few higher-curvature interactions modifying the two-derivative description of certain string theory constructions in ten, nine, eight and four spacetime dimensions, focusing on those terms whose moduli dependence can be computed exactly. These consist in BPS-protected operators in maximal supergravity, as well as certain F-terms in 4d $\mathcal{N}=2$ theories. Such local operators are moreover (marginally) relevant in the Wilsonian sense, and indeed probe the ultra-violet nature of quantum gravity in a non-trivial fashion, therefore having potentially great impact on the infra-red physics via e.g., (small) black holes or rather through entropy considerations (see e.g., \cite{Calderon-Infante:2023uhz,Cribiori:2023ffn}). 
In all the aforementioned cases, the moduli-dependent function controlling their associated Wilson coefficients was seen to match asymptotically with the usual species counting computations. This means that, starting from any dual frame, one can retrieve in principle the fundamental quantum gravity scale at any other corner of the duality web in the theory by looking at how these operators change when moving around the moduli space. Our analysis here moreover extends the results from refs. \cite{vandeHeisteeg:2022btw,vandeHeisteeg:2023ubh}, which we reviewed and broadened in Section \ref{s:4dN=2}. 

On the other hand, despite this encouraging agreement with the expectations described in Chapter \ref{ch:SpeciesIntro}, several relevant observations were made. First, the simple expansion in power series of the inverse cut-off was seen to appear only \emph{asymptotically}. Hence, even though the different gravitational operators are suppressed by the same energy scale close to infinite distance degenerations, this may be no longer the case when venturing towards the bulk of moduli space. We believe that this has to do with the fact that the expansion in \eqref{eq:scalargravDlag} should be taken as some sort of approximation strictly valid close to the boundaries, where weak coupling behaviours arise and the perturbative series rapidly converge. There, the classical dimension of the different operators in the EFT expansion becomes reliable, whilst in the interior large quantum corrections (both perturbative and non-perturbative) may occur, thus giving rise to big anomalous dimensions. Second, we found that certain non-leading gravitational operators in the theory, given by higher derivatives of four Riemann tensors in maximal supergravity \cite{Green:2010kv,Green:2010wi} or higher-dimensional F-terms of the form $\mathcal{R}^2\, F^{2g-2}$ in the 4d set-up \cite{Bershadsky:1993ta, Bershadsky:1993cx,Antoniadis:1993ze,Antoniadis:1995zn}  --- with $F_{\mu \nu}$ denoting the graviphoton field strength, do not seem to exactly follow the pattern proposed in \eqref{eq:scalargravDlag}, but are instead controlled by different integer powers of the lightest tower scale. Interestingly though, once we decompactify the theory and look at the same higher-curvature operator, the behaviour observed is in perfect accordance with the predictions made in Chapter \ref{ch:SpeciesIntro}. Hence, a more clear-cut statement would be that indeed the species cut-off seems to control the effective field theory expansion in gravitational theories, and this can be seen upon measuring the corresponding Wilson coefficients at energy scales close to the cut-off itself. 


Finally, let us point out certain issues that are raised by our findings here. First, it would be interesting to try to extend this analysis to other set-ups both in different number of spacetime dimensions and supersymmetries.\footnote{See \cite{
vandeHeisteeg:2023dlw} for recent progress along these lines.} This is particularly pressing in the case of 16 supercharges or less, where certain infinite distance limits probe running-solutions (i.e., not vacua) \cite{Etheredge:2023odp} and where the computation of the species scale seems challenging with our current techniques (see Section \ref{ss:het s1} for more on this). Moreover, one could try to study further implications of these considerations within the Swampland program, as well as to revisit certain naturalness/fine tuning arguments that are commonly employed in model building scenarios.

\chapter{The Emergence Proposal}\label{ch:Emergence} 
The four-dimensional Universe we observe seems to be in a weakly-coupled phase, although not very far away from strong coupling. For instance, the electro-weak coupling (c.f. discussion after eq. \eqref{eq:fermilagrangian}) is measured to be around $g \approx 0.7$ at the electro-weak scale itself. One could then envisage a situation in which the underlying fundamental theory is actually strongly coupled in the ultra-violet, whilst the perturbatively weak interactions arise only in the infra-red after following the renormalization group flow below some fundamental UV scale. Thus, in a sense, one could say that the corresponding kinetic terms would be {\it emergent}. This may happen e.g., if the theory is not asymptotically free below the aforementioned ultra-violet scale due to the presence of a large number of very massive vector-like particles. Alternatively, the fact that the fundamental theory could be strongly coupled can be effectively reformulated by postulating vanishing (or at least very small) kinetic terms for all the light states in the theory. One could then hope to generate such kinetic functions deep in the IR by quantum corrections involving heavy particles. However, as appealing as this idea may seem, to actually make it work is not as easy as it sounds, since loop contributions to the field metrics are in general divergent and hence cut-off dependent. 
Counterterms for these divergences at the cut-off scale would in principle be needed --- if one tries to extend the theory up to the continuum limit, against the original assumption that no significant kinetic terms were present at the fundamental UV scale.
	
Nonetheless, the possibility of fully generating the relevant kinetic terms in low energy effective field theories has been recently reconsidered in the context of quantum gravity. The motivation for this is purely theoretical at this point, and stems from the necessity of background independence in any underlying theory of quantum gravity. In a similar vein, the holographic principle \cite{tHooft:1993dmi,Susskind:1994vu,Bousso:1999xy, Bousso:1999cb} (see also \cite{Bousso:2002ju} for a review) posits that the maximum information content associated to any given spacetime region is encoded into its adjacent area, therefore limiting the number of fundamental degrees of freedom in quantum gravity as compared to usual local field theories, which allow for a volume-law growth. This latter point of view is also supported by explicit examples in holography such as the AdS/CFT correspondence \cite{Maldacena:1997re,Witten:1998qj}, where local degrees of freedom in the bulk --- and even spacetime itself --- are believed to emerge non-trivially from the boundary data. Furthermore, it has been argued that an emergence principle for both moduli fields and gauge bosons in theories of quantum gravity may provide for a microscopic understanding of some of the most prominent Swampland criteria, namely the Weak Gravity and the Distance conjectures (see Section \ref{s:SwamplandProgram} for details). Such a phenomenon in this specific context is usually referred to as the {\it Emergence Proposal} (see \cite{Harlow:2015lma, Grimm:2018ohb,Corvilain:2018lgw,Heidenreich:2017sim, Heidenreich:2018kpg,Palti:2019pca} for the original works and \cite{Castellano:2022bvr, Hamada:2021yxy,Marchesano:2022axe,Castellano:2023qhp,Blumenhagen:2023yws,Kawamura:2023cbd,Seo:2023xsb,Blumenhagen:2023tev,Blumenhagen:2023xmk,Calderon-Infante:2023uhz,Hattab:2023moj,Casas:2024ttx,Blumenhagen:2024ydy,Hattab:2024thi,Blumenhagen:2024lmo} for multiple follow-ups). In spite of these surprising connections, almost all the evidence in favour of the proposal has been developed in simple toy model constructions, with the remarkable exception of certain studies in the large complex structure point of 4d $\mathcal{N}=2$ theories obtained from Type IIB compactified on a Calabi--Yau three-fold \cite{Grimm:2018ohb}. Hence, what one would like to know is whether Emergence can be formulated as a general phenomenon in quantum gravity, as well as to study how the kinetic terms in bona-fide low energy $d$-dimensional EFTs may emerge, in practice. In addition, one would like to address the question of whether other terms in the effective lagrangian such as scalar potentials or even higher-derivative and higher-curvature interactions, could also appear fully as an infra-red effect.

In order to answer properly these questions it is crucial to understand what is the precise regime of validity of the low energy \emph{emergent} descriptions. In other words, what is the energy scale beyond which the local fields in the EFT no longer provide for the relevant degrees of freedom of the theory. This, in turn, sets the required renormalization group boundary condition for the latter, hence imposing vanishing kinetic terms at that scale, whilst the non-trivial kinematics would arise in the infra-red after integrating out the dual massive degrees of freedom. On the other hand, from our analysis in Chapters \ref{ch:SpeciesIntro} and \ref{ch:Higherdimops}, we know that such energy cut-off should be identified with the species cut-off $\LSP$, which sets the energy scale at which strong quantum-gravitational effects must be taken into account, therefore invalidating any low-energy effective field theory description. Based on this, we expect a strong interconnection between the idea of Emergence in quantum gravity and the concept of the species scale, which is what we investigate in the following.

The chapter is organized as follows. In Section \ref{s:CPNEmergence} we review the $\mathbb{P}^{\,\mathsf{N}}$--\,model, which provides for a toy model of the emergence mechanism in field theory. Then, in Section \ref{s:EmergenceQG} we discuss certain general aspects of the Emergence Proposal in quantum gravity. In particular, we draw the connection with the Swampland program, elaborating on its relation with the Weak Gravity as well as the Distance conjectures. A crucial role within this story is played by quantum (loop) computations, and in Section \ref{s:selfenergybosons} we present detailed calculations that will be necessary later on in order to check the proposal in realistic string theory vacua. Subsequently, in Section \ref{s:EmergenceStringTheory} we test these ideas within string theory compactifications in diverse spacetime dimensions and preserving different amounts of supersymmetry. Finally, in Section \ref{s:emergenceinteractions} we comment briefly on how the proposal can account for the generation of other operators in the effective action different than the kinetic terms, including the case of higher-dimensional and higher-derivative operators. 

The material presented hereafter is based on the publication \cite{Castellano:2022bvr}, which has been slightly adapted and broadened so as to better fit with the rest of this thesis. (See also \cite{Castellano:2023qhp} for interesting phenomenological applications of the Emergence Proposal.)

\section{Emergence in quantum field theory}
\label{s:CPNEmergence}

The idea of emergent gauge fields in quantum field theory is rather old, having a long history behind. First, in \cite{DAdda:1978vbw} it was pointed out how composite gauge bosons become dynamical in $\mathbb{P}^{\,\mathsf{N}}$ sigma-models. Subsequently, inspired by this work and in the context of $\mathcal{N}=8$ supergravity, Cremmer and Julia \cite{Cremmer:1979up} suggested that composite gauge fields transforming in the adjoint of $\mathsf{SU(8)}$ could acquire non-trivial kinetic terms and even be used as a unification group. This idea was further explored in \cite{Ellis:1980cf,Ellis:1980tf}, where the possibility of embedding $\mathsf{SU(5)}$ Grand Unified Theories within this context was scrutinized. In the end, however, these considerations did not work as expected due to several reasons, most notably the endemic presence of anomalies. In the following, we will revisit these models so as to illustrate in a simple example how the phenomenon of Emergence works in field theory. We will closely follow the discussion in \cite{Palti:2019pca} (see also \cite{Rabinovici:2011jj,Milekhin:2012ca,Witten:1978bc,Gross:1974jv}), and we refer the reader interested in the details to the original references.

\subsubsection*{The $\mathbb{P}^{\,\mathsf{N-1}}$ model}

Here we consider the following sigma-model for $N$ complex scalar fields $\{ z^i \}$ in four spacetime dimensions, whose lagrangian reads
\beq
\mathcal{L} = -\partial_{\mu} \bar{z}^i \partial^{\mu} z_i  + \left( \bar{z}_i \partial_{\mu} z^i\right) \left( \bar{z}_j \partial_{\mu} z^j\right) \, ,
\label{eq:CPnlagrangianI}
\eeq
and which are subject to the non-linear constraint $\bar{z}^i z_i = \frac{N}{g^2}$, where the physical meaning of $g$ will become more clear later on. This theory enjoys a $\mathsf{U(1)}$ gauge symmetry which acts on the scalars as $z^j \to e^{\i \alpha (x)} z^j$, thus leaving the lagrangian \eqref{eq:CPnlagrangianI} invariant, as it may be easily checked.\footnote{The set $\{ z^i\}$ can be thought of as local inhomogeneous coordinates on $\mathbb{P}^{\,\mathsf{N-1}}$, which topologically is described as $\mathbf{S}^{\mathsf{2N}}/\mathsf{U}(1)$. Moreover, the kinetic term in \eqref{eq:CPnlagrangianI} arises from the natural non-linear K\"ahler metric of the Fubini-Study type.} Such invariance can be made manifest upon introducing an auxiliary 1-form field $A_{\mu}$ as follows
\beq
\mathcal{L} = -\overline{D_{\mu} z^i} D^{\mu} z_i  - \sigma \left( \bar{z}^i z_i - \frac{N}{g^2}\right) \, ,
\label{eq:CPnlagrangianII}
\eeq
where $D_{\mu} = \partial_{\mu} - \i A_{\mu}$ is the familiar covariant derivative and we have also introduced a lagrange multiplier $\sigma$ implementing the constraint mentioned above. Notice that the 1-form field is \emph{classically} non-dynamical --- i.e. it has no explicit kinetic term in \eqref{eq:CPnlagrangianII} --- and transforms under the $\mathsf{U(1)}$ gauge symmetry as usual, namely $A_{\mu} \to A_{\mu} + \partial_{\mu} \alpha$. Of course, what we have done is nothing but a relabelling and the original lagrangian can be retrieved upon integrating out the non-dynamical gauge field, i.e. upon imposing its associated equation of motion:
\beq
A_{\mu} = \frac{\i g^2}{2 N} \left( z_i \partial_{\mu} \bar{z}^i-\bar{z}_i \partial_{\mu} z^i\right)\, ,
\label{eq:gaugefieldEOM}
\eeq
which ensures that $A_{\mu}$ indeed transforms under the gauge symmetry as stated before. On the other hand, when considering the associated quantum theory, it turns out to be more convenient to redefine the scalar fields as $z^i \to \frac{\sqrt{N}}{g} z^i$, yielding the following lagrangian \cite{Harlow:2015lma}
\beq
\mathcal{L} = -\frac{N}{g^2} \left[ \overline{D_{\mu} z^i} D^{\mu} z_i  + \sigma \left( \bar{z}^i z_i - 1\right) \right] \, ,
\label{eq:CPnlagrangianIII}
\eeq
since that way one can perform a perturbative expansion in $1/N$, such that the large $N$ limit may be interpreted as the classical regime of the theory. It is now intuitively clear as well that the parameter $g$ can be thought of as some sort of coupling constant for the non-linear sigma-model.

Classically, the scalar fields $\{ z^i \}$ are exactly massless and their v.e.v.s parametrize the vacuum manifold of the theory,\footnote{Note that the $\mathbb{P}^{\,\mathsf{N-1}}$ model also enjoys a $\mathsf{SU(N)}$ global symmetry which precisely relates the different classical vacua mentioned in the main text.} where also $\sigma = A_{\mu} = 0$ on-shell. Quantum-mechanically, however, the story gets a bit more interesting, and in fact the aforementioned degeneracy of vacua is lifted by loop effects. This can be seen upon integrating out the scalar fields in the path integral, which is moreover one-loop exact (at the perturbative level) and leads to the following euclidean quantum effective action (c.f. eq. \eqref{eq:EFTpathintegral})
\begin{align}
\Gamma_{\Lambda} \left[ A_{\mu}, \sigma \right] &= - N \left( \log \det \left( - D^2 + \sigma\right) + \frac{1}{g^2} \int \dd^4 x\, \sigma\right) \notag\\
&=-N \int \dd^4 x \Bigg \lbrace \int \dd^4p\, \log \left[ \left( p-A(x)\right)^2  + \sigma \right] -\frac{1}{g^2} \sigma \Bigg\rbrace\, ,
\label{eq:CPnquantumeffectiveaction}
\end{align}
where in the second step we have switched to the momentum parametrization and we denote by $\Lambda$ the UV cut-off of the 4d theory. From \eqref{eq:CPnquantumeffectiveaction} we can determine what are the vacua that survive at the quantum level. This amounts to compute the effective scalar potential $V_{\rm eff} (\sigma)$, which reads
\begin{align}
V_{\rm eff} (\sigma) = -N \left[ \int \dd^4p\, \log \left( p^2  + \sigma \right) -\frac{1}{g^2} \sigma \right]\, ,
\label{eq:CPnpotential}
\end{align}
and find its minima, which are determined by the following implicit equation
\begin{align}
\frac{1}{g^2} = \int_{p^2 \leq \Lambda^2} \dd^4p\ \frac{1}{p^2 + \braket{\sigma}}\, .
\label{eq:CPnminimumpotential}
\end{align}
It is easy to show that for $g > \left(\int_{p^2 \leq \Lambda^2} \dd^4p\ \frac{1}{p^2} \right)^{-1/2}$ one can find a physical real solution for $\braket{\sigma}$ which moreover satisfies $\braket{\sigma} >0$.\footnote{More precisely, one finds that if $g^{-2} = \beta \int_{p^2 \leq \Lambda^2} \dd^4p\ p^{-2}$ with $1 \gtrsim \beta \gtrsim 0.3069$ there is a unique real solution for $\braket{\sigma}$ in \eqref{eq:CPnminimumpotential} which reads $\braket{\sigma} = \alpha \Lambda^2$, where $\alpha^{-1}= -\left(1-\beta\right)^{-1} W_{-1} \left( -\left(1-\beta\right) e^{-\frac{1}{1-\beta}}\right) -1 > 1$, c.f. footnote \ref{fn:Lambert}.} This implies, via eq. \eqref{eq:CPnlagrangianIII}, that the complex fields $\{ z^i \}$ acquire some positive mass $m_z = \sqrt{\braket{\sigma}}$ in the quantum vacuum, such that it is actually meaningful to integrate them out and construct an effective field theory description for energies below $m_z$. Doing so leaves us with an EFT for the massless field $A_\mu$, whose Wilsonian effective action is computed to be 
\begin{align}
\Gamma_{m_z} \left[ A_{\mu}\right] &= - N \log \det \left( - D^2 + m_z^2 \right) = - \frac{1}{4e^2} \int \dd^4 x\ F^2\ +\ \mathcal{O} \left( \frac{1}{m_z^2}\right)\, ,
\label{eq:effectiveactionphoton}
\end{align}
where $F^2 = F_{\mu \nu} F^{\mu \nu}$ is the field strength squared and we have expanded the (logarithm of the) determinant of $ - D^2 + m_z^2$ above up to leading order in $1/m_z^2$. Crucially, notice that contrary to the original lagrangian \eqref{eq:CPnlagrangianIII}, the 1-form gauge field now exhibits non-trivial dynamics in the infra-red, with a gauge coupling $e$ which reads  
\begin{align}
\frac{1}{e^2} = \frac{N}{12 \pi^2} \log \frac{\Lambda}{m_z}\, ,
\label{eq:gauglecouplingphoton}
\end{align}
that follows precisely the familiar behaviour of the renormalization group running in scalar QED, with the particularity of having a vanishing kinetic term in the ultra-violet, namely
\begin{align}
\frac{1}{e^2} \bigg |_{\rm UV} = 0\, .
\label{eq:gauglecouplingphotonUV}
\end{align}
From this perspective, one may argue that the low energy dynamics of the massless gauge boson \emph{emerges} completely from the quantum contributions associated to the charged massive scalar fields $\{ z^i \}$, which provide the fundamental degrees of freedom for energies above $m_z$. On the other hand, below that energy scale the physics is governed by the $\mathsf{U}(1)$ gauge boson instead, which would be a composite field from the UV point of view (c.f. eq. \eqref{eq:gaugefieldEOM}). Let us note, in passing, that the higher-order terms in \eqref{eq:effectiveactionphoton} --- which include operators of the form $F^4$ or even $F_{\mu \nu} \partial^2 F^{\mu \nu}$ --- also emerge from the non-trivial quantum corrections associated to the scalar fields, and moreover match the corresponding quantities in scalar QED (with trivial renormalization boundary conditions in the ultra-violet). We will discuss similar matters in the context of quantum gravity later on in Section \ref{s:emergenceinteractions}.

\section{Emergence in quantum gravity}
\label{s:EmergenceQG}

The previous discussion nicely illustrates how non-trivial dynamics for certain low energy field-theoretic degrees of freedom can be fully generated at the quantum level from (a finite number of) very massive fields. Furthermore, we learned that the fundamental degrees of freedom in the ultra-violet are, in a sense, dual to the dominant ones in the infra-red regime, which may be seen as composite and in fact describe the relevant physics that survives up to very low energies.

Therefore, a natural question that one may ask at this point is whether such an emergence phenomenon may happen as well in the context of gravity. In trying to address this point we face several new problems that were absent in the $\mathbb{P}^{\,\mathsf{N-1}}$ example from Section \ref{s:CPNEmergence}. On the one hand, it is clear that the gravitational field cannot emerge from a simple underlying local and Lorentz-invariant quantum field theory, as per the Weinberg-Witten theorem \cite{Weinberg:1980kq}. This does not imply, however, that gravity cannot be emergent at all, but rather that whatever the underlying theory is, it must be of fundamentally different nature and possibily non-local --- such as string theory. (Notice that this issue does not arise in the field theory example since the emergent gauge field $A_{\mu}$ carries no charge under the conserved $\mathsf{U}(1)$ 1-form current.) Relatedly, the purported non-locality of the fundamental theory implies that the dual ultra-violet modes from which the graviton may emerge cannot be arranged into a finite family of degrees of freedom --- as in the $\mathbb{P}^{\,\mathsf{N-1}}$ model --- and in fact it is very natural to expect the same role to be played now by an \emph{infinite} number of those. In practice, however, this is bad news for us since it complicates tremendously a general study of the emergence mechanism in generic gravitational EFTs. Nevertheless, in certain circumstances we may be able to perform some zero-th order analysis and even use the more familiar tools of quantum field theory via the renormalization group flow, as explained below in more detail.
 
More generally, there exists a serious proposal in quantum gravity, dubbed \emph{Emergence} \cite{Palti:2019pca,Harlow:2015lma,Grimm:2018ohb,Corvilain:2018lgw,Heidenreich:2017sim,Heidenreich:2018kpg}, which holds that the dynamics of all massless (or at least very light) degrees of freedom entering into the effective low energy description of semi-classical gravity emerges from integrating out certain towers of states below some particular energy cut-off, that is bounded above by the Planck scale. Thus, to first approximation, the Emergence Proposal may be stated as follows:

\newpage

\begin{center}
	\textbf{Emergence Proposal (Strong)}: \textit{In a consistent theory of quantum gravity all light particles in a perturbative regime  have no kinetic terms in the UV. The required kinetic terms appear as an IR effect upon integrating out a tower of asymptotically massless states.} 
\end{center}	
	
Furthermore, it has been argued \cite{Palti:2019pca,Harlow:2015lma,Grimm:2018ohb} that the condition of vanishing kinetic terms for the light degrees of freedom in the ultra-violet regime of gravity could perhaps suggest the existence of an underlying \emph{topological} fundamental theory, wherein these particles do not propagate. 
Hence, there would be couplings for these non-propagating fields --- as in the $\mathbb{P}^{\,\mathsf{N-1}}$ model --- but no geometric objects to start with, i.e. no kinetic terms (see also \cite{Harlow:2015lma,Agrawal:2020xek}). 

Alternatively, one could be slightly more conservative and formulate the following general (but less ambitious) version of the Emergence Proposal \cite{Heidenreich:2017sim,Heidenreich:2018kpg}:

\begin{center}
	\textbf{Emergence Proposal (Weak)}: \textit{In a consistent theory of quantum gravity, for any singularity located at infinite distance in moduli space, there is an associated infinite tower of states becoming massless which induce quantum corrections to the metrics matching the `tree-level' singular behavior}. 
\end{center}
	
Notice that the two formulations are closely related to each other, although they depart in a conceptual but important way. In practice, when analyzing (if possible) the behaviour induced by the infinite towers of states on the kinetic terms of the massless modes, the two statements only differ by the UV boundary condition, namely whether the metric evaluated at the quantum gravity cut-off vanishes (\emph{strong}) or rather it provides for a contribution which is of the same order or subleading (\emph{weak}). Conceptually, however, this difference has a tremendous impact on our understanding of the underlying physics, since it means that in the former case one can really say that e.g., the graviton dynamics completely emerges from that of the ultra-violet fundamental degrees of freedom, whilst for the latter such an statement would be rather inaccurate.

Let us stress here that the proposal, as currently formulated, is still highly speculative. Despite the recent progress towards our understanding of the latter, there remains important open questions that must be addressed before claiming victory. However, both its simplicity as well as the good amount of evidence in favour of the conjecture that we present in this chapter (see also \cite{Marchesano:2022axe, Blumenhagen:2023yws,Kawamura:2023cbd,Seo:2023xsb,Blumenhagen:2023tev,Blumenhagen:2023xmk,Calderon-Infante:2023uhz,Hattab:2023moj,Casas:2024ttx,Blumenhagen:2024ydy,Hattab:2024thi,Blumenhagen:2024lmo} for subsequent works) suggest that these ideas may be indeed pointing towards some interesting physics in quantum gravity. Moreover, there are also various heuristic arguments that support this claim. For instance, the philosophy of the gravitational effective field theory already capturing the dynamics of non-perturbative massive (and dual) states is in agreement with the more familiar example of the conifold singularity in Type IIB string theory compactified on a Calabi--Yau three-fold \cite{Candelas:1990rm,Candelas:1989ug,Candelas:1989js, Aspinwall:1993nu}. In that case, the resolution of the pathological behaviour exhibited by the kinetic terms in the EFT can be understood as fully generated by integrating out a single D3-brane BPS mode that becomes massless precisely at the conifold locus \cite{Strominger:1995cz}. Still, this set-up is both qualitatively and quantitatively different in the sense that in the Emergence cases here described an infinite number of states become asymptotically massless in the limit, yielding an infinite distance singularity, as opposed to the finite number of states for the conifold, which give rise to a finite distance one. Another heuristic motivation for the emergence prescription comes from holography. Indeed, in AdS/CFT \cite{Maldacena:1997re,Witten:1998qj} the graviton bulk dynamics arises, when seen from the boundary perspective, as correlation functions involving the stress-energy momentum-tensor $T_{i j}$, which may be thought of as a \emph{derived} quantity in terms of the `fundamental' fields of the 4d $\mathcal{N}=4$ super-Yang-Mills theory --- as it was also the case for the 1-form gauge field in the $\mathbb{P}^{\,\mathsf{N-1}}$ model, c.f. eq. \eqref{eq:gaugefieldEOM}. For example, one may identify \cite{Mueck:1998ug,Liu:1998bu}
\begin{align}
\braket{h_{i j} (\mathbf{x}) h_{r s} (\mathbf{y})}_{\rm bulk} &= \braket{T_{i j} (\mathbf{x}) T_{r s} (\mathbf{y})}_{\rm boundary} \notag\\
&= \frac{20}{\pi^2 \left| \mathbf{x}-\mathbf{y}\right|^8} \left[ \frac12 \left(J_{ir} (\mathbf{x}-\mathbf{y}) J_{js} (\mathbf{x}-\mathbf{y}) + \left( i \leftrightarrow j \right) \right) -\frac14 \delta_{ij} \delta_{rs}\right]\, ,
\label{eq:2pointgravAdS}
\end{align}
where $h_{i j} (\mathbf{x})$ denotes the fluctuation in the gravitational field around the AdS vacuum, the points $\{\mathbf{x}, \mathbf{y} \}$ belong to the (conformal) boundary of AdS and we have defined the quantity
\begin{align}
J_{ij} (\mathbf{x}) = \delta_{ij} - 2 \frac{x_i x_j}{\mathbf{x}^2}\, .
\end{align}
This does not constitute a proof for the emergence of the graviton, but instead can be regarded as circumstantial evidence for its reformulation in terms of more `fundamental' degrees of freedom, in this case the conformal field theory living at the boundary. 
 
In any event, the aim of this chapter will be to present further compelling evidence in favour of the Emergence Proposal within string theory set-ups. Strictly speaking, in this work we restrict ourselves to the more conservative weaker version of the proposal, even though the results may also be fully compatible with the stronger one.

\subsection{Emergence of the graviton and the species cut-off}
\label{ss:Emergencegraviton}

\begin{figure}[tb]
		\begin{center}
			\includegraphics[scale=1.3]{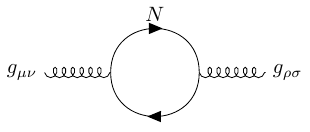}
			\caption{Contribution of the one-loop determinant of $N$ light species to the graviton self-energy. This can be related to the emergence of the gravitational field, as explained in the text.}
			\label{fig:oneloopgraviton}
		\end{center}
\end{figure}

One interesting observation that can be drawn at this point concerns the relation between the \emph{perturbative} arguments in favour of the species scale as the quantum gravity cut-off (see Section \ref{ss:perturbative} for details) and the emergence --- in the weaker sense --- of the graviton kinetic term. Indeed, as we elaborated on in Chapter \ref{ch:SpeciesIntro}, any state carrying energy and momentum couples to the gravitational field and in particular it can renormalize the Einstein-Hilbert action, as shown in Figure \ref{fig:oneloopgraviton}. In the following, we would like to make this statement more precise by considering how (towers of) states with different spin --- such as scalars, fermions, vectors, etc. --- contribute to the graviton kinematics at one-loop. For concreteness, let us assume that there is an infinite tower of species with masses $m_1 \leq m_2 \leq \ldots \leq m_N \leq \LSP$ weakly coupled to gravity. For energies well below $m_1$, we can integrate out the tower up to the quantum gravity cut-off, thus obtaining certain threshold corrections to the `bare' Einstein-Hilbert term, which reads as
\begin{equation}\label{eq:EH}
	S_{\text{EH},\, \LSP} = \frac{\LSP^{d-2}}{2} \int \dd^{d}x\, \sqrt{-g}\,  \mathcal{R}\, .
\end{equation}
Note that in the previous expression we have assumed the reduced Planck mass measured at energies around the cut-off $\LSP$ to be given precisely the species scale itself. Therefore, upon using the worldline formalism\cite{Schubert:2001he, Vassilevich:2003xt, Bastianelli:2008cu,Bastianelli:2005rc} --- which is manifestly gauge invariant --- one finds (see Appendix \ref{ap:heatkernel} for details)
\begin{equation}\label{eq:quantumEH}
	S_{\text{EH},\, \text{eff}} = \frac{\LSP^{d-2}}{2} \int \dd^{d}x\, \sqrt{-g}\,  \mathcal{R} \left( 1+\sum_{n=1}^N \gamma_n \right)\, ,
\end{equation}
where $\{\gamma_n\}$ comprise some positive $\mathcal{O}(1)$ factors depending both on the statistics and spin of the corresponding field (c.f. eqs. \eqref{eq:dewittcoeffscalar} and \eqref{eq:dewittcoefffermion}). Moreover, if we assume the total number $N$ of such particles to be very large, we can then take the summation to be roughly proportional to the number of species, such that
\begin{equation}\label{eq:finalquantumEH}
	S_{\text{EH},\, \text{eff}}\, =\, \int \dd^{d}x\, \sqrt{-g}\,  \left(\frac{N\, \LSP^{d-2}}{2}\, \mathcal{R}\ +\ \mathcal{O}\left(\frac{1}{N} \right)\right)\, ,
\end{equation}
which thus defines \emph{a posteriori} the Planck mass measured in the infra-red to be given by $\Mpd^{d-2} := \LSP^{d-2}\, N$. Notice that in order to arrive at \eqref{eq:finalquantumEH} it is indeed crucial that all fields entering into the one-loop computation contribute \emph{positively} to the latter, irrespective of their spin or statistics. This amounts to asking for no `anti-screening' phenomenon to happen in gravitational theories, and is to be expected based on general physical grounds \cite{Anber:2011ut,Donoghue:1994dn, Han:2004wt}.\footnote{Heuristically, the reason for this stems from the fact that the graviton couples to energy, which is positive definite. Furthermore, the modern perspective based on unitary methods (see e.g., \cite{Elvang:2013cua,Kruczenski:2022lot} and references therein) allows one to reconstruct generic loop corrections from tree-level diagrams, which would carry the positivity of the \emph{on-shell} coupling between gravity and energy-momentum \cite{Donoghueprivate}.} Furthermore, let us mention that even in the extreme case where $m_n \approx \LSP$ for all $n \leq N$, one obtains a correction of the form (c.f. eq. \eqref{eq:integratedoneloop})
\begin{equation}\label{eq:extremequantumEH}
	\mathcal{L}_{\text{EH},\, \text{eff}}\, \sim\, \sqrt{-g} \left( N\, \LSP^{d-2} E_{\frac{d}{2}} (1)\, \mathcal{R} \right)\, \sim\, \sqrt{-g} \left(\Mpd^{d-2}\, \mathcal{R} \right)\, ,
\end{equation}
where $E_{k}(z)$ is the exponential integral function. This means that even if all $N$ species have a mass of order of the quantum gravity cut-off, we still get the desired result \eqref{eq:finalquantumEH}.

Let us stress at this point that, strictly speaking, one does not need to assume the boundary condition \eqref{eq:EH} at the cut-off scale in order to obtain the correct asymptotic dependence for the Einstein-Hilbert action in the infra-red. Consequently, this would allow for the possibility of fully generating the graviton kinematics via quantum corrections induced by the infinite towers of states, as in the stronger versions of the Emergence Proposal. 

\subsection{Relation to the Swampland conjectures}
\label{ss:Emergence&Swampland}
	
One attractive feature of the Emergence Proposal is that it provides us with a very simple microscopic rationale for the understanding the existence of both the Weak Gravity and the Distance conjectures \cite{Arkani-Hamed:2006emk,Heidenreich:2015nta,Heidenreich:2016aqi,Montero:2016tif,Andriolo:2018lvp,Ooguri:2006in, Etheredge:2022opl}.\footnote{See also \cite{Stout:2021ubb,Stout:2022phm} for alternative approaches to explain the Distance Conjecture based on information theory.} In order to give a flavour of why this is so, let us consider here a simple toy model in $d > 4$ spacetime dimensions with a BPS-like spectrum of charged particles labeled both by their $\mathsf{U(1)}$ gauge charge $n \in \mathbb{Z}$ and mass $m_n=|n|\, \Mt$, with $|n| \leq N$, where $N$ is effectively very large. We also assume that these massive states couple to a single real modulus $\phi$ through its field-dependent mass, namely $m_n = m_n (\phi)$.

\begin{figure}[t]
		\begin{center}
			\subfigure[]{
				\includegraphics[height=4.0cm]{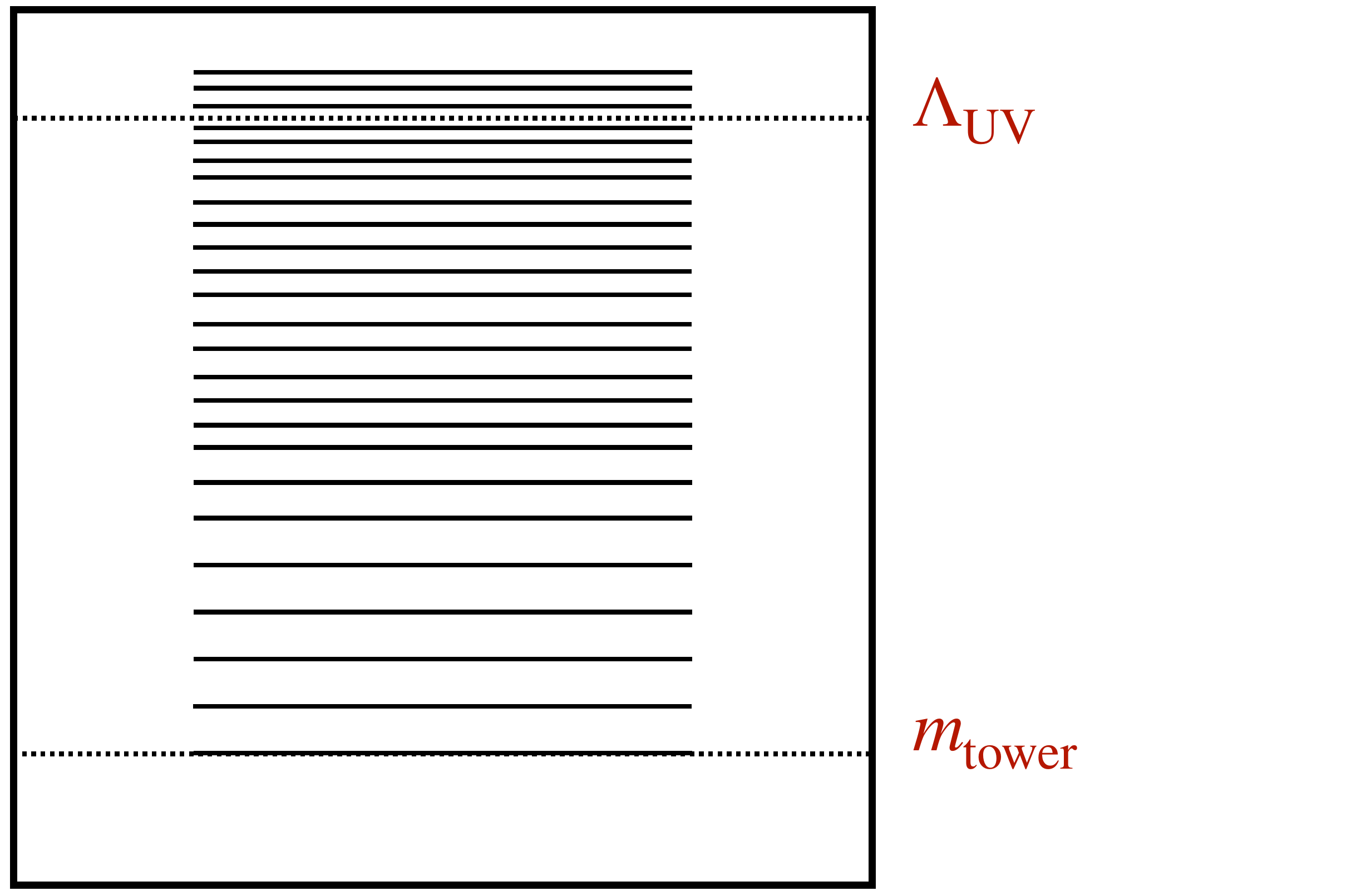} 
				\label{torresa}
			}
			\subfigure[]{
				\includegraphics[height=4.0cm]{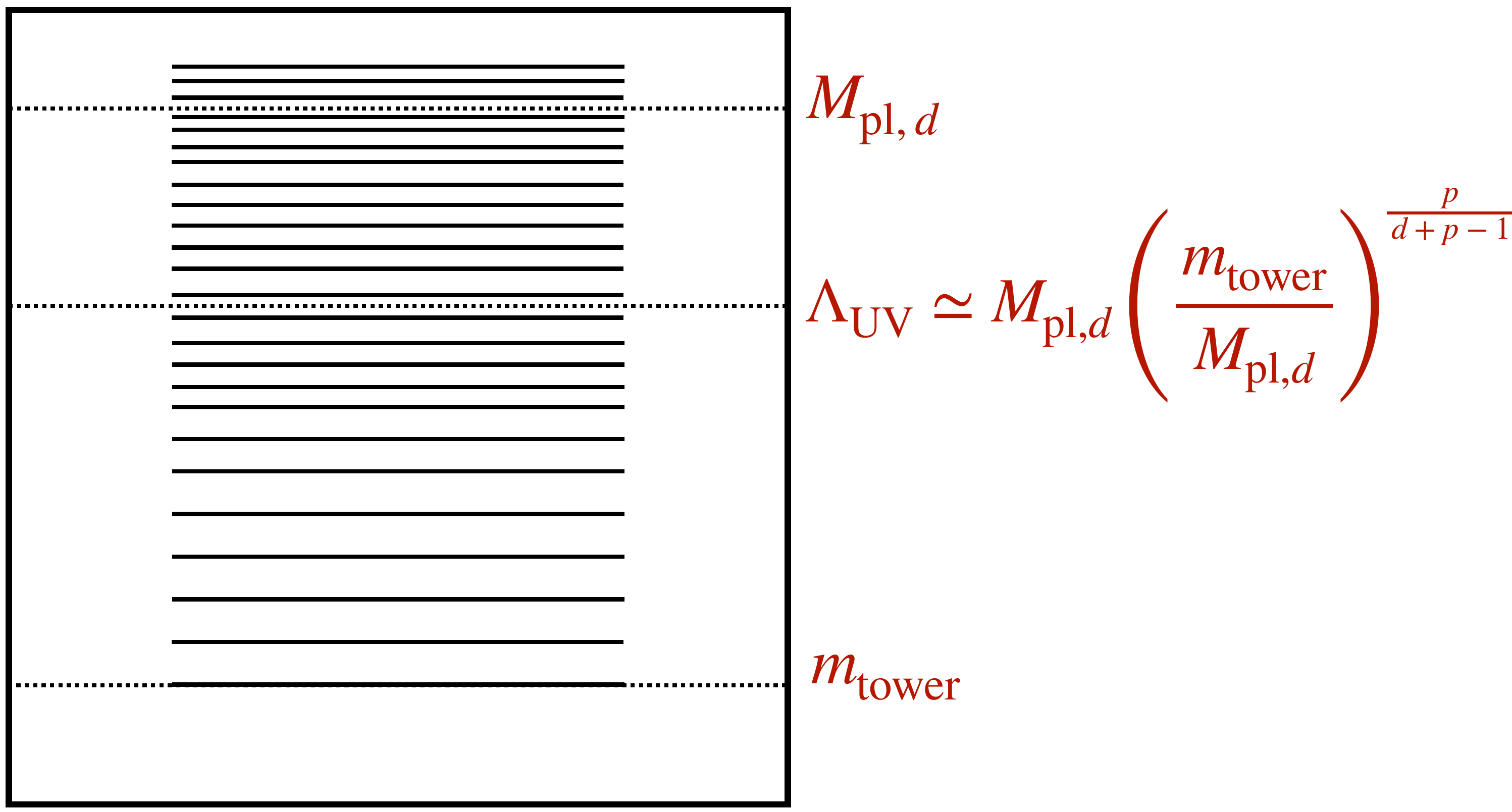}
				\label{torresb}
			}
			\caption{Emergence of metrics from a tower of states. \textbf{(a)} In the absence of gravity they are sensitive to the cut-off $\Lambda_{\text{UV}}$. \textbf{(b)} In the presence of gravity the UV cut-off must be identified with the species scale. The latter is related to the mass scale of the tower $\Mt$ such that in the end, only $\Mt$ and the Planck mass, $\Mpd$, appear in the EFT.}			
		\label{torres}
		\end{center}
\end{figure}

On the one hand, if gravity was no present we could still compute the one-loop contribution to the modulus kinetic term, which we denote by $g_{\phi\phi}$ (see Figure \ref{torresa}). For concreteness, we take the infinite tower to be comprised by e.g., massive fermionic fields $\Psi_n$, whose moduli-dependent mass is controlled by some Yukawa interaction of the type $(\partial_\phi \Mt)\, \phi \ \overline{\Psi^{(n)}} \Psi^{(n)}$. Therefore, upon summing over the whole spectrum we find\footnote{We ignore for the moment some subtleties associated to the precise loop computations which will be explained in upcoming sections. Let us stress though, that the results presented here are essentially unchanged.}
\beq \label{eq:metricemergence}
	\delta g_{\phi\phi}\, \sim\, \sum_{n=1}^N n^2 \Lambda_{\text{UV}}^{d-4}  (\partial_\phi \Mt)^2\, \sim\, N \Lambda_{\text{UV}}^{d-2} \left(\frac {\partial_\phi \Mt}{\Mt}\right)^2\, ,
\eeq
where we have not kept track of the $\mathcal{O}(1)$ factors that arise either from the loop diagram nor upon performing the finite sum, which has been cut off at an energy scale $\Lambda_{\text{UV}} = N\, \Mt$ beyond which our effective description breaks down. Indeed, we see that a kinetic term is obtained at the quantum level, but it is in principle divergent if one naively insists on taking the continuum limit, i.e. $\Lambda_{\text{UV}} \to \infty$. However, the renormalization prescription will force us to have some (divergent) kinetic term already at the UV scale so as to be able to make finite  physical predictions at low energies. Thus, one cannot simply claim that kinematics is completely induced in the infra-red. 

Similarly, one can easily compute the quantum contributions to the wave-function of the $\mathsf{U(1)}$ gauge boson induced by the tower of charged fermions. As is well-known, this captures the renormalization of the (inverse) gauge coupling, which reads
\beq
	\delta \left(\frac {1}{g^2}\right)\, \sim\,  \sum_{n=1}^N n^2 \Lambda_{\text{UV}}^{d-4}\, \sim\, N \Lambda_{\text{UV}}^{d-2} \frac {1}{\Mt^2}\, ,
	\label{eq:gaugecouplingemergence}
\eeq
where we are again ignoring any numerical prefactor in \eqref{eq:gaugecouplingemergence}. Notice that, as it was also the case for the scalar field, strictly speaking there is no emergence phenomenon whatsoever, since large kinetic terms must be already present in the UV regime.  

Let us now reconsider the previous analysis but now in the context of quantum gravity (see Figure \ref{torresb}). In this case, as already argued in Part \ref{part:QGscale} of the thesis, it makes sense to identify the ultra-violet cut-off $\Lambda_{\text{UV}}$ with the species scale, c.f. eq. \eqref{species}. By doing so, the explicit dependence on the momentum cut-off disappears and one is left with 
\beq\label{eq:emergencegeneralQG}
	g_{\phi\phi}\, \lesssim\,  \Mpd^{d-2}\, \left(\frac {\partial_\phi \Mt}{\Mt}\right)^2\, , \qquad \frac {1}{g^2}\, \lesssim\, \Mpd^{d-2}\, \frac {1}{\Mt^2}\, .
\eeq
Consider for the moment the one-loop contribution to the gauge kinetic term given by the second expression above. One thus finds
\beq
	\Mt^2\, \sim\, g^2\, \left(N\LSP^{d-2}\right)\, \lesssim\, g^2 \Mpd^{d-2}\, ,
\eeq
which exhibits the same qualitative structure as the Weak Gravity Conjecture for a $\mathsf{U(1)}$ gauge field \cite{Arkani-Hamed:2006emk}. In this sense, the emergence of the gauge kinetic term together with the species scale imply the WGC.

Concerning the modulus scalar, once we found the field space metric we can easily compute the distance in moduli space between any two given points, $\phi_a$ and $\phi_b$ (provided of course they lie ultimately at infinite distance from one another, which is where our computations have been done reliably). By doing so, we arrive at 
\beq
\label{eq:distancegeneral}
	\kappa_d\, \Delta\phi_{ab}\, =\,  \kappa_d\, \int_{\tau_a}^{\tau_b} \text{d}\tau \sqrt{g_{\phi \phi}\, \dot{\phi}^2}\, \sim \, \int_{\phi_a}^{\phi_b} \frac {\partial_\phi \Mt}{\Mt}\, d\phi\, \sim \, \log \left(\frac {\Mt (\phi_b)}{\Mt (\phi_a)}\right)\, ,
\eeq
where we denote $\dot{\phi}=d \phi/d\tau$, with $\tau$ being some affine parameter, and we have substituted $\kappa_d^2\, g_{\phi \phi} \sim (\partial_\phi \Mt/\Mt)^2$ in eq. \eqref{eq:distancegeneral} above. From here one obtains the sought-after exponential behaviour for the mass scale of the tower, namely
\beq
	\Mt (\phi_b)\, \sim\, \Mt(\phi_a)\, e^{-\lambda \kappa_d \Delta\phi_{ab}}\, , \qquad \text{with}\ \lambda=\mathcal{O}(1)\, .
\eeq
This is precisely the content of the Distance Conjecture \cite{Ooguri:2006in}.
	
Let us mention that even though the above analysis is framed within a very simple $d$-dimensional example --- with $d>4$, we will see that similar results are attained for more complicated tower structures and different spacetime dimensions, including also the case of string oscillator modes. The take-home message is that the concept of Emergence is intimately related to the Weak Gravity and the Distance conjectures and that in this connection it is crucial that the cut-off in the effective field theory is taken to be the species scale. However, the final expression for the metrics may be written in a way which is independent of the quantum gravity scale (equivalently $N$) and depends explicitly only on `infra-red' data, such as the value of $\Mt$ as well as the Planck mass.

\subsection{Classical metrics from quantum effects}\label{ss:classicalfromquantum}

As an aside, let us take the opportunity to discuss a subtle point that is raised by our analysis here. This has to do with the fact that the emergence prescription in principle tells us that seemingly \emph{classical} results, such as the tree-level graviton term (c.f. eq. \eqref{eq:EinsteinHilbertaction}), can be generated by summing instead over an infinite number of \emph{quantum} contributions. This naively contradicts our quantum field theory intuition, where the Wilsonian effective action --- or rather any physical quantity derived from it --- can be organized as a perturbative series in $\hbar$, thereby separating classical effects from purely quantum corrections. 

The resolution to this puzzle lies on the very definition of the species cut-off, which is a quantity that is quantum in nature. This can be seen both from the perturbative and non-perturbative analyses presented in Section \ref{s:speciesmotivation}. In any event, after carefully keeping track of the relevant powers of $\hbar$ we arrive at the relation
\beq
\begin{aligned}
    \ell_{\rm sp}^{d-2} = \ell_d^{d-2}\, \left(\frac{N}{4\pi} \right) = 8\pi G_N \hbar^{d-3} N\, ,
\end{aligned}
\eeq
where crucially $\ell_{\rm sp}$ ends up depending explicitly on Planck's fundamental constant. Therefore, by performing the loop integrals as outlined in Sections \ref{ss:Emergencegraviton} and \ref{ss:Emergence&Swampland} and upon imposing $\LSP$ to be the physical cut-off, we actually obtain fully consistent results (at least from this perspective) where the overall normalization of the kinetic terms is controlled by a factor of $\kappa_d^{-2}$, which is independent of $\hbar$. Hence, even though the Emergence Proposal is quantum in  nature, the species scale comes itself from a purely quantum computation such that both effects ultimately cancel each other, yielding a seemingly classical result.\footnote{The idea that loop corrections in gravity can lead to classical effects in the infra-red is a well-known fact in the literature, see e.g., the recent review \cite{Donoghue:2022eay} and references therein.}

\section{Kinetic terms from one-loop corrections}
\label{s:selfenergybosons}	

In this section we discuss the  computation of the wave-function renormalization that an infinite tower of bosonic and/or fermionic particles induce on a given modulus, $p$-form gauge field and massless fermion. For concreteness, we perform such computations for towers comprised by spin-0 scalars and spin-$\frac{1}{2}$ Dirac fields, keeping in mind that we use them as a proxy to estimate the contribution of general towers of scalars and/or fermions to the quantum loops. We outline the basic logic and main results here, leaving the detailed calculations for Appendix \ref{ap:Loops}. In particular, we perform all computations in general $d$ spacetime dimensions, and comment on qualitative differences that arise depending on the latter.

\subsection{Emergence of moduli metrics}
\label{ss:Emergencemodulus}
	
\subsubsection{Self-energy of a modulus field}
\label{sss:selfenergymodulus}
\begin{figure}[t]
		\begin{center}
			\subfigure[]{
				\includegraphics[scale=1.2]{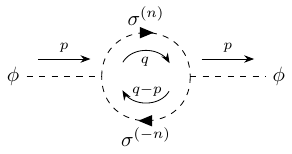} 
				\label{fig:scalarloopscalar}
			}\qquad \qquad
			\subfigure[]{
				\includegraphics[scale=1.2]{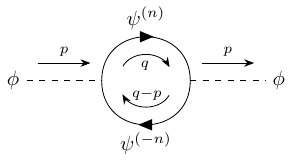}
				\label{fig:scalarloopfermion}
			}
			\caption{ Contributions to the wave-function renormalization for massless scalars from \textbf{(a)} scalar and \textbf{(b)} fermion loops.}			
		\label{fig:scalarpropagator}
		\end{center}
\end{figure} 

We begin by considering a real modulus field, $\phi$, coupled to a tower of massive scalars $\{\sigma^{(n)}\}$ or fermions $\{\psi^{(n)}\}$, through their field-dependent masses. The relevant piece of the action can be found in eqs. \eqref{eq:Skinphi}-\eqref{eq:Spsin}. On the other hand, the specific trilinear couplings that enter into the Feynman diagrams contributing to the process are shown in Figure \ref{fig:scalarpropagator}, and can be obtained by expanding the mass term of the states running in the loop up to linear order in the (fluctuation of the) modulus. Their strength is given by
\begin{equation}
		\label{eq:scalarcouplings}
		\lambda_n=2m_n(\partial_\phi m_n)\,,  \qquad \mathrm{and} \qquad \mu_n=\partial_\phi m_n \, ,
\end{equation}
for massive scalars and fermions, respectively.
	
In the context of Emergence, we are interested in computing the wave-function renormalization of the modulus due to scalar and fermionic loops (see also \cite{Heidenreich:2018kpg, Hamada:2021yxy} for related computations). The idea would be thus to extract the momentum-dependent part of the exact propagator of $\phi$ at one loop, which takes the following form (see Appendix \ref{ap:Loopsscalar} for details)
\beq
	\label{eq:scalarpropagator}
	D(p^2)=\frac{1}{p^2-\Pi(p^2)}\, .
\eeq
Here, $\Pi(p^2)$ corresponds to the self-energy of the msassless field $\phi$. In the following, we will content ourselves with computing $\Pi(p^2)$ up to $\mathcal{O}(\hbar)$ in the effective action, such that we concentrate on the (amputated) one-loop graph displayed in Figure \ref{fig:scalarpropagator}. (Recall that at tree-level $\Pi_0(p^2)\, = \, 0$.) Thus, the correction to the propagator --- i.e. the modulus metric --- is given by the term in $\Pi(p^2)$ proportional to $p^2$, namely
\begin{equation}\label{eq:metriccorrection}
		\delta g_{\phi \phi}= \frac {\partial \Pi(p^2)}{\partial p^2} \bigg\rvert_{p=0}\, .
\end{equation}

\subsubsection*{Scalar loop}
	
Let us consider first the contributions coming from massive real scalar fields $\{\sigma^{(n)}\}$ corresponding to the Feynman diagram displayed in Figure \ref{fig:scalarloopscalar}, which reads
\beq
	\Pi_n(p^2) \ = \frac{\lambda_n^2}{2} \int \frac {\text{d}^dq}{(2\pi)^d} \frac {1}{(q^2+m_n^2)} \frac {1}{((q-p)^2+m_n^2)}\, ,
	\label{eq:selfenergyscalar}
\eeq
with the coupling $\lambda_n$ defined in \eqref{eq:scalarcouplings}. In order to compute $\delta g_{\phi \phi}$ we need to extract the term linear in $p^2$ from the expression above, which leads to the following integral in momentum space
\beq
	\frac {\partial \Pi_n(p^2)}{\partial p^2} \bigg\rvert_{p=0}  = - \frac{\lambda_n^2}{2} \int \frac {\text{d}^dq}{(2\pi)^d} \frac {1}{(q^2+m_n^2)^3}\, .
	\label{eq:sigmaa}
\eeq
From this one can already anticipate the different behaviour in terms of convergence of the momentum integral depending on whether $d$ is equal, lower than, or greater than 6. Even though for $d<6$ the loop integral is convergent for large $q$, we introduce here a momentum cut-off (that will be identified ultimately with the species scale), since this is required for later consistency once we fix the energy scale up to which we include the contribution from the states in the tower that is integrated out. The exact solution of \eqref{eq:sigmaa} is given in terms of hypergeometric functions (c.f. eq. \eqref{eq:scalarloopscalarexact}), but we will focus here on the dependence of the \emph{leading} term on the relevant energy scale (i.e. either the mass of the particle, $m_n$, or the UV cut-off, $\Lambda$), since these are the ones from which we will eventually extract the field dependence of the emergent kinetic terms. In particular, there are two relevant cases in which we are interested: \emph{(i)} The limit $\Lambda \gg m_n$, in which the mass of the particle running in the loop is negligible with respect to the UV cut-off, and \emph{(ii)} $\Lambda\simeq m_n$, where both are roughly of the same order. The former is relevant for most states of KK-like towers in asymptotic regimes, since as we saw in Chapter \ref{ch:SpeciesIntro} the mass scale of the tower is typically parametrically lighter than the species scale. The latter is relevant for the highest modes in KK-like towers and for most states in stringy-like towers, since the masses of the corresponding particles coincide asymptotically with the species scale. Luckily, both limits give rise to the same functional dependence on either $\Lambda$ or $m_n$ (c.f. Appendix \ref{ap:Loopsscalar}), where the expressions only differ in the numerical prefactors.\footnote{To be precise, the numerical coefficients shown in Appendix \ref{ap:Loops} can be thought of as upper and lower bounds on the contributions from each particle in the loop, since they include the two limiting cases for the mass of the relevant particles.} From the results summarized in Tables \ref{tab:scalarloopscalarLambda>>m} and \ref{tab:scalarloopscalarLambda=m} we can thus extract the following asymptotic dependence for the dominant contribution to the 2-point function of the massless scalar $\phi$
\begin{equation}\label{eq:scalarloopscalarssummary}
		\frac {\partial \Pi_n(p^2)}{\partial p^2} \bigg\rvert_{p=0}\,   \sim 
		\left\{\begin{array}{lr}
			-  \dfrac{\lambda_n^2}{m_n^{6-d}} & \qquad\text{for } d< 6\, ,\\ \\ 
			-\lambda_n^2 \ \log \left( \dfrac{\Lambda^2}{m_n^2}\right) &\qquad \text{for } d= 6\, ,\\ \\ 
			-\lambda_n^2 \ \Lambda^{d-6}&\qquad \text{for } d>6\, ,
		\end{array}\right.
\end{equation}
where the precise meaning of $\sim$ is that we keep track of all the quantities that are field dependent and neglect only the numerical prefactors. 
	
Before proceeding with the fermionic loop, let us make some comments about possible natural generalizations of the scalar case just discussed. First, as typically happens in supersymmetric theories, one could consider a tower of \emph{complex} scalar fields coupled to the modulus $\phi$ through its mass. This scenario reduces essentially to the one described here, since one can always write a complex field in terms of its real and imaginary parts
\begin{align}
    \chi^{(n)}=\frac{\sigma^{(n)}_1+\i\sigma^{(n)}_2}{\sqrt{2}}
\end{align}
which both share the same moduli-depenedent mass and thus contribute to the self-energy \eqref{eq:metriccorrection} as summarized in eq. \eqref{eq:scalarloopscalarssummary} above (with an extra factor of 2). Second, one could even study the case in which the modulus itself is complexified, namely $\phi \to \Phi = \text{Re}\, \Phi + \i\, \text{Im}\, \Phi$. The scalar charges \eqref{eq:scalarcouplings} would now be complex-valued, and similar considerations would lead to the one-loop generated metric $\delta g_{\Phi \bar \Phi}$.\footnote{Notice that the fact that one obtains a hermitian metric at one loop is due to the pseudo-scalar nature of the imaginary part of the modulus, which prevents a term of the form $(\partial \text{Re}\, \Phi)(\partial \text{Im}\, \Phi)$ from appearing in the effective action.}

\subsubsection*{Fermionic loop}
	
We consider now the contribution to the propagator from a loop of massive fermions $\{\psi^{(n)}\}$ with trilinear couplings $\mu_n$ as defined in \eqref{eq:scalarcouplings}. The relevant Feynman diagram, shown in Figure \ref{fig:scalarloopfermion}, reads as follows
\begin{equation}
		\label{eq:selfenergyfermion}
		\Pi_n(p^2) \, = \, -\mu_n^2  \int \frac {\text{d}^dq}{(2\pi)^d}\ \text{tr} \left (\frac {1}{\i \slashed{q}+m_n}\ \frac {1}{\i (\slashed{q}-\slashed{p})+m_n} \right) \, .
\end{equation}
After performing the relevant traces and rearranging terms (c.f. around eq. \eqref{eq:scalarloopfermionstraces} for details), we get the following  expression for the one-loop contribution to the wave-function renormalization 
\begin{equation} \label{eq:wfrfermionloop}
		\frac{\partial \Pi_n(p^2)}{\partial p^2} \bigg\rvert_{p=0} \, = \,   -\mu_n^2\, \fdim  \int \frac {\text{d}^dq}{(2\pi)^d} \frac{1}{(q^2+m_n^2)^2} \ + \ 2 m_n^2 \, \mu_n^2\ \fdim \int \frac {\text{d}^dq}{(2\pi)^d} \frac{1}{(q^2+m_n^2)^3} \, .
\end{equation}
The first term is negative and looks very analogous to the scalar contribution, with the difference that it naively diverges for $d\geq 4$ instead of $d\geq 6$. Performing a similar analysis as for the scalar loop, we obtain akin results (see Appendix \ref{ap:Loopsscalar}). Namely, for the two limits of interest, $\Lambda \gg m_n$ and $\Lambda \simeq m_n$, we get the same functional dependence on the coupling constants and the pertinent energy scales. The only difference being the numerical prefactors, that play no role in determining the field-dependent part of the emergent metric. The detailed results are summarized in Tables \ref{tab:scalarloopfermionLambda>>m} and \ref{tab:scalarloopfermionLambda=m}, which take the schematic form
\begin{equation}\label{eq:scalarloopfermionssummary}
		\frac {\partial \Pi_n(p^2)}{\partial p^2} \bigg\rvert_{p=0}\,   \sim 
		\left\{\begin{array}{lr}
			-  \dfrac{\mu_n^2}{m_n^{4-d}} & \qquad\text{for } d< 4\, ,\\ \\ 
			-\mu_n^2\ \log \left( \dfrac{\Lambda^2}{m_n^2}\right) &\qquad \text{for } d= 4\, ,\\ \\ 
			-\mu_n^2\ \Lambda^{d-4}&\qquad \text{for } d>4\, .
		\end{array}\right.
\end{equation}
On the other hand, the second term in \eqref{eq:wfrfermionloop} is of the same form as the scalar contribution \eqref{eq:sigmaa}, with $\lambda_n= 2 m_n (\partial_\phi m_n)=2 m_n \mu_n$, including also a prefactor of $\fdim$ which takes into account the number of fermionic degrees of freedom in $d$ spacetime dimensions. Moreover, it has the opposite sign as the scalar contribution, so for supersymmetric theories both terms cancel and the leading contribution to the emergent metric comes from the first term in \eqref{eq:wfrfermionloop}. For instance, in 4d a single Dirac fermion seems to cancel the renormalization due to two complex --- or four real scalar fields, so that e.g., in 4d $\mathcal{N}=2$ a hypermultiplet only contributes to the modulus metric through the fermion loop.\footnote{A similar cancellation (this time exact) occurs for the would-be mass term generated for the modulus in case supersymmetry is preserved in our theory, which can be checked explicitly upon using our formulae, although one may need to consider some extra loop diagrams not contributing to the wave-function renormalization and therefore not displayed in Figure \ref{fig:scalarpropagator}.} This suggests that following the leading order contribution coming from fermionic towers may be a good proxy for keeping track of the behaviour exhibited by the emergent kinetic terms.

\subsubsection{Generating moduli metrics}
\label{sss:emergencemodulimetric}
	
Armed with the above results, we study now the emergence of the kinetic term for real scalar moduli in the two relevant scenarios of decompactification and emergent string limits. For the latter, given the lack of a manifestly off-shell formulation of string theory as of today,\footnote{See \cite{Ahmadain:2022tew,Ahmadain:2022eso} though for recent proposals to alleviate this problem.} we will perform the computation using a purely field-theoretic approach, following the perturbative discussion of Section \ref{sss:stringtowersspecies}. Hence, the results there should be taken with a grain of salt. 
	
\subsubsection*{Moduli metrics from  Kaluza-Klein towers}
	
Let us consider a KK-like tower of scalars with mass spectrum and associated species scale given by
\beq\label{eq:@@@}
	m_n\, =\, n^{1/p}\, \Mt \, , \qquad   \LSP\,  \simeq\, N^{1/p}\, \Mt \, ,
\eeq
so that e.g., a single KK tower would correspond to $p=1$ (see Section \ref{ss:MultipleTowers} for precise definitions). Note that an analogous analysis can be performed with fermionic towers and similar results are obtained. The  contribution of a single scalar to the wave-function renormalization of a modulus field is given in \eqref{eq:scalarloopscalarssummary}, and for $d>6$ it takes the form
\beq
	\delta g_{\phi\phi}^{(n)}\, \sim\,  \mathcal{A}_d\, \lambda_n^{2}\, \Lambda^{d-6}\, \sim\, 4\, \mathcal{A}_d\, m_n^2\, (\partial_\phi m_n)^2\, \Lambda^{d-6}\, ,
\eeq
with $\mathcal{A}_d$ being a numerical prefactor depending only on $d$ that is not explicitly included in eq. \eqref{eq:scalarloopscalarssummary}. Its precise value for the two relevant limits is displayed in Tables \ref{tab:scalarloopscalarLambda>>m} and \ref{tab:scalarloopscalarLambda=m}. After adding up the contribution from the states of the tower below the cut-off $\Lambda = \LSP $, we get
\beq \label{eq:KKloopscalarmetricd>6}
	\delta g_{\phi\phi} \,=\, \sum_{n=1}^{N} \delta g_{\phi \phi}^{(n)}  \, \sim\, \frac {4 p \mathcal{A}_d}{p+4}\,  N^{\frac{4}{p}+1}\,  (\partial_\phi \Mt)^2\, \Mt^2\, \LSP^{d-6}\, \sim\, \frac {4 p \mathcal{A}_d}{p+4}\, \Mpd^{d-2} \left( \frac{\partial_\phi \Mt}{\Mt}\right)^2\, ,
\eeq
where we have used $N \simeq\Mpd^{d-2}/ \LSP^{d-2}$ as well as the second relation in \eqref{eq:@@@}. Thus, the non-trivial dependence on the characteristic mass of the tower, namely $\delta g_{\phi\phi} \sim 1/\Mt^2$, is recovered for any dimension, leading to the structure needed for the Distance Conjecture to hold a posteriori. Notice that the dependence on the tower density parameter, i.e. $p \in \mathbb{R}$, only enters through the numerical prefactor in \eqref{eq:KKloopscalarmetricd>6}, so that it is irrelevant for our purposes here.

For $d<6$, on the other hand, the correction associated to one real massive scalar in the loop integral is of the form
\beq
	\delta 
	g_{\phi\phi} ^{(n)}\, \sim\, \mathcal{B}_d\, \lambda_n^2\,  m_n^{d-6}\, \sim\, 4\, \mathcal{B}_d\, n^{\frac{d-2}{p}}\, \Mt^{d-4}\, (\partial_\phi \Mt)^2\, .
\eeq
Again, $\mathcal{B}_d$ is a $d$-dependent numerical prefactor whose precise value is bounded by the ones shown in Tables \ref{tab:scalarloopscalarLambda>>m} and  \ref{tab:scalarloopscalarLambda=m}. The total contribution from a tower up to the species scale therefore reads
\beq 
	\delta g_{\phi\phi}\,=\, \sum_{n=1}^{N} \delta g_{\phi \phi}^{(n)}\, \sim\, \frac {4p\mathcal{B}_d}{d-2+p}\, N^{\frac{d-2+p}{p}}\, \Mt^{d-4}\, (\partial_\phi \Mt)^2\, \sim\, \frac {4p\mathcal{B}_d}{d-2+p}\, \Mpd^{d-2} \left( \frac{\partial_\phi \Mt}{\Mt}\right)^2\, ,
\eeq
so that essentially we arrive at the same expression as for $d>6$, with a different numerical coefficient. A similar behaviour is obtained for the marginal case $d=6$ up to a numerical prefactor.
	
\subsubsection*{Moduli metrics from stringy towers}
As a second example, we consider this time the coupling of a real modulus $\phi$ to the oscillator modes of some fundamental string. For concreteness, we choose to do the computation for the fermionic modes only, although the same analysis may be repeated using the bosons yielding similar results as well. Let us study first the case $d>4$. Consider the spectrum of a string with masses and degeneracies for fermionic excitations given by
\beq
	m_n^2 = 16 \pi^2 (n-1) \Ms^2\, , \qquad  d(n)\, \sim\,  n^{-b}e^{a\sqrt{n}}\, ,
\eeq
where $a$ and $b$ are constants characteristic of each string, see e.g., eq. \eqref{eq:exactleveldensitystrings}. For simplicity, we will approximate in this section $d(n) \sim e^{\sqrt{n}}$, since this is enough for our purposes and the results are already reliable up to log corrections. We moreover assume the string scale $\Ms$ to depend on $\phi$ when measured in Planck units, as indeed happens in string theory (c.f. eq. \eqref{eq:ddimdilaton}). Then the total contribution for the metric arising from fermion loops is 
\beq
	\delta g_{\phi \phi}\, \sim\, \sum_{n=1}^{\Ns}\mu_n^2\, d(n)\, \LSP^{d-4} \sim \sum_{n=1}^{\Ns} (\partial_\phi \Ms)^2\, n\, d(n)\, \LSP^{d-4}\, .
\eeq
Recalling as well that $\LSP ^2 \simeq \Ns\, \Ms^2$, one finds
\beq
	\delta g_{\phi \phi}\, \sim\, \Ms^{d-4}\Ns^{\frac{d-4}{2}} (\partial_\phi \Ms)^2\int^{\Ns}_1 dn\, n\;  e^{\sqrt{n}} \sim  \Ms^{d-4} \Ns^{\frac{d-4}{2}} (\partial_\phi \Ms)^2  \Ns^{3/2}e^{\sqrt{\Ns}}\, .
\eeq
Using now the expression \eqref{eq:maxstringlevel}, we finally get
\beq\label{eq:modulusWFstringtow}
	\delta g_{\phi \phi}\, \sim\, \frac {(\partial_\phi \Ms)^2}{\Ms^2}\Mpd^{d-2}\, ,
\eeq
which is again the expected asymptotic behaviour. Note that the explicit dependence on $\Ns$ drops out, and the result only depends on the mass of the lightest string excitation $\Ms$, as it happened with the Kaluza-Klein towers. It is easy to check that \eqref{eq:modulusWFstringtow} is also reproduced for $d\leq 4$. 

 \subsection{Emergence of $\mathsf{U(1)}$ gauge kinetic terms}\label{ss:EmergenceU(1)gauge}
	
\subsubsection{Self-energy of a gauge 1-form}
\label{sss:selfenergy1form}
\begin{figure}[t]
		\begin{center}
			\subfigure[]{
				\includegraphics[scale=1.2]{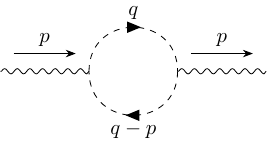} 
				\label{fig:1-formloopscalar}
			}\qquad \quad 
			\subfigure[]{
				\includegraphics[scale=1.2]{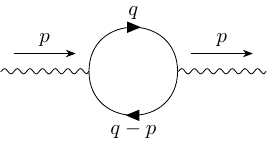}
				\label{fig:1-formloopfermion}
			}
			\caption{Wave-function renormalization at one-loop for $\mathsf{U(1)}$ gauge bosons due to \textbf{(a)} charged scalars and \textbf{(b)} fermionic fields.}			
			\label{fig:1-formpropagator}
		\end{center}
\end{figure} 	
Let us now consider the contribution coming from loops of complex scalar fields $\{\chi^{(n)}\}$ and Dirac fermions $\{\psi^{(n)}\}$ to the propagator of a gauge 1-form, denoted $A_1$. In particular, we take the mass and charge of the $n$-th scalar/fermion to be given by $m_n$ and $q_n$, respectively, whereas the gauge coupling associated to the 1-form is denoted by $g$ (c.f. eqs. \eqref{eq:SkinA1}-\eqref{eq:SpsinA1} for details on the conventions and the precise spacetime action).
	
As in the modulus scenario analyzed in Section \ref{sss:selfenergymodulus}, in order to compute the emergent gauge kinetic terms we must study the relevant Feynman diagrams that contribute to the wave-function renormalization of $A_1$. Subsequently, we need to extract the momentum-dependent part of the exact propagator, after taking into account the one-loop corrections. We choose to use the Lorenz gauge $\partial_\mu A^{\mu}\, =\, 0$ for convenience, since it can be easily generalized to higher $p$-forms as $\partial_\mu A^{[\mu \nu_1 \ldots \nu_{p-1}]}\, = \, 0$. In this gauge, the propagator for the 1-form can be written as
\begin{equation}
		\label{eq:A1propagator}
		D^{\mu \nu} (p^2) \, = \, \left( \dfrac{p^2}{g^2} \delta ^{\mu \nu} - \Pi^{\mu \nu}(p^2) \right)^{-1}\, ,
\end{equation}
where $\Pi^{\mu \nu}(p^2)$ vanishes at tree-level since it is the amputated Feynman diagram coming from the loops shown in Figure \ref{fig:1-formpropagator}. By imposing again our gauge choice, we can extract the tensorial dependence on
\begin{equation}
		\label{eq:A1loopamplitude}
		\Pi^{\mu \nu} (p^2) \, = \, \Pi(p^2) \delta ^{\mu \nu} \, ,
\end{equation}
so that we will be interested in the term linear in $p^2$ within $\Pi(p^2)$, as it provides the correction to the propagator and thus to the gauge coupling itself, namely
\begin{equation}\label{eq:gaugecouplingcorrection}
		\delta\left(\dfrac{1}{g^2} \right) = \frac {\partial \Pi(p^2)}{\partial p^2} \bigg\rvert_{p=0}\, .
\end{equation}
In the following, we analyze the problem both for the case in which the particle running in the loop is bosonic and fermionic.
	
Before going into the systematics of the loop calculations, let us remark that for a general gauge $p$-form, we would have a propagator with Lorentz indices $D^{\mu_1 \ldots \mu_p}_{{\nu_1 \ldots \nu_p}}$ instead of $D^\mu_\nu$ in \eqref{eq:A1propagator}. Then, upon working in Lorenz gauge, we would only need to replace ${\delta^{\mu}_\nu \, \to \, p! \, \delta^{[\mu_1}_{[\nu_1}\ldots \delta^{\mu_p]}_{\nu_p]}}$ in the previous equations in order to obtain the correct Lorentz structure.
	
\subsubsection*{Scalar loop}
	
We start with the contribution due to complex charged scalars $\{\chi^{(n)}\}$, with mass $m_n$ and charge $q_n$, given by the Feynman diagram shown in Figure \ref{fig:1-formloopscalar}, which reads
\beq
	\Pi^{\mu \nu}_n(p^2) \, =\,   g^2 \, q_n^2 \int \frac {\text{d}^dq}{(2\pi)^d} \frac {(2q-p)^{\mu} (2q-p)^{\nu}}{(q^2+m_n^2)\left( (q-p)^2+m_n^2\right)} \, .
	\label{eq:A1scalar(ap)}
\eeq
Hence, the one-loop correction to the gauge field propagator is given by (see Appendix \ref{ap:Loops1-form} for details)
\begin{equation}
		\frac{\partial \Pi^{\mu \nu}_n(p^2)}{\partial p^2} \bigg\rvert_{p=0} \, = \, -g^2\,  q_n^2\,   \frac{4}{d} \, \delta^{\mu\nu} \, \int \dfrac{\dd^d q}{(2\pi)^d} \dfrac{q^2}{(q^2+m_n^2)^3} \, .
\end{equation}
The above expression is expected to diverge for $d\geq 4$. However, we will introduce a UV cut-off for any $d$, since the goal is to identify it with $\LSP$ once we integrate out the different states in the relevant towers up to that energy scale. The exact expression for the amplitude is computed in eq. \eqref{eq:1-formloopscalarexact}, but since we are interested here just in the asymptotic dependence with the mass, the cut-off, and the charges, we will only retain the leading order terms. The two relevant limits are again $\Lambda \gg m_n$ and $\Lambda \simeq m_n$, which essentially correspond to the KK and the stringy cases, respectively. In both settings, as discussed in detail in Appendix \ref{ap:Loops1-form} (see in particular Tables \ref{tab:1-formloopscalarLambda>>m} and \ref{tab:1-formloopscalarLambda=m}), the leading dependence on the aforementioned quantities is the same for any $d$, and only the numerical prefactors change. Hence, we can summarize the relevant part of the leading correction to the propagator as follows
\begin{equation}\label{eq:1-formloopscalarssummary}
		\frac {\partial \Pi_n(p^2)}{\partial p^2} \bigg\rvert_{p=0}\,   \sim 
		\left\{\begin{array}{lr}
			-  \dfrac{g^2\,  q_n^2}{m_n^{4-d}} & \qquad\text{for } d< 4\, ,\\ \\ 
			-g^2\,  q_n^2 \ \log \left( \dfrac{\Lambda^2}{m_n^2}\right) &\qquad \text{for } d= 4\, ,\\ \\ 
			-g^2\,  q_n^2\ \Lambda^{d-4}&\qquad \text{for } d>4\, .
		\end{array}\right.
\end{equation}

\subsubsection*{Fermionic loop}
	
Including now the charged Dirac fermions $\{\psi^{(n)}\}$ --- with masses $\{m_n\}$ and charges $\{q_n\}$ --- in the loop given by Figure \ref{fig:1-formloopfermion} we get
\begin{equation}
		\Pi^{\mu\nu}_n(p^2) \ = - (\i g)^2 \, q_n^2  \int \frac {\text{d}^dq}{(2\pi)^d}\ \text{tr} \left (\frac {1}{\i \slashed{q}+m_n}\ \gamma^\mu \ \frac {1}{\i (\slashed{q}-\slashed{p})+m_n} \ \gamma^\nu  \right)\, .
\end{equation}
After performing the traces and selecting the terms that survive the angular integration (see discussion around eq. \eqref{eq:traces1-formloop}) we arrive at the following result for the relevant piece that corrects the propagator
\begin{equation}\label{eq:1-formloopfermions2terms}
 \begin{aligned}
	\frac{\partial \Pi^{\mu \nu}_n(p^2)}{\partial p^2} \bigg\rvert_{p=0}  =& \, -\fdim\, g^2\,  q_n^2 \, \delta^{\mu\nu}  \int \dfrac{\dd^d q}{(2\pi)^d} \dfrac{1}{(q^2+m_n^2)^2}\\
    & + \,  \fdim\,  g^2\,  q_n^2\,   \frac{2}{d} \, \delta^{\mu\nu}  \int \dfrac{\dd^d q}{(2\pi)^d} \dfrac{q^2}{(q^2+m_n^2)^3}\, .
 \end{aligned}
\end{equation}
The first piece can be computed after introducing the UV cut-off $\Lambda$ and it gives the exact result presented in \eqref{eq:1-formloopfermionexact}. For the two relevant limits, namely when $\Lambda \gg m_n$ or $\Lambda \simeq m_n$, it can be seen that the functional dependence with $\{g,\, q_n,\, m_n,\, \Lambda\}$ is again the same for all $d$, where only the numerical coefficients are different (see Tables \ref{tab:1-formloopfermionLambda>>m} and \ref{tab:1-formloopfermionLambda=m}). The pertinent leading expressions take the form
\begin{equation}\label{eq:1-formloopfermionssummary}
		\frac {\partial \Pi_n(p^2)}{\partial p^2} \bigg\rvert_{p=0}\,   \sim 
		\left\{\begin{array}{lr}
			-  \dfrac{g^2\, q_n^2}{m_n^{4-d}} & \qquad\text{for } d< 4\, ,\\ \\ 
			-g^2\, q_n^2\ \log\left( \dfrac{\Lambda^2}{m_n^2}\right) &\qquad \text{for } d= 4\, ,\\ \\ 
			-g^2\, q_n^2\ \Lambda^{d-4}&\qquad \text{for } d>4\, .
		\end{array}\right.
\end{equation}
On the other hand, the second term in \eqref{eq:1-formloopfermions2terms} can be seen to be equal --- but with opposite sign --- to the scalar contribution, up to a relative prefactor that accounts for number of degrees of freedom, given by $2^{\lfloor d/2 \rfloor-1}$. In parallel to the modulus case, this indicates that in the presence of unbroken supersymmetry the contribution from bosons would cancel against this second term coming from the fermionic loop, such that the first contribution in \eqref{eq:1-formloopfermions2terms} seems to be again a good proxy for keeping track of the leading order correction to the kinetic terms of the $\mathsf{U(1)}$ gauge field.

\subsubsection{Generating gauge kinetic terms}
\label{sss:emergenceU(1)}

Having computed the general correction to the kinetic term of a gauge 1-form, we describe in what follows how this can be employed to generate via emergence the abelian gauge kinetic function in the two relevant cases of Kaluza-Klein and stringy towers. Again, in the latter case we will adopt a naive field-theoretic approach based on our discussion in Section \ref{sss:stringtowersspecies}.

\subsubsection*{Gauge kinetic function from Kaluza-Klein towers}

Let us consider a tower of the general form
\beq
	m_n\, =\, n^{1/p}\, \Mt\, , \qquad  \LSP\, \simeq\, N^{1/p}\, \Mt\, ,
	\label{eq:abelianmasstower}
\eeq
where we assume the particles to have some quantized charges $\{q_n\}$ under certain $\mathsf{U(1)}$ gauge field. Here, the functional form of the charges with respect to the integer $n$ can be somewhat model-dependent. In order to be as general as possible, we will consider some useful parameterization given by\footnote{In this section we normalize the vector fields $A_{\mu}$ so that they have mass dimension one (c.f. eq. \eqref{eq:SkinA1}). Later on, in Section \ref{s:EmergenceStringTheory} we adopt the usual conventions in supergravity/string theory where all fields are dimensionless. Thus, upon doing so, eq. \eqref{eq:abelianquantizedcharges} catches some extra factors involving the appropriate Planck/string length.}
\beq
	q_n =  n^{1/r}\, ,
	\label{eq:abelianquantizedcharges}
\eeq
with $r=1,\infty$. For $r=1$ one recovers the charge spectrum of a standard Kaluza-Klein tower, whereas $r \to \infty$ corresponds to the case in which all states in the tower present the same (constant) charge under the $\mathsf{U(1)}$ field. String theory realizations of both kind towers will be discussed in concrete examples later on, see Section \ref{s:EmergenceStringTheory}. 
We have just seen that e.g., a single fermion of mass $m_n$ contributes to the gauge kinetic function of a $\mathsf{U(1)}$ gauge field as follows (for $d>4$)
\beq
	\delta \left(\frac {1}{g^2}\right) \bigg\rvert_{n-\text{th}} \sim\,  q_n^2\,  \mathcal{C}_d\, \LSP^{d-4}\, ,
\eeq
with $\mathcal{C}_d$ being a numerical prefactor which only depends on $d$. Its precise value for the two limits of interest is displayed in Tables \ref{tab:1-formloopfermionLambda>>m} and \ref{tab:1-formloopfermionLambda=m}. Subsequently, performing the sum over the full tower leads to
\beq\label{eq:gaugeemergenceddimensions}
	\delta \left(\frac {1}{g^2}\right)\, \sim\, \sum_{n=1}^N n^{2/r} \mathcal{C}_d\, \LSP^{d-4}\, \sim\, \frac {\mathcal{C}_d\, r}{2+r}\, \left(\frac {\Mpd^{d-2}} {\Mt^2}\right)^{\alpha_{d,r}/\alpha_{d,p}}\, ,
\eeq
where we have used eqs. \eqref{eq:abelianmasstower} and \eqref{eq:abelianquantizedcharges} as well as the species bound, with
\beq
	\alpha_{d,p} = \frac {d-2+p}{2p(d-1)}\, , \qquad  \alpha_{d,r} = \frac {d-2+r}{2r(d-1)}\, .
	\label{eq:dosalphas}
\eeq
Note that for $r=p$ one finds $\alpha_{d,p} =\alpha_{d,r}$ and $1/g^2\sim 1/\Mt^2$, as expected for the particular case of a BPS tower. 

It is interesting to consider as well the case in which only a sub-lattice of the full charge lattice is realized in the spectrum. Thus, let us assume to have an infinite set of particles labeled by $n\in \mathbb{Z}$ with charges given by $q_n=kn$, such that $k$ is some fixed positive integer. Then the above analysis for $r=p=1$ still applies except for the replacement $1/g^2\rightarrow k^2/g^2$, so that one rather finds
\beq
  m^2\, \lesssim\, k^2 g^2 \Mpd^{d-2}\, .
\eeq
Hence, we see that in this case the WGC is slightly weakened, in agreement with the $\mathbb{Z}_k$--Weak-Gravity-Conjecture as formulated in \cite{Buratti:2020kda}.

\subsubsection*{Gauge kinetic function from stringy towers}

Concerning the quantum correction to the gauge kinetic function induced by the oscillation modes of a critical string, the analysis will of course depend on which type of string is involved in the limit as well as the precise origin of the $\mathsf{U(1)}$ field whose renormalization we care about. In practice, however, this means that it is difficult to provide completely general results at this point. Still, we discuss in the following a particular structure which turns out to appear in large classes of e.g., Heterotic vacua in $d<10$ spacetime dimensions. Consider, for concreteness, a tower of fermionic fields with integer-valued charges $\{q\}$ at the $n$-th oscillator level of a given string theory compactification in $d>4$. We can try to estimate their contribution to the gauge kinetic function at the one-loop level as follows
\beq\label{eq:oneloopstringtowergauge}
	\delta \left(\frac {1}{g^2}\right)\bigg\rvert_{n-\text{th}} \sim\, \sum_{q=1}^{q_{\text{max}}} q^2\, d_{q,n}\, \LSP^{d-4} .
\eeq
Here the function $d_{q,n}$ parameterizes the degeneracy of each charge $q\in \mathbb{Z}$ present at the $n$-th oscillator mode of the emergent string. However, since we know (c.f. Section \ref{s:speciesscale}) that the \emph{total} degeneracy at each level $n$ behaves roughly as $d_n \sim e^{\sqrt{n}}$ (for large $n$), we will propose as an anzatz for $d_{q,n}$ the following form factor 
\beq
\label{eq:stringytowerdensity}
	d_{q,n}\, \sim\, f(q)\, e^{\sqrt{n}}\, ,
\eeq
where $f(q)$ is some \emph{polynomial} function of the charges, normalized so as to have total degeneracy equal to $d_n \sim e^{\sqrt{n}}$, namely $\sum_{q=1}^{q_{\text{max}}} f(q) = 1$. Notice that we assumed in \eqref{eq:oneloopstringtowergauge} that there exists a maximum charge $q_{\text{max}}$ at each oscillator level, which in principle could be some arbitrary function of $n$. However, in the following we take $q_{\text{max}} \sim \sqrt{n}$, a particular choice that is motivated from our experience with the Heterotic string.\footnote{This ultimately arises from the level matching condition on the left- and right-handed movers of the Heterotic string, namely $\alpha'm^2/4=N_R =N_L+\textbf{Q}^2/2-1$ (see Appendix A of \cite{Castellano:2022bvr} for more details on this issue).}
Hence, we obtain for each oscillator level the following one-loop contribution
\beq
	\delta \left(\frac {1}{g^2}\right)\bigg\rvert_{n-\text{th}} \sim\, n\, e^{\sqrt{n}}\, \LSP^{d-4}\, ,
\eeq
where we have approximated the sum over the gauge charges $q$ in \eqref{eq:oneloopstringtowergauge} by an integral $\int_0^{q_{\text{max}}} \text{d}q\ f(q)\ q^2 \sim n$. Summing now over all levels in the tower up to $n_{\text{max}} = \Ns$, one finds
\beq
	\delta \left(\frac {1}{g^2}\right)\, \sim\, \LSP^{d-4}  \int^{\Ns}_1 \text{d}n\, n\, e^{\sqrt{n}}\, \sim\, \LSP^{d-4}\, \Ns^{3/2}e^{\sqrt{\Ns}}\, \sim \Ms^{d-4}\, \Ns^{\frac{d-1}{2}}\, e^{\sqrt{\Ns}}\, ,
\eeq
where we made use of the relation $\LSP \simeq \sqrt{\Ns}\, \Ms$. Finally, upon substituting \eqref{eq:maxstringlevel} into the expression above, we get
\beq\label{eq:1-formemergencestringytowersgeneral}
	\delta \left(\frac {1}{g^2}\right)\, \sim\, \frac {\Mpd^{d-2}}{\Ms^2}\, ,
\eeq 
for the resummed wave-function renormalization induced by the tower of string modes. This implies that $1/g_{\text{IR}}^2 \sim 1/\Ms^2$ (in Planck units), similarly to what happened in the Kaluza-Klein case. One also recovers a WGC-like expression applied to this stringy set-up, with $\Ms^2 \sim g^2\, \Mpd^{d-2}$. Let us remark that the example here analyzed should be regarded as a string-theory-inspired toy model. We will discuss in more detail in Section \ref{ss:emergence6d} a particular realization of this scenario associated to some 6d (emergent) Heterotic string obtained from a singular limit within the K\"ahler moduli space of F-theory compactified on an elliptic Calabi--Yau three-fold. 
	
\subsection{Emergence of fermion kinetic terms}
\label{ss:kineticfermions}
	
In Sections \ref{ss:Emergencemodulus}-\ref{ss:EmergenceU(1)gauge} we studied the emergence phenomenon of kinetic terms associated to fields with \emph{bosonic} spin-statistics, namely scalars and 1-form gauge fields. We now turn to the generation of kinetic terms for light \emph{fermions} in quantum gravity. In fact, this is an important case from the phenomenological point of view, since most of the Standard Model (SM) particles are indeed of this kind.
	
\begin{figure}[t]
		\begin{center}
			\subfigure[]{
				\includegraphics[scale=1.2]{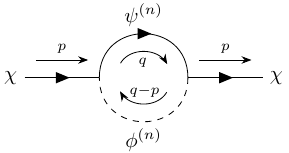} 
				\label{fig:fermionloopscalar}
			}\qquad \qquad 
			\subfigure[]{
				\includegraphics[scale=1.2]{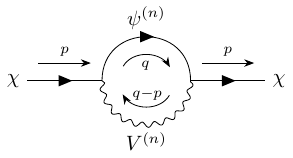}
				\label{fig:fermionloopvector}
			}
			\caption{One-loop diagrams contributing to the wave-function renormalization of light fermion fields in our theories.}
			\label{fig:kineticfermionsbas}
		\end{center}
\end{figure}

As first discussed in ref. \cite{Palti:2020tsy}, the fermionic set-up is different due to two main reasons. To start with, the loop graphs generically involve more than one different tower: one comprised by fermions and at least a second one with bosons, see Figure \ref{fig:kineticfermionsbas}. Such towers are moreover independent from each other in general, and may even have different mass gaps $\{ m_{\mathrm{b}}$, $m_{\mathrm{f}}\}$ --- as well as different structures. Secondly, the coupling between the light fermions and the particle states in the towers $\{\mathcal{Y}_n \}$ appears to be model-dependent. Thus, recall that in the case of the kinetic terms for moduli scalars or gauge bosons, those couplings arose naturally either from the moduli-dependent mass of the heavy fields or were rather determined by their gauge charge $q_n$, respectively. With light fermions, however, the analogous couplings are only fixed whenever they are related by some symmetry --- e.g., supersymmetry --- to the aforementioned bosonic fields. More generally, one needs to specify the structure of towers involved as well as their interactions, or perhaps make some simplifying assumptions so as to be able to proceed any further.
	
In any event, before studying the emergence mechanism in the present set-up we need to first compute the relevant one-loop diagrams. We will always have in mind in what follows the 4d case, which has possible direct applications to the real world \cite{Castellano:2023qhp}, even though the computations will be performed in $d$ spacetime dimensions for completeness. We will moreover consider the case in which $\chi$ is \emph{chiral} (i.e. we restrict to even-dimensional spacetimes). This is to ensure that there is no mass term generated at the quantum level for the light fermion, even in the absence of supersymmetry. The natural couplings would be thus of the general form
\begin{equation}\label{eq:fermion&bosonYukawas}
		\mathcal{Y}_n\ \overline{\phi^{(n)}} \left(\psi^{(n)}\chi \right)\, , \qquad  \tilde{\mathcal{Y}}_n \left(\psi^{(n)} \sigma^{\mu} \overline \chi \right) V_{\mu}^{(n)}\, ,   
\end{equation}
where the $\sigma$-matrices belong to the appropriate irreducible spinor representation of the Lorentz group in $d=2k$ dimensions acting on Weyl spinors, and $n \in \mathbb{Z}$ labels the states in the tower. Such towers present, within the bosonic sector, either massive complex scalars or vector fields $\{\phi^{(n)}, V_{\mu}^{(n)} \}$, respectively. The Weyl fermions $\{\psi^{(n)}\}$, on the other hand, pair up with their charge conjugate (say the one labeled by $-n$) so as to form massive Dirac fermions, i.e. $\Psi^{(n)}=\left(\psi^{(n)}, \overline{\psi^{(-n)}} \right)^{\text{T}}$. Furthermore, both towers will be assumed to be independent,\footnote{In principle, one could consider two  independent  indices $(m, n) \in \mathbb{Z}^2$ to label the fields $\lbrace \phi^{(m)} \psi^{(n)} \rbrace$, as well as the interactions in \eqref{eq:fermion&bosonYukawas}, i.e. $\mathcal{Y}_{m, n}$. For simplicity, however, we choose to have only diagonal couplings, namely those between fields with $m=n$, as happens e.g., when there is some conserved charge or when the towers involved are supersymmetric.} meaning that their mass spectrum is not a priori correlated, with the Yukawa-like couplings $\lbrace \mathcal{Y}_n, \tilde{\mathcal{Y}}_n \rbrace$ not having any further restriction.
	
It is interesting to point out that the  necessity of having both fermionic and bosonic towers at the same time in order to account for such an emergence mechanism in the present case is consistent with the existence of some sort of \emph{symmetry} relating both kinds of fields \cite{Palti:2020tsy}, such as (misaligned) supersymmetry \cite{Dienes_1995}. Hence, the mere presence of light fermionic degrees of freedom in the gravitational EFT points (when viewed from the Emergence prism) towards the existence of towers of massive states with different spin-statistics, given that it seems difficult to generate their corresponding kinetic terms otherwise.
Relatedly, in a theory with a supersymmetric spectrum, the on-shell bosonic content equals that of the fermions, such that one would expect the towers to come along with their respective superpartners. Thus, if e.g., $\chi$ in Figure \ref{fig:fermionloopscalar} is a massless gaugino in a 4d $\mathcal{N}=1$ gauge theory, it would couple to both supersymmetric fields belonging to the same charged chiral multiplet. Moreover, this kind of Yukawa couplings associated to supersymmetrized gauge interactions involve the gauge charges precisely like their vector partners do\cite{Wess:1992cp}, namely the Yukawa couplings are of the form $\mathcal{Y}_n \propto \i n$. 

\subsubsection{Self-energy of a Weyl fermion}
\label{sss:selfenergyfermion}	
Let us compute the wave-function renormalization induced on a chiral fermion $\chi$ by the tower of bosonic and fermionic particles interacting with the former through Yukawa-like couplings of the form discussed in \eqref{eq:fermion&bosonYukawas}. In what follows, we will work with Dirac fermions, such that in order to take into account that the massless spin-$\frac{1}{2}$ field $\chi$ is chiral, it is thus necessary to introduce the  chirality projector $P_{-}=\frac{1}{2}(1-\gamma^{d+1})$, where $\gamma^{d+1}$ is the proper generalization of the four-dimensional $\gamma^5$ to $2k$ dimensions (see Appendix \ref{ap:conventions} for conventions). This way, one can directly work with a Dirac spinor $\mathcal{X}$, whereas the original chiral field can be recovered upon projecting with $P_-$. With this in mind, the interactions in \eqref{eq:fermion&bosonYukawas} can be rewritten in terms of $\{\Psi^{(n)}, \mathcal{X} \}$ as follows
\begin{equation}\label{eq:interactionsfermion2}
		\mathcal{Y}_n\ \phi^{(n)} \left( \overline{\Psi^{(n)}} P_{-} \mathcal{X} \right) + \text{h.c.}\, , \qquad  \tilde{\mathcal{Y}}_n\ \left(\overline{\Psi^{(n)}} \gamma^{\mu} P_{-} \mathcal{X} \right) V_{\mu}^{(n)} + \text{h.c.}\, .  
\end{equation}
The idea then is to extract the momentum-dependent part of the exact propagator of our massless fermion $\chi$ at one-loop after integrating out such heavy particles (see Appendix \ref{ap:LoopsWeylfermion} for details)
\beq\label{eq:Euclexactpropagator}
	S(\slashed{p})= \frac{1}{\i\slashed{p}}\, P_{-} + \frac{1}{\i\slashed{p}}\, P_{-}\, \left(\i\Sigma(\slashed{p})\right)\, \frac{1}{\i\slashed{p}}\, P_{-} + \ldots\ ,
\eeq
$\i \Sigma(\slashed{p})$ being the (amputated) Feynman diagram depicted in Figure \ref{fig:kineticfermionsbas}. Notice that this is nothing but the fermionic analogue of $\Pi(p^2)$ shown in eq. \eqref{eq:scalarpropagator}. Therefore, if one wants to isolate the quantum corrections to the kinetic term arising from the loops, one then needs to take a derivative of $\i \Sigma(\slashed{p})$ with respect to the external momentum $p^{\mu}$, and subsequently set $p$ equal to zero.
	
\subsubsection*{The loop computation}
In the following, we will particularize to the case in which the bosonic particle running in the loop is a spin-0 field, as in Figure \ref{fig:fermionloopscalar}, although a similar analysis could be performed for the massive 1-form case. The computation thus involves a tower of Dirac fermions $\{\Psi^{(n)}\}$ of masses $\{m_n^{{\text{f}}}\}$ as well as another one made of complex bosonic scalars $\{\phi^{(n)}\}$ with masses given by $\{m_n^{{\text{b}}}\}$. Hence, the self-energy integral we need to examine for each step $n$ in the infinite tower is the following
\beq
	\i \Sigma_n(\slashed{p}) \ = |\mathcal{Y}_n|^2 \int \frac{\text{d}^dq}{(2\pi)^d} \frac{P_{-} \left( -\i \slashed{q} + m_n^{\text{f}}\right) P_{+}}{q^2+(m_n^{\text{f}})^2} \frac {1}{(q-p)^2+(m_n^{{\text{b}}})^2}\ ,
	\label{eq:selfenergychi}
\eeq
whilst the relevant contribution to the fermionic propagator reads (see Appendix \ref{ap:LoopsWeylfermion})
\beq
	\begin{aligned}\label{eq:Euclwavefunctionfermion}
		\frac{\partial \Sigma_n(\slashed{p})}{\partial p^{\mu}} \bigg\rvert_{p=0} = \frac{-2 |\mathcal{Y}_n|^2 \delta_{\mu \nu} \gamma^{\nu}\ P_{+}}{d} \int \frac{\text{d}^dq}{(2\pi)^d} \frac{q^2}{\left[ q^2 + (m_n^{{\text{f}}})^2 \right] \left[ q^2 + (m_n^{{\text{b}}})^2 \right]^2}\ .
	\end{aligned}
\eeq
Notice that by carefully evaluating $\i\Sigma_n(\slashed{p})$ at zero external momentum one can see that \eqref{eq:selfenergychi} vanishes, such that no mass term is generated quantum mechanically at one-loop, as it should be. Additionally, the self-energy includes the projector $P_{+}$, since it is associated to the \emph{chiral} field $\chi$. Let us also stress that the behaviour of the above momentum integral strongly depends on the dimension  of our spacetime. In particular, it can be seen to diverge depending on whether $d\geq4$ or not. In any event, since we are interested in its consequences for the Emergence Proposal, we will impose again some UV cut-off which ultimately will be identified with the species scale, rendering the previous integral finite.
	
Now, in order to study with more detail the kind of corrections induced by the previous diagrams and for future reference, we will first specialize to the easier case in which both towers present identical mass gaps, namely we take $m_n^{{\text{b}}} = m_n^{{\text{f}}} = m_n$. For concreteness, let us show in here the explicit results for the case in which $\Lambda/m_n \gg 1$. The pertinent leading expressions take the form
\begin{equation}\label{eq:fermionloopsummary}
		\frac{\partial \Sigma_n(\slashed{p})}{\partial p^{\mu}} \bigg\rvert_{p=0}\,   \sim 
		\left\{\begin{array}{lr}
			-|\mathcal{Y}_n|^2 \gamma_{\mu}\ P_{+}\ \frac{1}{m_n^{4-d}} & \qquad\text{for } d< 4\, ,\\ \\ 
			-|\mathcal{Y}_n|^2 \gamma_{\mu}\ P_{+}\ \log\left( \dfrac{\Lambda^2}{m_n^2}\right) &\qquad \text{for } d= 4\, ,\\ \\ 
			-|\mathcal{Y}_n|^2 \gamma_{\mu}\ P_{+}\ \Lambda^{d-4}&\qquad \text{for } d>4\, .
		\end{array}\right.
\end{equation}
These results are analogous to those found for the wave-function renormalization of gauge bosons in eqs. \eqref{eq:1-formloopscalarssummary} and \eqref{eq:1-formloopfermionssummary}. In any case, the more general expression for the one-loop contribution to the fermion self-energy (c.f. eq. \eqref{eq:Euclwavefunctionfermion}) in which the towers have different spins as well as different masses can be found at the end of Appendix \ref{ap:LoopsWeylfermion}.
	
\subsubsection{Generating fermion kinetic terms in 4d }
\label{sss:emergencefermion}
		
As stated above, we cannot give a general expression for the emergent kinetic term of a light fermion field without further specifying the structure of towers and couplings involved in the problem. Therefore, in this section we will restrict ourselves to the case of four spacetime dimensions with the aim of illustrating with a simple toy model how the generation of kinetic terms may arise for massless fermions.\footnote{See \cite{Casas:2024ttx} for recent applications of these ideas in 4d $\mathcal{N}=1$ set-ups arising from string theory.} In order to be as general as possible, we consider here a model which does not impose identical mass gaps for the towers. In particular, we start with two different sets of fermions and bosons exhibiting a structure of the form
\beq
	m_n^{\mathrm{b}}\, =\, n\, m_{\mathrm{b}}\, , \qquad m_n^{\mathrm{f}}\, =\, n^{1/p}\, m_{\mathrm{f}}\, , \qquad  \LSP\, \simeq\, N\, m_{\mathrm{b}}\, \simeq\, N^{1/p}\, m_{\mathrm{f}}\, ,
\label{eq:dostorres}
\eeq
where we have allowed for $m_{\mathrm{f}}$ to be a priori different from $m_{\mathrm{b}}$. For concreteness, we assume $m_{\mathrm{f}}$ to be greater than $m_{\mathrm{b}}$, but notice that we could alternatively start with the reversed situation in which $m_{\mathrm{f}} \leq m_{\mathrm{b}}$, yielding identical results --- upon exchanging $m_{\mathrm{f}}$ and $m_{\mathrm{b}}$. The case with $p=1$ is indeed quite interesting, since depending on the precise value for $\mathcal{Y}_n$ one can recover the wave-function renormalization of the fermionic components either in 4d $\mathcal{N}=1$ chiral or vector multiplets. 
	
On the other hand, for $p\rightarrow \infty$, one has a tower in which the mass gap between two adjacent levels is much smaller than the scale of the tower itself, namely $\Delta m^{\mathrm{f}}_n \ll m_{\mathrm{f}}$. Notice that with the above parameterization we can actually describe towers with different mass scales, for more generality. We moreover assume the massless fermion $\chi$ to couple to the states comprising both towers through the Yukawa-like couplings displayed in \eqref{eq:fermion&bosonYukawas}, thus leading to the one-loop diagrams shown in Figure \ref{fig:kineticfermionsbas}. However, as already mentioned, the dependence of the Yukawa couplings  $\{\mathcal{Y}_n$, $\tilde{\mathcal{Y}}_n\}$ on the state labelling $n$ is, in general, not fixed by any gauge principle. Hence, in order to accommodate different possibilities we consider here Yukawas of the general form\footnote{There is also the possibility of having $m_{{\text{b}}}=m_{{\text{f}}}$ and $\mathcal{Y}_n = \mu_n = \partial_{\Phi} m_n$, which indeed captures the couplings of a supersymmetric tower to the fermionic component within some massless $\mathcal{N}=1$ chiral multiplet, leading to similar results as in Section \ref{sss:emergencemodulimetric}.}
\beq
	\mathcal{Y}_n\, \propto\, n^{1/r}\, .
\eeq
For $r=p=1$ one indeed recovers a situation in which both towers share the same mass gap (i.e. $m_{\mathrm{f}}=m_{\mathrm{b}}$ ) and the Yukawa charges grow like $\mathcal{Y}_n \propto n$. 
On the other hand, if $r\rightarrow \infty$, the Yukawa coupling does not depend on the state of the tower and it is just some fixed constant. Now, upon using the third relation in \eqref{eq:dostorres}, one sees that the species scale is related to the masses of the towers as follows
\beq
	\LSP^4\, \simeq\, N^{1/p}\, m_{\mathrm{b}}\, m_{\mathrm{f}}\, \Mpf^2\, ,
\eeq
where in four dimensions, the species scale is defined as $\LSP^2 \simeq \Mpf^2/N$. Thus, we obtain
\beq
	g_{\chi \chi}\, \sim\,  \sum^{N} n^{2/r}\, \sim\, N^{\frac{2}{r}+1}\, \sim\, \left(\frac {(m_{\mathrm{f}}\Mpf^2)^{1/3}}{m_{\mathrm{b}}}\right)^{\gamma_r}\, ,
\label{eq:metricafermiongeneral}
\eeq
where
\beq
	\gamma_{r,p}\, =\, \frac {3p(2+r)}{r(4p-1)}\, .
\eeq
 Note that by considering different values for $\{r,p\}$ we find the $\gamma$-parameter to lie in the range $1\leq \gamma \leq 3$. For the particular case of $r=p=1$ one gets $m_{\mathrm{b}}=m_{\mathrm{f}}=\Mt$  and $\gamma_{1,1}=3$ yielding\cite{Palti:2020tsy}
\beq
	g_{\chi\chi}\, \sim\, \left( \frac {\Mpf}{\Mt}\right)^2\, ,
\eeq
which is analogous to the structure found for the gauge kinetic functions. (In particular, if we take $\mathcal{Y}_n = -\i \sqrt{2}\ q_n g$ \cite{Wess:1992cp}, as in the gaugino case, one obtains agreement with the 1-form wave-function renormalization discussed in Section \ref{sss:emergenceU(1)}.) For towers/charges with a different structure the more general result, i.e. eq. \eqref{eq:metricafermiongeneral}, applies. Still, the take-home message is that fermions may get large wave-function renormalization if at least one of the towers running in the loop is sufficiently light. 

\subsubsection{A Weak-Gravity-like conjecture for  Yukawa couplings}
\label{sss:WGCYukawa}

The stronger versions of the Weak Gravity Conjecture propose that in the presence of a $\mathsf{U(1)}$ gauge field along with charged particles, the limit $g\rightarrow 0$ is singular and should be accompanied by (infinite) towers of states becoming light. A similar question that can arise is what happens with Yukawa couplings, namely when some of these go to zero. In the present discussion we would like to argue that, at least in the context of Emergence, the answer to this question is positive (see \cite{Palti:2020tsy} for similar ideas and \cite{Casas:2024ttx} for recent non-trivial tests in string theory).
	
In the emergence picture one expects every non-vanishing Yukawa coupling already present in the UV (if any) to be generically of order one, up to exponentially suppressed instanton corrections. In fact, this is what is generically found in specific string theory constructions, see e.g., \cite{Ibanez:2012zz}. Thus, couplings involving three fields in four dimensions will have typically the form $\mathcal{Y}^{\text{UV}}_{ijk}\simeq \mathcal{S}_{ijk}$, with all entries in the matrix $\mathcal{S}$ being essentially of $\mathcal{O}(1)$, such that no hierarchies of Yukawa interactions would appear at this level. Therefore, any hierarchical behaviour (like the ones existing in the SM) would appear as an infra-red effect. Indeed, the above considerations about the generation of kinetic terms for fermions may give us a hint regarding how this issue. We will frame the present discussion for concreteness in an 4d $\mathcal{N}=1$ supersymmetric setting, but most of the results should still be amenable to generalization to non-supersymmetric set-ups with minimal changes. Thus, in such a supersymmetric scenario with a \emph{renormalizable} 4d superpotential, $W (\Phi)= \mathcal{S}_{ijk}\Phi^i\Phi^j\Phi^k$, the properly normalized Yukawa couplings would have the form
\beq
	\mathcal{Y}_{i j k}\, =\, \mathcal{S}_{ijk}\,  (g_{i \bar i}g_{j \bar j} g_{k \bar k})^{-1/2}\, .
\eeq
where $g_{i \bar i}$ are the K\"ahler metrics of the corresponding (massless) chiral multiplets, which we have taken here to be diagonal for simplicity. We now assume that all these metrics for the massless fields are obtained via the emergence mechanism such that, according to the discussion above, we find
\beq
	\mathcal{Y}_{i j k}\,  \gtrsim \,  \mathcal{S}_{ijk} \Mpf^{-\gamma}\left[ \left(\frac {m_{l,i}}{m_{h,i}^{1/3}}\right)^{\gamma_i/2}
	\left(\frac {m_{l,j}}{m_{h,j}^{1/3}}\right)^{\gamma_j/2}
	\left(\frac {m_{l,k}}{m_{h,k}^{1/3}}\right)^{\gamma_k/2} \right]\, ,
\eeq
where $\gamma =\sum_i\gamma_i/3$ and $\gamma_i=\frac{3p_i(2+r_i)}{r_i (4p_i-1)}$. Here the subindices $\{h,l\}$ stand for the heaviest and lightest masses in each of the loops, respectively. Now, starting with $\mathcal{S}_{ijk}$ being $\mathcal{O}(1)$, if we take some entry of $\mathcal{Y}_{i j k}$ to approach zero it will imply that some (or all) of the towers have to become nearly massless in Planck units. Hence, the above expression is, in some sense, the Yukawa analogue of the magnetic WGC. Notice  also that, depending on how light each of the $\{i,j,k\}$ towers (which need not be different) is, hierarchies in the Yukawa couplings could naturally arise, even though all components in the UV matrix --- i.e. the $\mathcal{S}_{ijk}$ --- were originally of order one. Thus, one could argue that hierarchies in the Yukawa interactions may be a generic effect in \emph{emergent} EFTs, with essentially the same loops inducing the kinetic terms also providing for the hierarchical behaviour of the set $\{ Y_{i j k} \}$ \cite{Castellano:2023qhp}.
	
If the bosonic and fermionic towers present identical mass gaps and $g_{i \bar i}\sim 1/m_i^2$, as in the case in which they are supersymmetric partners of one another, one has $\gamma_i=\gamma=3$ and 
\beq
	\mathcal{S}_{ijk}\, m_im_jm_k\, \lesssim\, \mathcal{Y}_{i j k} \, \Mpf^3\, ,
\eeq	
which is again reminiscent of a WGC-like expression for the Yukawa couplings. In particular, it was shown in \cite{Heidenreich:2017sim} that in a 4d $\mathsf{U(1)}$ gauge theory coupled to gravity, the cut-off should be bounded as $\LSP \lesssim e^{1/3}\Mpf$, with $e$ denoting the gauge coupling. Hence, for any charged matter particle with mass $m$ below the cut-off one finds
\beq
	m^3\, \lesssim\, \LSP^3\, \lesssim\, e\, \Mpf^3\, ,
\eeq
which is a `gauge counterpart' of the above Yukawa constraint for the case of a single tower of states with mass parameter given by $m$. 
	
We end this section with an important comment concerning the gauge coupling renormalization. One could naively argue that by applying this idea to the interaction terms coupling the vector boson to some (conserved) current as in e.g., $({\bar \psi}\gamma^\mu A_\mu \psi)$, gauge interactions could also become hierarchical due to possible large corrections to the wave-function renormalization of the charged fermions. Of course, this cannot be the case since the Ward-Takahashi identities associated to the gauge field relate precisely the vertex and the wave-function factors, such that once we sit in a frame with canonically normalized fermions, the apparent hierarchies disappear. On the other hand, there are no a priori Ward-Takahashi identities for Yukawa couplings (unless they come from a gaugino interaction) and thus hierarchies can arise in general due to possibly large anomalous dimensions. 

\section{Emergence in string theory}\label{s:EmergenceStringTheory}

In this section we test the Emergence prescription introduced in Section \ref{s:EmergenceQG} within several concrete string theory constructions. In particular, we consider theories with 8 supercharges in 4d (Section \ref{ss:emergence4d}) and 6d (Section \ref{ss:emergence6d}) arising from Type II and F-theory Calabi--Yau three-fold compactifications, respectively; as well as 7d theories preserving 16 supercharges obtained from M-theory on $K3$, see Section \ref{ss:emergence7dN=1}. As we will see, each of these examples exhibits different singularity structures that may be ultimately resolved by different kinds of towers of states becoming asymptotically light (in Planck units). The computations that we perform rely heavily on the machinery and formulae presented in Section \ref{s:selfenergybosons}, so that we refer oftentimes to the material presented in that section for the technical details. 

\subsection{Emergence in 4d $\mathcal{N}=2$ theories}
\label{ss:emergence4d}

The first realistic set-up where the Emergence mechanism was studied arises in the context of 4d $\mathcal{N}= 2$ theories as obtained from Type II Calabi--Yau compactifications, see Section \ref{s:4dN=2} for details. The focus was originally placed on the large volume/large complex structure singularity \cite{Corvilain:2018lgw,Grimm:2018ohb,Gendler:2020dfp}, since the associated infinite distance degeneration corresponds to a simple circle decompactification to M-theory on the same three-fold (see Section \ref{sss:5dMtheory} for a detailed discussion of the associated effective 5d $\mathcal{N}=1$ supergravity theory).

In the following, we revisit this analysis in light of our considerations from Section \ref{s:selfenergybosons} above. We will pay special attention to the infinite set of states that are necessary so as to generate (via Emergence) the full gauge kinetic function close to the large volume point. Later on, in Section \ref{ss:generalizations} we turn to other infinite distance degenerations probing instead different duality frames of the theory. Finally, in Section \ref{ss:hypermultiplet4d} we briefly discuss the hypermultiplet moduli space of 4d $\mathcal{N}=2$ theories within the present context.

\subsubsection{Preliminary remarks} 

In this section we consider Type IIA string theory on a CY three-fold $X_3$, and we sit ourselves close to the large volume point, i.e. $\mathcal{V}_{X_3} \to \infty$. This corresponds to an infinite distance singularity, where a number of towers of states become exponentially light with the traversed geodesic distance \cite{Font:2019cxq}. Indeed, notice that for a given BPS particle with $n_{2,a}\in \mathbb{Z}$ units of D2-brane charge --- where the subindex $a$ indicates the holomorphic 2-cycle wrapped by the brane --- and $n_0\in \mathbb{Z}$ the corresponding D0-brane charge, the moduli dependence of the mass measured in 4d Planck units is encapsulated by the normalized $\mathcal{N}=2$ central charge
\beq\label{eq:centralchargeI}
		\frac{m_{n_{2p}}}{\Mpf} = \sqrt{8\pi } e^{K_{\text{ks}}/2} |\mathcal{Z}_{\text{IIA}}| = \sqrt{\frac{\pi}{ \mathcal{V}_{X_3}}}\, |n_0+n_{2,a} z^a|\, .
\eeq
Therefore, for D2-branes wrapping 2-cycles whose volume grows no faster than $\mathcal{V}_{X_3}^{1/2}$, the above central charge vanishes asymptotically, in agreement with our previous claim. Furthermore, using the duality map between the moduli coordinates of Type IIA string theory on $X_3$ and M-theory on $X_3 \times \mathbf{S}^1$ (see Section \ref{sss:IIA/Mthy} below for details on this), we can relate the 4d states \eqref{eq:centralchargeI} with their five-dimensional counterparts
\begin{equation}\label{eq:massesD0D2Emergence}
	 \begin{aligned}
				\frac{m_{\text{D0}}}{\Mpf}\, &\, \sim g_s^{-1} e^{\varphi_4}\, \sim\, \frac{1}{\mathcal{V}_{X_3}^{1/2}}\, \sim\, \frac{m_{\text{KK},\, 5}}{\Mpf}\, ,\\
				\frac{m_{\text{D2}}}{\Mpf}\, & \sim\, e^{K_{\text{ks}}/2} |t^a|\, \sim\, \frac{\tilde{t}^a}{\mathcal{V}_{X_3}^{1/6}}\, \sim\, \frac{m_{\text{M2}}}{\Mpf}\, ,
	 \end{aligned}
\end{equation}
where in the last expression we have considered a single D2-brane wrapping some 2-cycle and we have set the corresponding axion v.e.v. $b^a$ to zero. Notice that in order to identify the mass of the D2-particles with that of the M2-branes wrapping the same cycles in the Calabi--Yau, we have split the K\"ahler coordinates into the overall volume $\mathcal{V}_{X_3}$, and rescaled moduli $\tilde{t}^a=t^a/\mathcal{V}_{X_3}^{1/3}$, which are subject to the very special geometry constraint (c.f. eq. \eqref{eq:veryspecialgeometry})
\beq
 \mathcal{K}_{a b c} \tilde{t}^a \tilde{t}^b \tilde{t}^c \stackrel{!}{=} 6\, .
\eeq
Now, given that we are dealing with a decompactification limit on a circle, and that the Kaluza-Klein scale is identified with the mass of the D0-branes, one can readily see that the species scale associated to that tower alone already coincides with the 5d Planck scale, i.e.
\beq \label{eq:D0towerEmergence}
		\frac{\LSP}{\Mpf}\, \simeq\, \left( \frac{m_{\text{D0}}}{M_{\text{Pl;}\, 4}} \right)^{1/3}\, \sim\, \frac{1}{\mathcal{V}_{X_3}^{1/6}}\, \sim\, \frac{M_{\text{Pl;}\, 5}}{\Mpf}\, ,
\eeq
which verifies $\LSP/\Mpf \to 0$, as expected. Nonetheless, following the general procedure described in Section \ref{ss:MultipleTowers} so as to compute the species scale in the presence of multiple towers, once we have calculated the scale associated to the lightest set of states, we should compare it with the characteristic mass of the subsequent lightest tower. In the large volume limit, the next-to-leading one would correspond to the aforementioned set of D2-D0-particles. However, as can be seen from eq. \eqref{eq:massesD0D2Emergence} above, the mass of a wrapped D2-brane is already of the order of the species scale --- if we keep the $\{ \tilde{t}^a\}$ fixed and finite. This means, in turn, that the condition \eqref{eq:algorithmstops} is saturated, such that accounting for this extra set of states as well does not change significantly the previously computed species cut-off, since both towers behave additively. 
		
\subsubsection*{The D-brane field content}
		
Having discussed the objects that must be included in the loop computation when exploring the large volume limit, let us now turn our attention to their corresponding field-theoretic content. The basic claim here is that we need to consider both the tower of BPS bound states of D0-branes, whose particle content corresponds to that of massive Kaluza-Klein replicas within a circle reduction from 5d $\mathcal{N}=1$ supergravity; as well as certain fields with spin strictly smaller than 2, which arise from bound states between a single wrapped D2-brane and $n$ D0 modes. 
		
There are in fact several arguments that support these claims, which stem either from the duality between Type IIA string theory and M-theory (c.f. Section \ref{ss:dualitieswithhighersusy}), or rather from a more concrete super-quantum mechanical analysis. Here we will only review the former, whilst a detailed discussion of the super-quantum mechanics of the D0-branes can be found in Appendix C of \cite{Castellano:2022bvr}. Therefore, recalling that Type IIA string theory on $X_3$ is dual to M-theory compactified on $X_3 \times \mathbf{S}^1$, we deduce that the spectrum of (supersymmetric) states in four dimensions must necessarily include the Kaluza-Klein replica of every light field already present in the 5d theory. On the one hand, this implies that there should exist massive KK modes associated to the 5d massless fields, namely a spin-2 multiplet, ($h^{1,1}-1$) vector multiplets, and ($h^{2,1}+1$) hypermultiplets \cite{Cadavid:1995bk}. From the four-dimensional perspective, these states are nothing but the BPS tower of D0-brane bound states. On the other hand, regarding the spectrum associated to the D0-D2 tower, it is clear from the above picture that it should correspond to KK replica of some massive supermultiplet in 5d arising from an M2-brane wrapping the corresponding 2-cycle. There is, however, an extra difficulty due to the fact that its field-theoretic content strongly depends on the moduli space of the supersymmetric cycle wrapped by the 2-brane --- along with its possible flat connections. In any event, for the simplest case in which such moduli space is trivial (i.e. just a point), one can readily see that each step in the tower is associated to a 4d massive hypermultiplet.

\subsubsection{Emergence of the gauge kinetic function}
\label{ss:4doneloop}
Once we identify the relevant physics taking place in the asymptotic limit of interest, as well as the corresponding towers which should be responsible for such transition, one can proceed by studying whether or not Emergence can work in practice. One possible route would be to use the general results from Section \ref{sss:emergencemodulimetric} to try to reproduce the moduli space metric shown in \eqref{eq:kahlersectormetric}. Here, we focus instead on the more interesting case corresponding to the emergence of the gauge kinetic function, even though both computations are intimately related due to the underlying $\mathcal{N}=2$ structure of the theory. To do so, we hence calculate the one-loop wave-function renormalization of the gauge fields and show that upon integrating out the tower of BPS states (up to the species cut-off) we recover precisely the functions displayed in \eqref{eq:gaugekineticmatrix}.
		
For concreteness, we restrict ourselves to one-dimensional K\"ahler spaces --- i.e. those with $h^{1,1}=1$, where we denote by $z=b+\i t$ the corresponding K\"ahler modulus, whereas the overal volume reads as $\mathcal{V}_{X_3}=t^3$. In that case, we are left with just two $\mathsf{U(1)}$ gauge fields: the graviphoton, and the one belonging to the unique vector multiplet. Still, let us stress here that in the more general multi-moduli scenario, one can similarly probe the large volume regime by splitting the K\"ahler moduli into the volume scalar and rescaled coordinates, see Section \ref{sss:IIA/Mthy} for details. Thus, we conclude that the physics of the $\mathcal{V}_{X_3}\to \infty$ limit is already captured by our current simplified set-up.
		
Therefore, the relevant gauge kinetic matrix that we aim to reproduce takes the following simple form
\begin{equation} \label{eq:matrixonemodulus}
	\text{Im}\, \mathcal{N}\, = \frac{\mathcal{K}}{6} \, \left(
		\begin{array}{cc}
			1+3(b/t)^2 & -3b/t^2  \\
			-3b/t^2 & 3/t^2  \\
		\end{array}
	\right) \, ,
\end{equation}
where we have substituted the K\"ahler metric 
\beq
		G_{z\bar z}= \frac{1}{4t^2}\, ,
\eeq
in eq. \eqref{eq:gaugekineticmatrix}. Let us remark two important features exhibited by \eqref{eq:matrixonemodulus} above. First, due to the off-diagonal terms, it implements the phenomenon of gauge kinetic-mixing\cite{Holdom:1985ag,delAguila:1988jz}. Secondly, in the large volume limit (i.e. $t \to \infty$) every entry blows up, such that both vectors become weakly coupled.\footnote{Strictly speaking, when talking about weakly/strongly coupled gauge fields, one refers to the \emph{physical} fields, that is those for which the kinetic functions adopt the canonical form and are thus diagonal. In our example it can be checked that the corresponding eigenvalues of the physical gauge couplings are also vanishing in the $t \to \infty$ limit.} Notice as well that the polynomial dependence on the divergent K\"ahler modulus is different between those terms involving the axion $b$ and those in which it is absent. In the following, we explain how these facts arise naturally when integrating out the tower of D0-branes and D0-D2 particles discussed before.
		 
\begin{figure}[tb]
			\begin{center}
				\includegraphics[scale=1.2]{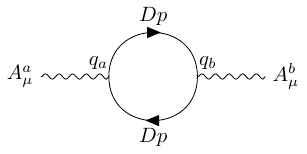}
				\caption{One-loop diagram with BPS states circulating and giving rise to an effective low-energy kinetic mixing between the different $\mathsf{U(1)}$ vector bosons.}
				\label{kineticmixing}
			\end{center}
		\end{figure}

To show this, we make use of the  renormalization group flow, which induces kinetic mixing when we allow for heavy multi-charged particles to run in the loop, as displayed in Figure \ref{kineticmixing}. Indeed, the gauge kinetic function $f_{AB}^{\text{UV}}$ at a certain UV scale $\Lambda_{\text{UV}}$, is related to the low-energy gauge couplings (for energies well below the masses of the particles) as follows
		\beq \label{eq:oneloopmixing}
		f_{AB}^{\text{UV}}\, =\, f_{AB} - \sum_i \frac{\beta_i}{8\pi^2 \kappa_4^2}\ q_{A}^{(i)} q_{B}^{(i)}\ \log \left(\frac{\Lambda_{\text{UV}}}{m_i}\right)\, ,
		\eeq
where $m_i$ and $q_{A}^{(i)}$ denote the mass and (dimensionless) charge of the $i$-th particle, whereas $\beta_i$ is the corresponding $\beta$-function coefficient. Notice that the kinetic matrix $f_{AB}$ defined here includes a factor of $\kappa_4^{2}$ in the denominator that was explicitly extracted in the definition of $\text{Im}\, \mathcal{N}_{AB}$ within the supergravity action \eqref{eq:IIAaction4d}. This arises since in our conventions the gauge fields are taken to be dimensionless. Additionally, the value of $\beta_i$ depends on the character and degeneracy of the particles running in the loop. For BPS states corresponding to e.g., the D0-D2 tower --- in the simplest one-modulus scenario --- we find $\beta_i= 1$, since the tower is made of charged $\mathcal{N}=2$ hypermultiplets.

Before proceeding with the computation, let us make an important remark about gauge charges, since their precise value depends on the basis of vector bosons that is used. In particular, the basis defined in eq. \eqref{eq:rotatedbasisvectorfields4d} is better adapted for the present analysis, since the associated field strengths are (locally) exact and therefore well-suited for the renormalization of the propagator. Relatedly, the charges of BPS states running in the loop are quantized (they essentially count the number of D$p$-particles comprising each bound-state), and hence moduli-independent. This is to be contrasted with the usual supermultiplet eigenstates basis (i.e. the graviphoton and the orthogonal gauge fields in the vector multiplets), which have neither locally exact field strengths, nor integrally quantized charges.\footnote{The moduli-dependent shifts in the charge vectors can be understood as e.g., induced D0-charges within the 2-brane worldvolume due to a non-trivial $B_2$-flux along the wrapped 2-cycle, c.f. eq. \eqref{eq:CSWZactionTypeIIA}.}
		
\subsubsection*{The case $b=0$}
		
We start here with the simpler case in which the axion v.e.v. is set to zero, i.e. $b=0$. The gauge kinetic functions \eqref{eq:matrixonemodulus} become thus diagonal, so that there is no kinetic mixing anymore, and the dependence with respect to the saxion field is different for each boson. Let us explain how this matrix appears via Emergence, i.e. by integrating out the tower of D2-D0 bound states with one unit D2-charge and $n$ D0-charge.\footnote{In the notation of Section \ref{sss:emergenceU(1)}, the KK-photon coupling to the D0 tower corresponds to $r=p=1$ and that of the D2-D0 to $r=1,\, p \to \infty$. For the other $\mathsf{U(1)}$, the D2-D0 tower corresponds to $r=p \to \infty$.} Consider first the diagonal terms. Hence, from \eqref{eq:oneloopmixing} we get the following kinetic term for $A^0$ 
\beq \label{eq:f00}
		f_{00}\, =\, f_{00}^{\text{UV}} + \frac{\beta}{16 \pi^2} \sum_{-n_{\text{max}}}^{n_{\text{max}}}  \left( \pi \Mpf^2 n^2 \right)\ \log \left(\frac{t^3 \,  \LSP^2}{\pi  \, (n^2+t^2) \, \Mpf^2}\right)\, ,
\eeq
where we have substituted eq. \eqref{eq:centralchargeI} for the BPS masses, whereas the quantized charges under $A_0$ are given by $q_0^{(n)}\, =\, n  \sqrt{\pi} \Mpf$. Note that $n_{\text{max}}$ is such that the mass of the heaviest state considered in the tower agrees with the cut-off scale, namely $m_{n_{\text{max}}}^2 \sim \LSP^2 \sim 2 \pi \Mpf^2/ t$, thus implying that $n_{\text{max}} \sim t$. Therefore, by approximating the sum with an integral (which is justified in the asymptotic limit) one recovers the following behaviour for the gauge kinetic function
\beq 
		f_{00}\, \simeq\, f_{00}^{\text{UV}} +  \beta \,  \Mpf^2 \, \left(\dfrac{3 \pi - 8}{144 \pi } \right)
		t^3 \sim  \,  t^3 \, \Mpf^2 \, ,
\eeq
where in the last step we have assumed (following the Emergence prescription) that the UV-contribution is at most as large as the one-loop piece. Hence, we are able to generate the right asymptotic behavior in the first diagonal entry of eq. \eqref{eq:matrixonemodulus}. Similarly, including the contribution from the tower of D0-particles --- which couples only to $A^0$, for which the mass in the denominator should be substituted as $(n^2+t^2) \, \to n^2$, the same asymptotic dependence $\sim t^3$ is generated.
		
Analogously, we can compute the one-loop correction to the kinetic term associated to the second gauge field, $A^1$, induced by the D0-D2 particles (the D0-branes do not couple to $A^1$). In this case, the mass of the states is the same as before, but the charge is now constant $q_1^{(n)}\, =\,  \sqrt{\pi} \Mpf$ for every mode in the tower. The relevant contribution thus takes the form
\beq \label{eq:f11} 
		\begin{aligned} f_{11}\, &=\, f_{11}^{\text{UV}} + \frac{\beta}{16 \pi^2} \sum_{-n_{\text{max}}}^{n_{\text{max}}}  \left( \pi \Mpf^2 \right)\ \log \left(\frac{t^3 \,  \LSP^2}{\pi  \, (n^2+t^2) \, \Mpf^2}\right)\, , \\ &\simeq\, f_{11}^{\text{UV}} + \beta \Mpf^2 \left( \frac{4-\pi}{16 \pi }\right) t \sim   t \, \Mpf^2 \, ,
		\end{aligned}
\eeq
where in the last step we have assumed the UV piece to be again subleading. Interestingly, the saxion dependence of the second diagonal component of the gauge kinetic matrix   \eqref{eq:matrixonemodulus} presents the same asymptotic behaviour as the one obtained via Emergence.
		
Let us now turn to the off-diagonal contributions. This is arguably the most interesting piece of the discussion, since it gives a couple of nice insights. First, recall the gauge kinetic matrix is diagonal for $b=0$. However, in principle, since the D0-D2 tower is charged with respect to both $\mathsf{U(1)}$ fields, kinetic mixing may be generated via loops. Thus, if the Emergence mechanism works in this case, it should produce a vanishing one-loop correction when the full tower is included. Indeed, the computation yields
\beq \label{eq:f01}
		f_{01}\, =\, f_{01}^{\text{UV}} + \frac{\beta \, \Mpf^2}{4\pi} \sum_{-n_{\text{max}}}^{n_{\text{max}}}    n\ \log \left(\frac{t^3 \,  \LSP^2}{\pi  \, (n^2+t^2) \, \Mpf^2}\right) = f_{01}^{\text{UV}}\,  ,
\eeq
where the exact cancellation of the loop terms is due to the fact that we sum from $-n_{\text{max}}$ to $n_{\text{max}}$. 
Therefore, to match with the exact result \eqref{eq:matrixonemodulus}, we need to impose $f_{01}^{\text{UV}}=0$ in eq. \eqref{eq:f01} above. Crucially, this is precisely the expected UV boundary condition, since the decompactified 5d $\mathcal{N}=1$ theory obtained by M-theory on the Calabi--Yau contains no axion-like particles (they arise in 4d as Wilson lines of the 5d vectors along the M-theory circle), and moreover presents no kinetic mixing between the 5d graviphoton and the five-dimensional Einstein-Hilbert term (from which the four-dimensional KK-photon descends).  
		
\subsubsection*{The case $b\neq0$}
		
Let us now turn on the axion v.e.v. and discuss how the preceding results are modified. Notice that once we have non-vanishing axion fields in the vacuum, the gauge kinetic functions include off-diagonal terms, and therefore there is some amount of kinetic mixing between the two vector fields. This, together with the fact that also the $A^0$ gauge coupling is shifted by an axion-dependent amount, are the two main points of the present analysis, and as we will see they can be nicely accounted for in the framework of Emergence. Concerning the light states, the main difference with respect to the $b=0$ case is the $b$-dependent shift in the mass of the D0-D2 tower. More concretely, using \eqref{eq:centralchargeI}, we can see that this is taken into account by the replacement
\begin{equation}
\label{eq:shifbneq0}
			m_n^2\, =\, \dfrac{\pi \Mpf^2}{t^3}\left[ n^2 +t^2 \right] \, \longrightarrow \dfrac{\pi \Mpf^2}{t^3}\left[ (n+b)^2 +t^2 \right] ,
\end{equation}
which basically means $n \to n+b$ in the denominators inside the logs of eqs. \eqref{eq:f00}-\eqref{eq:f01}. On the contrary, the charges with respect to the fields $\{A^0, A^1\}$ are left unmodified. This seemingly innocuous change is actually at the core of the generation of the axion-dependent terms in the gauge kinetic matrix, as we explain in what follows. 
		
Notice that the structure of the tower is very similar to the one we had before except from a key difference regarding which states lie below $\LSP$. In particular, if we now compute the values of $\{n_{\text{min}}, n_{\text{max}}\}$ so that $m_{n_{\text{min}}}^2 \simeq \LSP^2 \simeq m_{n_{\text{max}}}^2$, we obtain the shifted quantities $n \in [-t-b, t-b]$, such that indeed $n_{\text{min}} \neq -n_{\text{max}}$.\footnote{Notice that CPT does not prevent this asymmetry in the D0-brane charge from happening, given that the would-be hypermultiplets inside the D0-D2 towers are already CPT invariant, such that for a given D0-brane charge $n$, the opposite one appears along with the anti-D2-brane wrapping the same 2-cycle.} Taking this into account, as well as the shift discussed in \eqref{eq:shifbneq0}, we finally obtain
\begin{align}\label{eq:4dEmergenceaxion}
			f_{11}&\, \simeq\, f_{11}^{\text{UV}} + \beta \Mpf^2 \left( \frac{4-\pi}{16 \pi }\right) t\, \sim\,   t \, \Mpf^2 \, ,\\ 
			f_{01}&\, \simeq\, f_{01}^{\text{UV}} - \beta \Mpf^2 \left( \frac{4-\pi}{16 \pi }\right) b t\, \sim \, - b  t\, \Mpf^2 \, ,\\
			\begin{split}
				f_{00}&\, \simeq\, f_{00}^{\text{UV}} + \frac{\beta \Mpf^2}{144 \pi } \left[  \left( 3\pi -8  \right) \, t^3 + \left( 4 -\pi \right)\,  b^2 t \right]\, \sim\, (\, C \, t^3 +  \,  b^2 t)\, \Mpf^2\, .
			\end{split}
\end{align}
For $f_{11}$ we recover the same result as in the $b=0$ case, as expected. For the mixed term $f_{01}$, however, we get instead a non-vanishing result which is essentially $(-b)$-times the analogous contribution to $f_{11}$, and indeed reproduces \eqref{eq:matrixonemodulus}. (We have set again $f_{01}^{\text{UV}}=0$ in eq. \eqref{eq:4dEmergenceaxion}, since as already discussed there is no such kinetic mixing in the 5d parent theory.) Finally, for the $f_{00}$ component we seem to get the right structure of the gauge kinetic matrix, with a subleading --- but nevertheless diverging --- contribution depending on the axion v.e.v. Notice that we are unable to fix the relative coefficient $C$,  between the two terms in $f_{00}$, since the $t^3$ contribution also receives corrections from the tower of D0-branes alone. 
Barring this subtlety, we seem to recover the correct asymptotic structure exhibited by the gauge kinetic matrix \eqref{eq:matrixonemodulus} in this set-up as well.

\subsubsection{Other possible infinite distance limits in 4d}
\label{ss:generalizations}	

In Section \ref{ss:4doneloop} we restricted ourselves to one-dimensional K\"ahler moduli spaces, in which the relevant infinite distance singularity was identified with the large volume point. Moreover, the asymptotically massless tower of states giving rise to the singularity effectively implemented a circle decompactification to 5d M-theory on the same Calabi--Yau three-fold. However, as already commented around eq. \eqref{eq:matrixonemodulus}, the discussion there seems to apply equally well to higher dimensional moduli spaces, as long as we fix the rescaled K\"ahler parameters, $\tilde{t}^a = t^a/\mathcal{V}_{X_3}^{1/3}$, to be finite (and non-vanishing) whilst the volume modulus, $\mathcal{V}_{X_3}$, is taken to infinity.
		
Therefore, one could also ask at this point what happens now if we additionally move close to infinite distance boundaries along the $\tilde{t}^a$ directions as well. Indeed, what one expects is to approach another kind of singular limit, exhibiting a different nearly massless tower dominating the infinite distance regime and thus implementing some other gravitational phase transition. Notice that given that we essentially explore the large volume phase of Type IIA string theory on $X_3$, any other infinite distance singularity that we may encounter there should be present already in the vector multiplet moduli space of M-theory on the same three-fold \cite{Witten:1996qb}. In what follows we will concentrate on two other possible infinite distance boundaries, as studied first in \cite{Lee:2019wij, Corvilain:2018lgw}. These correspond to having a universal fibration structure, where the fibre shrinks (in rescaled coordinates) and is given by an elliptic curve or a K3 surface. These limits were shown to be originally in tension with the Emergence Proposal, and in the following we revisit them, elaborating on several important points which solve all the problems encountered by the authors in \cite{Grimm:2018ohb}.
		
\subsubsection*{The F-theory limit}
\label{sss:IIA-Ftheorylimit}		
Let us start with the $\mathbf{T}^2$-limits of \cite{Lee:2019wij}, in which apart from having $\mathcal{V}_{X_3}\to \infty$, 
the Calabi--Yau exhibits asymptotically certain elliptic fibration
\begin{equation}
			\begin{aligned}\label{eq:ellipticcaseEmergence}
				\pi: \qquad \mathbf{T}^2 \hookrightarrow &\;X_{3} \\
				&\;\; \downarrow \qquad , \\ &\;B_{2}
			\end{aligned}
\end{equation}
where $\mathbf{T}^2$ denotes the elliptic fibre whose associated rescaled K\"ahler modulus, denoted by $\tilde{t}^0$ here, vanishes asymptotically. Notice that this kind of degenerations correspond to Type III singularities in Mixed Hodge Structure (MHS) language, as discussed in \cite{Grimm:2018ohb, Corvilain:2018lgw} (see also Table \ref{tab:limitsN=2} below).
		
It is easy to see that along the aforementioned class of infinite distance limits, both the D0- and D2-particles present the same asymptotic mass scale (see discussion around eq. \eqref{eq:ellipticmass}), contrary to what happens in the large volume point. Moreover, the D2-branes wrapping the elliptic fibre provide now for an \emph{infinite} number of distinct BPS states, which geometrically is reflected in the fact that the Gopakumar-Vafa invariants associated to the $\mathbf{T}^2$ fibre are (generically) non-zero and constant\cite{Klemm:2012sx, Klemm:1996hh} for each $k\in \mathbb{Z}_{\geq 0}$ (c.f. eq. \eqref{prepotentialIIA})
\beq
		n_{\mathbf{k}}^{(0)} = \chi(X_3) = 2 \left ( h^{1,1} (X_3) -  h^{2,1} (X_3) \right)\, ,
\eeq
where $k$ denotes the wrapping number of the D2-brane along the elliptic cycle. Additionally, these two towers can mix, forming bound states with D0 and D2 charges which can in principle run independently as long as the total mass remains below the species scale, $\LSP$. Hence, the present scenario corresponds to the case of multiplicative species (with $p=2$) discussed in Section \ref{ss:MultipleTowers}, such that upon using our formulae \eqref{eq:massmixedsprectra}-\eqref{eq:NtotLQGeff}, one indeed reproduces the behaviour exhibited by the exact gauge kinetic matrix, as we show here. Denoting the divergent modulus by $t$, the mass of the D0-branes --- as well as that of D2-branes wrapping the elliptic fibre --- behave as $m_{\text{D}0}\sim m_{\text{D}2}\sim \Mpf/t$. Using the general formula \eqref{eq:NtotLQGeff} for $d=4$ and $p=2$ we obtain the following behaviour for the species scale and the total number of species
\begin{equation}
    \LSP\, \sim\,  \dfrac{\Mpf}{t^{\frac{1}{2}}}\, , \qquad \Ntot\, \simeq \, N_{\mathrm{D}0}\,  N_{\mathrm{D}2} \sim t\, .
\end{equation}
Performing now a similar calculation to the one shown in \eqref{eq:f00}, where we compute the contribution of the full tower of D2-D0 particles to the kinetic term of the Kaluza-Klein photon $A^0$, we get
\beq \label{eq:f00F-theorylimit}
		f_{00}\, \sim\, \sum_{n_{\text{D}2}}^{N_{\mathrm{D}2}} \sum_{n_{\text{D}0}}^{N_{\mathrm{D}0}}  \left(  \Mpf^2\,  n_{\text{D}0}^2 \right)\, \log \left(\dfrac{t}{n_{\text{D}0}^2 + n_{\text{D}2}^2}\right)\, \sim \,  \Mpf^2 \, t^2\,  ,
\eeq
where we used that $N_{\mathrm{D}0} \, \sim \, N_{\mathrm{D}2} \sim t^{1/2}$. Similarly, for the kinetic term of the 1-form $A^1$ that couples to the D2-particles becoming light, we obtain the same parametric behaviour, namely
\beq \label{eq:f11F-theorylimit}
		f_{11}\, \sim\, \sum_{n_{\text{D}2}}^{N_{\mathrm{D}2}} \sum_{n_{\text{D}0}}^{N_{\mathrm{D}0}}  \left(  \Mpf^2\,  n_{\text{D}2}^2 \right)\, \log \left(\dfrac{t}{n_{\text{D}0}^2 + n_{\text{D}2}^2}\right)\, \sim \,  \Mpf^2 \, t^2\, ,
\eeq
whereas for the mixed terms (in the vanishing axion case) as well as the remaining $\mathsf{U(1)}$ fields --- under which the light D2-D0-particles are not charged --- we obtain vanishing contributions from the quantum corrections. Note that eqs. \eqref{eq:f00F-theorylimit} and \eqref{eq:f11F-theorylimit} indeed reproduce the right field dependence of the gauge kinetic matrix \eqref{eq:gaugekineticmatrix} on the divergent modulus $t$ \cite{Grimm:2018ohb} (see also \cite{Marchesano:2022axe} for a complementary approach).

To finish this example, let us just add a few relevant comments about the resolution of the infinite distance singularity. Indeed, as already stressed in \cite{Corvilain:2018lgw}, one expects some partial decompactification to happen upon approaching this Type III limit. The main intuition comes from Type IIA/M-theory duality first, and then from M-/F-theory duality (see Section \ref{ss:dualitieswithlowersusy} for details). Thus, recall that upon taking the $\mathcal{V}_{X_3}\to \infty$ limit and on top of that exploring singularities along the $\tilde{t}^a$ directions, what we are effectively doing is probing the 5d moduli space \cite{Witten:1996qb}. Therefore, since a $\mathbf{T}^2$ limit there leads to a circle decompactification to 6d F-theory \cite{Lee:2019wij}, the natural conclusion here would be to identify this Type III singularity with a nested limit from 4d Type IIA to 6d F-theory. One non-trivial check that can be performed so as to provide evidence for the previous conclusion is to employ the super-quantum mechanical machinery so as to deduce the field-theoretic content associated to the D2-branes wrapping the elliptic fibre (see e.g., Appendix C in \cite{Castellano:2022bvr}). Upon doing so, what one finds is that the moduli space of the elliptic fibre $\mathbf{T}^2$ together with its flat connections is again an elliptically-fibered K\"ahler three-fold, such that the D2-particles contain --- for each $n_{\text{D}2}\in  \mathbb{Z} \setminus \lbrace 0 \rbrace$ --- precisely one massive spin-2 multiplet, $h^{1,1} (B_2)=h^{1,1} (X_3)-1$ massive vector multiplets and $h^{2,1} (X_3)+1$ massive hypermultiplets.
		
\subsubsection*{The emergent Heterotic string limit}
\label{sss:IIA-heteroticlimit}		
Finally, we discuss the so-called $K3$-limits introduced in ref. \cite{Lee:2019wij}, which only exist at large volume (due to large $\alpha'$-corrections) and are characterized by the fact that the three-fold presents an asymptotic fibration structure of the form
\begin{equation}\label{eq:K3fibration}
			\begin{aligned}
				\rho: \qquad K3 \hookrightarrow &\;X_{3} \\
				&\;\; \downarrow \qquad , \\ &\;\mathbb{P}^1
			\end{aligned}
\end{equation}
where again it is the fibre the one shrinking the fastest in rescaled coordinates, upon taking the singular limit. In MHS language, this set-up corresponds to the Type II singularities discussed in \cite{Corvilain:2018lgw}.
		
The crucial difference here with respect to the previous case is that the leading tower of states now comes from a \emph{critical} emergent Heterotic string \cite{Lee:2019wij}. This string arises from a NS5-brane wrapping the $K3$ fibre, whose effective world-sheet theory precisely contains the spectrum associated to a would-be Heterotic string compactification on $\widehat{K3} \times \mathbf{T}^2$ (or a free quotient thereof)\cite{Harvey:1995rn}. Therefore, upon assuming this limit to correspond in general to an emergent Heterotic string limit (c.f. eq. \eqref{eq:IIA/HETduality4d}) we can run our computations for critical string limits described in Section \ref{sss:emergenceU(1)} (more precisely around eq. \eqref{eq:oneloopstringtowergauge}) in order to reproduce the behaviour of the gauge kinetic matrix \eqref{eq:gaugekineticmatrix} within the present case. In particular, using the general formula for the emergent gauge coupling \eqref{eq:1-formemergencestringytowersgeneral} with the mass scale for the string oscillator modes in the case at hand, which reads $m_{\rm h}\sim \Mpf/t^{1/2}$, we obtain
\beq
	\delta \left(\frac {1}{g^2}\right)\, \sim\, \frac {\Mpf^4}{m_{\rm h}^2} \, \sim \, \Mpf^2\, t \, .
\eeq 
As expected, this gives rise to the correct asymptotic dependence --- for vanishing axion v.e.v.s, see e.g., \cite{Grimm:2018ohb}. 

\subsubsection{Emergence in the hypermultiplet sector}
\label{ss:hypermultiplet4d}
		
Let us now turn our attention to the other (independent) sector of the moduli space of 4d $\mathcal{N}=2$ EFTs coming from Type II Calabi--Yau three-fold compactifications. In particular, our aim here will be to highlight some interesting points about the consistency of the Emergence Proposal with a ubiquitous phenomenon pertaining to these theories, namely that of instanton corrections (see also \cite{Hamada:2021yxy} for complementary discussions about this point).
		
Due to $\mathcal{N}=2$ supersymmetry, the moduli space of the 4d theory factorizes at the two-derivative level (c.f. eq. \eqref{productmoduli}) between the vector multiplet sector $\mathcal{M}_{\rm VM}$, that is a projective special K\"ahler manifold, and the hypermultiplet sector ${\cal M}_{\rm HM}$, which is a quaternionic K\"ahler manifold instead. This can be found in Section \ref{sss:4dN=2basics}, so that we refer the reader interested in the details to that section. 
One of the features that makes the hypermultiplet moduli space qualitatively different from its vector counterpart is that it receives a plethora of perturbative and non-perturbative quantum corrections. This is so because it includes in e.g., the Type IIA case, both $\varphi_4$ and the complex structure moduli $\{z^i\}$, which control the action of euclidean D2-branes (i.e. 4d instantons) wrapping minimal 3-cycles in the internal geometry. Therefore, the associated effective 4d action can be heavily quantum-corrected depending on where we sit in moduli space, as summarized in Appendix \ref{ap:hypermetric}. In particular, it was explained in \cite{Marchesano:2019ifh,Baume:2019sry} that a would-be infinite distance point which is present at tree-level in the moduli space geometry can get obstructed or excised from the \emph{exact} $\mathcal{N}=2$ quantum moduli space precisely due to infinitely many instanton corrections that become relevant along the limit.
		
%
\begin{figure}[t]
		\begin{center}
			\subfigure[]{
				\includegraphics[height=4.5cm]{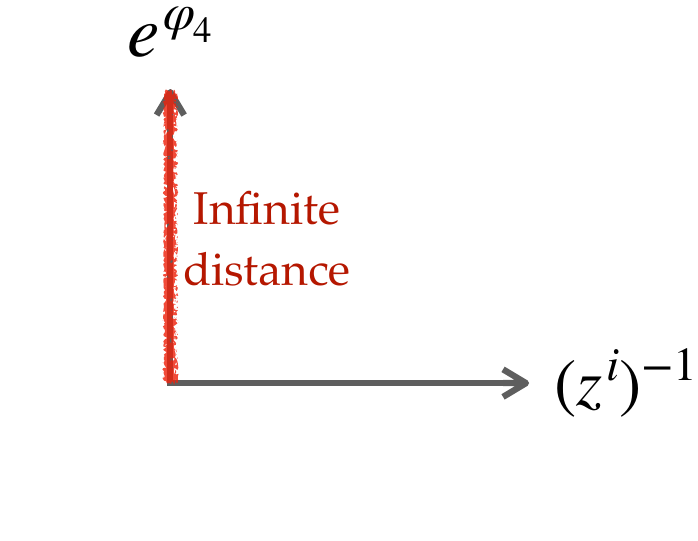} 
				\label{fig:Instantoncorrections1}
			}\qquad \qquad\qquad
			\subfigure[]{
				\includegraphics[height=4.5cm]{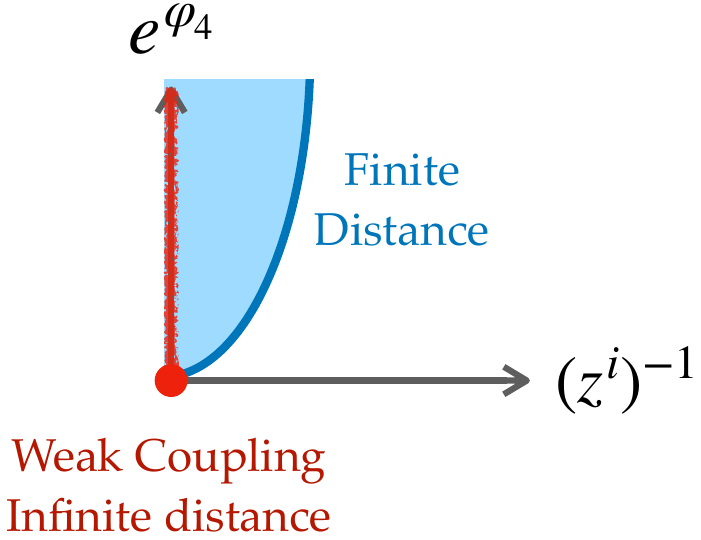}
				\label{fig:Instantoncorrections2}
			}
			\caption{\textbf{(a)} Classical infinite distance limit as $z^i \, \to \,\i \infty$. \textbf{(b)} Including the instanton corrections that become relevant along the limit actually obstructs the would-be classical singularity and leaves only the weak coupling point as an infinite distance boundary.}			
		\label{fig:Instantoncorrections}
		\end{center}
\end{figure} 
Let us briefly review one simple instance in which such obstruction takes place, and try to understand it from the prism of Emergence. We will concentrate on the infinite distance singularity in the mirror symmetric analogue of the D1-limit of \cite{Baume:2019sry}, which was recently investigated in \cite{Alvarez-Garcia:2021pxo}. Hence, consider a geodesic trajectory moving only along the Type IIA hypermultiplet moduli space. The first such trajectory that may come to mind is the one that reaches the large complex structure (LCS) point of the mirror dual Type IIB vector multiplet space, embedded via the c-map within the Type IIA side (we thus keep the 4d dilaton fixed). In the mirror Type IIB compactification, the aforementioned singularity can be understood from Emergence by integrating out D3-branes wrapping some combination of supersymmetric 3-cycles \cite{Grimm:2018ohb, Grimm:2018cpv}. However, in the Type IIA frame, there is no such D3-brane particle states whose mass spectrum is controlled by the complex structure moduli, so that it seems difficult to find an infinite tower which could give rise by quantum corrections to the singular behaviour of the tree-level 4d effective action. 
		
At this point, one could be tempted to conclude that the Emergence Proposal (and even the Distance Conjecture) does not seem to hold in the present case. This is where the analysis of \cite{Baume:2019sry,Alvarez-Garcia:2021pxo} comes to our rescue. It can be seen that, even though we do not have the D3-particles here, there are indeed solitonic strings arising from wrapped D4-branes on non-trivial 3-cycles, together with euclidean D2-instantons, whose tension is controlled precisely by the complex structure moduli.\footnote{As an aside, both strings and instantons happen to be the c-duals of the same D3-particles in the Type IIB side, hence explaining why their tension and action present the same moduli dependence.} The upshot is that along the limit at hand, the action of infinitely many euclidean D2-instantons decreases asymptotically (c.f. eq. \eqref{eq:strongorrections}), such that the tree-level geometry is heavily corrected and, in particular, the infinite distance point at fixed 4d dilaton is excised from the exact quantum moduli space, as schematically displayed in Figure \ref{fig:Instantoncorrections}. Crucially, this is precisely what we would have expected from the Emergence perspective, since we could not find any tower that could generate the tree-level singularity from the beginning. Thus, we can reconcile the absence of a tower with the fact that the infinite distance singularity is actually not present in the exact moduli space.
		
Interestingly enough, even though the LCS singularity at fixed 4d dilaton does not belong to ${\cal M}_{\rm HM}$, one can see that it is actually traded for another (very different) infinite distance degeneration, since the instanton corrections force the dilaton coordinate itself to run towards weak coupling if we insist into approaching the LCS point. The upshot, as explained in \cite{Alvarez-Garcia:2021pxo}, is that an emergent S-dual Type IIA string arising from a D4-brane wrapping the certain shrinking 3-cycle becomes asymptotically tensionless (c.f. eq. \eqref{eq:D4SYZSdual}), and in particular it does so faster than any other critical string in the theory, including the original fundamental one we started with. Therefore, at the end of the day one reaches a new kind of emergent string limit \cite{Lee:2019wij}, which can be reproduced by integrating out the oscillator modes associated to the D4-string --- along with its KK-replicas, in agreement with our expectations coming from Emergence.

\subsubsection*{Instanton corrections $\longleftrightarrow$ Emergence loop corrections}
		
In the previous discussion we have used the Emergence Proposal to predict, in some sense, the absence of an infinite distance singularity based on the lack of any infinite tower of particles which could give rise to the latter after being integrated out. To do so, we built upon several results from previous works so as to interpret the would-be infinite distance singularity to be precluded by the presence of an infinite tower of instanton corrections. Our aim in the following is to turn this logic around and use such corrections (which are quantum in nature) to provide more evidence for the picture advocated in the Emergence Proposal.
		
The strategy is to employ the c-map \cite{Cecotti:1988qn,Ferrara:1989ik} so as to relate the D2-instanton corrections in e.g., the Type IIA hypermultiplet sector to one-loop contributions from the infinite tower of D3-particles, in its type IIB vector multiplet counterpart. For concreteness, we skip the technical details of such construction and focus on summarizing the logic behind it. Concretely, the c-map can be understood as some \emph{dressed} T-duality (see Section \ref{ss:dualitieswithhighersusy}) which relates the vector multiplet moduli space of Type IIA on $X_3$ with the hypermultiplet space of Type IIB on the same three-fold, and the vice versa. (This requires from a further $\mathbf{S}^1$-compactification of both theories, and then performing the usual T-duality transformation along said 1-cycle.)
		
What is interesting about the c-map in this context is that it matches quantum contributions to the partition function of the theory as coming from D-instanton sums in one side of the duality, to one-loop corrections associated to integrating out a tower of D-particles on the other. In particular, one can match the instanton number in one theory with the winding number of the particle worldline along the extra $\mathbf{S}^1$ on the dual theory, being the latter moreover related upon Poisson resummation with the Kaluza-Klein modes along the circle. 
This would naturally explain why starting from the Type IIA hypermultiplet sector, where we find that quantum corrections associated to a tower of euclidean D2-instantons smooth out the moduli space around a given point in ${\cal M}_{\rm HM}^{\rm IIA}$, one should analogously obtain an infinite number of D3-particles that resolve the singularity in ${\cal M}_{\rm VM}^{\rm IIB}$, as required by the Emergence Proposal.
		
To close this section, let us mention that this is nothing but a generalization to infinite distance of a well-understood phenomenon in the finite distance case. Namely, as shown in \cite{Ooguri:1996me}, by using the c-map one can understand the resolution of the conifold singularity within Type II string theory either by invoking the appearance of massless BH-like states at that point in the vector multiplet moduli space (see \cite{Strominger:1995cz}), or by a smoothing-out procedure due to a \emph{single} D-instanton in the hypermultiplet moduli space of the c-dual theory.

\subsection{Emergence in 6d $\mathcal{N}=(1,0)$ theories}
\label{ss:emergence6d}
		
The purpose of this section is to test the general analysis performed in Section \ref{sss:emergenceU(1)} concerning the wave-function renormalization induced by a tower of string oscillator modes on some $\mathsf{U(1)}$ gauge field under which they are charged. To do so, we consider certain finite volume infinite distance limits in F-theory compactifications on an elliptically fibered three-fold \cite{Vafa:1996xn,Morrison:1996na,Morrison:1996pp}, as studied originally in \cite{Lee:2018urn}, where some gauge coupling tends to zero as well. There, it was shown that any such limit is indeed equi-dimensional (c.f. footnote \ref{fnote:equidimensional}), such that the light spectrum of the theory is dominated by a critical Heterotic string arising from a D3-brane wrapping some vanishing holomorphic 2-cycle. Moreover, (part of) the tower of excitation states associated to this fundamental dual string is the one satisfying both the Distance and the sub-Lattice Weak Gravity conjectures (see Section \ref{s:SwamplandProgram} for details), thus providing the relevant modes for Emergence.  

Let us summarize first the main results of \cite{Lee:2018urn}, which will be thus necessary so as to study the Emergence phenomenon in this scenario. On the one hand, recall that the 6d Planck scale is fixed by the internal geometry in F-theory, in particular by the volume\footnote{In this section we measure all dimensionful quantities in Type IIB 10d Planck units.} of the complex K\"ahler surface $B_2$ which serves as the base for the elliptically-fibered CY three-fold, namely
\beq
\label{eq:Planckmass6d}
	M_{\text{Pl};\, 6}^4 = 4 \pi \mathcal{V}_{B_2}\, \Mpt^4\, .
\eeq
On the other hand, the relevant $\mathsf{U(1)}$ field for us arises from 7-branes wrapping certain complex curves $\mathcal{C}$ within $B_2$ (i.e. the discriminant locus $\Delta$ of the elliptic fibration), whose volume sets the corresponding six-dimensional gauge coupling as follows
\beq
\label{eq:gaugecoupling6d}
	\frac{1}{g_{\text{IR}}^2} = 2 \pi \mathcal{V}_{\mathcal{C}}\, \Mpt^4\, .
\eeq
Therefore, as shown in \cite{Lee:2018urn, Lee:2018spm}, the weak coupling limit for such $\mathsf{U(1)}$ gauge field and with gravity kept dynamical --- which is attained when we blow up the curve wrapped by the 7-brane while keeping the overall volume of the complex surface $B_2$ fixed and finite, require that $B_2$ contains some other rational curve (of vanishing self-intersection), $\mathcal{C}_0$, which moreover intersects $\mathcal{C}$ and whose volume goes to zero as $\mathcal{V}_{\mathcal{C}} \to \infty$. Consequently, a D3-brane wrapping such rational curve gives rise to a solitonic string in the six non-compact dimensions, whose tension is controlled by 
\beq
\label{eq:stringtension6d}
	T_{\rm D3,\, str} = 2 \pi \mathcal{V}_{\mathcal{C}_0}\ \Mpt^2\, \sim\, 2 \pi \frac{\mathcal{V}_{B_2}}{\mathcal{V}_{\mathcal{C}}}\ \Mpt^2 \, ,
\eeq
and which can be shown to correspond to a weakly coupled dual Heterotic string (compactified on some $K3$ surface), thus having excitation states charged under the 7-brane gauge theory due to the fact that $\mathcal{C}_0$ and $\mathcal{C}$ non-trivially intersect.
		
Our aim here will be to relate the functional form of the gauge coupling in eq. \eqref{eq:gaugecoupling6d} to the appearance of the tower of asymptotically massless charged string states, in the spirit of the Emergence Proposal. In order to do this, we need two basic ingredients: the 6d species scale for this set-up, together with its relation to the characteristic data of the tensionless Heterotic string; and second, the one-loop renormalization of the inverse gauge coupling as well as the spectrum of oscillator modes which are charged under the $\mathsf{U(1)}$ gauge field.
		
To address the first part of the relevant information, we follow the field-theoretic approach introduced in Section \ref{sss:stringtowersspecies}. In particular, evaluating eqs. \eqref{eq:speciescale}-\eqref{eq:Ns} for $d=6$ leads to an expression of the form
\beq
		\left ( \frac{M_{\text{Pl};\, 6}}{m_{\text{h}}}\right )^4 =\, \mathsf{N}_{\text{h}}^2 \sum_{n=1}^{\mathsf{N}_{\text{h}}} \text{exp}(\sqrt{n}) \, \sim 2\ \mathsf{N}_{\text{h}}^{5/2} e^{\sqrt{\mathsf{N}_{\text{h}}}}\, ,
\label{eq:maxstringlevel6d}
\eeq
where the quantities with the subscript `h' are associated to the emergent critical Heterotic string. In particular, since it is obtained from the wrapped D3-brane discussed above, we have $m_{\text{h}}^2\, =\, T_{\rm D3,\, str}/2 \pi$. Recall also that for such a string tower $\LSP^2 \, \simeq \, \mathsf{N}_{\text{h}} \, m_{\text{h}}^2$.
		
Additionally, the one-loop gauge coupling renormalization at leading order reads as follows
\beq
		\label{eq:runninggaugecoupling6d}
		\frac{1}{g_{\text{IR}}^2} = \frac{1}{g_{\text{UV}}^2} + \beta \Mpt^2 \sum_{k=1}^{k_{\text{max}}} q_k^2\, \LSP^2\, ,
\eeq
where $\beta$ is a positive numerical prefactor that depends on the specific type of state considered (e.g., whether it is a fermion or a boson, its tensor structure, etc.), and the sum runs over every oscillator state (labelled by the collective index $k$) charged under the $\mathsf{U(1)}$ with quantized charge $q_k$ that appears below the species scale.\footnote{Notice from eq. \eqref{eq:runninggaugecoupling6d} that we have extracted a factor of $\Mpt$ from the gauge charges associated to the oscillator modes of the wrapped D3-brane, so as to have the $q_k$ dimensionless. One can see that this is indeed the correct normalization from e.g., anomaly inflow in the worldsheet of the resulting EFT string\cite{Callan:1984sa,Heidenreich:2021yda}.}  In order to see how the above computation reproduces the expected divergence in \eqref{eq:gaugecoupling6d} upon sending $\mathcal{V}_{\mathcal{C}} \to \infty$, we need a way to parameterize what gauge charges appear for each oscillator level $n$, which in turn controls the mass of the states by the usual linear relation $m_n^2 = 8\pi T_{\rm D3,\, str} (n-1) \,$, as well as their degeneracy. Hence, we borrow some useful results from  \cite{Lee:2018urn}. First, one can estimate the highest charge  appearing at each excitation level $n$ by $q_{\text{max}}(n)= \sqrt{n}$, as frequently happens for Heterotic compactifications.\footnote{\label{fn:modularity}This is essentially fixed by the modular properties of the Heterotic string partition function on the worldsheet torus, see \cite{Lee:2018urn} and references therein.} Secondly, we notice that in concrete examples, such as e.g., F-theory on an elliptic three-fold with base given by a Hirzebruch surface, $B_2=\mathbb{F}_1$, at a given $n$ each value of the charges with $|q| < q_{\text{max}}(n)$ is indeed populated by string states. Taking this kind of behaviour as representative, we need to parameterize the degeneracy $d_{k,n}$ of charged states with charge $k \in \mathbb{Z}$ within each oscillator level $n$. In principle, one would need to extract the precise dependence by studying physical quantities such as the partition function of the tensionless string. Here, however, we take a (possibly oversimplified) parameterization of the form
\beq
\label{eq:dmn}
	d_{k,n} \sim f(k)\, e^{\sqrt{n}} \ ,
\eeq
with $f(k)$ being any polynomial function with the only restriction that summing over all charges up to $q_{\text{max}}(n)$ within a given step in the tower reproduces the level density of states, i.e. $d_n \sim e^{\sqrt{n}}$. For instance, a simple example for $f(k)$ would be a monomial of the form $f(k) \sim k^p/n^{(p+1)/2}$, with $p \in \mathbb{R}_+$. With these ingredients at hand, we can then perform the summation implicit in eq. \eqref{eq:runninggaugecoupling6d}
\beq
\begin{aligned}
			\frac{1}{g_{\text{IR}}^2} & \sim \,  \Mpt^2 \,  \LSP^2 \sum_{n=1}^{\mathsf{N}_{\text{h}}} \sum_{k=0}^{\sqrt{n}} k^2 d_{k,n}\, \sim\, \Mpt^2 \,  T_{\rm D3,\, str} \,  \mathsf{N}_{\text{h}} \sum_{n=1}^{\mathsf{N}_{\text{h}}} n\ \text{exp} (\sqrt{n})\\
			&\sim\, \Mpt^2 \, T_{\rm D3,\, str} \, \mathsf{N}_{\text{h}}^{5/2} \text{exp} (\sqrt{\mathsf{N}_{\text{h}}})\, .
		\end{aligned}
\eeq
Next, upon substituting eqs. \eqref{eq:maxstringlevel6d}-\eqref{eq:stringtension6d}, together with $m_{\text{h}}^2\, =\, T_{\rm D3,\, str}/2\pi$, and using also the definition of the 6d Planck mass given in \eqref{eq:Planckmass6d}, we find that the exact same dependence for the $\mathsf{U(1)}$ gauge coupling as in eq. \eqref{eq:gaugecoupling6d} is recovered. 
		
\subsection{Emergence in 7d $\mathcal{N}=1$ theories}
\label{ss:emergence7dN=1}
		
As our final example, we consider M-theory compactified on a $K3$ surface. The resulting low energy effective field theory (see Section \ref{sss:MtheoryonK3} for a detailed discussion) preserves minimal supersymmetry in seven dimensions, and its bosonic sector is described by the action \eqref{eq:7dMthyK3}. We moreover restrict ourselves to the attractive $K3$ case, since the analysis is enormously simplified and it already serves to prove our point. In that scenario, one can argue that all integral curve classes in $H_2(K3,\mathbb{Z})$ admit holomorphic representatives (see discussion around eq. \eqref{eq:7dMthyattractive}), and therefore they can all give rise to distinguished BPS states from wrapped M2-branes. 

For concreteness, let us reproduce here the scalar-vector-tensor sector of the 7d action, which reads\footnote{\label{fnote:Weylrescaling7d}In order to arrive at the Einstein-framed action \eqref{eq:7dMthyK3scalarvectortensor} after compactifying M-theory on $K3$ one needs to perform the Weyl rescaling $g_{\mu \nu} \to \mathcal{V}_{K3}^{-2/5} g_{\mu \nu}$.}
\begin{equation}\label{eq:7dMthyK3scalarvectortensor}
	\begin{aligned}
		S_\text{M-th}^{\text{7d}}\, \supset\, & \frac{1}{2\kappa^2_7} \int \dd^{7}x\, \sqrt{-g}\,  \left( \mathcal{R} - \frac{9}{20} \left(\frac{\partial \mathcal{V}_{K3}}{\mathcal{V}_{K3}} \right)^2 + \mathsf{G}_{a b}\, \partial \tilde{t}^a \cdot \partial \tilde{t}^b\right) -\frac{1}{4\kappa_{7}^2} \int \mathfrak{g}_{a b}F^a \wedge F^b\, ,
	\end{aligned}
\end{equation}
where we recall that $J=t^a \omega_a$ denotes the K\"ahler 2-form and $\tilde{t}^a =t^a/\mathcal{V}_{K3}^{1/2}$ are rescaled K\"ahler coordinates. In the following, we will measure every dimensionful quantity in Planck units, keeping in mind that the 7d Planck scale simply reads $M_{\text{Pl};\, 7}^5= \frac{4 \pi}{\ell^5_{7}}$ (see Appendix \ref{ap:conventions} for conventions).

On the other hand, the gauge kinetic matrix $\mathfrak{g}_{a b}$ depends on the K\"ahler moduli as follows		
\begin{align}\label{eq:matrix1forms7dMtheory}
			\mathfrak{g}_{a b}= \frac{1}{\mathcal{V}_{K3}^{3/5}} \mathsf{G}_{a b}= \frac{1}{{\mathcal{V}_{K3}^{8/5}}}t_a t_b- \frac{1}{\mathcal{V}_{K3}^{3/5}}\eta_{a b}\, ,    
\end{align}
with $\eta_{a b}$ being the intersection form of the corresponding two-fold.

\subsubsection*{F-theory limit}
		
Let us consider first the corner of the 7d moduli space that corresponds to the F-theory limit, where the theory effectively decompactifies to 8d $\mathcal{N}=1$ supergravity. We closely follow the analysis performed in ref. \cite{Lee:2019xtm}, where such limit was carefully studied.

Therefore, suppose that we move in K\"ahler moduli space while keeping the overall $K3$ volume constant, so as to maintain 7d gravity dynamical. The crucial point is that in order to have such an infinite distance, weak coupling point, one of the entries in $\mathfrak{g}_{a b}$ has to blow-up. Indeed, as demonstrated in \cite{Lee:2019xtm}, this requires that the K\"ahler form must behave asymptotically as follows
\begin{align}\label{eq:KahlerFtheorylimit}
			J= t^0 \omega_0 + \sum_i \frac{a^i}{2t^0} \omega_i\, , \qquad t^0 \to \infty\, ,    
\end{align}
where $\{\omega_0, \omega_i\}$ are generators of the K\"ahler cone, $\{a^i\}$ are some constant numerical factors and $t^0$ is the K\"ahler modulus that scales to infinity in the limit. Additionally, the finite volume requirement imposes the following restrictions on the generators $\{\omega_0, \omega_i\}$ \cite{Lee:2019xtm}
\begin{align}
			\omega_0 \cdot \omega_0 = 0\, , \qquad \sum_i \frac{a_i}{2} \omega_i \cdot \omega_0= \mathcal{V}_{K3} + \mathcal{O} (1/(t^0)^2)\, .    
\end{align}
Therefore, upon taking as a basis for $H^{1,1}(K3)$ precisely these 2-forms, $\lbrace \omega_a \rbrace = \lbrace \omega_0, \omega_i \rbrace$, one can see that the vector which becomes weakly coupled when we take $t^0 \to \infty$ is precisely $A^0$, whose kinetic term behaves as follows
\begin{align}\label{eq:gaugecoupling7d}
			\mathfrak{g}_{0 0} = \frac{(t^0)^2}{\mathcal{V}_{K3}^{8/5}} - \frac{1}{\mathcal{V}_{K3}^{3/5}}\mathcal{C}_0 \cdot \mathcal{C}_0\, .    
\end{align}
where the second term is independent of $t^0$. Moreover, it is easy to see that neither $\mathfrak{g}_{i j}$ nor $\mathfrak{g}_{0 i}$ blow up along the limit \eqref{eq:KahlerFtheorylimit}, and hence become negligible when compared to $\mathfrak{g}_{0 0}$. Similarly, since the kinetic terms for the scalar fields $\{ \tilde{t}^a\}$ behave (up to a constant volume prefactor) precisely as the gauge kinetic functions $\mathfrak{g}_{a b}$ (c.f. eq. \eqref{eq:matrix1forms7dMtheory}), one concludes that it is the massless mode parametrized by $\tilde{t}^0$ the one associated to the infinite distance singularity.

Note that in eq. \eqref{eq:gaugecoupling7d} we have introduced a curve class $\mathcal{C}_0$ which belongs the set $\lbrace \mathcal{C}_0, \mathcal{C}_i \rbrace$ dual to $\lbrace \omega_0, \omega_i \rbrace = \lbrace \mathcal{C}^0, \mathcal{C}^i \rbrace$, namely it satisfies
\begin{align}
			\mathcal{C}_0 \cdot \mathcal{C}^0 = 1\, , \qquad \mathcal{C}_0 \cdot \mathcal{C}^i = 0\, , \qquad \mathcal{C}_i \cdot \mathcal{C}^0 = 0\, , \qquad \mathcal{C}_i \cdot \mathcal{C}^j = \delta_i^j\, .    
\end{align}
Furthermore, it can be shown that the divisor class defined by $\mathcal{C}^0=\omega_0$ contains necessarily a genus-one holomorphic curve whose volume tends to zero along the limit $t^0 \to \infty$ at a rate $\mathcal{V}_{\mathcal{C}^0} = \mathcal{V}_{K3}/t^0 + \mathcal{O} (1/(t^0)^3)$. This means, in particular, that the $K3$ surface admits an elliptic fibration --- over a $\mathbb{P}^1$-base --- of the form depicted in eq. \eqref{eq:fibration}.
		
Additionally, one obtains a tower of asymptotically light BPS particles arising from wrapped M2-branes on the shrinking curve $\mathcal{C}^0$. In fact, due to its $\mathbf{T}^2$-topology, there are indeed bound states of $n$ M2-particles for each $n \in \mathbb{Z}$, such that we obtain a tower of asymptotically light particles with constant degeneracy at each mass level in the spectrum (see \cite{Katz:1999xq} for a more rigorous analysis of these matters). These constitute precisely the Kaluza-Klein replica that effectively implement the circle decompactification along the F-theory limit. Their mass and charge can be computed from the dimensional reduction of the Nambu-Gotto plus Chern-Simons action associated to the M2-brane, yielding (c.f. eq. \eqref{eq:massM2branes})
\begin{align}\label{eq:M2particlemassafterWeylrescaling}
			m_{\text{M2}} = \frac{2 \pi}{\ell_{7}} \mathcal{V}_{\mathcal{C}^0}\, \mathcal{V}_{K3}^{-1/5} \sim \frac{2 \pi}{\ell_{7}} \frac{\mathcal{V}_{K3}^{4/5}}{t^0}\, , \qquad \text{as}\ \ t^0 \to \infty\, .    
\end{align}
With this, we can now use our general formulae for the one-loop contribution to both the gauge and scalar metrics in the presence of a tower of charged light particles. Thus, starting from eqs. \eqref{eq:KKloopscalarmetricd>6} and \eqref{eq:gaugeemergenceddimensions} and specializing to the seven-dimensional case at hand, we obtain
\beq\label{eq:oneloopgaugecoupling7d}
 \begin{aligned}
     \mathsf{G}_{00} &\sim M_{\text{Pl}; 7}^5 \left(\partial_{\tilde{t}^0} \log m_{\text{M2}} \right)^2 \sim M_{\text{Pl}; 7}^5\, \frac{(t^0)^2}{\mathcal{V}_{K3}}\, ,\\
     \frac{1}{g_{\text{IR}}^2} &= \mathfrak{g}_{0 0} \sim M_{\text{Pl}; 7}^5 \left( \frac{2 \pi}{\ell_{7}\, m_{\text{M2}}} \right)^2 \sim M_{\text{Pl}; 7}^5\, \frac{(t^0)^2}{\mathcal{V}_{K3}^{8/5}}\, .
 \end{aligned}
\eeq
Above we have used the fact that the M2-particle charges are given by $q_n\, = \,  \frac{2 \pi n}{\ell_{7}}$, as well as the definition of the species scale \eqref{species}. Thus, the infinite distance/weak coupling singularity in the F-theory limit \eqref{eq:KahlerFtheorylimit} can in principle be reproduced via the Emergence mechanism.

\subsubsection*{Emergent string limit}

For completeness, let us also analyze another interesting infinite distance limit that is exhibited by the present theory. It corresponds to the \emph{small radius} point, where the overall $K3$ volume goes to zero size while keeping the set $\{ \tilde{t}^a\}$ fixed and non-degenerate. Indeed, one can argue that this singularity actually corresponds to an emergent string limit \cite{Lee:2019wij}, where an asymptotically tensionless and weakly coupled Heterotic string emerges at infinite distance (see Section \ref{ss:dualitieswithlowersusy} for more on this duality). Such string can be constructed by wrapping the M5-brane on the whole $K3$ surface \cite{Cherkis:1997bx,Park_2009}, with a tension in 7d Planck units which reads as
\begin{equation}
	T_{\text{M5,\, str}} = \frac{2 \pi}{\ell_{7}^2}\mathcal{V}_{K3}^{3/5}\, .
\end{equation}
It is moreover $\frac12$-BPS and thus couples to some 2-form gauge field $B_2$ which is dual to the massless 3-form potential in \eqref{eq:7dMthyK3}. In fact, upon dualizing the latter following the usual lagrange multiplier procedure, one arrives at an action of the form
\begin{equation}\label{eq:7dMthyK3dual}
\begin{aligned}
	S_\text{M-th}^{\text{7d}}\, =\, & \frac{1}{2\kappa^2_7} \int \dd^{7}x\, \sqrt{-g}\,  \left( \mathcal{R} - \frac{9}{20} \left( \partial \log \mathcal{V}_{K3} \right)^2 - G_{i j} \partial \phi^i \cdot \partial \phi^j\right)\\
       &  -\frac{1}{4\kappa_{7}^2} \int \mathcal{V}_{K3}^{-6/5}\, H_3 \wedge \star H_3 + \mathcal{V}_{K3}^{-3/5} \mathsf{G}_{a b} F^a \wedge F^b\, ,
\end{aligned}
\end{equation}
where $H_3$ is the field strength of the 2-form to which the M5-string couples, which is given by
\begin{equation}\label{eq:H3fieldstrength}
	H_3= dB_2- \frac{1}{3}\eta_{a b} A^a \wedge F^b\, .
\end{equation}
Notice that the Chern-Simons term has disappeared in eq. \eqref{eq:7dMthyK3dual} above. This is indeed a usual phenomenom, where upon dualization Chern-Simons couplings get exchanged with transgression terms in the $p$-form field strengths. In the present case, this transgression term shows up in the Bianchi identity of the $B_2$-field, namely
\begin{equation}
	dH_3= - \frac{1}{3}\eta_{a b} F^a \wedge F^b\, .
\end{equation}
Therefore, performing an exactly analogous analysis as in Section \ref{ss:emergence6d}, one finds an asymptotic behaviour of the form
\beq
 \begin{aligned}
     G_{\mathcal{V}_{K3} \mathcal{V}_{K3}} &\sim M_{\text{Pl}; 7}^5 \left(\frac12 \partial_{\mathcal{V}_{K3}} \log T_{\text{M5,\, str}} \right)^2 \sim M_{\text{Pl}; 7}^5\, \frac{1}{\mathcal{V}_{K3}^2}\, ,\\
     \mathfrak{g}_{a b} &\sim \frac{M_{\text{Pl}; 7}^7}{T_{\rm M5}} \mathsf{G}_{a b} \sim \frac{M_{\text{Pl}; 7}^7}{\mathcal{V}_{K3}^{3/5}} \mathsf{G}_{a b}\, .
 \end{aligned}
\eeq
for the one-loop contribution to the scalar kinetic term and inverse gauge coupling, respectively.

\section{Emergence of higher-dimensional terms}
\label{s:emergenceinteractions}

Up to now we have only considered the generation of kinetic terms in quantum gravitational field theories. However, as we already know, the non-renormalizability of gravity (c.f. Section \ref{ss:basics}) requires necessarily from the presence of higher-dimensional and higher-derivative operators in the effective action. Therefore, a natural question at this point concerns whether those can also be accounted for in the framework of Emergence.

The message we want to convey in this section is that this may well be also the case, for various reasons (see also \cite{Lee:2021qkx} for a complementary discussion). First, as it is familiar from our quantum field theory experience, loop corrections involving massive states generate non-local contributions to the Wilsonian effective action, which can be usually expanded as a power series of local operators of increasing number in derivatives and dimension. Moreover, as argued in Section \ref{ss:classicalfromquantum}, the contribution due an infinite tower of massive modes can yield seemingly classical terms in the effective action once we cut off the sum at the species scale $\LSP$. Hence, a priori tree-level corrections for higher-dimensional operators in string theory could be accounted for upon integrating out the relevant towers of states (see e.g., \cite{Blumenhagen:2024ydy} for a recent analysis regarding the $\mathcal{R}^4$--\,term in maximally supersymmetric theories with $7 \leq d < 11$). Secondly, even operators within the two-derivative lagrangian not corresponding to kinetic terms can be reproduced via Emergence, in principle. Indeed, a particularly interesting instance is that of scalar potentials, which play a key role in determining the possible vacua of the theory. Crucially, it turns out that the generation of such terms can be reformulated as the question of whether non-propagating $(d-1)$-forms (in $d$ spacetime dimensions) \cite{Herraez:2018vae,Bielleman:2015ina,Carta:2016ynn} can get emergent kinetic terms. In this regard, one can argue that e.g., the saxionic dependence of the flux potential in Type IIA Calabi--Yau compactifications can emerge upon integrating out the relevant D$p$-brane states becoming light along the limit of interest (see Section 6 of \cite{Castellano:2022bvr} for details on this procedure).

Here, we choose to focus on higher-derivative operators instead, since their analysis within the Emergence point of view already serves to illustrate various important points.

\subsubsection*{Example: the $\mathcal{R}^2$--\,operator in 4d $\mathcal{N}=2$ theories}


Let us consider the BPS operator involving two curvature tensors in the 4d efective action obtained from Type IIA on a CY threefold $X_3$. As discussed in Section \ref{s:4dN=2}, this term reads as follows
\beq
\label{eq:R2termEmergence}
	S_{\text{IIA}}^{\text{4d}} \supset \int \dd^4x\, \sqrt{-g}\, \mathcal{F}_1(X^A)\, \mathcal{R}_+^2\ +\ \text{h.c.}\, ,
\eeq
where $\mathcal{F}_1$ can be identified with the topological free energy at genus one (see Section \ref{ss:threshold4d} for a list of its main properties). For concreteness, we focus here on partial decompactification limits to 6d, where the three-fold exhibits some elliptic fibration of the form shown in \eqref{eq:ellipticcaseEmergence}. Our aim will be to reproduce via certain one-loop computation the relevant functional dependence displayed by $\mathcal{F}_1$ in eq. \eqref{eq:R2termEmergence} above along this set of limits, including subleading (i.e. non-divergent) contributions as well.

Therefore, we recall that in this scenario one finds two infinite towers of asymptotically light states, namely D0 bound states and D2-branes wrapping the elliptic fibre, whose charges run arbitrarily over the integers. Hence, the one-loop contribution to the operator $\mathcal{R}^2$ due to the aforementioned particles is given by (c.f. eq. \eqref{eq:generatingseries})
\begin{align}
\label{eq:F1ell}
	\mathcal{F}^{\rm ell}_{1} (z,\bar z)&= -\frac{1}{12} \sum_{(\omega, n) \in \mathbb{Z}^2} \int_{\varepsilon}^{\infty} \frac{\dd\tau}{\tau}\, e^{-\tau m_{\rm D0}^2\left[ (\omega t)^2+(n+ \omega b)^2\right]}\, ,
\end{align}
where $z=b+ \i t$ is the K\"ahler modulus associated to the elliptic fibre. Note that we have included in eq. \eqref{eq:F1ell} the contribution due to the zero modes as well, namely those states with $n=\omega=0$ corresponding to the massless particles in the 4d EFT. 
Moreover, the exponent in the one-loop integral above can be written in a manifestly quadratic form on the integers $\mathbf{n}=(n, \omega)$ as follows
\begin{align}
	m_{\rm D0}^2 \left[(\omega t)^2+(n+ \omega b)^2\right] = G^{ij}n_i n_j\, , 
\end{align}
where $G$ is a $2\times 2$ matrix with entries
\begin{align}
	 G= \frac{1}{(m_{\rm D0}\, t)^2} \left( \begin{array}{cc}
				|z|^2 & -b  \\
				-b & 1  \\
			\end{array} \right)\, , \qquad G^{-1}= m_{\rm D0}^2
 			\left( \begin{array}{cc}
				1 & b  \\
				b & |z|^2  \\
			\end{array} \right)\, ,
\end{align}
whilst $G^{-1}$ denotes its inverse. This allows us to rewrite \eqref{eq:F1ell} as
\begin{align}
	\mathcal{F}^{\rm ell}_{1}&= -\frac{1}{12} \sum_{\mathbf{n}\in \mathbb{Z}^2} \int_{\varepsilon}^{\infty} \frac{\dd\tau}{\tau}\, e^{-\tau G^{ij}n_i n_j} = -\frac{\pi}{12} \frac{1}{\sqrt{\det G^{-1}}} \sum_{\mathbf{k}\in \mathbb{Z}^2} \int_{\varepsilon}^{\infty} \frac{\dd\tau}{\tau^2}\, e^{-\frac{\pi^2}{\tau} G_{ij}k^i k^j}\, ,
\end{align}
where in the second step we have performed a Poisson resummation on the integer-valued vector $\mathbf{n}$ (c.f. footnote \ref{fnote:Poissonresummation}). Next, upon changing the integration variable to $\hat \tau=\tau^{-1}$, we find
\begin{align}
	\mathcal{F}^{\rm ell}_{1}&= -\frac{\pi}{12} \frac{1}{m_{\rm D0}^2\, t} \sum_{\mathbf{k}\in \mathbb{Z}^2} \int_0^{\varepsilon^{-1}} \dd \hat \tau\, e^{-\hat \tau \frac{\pi^2}{(m_{\rm D0}\, t)^2} \left| k^1 -k^2 z\right|^2}\, ,
\end{align}
where we used that
\begin{align}
	G_{ij}k^i k^j= \frac{1}{(m_{\rm D0}\, t)^2} \left[(k^1-b k^2)^2+(k^2 t)^2\right] = \frac{1}{(m_{\rm D0}\, t)^2} \left| k^1 -k^2 z\right|^2\, .
\end{align}
Let us first study the UV-divergent term, which is associated to zero-winding number, i.e. $\mathbf{k}=0$. It provides a contribution to $\mathcal{F}^{\rm ell}_{1}$ of the form
\begin{align}\label{eq:numberspeciesEmergenceelliptic}
	\mathcal{F}^{\rm ell}_{1}&\, \supset\, -\frac{\pi}{12} \frac{1}{m_{\rm D0}^2\, t} \varepsilon^{-1} = -\frac{\pi}{12} \frac{\Lambda_{\rm sp}}{m_{\rm D0}} \frac{\Lambda_{\rm sp}}{m_{\rm D0}\, t} \sim N_{\rm D0} N_{\rm D2} \sim N_{\rm sp}\, ,
\end{align}
where we have substituted $\varepsilon=\LSP^{-2}$ in the second step, and the final result depends on the total number of light species in the theory, as expected from the usual Emergence computations.\footnote{The authors in \cite{vandeHeisteeg:2022btw} identify $\mathcal{F}_1$ with the number of light species $N$ over the entire moduli space, giving various motivations for this. Note that this is precisely suggested by the Emergence computation here performed, c.f. eq. \eqref{eq:numberspeciesEmergenceelliptic}.} The second piece leads to some UV-convergent contribution to $\mathcal{F}_1$, which reads as
\begin{align}
	\mathcal{F}^{\rm ell}_{1}& \supset -\frac{1}{12 \pi} \sideset{}{'}\sum_{\mathbf{k} \in \mathbb{Z}^2} \frac{t}{\left| k^1+k^2 z\right|^{2}}\, ,
\end{align}
that is moreover modular invariant. Indeed, it can be seen to correspond to the non-holomorphic Eisenstein series of order one (see Appendix \ref{s:SL2Waveforms} for details), and it moreover presents a logarithmic divergence. In this context, though, such divergence is associated to an infra-red effect arising from the inclusion of the massless modes within the sum \eqref{eq:F1ell}. Hence, upon regularization, and requiring modular invariance to be preserved, one finally obtains
\begin{align}
	\mathcal{F}^{\rm ell}_{1} (z,\bar z)& = -\frac{\pi}{12} \frac{\Lambda_{\rm sp}}{m_{\rm D0}} \frac{\Lambda_{\rm sp}}{m_{\rm D0}\, t} + \frac{1}{12} \log \left( t \left| \eta(z)\right|^4\right)\, ,
\end{align}
which is close to the exact result, see e.g., \cite{Bershadsky:1993ta, vandeHeisteeg:2023dlw}.

\section{Summary}

In the present chapter we have taken the first steps toward a systematic analysis of the idea of Emergence in quantum gravity, as discussed at the beginning of Section \ref{s:EmergenceQG}. A crucial ingredient within this proposal deals with the maximum regime of validity of any gravitational effective field theory, which is captured by the quantum gravity cut-off $\LQG$. This was identified in Chapter \ref{ch:SpeciesIntro} with the species scale $\LSP$, and serves as the physical energy where loop computations must be cut off in gravity. In order to test this, we studied several string theory constructions in diverse dimensions and with different amounts of supersymmetry, thus providing strong evidence for the overall picture advocated in this thesis. 
 
One-loop self-energy computations play an important role in any practical calculation of Emergence, and we addressed them in Section \ref{s:EmergenceQG}. In particular, we analyzed the general case in $d$-dimensions of moduli fields, gauge bosons and (chiral) fermions. As already mentioned, the crucial ingredient so as to get emergent kinetic terms is to use the species scale as the physical cut-off, as well as to sum over all states belonging to the relevant tower(s) becoming light. Remarkably, even though each individual contribution to the kinetic functions gives a quantum loop effect, the final resummed metric turns out to be independent of $\hbar$, due to this very special choice of ultra-violet cut-off. This suggests that perhaps it may not make sense to actually distinguish between classical and quantum gravity, given that the very concept of what is \emph{classical} and what is \emph{quantum-mechanical} is not duality invariant (see e.g., \cite{Vafa:1995fj,Strominger:1995qi}). Furthermore, with this analysis we obtain a number of interesting results. First, we show explicitly that in simple examples one can reproduce the correct leading-order dependence for the known non-linear sigma model metrics of several string theory constructions, in agreement with predictions by the Distance and Weak Gravity conjectures, both in the presence of either Kaluza-Klein or string towers. (In the latter case an approximate field theory inspired approach was employed to treat the higher-spin modes, which should be properly modified using manifestly off-shell string theory techniques, such as e.g., string field theory.) Secondly, we also find that fermion fields can indeed get large contributions to their wave-function renormalization, a fact that may be important for the application of quantum gravity ideas to phenomenology, since most of the Standard Model particles are actually described by spin-$\frac{1}{2}$ fields \cite{Castellano:2023qhp}.  
 
Therefore, we have tested the Emergence Proposal in various selected examples that display different features both in terms of the number of dimensions where the EFT lives, the amount of supersymmetry preserved and the structure of relevant towers appearing along the infinite distance limits. In particular, we have revisited 4d $\mathcal{N}=2$ theories arising from Type IIA Calabi--Yau three-fold compactifications, performing a detailed analysis of the Emergence mechanism in this set-up. In addition, we have studied higher-dimensional examples, including the case of 6d  F-theory compactifications in which the kinetic term for the gauge bosons arise from a solitonic tensionless Heterotic string, as well as 7d theories from M-theory on (attractive) $K3$ surfaces. All the top-down constructions checked so far seem to be naively consistent with Emergence, at least at the level of properly generating the leading (divergent) field dependence, thus suggesting that generic infinite distance singularities may arise non-trivially in quantum gravity as an intrinsic infra-red phenomenon, as originally proposed in \cite{Palti:2019pca,Harlow:2015lma,Grimm:2018ohb,Corvilain:2018lgw,Heidenreich:2017sim,Heidenreich:2018kpg}.

Furthermore, in Section \ref{s:emergenceinteractions} we briefly considered the possibility of generating terms in the effective lagrangian other than the kinetic ones, including scalar potentials and higher-dimensional/derivative operators. The former were discussed in \cite{Castellano:2022bvr} within the context of Type IIA CY compactifications with fluxes, and we refer the reader interested in the details to that reference. The latter also exhibit interesting features that are currently under investigation, see e.g., \cite{Blumenhagen:2023xmk}. To illustrate this point we revisited the behaviour exhibited by the $\mathcal{R}^2$--\,operator in 4d $\mathcal{N}=2$ theories along partial decompactification limits, finding perfect agreement with exact results already present in the literature. 
 
Many important questions remain open before claiming success, so let us finally comment on some of these. First, it would be very interesting to study in more detail how Emergence applies in general for mixed limits where the resolution of the singularity involves a nested chain of decompactifications and/or emergent string limits. Partial progress along this direction has been accomplished recently in \cite{Blumenhagen:2023yws}, where several relevant observations and issues were raised. More generally, understanding how this idea could be realized in AdS spacetimes --- upon using the AdS/CFT correspondence \cite{Maldacena:1997re}, also seems particularly interesting. Moreover, assuming Emergence to pass every test one may think of, the important question arises of what is the fundamental origin of this property as well as its role in the context of the Swampland program. Indeed, the fact that various conjectures lying at the core of the program could be explained via this simple mechanism would point towards some sort of unifying picture in which Emergence, together with the concept of the species scale, could play a major role within the latter. On the other hand, if the strong formulation of Emergence were to be true, it would be necessary to get some insights about the underlying theory of quantum gravity as well as more serious tests in favour of this rather counter-intuitive phenomenon.\footnote{See \cite{Blumenhagen:2023tev,Blumenhagen:2023xmk, Hattab:2023moj,Hattab:2024thi} for recent progress along this direction.}


\part{\scshape  \textcolor{MyDarkRed}{Universal constraints on the Species Scale}} 
\label{part:pattern}

\chapter{Bounds on Asymptotic Decay Rates}\label{ch:bounds} In Chapter \ref{ch:SpeciesIntro} we explained the central role of the species cut-off $\LSP$ as the energy scale controlling the maximum regime of validity of any effective field theory weakly coupled to Einstein gravity. Therefore, an interesting avenue to infer generic properties of quantum gravity, as seen from the low energy realm, involves systematically studying the behaviour of this energy scale across the landscape of consistent EFTs. In addition, this strategy would perhaps even allow us to obtain non-trivial information about yet unknown consistency conditions that must be generically satisfied in quantum gravity, and formulate them in the form of new universal constraints. To do so, a good starting point would be to analyze the behaviour of the QG cut-off close to infinite distance boundaries of moduli space, where many Swampland criteria have been already proposed and thoroughly tested (see Section \ref{s:SwamplandProgram} for details). In particular, the connection between the Swampland program and the species scale becomes very apparent when focusing on those consistency conditions which are formulated as continuous statements, such as the (sub-)Lattice/tower Weak Gravity Conjecture \cite{Heidenreich:2015nta,Heidenreich:2016aqi,Montero:2016tif,Andriolo:2018lvp} or the Distance Conjecture \cite{Ooguri:2006in, Etheredge:2022opl}. Indeed, these criteria usually deal with extreme regimes in the parameter (or moduli) space of low energy EFTs, where certain gauge couplings are taken to be close to zero value or rather we allow for enormous vacuum expectation values in the scalar fields; and the way in which these are censored is via the appearance of an infinite number of light states. This latter fact implies, in turn, that there should be a significant decrease of the species cut-off $\LSP$, such that the regime of validity of our starting effective description shrinks to the point in which it actually becomes completely invalidated (i.e. when sitting precisely at the infinite distance boundary).

The purpose of the present chapter is to revisit these considerations so as to propose, based several bottom-up arguments, a very sharp and universal \emph{lower bound} on the exponential decay rate, $\lambda_{\rm sp}$, of the quantum gravity cut-off (see Section \ref{s:convexhull} below for a precise definition of $\lambda_{\rm sp}$). More concretely, such constraint would read as follows
\begin{equation}\label{eq:lowerboundspecies}
  \lambda_{\rm sp} \geq \frac{1}{\sqrt{(d-1)(d-2)}}\, ,
\end{equation}
where $d$ denotes the spacetime dimension of our theory, and it should hold when venturing towards the boundaries of moduli space, regardless of the particular infinite distance limit that we choose to sample. We will thus first motivate the existence of the lower bound \eqref{eq:lowerboundspecies} in Section \ref{s:convexhull}, using our experience with string theory to guide us.\footnote{See also \cite{Calderon-Infante:2023ler} for complementary bottom-up arguments.} In practice, it may become highly involved to check the proposed bound, especially in the presence of several scalar fields, where the number of infinite distance singularities and relevant towers of states can easily proliferate. To sidestep this difficulty, we can reformulate the constraint \eqref{eq:lowerboundspecies} in terms of a convex hull condition, which is considerably easier to analyze. Later on, in Section \ref{s:consistencydimreduc}, we will study the stability (or consistency) of the lower bound under dimensional reduction. This analysis will moreover highlight the role of the saturating value $\lambda_{\rm sp,\, min} = \frac{1}{\sqrt{(d-1)(d-2)}}$, which actually arises in explicit examples such as $\mathbf{S}^1$--\,compactifications, as being the only one which is both preserved and consistent under the compactification process. Finally, we present top-down evidence for the bound \eqref{eq:lowerboundspecies} in Section \ref{s:examplesbound}, restricting to the case of maximal supergravity in $d\geq4$. Further non-trivial evidence for the latter in set-ups preserving less amount of supersymmetry will be provided in Chapter \ref{ch:pattern} below.

This chapter is based on the publication \cite{Calderon-Infante:2023ler}, which has been adapted to better fit in the broader context of this thesis.

\section{A convex hull condition for the species scale}
\label{s:convexhull}

We consider in the following a $d$-dimensional EFT containing a set of massless scalar fields (i.e. moduli), weakly coupled to Einstein gravity as follows
\begin{equation}\label{eq:action}
	\mathcal{L}_{\text{EFT}} \supset \dfrac{1}{2\kappa_d^2}\,  \sqrt{- g} \left(\mathcal{R} + G_{i j} (\phi)\, \partial \phi^i \cdot \partial \phi^j\right)\, ,
\end{equation}
where $G_{ij}(\phi)$ is the metric tensor associated to the moduli space, $\mathcal{M}_{\phi}$, spanned by the vacuum expectation values (v.e.v.s) of the massless scalars. According to the Distance Conjecture (c.f. Section \ref{s:SDC}), we should have an infinite tower of states becoming exponentially light at every infinite distance boundary within $\mathcal{M}_{\phi}$. Therefore, in terms of the traversed \emph{geodesic} distance, which is defined by
\begin{equation}\label{eq:modspacedist}
	\Delta_{\phi} = \int_{\gamma} \dd\sigma \sqrt{G_{i j} (\phi) \frac{d \phi^i}{d \sigma} \frac{d \phi^j}{d \sigma}}\, ,
\end{equation}
with $\gamma$ denoting some geodesic path and $\sigma$ an affine parameter, there should exist a tower whose mass scale decreases as $m\sim e^{-\lambda \Delta_{\phi}}$ for $\Delta_{\phi} \gg 1$ (in Planck units). Moreover, the asymptotic \emph{decay rate} $\lambda$ should be given by some $\mathcal{O}(1)$ factor.

On the other hand, in the presence of several moduli, it is convenient to define a $\zeta$-vector for every tower becoming light asymptotically, whose components read
\begin{equation}\label{eq:chargetomass}
	\zeta^i := - G^{i j} \frac{\partial}{\partial \phi^j} \log m= -\partial^i \log m\, .
\end{equation}
These are referred to as \emph{scalar charge-to-mass vectors} \cite{Calderon-Infante:2020dhm,Etheredge:2022opl,Etheredge:2023odp},\footnote{The name originates from the Scalar Weak Gravity Conjecture \cite{Palti_2017}, as these vectors measure the strength of the scalar force induced by the moduli in comparison to the gravitational interaction, c.f. eq. \eqref{eq:scalarcouplings}.} and they precisely encode the information about how fast the associated tower of states becomes light (see Appendix \ref{ap:generalities} for details). In particular, for any given asymptotically geodesic trajectory in moduli space characterized by some normalized tangent vector $\hat{T}$, the decay rate of the tower can be determined as the projection
\begin{equation}\label{eq:decayrate}
  \lambda=\vec{\zeta} \cdot \hat{T} = G_{ij}\zeta^i \hat{T}^j\, .
\end{equation}
In practice, however, the metrics $G_{ij}(\phi)$ we have to deal with tend to be rather complicated, which stems from the fact that the moduli space geometry is usually very non-trivial. Therefore, it becomes useful to define some orthonormal frame at each point in $\mathcal{M}_{\phi}$,\footnote{Generally speaking, one can only define a flat frame in a local fashion. However, in some cases, when focusing on the slice of moduli space parametrized by the non-compact scalar fields, it becomes possible to extend such definition globally. This is the case in e.g., maximal supergravity and even certain examples with lower supersymmetry, see Chapter \ref{ch:pattern} for more on this.} in terms of which the $\zeta$-vector components read as follows
\begin{equation} \label{eq:zeta-vectors}
  \zeta^{a} = e^{a}_{i} \zeta^i\, , 
\end{equation}
where $e^a_i(\phi)$ defines some vielbein in field space, thus satisfying $\delta_{ab} e^a_i(\phi) e^b_j(\phi) = G_{ij}(\phi)$. Hence, when using such a (local) flat frame, the inner product in eq. \eqref{eq:decayrate} is simply taken with respect to the Cartesian metric $\delta_{ab}$.
	
Crucially, note that the decay rate $\lambda$ for any given tower strongly depends on the geodesic trajectory taken, and thus it is not an intrinsic property of the tower itself, whereas the $\zeta$-vectors are. Consequently, given the set of all possible towers becoming light, we will denote by $m_{\rm t}$ the one that does so at the fastest rate --- i.e. $\lambda_{\rm t} = \vec{\zeta}_{\rm t} \cdot \hat T$ is the largest exponent.

Regarding the allowed values for $\lambda_{\rm t}$, it is strongly believed that there exists a lower bound for the latter, given by 
\begin{equation}\label{eq:sharpenedDistConj}
  \lambda_{\rm t} \geq \frac{1}{\sqrt{d-2}}\, ,
\end{equation}
which was originally proposed in \cite{Etheredge:2022opl} and thoroughly tested since then in a number of works, see e.g., \cite{Etheredge:2023odp} for a recent non-trivial check in 9d $\mathcal{N}=1$ set-ups arising from Heterotic string theory on $\mathbf{S}^1$. That being said, the evidence for the precise saturating value $\lambda_{\rm t,\, min} = \frac{1}{\sqrt{d-2}}$ is mostly empirical, and comes from the fact that the typical decay rates for the towers arising in string theory, namely Kaluza-Klein or string oscillation modes \cite{Lee:2019wij}, behave in the following way (c.f. eqs. \eqref{eq:KKscale} and \eqref{eq:stringmassdependence})\footnote{\label{fnote:exactpreservation}The particular values for the $\zeta$-vectors exhibited in eq. \eqref{eq:zetaveconemodulus} are special since their functional form is preserved under dimensional reduction. On top of this, the precise lower bound appearing in \eqref{eq:sharpenedDistConj} is selected as being the only one that is \emph{exactly} preserved, in the sense that saturation in $d+1$ dimensions leads to saturation in the lower dimensional theory \cite{Calderon-Infante:2023ler}.}
\beq\label{eq:zetaveconemodulus}
	\zeta_{{\rm KK},\, n} = \sqrt{\frac{d+n-2}{n (d-2)}}\, , \qquad \zeta_{\rm osc}= \frac{1}{\sqrt{d-2}}\, ,
\eeq
where $n$ counts the number of decompactifying dimensions. Hence, since we always consider $d\geq 4$, we deduce that the minimum value for $\lambda_{\rm t}$ is attained in the string case, thus saturating \eqref{eq:sharpenedDistConj}.

Notice that a simple and straightforward procedure in order to check the bound \eqref{eq:sharpenedDistConj} at any given asymptotic regime in moduli space involves plotting the relevant $\zeta$-vectors (once they have been canonically normalized, as per \eqref{eq:zeta-vectors}), and verify whether the \emph{convex hull} determined by the latter contains the extremal ball of radius $\frac{1}{\sqrt{d-2}}$. This boils down to the fact that the asymptotic direction with minimum $\lambda$--\,parameter between two competing towers with different $\zeta$-vectors, precisely coincides with that of the closest point to the origin of the convex hull determined by the latter. Hence, satisfying \eqref{eq:sharpenedDistConj} is equivalent to ask for the convex hull to include the `extremal region'. Note that this statement resembles that of the Convex Hull Distance Conjecture proposed in \cite{Calderon-Infante:2020dhm}, with the crucial difference that the latter assumes this convex hull diagram to be always globally defined, whilst in general it is known to be able to change upon exploring different asymptotic regions \cite{Etheredge:2023odp}.

Let us remark here the importance of knowing exactly which values for $\vec{\zeta}$ are allowed in any consistent theory of quantum gravity. The reason for this is twofold: First, as we will see in more detail below, their norms can be sometimes mapped in a one-to-one fashion with the asymptotic physics encountered at the infinite distance boundary --- i.e. the QG resolution. Second, knowing what are the lowest possible values for $\lambda_{\rm t}$ (if any) may in principle restrict the maximum (geodesic) variation of the scalar fields that can be accommodated within the original EFT, which is computed as
\beq
	\Delta_{\phi}\, \lesssim\, \frac1{\lambda_{\rm t}}\log{\frac{\Mpd}{m_{\rm t}}}\, ,
\eeq
and can be used in turn to place strong bounds of phenomenological interest for cosmological models of inflation or quintessence \cite{Scalisi:2018eaz}, as well as other dynamical proposals to explain the electro-weak hierarchy problem such as cosmological relaxation \cite{Graham_2015}.

Now, given that in this thesis we are most interested in the characterization of the possible behaviours of the quantum gravity cut-off, namely the species scale, a natural question to ask at this point is whether one could formulate an analogous condition for the latter. To do so, we first need to recall how to properly compute such quantity in the presence of several infinite sets of states, since $\LSP$ may receive contributions from towers other than the lightest one. The details of the calculation strongly depend on how the towers relate to each other, i.e. whether they are additive or multiplicative (c.f. Chapter \ref{ch:SpeciesIntro}). We review the latter case in the following both for the ease of reading and since it will be crucial for our purposes in this chapter. 

Let us therefore consider a spectrum of mixed states with quantum numbers $(j, k) \in \mathbb{Z}^2\,$, associated to two different infinite towers. For definiteness, we take their mass dependence to be of the form
\begin{equation} \label{eq:massmultiplicativetowers}
  m_{j,k}^{2} = j^{2/p_1} m_{\text{tow}}^2 + k^{2/p_2} m_{\text{tow}'}^2 \, ,
\end{equation}
with $m_{\text{tow}}\leq m_{\text{tow}'}$ without any loss of generality. A useful way to think of this spectrum is as if it was coming from two distinct \emph{multiplicative} towers with mass scales $\{ m_{\text{tow}},\, m_{\text{tow}'} \}$ and density parameters $p_1$ and $p_2$, respectively. One canonical example would be that of a pair of Kaluza-Klein towers corresponding to two compact internal directions of radius $\{ R_1, R_2\}$, with masses $m_{\text{KK},\, 1}=1/R_1$ and $m_{\text{KK},\, 1'}=1/R_2$. (Henceforth we denote by $m_{\text{KK},\, n}$ the mass scale associated to a KK-like tower with density parameter $p=n$.)

To each of these separate towers one can associate a would-be species scale as follows
\begin{equation}\label{eq:powerlikespecies}
  \Lambda_{\text{tow}}\, \sim\, m_{\text{tow}}^{\frac{p_1}{d-2+p_1}} \Mpd^{\frac{d-2}{d-2+p_1}}\, , \qquad \Lambda_{\text{tow}'}\, \sim\, m_{\text{tow}'}^{\frac{p_2}{d-2+p_2}}\Mpd^{\frac{d-2}{d-2+p_2}}\, ,
\end{equation}
where each of them is computed by accounting just for the subset of states associated to the corresponding tower and ignoring those arising from the remaining one.

On the other hand, in general, one should account for states with mixed quantum numbers $(j,k)$, thus considering the combined effect of the two aforementioned towers. The algorithmic procedure was explained in Section \ref{ss:MultipleTowers}, so that we state here only the final result and refer the reader interested in the details to that section. Upon doing so, one finds
\begin{equation}\label{eq:effspeciesscale}
  \Lambda_{\text{eff}}\, \sim\, m_{\text{eff}}^{\frac{p_{\text{eff}}}{d-2+p_{\text{eff}}}}\Mpd^{\frac{d-2}{d-2+p_{\text{eff}}}}\, ,
\end{equation}
where we have defined (geometric) `averaged' quantities as follows\footnote{Recall that such `effective' towers, together with their averaged mass scale and density parameters, are just book-keeping devices that allow us to easily compute the total number of species and quantum gravity cut-off via e.g., eq. \eqref{eq:effspeciesscale}.}
\begin{equation}\label{eq:masseffectivetower}
  m_{\text{eff}}\, \sim\, \left( m_{\text{tow}}^{p_{1}}\, m_{\text{tow}'}^{p_{2}} \right)^{1/p_{\text{eff}}}\, , \qquad p_{\text{eff}} = p_1 + p_2 \, .
\end{equation}
Furthermore, the main reason why it is useful to divide such computation into two steps is because, depending on the asymptotic direction in moduli space $\hat T$ that we explore, the states associated to either one of the two towers can become arbitrarily lighter than those coming from the second one. Therefore, in certain circumstances it may be enough to consider particle states arising from just one of them in order to compute $\LSP$. However, oftentimes it may still be necessary to consider mixed states of the form \eqref{eq:massmultiplicativetowers} --- even if one of the two scales (say $m_{\text{tow}}$) becomes parametrically lighter than the other, which happens precisely when
\begin{equation}\label{eq:conforeffectivetower}
  m_{\text{tow}'}\, \lesssim\, \Lambda_{\text{tow}} \iff \Lambda_{\text{eff}}\, \lesssim\, \Lambda_{\text{tow}}\, .
\end{equation}
This means, in particular, that the true species cut-off for any definite infinite distance limit is simply given by the \emph{smallest} scale out of the set $\{\Lambda_{\text{tow}},\, \Lambda_{\text{tow}'},\, \Lambda_{\text{eff}}\}$. Moreover, the asymptotic moduli-dependence of $\LSP$ is typically exponential, as per \eqref{eq:powerlikespecies}, namely $\LSP \sim e^{-\lambda_{\rm sp} \Delta_{\phi}}$ with $\lambda_{\rm sp}$ being some $\mathcal{O}(1)$ factor. Hence, in complete analogy with the case of the towers discussed around eq. \eqref{eq:chargetomass}, one may define some \emph{species vectors}, which in a flat orthonormal frame are computed as follows
\begin{equation}\label{eq:defspeciesvectors}
  \mathcal{Z}^{a} = - \delta^{ab} e^{i}_{b} \,  \partial_{i} \log\Lambda_{\text{sp}}\, ,
\end{equation}
and whose projection gives, for every asymptotic direction $\hat T$, the previously defined exponential decay rate (c.f. eq. \eqref{eq:decayrate})
\begin{equation}\label{eq:decayratespecies}
  \lambda_{\text{sp}}=\vec{\mathcal{Z}} \cdot \hat{T}\, .
\end{equation}
With this in mind, we can now ask whether there could be some non-trivial constraints on the possible allowed values for the parameter $\lambda_{\text{sp}}$. More concretely, one would like to know if a precise lower bound can be proposed for the latter, similarly to what the sharpened Distance Conjecture holds, c.f. eq. \eqref{eq:sharpenedDistConj}. The idea that we want to put forward here is that there seems to be an exact analogous statement, and one can actually find such lower bound: 
\begin{equation}\label{eq:lowerboundspecies2}
  \lambda_{\rm sp} \geq \frac{1}{\sqrt{(d-1)(d-2)}}\, ,
\end{equation}
which again should be verified for any asymptotic trajectory exploring infinite distance in field space. Nonetheless, at this point, the main motivation for \eqref{eq:lowerboundspecies2} comes entirely from our string theory experience. Indeed, in the usual infinite distance limits arising in the moduli spaces of QG theories, the species cut-off always corresponds to either some higher-dimensional Planck mass or rather to some fundamental string scale. This yields the following species vectors
\beq\label{eq:speciesveconemodulus}
	\mathcal{Z}_{{\rm KK},\, n} = \sqrt{\frac{n}{(d+n-2) (d-2)}}\, , \qquad \mathcal{Z}_{\rm osc}= \frac{1}{\sqrt{d-2}}\, ,
\eeq
which in turn verify the condition \eqref{eq:lowerboundspecies2} and saturate the latter in the particular case of a decompactification of one extra dimension (i.e. when $n=1$). 

Several comments are in order. First, note that the very existence of such a constraint implies, when taken seriously, that indeed the asymptotic field-space behaviour of both the quantum gravity cut-off and the number of species (equivalently the minimal black hole entropy, see Chapter \ref{ch:SpeciesIntro}) must be \emph{at least} exponential. Therefore, if combined with some additional upper bound for the parameter $\lambda_{\rm sp}$ --- as recently argued in \cite{vandeHeisteeg:2023ubh} based on EFT arguments and semi-classical black hole considerations --- the condition \eqref{eq:lowerboundspecies2} forces the asymptotic behaviour of the species scale to be \emph{exactly} exponential. Second, in analogy with the convex hull condition (CHC) for the $\zeta$-vectors associated to the towers, one can reformulate the lower bound \eqref{eq:lowerboundspecies2} as the following equivalent statement

\begin{center}
	\textbf{Convex Hull Condition for the Species Scale}: \textit{The convex hull of species vectors $\{ \vec{\mathcal{Z}}\}$ defined at infinity should contain the ball of radius $\lambda_{\text{sp}, \, \text{min}}= \dfrac{1}{\sqrt{(d-1)(d-2)}}$.} 
\end{center}

Finally, let us also mention that there exists a very simple and practical algorithm which translates the convex hull for the towers into that associated to the species vectors. Indeed, starting with the $\zeta$-vectors defined for each, say multiplicative, tower in the theory, $\{\vec{\zeta}_{\alpha}\}$ with $\alpha=1, \ldots, N$, as well as their density parameters $\{p_{\alpha}\}$, one may compute the species vector of any effective combination thereof as follows
\begin{equation} \label{eq:eff-vector}
    \vec{\mathcal{Z}}_{\text{eff},\, p_{\rm eff}} = \frac{1}{d-2+ \sum_\alpha p_{\alpha}} \, \sum_{\alpha=1}^{N} p_{\alpha} \, \vec{\zeta}_{\alpha}\, ,
\end{equation}
where we use a similar notation than that of the $\zeta$-vectors and indicate the density parameter of the effective tower with a subscript $p_{\rm eff}$. In addition, one can analogously define the scalar charge-to-mass vector associated to the effective tower introduced in eq. \eqref{eq:masseffectivetower} above by 
\begin{equation}
    \vec{\zeta}_{\text{eff},\, p_{\rm eff}} = \frac{1}{\sum_\alpha p_{\alpha}} \, \sum_{\alpha} p_{\alpha} \, \vec{\zeta}_{\alpha}\, ,
\end{equation}
which allows us to rewrite \eqref{eq:eff-vector} simply as $\vec{\mathcal{Z}}_{\text{eff},\, p_{\rm eff}}= \frac{p_{\text{eff}}}{d-2+p_{\text{eff}}}\, \vec{\zeta}_{\text{eff},\, p_{\rm eff}}$.

In the upcoming sections we will both present explicit examples of convex hulls constructed out of the $\mathcal{Z}$-vectors computed in certain quantum gravity theories, as well as analyze the consistency of the CHC under dimensional reduction. Moreover, we will remark at various points how important it is to include the effective towers into the analysis since, without some of them, the convex hull would not capture the relevant underlying physics, and the condition \eqref{eq:lowerboundspecies2} would be moreover immediately violated. 

\section{Consistency of the bound under dimensional reduction}
\label{s:consistencydimreduc}

Before testing the lower bound for the species scale decay rate proposed in \eqref{eq:lowerboundspecies2} we would like to investigate whether such a requirement can be consistently formulated in the first place, which amounts to showing that indeed it is preserved (or at least not immediately violated) under the remormalization group (RG) flow. This makes sense, since any bona-fide Swampland condition should be regarded as an IR constraint and thus must take into account all possible deformations of the theory due to quantum effects. Furthermore, given that in quantum gravity dynamical processes involving topology change are to some extent allowed (see e.g., \cite{McNamara:2019rup} and references therein), one is also prompted to check how the bound \eqref{eq:lowerboundspecies2} behaves when considering different compactification backgrounds. In fact, this latter strategy has proven to be a very fruitful avenue in the past \cite{Heidenreich:2015nta,Rudelius:2021oaz,Etheredge:2022opl}, leading to some important insights as well as allowing us to improve or `sharpen' certain Swampland criteria upon imposing their consistency under dimensional reduction.

In what follows we will thus analyze the behaviour of the convex hull condition just proposed during the compactification process. Accordingly, in Section \ref{ss:field-theory} we show that the constraint \eqref{eq:lowerboundspecies2} is well-defined and indeed stable under dimensional reduction, at least within the realm of effective field theory. In fact, we find that it is the \emph{strongest} bound on the exponential rate $\lambda_{\text{sp}}$ yet compatible with this procedure, when seen from the bottom-up perspective. To do so, we restrict ourselves to those asymptotic directions in moduli space where a purely field-theoretical approach suffices to determine the relevant spectrum of towers, namely without the need to include the presence of additional extended objects. Subsequently, in Section \ref{ss:compactificationstring} we go beyond field theory, also allowing for the possibility of having lower-codimension objects such as strings (together with their winding modes, etc.). Crucially, using a simple yet instructive toy model, we find that for theories living in less than ten dimensions, the CHC generically requires from the existence of extra non-perturbative states for it to be satisfied after the compactification process.

\subsection{Field theoretic considerations} 
\label{ss:field-theory}

Let us start by studying how eq. \eqref{eq:lowerboundspecies2} behaves under generic RG flows. First, notice that the bound becomes strictly \emph{stronger} when reducing the number of non-compact dimensions $d$ of our theory. This is in contrast to what happens for instance in the case of the WGC (c.f. Section \ref{s:WGC}), which naively becomes monotonically weaker in the infrared, even if we allow for the radion modes to be stabilized by e.g., quantum effects \cite{Heidenreich:2015nta}. Therefore, if we imagine starting with some $d$-dimensional EFT which satisfies the CHC and then compactify the theory on a circle whose physical radius is dynamically fixed, then the same species cut-offs already existing in $d$ dimensions will no longer verify \eqref{eq:lowerboundspecies2}. As a consequence, one would thus be tempted to conclude that the bound is automatically violated. Notice that the exact same complaint could be raised for the analogous lower bound in the Distance Conjecture parameter, c.f. eq. \eqref{eq:sharpenedDistConj}, which also becomes monotonically stronger when reducing the number of dimensions (as long as $d>2$). 

The resolution to this puzzle involves realizing that, in fact, one should \emph{(i)} either still consider log-derivatives with respect to the radion field, thus adding some extra component to the $\zeta$- and $\mathcal{Z}$-vectors which precisely compensates for the increase in the value of $\lambda_{\rm t,\, min}$ (correspondingly $\lambda_{\rm sp,\, min}$); or \emph{(ii)} rather the new decompactifying directions 
are essentially obstructed by the stabilizing potential, which tells us that the two set-ups are no longer dynamically related to each other and thus it does not make sense to retrieve information from one using the other setting, and viceversa. Note that the former scenario would apply to those cases in which, despite the radion being in general massive, there is a sense in which the large radius limit can be faithfully taken within the EFT --- perhaps by tuning some discrete flux, and hence it is still meaningful to compute the variation of the masses of the towers as the physical radius changes. In the following, we will discuss in detail how, upon taking into account the additional contribution due to the radion field, the bound \eqref{eq:lowerboundspecies2} turns out to be non-trivially satisfied along all asymptotic directions which can be described within field theory. Along the way, we will argue that the saturation value, namely $\frac{1}{\sqrt{(d-1)(d-2)}}$, is selected as special by the dimensional reduction process.

\subsubsection{Testing the CHC in $\mathbf{S}^1$--\,compactifications}

We consider in what follows some effective field theory in $D=d+1$ dimensions with a single canonically normalized field that we denote by $\hat \phi$ in the following. Note that, despite the one-modulus simplification, one could interpret this as parametrizing \emph{any} geodesic trajectory within a multi-moduli theory without loss of generality. Our starting point will be to assume that the bound \eqref{eq:lowerboundspecies2} is satisfied along this trajectory, so that we need to introduce some tower of states with density parameter $p$ and exponential rate $\lambda_{\text{t}}$, thus verifying (c.f. eq. \eqref{eq:effspeciesscale})
\begin{equation}
    \lambda_{\text{sp}} = \frac{p}{D-2+p} \lambda_{\text{t}} \geq \frac{1}{\sqrt{(D-1)(D-2)}} \, .
\end{equation}
The strategy here will consist in dimensionally reducing this theory on a circle in order to see under what conditions the CHC is still fulfilled in $d$-dimensions. Furthermore, to be as general as possible, we consider the pair $\{\lambda_{\text{t}},p\}$ (equivalently $\{\lambda_{\text{sp}},p\}$) as independent and free parameters.\footnote{Note that, even though these two parameters do not depend a priori on each other, in explicit string theory examples they always end up being correlated. We elaborate more on this in Chapter \ref{ch:pattern}, where a universal pattern at infinite distance is introduced and thoroughly discussed. This condition implies that the values of $|\vec{\zeta}|$ and $p$ are directly connected, at least in simple set-ups.} Notice that this assumption also fits well with the interpretation of $\hat \phi$ as encoding any possible geodesic trajectory, since depending on the latter, the same tower may present different exponential rates (see discussion after eq. \eqref{eq:zeta-vectors}). Upon doing so, we end up with a $d$-dimensional EFT featuring two extra ingredients: a new modulus field as well as an additional tower. These correspond to the (canonically normalized) radion $\hat \sigma$ and the KK tower. The relevant scalar charge-to-mass vectors in $d$ dimensions read (see Appendix \ref{ap:generalities})
\begin{equation}\label{eq:zvectorafterdimreduction}
	\vec{\zeta}_{\text{KK}} = \left( 0 \ , \ \sqrt{\frac{d-1}{d-2}} \right) \, , \quad \vec{\zeta}_{\text{t}} = \left( \lambda_{\text{t}} \ ,\ \frac{1}{\sqrt{(d-1)(d-2)}} \right) \, .
\end{equation}
Per the relation \eqref{eq:eff-vector}, we can easily translate these tower vectors into their species analogues
\begin{equation} \label{vectors}
\begin{split} 
	&\vec{\mathcal{Z}}_{\text{KK}} = \left( 0 \ ,\ \frac{1}{\sqrt{(d-1)(d-2)}} \right) \, ,\\
	&\vec{\mathcal{Z}}_{\text{t}} = \left( \frac{d-1+p}{d-2+p} \ \lambda_\text{sp} \ ,\ \frac{p}{(d-2+p)\sqrt{(d-1)(d-2)}} \right) \, ,\\
	&\vec{\mathcal{Z}}_{\text{KK-t},\, p+1} = \frac{1}{d-1+p} \left( \vec{\zeta}_{\text{KK}} + p\, \vec{\zeta}_{\text{t}} \right) = \left( \lambda_\text{sp} \ ,\ \frac{1}{\sqrt{(d-1)(d-2)}} \right) \, ,
\end{split}
\end{equation}
where we have used that the KK tower of the $\mathbf{S}^1$ has density parameter equal to one. In addition, we also expressed everything in terms of $\lambda_{\text{sp}}$ instead of $\lambda_{\text{t}}$, since the former is the relevant parameter for us.

With this information, we can now test the CHC in the resulting $d$-dimensional theory. The first thing to note is that the presence of the aforementioned towers is not enough so as to verify the conjecture for every geodesic trajectory in moduli space. This is, nonetheless, not as bad as it sounds, since even the Distance Conjecture itself typically requires from elements beyond field theory to be satisfied after compactification on a circle (e.g., in the limit of small radius). Crucially, the latter `predicts' the existence of extended objects already in the parent theory which can wrap along the internal $\mathbf{S}^1$, such as fundamental strings together with their winding modes. Consequently, in this section we focus on those asymptotic directions in which the $\mathcal{Z}$-vectors \eqref{vectors} are enough to build the relevant convex hull, thus remaining agnostic about new extra ingredients that would be needed if exploring other possible infinite distance limits (see Section \ref{ss:compactificationstring} for more on this). An example of this restriction as well as the portion of the convex hull generated by the species vectors \eqref{vectors} is shown in Figure \ref{fig:ch-example} below. There we see that the boundary of the polytope is given by two edges joining $\vec{\mathcal{Z}}_{\text{KK}}$ with $\vec{\mathcal{Z}}_{\text{KK-t},\, p+1}$ and $\vec{\mathcal{Z}}_{\text{KK-t},\, p+1}$ with $\vec{\mathcal{Z}}_{\text{t}}$, respectively. In the following, we discuss separately the implications of each of these two lines for the CHC.

\begin{figure}[htb]
\begin{center}
\includegraphics[width=0.45\textwidth]{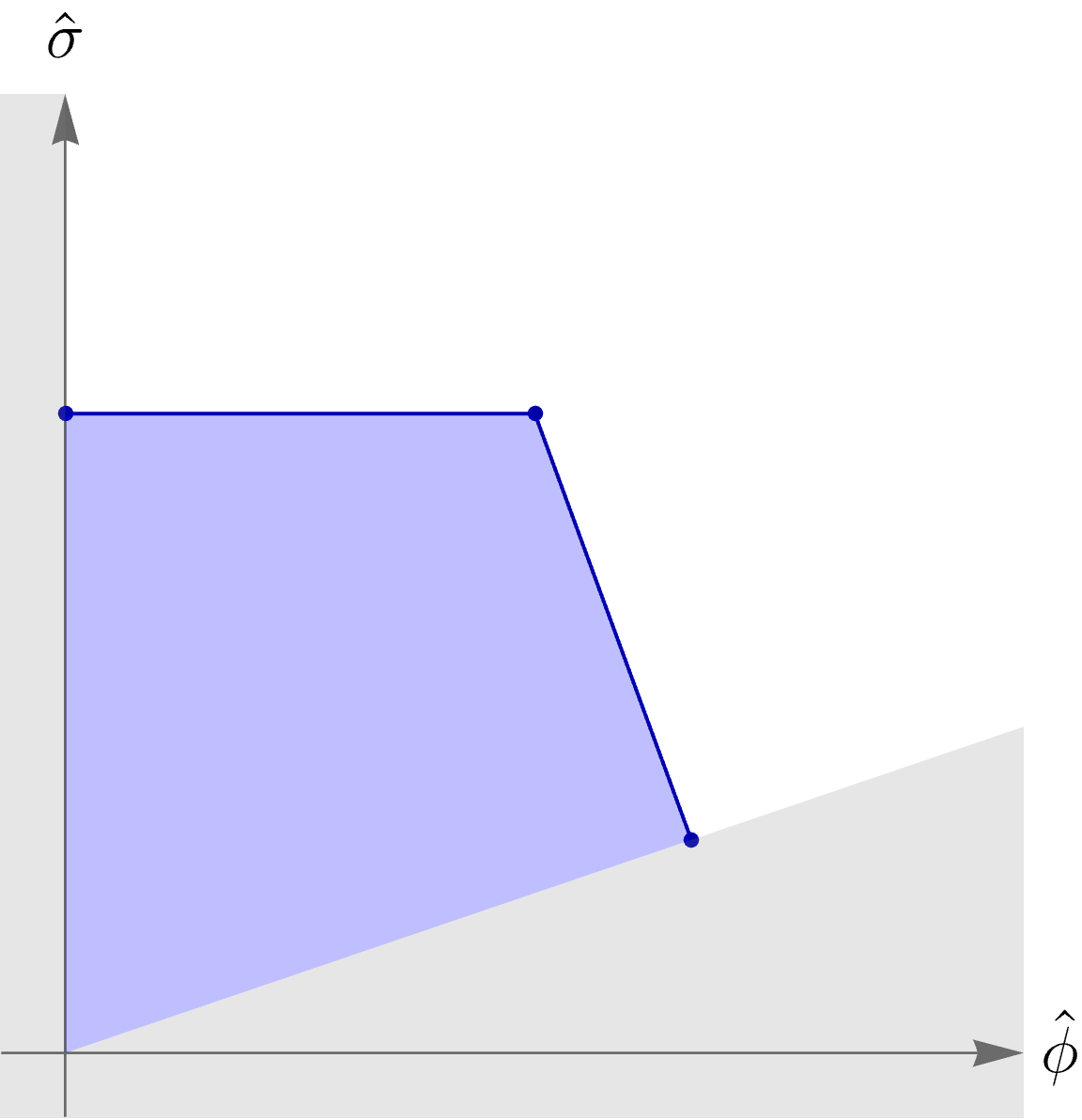}
\caption{\small Example of the convex hull generated by the triplet of vectors in \eqref{vectors}, restricted to the region in which the towers of states are enough to build it. In the region shaded in grey, new towers potentially arising from extended objects in $D$-dimensions are expected to become relevant and complete the rest of the diagram.} 
\label{fig:ch-example}
\end{center}
\end{figure}

Let us start with the first boundary, namely the edge joining $\vec{\mathcal{Z}}_{\text{KK}}$ with $\vec{\mathcal{Z}}_{\text{KK-t},\, p+1}$. As one can clearly see, this line is horizontal for any value of $\lambda_{\text{sp}}$ and $p$. Furthermore, the distance from this edge to the origin is always given by the saturating value
\begin{equation}\label{eq:lspmin}
    \lambda_{\text{sp}, \, \text{min}}\, =\, \frac {1}{\sqrt{(d-1)(d-2)}} \, ,
\end{equation}
again irrespective of the values of $\lambda_{\text{sp}}$ and $p$. Therefore, we conclude that \eqref{eq:lowerboundspecies2} is indeed the \emph{strongest} possible bound yet compatible with dimensional reduction on a circle (see also the discussion after eq. \eqref{eq:Zspeciescompactification}). The fact that the KK tower associated to some internal $\mathbf{S}^1$ typically yields this particular value was pointed out as well in \cite{vandeHeisteeg:2023uxj} as a bottom-up argument for the latter. Moreover, their results regarding the global behaviour of the species cut-off in 4d $\mathcal{N}=2$ theories suggest that \eqref{eq:lowerboundspecies2} could be extended slightly beyond the strictly asymptotic regime, due to the positivity of the generic leading-order logarithmic corrections found therein. Here we also uncover that, if the $(d+1)$-dimensional theory has some non-trivial moduli space, it is crucial to take into account the effective towers for the CHC to be satisfied in the lower-dimensional EFT. In particular, the saturation featured by the vector $\vec{\mathcal{Z}}_{\text{KK}}$ towards the $\hat \sigma \to \infty$ limit would actually present a pressing problem for the bound to be satisfied along other neighbouring directions, unless some additional species vectors were present. These must be, in addition, tightly constrained, since otherwise the resulting convex hull would not contain the ball of radius $\frac {1}{\sqrt{(d-1)(d-2)}}$. Fortunately, we are able to find a vector --- namely $\vec{\mathcal{Z}}_{\text{KK-t},\, p+1}$ --- that precisely accomplishes this, regardless of the details of the tower already existing in $D$-dimensions. Let us also stress that, since we interpret $\hat \phi$ as parametrizing any geodesic trajectory in the moduli space of the parent $D$-dimensional theory, the above conclusion is in fact not limited to theories with just one single field. 

\begin{figure}[htb]
\begin{center}
	\subfigure[\label{sfig:4dp=0.2}$d=4$ and $p=0.2$]{\includegraphics[width=0.32\textwidth]{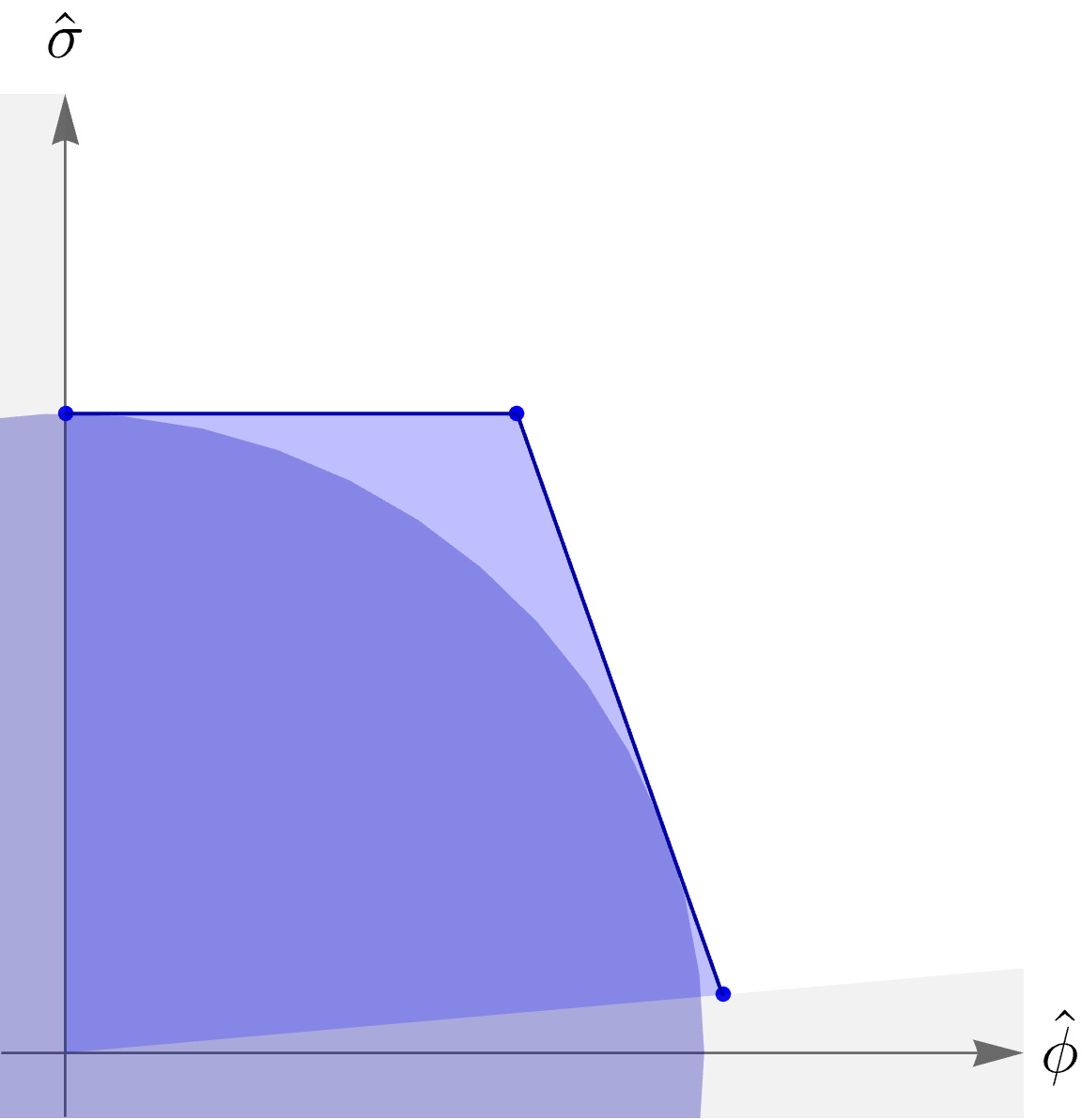}}
	\subfigure[\label{sfig:4dp=1}$d=4$ and $p=1$]{\includegraphics[width=0.32\textwidth]{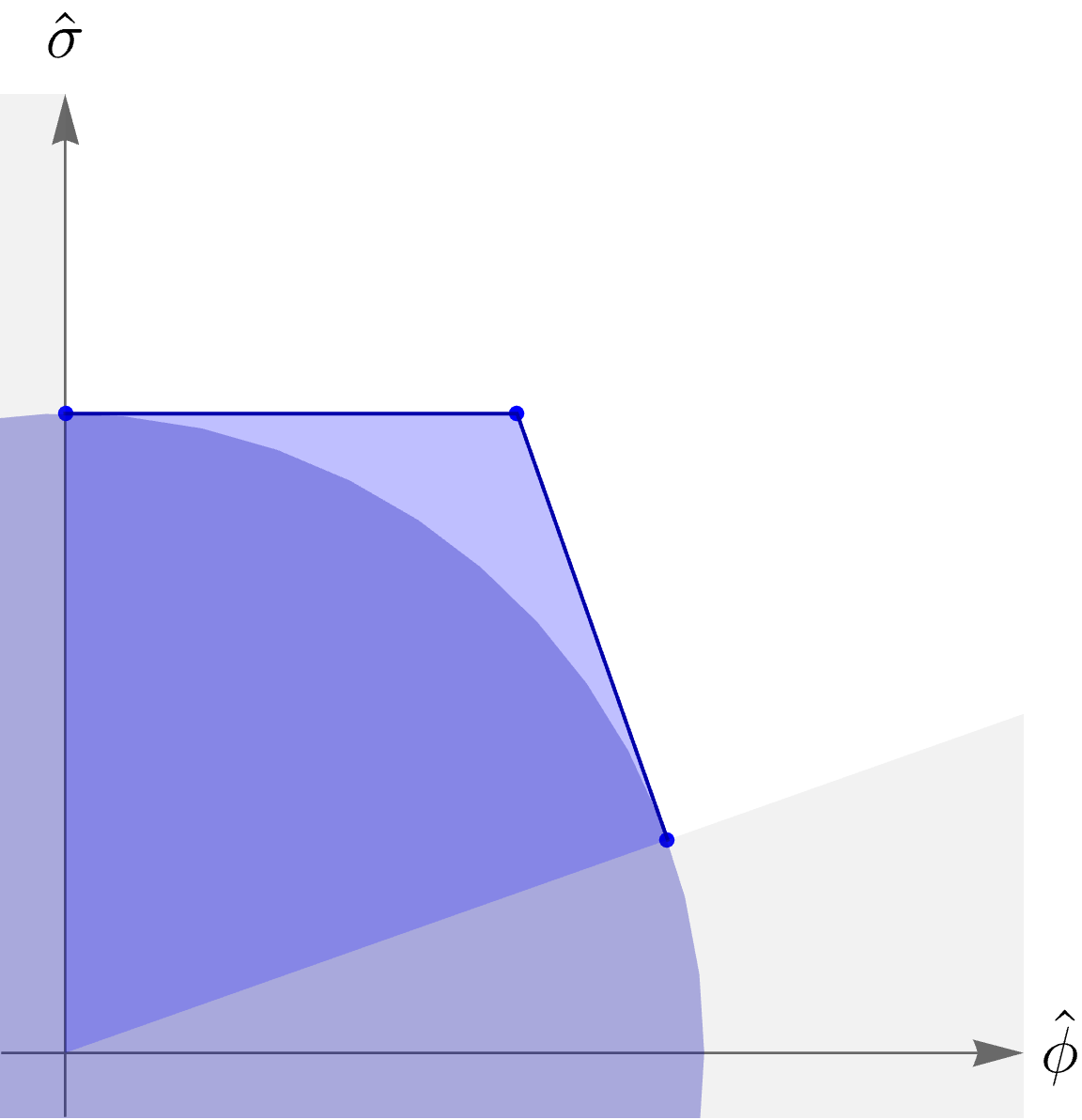}}
	\subfigure[\label{sfig:4dp=5}$d=4$ and $p=5$]{\includegraphics[width=0.32\textwidth]{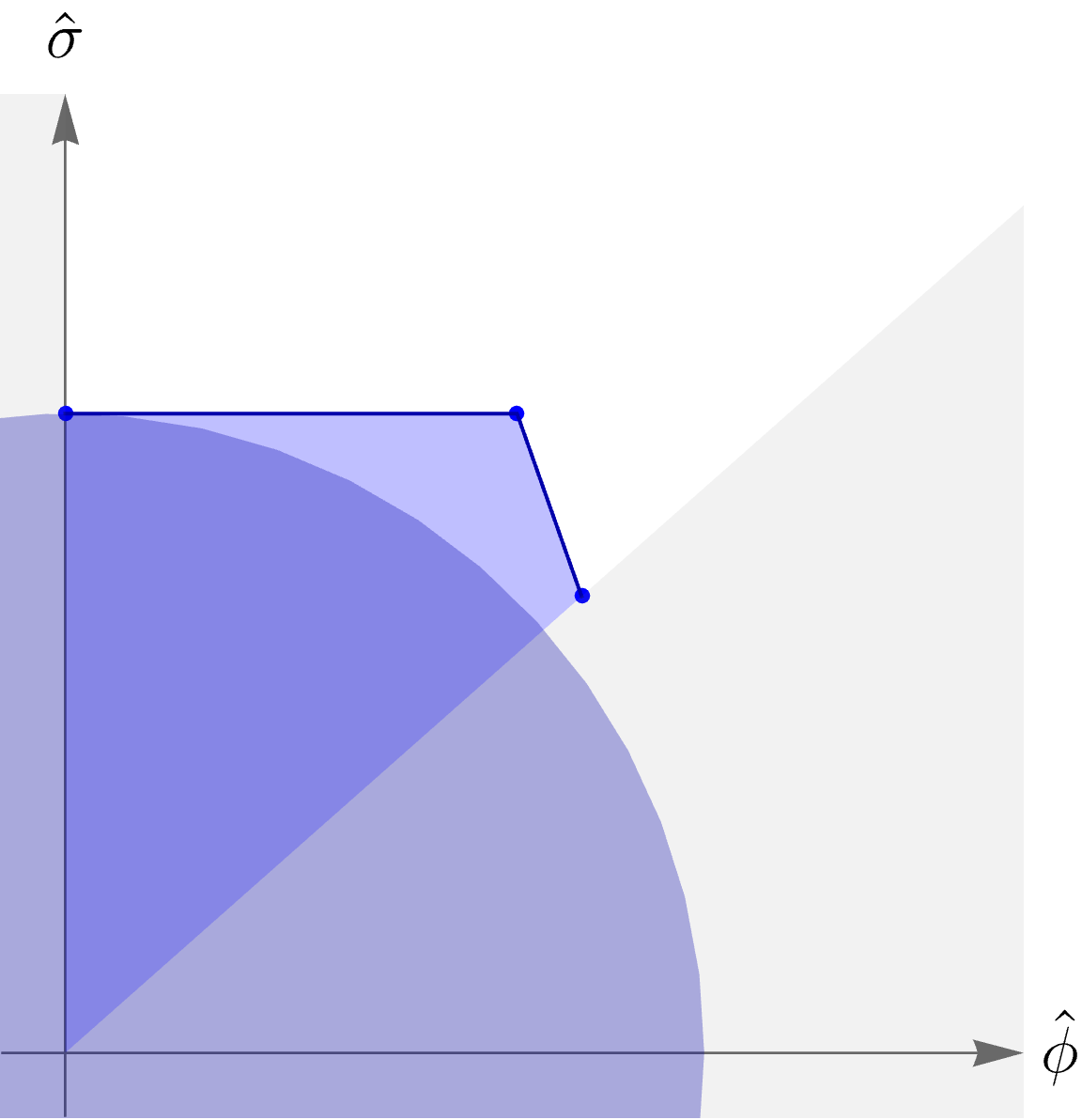}}
	\caption{Convex hull diagrams for the species scale vectors in four dimensions. We assume that the parent 5-dimensional theory is such that $\lambda_{\text{sp}}$ saturates \eqref{eq:lowerboundspecies2}, and show the resulting plot for three different values of $p$.}
	\label{fig:dim-red}	
\end{center}
\end{figure}

For the remaining boundary, namely the line determined by the vectors $\vec{\mathcal{Z}}_{\text{KK-t},\, p+1}$ and $\vec{\mathcal{Z}}_{\text{t}}$, we first note that at fixed $\lambda_{\text{sp}}$, varying $p$ only modifies the length of the corresponding edge but not its slope. This means, in practice, that the distance from the origin to this facet of the polytope is bounded from below by that associated to its infinite length extension. Second, we realize that $\lambda_{\text{sp}}$ only appears in the first components of the vectors \eqref{vectors}, and it does so in such a way that the bigger $\lambda_{\text{sp}}$, the farther away from the origin this edge will appear. In physical terms, this means that the closer we are to violate \eqref{eq:lowerboundspecies2} in $D$-dimensions, the easier it gets to do so after dimensional reduction as well. As a consequence, we deduce that the most dangerous situation happens precisely when the bound is saturated in the higher-dimensional theory. Nevertheless, even in this scenario there is no violation whatsoever of the CHC, regardless of the particular value of $p$ (see Figure \ref{fig:dim-red}), and in fact one finds again saturation when $p=1$. 

In conclusion, we see that the bound \eqref{eq:lowerboundspecies2} is preserved under dimensional reduction on a circle, at least for those asymptotic directions where field-theoretic considerations suffice to determine the convex hull diagram. In fact, it can be regarded as the strongest bound on the exponential decay rate of the species scale yet compatible with this procedure. This moreover involves a pretty robust mechanism, where the CHC is crucially protected by the KK replicas of the tower already existing in the higher-dimensional theory. 

\subsubsection*{Generalization to $D=d+n$ dimensions}

It is worth emphasizing that the previous analysis also extends to the more general case of compactification on a $n$-dimensional Ricci-flat manifold $\mathcal{X}_n$ (e.g., a $\mathbf{T}^n$). To see this, we start by reducing the original $D$-dimensional EFT on $\mathcal{X}_n$, such that the relevant dynamics in $d$ dimensions is described by the action (see Appendix \ref{ap:generalities} for details)
\begin{equation}
	S_{d} \supset \int \dd^{d}x\, \sqrt{-g}\,  \left[ \frac{1}{2\kappa_{d}^2} \left(\mathcal{R} - \frac{d+n-2}{n (d-2)} \left(\partial \log \mathcal{V}_n \right)^2 \right)- \frac{1}{2} \left(\partial \hat \phi \right)^2 \right]\, ,
\end{equation}
where $\mathcal{V}_n$ denotes the overall volume modulus measured in $D$-dimensional Planck units and we have retained only the massless scalar-tensor sector of the theory. Hence, upon canonically normalizing the volume modulus
\begin{equation}
	\hat \sigma = \frac{1}{\kappa_d}\sqrt{\frac{d+n-2}{n(d-2)}} \log \mathcal{V}_n\, ,
\end{equation}
one finds again two competing scalar charge-to-mass vectors, namely the one associated to the isotropic KK tower (with density parameter $p_{\text{KK}}=n$) and the original $D$-dimensional tower, which read
\begin{equation}\label{eq:zvectornmfd}
	\vec{\zeta}_{\text{KK},\, n} = \left( 0 \ ,\ \sqrt{\frac{d+n-2}{n(d-2)}} \right) \, , \quad \vec{\zeta}_{\text{t}} = \left( \lambda_{\text{t}} \ ,\ \sqrt{\frac{n}{(d+n-2)(d-2)}} \right) \, .
\end{equation}
From these one obtains three species vectors
\begin{equation}
\label{eq:Zspeciescompactification}
\begin{split} 
	&\vec{\mathcal{Z}}_{\text{KK},\, n} = \left( 0 \ ,\ \sqrt{\frac{n}{(d+n-2)(d-2)}} \right) \, ,\\
	&\vec{\mathcal{Z}}_{\text{t}} = \left( \frac{d-1+p}{d-2+p} \ \lambda_{\text{sp}} \ ,\ \frac{p}{d-2+p} \sqrt{\frac{n}{(d+n-2)(d-2)}} \right) \, ,\\
	&\vec{\mathcal{Z}}_{\text{KK-t},\, p+n}  = \left( \frac{d-1+p}{d-2+n+p} \ \lambda_{\text{sp}} \ ,\ \sqrt{\frac{n}{(d+n-2)(d-2)}} \right) \, .
\end{split}
\end{equation}
Notice that, as in the previous $\mathbf{S}^1$--\,compactification, we see that the boundary of the convex hull determined by $\vec{\mathcal{Z}}_{\text{KK},\, n}$ and $\vec{\mathcal{Z}}_{\text{KK-t},\, p+n}$ is again horizontal. Strictly speaking, though, this is not needed anymore so as to protect the CHC, since $\vec{\mathcal{Z}}_{\text{KK},\, n}$ does actually satisfy the bound with room to spare when $n>1$. In any event, the conclusion remains unchanged, thus confirming the expectation that the most constraining compactification for the exponential rate of the species scale corresponds to $n=1$, i.e. a circle/interval reduction.

Similarly, it is straightforward to check that the slope of the edge connecting $\vec{\mathcal{Z}}_{t}$ and $\vec{\mathcal{Z}}_{\text{KK-t},\, p+n}$ is also independent of $p$. Moreover, the larger $\lambda_{\text{sp}}$ the farther away from the origin that the corresponding edge gets. In fact, in the most dangerous situation, i.e. when \eqref{eq:lowerboundspecies2} is saturated in $D=d+n$ dimensions, this line still preserves the CHC in the lower dimensional theory, leading again to saturation if $p = 1$. 
All in all, the claim that the bound \eqref{eq:lowerboundspecies2} is preserved under dimensional reduction (within the directions in which field theory is enough to determine the convex hull), remains true for $n>1$ compactifications as well.    

\subsection{Beyond field theory}
\label{ss:compactificationstring}

The aim of this subsection is to go beyond our previous field-theoretical considerations and study the fate of the bound \eqref{eq:lowerboundspecies2} after the inclusion of genuine quantum-gravitational ingredients. To illustrate this, we consider a very simple toy model which is clearly inspired by Type IIB string theory, thus featuring two fundamental strings in $D$ spacetime dimensions. These strings may become asymptotically tensionless when exploring two different infinite distance regimes. For simplicity, we will again restrict ourselves to one-dimensional moduli spaces, where the previous emergent string limits arise upon taking $\hat \phi \to \pm \infty$. Therefore, we assume the associated string excitation modes to become exponentially massless with some constant decay rate $\lambda_{\text{osc}}$ (correspondingly $\lambda_{\text{osc}'} =-\lambda_{\text{osc}}$), which in this case coincides with the exponential rate for the species scale. 

The dimensional reduction analysis of the aforementioned theory on a circle proceeds exactly as in Section \ref{ss:field-theory}, with the important difference that we can have additional towers in $d$ dimensions arising from the winding modes of the extended strings. The relevant scalar charge-to-mass vectors read \cite{Etheredge:2022opl}
\begin{equation}\label{eq:zvectorsp=1}
\begin{split} 
	\vec{\zeta}_{\text{KK}} =& \left( 0 \ ,\ \sqrt{\frac{d-1}{d-2}} \right)\, , \qquad \vec{\zeta}_{\text{w}} = \left( 2 \lambda_{\text{osc}} \ ,\ - \frac{d-3}{\sqrt{(d-1)(d-2)}} \right) \, ,\\
	\vec{\zeta}_{\text{w}'} =& \left( -2 \lambda_{\text{osc}} \ ,\ - \frac{d-3}{\sqrt{(d-1)(d-2)}} \right) \, ,
\end{split}
\end{equation}
for the KK-like towers, whilst for the string oscillator modes one rather finds
\begin{equation}\label{eq:zvectorsStrings}
  \vec{\zeta}_{\text{osc}} =\left( \lambda_{\text{osc}} \ ,\ \frac{1}{\sqrt{(d-1)(d-2)}} \right) \, , \qquad 
  \vec{\zeta}_{\text{osc}'} = \left( - \lambda_{\text{osc}} \ ,\ \frac{1}{\sqrt{(d-1)(d-2)}} \right) \, .
\end{equation}
Moreover, upon using eq. \eqref{eq:eff-vector} above, we can translate the set \eqref{eq:zvectorsp=1} into the following species scale vectors
\begin{equation}\label{vectors2}
\begin{split} 
	\vec{\mathcal{Z}}_{\text{KK}} =& \left( 0 , \frac{1}{\sqrt{(d-1)(d-2)}} \right)\, , \qquad \qquad \qquad \ \ \vec{\mathcal{Z}}_{\text{w}} = \left( \frac{2 \lambda_{\text{osc}}}{d-1} , - \frac{d-3}{(d-1)^{3/2}\sqrt{(d-2)}} \right) \, ,\\
	\vec{\mathcal{Z}}_{\text{w}'} =& \left( -\frac{2 \lambda_{\text{osc}}}{d-1} , - \frac{d-3}{(d-1)^{3/2}\sqrt{(d-2)}} \right) \, , \quad \vec{\mathcal{Z}}_{\text{w}, \,2} = \left( 0 , - \frac{2(d-3)}{d\sqrt{(d-1)(d-2)}} \right) \, .
\end{split}
\end{equation}
Here we have taken into account several things. First, notice that even though strings can form effective towers together with some additional KK-like spectrum, their degeneracy is so strong that they already dominate the state counting and thus saturate the species scale alone.\footnote{Recall that this claim can be intuitively understood upon taking $p\to\infty$ in eq. \eqref{eq:eff-vector}, which forces the species scale to be given precisely by the fundamental string scale.} Additionally, we have assumed the two towers of winding modes to be multiplicative, thus giving rise to the effective species vector $\vec{\mathcal{Z}}_{\text{w}, \,2}$, which has $p=2$. In Type IIB string theory this comes about from having not only the fundamental and the D1-string, but also the spectrum of $(p,q)$ bound states thereof \cite{Witten:1995im}. We will see later on how this fits nicely with the results. Finally, in spite of the fact that one could also a priori consider winding and KK modes to be multiplicative amongst each other, their effective combination ends up being, to all effects, irrelevant for testing the CHC, as one may readily check.

\begin{figure}[htb]
\begin{center}
\includegraphics[width=0.45\textwidth]{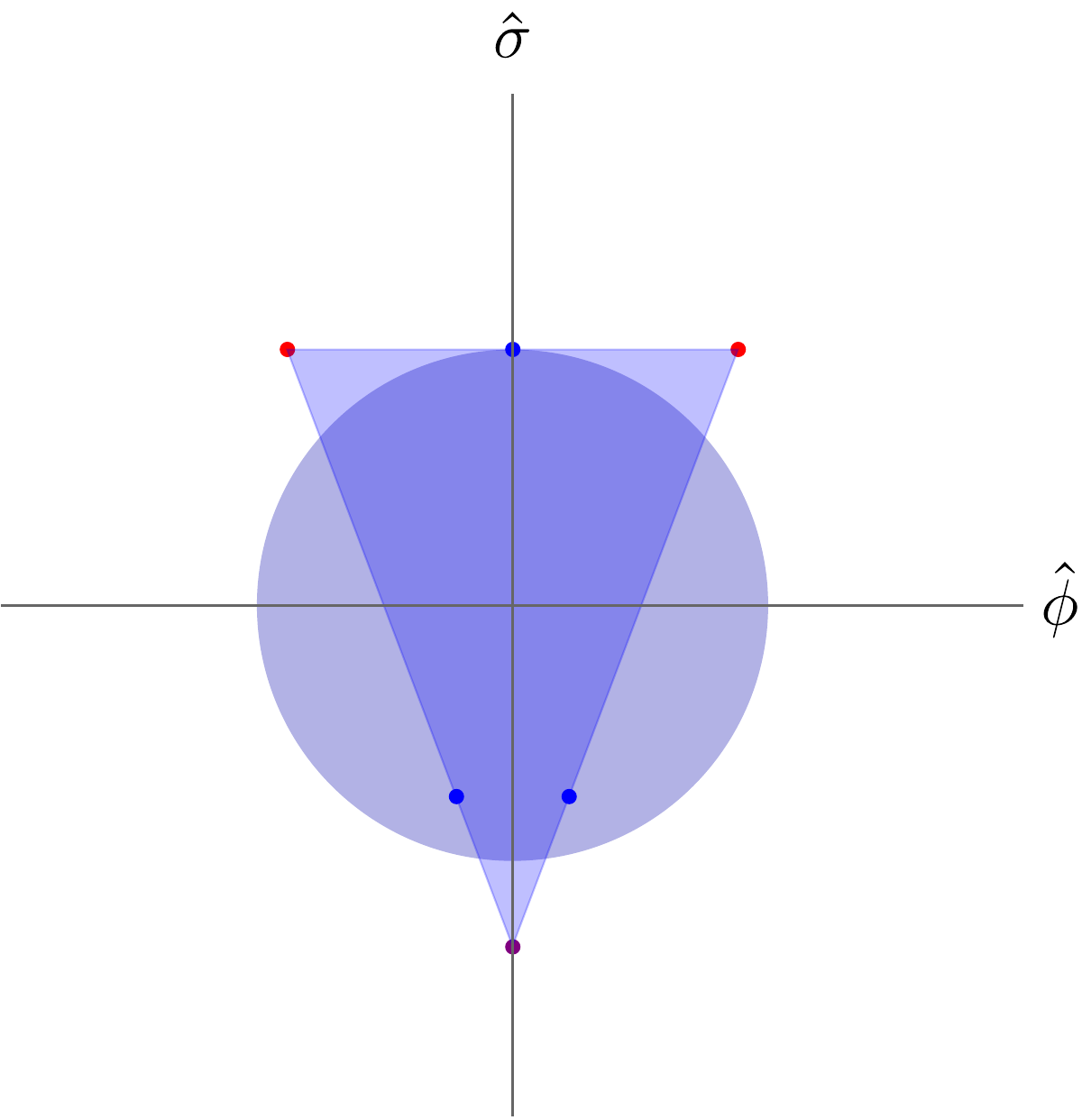}
\caption{\small Convex hull diagram for a theory featuring two fundamental strings (red dots) that saturate the CHC \eqref{eq:lowerboundspecies2} in $D=10$ dimensions, after compactification on $\mathbf{S}^1$. The blue dots are associated to KK-like towers with $p=1$, whilst the purple one represents the effective winding tower of $p=2$. Finally, the axes correspond to the radius modulus $\hat \sigma$ as well as the $D$-dimensional dilaton, $\hat \phi$.} 
\label{fig:ch3}
\end{center}
\end{figure}

With this information, we are now ready to study the behavior of the bound \eqref{eq:lowerboundspecies2} under dimensional reduction within the present toy model. As a first interesting observation, let us note that $\lambda_{\text{osc}}$ only appears in the first components of (some of) the vectors in eqs. \eqref{eq:zvectorsStrings} and \eqref{vectors2}, in such a way that for larger values of $\lambda_{\text{osc}}$, these vectors get farther away from the origin, similarly to what happened in Section \ref{ss:field-theory}. Following the same logic as in there, one can then try to fix $\lambda_{\text{osc}}$ so as to saturate the CHC in the parent $D$-dimensional theory, namely
\begin{equation}
    \lambda_{\text{osc}} \stackrel{!}{=} \frac{1}{\sqrt{(D-1)(D-2)}} = \frac{1}{\sqrt{d(d-1)}} \, .
\end{equation}
Plugging this into the previous set of $\mathcal{Z}$-vectors and drawing the convex hull, one realizes that there seems to be an unavoidable violation of the bound \eqref{eq:lowerboundspecies2} for any spacetime dimension $d$. (An example of this is shown in Figure \ref{fig:ch3} for the $d=9$ case.) Let us stress, however, that this by itself does \emph{not} imply a violation of the CHC for the species scale within quantum gravity in general, since we are just considering a very crude toy model in which there exists some critical string with decay rate $\lambda_{\text{osc}}$ that is moreover taken to saturate \eqref{eq:lowerboundspecies2} in $D$-dimensions. In fact, as we know from our experience with string theory, critical strings never feature this particular value for the exponential decay rate along their own gradient flow, but instead a much \emph{larger} one. In a sense, what this toy model tells us is that the consistency of the CHC prevents the strings from saturating alone the lower bound \eqref{eq:lowerboundspecies2}. This resembles prior studies in the literature of certain Swampland criteria, which under some circumstances become stronger when imposing its consistency under dimensional reduction (see e.g., \cite{Heidenreich:2015nta}).

\begin{figure}[htb]
\begin{center}
	\subfigure[\label{sfig:5dto4d}$D=5 \, \rightarrow\, d=4$]{\includegraphics[width=0.45\textwidth]{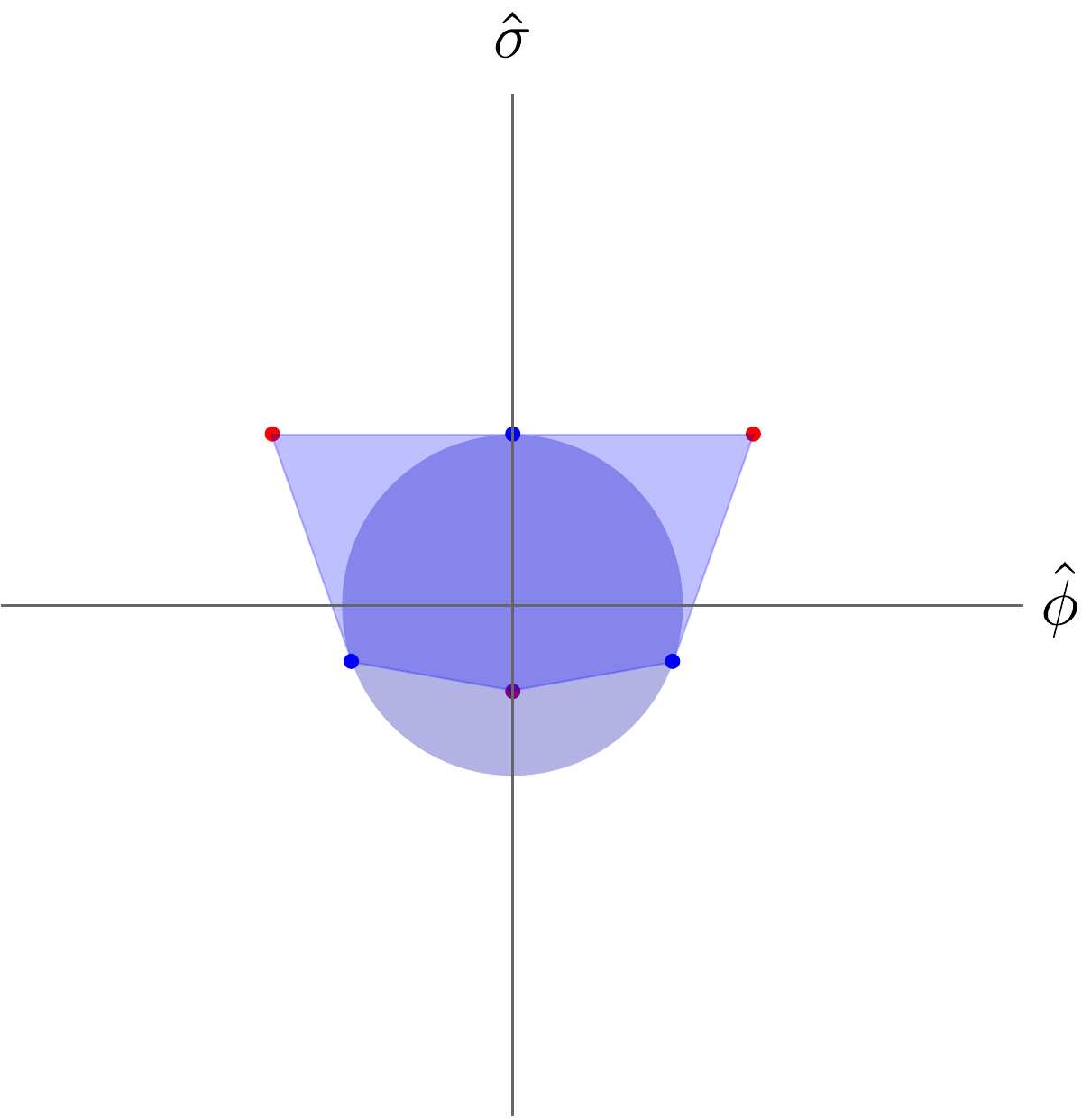}}\qquad 
	\subfigure[\label{sfig:9dto8d}$D=9 \, \rightarrow\, d=8$]{\includegraphics[width=0.45\textwidth]{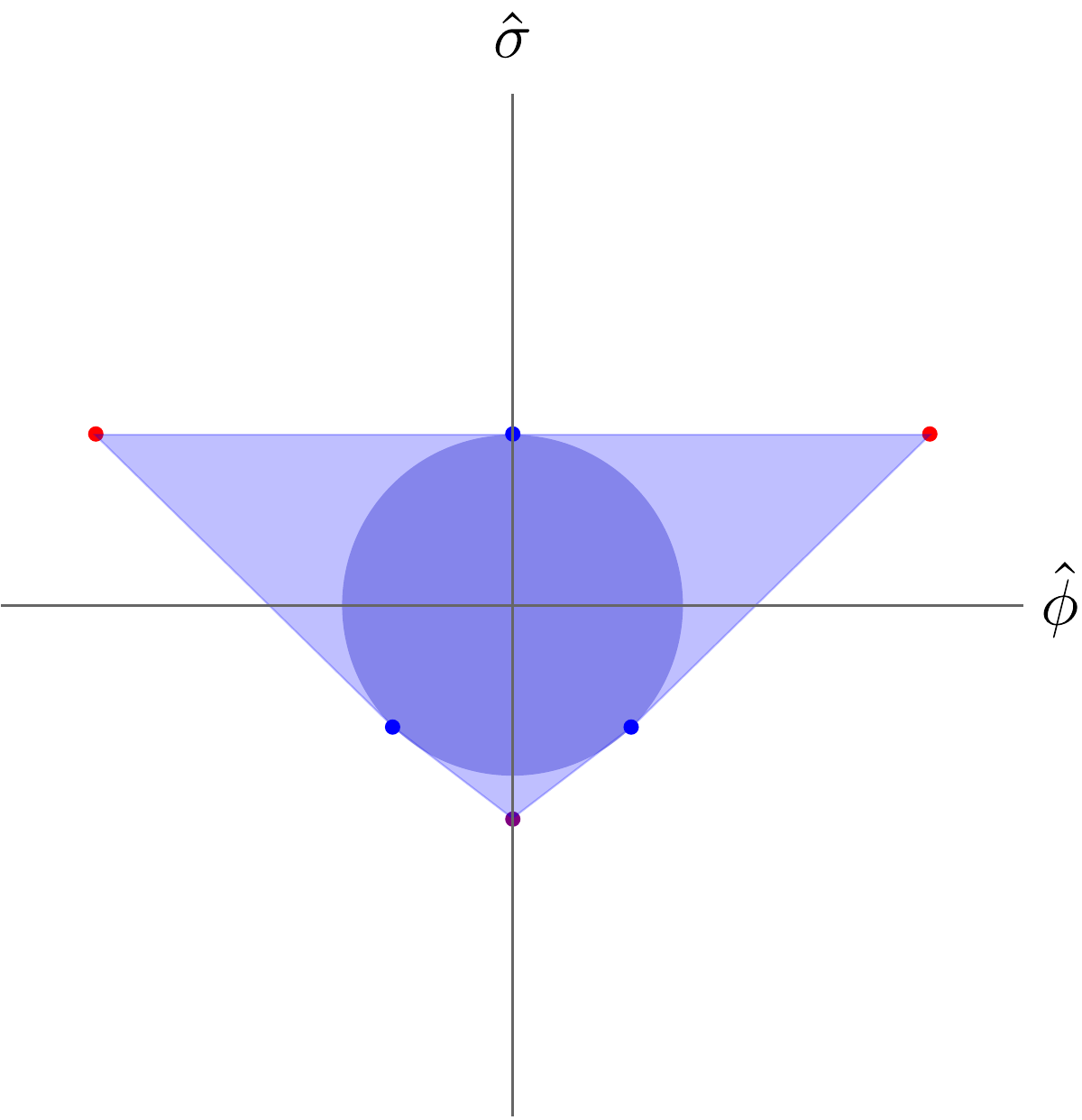}}
	\caption{Convex hull diagram of a theory of strings with $|\lambda_{\rm osc}|=\frac{1}{\sqrt{D-2}}$ in \textbf{(a)} $D=5$ and \textbf{(b)} $D=9$ dimensions compactified on a circle. Although difficult to spot by eye, one may easily check that the 9d $\to$ 8d case also violates the CHC for asymptotic directions centered around the two lower blue dots, corresponding to T-dual decompactifications of one extra dimension.}
	\label{fig:ch4}	
\end{center}
\end{figure}

Therefore, following our discussion in the previous paragraph, a better justified value for $\lambda_{\text{osc}}$ would be (c.f. eq. \eqref{eq:zetaveconemodulus})
\begin{equation}
    \lambda_{\text{osc}} = \frac{1}{\sqrt{D-2}} = \frac{1}{\sqrt{d-1}} \, .
\end{equation}
After plugging this into eqs. \eqref{eq:zvectorsStrings} and \eqref{vectors2} so as to draw the convex hull, we find that the CHC is a priori satisfied only for $D \geq 10$, see Figure \ref{fig:ch4} for the particular cases of $D=5$ and $D=9$. Furthermore, we find saturation happening precisely in the ten-dimensional case. 
Of course, this is not a coincidence, since in this case our toy model precisely reproduces Type IIB string theory on $\mathbf{S}^1$, which is known to be dual to M-theory on $\mathbf{T}^2$, see Section \ref{ss:MthyT2SSDC} below.

\begin{figure}[htb]
\begin{center}
\includegraphics[width=0.45\textwidth]{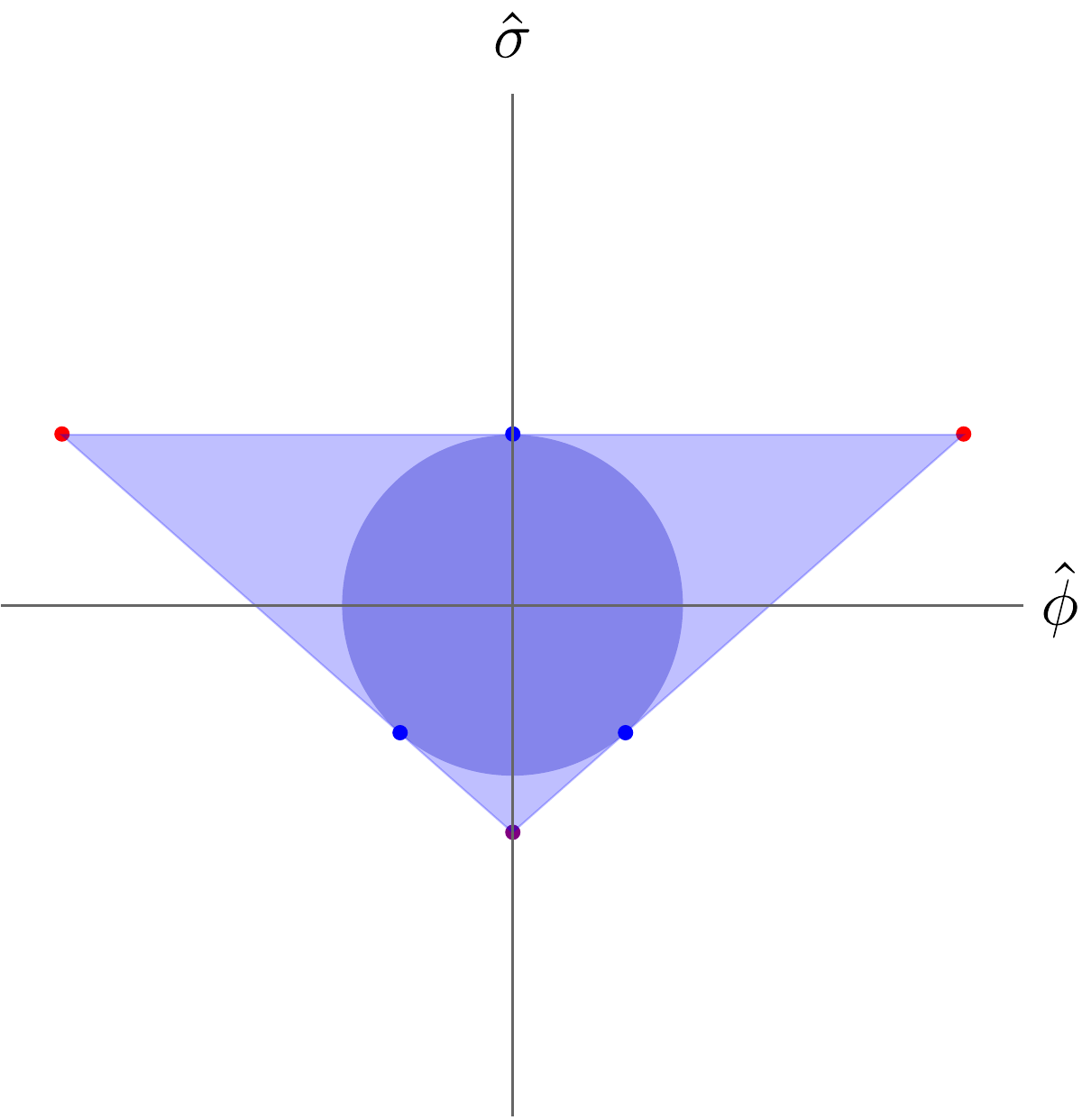}
\caption{\small Convex hull diagram of a theory of strings with $|\lambda_{\rm osc}|=\frac{1}{\sqrt{D-2}}$ in $D=10$ dimensions compactified on a circle. It can be related with the corresponding picture in M-theory on $\mathbf{T}^2$ after a $\pi/2$ rotation, c.f. Figure \ref{fig:ch1}.} 
\label{fig:ch5}
\end{center}
\end{figure}

Very remarkably, we conclude that the present toy model, which presents only one-dimensional strings to start with, does not satisfy the bound \eqref{eq:lowerboundspecies2} in $D\leq 9$ after dimensionally reducing on a circle. This is actually consistent with our intuition from string theory, since upon compactifying down to $d\leq8$ dimensions, not only strings and winding modes arise, but also higher-dimensional non-perturbative objects do, such as D$p$-branes. Such extended objects potentially give rise --- when wrapped along some internal cycle of the compact geometry --- to new infinite towers of states. Indeed, an explicit realization of these matters will be presented later on in Section \ref{ss:MthyT3SSDC}, where we consider M-theory compactified on $\mathbf{T}^3$. There, we will not only verify the CHC but we will also be able to construct a picture that is very reminiscent of the one shown in Figure \ref{sfig:9dto8d}, which features certain additional towers that indeed protect the lower bound \eqref{eq:lowerboundspecies2} in a non-trivial manner.

Before closing this subsection, let us compare our results with the ones obtained for the sharpened Distance Conjecture \cite{Etheredge:2022opl}. There it was shown, by considering the exact same toy model, that the winding modes associated to the critical strings were in fact sufficient so as to ensure that the bound \eqref{eq:sharpenedDistConj} holds in $d\geq5$. Additionally, it was argued that a tower of KK monopoles prevents this condition from being violated in $d=4$. Here we find that the analogous condition for the species scale decay rate requires from more than this. Indeed, as argued before, the presence of higher-dimensional objects in $D$-dimensions seems to be crucial for the CHC to be satisfied in lower dimensional (Minkowski) vacua as soon as we get down to nine non-compact dimensions. 

\section{String theory evidence}
\label{s:examplesbound}

In this section we present non-trivial evidence in favour of the proposed lower bound for the asymptotic decay rate of the species scale in theories of quantum gravity. We will restrict ourselves to maximally supersymmetric set-ups arising from toroidal compactifications of Type II/M-theory, where the verification of \eqref{eq:lowerboundspecies2} becomes highly intricate even in the most simple cases. In particular, in Sections \ref{ss:MthyT2SSDC} and \ref{ss:MthyT3SSDC} we study the problem very explicitly in maximal supergravity in nine and eight spacetime dimensions. Later on, in Section \ref{ss:MthyTk}, we present a brief argument extending the proof to lower-dimensional toroidal compactifications as well. For completeness, let us mention that the convex hull diagrams here presented can be regarded as `boundary conditions' that any globally defined species scale function must agree with, see Chapter \ref{ch:Higherdimops} for details on this. 

\subsection{M-theory on $\mathbf{T}^2$}
\label{ss:MthyT2SSDC}

We consider first a 9d $\mathcal{N}=2$ example arising from compactifying M-theory on a two-dimensional torus. The bosonic action for the scalar and gravitational sectors reads
\begin{equation}\label{eq:9d}
	S^{\text{9d}}_{\text{M-th}} \supset \frac{1}{2\kappa_9^2} \int \dd^{9}x\, \sqrt{-g}\,  \left( \mathcal{R} - \frac{9}{14} \left( \frac{\partial \mathcal{V}_2}{\mathcal{V}_2} \right)^2 -\frac{\partial \tau \cdot \partial \bar \tau}{2 \left(\text{Im}\, \tau\right)^2} \right)\, ,
\end{equation}
where $\tau=\tau_1+ \i \tau_2$ and $\mathcal{V}_2$ are the complex structure and the volume of the $\mathbf{T}^2$, respectively (see Section \ref{ss:9dmaxsugra} for details). As discussed in Section \ref{s:dualities}, this theory enjoys a non-perturbative $\mathsf{SL(2, \mathbb{Z})}$ duality symmetry, whose origin is clearly geometric from this perspective since it is associated to the group of large diffeomorphisms of the internal torus\cite{Schwarz:1995dk,Aspinwall:1995fw}. 

Our goal here will be to check the convex hull condition for the species scale, namely the requirement
\begin{equation} \label{eq:bound9d}
  \lambda_{\text{sp}} \geq \frac{1}{\sqrt{(d-1)(d-2)}} \stackrel{\text{9d}}{=} \frac{1}{\sqrt{56}}\, ,
\end{equation}
at any boundary of moduli space. Moreover, since this condition should hold for any locally geodesic trajectory within the latter as it explores infinite distance, the first step is to properly characterize the latter. For this, we recall that the 9d moduli space can be identified with group coset
\begin{equation}\label{eq:9dmodspaceSSDC}
 \mathcal{M}_{\text{9d}}=\mathsf{SL(2, \mathbb{Z})}\backslash \mathsf{SL(2, \mathbb{R})}/\mathsf{U(1)} \times \mathbb{R}_+\, ,
\end{equation}
where we have already modded out by the modular duality group. Furthermore, as we discuss in more detail below, all geodesic trajectories reaching infinity are such that $\tau_1 \to \text{const.}$ asymptotically. This allows us to restrict our discussion to the slice of $\mathcal{M}_{\text{9d}}$ parametrized by the non-compact directions $\{ \mathcal{V}_2, \text{Im}\, \tau \}$, which corresponds to the subspace of asymptotically geodesic tangent vectors introduced in ref. \cite{Calderon-Infante:2020dhm}.

In a next step, one needs to account for the relevant towers of states, as well as compute all possible quantum gravity cut-offs that could arise depending on the infinite distance singularity that we probe. We start by considering $\frac{1}{2}$-BPS strings, which arise from wrapped M2-branes on any $(p,q)$ 1-cycle of the internal geometry. Their tension can be computed to be
\begin{equation}\label{eq:pqstrings9d}
	T_{p,q} = \frac{2\pi}{\ell_{9}^2} \frac{|p+q\tau|}{\sqrt{\tau_2}} \mathcal{V}_2^{\frac{3}{14}}\, ,
\end{equation}
where $\ell_{9}$ denotes the 9d Planck length. In the following, we will fix the axion v.e.v. to zero for simplicity (see however the discussion after eq. \eqref{eq:chargetomasspq}), and only keep track of the saxionic dependence of the relevant masses involved. We moreover define canonically normalized fields $\hat U$ and $\hat \tau$ as follows
\begin{equation} \label{eq:canonicalnormalization}
  \log \mathcal{V}_2 =  \kappa_9 \sqrt{\frac{14}{9}}\, \hat U \, , \quad \tau_2 = \kappa_9\, e^{\sqrt{2} \, \hat\tau} \, ,
\end{equation}
in terms of which the mass scale of the oscillation modes of the $(p,q)$-strings reads as
\begin{equation} \label{strings}
  m^{(\text{osc})}_{p,q} = \sqrt{T_{p,q}} =  \frac{(4\pi)^{5/14} M_{\text{Pl};\, 9}}{\sqrt{2}} \left( p^2 e^{-\sqrt{2} \,\hat\tau} + q^2 e^{\sqrt{2} \,\hat\tau}\right)^{1/4} e^{\frac{1}{2 \sqrt{14}} \, \hat U} \, .
\end{equation}
Note that the above expression presents two different asymptotic behaviors depending on which infinite distance limit is probed and whether $p$ and $q$ are non-vanishing. As it is to be expected, any infinite distance limit --- at fixed $\tau_1=0$ --- is dominated by either the $q=0$ or $p=0$ cases, which are associated to the fundamental and S-dual Type II strings, respectively. These lead to two relevant asymptotic scales for the QG cut-off
\begin{equation} \label{string-species}
  \frac{\Lambda_{\text{osc}}}{M_{\text{Pl};\, 9}}\, \sim\, e^{\frac{1}{2 \sqrt{14}} \, \hat U-\frac{1}{2\sqrt{2}} \,\hat\tau } \, , \qquad \frac{\Lambda_{\text{osc'}}}{M_{\text{Pl};\, 9}}\, \sim\, e^{\frac{1}{2 \sqrt{14}} \, \hat U + \frac{1}{2\sqrt{2}} \,\hat\tau } \, ,
\end{equation}
where we are using the fact that the species scale associated to a critical string is given at leading order by its own mass. One thus obtains the following relevant species vectors
\begin{equation}
\label{eq:speciesscalevectorsstringsT2}
  \vec{\mathcal{Z}}_{\text{osc}} = \left( -\frac{1}{2 \sqrt{14}},\frac{1}{2\sqrt{2}} \right) \, , \qquad 
  \vec{\mathcal{Z}}_{\text{osc'}} = \left( -\frac{1}{2 \sqrt{14}},- \frac{1}{2\sqrt{2}} \right) \, ,
\end{equation}
where the notation is $\vec{\mathcal{Z}} = \left(\mathcal{Z}_{\hat U}, \mathcal{Z}_{\hat \tau} \right)$.

On the other hand, the 9d theory also presents some particular spectrum of $\frac{1}{4}$-BPS particles, whose masses depend on where we sit in moduli space and are given by \cite{Obers:1998fb}
\begin{equation}\label{eq:pqparticles9d}
	m^{(\text{part})}_{p,q,w} = \frac{2\pi}{\ell_{9}} \left[ \frac{|p+q\tau|}{\sqrt{\tau_2}} e^{-\frac{9}{14}U} + |w| e^{\frac{6}{7}U} \right]\, .
\end{equation}
Setting again the axion v.e.v. to zero and re-expressing everything in terms of the canonically normalized fields \eqref{eq:canonicalnormalization}, we get 
\begin{equation} \label{particles}
  m^{(\text{part})}_{p,q,w} = \frac{(4\pi)^{6/7} M_{\text{Pl};\, 9}}{2} \left[\left( p^2 e^{-\sqrt{2} \,\hat\tau} + q^2 e^{\sqrt{2} \,\hat\tau}\right)^{1/2} e^{-\frac{3}{\sqrt{14}} \, \hat U} + |w| e^{\sqrt{\frac{8}{7}} \, \hat U} \right]\, .
\end{equation}
These particles arise as bound states of Kaluza-Klein modes along the compact directions (with charges $p, q \in \mathbb{Z}$) and non-perturbative states obtained by wrapping an M2-brane $w \in \mathbb{Z}$ times along the internal 2-cycle.\footnote{Alternatively, the M2-particles may be viewed as winding modes of the critical Type IIA strings described in \eqref{strings}.} For us, it turns out to be enough to focus on towers comprised by $\frac{1}{2}$-BPS states, since only these become light and dense enough asymptotically so as to saturate the species scale at some infinite distance corner of the 9d moduli space \eqref{eq:9dmodspaceSSDC}. Heuristically, this may be understood from the fact that any other state necessarily contains fields with spin higher than 2, and thus whenever they become nearly massless we expect some other critical string to dominate the asymptotic physics. Therefore, we may divide the spectrum into two sectors, corresponding to either $w=0$ or $p=q=0$. 

Consider first the $w=0$ sector. It does behave as two multiplicative towers of Kaluza-Klein type --- indeed they are the KK modes corresponding to either one of the two 1-cycles of the torus --- with mass scales behaving in Planck units as
\begin{equation} \label{KK-mass}
  m_{\text{KK},\, 1} = \frac{(4\pi)^{6/7} M_{\text{Pl};\, 9}}{2} e^{-\frac{3}{\sqrt{14}} \, \hat U - \frac{1}{\sqrt{2}} \,\hat\tau } \, , \qquad 
  m_{\text{KK},\, 1'} = \frac{(4\pi)^{6/7} M_{\text{Pl};\, 9}}{2}  e^{-\frac{3}{\sqrt{14}} \, \hat U + \frac{1}{\sqrt{2}} \,\hat\tau } \, ,
\end{equation}
and density parameter $n=1$. Their associated species cut-offs can be easily computed:
\begin{equation}
  \frac{\Lambda_{\text{KK},\, 1}}{M_{\text{Pl};\, 9}}\, \sim\, e^{-\frac{3\sqrt{14}}{112} \, \hat U - \frac{\sqrt{2}}{16} \,\hat\tau } \, , \quad 
  \frac{\Lambda_{\text{KK},\, 1'}}{M_{\text{Pl};\, 9}}\, \sim\, e^{-\frac{3\sqrt{14}}{112} \, \hat U + \frac{\sqrt{2}}{16} \,\hat\tau } \, , \quad
   \frac{\Lambda_{\text{KK},\, 2}}{M_{\text{Pl};\, 9}}\, \sim\, e^{-\frac{\sqrt{14}}{21} \, \hat U} \, ,
\end{equation}
where the last one corresponds to the effective combination of the first two, c.f eq. \eqref{eq:effspeciesscale}. Thus, their species vectors become
\begin{equation}
  \vec{\mathcal{Z}}_{\text{KK},\, 1} = \left( \frac{3\sqrt{14}}{112},\frac{\sqrt{2}}{16} \right)\, , \quad
  \vec{\mathcal{Z}}_{\text{KK},\, 1'} = \left( \frac{3\sqrt{14}}{112},-\frac{\sqrt{2}}{16}\right) \, , \quad
  \vec{\mathcal{Z}}_{\text{KK},\, 2} = \left( \frac{\sqrt{14}}{21}  , 0\right) \, .
\end{equation}

For the $p=q=0$ sector, eq. \eqref{particles} tells us that the tower behaves essentially as some sort of KK spectrum. This is easy to understand, since they are nothing but the Kaluza-Klein replica of the 10d fields implementing the M-/F-theory duality, c.f. Section \ref{ss:dualitieswithlowersusy}. Their mass scale is thus given by
\begin{equation}\label{eq:Ftheorytower}
  m_{\text{M}2} = \frac{(4\pi)^{6/7} M_{\text{Pl};\, 9}}{2}\,  e^{\sqrt{\frac{8}{7}} \, \hat U } \, ,
\end{equation}
and its associated species scale and charge-to-mass ratio read
\begin{equation}
  \frac{\Lambda_{\text{M}2}}{M_{\text{Pl};\, 9}}\, \sim\, e^{\frac{1}{2\sqrt{14}} \, \hat U } \quad \Longrightarrow \quad \vec{\mathcal{Z}}_{\text{M}2} = \left( -\frac{1}{2\sqrt{14}} , 0\right) \, .
\end{equation}
A priori, one should also include a pair of vectors obtained from considering effective towers comprised by BPS states with non-vanishing charges $(p, \omega)$ (or rather $(q, \omega)$). However, it is easy to convince ourselves that these vectors will not modify the convex hull diagram in any way, since the corresponding scales turn out to be always above the mass scale of the strings in eq. \eqref{string-species}. Relatedly, if one were to include the species scale associated to an effective tower including all three KK-like charges, the result would be that it is always above the ones just considered, which follows from the fact that the states in \eqref{eq:pqparticles9d} with all $p,q,w\neq 0$ never become asymptotically massless and thus never help in lowering the QG cut-off. 

\subsubsection*{Plotting the convex hull}

\begin{figure}[htb]
		\begin{center}
			\subfigure[]{
				\includegraphics[width=0.45\textwidth]{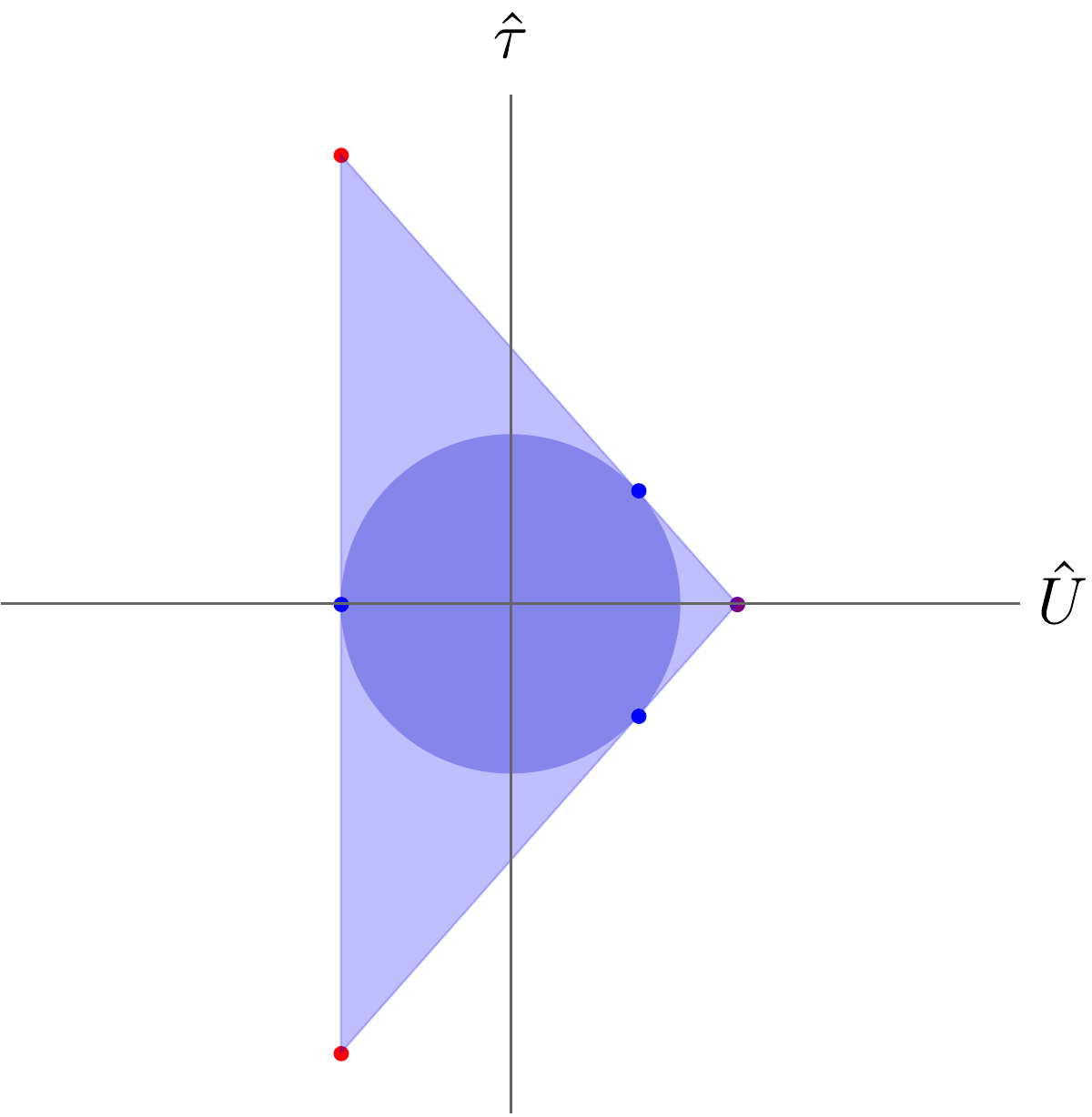}\label{sfig:9dCH}
			}
            \quad
			\subfigure[]{
				\includegraphics[width=0.45\textwidth]{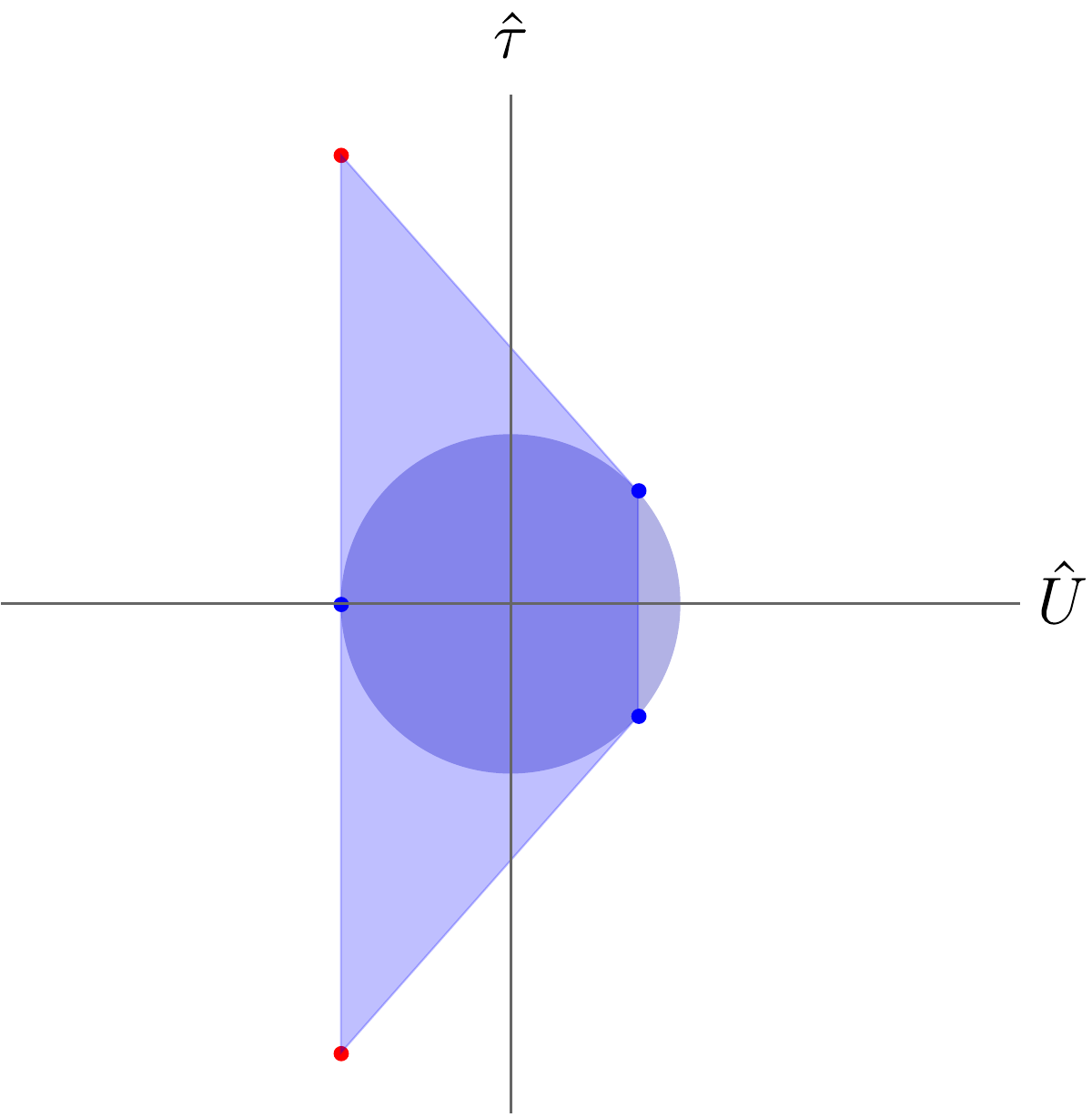}\label{sfig:9dCH-2}
			}
			\small \caption{\textbf{(a)} Convex hull condition for the species scale in M-theory on $\mathbf{T}^2$. \textbf{(b)} Convex hull condition for the species scale in M-theory on $\mathbf{T}^2$, if the effective double KK tower would not exist. The blue dots appearing in the edges of the convex hull are single KK towers, whereas the red and purple dots arising at the vertices represent stringy and double KK towers, respectively.}
			\label{fig:ch1}
		\end{center}
\end{figure}

With this we already have all the necessary ingredients in order to draw the convex hull associated to the species vectors $\vec{\mathcal{Z}}$. The result is shown in Figure \ref{sfig:9dCH}, where we additionally include the extremal radius corresponding to $\lambda_{\text{sp,\, min}}=1/\sqrt{56}$ in the present set-up. For comparison, we also depict in Figure \ref{sfig:9dCH-2} how the convex hull would look like in the absence of the effective tower. As advocated, the latter is indeed crucial for capturing the underlying asymptotic physics, also ensuring that the bound \eqref{eq:bound9d} is non-trivially satisfied. 

Notice that there is a  $\mathbb{Z}_2$-symmetry relating the upper and lower halves of the convex hull in Figure \ref{sfig:9dCH}, which is nothing but a manifestation of the $\mathsf{SL(2, \mathbb{Z})}$ duality group of this theory. On top of this, the diagram presents lots of structure that nicely encodes the physics at the different asymptotic limits, corresponding to different directions within the diagram. We discuss each of these in turn.

Let us start with the vertices. Indeed, upon pointing towards any one of the red dots in Figure \ref{sfig:9dCH}, the species scale is dominated by a (critical) string tower. Therefore, the associated regime turns out to be an emergent string limit \cite{Lee:2019wij}. Similarly, for the direction determined by the purple dot, the species counting is saturated by the double KK tower, thus signalling towards full decompactification to 11d M-theory. In fact, this also holds upon exploring any intermediate direction between the blue dots in Figure \ref{sfig:9dCH}, since despite one KK tower being parametrically lighter than the other, the species scale is yet saturated by accounting for mixed states thereof (see discussion around eq. \eqref{eq:masseffectivetower}).

Finally, let us discuss the directions associated to the blue dots themselves. Notice that these vectors are always orthogonal to some edge of the convex hull. In fact, were not this the case, the condition \eqref{eq:bound9d} would be automatically violated since such single KK vectors precisely saturate the bound. This means, incidentally, that all three potential species scales lying at each edge of the convex hull fall at the same rate along the aforementioned limits, with their (finite) ratios not encoded in the convex hull and thus depending on the values of the moduli that are not sent to infinity. This fact has a nice physical interpretation: The species scale of the single Kaluza-Klein tower we are pointing to signals towards decompactification of one extra dimension, whereas the remaining pair of scales correspond to towers that are already present in the higher-dimensional theory. In fact, we observe that the edges precisely reproduce the (one-dimensional) convex hull of the decompactified theory. Indeed, the vertical line on the left-hand side of the diagram corresponds to the convex hull of 10d Type IIB string theory, with the F1 and D1-strings becoming light at weak and strong coupling, respectively. Similarly, the other two edges correspond to the convex hull of 10d Type IIA, with the fundamental string and the tower of D0-branes becoming light analogously at weak and strong coupling. 

\begin{figure}[htb]
\begin{center}
\includegraphics[width=0.45\textwidth]{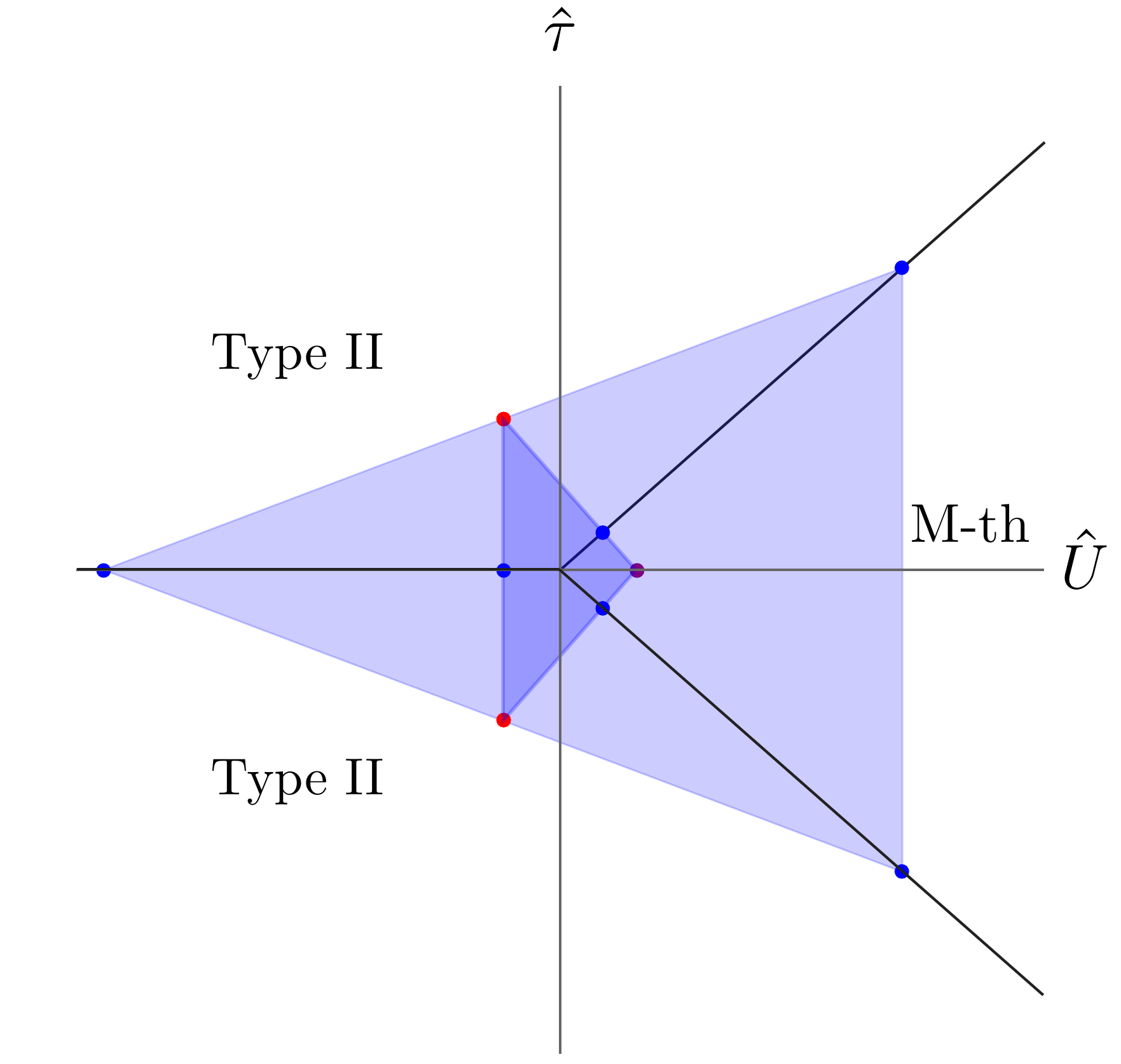}
\caption{\small Convex hulls superimposed for both the species scale and mass scales of the leading towers in M-theory on $\mathbf{T}^2$. The black lines separate the different duality frames within the 9d set-up, corresponding to Type II string theory and 11d M-theory. The self-dual line, at $\hat{\tau}=0$, is fixed under the $\mathbb{Z}_2$ remnant symmetry.} 
\label{fig:ch2}
\end{center}
\end{figure}

In comparison with \cite{Etheredge:2022opl}, we also note that the roles of vectors saturating/protecting the convex hull condition are exchanged between strings and KK towers for the case of the species scale. This can be easily seen upon superimposing the diagrams for the $\zeta$- and $\mathcal{Z}$-vectors, as shown in Figure \ref{fig:ch2}. This moreover allows us to appreciate some hidden duality-like symmetry relating both convex hulls, which will be further investigated in Chapter \ref{ch:pattern} of this thesis.

\subsubsection*{Revisiting the axion}
\label{sss:axions}

Before turning to our next string theory example, let us briefly reconsider the role of the axion $\tau_1$ in our previous discussion. This simple exercise proves to be instructive and teaches us a generic lesson on the significance of compact scalar fields in our analysis of the bound \eqref{eq:lowerboundspecies2}.

Notice from eq. \eqref{eq:9d} that the two scalar sectors of the theory are decoupled (at the two-derivative level) and only the modular one, which describes the complex structure of the internal $\mathbf{T}^2$, contains the axion. Such complex-valued field parametrizes the manifold $\mathsf{SL(2, \mathbb{Z})}\backslash \mathsf{SL(2, \mathbb{R})}/\mathsf{U(1)}$, with $\mathsf{SL(2, \mathbb{Z})}$ being the U-duality group of the 9d theory.

Restricting ourselves to the fundamental domain, $\mathscr{F}$, of the aforementioned moduli space (see Figure \ref{sfig:funddom} below) it is transparent that there is just one infinite distance singularity, namely $\tau \to \i \infty$. This corresponds to a weak coupling limit for the fundamental (F1) Type IIA string, whose tension reads
\begin{equation}\label{eq:IIAtension}
	T_{\text{F1}} = \frac{2\pi}{\ell_{9}^2} \frac{1}{\sqrt{\tau_2}}  \mathcal{V}_2^{\frac{3}{14}}\, ,
\end{equation}
leading to the species scale vector $\vec{\mathcal{Z}}_{\text{osc}} = \left(\left(\mathcal{Z}_{\rm osc}\right)_{U}, \left(\mathcal{Z}_{\rm osc}\right)_{\tau_2}, \left(\mathcal{Z}_{\rm osc}\right)_{\tau_1} \right)$ given by the components displayed in the first vector of equation \eqref{eq:speciesscalevectorsstringsT2}, together with an additional axionic component of the form
\begin{equation} \label{eq:chargetomassF1}
 \left(\mathcal{Z}_{\rm osc}\right)_{\tau_1} = -\frac{1}{2}\sqrt{G^{\tau_1 \tau_1}}\, \partial_{\tau_1} \log \left(T_{\text{F1}}\right) = 0 \, .
\end{equation}
The fact that this extra component vanishes --- since the tension of the F1 does not depend on the axion --- implies that the decay rate parameter $\lambda_{\text{sp}}$ of the species scale along directions fulfilling $\tau \to \i \infty,\, U \to \infty$ is controlled just by the saxionic dependence of the species vector, regardless of the particular trajectory we follow so as to approach the weak coupling point. Furthermore, geodesics reaching infinity within $\mathscr{F}$ are given by vertical straight lines --- i.e. they satisfy $\dot \tau_1=\frac{d\tau_1}{d\sigma}=0$ with $\sigma \in \mathbb{R}$ some affine parameter --- such that we can effectively forget about the axionic direction $\tau_1$.

\begin{figure}[htb]
		\begin{center}
			\subfigure[]{
				\includegraphics[width=0.5\textwidth]{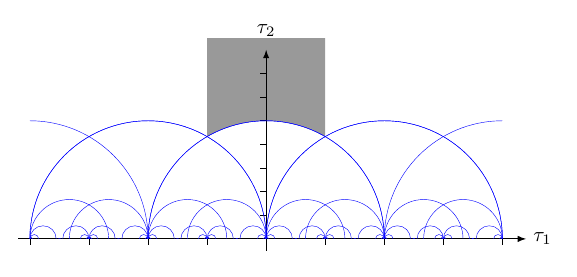}\label{sfig:funddom}
			}
            \quad
			\subfigure[]{
				\includegraphics[width=0.425\textwidth]{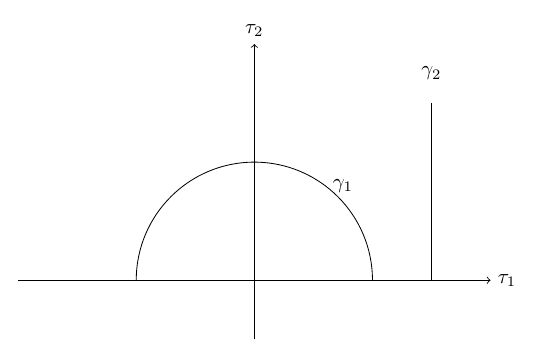}\label{sfig:geodesics}
			}
			\caption{\small \textbf{(a)} The upper half-plane $\mathbb{H}$ and the different $\mathsf{SL(2,\mathbb{Z})}$ domains one can consider, namely the `triangular' regions. The fundamental one $\mathscr{F}$ is shown shaded in grey. \textbf{(b)} Geodesics in the hyperbolic plane are given by vertical straight lines as well as half-circles intersecting the real axis at right angles.}
			
		\end{center}
\end{figure} 

At this point, one could complain that the above conclusion might be an artifact of restricting ourselves to the fundamental domain, since the theory presents a whole plethora of $\frac{1}{2}$-BPS $(p,q)$-strings whose tension in Planck units is given by eq. \eqref{eq:pqstrings9d} above. Moreover, the latter actually provide for the dominant tower of light states required by the Distance Conjecture upon approaching different boundaries of the 9d moduli space, and are characterized by the following species vectors
\begin{equation} \label{eq:chargetomasspq}
\left(\mathcal{Z}_{p,q}\right)_{U} = -\frac{1}{2 \sqrt{14}} \, , \qquad	
 \left(\mathcal{Z}_{p,q}\right)_{\tau_2} = \frac{(p+q\tau_1)^2-q^2\tau_2^2}{2\sqrt{2} |p+q\tau|^2}\, , \qquad \left(\mathcal{Z}_{p,q}\right)_{\tau_1} =  -\frac{q \tau_2 (p+q\tau_1)}{\sqrt{2} |p+q\tau|^2}\, .
\end{equation}
In fact, it is easy to check --- taking also into account the $\tau_1$--\,direction --- that these vectors densely fill a circumference of radius $1/\sqrt{7}$, regardless of the particular point in moduli space we are sitting at \cite{Etheredge:2022opl}.\footnote{This point becomes subtle if one approaches an infinite distance boundary by taking e.g., $\tau_2 \to \infty$, since to actually fill the circumference one needs to consider $(p,q)$ bound states with increasingly bigger F1-string charge, namely $p \sim k \tau_2$ for $k\in \mathbb{Z}$, since otherwise their charge-to-mass vectors collapse to that of the F1 or D1-strings. Upon doing so, eq. \eqref{eq:chargetomasspq} leads to $\vec{\mathcal{Z}}_{p,q} = \frac{1}{\sqrt{8}} \left(\frac{1}{\sqrt{7}}, \cos 2\theta, -\sin 2\theta\right)$, with $\cos \theta =k/\sqrt{k^2+q^2}$.} This suggests that perhaps one might need to include the axionic component $\left(\mathcal{Z}_{p,q}\right)_{\tau_1}$ into our analysis, since it is non zero in general depending crucially on the value of $\braket{\tau_1}$. However, as we argue in what follows, the latter turns out to be irrelevant whenever we reach an asymptotic boundary of $\mathcal{M}_{\text{9d}}$. The reason being that for geodesic paths reaching an infinite distance singularity other than $\tau_2 \to \infty$, namely any point with $\tau_2=0$ and $\tau_1 \in \mathbb{Q}$ in Figure \ref{sfig:funddom}, one finds again both $\dot \tau_1 \to 0$ asymptotically as well as $\left(\mathcal{Z}_{p,q}\right)_{\tau_1} \to 0$ for the \emph{lightest} $(p,q)$-string. Indeed, the geodesic trajectories reaching such infinite distance points are given by half-circles orthogonal to the real axis with endpoint $(\tau_1, \tau_2)=(-p/q,0)$, see Figure \ref{sfig:geodesics}. Hence, upon substituting these values for $\tau_2$ and $\tau_1$ we obtain $\vec{\mathcal{Z}}_{p,q} \to \left(-\frac{1}{2 \sqrt{14}}, \frac{1}{2\sqrt{2}}, 0 \right)$, matching those from \eqref{eq:speciesscalevectorsstringsT2} and \eqref{eq:chargetomassF1}. As expected, this is merely a manifestation of the fact that an $\mathsf{SL(2, \mathbb{Z})}$ transformation relating F1 to any other $(p,q)$-string (i.e. their tensions) maps the vertical lines reaching $\i\infty$ to the circles intersecting the real axis at $\tau_1=-p/q$.

Therefore, the take-home message is that whenever we find some modular sector within the moduli space of our theory and we want to check whether the bound \eqref{eq:lowerboundspecies2} is satisfied \emph{asymptotically}, it is sufficient to restrict ourselves to the fundamental domain and focus just on the saxionic components of the species scale vectors. This follows since any other infinite distance path associated to a different $(p,q)$ tower can be effectively translated --- via some modular transformation --- to this simplified set-up. Notice that restricting to the fundamental domain means, in turn, that when plotting the convex hull in e.g., Figure \ref{sfig:9dCH} it is not necessary to consider directions exploring $\tau_2 \to 0$, since those are already accounted for once we sit in the appropriate duality frame.

\subsection{M-theory on $\mathbf{T}^3$}
\label{ss:MthyT3SSDC}

Let us now consider M-theory compactified on $\mathbf{T}^3$, yielding a 8d $\mathcal N=2$ supergravity effective field theory. The gravitational and scalar sectors are described by the following Einstein-frame action
\begin{align}\label{eq:8d}
	S_\text{M-th}^{\text{8d}}\, \supset\, &\frac{1}{2\kappa_8^2} \int \dd^{8}x\, \sqrt{-g}\,  \left( \mathcal{R} + \frac{1}{4} \text{tr} \left( \partial \tilde{g} \cdot \partial \tilde{g}^{-1} \right) -\frac{\partial \mathcal{T} \cdot \partial \bar{\mathcal{T}}}{2 \mathcal{T}_2^2} \right)\, ,
\end{align}
where $\tilde{g}_{m n}$ is related to the internal metric of the $\mathbf{T}^3$ and $\mathcal{T}=C_{123}^{(3)} + \text{i} \mathcal{V}_3$ is a complex field containing in particular the overall volume modulus (see Section \ref{ss:8dmaxsugra} for details). In this case, the U-duality group of the theory is enhanced to $\mathsf{SL(2, \mathbb{Z})} \times \mathsf{SL(3, \mathbb{Z})}$, with the above fields transforming in the adjoint representation of each of these factors. They moreover parametrize the group coset 
\begin{equation}\label{eq:8dmodspaceSSDC}
 \mathcal{M}_{\text{8d}}=\mathsf{SL(2, \mathbb{Z})}\backslash \mathsf{SL(2, \mathbb{R})}/\mathsf{U(1)} \times \mathsf{SL(3, \mathbb{Z})}\backslash \mathsf{SL(3, \mathbb{R})}/\mathsf{SO(3)}\, ,
\end{equation}

see discussion around eq. \eqref{eq:8dsl3}. The goal of this section is twofold: First we want to check that the convex hull condition 
\begin{equation} \label{eq:bound8d}
  \lambda_{\text{sp}} \geq \frac{1}{\sqrt{(d-1)(d-2)}} \stackrel{\text{8d}}{=} \frac{1}{\sqrt{42}}\, ,
\end{equation}
is indeed satisfied in the present set-up; and second we would like to understand how quantum gravity avoids the naive violation of the bound \eqref{eq:lowerboundspecies2} for the species scale in $D\leq9$, as discussed in our simple toy model from Section \ref{ss:compactificationstring} above. Of course, the mechanism by which this happens is precisely the appearance of new towers of light states along certain asymptotic directions, as will be verified later on. 

In principle, in order to check the condition \eqref{eq:bound8d} in full generality, one should take into account the dependence on the compact scalar fields (e.g., $ C_{123}^{(3)}= \text{Re}\, \mathcal{T}$) of the mass scale and $\zeta$-vectors associated to the infinite towers of states. However, for simplicity and in light of our analysis in nine dimensions, we will henceforth freeze all the axion fields in the theory, thus exploring geodesics in moduli space which move just along the non-compact directions. In any event, by making use of $\mathsf{SL(2, \mathbb{Z})} \times \mathsf{SL(3, \mathbb{Z})}$ duality, it is clear one can actually relate such trajectories to analogous ones exploring some other equivalent infinite distance singularity.

Therefore, following the logic of Section \ref{ss:compactificationstring}, let us rewrite the scalar lagrangian \eqref{eq:8d} as if it was obtained by circle-reduction from the 9d theory \eqref{eq:9d}, instead of compactifying M-theory directly on $\mathbf{T}^3$. Hence, we take the 9d metric in \eqref{eq:11dmetric} and propose the following ansatz
\begin{equation}\label{eq:9dmetric}
	ds^2_{9} = e^{-\sqrt{1/21}\,\rho} ds_8^2 + e^{\sqrt{12/7}\,\rho} \left(dy^3 \right)^2\, ,
\end{equation}
with the radion field $\rho$ being related to the extra radius by $R_3=e^{\sqrt{3/7}\,\rho}$. After dimensional reduction, one arrives at (c.f. eq. \eqref{eq:8dalternativeaction} for the full scalar action) 
\begin{equation}\label{eq:8dcanonical}
	S^{\text{8d}}_{\text{M}} \supset \frac{1}{2\kappa_8^2} \int \dd^{8}x\, \sqrt{-g}\,  \left( \mathcal{R} - \left( \partial \hat U \right)^2 - \left(\partial \hat \tau\right)^2 - \left( \partial \hat \rho \right)^2 \right) + \left( \text{axions} \right)\, ,
\end{equation}
where we have introduced the canonically normalized radion $\hat \rho=\frac{\rho}{\kappa_8 \sqrt{2}}$ as well as those defined in \eqref{eq:canonicalnormalization}. Note that the compactification process just described can be equivalently seen as a $\mathbf{T}^2$-fibration over $\mathbf{S}^1$.

With this, we can now start computing the mass scales of the infinite tower of states, their charge-to-mass vectors $\vec{\zeta}$, as well as their (combined) species scales, similarly to what we did in the 9d setting. The relevant towers are shown in Table \ref{tab:BPSstates}. 
\begin{table}[tb]\begin{center}
\renewcommand{\arraystretch}{2.00}
\begin{tabular}{|c|c|c|c|}
\hline BPS states & Microscopic interpretation & Tension & Multiplicity \\
\hline \hline 
strings &  M2 on $\mathbf{S}^1_i$  &    $T = \frac{2 \pi}{\ell_{11}^3} (2\pi\, \mathsf{R}_i)$  &   $3$ \\
particles (non-pert.) & M2 on $\mathbf{S}^1_i \times \mathbf{S}^1_j$  &    $m_{\text{M2}} = \frac{2 \pi}{\ell_{11}^3} (4\pi^2\, \mathsf{R}_i\, \mathsf{R}_j)$  &   $3$ \\
particles (pert.) &  KK from $\mathbf{S}^1_i$  &    $m_{\text{KK}}= \frac{1}{\mathsf{R}_i}$  &   $3$\\
\hline
\end{tabular}
\caption{Relevant towers of $\frac{1}{2}$-BPS states in M-theory on $\mathbf{T}^3$. Their mass/tension is computed in terms of the dimensionful radii $\mathsf{R}_i$ and the 11d Planck length $\ell_{11}$.}
  \label{tab:BPSstates}
  \end{center}\end{table} 

%
%

Let us first study the case of solitonic critical strings. It is easy to see that apart from the set of $(p,q)$-strings which were already present in nine dimensions, we get an additional one by wrapping the M2-brane along the extra $\mathbf{S}^1$, whose tension is given by
\begin{equation}\label{eq:8dstring}
	T_{\text{str}''} =  \frac{2\pi}{\ell_8^2} e^{-2\hat U/\sqrt{14}}\, e^{2\sqrt{2}/\sqrt{21}\, \hat \rho} \quad \Longrightarrow \quad \frac{m_{\text{osc}''}}{M_{\text{Pl};\, 8}}  \sim e^{-\frac{1}{\sqrt{14}}\hat U}\, e^{\sqrt{\frac{2}{21}}\, \hat \rho}\, .
\end{equation}
Therefore, upon taking this extra M2-string into account, we arrive at the following species vectors associated to towers of oscillator modes
\begin{equation} \label{eq:stringvectors}
\begin{split} 
	\vec{\mathcal{Z}}_{\text{osc}} &= \left( \frac{1}{2\sqrt{2}} , \frac{1}{\sqrt{42}}, -\frac{1}{2 \sqrt{14}} \right)\, , \qquad \vec{\mathcal{Z}}_{\text{osc}'} = \left( -\frac{1}{2\sqrt{2}} , \frac{1}{\sqrt{42}}, -\frac{1}{2 \sqrt{14}} \right) \, ,\\
	\vec{\mathcal{Z}}_{\text{osc}''} &= \left( 0 , -\sqrt{\frac{2}{21}}, \frac{1}{\sqrt{14}} \right) \, ,
\end{split}
\end{equation}
where we have adopted the notation $\vec{\mathcal{Z}} = \left(\mathcal{Z}_{\hat \tau}, \mathcal{Z}_{\hat \rho}, \mathcal{Z}_{\hat U} \right)$. To obtain the first two we made use of eq. \eqref{string-species} together with the second relation in \eqref{eq:zvectorafterdimreduction}. Notice that they satisfy $|\vec{\mathcal{Z}}_{\text{osc}}|^2=1/(d-2)=1/6$, as expected. Physically, these strings can be interpreted as a fundamental Type IIA string and S-duals thereof by choosing one reference 1-cycle as the M-theory circle.

For the Kaluza-Klein towers, we proceed analogously to the string case above, namely we borrow the results from the 9d set-up and then compute the additional KK scale associated to the extra circle. One then finds
\begin{equation} \label{eq:KKvectors}
\begin{split} 
	\vec{\mathcal{Z}}_{\text{KK},\, 1} &= \left( \frac{1}{7\sqrt{2}} , \frac{1}{7 \sqrt{42}}, \frac{3}{7 \sqrt{14}} \right) \, , \qquad \vec{\mathcal{Z}}_{\text{KK},\, 1'} = \left( -\frac{1}{7\sqrt{2}} , \frac{1}{7 \sqrt{42}}, \frac{3}{7 \sqrt{14}} \right) \, ,\\
	\vec{\mathcal{Z}}_{\text{KK},\, 1''} &= \left( 0 , \frac{1}{\sqrt{42}}, 0 \right) \, .
\end{split}
\end{equation}
Note that all these vectors verify $|\vec{\mathcal{Z}}_{\text{KK}}|^2=1/(d-1)(d-2)=1/42$, thus saturating the bound \eqref{eq:bound8d}. In string theory language, they become two distinct KK towers and (bound states of) D0-branes upon choosing --- without loss of generality --- some $\mathbf{S}^1$ as the M-theory circle. Moreover, the first two vectors, which were already present in nine dimensions, have some $\lambda_{\text{sp}}$--\,parameter whose functional form is preserved upon dimensional reduction, in agreement with our claims from Section \ref{s:consistencydimreduc}.

Lastly, the other relevant set of $\frac{1}{2}$-BPS particles arises from M2-branes wrapping any 2-cycle of the internal manifold, as shown in Table \ref{tab:BPSstates}. These can be either seen as winding modes associated to each one of the three critical strings displayed in \eqref{eq:stringvectors}, or alternatively, as the 8d analogues of the F-theory tower discussed around eq. \eqref{eq:Ftheorytower}. In any event, it is clear that they are nothing but KK towers associated to decompactification limits in a dual Type IIB frame. Their species vectors can be easily computed to be
\begin{equation} \label{eq:M2vectors}
\begin{split} 
	\vec{\mathcal{Z}}_{\text{M},\, 1} &= \left( \frac{1}{7\sqrt{2}}, -\frac{5}{7 \sqrt{42}}, -\frac{1}{7 \sqrt{14}} \right) \, , \qquad \vec{\mathcal{Z}}_{\text{M},\, 1'} = \left( -\frac{1}{7\sqrt{2}}, -\frac{5}{7 \sqrt{42}}, -\frac{1}{7 \sqrt{14}} \right) \, ,\\
	\vec{\mathcal{Z}}_{\text{M},\, 1''} &= \left( 0, \frac{1}{7 \sqrt{42}}, -\frac{\sqrt{8}}{7 \sqrt{7}} \right)\, .
\end{split}
\end{equation}
which verify $|\vec{\mathcal{Z}}_{\text{M}}|^2=1/(d-1)(d-2)=1/42$ as well. 

We are not done yet, though. As we learned from the general analysis of Section \ref{s:consistencydimreduc} and the 9d example above, one needs to take into account effective combinations of towers of states, whose associated species scale can be lower than naively expected. For the case at hand, the ones that will be relevant are those formed by bound states involving only Kaluza-Klein replica, those comprised by M2-particles alone and a mixture of these two sectors. The KK bound states lead to the following set of vectors
\begin{equation} \label{eq:KKvectorscombined}
\begin{split} 
	\vec{\mathcal{Z}}_{\text{KK},\,2} &=  \left( 0, \frac{1}{4 \sqrt{42}}, \frac{3}{4 \sqrt{14}} \right) \, , \qquad \qquad \, \vec{\mathcal{Z}}_{\text{KK},\,2'} =  \left( \frac{1}{8 \sqrt{2}}, \frac{1}{ \sqrt{42}}, \frac{3}{8 \sqrt{14}} \right) \, ,\\
	\vec{\mathcal{Z}}_{\text{KK},\,2''} &=  \left( -\frac{1}{8 \sqrt{2}}, \frac{1}{ \sqrt{42}}, \frac{3}{8 \sqrt{14}} \right) \, , \qquad \vec{\mathcal{Z}}_{\text{KK},\,3} =  \left( 0, \frac{1}{ \sqrt{42}}, \frac{2}{3 \sqrt{14}} \right) \, .
\end{split}
\end{equation}
whilst for the M2-particles one finds instead
\begin{equation} \label{eq:M2vectorscombined}
\begin{split} 
	\vec{\mathcal{Z}}_{\text{M},\, 2} &=  \left( 0, -\frac{5}{4 \sqrt{42}}, -\frac{1}{4 \sqrt{14}} \right) \, , \qquad \qquad \ \ \vec{\mathcal{Z}}_{\text{M},\, 2'} =  \left( \frac{1}{8 \sqrt{2}}, -\frac{1}{ 2 \sqrt{42}}, -\frac{5}{8 \sqrt{14}} \right) \, ,\\
	\vec{\mathcal{Z}}_{\text{M},\, 2''} &=  \left( -\frac{1}{8 \sqrt{2}}, -\frac{1}{ 2 \sqrt{42}}, -\frac{5}{8 \sqrt{14}} \right) \, , \qquad \vec{\mathcal{Z}}_{\text{M},\, 3} =  \left( 0, -\frac{1}{ \sqrt{42}}, -\frac{2}{3 \sqrt{14}} \right) \, ,
\end{split}
\end{equation}
Finally, one can construct effective BPS towers of M2-particles with non-trivial KK momentum along the 1-cycle they do not wrap. These will be denoted as
\begin{equation} \label{eq:KK&M2vectorscombined}
\begin{split} 
	\vec{\mathcal{Z}}_{\text{KK-M},\, 2} &=  \left( 0, \frac{1}{\sqrt{42}}, -\frac{1}{2 \sqrt{14}} \right) \, , \qquad \vec{\mathcal{Z}}_{\text{KK-M},\, 2'} =  \left( -\frac{1}{4 \sqrt{2}}, -\frac{1}{2 \sqrt{42}}, \frac{1}{4 \sqrt{14}} \right) \, ,\\
	\vec{\mathcal{Z}}_{\text{KK-M},\, 2''} &=  \left( \frac{1}{4 \sqrt{2}}, -\frac{1}{2 \sqrt{42}}, \frac{1}{4 \sqrt{14}} \right) \, .
\end{split}
\end{equation}
The physical interpretation of the species scales associated to the vectors \eqref{eq:KKvectorscombined}-\eqref{eq:KK&M2vectorscombined} is straightforward: The first three of each set can be seen to be effective towers with density parameter $p=2$, such that they satisfy $|\vec{\mathcal{Z}}_{\rm sp}|^2=1/24$ and moreover implement some double decompactification to ten dimensions (possibly in a dual frame). On the other hand, the last vector of both eqs. \eqref{eq:KKvectorscombined} and \eqref{eq:M2vectorscombined} take us back to 11d M-theory, and as such verify $|\vec{\mathcal{Z}}_{\text{KK},\,3}|^2=|\vec{\mathcal{Z}}_{\text{M},\,3}|^2=1/18$. 

\subsubsection*{Plotting the convex hull}

\begin{figure}[htb]
		\begin{center}
			\subfigure[]{
				\includegraphics[width=0.45\textwidth]{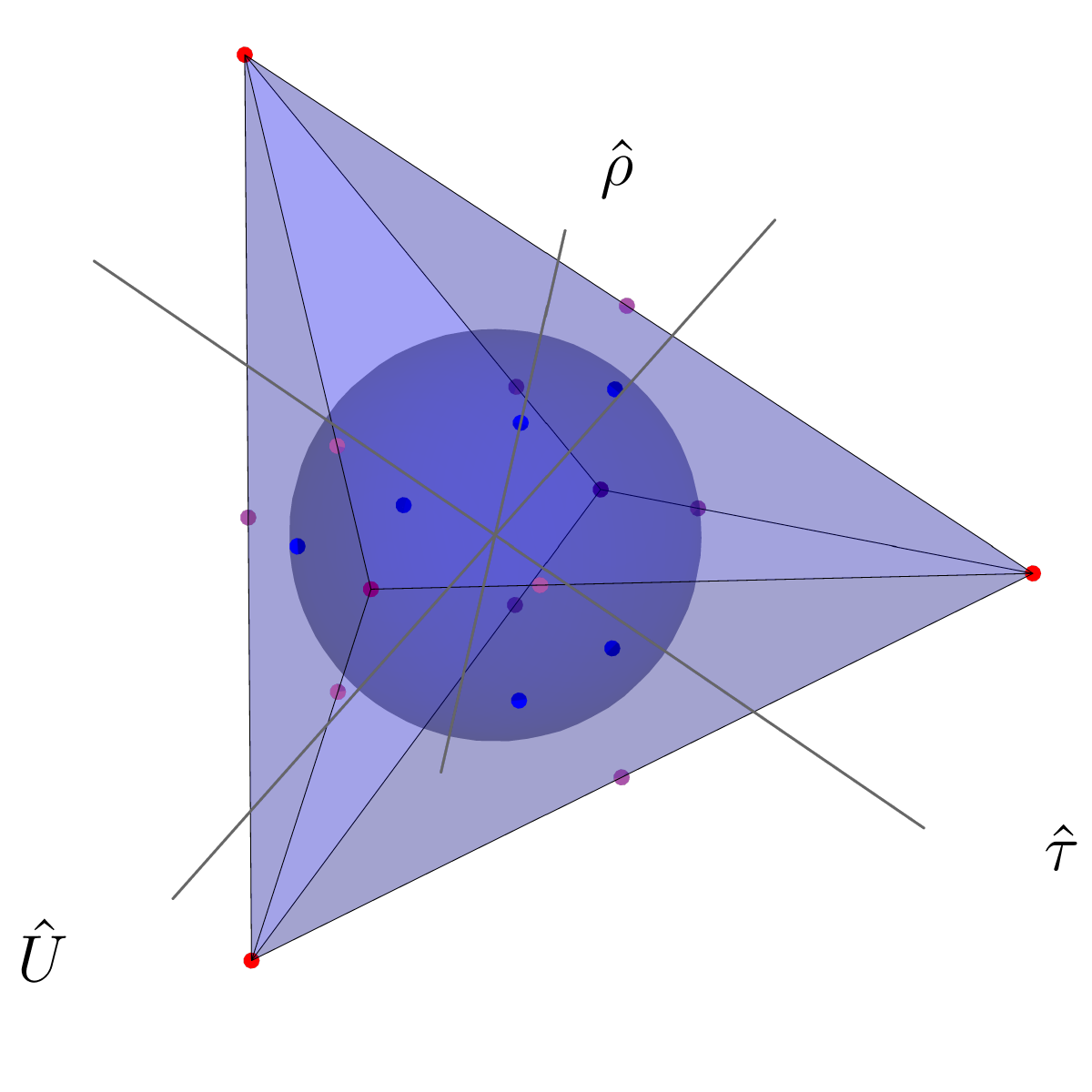}\label{sfig:slice3dCH}
			}
			\subfigure[]{
				\includegraphics[width=0.45\textwidth]{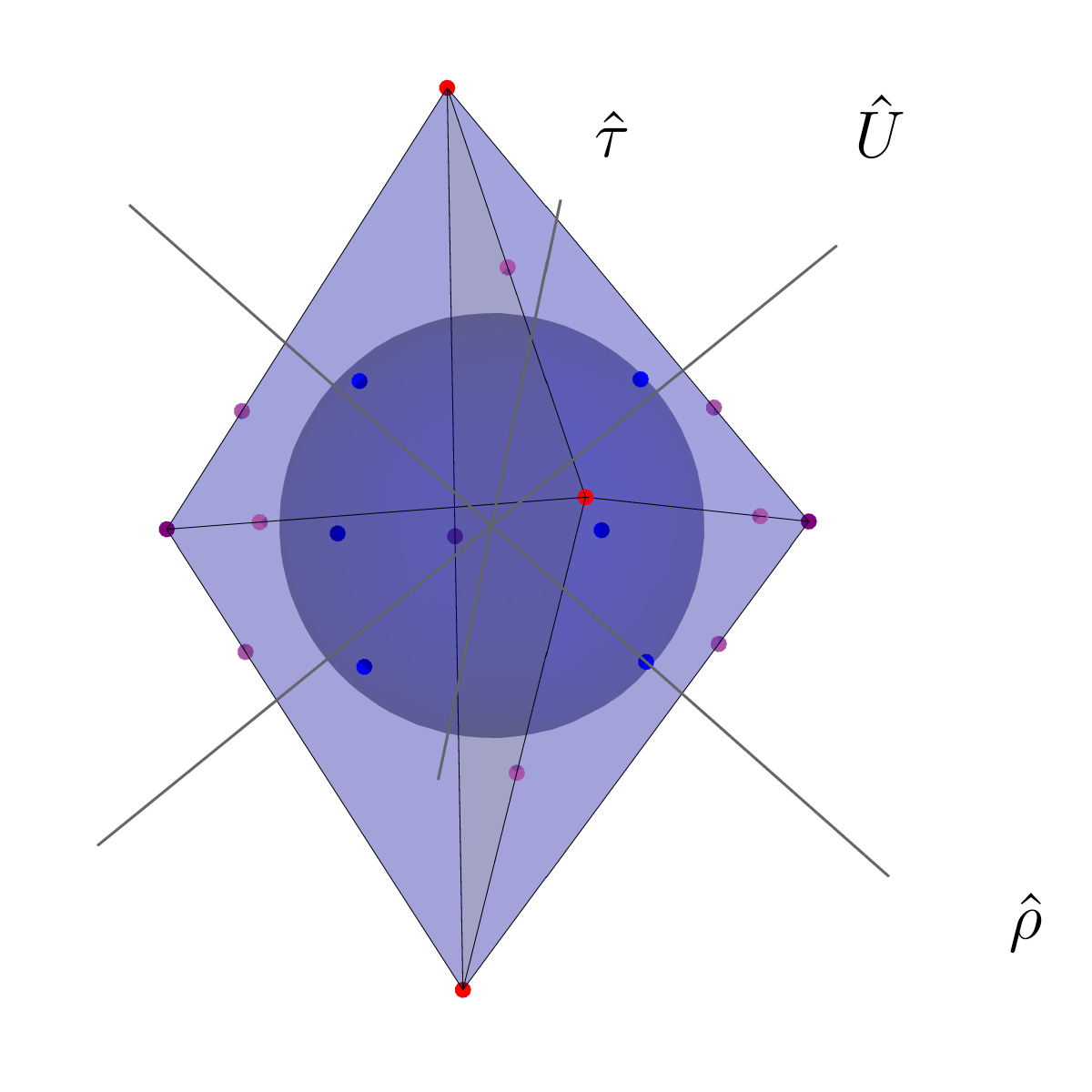}\label{sfig:2ndslice3dCH}
			}
			\caption{\small Convex hull for the species scale in M-theory on $\mathbf{T}^3$ with the axions set to a constant value, as seen from two different angles. The blue dots in the faces of the resulting polyhedron correspond to single KK towers ($p=1$), the light purple dots in the edges indicate double KK towers ($p=2$) and the dark purple and red dots in the vertices correspond to triple KK towers ($p=3$) and string towers, respectively.}
			\label{fig:ch8dgeneric}
		\end{center}
\end{figure}

Once we have all the $\mathcal{Z}$-vectors associated to the individual species scales, we can plot them in a 3d graph to check whether the CHC is satisfied or not. This is shown in Figure \ref{fig:ch8dgeneric} from two different perspectives, where one can see very explicitly that the convex hull for the present 8d example contains the ball of radius $\frac{1}{\sqrt{42}}$, thus fulfilling the bound \eqref{eq:lowerboundspecies2}. 

Interestingly, and similarly to what happened in the 9d setting discussed in Section \ref{ss:MthyT2SSDC}, the convex hull diagram for the species vectors features various nice properties capturing both the symmetries of the quantum theory as well as the relevant physics associated to the infinite distance boundaries in $\mathcal{M}_{\text{8d}}$. Indeed, from Figure \ref{sfig:2ndslice3dCH} (see also Figure \ref{fig:ch8dSL3} below) one can easily spot a surviving $\mathsf{S}_3 \times \mathsf{S}_2$ symmetry, where $\mathsf{S}_n$ denotes the permutation group of $n$ elements. Such discrete group can be thought of as a remnant of the U-duality group existing in the full eight-dimensional theory (see footnote \ref{fnote.Weyl} for more on this). In fact, this implies that the convex hull for the species scale is completely encoded within some \emph{fundamental domain} $\mathscr{F}_{8}$, which is replicated by acting on it with different elements of the discrete symmetry group. This latter observation will be crucial later on in Section \ref{ss:MthyTk} so as to extend the present analysis to set-ups with maximal supersymmetry in $d<8$.

\begin{figure}[htb]
\begin{center}
\includegraphics[scale=.4]{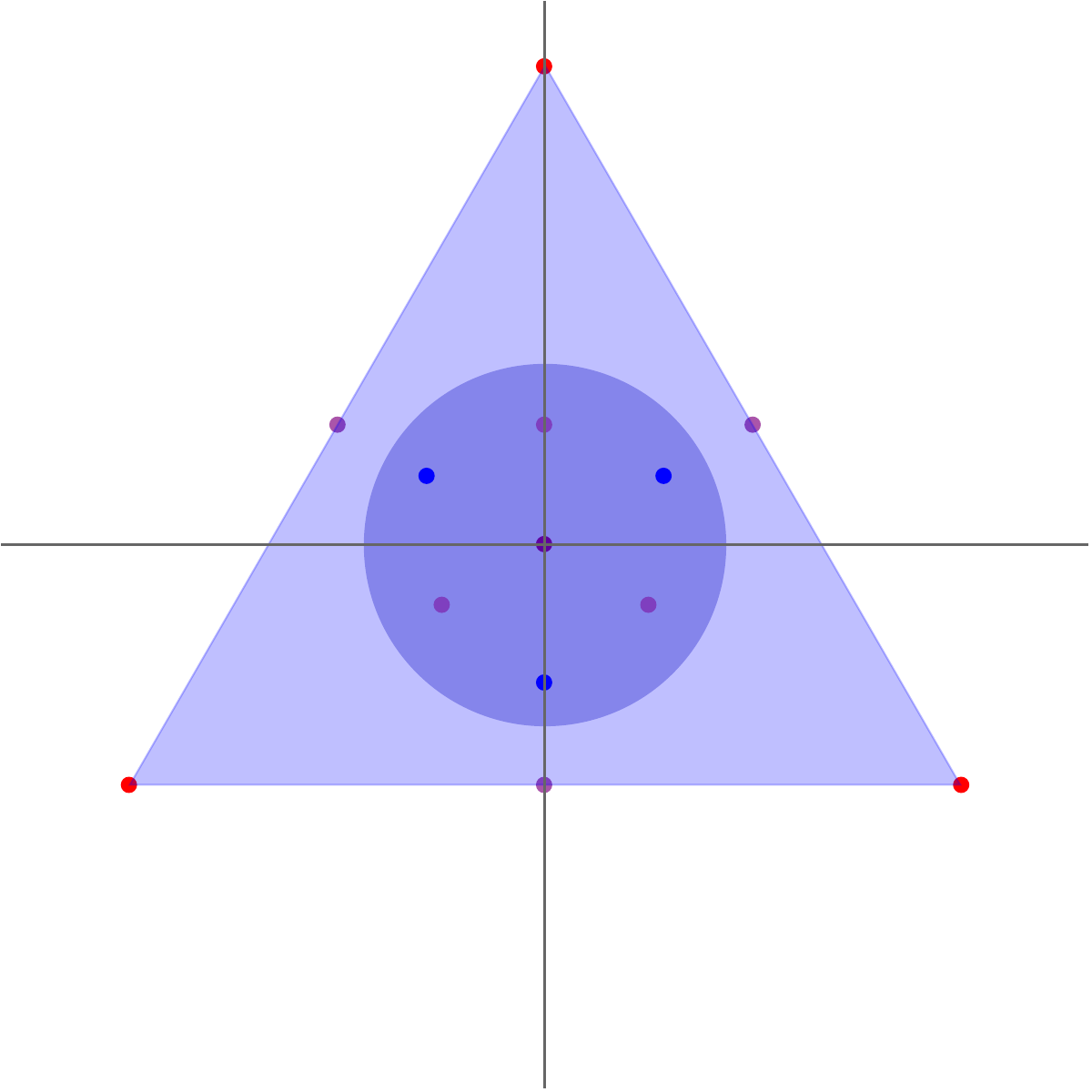}
\caption{\small Two-dimensional projection of the convex hull for the species scale in M-theory on $\mathbf{T}^3$. The slice is chosen with respect to the normal vector $\hat T = \frac{\vec{\mathcal{Z}}_{\text{KK},\, 3}}{|\vec{\mathcal{Z}}_{\text{KK},\, 3}|}$, making thus manifest the discrete symmetry remnant of the $\mathsf{SL(3, \mathbb{Z})}$ duality (sub-)group.} 
 \label{fig:ch8dSL3}
 \end{center}
 \end{figure}

Regarding the structure exhibited by the diagram, let us first notice that the convex hull is fully generated again by the species vectors associated to either emergent string limits (the red dots in Figure \ref{fig:ch8dgeneric}) or full decompactification to 11d M-theory (the purple dots). On the other hand, the vectors marked by blue dots, corresponding to $p=1$ towers, saturate the constraint \eqref{eq:bound8d} and appear precisely at the faces of the convex polyhedron, being moreover perpendicular to the latter. In fact, these faces turn out to be nothing but the convex hull diagram of the `parent' 9d theory (see Figure \ref{sfig:9dCH-2}), which in turn include at its edges the convex hull for the different limiting 10d theories. Therefore, we find an inductive sequence of the form
\beq
    {\rm Hull}\left(\{\vec{\mathcal{Z}}_I\}\right)\big\rvert_{\rm 8d} \supset {\rm Hull}\left(\{\vec{\mathcal{Z}}_I\}\right)\big\rvert_{\rm 9d} \supset {\rm Hull}\left(\{\vec{\mathcal{Z}}_I\}\right)\big\rvert_{\rm 10d} \supset {\rm Hull}\left(\{\vec{\mathcal{Z}}_I\}\right)\big\rvert_{\rm 11d}\, ,
\eeq
informing us about all possible infinite distance limits that can be explored either directly from the lower dimensional perspective, or rather by passing first through some intermediate higher-dimensional frame.  

\subsubsection*{Comparison with the toy model}

Finally, in order to appreciate the crucial role played by the effective towers of states so as to ensure that the convex hull condition to be verified, let us study one concrete 2d slice of Figure \ref{fig:ch8dgeneric}, namely the one spanned by the $\{\hat \rho, \hat \tau \}$ directions. Hence, upon projecting the $\mathcal{Z}$-vectors down to a plane characterized by its normal $\hat T = \partial_{\hat{U}}$ as follows
\begin{equation}\label{eq:2dprojection}
	\vec{\mathcal{Z}}_{\hat T} = \vec{\mathcal{Z}} - \left( \hat T \cdot \vec{\mathcal{Z}}\right) \hat T\, ,
\end{equation}
one obtains precisely what is shown in Figure \ref{fig:ch-comparison-toymodel} below. The reason for choosing this particular slice is because it can be easily connected to the situation discussed previously in Section \ref{ss:compactificationstring}. There we showed that, upon compactifying a $D$-dimensional theory containing two S-dual strings on a circle, the CHC seemed to be naively violated whenever $D\leq9$. Indeed, the lower-dimensional vectors associated to the pair of strings together with their winding modes --- and effective combinations thereof, which are crucial to satisfy \eqref{eq:lowerboundspecies2} if $D=10$, were actually not sufficient to preserve the latter when starting from the non-critical dimension (c.f. Figure \ref{sfig:9dto8d}). On the other hand, if we consider M-theory on $\mathbf{T}^3$ instead, which of course provides for an eight-dimensional EFT coming from quantum gravity, there is indeed no violation of the CHC. In fact, the projected 2d slice of the hull depicted in Figure \ref{fig:ch-comparison-toymodel} turns out to have a similar structure to the one exhibited by the toy model, which is also shown for comparison. Indeed, in both cases there are two strings that correspond to the red dots appearing in the upper half of the image, which are connected by a line that is tangent to the critical ball of radius $\frac{1}{\sqrt{42}}$. The saturation happens precisely along the asymptotic direction where the KK tower associated to the extra $\mathbf{S}^1$ becomes relevant. In the opposite regime, namely along the negative $\hat \rho$ axis, where the (effective) tower of winding modes do not give rise to a convex hull containing the critical ball, a genuine 8d critical string appears in M-theory (i.e. the one in eq. \eqref{eq:8dstring}). It is precisely this new tower of oscillator modes, which was absent in the toy model, the one ensuring that the bound is preserved along every direction in the resulting 2d graph.


\begin{figure}[htb]
\begin{center}
	\subfigure[M-theory on $\mathbf{T}^3$]{\includegraphics[width=0.45\textwidth]{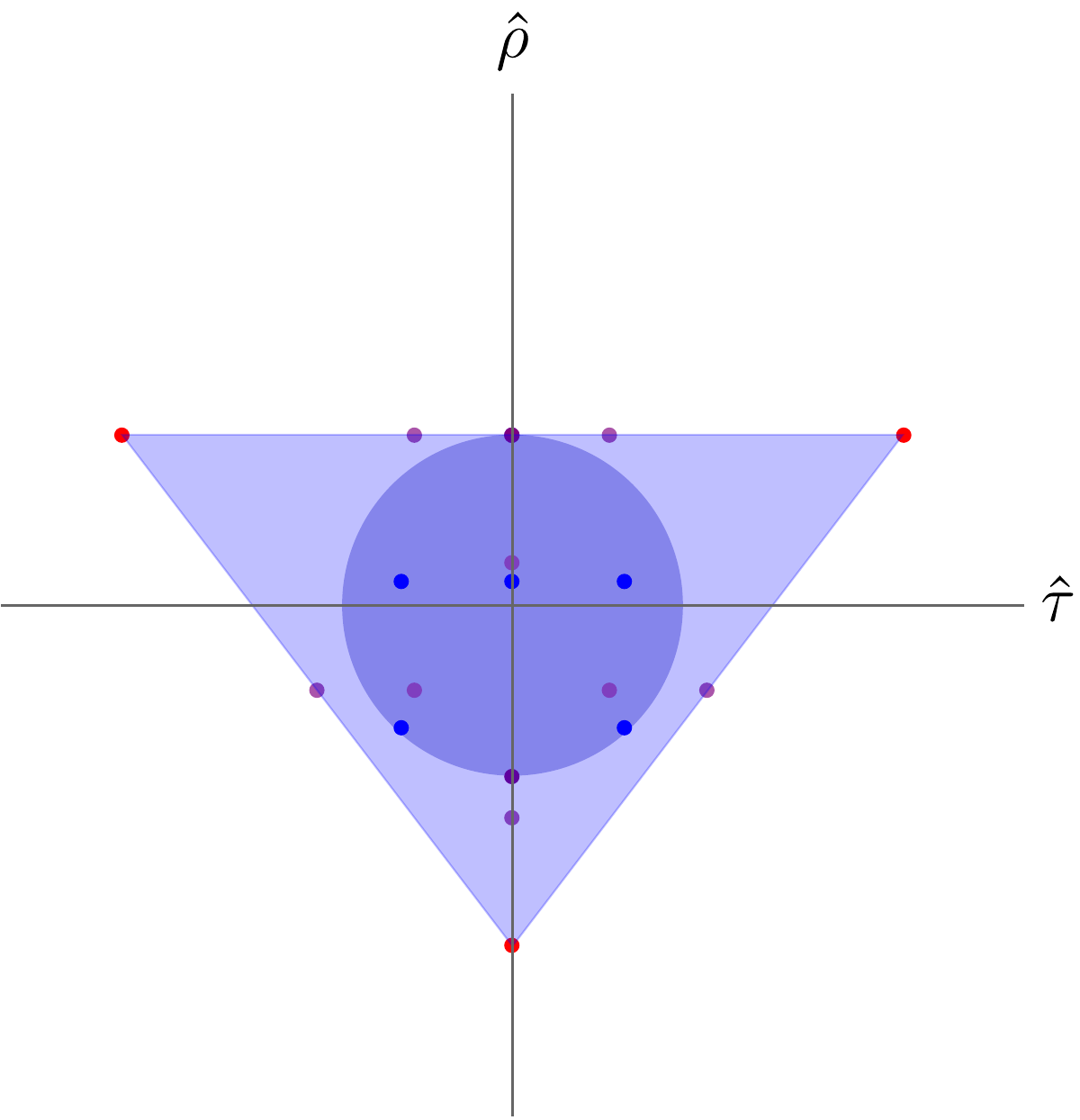}}\label{sfig:2dslice}
    \quad
	\subfigure[Toy model]{\includegraphics[width=0.45\textwidth]{CH-5.pdf}}\label{sfig:toymodel8d}
	\caption{\small \textbf{(a)} Two-dimensional projection of the convex hull for the species scale in M-theory on $\mathbf{T}^3$ along the plane perpendicular to $\hat T = \partial_{\hat{U}}$. \textbf{(b)} Convex hull diagram for the 9d $\to$ 8d compactification of the toy model discussed in Section \ref{ss:compactificationstring}.}
	\label{fig:ch-comparison-toymodel}	
\end{center}
\end{figure}

\subsection{M-theory on $\mathbf{T}^k$}
\label{ss:MthyTk}

In this subsection we extend the results from the previous examples in nine and eight dimensions to M-theory compactifications on $\mathbf{T}^k$, for $k>3$. This will provide strong evidence in favour of the bound \eqref{eq:lowerboundspecies2} as well as the idea that there seems to exist a minimum rate at which the species scale can fall-off at infinity.

Our argument proceeds inductively, relying heavily both on U-duality (c.f. Section \ref{s:dualities}) and the uniqueness of the supergravity theory for $d<9$ spacetime dimensions. Crucially, U-duality forces all particle states comprising infinite towers with $p=1$ to arrange themselves into a \emph{single} irreducible representation of the symmetry group (see Table \ref{tab:irreps} below).\footnote{This is actually not true in 9d, where the KK-like towers form a $\mathbf{2} \oplus \mathbf{1}$ representation of the $\mathsf{SL(2, \mathbb{Z})}$ duality group. The reason for this hinges on the fact that there are two different limiting theories in ten dimensions with 32 supercharges one can arrive at from 9d, depending on their chirality\cite{Hull:1994ys}.} Such orbit includes both perturbative (i.e. KK, winding modes, etc.) and non-perturbative states (wrapped $p$-branes, KK-monopoles, etc.), as seen from the original duality frame. For instance, consider M-theory compactified on $\mathbf{T}^4$ down to 7d. The U-duality group is identified with $\mathsf{SL(5, \mathbb{Z})}$ in this case, and the particle multiplet transforms as the $\mathbf{10}$ of $\mathsf{SL(5, \mathbb{Z})}$. These states may be understood microscopically as four Kaluza-Klein towers associated to the compact directions, as well as six additional infinite sets of wrapped M2-particles.

Hence, we start with a couple of insights provided by the examples analyzed in Sections \ref{ss:MthyT2SSDC} and \ref{ss:MthyT3SSDC}. There we saw that the states with $p=1$ saturate the lower bound \eqref{eq:lowerboundspecies2}, and they appear precisely at the facets of the corresponding convex hull polytope. Furthermore, as already noticed, these facets turn out to be nothing but the convex hull of the theory in one dimension higher. Intuitively, this is easy to understand, since upon dimensionally reducing the supergravity theory, all $\mathcal{Z}$-vectors give rise to analogous ones combined with the KK tower associated to the extra circle. The latter generate the same polytope already existing in the higher-dimensional theory, which is moreover orthogonal to the species vector associated to decompactifying the new compact dimension, $\vec{\mathcal{Z}}_{\text{KK}}$ (see Section \ref{ss:field-theory} for details). Finally, the general analysis of Section \ref{s:consistencydimreduc} showed that if the theory we start with satisfies the CHC then, upon $\mathbf{S}^1$--\,compactification, we end up with a set of species vectors which still verify \eqref{eq:lowerboundspecies2} along all intermediate asymptotic directions.

With this we are now ready to argue that the condition \eqref{eq:lowerboundspecies2} is indeed satisfied in $d$-dimensional maximal supergravity for $d\geq4$. We work by induction, such that we first assume the bound to hold for M-theory compactified on $\mathbf{T}^k$, with species vectors denoted by $\{\vec{\mathcal{Z}}_{\text{t}}\}$. In a next step, we dimensionally reduce the theory on a circle, leading to M-theory on $\mathbf{T}^k \times \mathbf{S}^1 \cong \mathbf{T}^{k+1}$.\footnote{We freeze the axions to zero v.e.v., since as we learned from the previous examples they play no role whatsoever in our analysis (see Section \ref{ss:MthyT2SSDC}).} Based on the general considerations from the previous paragraph we conclude that the (sub-)polytope spanned by the set of vectors $\lbrace \vec{\mathcal{Z}}_{\text{KK}}, \vec{\mathcal{Z}}_{\text{KK-t},\, p+1}\rbrace$ satisfies the CHC, saturating the bound precisely along the direction determined by $\vec{\mathcal{Z}}_{\text{KK}}$ (c.f. equation \eqref{vectors}). However, from U-duality we know that this is already enough to ensure that the CHC holds in every asymptotic direction, since upon acting with the discrete remnant of the symmetry group, one can completely reconstruct the rest of the diagram. The latter presents as many identical facets as the dimension of the representation into which the $p=1$ towers fit into, which follows from the uniqueness of the supergravity action, at least for $d<9$. Therefore, by noticing that the condition \eqref{eq:lowerboundspecies2} was already shown to be satisfied in M-theory compactified on $\mathbf{T}^k$ for $k=1,2,3$, we thus conclude that the same remains true for maximal supergravity in lower dimensions as well.

\section{Summary}\label{s:summarybounds}

Let us summarize the main findings extracted from this chapter. First of all, we were able to determine a lower bound for the exponential decay rate $\lambda_{\rm sp}$ (c.f. eqs. \eqref{eq:defspeciesvectors}-\eqref{eq:decayratespecies} for a precise definition) of the quantum gravity cut-off close to infinite distance boundaries in field space, which only depends on the spacetime dimension of our theory. This translates into having a minimum rate at which the species scale can fall off with respect to the Planck mass at infinity, thereby forcing the exponential behaviour predicted by the Distance Conjecture \cite{Ooguri:2006in} when combined with other bounds proposed in the literature \cite{vandeHeisteeg:2023ubh, Calderon-Infante:2023ler}. Moreover, we were able to formulate this constraint in a manifestly reparametrization invariant way, namely in terms of a convex hull condition.

On the other hand, several interesting properties associated to the bound \eqref{eq:lowerboundspecies} were uncovered. In fact, we argued that the precise lowest possible value $ \frac{1}{\sqrt{(d-1)(d-2)}}$ is selected both empirically and by consistency with dimensional reduction, since it is the only one whose saturation is \emph{exactly} preserved by the compactification process. However, as typically happens with Swampland criteria, field-theoretic considerations do not suffice in general to ensure that the bound is satisfied for any possible infinite distance limit. This typically requires from the inclusion of additional extended objects in the theory, such as strings or higher-dimensional $p$-branes, as we readily confirmed with the examples discussed in Section \ref{s:examplesbound}.

In addition, we provided strong evidence for the constraint \eqref{eq:lowerboundspecies} via explicit verification in toroidal compactifications of M-theory, leading to maximal supergravity constructions in $d \geq 4$. This allowed us to extract further general lessons that we believe go beyond this highly-constrained systems. In particular, we observed that the convex hull diagrams were fully generated by string or Kaluza-Klein towers corresponding to full decompactification (i.e. back to 11d M-theory). This seems to capture the idea that, at the end of the day, we always expect the species scale to encode either the ten-dimensional string scale or the eleven-dimensional Planck mass. On the contrary, the species vectors signalling decompactification of just one extra dimension --- which therefore saturate the bound \eqref{eq:lowerboundspecies} --- were seen to lie always at the point closest to the origin within the different facets comprising the convex polytope. 

During the course of this investigation, we realized that the convex hulls associated to each of the examples analyzed so far exhibited rich geometric and symmetry casuistics, which are related to the duality properties of the theory. Furthermore, upon comparison with the convex hull determined by the tower scales, i.e. the $\zeta$-vectors (see discussion around eq. \eqref{eq:chargetomass} for details), we uncovered a non-trivial relation between the two. Indeed, the role of saturating/protecting towers for the corresponding lower bounds in the asymptotic exponential decay rates seemed to be exchanged. This hinges on some interesting mathematical duality relating both convex hull diagrams that will be further explored in the upcoming chapter.


\chapter{A Universal Pattern at Infinite Distance}\label{ch:pattern} 

In the previous chapter we proposed and tested a very precise lower bound \eqref{eq:lowerboundspecies} for the exponential decay rate of the quantum gravity cut-off, $\lambda_{\rm sp}$. As argued there, such constraint ought to be satisfied for any given geodesic trajectory that explores infinite distance within the moduli spaces of QG theories, and it can be nicely reformulated as a convex hull condition for the so-called species vectors $\vec{\mathcal{Z}}$ (c.f. eq. \eqref{eq:defspeciesvectors}). Interestingly though, during the course of the investigation it was found that the convex hull diagrams determined by the potential candidates for (asymptotic) QG cut-offs were intimately related to those constructed just out of the individual towers. Indeed, as observed in the explicit examples from toroidal compactifications of M-theory (see Section \ref{s:examplesbound}), the vertices of one diagram appeared to be `dual' to the facets of the other, and viceversa. The main goal of this chapter will be to revisit this point and argue that, in fact, the aforementioned symmetry property relating both convex hull constructions can be encapsulated into a very sharp mathematical identity, which we dub \emph{the pattern}:
\beq \label{eq:patternmass}
	\frac{\vec\nabla m_{\text{t}}}{m_{\text{t}}} \cdot\frac{\vec\nabla \LSP}{\LSP}= \frac{\kappa_d^2}{d-2}\, ,
\eeq
where the product is taken using the metric in the moduli space, $d$ denotes again the spacetime dimension of our theory and $\kappa_d^2= \Mpd^{2-d}$ is the gravitational coupling constant. As we will see in the following, this pattern is non-trivially satisfied in all (up to now explored) string theory examples, which is a priori quite surprising given the rich casuistics that typically arise when checking different possible models in quantum gravity. Notice that, when written in terms of the number of light species (i.e. the number of weakly coupled fields whose mass falls at or below the species scale), \eqref{eq:patternmass} reduces to
\beq
\label{patternN}
	\frac{\vec\nabla m_{\text{t}}}{m_{\text{t}}} \cdot\frac{\vec\nabla N}{N}=-\kappa_d^2\, ,
\eeq
since $\LSP=M_{\text{Pl};\, d}\, N^{-1/(d-2)}$. The purported universality of the pattern, which becomes independent of the number of spacetime dimensions or the nature of the infinite distance/perturbative limit, is at the very least tantalizing, and suggests that there might be an underlying reason constraining the structure of the allowed infinite towers of states that can arise as per the Distance Conjecture. In addition, one should note that the relation \eqref{patternN} puts constraints on the variation on the density of states below the species scale and the rate at which they are becoming light. Roughly speaking, the more dense the spectrum gets, the faster the species scale goes to zero and therefore the slower the tower should become light.
	
On the other hand, since by definition $m_{\text{t}}\leq \LSP$, eq. \eqref{eq:patternmass} implies a definite bound on how slow the tower mass can go to zero asymptotically in comparison to the species scale. Indeed, as discussed below, from the pattern one may obtain a lower bound for the exponential rate of the tower given by $\frac{1}{\sqrt{d-2}}$, which reproduces precisely the bound proposed in the sharpened Distance Conjecture (c.f. \eqref{eq:sharpenedDistConj}). This is also closely related to the Emergent String Conjecture (see Section \ref{s:SDC}), as the bound is saturated by a tower of oscillator modes of a fundamental string, while Kaluza-Klein modes usually have larger exponential rates. Hence, understanding the pattern \eqref{patternN} from the bottom-up opens a new avenue to test the Emergent String Conjecture independently of string theory.
	
Therefore, in order to convince ourselves that the pattern \eqref{eq:patternmass} could be realized universally in quantum gravity, we provide strong evidence for the latter by checking multiple string theory constructions in different number of spacetime dimensions and with different amounts of supersymmetry. This includes maximal supergravity set-ups, as well as theories with sixteen or eight unbroken supercharges. For each different level of supersymmetry, we select a few representative examples to illustrate the realization of the pattern. Furthermore, in certain moduli spaces, we can even derive \eqref{eq:patternmass} in full generality. However, for the moment, it should be taken purely as an interesting observation, since we do not have a clear-cut argument that allows us to discern whether it is a lamppost effect or a general feature of quantum gravity. In any event, it is interesting either way, for in the former case, it provides at the very least an elegant and universal constraint that summarizes the casuistics of infinite distance limits observed in known string theory compactifications. In the latter case, it could be the definite criterion that characterizes the tower of the Distance Conjecture and constrains its exponential mass decay rate, providing therefore information about the QG cut-off of an EFT from the bottom-up perspective. 
	
The outline of the chapter is as follows. We start with an explanation of the pattern and its consequences in Section \ref{s:patternintro}, and provide compelling evidence for it within large classes of string theory compactifications in subsequent parts. Section \ref{s:maxsugra} is dedicated to set-ups with maximal supersymmetry, whilst Sections \ref{s:16supercharges} and \ref{s:8supercharges} analyze theories with sixteen and eight supercharges, respectively.\footnote{See also \cite{Rudelius:2023spc} for more top-down evidence in 5d $\mathcal{N}=1$ supergravity as well as \cite{Castellano:2023jjt} for 4d settings with minimal amount of supersymmetry preserved.} Finally, in Section \ref{s:bottomup}, we give the first steps towards providing a bottom-up rationale for the constraint \eqref{eq:patternmass} and identify some underlying sufficient conditions.

This chapter is based on the publications \cite{PhysRevLett.132.181601,Castellano:2023jjt} which have been adapted to fit in the broader context of this thesis.


\section{The pattern and its consequences}
\label{s:patternintro}

Our starting point here will be the exact same set-up as the one discussed in Section \ref{s:convexhull}. Thus, we consider some generic $d$-dimensional effective field theory, whose gravitational and scalar sectors are described by the lagrangian \eqref{eq:action}. This includes, in particular, a set of massless/light scalar fields $\{ \phi^i\}$, which parametrize some manifold --- dubbed moduli space $\mathcal{M}_{\phi}$ --- and whose kinetic term is controlled by a rank-2 symmetric tensor $G_{ij}(\phi)$. This latter quantity moreover allows us to define a very natural notion of distance within $\mathcal{M}_{\phi}$. Furthermore, per the Distance Conjecture (c.f. eq. \eqref{eq:masslesstower}), there should exist infinite towers of states whose masses decrease exponentially (in Planck units) with respect to the aforementioned distance at infinity, to which we can moreover associate certain vector-like quantities, usually referred to as scalar charge-to-mass vectors
\begin{equation}\label{eq:chargetomass2}
	\zeta_i = -\partial_i \log m\, .
\end{equation}
On the other hand, when approaching said infinite distance limits, the presence of the infinite towers of light states will inevitably force the original EFT to break down in a dramatic fashion. This process is physically captured by the behaviour exhibited by the quantum gravity cut-off, i.e. the species scale $\LSP$, above which it is not possible to have a semi-classical Einstein gravity description anymore. Moreover, its precise value strongly depends on the nature and masses of the towers becoming light, and it is given by
\beq
\label{LSP}
	\LSP\, =\, \frac{M_{\text{Pl};\, d}}{N^{\frac{1}{d-2}}}\, ,
\eeq
with $N$ denoting the number of (light) species. Such quantity may be defined (at least asymptotically) as the number of distinguishable weakly coupled light fields which fall at or below the species scale itself, namely
\beq
\label{NSP}
	N=\int^{\LSP}_0\dd m\, \rho(m)\, ,
\eeq
which is an implicit equation for both $N$ and $\LSP$, and $\rho(m)$ denotes the density of species per unit mass. Importantly, notice that since gravity couples to everything that carries energy-momentum, not only the leading but all light towers of states indeed matter when computing $\LSP$. 
	
Now, since the towers become massless in an exponential fashion, the species scale will similarly vanish at the infinite distance boundary, although at a different rate. To account for this, and following our discussion in Section \ref{s:convexhull}, we define the $\mathcal{Z} $-vectors
\begin{equation}\label{eq:speciescalechargetomass}
	\mathcal{Z}_i = -\partial_i \log \LSP\, ,
\end{equation}
which provide the rate at which the species scale goes to zero for any given asymptotically geodesic trajectory. Crucially, depending on the limit under consideration, we may have in principle a different microscopic interpretation both for the leading tower and the species scale, which is intimately tied to the value of their exponential rates. These are, however, a priori independent of the precise relation between $m_{\rm t}$ and $\LSP$. What we want to put forward in this chapter is a presumably universal relation between the variation of the mass of the leading tower and that of the species scale through the following simple constraint\footnote{Notice the disappearance of the factor $\kappa_d^2$ in the right-hand side of eq. \eqref{eq:pattern} with respect to \eqref{eq:patternmass}. This follows from our conventions for the scalar fields $\{ \phi^i\}$, which have been defined so as to have a kinetic energy proportional to $\Mpd^{d-2}$, c.f. eq. \eqref{eq:action}.}
\begin{equation}\label{eq:pattern}
	\vec{\zeta}_{\rm t} \cdot \vec{\mathcal{Z}}_{\text{sp}} = G^{ij} \left(\partial_i \log m_{\rm t}\right) \left(\partial_j \log \LSP\right)= \frac{1}{d-2}\; .
\end{equation}
This pattern, which is satisfied at least asymptotically, holds in all the string theory examples that we present here, regardless of the nature of the infinite distance limit and the microscopic interpretation of the light towers. 
Even more interestingly, using \eqref{LSP}, we can rewrite the pattern as
\beq\label{eq:patternN}
	G^{ij} \left(\partial_i \log m_{\rm t}\right) \left(\partial_j \log N\right)=-1\; ,
\eeq
which is moreover independent of the number of spacetime dimensions. This hints towards some deep universal relation between the density of states becoming light and their characteristic mass: The faster they become light as we approach the infinite distance limit, the less dense the towers can get, and viceversa. In some sense (that we will make more concrete later),  the variation of the mass and the number of states in the moduli space act as `dual variables'.

\subsubsection*{Derived bounds on exponential decay rates}
	
Notice that a relation like \eqref{eq:pattern} implies a lower bound for the scalar charge-to-mass ratio of the leading tower asymptotically, since the latter should be always lighter than the species scale, i.e. $m_{\rm t}\leq \LSP$. This consistency condition, together with the assumption of an exponential behavior for both scales \cite{Ooguri:2006in}, imply that $|\vec{\zeta}_{\rm t} \cdot \vec{\mathcal{Z}}_{\text{sp}}| \leq |\vec{\zeta}_{\rm t}|^2$ and, therefore,
\begin{equation}\label{eq:Rudelius}
	|\vec{\zeta}_{\rm t}|^2 \geq \frac{1}{d-2}\, ,
\end{equation}
which leads to the lower bound for the exponential rate of the leading tower (c.f. eq. \eqref{eq:sharpenedDistConj})
\beq
	\lambda_{\text{t}} = |\vec{\zeta}_{\rm t}| \geq \frac{1}{\sqrt{d-2}}\, .
\eeq
Analogously, in those cases (as it happens in all known examples) in which there exists a tower $\vec{\zeta}\propto\vec{\mathcal{Z}}_{\rm sp}$ satisfying \eqref{eq:pattern}, then one gets an upper bound on the exponential rate of the species scale since $|\vec{\mathcal{Z}}_{\rm sp}|\leq |\vec{\zeta}_{\rm t}| $, yielding
\beq\label{eq:Harvardbound}
	\lambda_{\text{sp}} = |\vec{\mathcal{Z}}_{\text{sp}}| \leq \frac{1}{\sqrt{d-2}}\, ,
\eeq
which matches the condition recently proposed in \cite{vandeHeisteeg:2023ubh}\footnote{Note that the pattern \eqref{eq:pattern}, in its present formulation, is only defined asymptotically, and this is why it is consistent that the constant in the right-hand side of \eqref{eq:Harvardbound} is fixed to $\frac{1}{\sqrt{d-2}}$. This might get modified when moving towards the interior of the moduli space, see \cite{Bedroya:2024uva} for more on this.} based both on EFT arguments and string theory evidence.
	
Notice that the above bounds are always saturated by the oscillator modes of a fundamental string. Hence, if we assume that Kaluza-Klein (KK) towers always have a larger exponential rate $\lambda_{\text{t}}$ (as indeed happens in all examples known so far), we are essentially recovering the Emergent String Conjecture (ESC) \cite{Lee:2019wij} as well, assuming that membranes decay at a slower rate than particles and strings, which is the case in all known string theory constructions (see also \cite{Alvarez-Garcia:2021pxo}). It would be interesting, though, to show that the only possible towers of states satisfying the pattern are indeed KK towers or oscillator string modes (as implied by the ESC) from a purely bottom-up perspective (see \cite{Basile:2023blg} for recent progress along this direction).
	
We want to remark that the pattern \eqref{eq:pattern} is much more concrete than previous analyses as it provides a sharp \emph{equality} relating the asymptotic behavior of the species scale and the leading tower of states, instead of just some bound on their respective decay rates. We expect that, upon further exploration, this may highly constrain the nature of the possible towers of states predicted by the Distance Conjecture.
	
Interestingly, we can also recover the bound \eqref{eq:lowerboundspecies2} for the exponential decay rate of the species scale, namely the condition
\begin{equation}\label{eq:lambdaspmin}
	\lambda_{\text{sp}} \geq \frac{1}{\sqrt{(d-1)(d-2)}}\, ,
\end{equation}
which was our main object of study in Chapter \ref{ch:bounds} of this thesis. More precisely, this follows upon assuming --- based on string theory evidence \cite{Etheredge:2022opl,Etheredge:2023odp} --- that the maximum possible value for the exponential rate of the leading tower is given by that of a KK tower decompactifying one (unwarped) extra dimension, i.e. $\lambda_{\text{t, max}} = \sqrt{\frac{d-1}{d-2}}$. In this regard, all the new examples analyzed in the present chapter can be equivalently seen to provide further evidence in favor of the bound \eqref{eq:lambdaspmin} (as well as its convex hull formulation, c.f. discussion around eq. \eqref{eq:eff-vector}).
	
\subsubsection*{First steps towards decoding the pattern}
	
Before getting into more complicated examples, let us first show how the pattern is satisfied for the case of a single modulus and a single tower of states becoming light. Let us consider two cases: either the leading tower is a KK tower or a tower of string oscillator modes, as dictated by the Emergent String Conjecture and as observed in all string theory examples so far. 
Recall that the species scale associated to a KK tower decompactifying $n$ (unwarped
) extra  dimensions  is given by the higher dimensional Planck mass
\beq
\label{Mp}
	\LSP\equiv M_{\text{Pl};\, d+n}=M_{\text{Pl};\, d}\, \left(\frac{m_{{\rm KK},\, n}}{M_{\text{Pl};\, d}}\right)^{\frac{n}{d+n-2}},
\eeq
as can be derived from applying \eqref{LSP} and \eqref{NSP} to an equi-spaced tower with $m_k=k^{1/n} m_{\text{KK},\, n}$, where $k=1,\ldots,\infty$. By dimensional reduction of the theory, it is also well-known that the exponential rates of the KK tower and the species scale read
\beq\label{eq:zeta&speciesveconemodulus}
	\zeta_{{\rm KK},\, n} = \sqrt{\frac{d+n-2}{n (d-2)}}\, , \qquad \mathcal{Z}_{{\rm KK}, \, n}=\sqrt{\frac{n}{(d+n-2) (d-2)}}\, ,
\eeq
where $\mathcal{Z}_{{\rm KK}, \, n}$ can be obtained from $\zeta_{{\rm KK},\, n}$ upon using \eqref{Mp}. It can be easily checked that this always reproduces the pattern \eqref{eq:pattern} independently of the number of dimensions that get decompactified
\beq\label{eq:patternKKn}
	\zeta_{{\rm KK},\, n} \cdot \mathcal{Z}_{{\rm KK},\, n}= \frac{1}{d-2}\, .
\eeq
Let us remark, though, that the above expressions for the exponential rates are valid when decompactifying to a higher dimensional \emph{vacuum}, since the story is more complicated when the theory decompactifies to a running solution instead, as recently shown in \cite{Etheredge:2023odp}. We will comment more on this in Section \ref{s:16supercharges}. 
	
The other relevant case is that of a tower of string oscillator modes. If these states arise from a fundamental string, we have
\beq\label{eq:zeta&speciesvecstring}
	\zeta_{\rm osc}= \frac{1}{\sqrt{d-2}}=\mathcal{Z}_{\rm osc}\, ,
\eeq
since the species scale coincides with the string scale (up to maybe logarithmic corrections that will not be relevant here) due to the exponential degeneracy of states at the string scale. It is then automatic that 
\beq\label{eq:patternstringsinglemodulus}
	\zeta_{\rm osc} \cdot \mathcal{Z}_{\rm osc} = \frac{1}{d-2}\, .
\eeq
In summary, for a single modulus, the pattern implies that the exponential rate of the species scale verifies
\beq
	\lambda_{\text{sp}}= \frac{\lambda_{\text{t}}^{-1}}{d-2}\, ,
\eeq
or, in other words, $\Lambda_{\text{sp}}\sim m_{\rm t}^{1/(d-2)\lambda_{\text{t}}^2 }$, which holds regardless of whether we consider KK or stringy towers. In the multi-moduli case, though, these vectors are no longer necessarily parallel to each other. Thus, the pattern is not giving a direct relation between the exponential rates along a given trajectory, but rather between the scalar charge-to-mass vectors $\vec\zeta_{\text{t}}$ and $\vec{\mathcal{Z}}_{\text{sp}}$ as we take some definite asymptotic limit. This is essential for the pattern to hold even in this more complicated scenario.
	
At this moment, one should be surprised by the claimed universality of the pattern, mainly for two reasons:
\begin{itemize}
	\item[$\circ$] The structure of the tower fixes the relation between $m_{\text{t}}$ and $\Lambda_{\text{sp}}$ at a given point of the moduli space. However, a priori, this relation is independent of the exponential decay rate of $m_{\text{t}}$ and $\Lambda_{\text{sp}}$ as we move in moduli space. The pattern implies that they are not independent but can be derived from each other, leading to a universal relation satisfied both for KK and string towers.
	\item[$\circ$] The pattern is verified even in the presence of multiple towers, when the species scale is not simply determined by the leading tower. For instance, we will see that there can be regions of the moduli space where e.g., the leading tower is a KK tower while the species scale corresponds to some string scale. Even then, the pattern is still satisfied as the angle between the vectors precisely compensates for the change in the magnitude, such that \eqref{eq:pattern} holds in a non-trivial manner. The same occurs when decompactifying to a larger number of dimensions than those associated to the leading tower, due to the presence of other subleading KK towers that change the value of the species scale.
\end{itemize}
	
Sometimes, it gets useful to define the convex hull determined by the $\zeta$- and species vectors of all light towers in a given asymptotic regime, since this provides us with useful information about which tower is dominating along each direction as well as the nature of the infinite distance limit, namely the quantum gravity theory above $\LSP$. Notice, though, that these convex hulls can only be defined if there is a region of moduli space in which the hull of the scalar charge-to-mass vectors does not change (see discussion after eq. \eqref{eq:zetaveconemodulus}). In that case, it follows from \eqref{eq:pattern} that both polytopes are dual to each other, as already hinted in Chapter \ref{ch:bounds}. This implies, in particular, that given any one of them one can simply retrieve the other upon imposing the aforementioned relation as a constraint. Therefore, both convex hulls contain the exact same information: It is then equivalent to keep track of all towers becoming light along a given trajectory (which allows one to compute the species scale), than to focus just on the leading tower along all asymptotic geodesics of a given asymptotic regime. Starting from a tower in some particular limit, we can then use the pattern to predict the nature of the towers in other asymptotic limits, and even reconstruct global information about how different limits (or even duality frames) glue together in moduli space \cite{Etheredge:2024tok}. 
	
In the upcoming sections we will test this pattern in the multi-moduli case within several familiar string theory vacua, differing in the number of spacetime dimensions, the amount of supersymmetry preserved, etc. We will see that it is always satisfied, independently of how complicated the tower structure may look like a priori.
 
\section{Deriving the pattern in maximal supergravity}\label{s:maxsugra} 

We begin by deriving the pattern in string theory compactifications with 32 supercharges, i.e. maximal supergravity set-ups arising from toroidal compactifications of M-theory. The advantage of these set-ups is that the $\zeta$-vectors associated to the leading towers of states take some very specific values that remain \emph{fixed} as we move within the moduli space \cite{Etheredge:2022opl}. This will allow us, in turn, to show that the pattern \eqref{eq:pattern} is verified in full generality at every infinite distance limit of the moduli space.
	
Due to the simplicity of these set-ups, we can basically summarize the results in two main scenarios that highlight the key features underlying the realization of the pattern. Hence, we will first explain the main points, and later on exemplify them in concrete examples of M-theory toroidal compactifications down to $d=9,8$. We finish the section by generalizing the discussion to any number of spacetime dimensions for the sake of completeness. 
	
\subsection{Summary of underlying key features}
\label{ss:summary}
	
Consider a $D$-dimensional theory compactified down to $d=D-n$ spacetime dimensions, both preserving maximal supersymmetry in flat space. As shown in \cite{Etheredge:2022opl}, such set-ups in Minkowski space satisfy the Emergent String Conjecture \cite{Lee:2019wij}, in the sense that every infinite distance limit corresponds either to an emergent string limit or to some decompactification. Hence, there are essentially two main scenarios, depending on whether the species scale associated to a given asymptotic regime corresponds to a higher dimensional Planck mass or to the fundamental string scale. In the following, we explain the underlying key features that make a relation like \eqref{eq:pattern} to be satisfied in these two cases, which we will later exemplify in some concrete examples. For a detailed derivation of the relevant formulae involved see Appendix \ref{ap:generalities}.

\begin{figure}
	\begin{center}
	\subfigure[]{
			\includegraphics[width=0.4\textwidth]{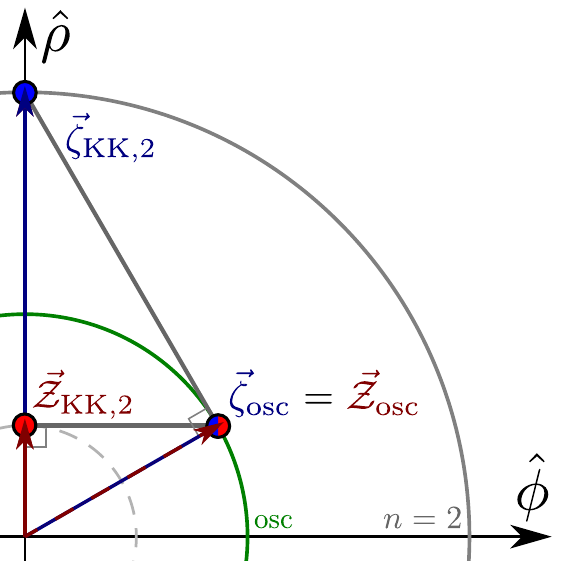}\label{sfig:KKstring}
		}
        \quad
	\subfigure[]{
			\includegraphics[width=0.4\textwidth]{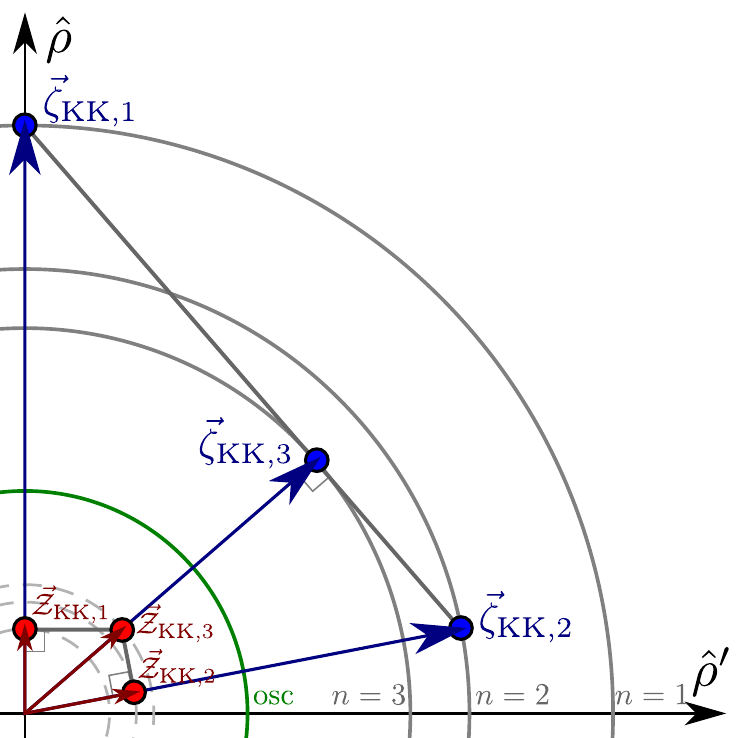}\label{sfig:twoKK}
		}
		\caption{\small Sketches depicting two possible scenarios in multi-field limits for maximal supergravity. For concreteness, we focus on $d=8$, with vectors associated to light towers in blue and to the species scale in red. \textbf{(a)} Decompactification of two internal dimensions and an emergent string limit. The species scale is controlled by the string scale unless we move along the pure decompactification direction, where it coincides with the ten-dimensional Planck mass. Here $\hat{\rho}$ and $\hat{\phi}$ denote the normalized radion and the ten-dimensional dilaton. \textbf{(b)} Two decompactification limits, of one and two internal dimensions, with towers $\vec{\zeta}_{{\rm KK},\, 1}$ and $\vec{\zeta}_{{\rm KK},\, 2}$ (as well as the total volume, $\vec{\zeta}_{{\rm KK},\, 3}$). Note that unless we decompactify a single cycle, the species scale is controlled by the eleven-dimensional Planck mass. The axes $\hat{\rho}$ and $\hat{\rho}'$ correspond to the normalized radions associated to decompactifying the 1- and 2-cycles, respectively.\label{fig:MT2radions} }
	\end{center}
\end{figure}

\subsubsection*{Perturbative string limit}
	
This first scenario is characterized by having the species scale equal to the string scale. Hence, the $\mathcal{Z}$-vector of the species scale is the same than the $\zeta$-vector associated to the tower of string oscillator modes. However, this does not necessarily mean that the tower of string modes is the leading one. As we already know, if we have both a KK and a string tower becoming light, the species scale will indeed correspond to the string scale (even if the KK tower is parametrically lighter) as long as the string scale remains below the species scale associated to the KK tower (i.e. the higher dimensional Planck mass). Hence, the most general scenario with $\LSP \simeq m_s$ can contain both KK and string modes below the species scale. For the sake of concreteness, let us focus on the KK tower associated to the overall volume of the compactification space and the oscillator modes arising from a fundamental string already existing in the higher dimensional theory. We can then restrict to a slice of the tangent space of the moduli space spanned by the overall volume modulus $\hat{\rho}$ and the string dilaton $\hat{\phi}$. The relevant $\zeta$-vectors for such towers within this subspace are (in the flat frame $\{\hat{\phi},\hat{\rho}\}$, c.f. eqs. \eqref{eq:higherDdim}-\eqref{eq:chargetomasstower}) 
\begin{equation}\label{eq:zetasstringlimit}
	\begin{split} 
		\vec{\zeta}_{{\rm KK},\, n} &= \left( 0 , \sqrt{\frac{d+n-2}{n (d-2)}} \right)\, ,\quad \vec{\mathcal{Z}}_{{\rm KK},\, n} = \left( 0 , \sqrt{\frac{n}{(d+n-2) (d-2)}} \right)\, ,\\
		\vec{\zeta}_{\text{osc}}=\vec{\mathcal{Z}}_{\text{osc}} &= \left( \frac{1}{\sqrt{d+n-2}} , \sqrt{\frac{n}{(d+n-2)(d-2)}} \right)\, .
	\end{split}
\end{equation}
These vectors are plotted in Figure \ref{sfig:KKstring}. The tangent vectors of asymptotic geodesics in this slice of the moduli space are radial vectors (i.e straight lines passing through the origin) \cite{Etheredge:2022opl}. As explained in Section \ref{s:convexhull}, to obtain the exponential rate $\lambda$ of a tower (or the species scale) along a given geodesic, we just need to compute the projection of the associated $\zeta$-vector (resp. $\mathcal{Z}$-vector) along such direction. The larger this projection is, the fastest the mass (or the species scale) goes to zero asymptotically. The leading (i.e. the lightest) tower of states is therefore the one with the largest projection of $\vec\zeta$ over such direction; and the same applies to the species scale, which will be the one with the largest projection of $\vec{\mathcal{Z}}$.
	
If we move parallel to $\vec{\zeta}_{{\rm KK},\, n} $, both the Planck scale and the string scale decay at the same rate, so we can simply take the species scale vector as $\vec{\mathcal{Z}}_{{\rm KK},\, n}$. Otherwise, for any other intermediate direction, $\LSP$ will be given by the string scale, as it always remains below the Planck scale, so we should take instead $\vec{\mathcal{Z}}_{\text{osc}}$. On the other hand, the leading tower is always the KK one, except if we move parallel to $\vec{\zeta}_{\text{osc}}$, where both towers present the same exponential rate.\footnote{Note that precisely in this case the limit qualifies as equi-dimensional, in the notation defined in \cite{Lee:2019wij}. Such limits probe gravitational theories in the same number of spacetime dimensions as the starting point of the (infinite distance) trajectory. The fact that there is a KK tower decaying at the same rate than the string tower along this direction is also expected from the Emergent String Conjecture.} It is clear from Section \ref{s:patternintro} that $\vec{\zeta}_{\text{KK,}\, n} \cdot \vec{\mathcal{Z}}_{\text{KK},\, n}=\frac1{d-2}$ and $\vec{\zeta}_{\text{osc}} \cdot \vec{\mathcal{Z}}_{\text{osc}}=\frac1{d-2}$ for each tower independently, but it is less obvious that the pattern will continue working when considering both towers simultaneously. We find here that even in the case in which the species scale is the string scale and the leading tower corresponds to the KK tower, the pattern still holds:
\beq\label{eq:KKstring}
	\vec{\zeta}_{{\rm KK},\, n} \cdot \vec{\mathcal{Z}}_{\text{osc}}=\frac1{d-2}\, .
\eeq
This can be easily understood geometrically from Figure \ref{sfig:KKstring} as follows. Since $\vec{\mathcal{Z}}_{\text{osc}}$ is perpendicular to the convex hull generated by $\vec{\zeta}_{{\rm KK},\, n}$ and $\vec{\zeta}_{\rm osc}$, it turns out that $\vec{\zeta}_{\rm osc}$ is the projection of $\vec{\zeta}_{{\rm KK},\, n}$ along the direction associated to $\vec{\mathcal{Z}}_{\text{osc}}$, so that the pattern holds in general. Alternatively, the projection of $\vec{\mathcal{Z}}_{\text{osc}}$ along the direction determined by $\vec{\zeta}_{{\rm KK},\, n}$ coincides with $\vec{\mathcal{Z}}_{\text{KK,}\, n}$ since the radion component of $\vec{\mathcal{Z}}_{\text{osc}}$ arises from changing the masses to lower dimensional Planck units and it is therefore equal to the $\hat{\rho}$ component of $\vec{\mathcal{Z}}_{\text{KK},\, n}$, as can be seen from \eqref{eq:zetasstringlimit}. 
	
\subsubsection*{Decompactification limit}
	
The second scenario occurs when all the light towers below the species scale are KK modes (possibly decompactifying to different number of dimensions), and we do not find any additional tower of string modes before reaching the lightest higher dimensional Planck mass. Hence, the species scale is a Planck scale in higher dimensions. For concreteness, let us focus on a two-dimensional slice spanned by two KK towers decompactifying to $d+n$ and $d+n'$ dimensions, respectively, with associated volume moduli $\hat{\rho}$ and $\hat{\rho}'$. The $\zeta$-vectors are given by \cite{Etheredge:2022opl} 
\begin{equation}\label{eq:n&n'zetas}
	\begin{split} 
		\vec{\zeta}_{{\rm KK},\, n} &= \left( 0 , \sqrt{\frac{d+n-2}{n (d-2)}} \right)\, ,\\
		\vec{\zeta}_{{\rm KK},\, n'} &= \left( \sqrt{\frac{d+n+n'-2}{n' (d+n-2)}} ,\, \sqrt{\frac{n}{(d+n-2)(d-2)}} \right)\, .
	\end{split}
\end{equation}
Depending on the infinite distance trajectory that we explore, the species scale will correspond to the Planck scale of decompactifying $n$, $n'$ or $n+n'$ extra dimensions. The associated $\mathcal{Z}$-vectors are
\begin{equation}
	\begin{split} 
		\vec{\mathcal{Z}}_{{\rm KK},\, n} &= \left( 0 , \sqrt{\frac{n}{(d+n-2) (d-2)}} \right) \, ,\\
		\vec{\mathcal{Z}}_{{\rm KK},\, n'} &= \left( \sqrt{\frac{n'(d+n+n'-2)}{(d+n'-2)^2 (d+n-2)}} ,\, \frac{n'}{d-2+n'} \sqrt{\frac{n}{(d+n-2) (d-2)}} \right) \, ,\\
		\vec{\mathcal{Z}}_{{\rm KK},\, n+ n'} &= \left( \sqrt{\frac{n'}{(d+n-2) (d+n+n'-2)}},\, \sqrt{\frac{n}{(d+n-2) (d-2)}} \right) \, \label{eq:combinedZ}.
	\end{split}
\end{equation}
All these vectors are represented in Figure \ref{sfig:twoKK}. The species scale corresponds to the lightest Planck scale along any chosen infinite distance trajectory. Hence, it will always be given by $\vec{\mathcal{Z}}_{{\rm KK},\,  n+ n'}$ in the entire asymptotic regime unless we move parallel to either $\vec{\zeta}_{{\rm KK},\, n}$ or $\vec{\zeta}_{{\rm KK},\, n'}\,$, in which case it reduces to $\vec{\mathcal{Z}}_{{\rm KK},\, n}$ or $\vec{\mathcal{Z}}_{{\rm KK},\, n'}\,$, respectively. However, the leading tower corresponds to decompactifying only $n$ or $n'$ extra dimensions unless we move precisely parallel to $\vec{\mathcal{Z}}_{{\rm KK},\,  n+ n'}$. The latter case would physically correspond to an isotropic decompactification of both $n$- and $n'$-dimensional internal cycles, with an effective KK tower of charge-to-mass vector given by (c.f. eq. \eqref{eq:effectivezeta})
\beq
	\vec{\zeta}_{{\rm KK},\, n+ n'} = \left( \sqrt{\frac{n' (d+n+n'-2)}{(d+n-2) (n+n')^2}}, \sqrt{\frac{n (d+n+n'-2)^2}{(n+n')^2(d+n-2) (d-2)}} \right) \, .
\eeq
Again, the pattern is clearly satisfied whenever we move along the asymptotic trajectories determined by any of the individual KK towers (due to \eqref{eq:patternKKn}), but it also nicely holds for intermediate directions within the asymptotic regime, since 
\beq\label{eq:doubleKK}
	\vec{\zeta}_{{\rm KK},\, n} \cdot \vec{\mathcal{Z}}_{{\rm KK},\, n+ n'}= \vec{\zeta}_{{\rm KK},\, n'} \cdot \vec{\mathcal{Z}}_{{\rm KK},\,n+ n'} = \frac1{d-2}\, .
\eeq
Notice that such relation may be easily understood from geometrical considerations as follows. The species vector $\vec{\mathcal{Z}}_{{\rm KK},\, n+n'}$ appears to be always perpendicular to the convex hull generated by $\vec{\zeta}_{{\rm KK},\, n}$ and $\vec{\zeta}_{{\rm KK},\, n'}$ (see Figure \ref{sfig:twoKK}), such that they both project to $\vec{\zeta}_{{\rm KK},\,  n+n'}$ along the direction determined by the former. Alternatively, $\vec{\mathcal{Z}}_{{\rm KK},\,  n+n'}$ projects to $\vec{\mathcal{Z}}_{{\rm KK},\, n}$ (analogously $\vec{\mathcal{Z}}_{{\rm KK},\, n'}$) along the direction determined by $\vec{\zeta}_{{\rm KK},\, n}$ (respectively $\vec{\zeta}_{{\rm KK},\, n'}$), which may be understood again as coming from a change of Planck units in both cases, given the commutativity of the compactification process (see Appendix \ref{ap:generalities}).
	
\subsubsection*{Summary}
	
What can be learned from the two scenarios above? The species scale vector $\vec{\mathcal{Z}}$ always happens to be perpendicular to the convex hull of the light towers of states. Conversely, the leading scalar charge-to-mass vector $\vec{\zeta}_{\rm t}$ is orthogonal to the convex hull generated by the species vectors. This is a feature that holds in general for M-theory toroidal compactifications, as we already observed in Chapter \ref{ch:bounds}. In fact, such constraints are restrictive enough so as to ensure that, once we assume that the pattern \eqref{eq:pattern} is verified by any pair of collinear vectors $\vec{\zeta}$ and $\vec{\mathcal{Z}}$ (i.e. when both are associated to one and the same tower of states), then the pattern extends automatically to any other asymptotic limit of the moduli space.\footnote{For instance, if $\vec{\zeta}_{\rm t}$ is orthogonal to the convex hull generated by $\vec{\mathcal{Z}}_{\rm sp}$ (the total specie scale) and $\vec{\mathcal{Z}}_{\rm t}$ (the one obtained only from considering the leading tower), then satisfying $\vec{\zeta}_{\rm t}\cdot \vec{\mathcal{Z}}_{\rm t}=\frac1{d-2}$ guarantees that $\vec{\zeta}_{\rm t}\cdot \vec{\mathcal{Z}}_{\rm sp}=\frac1{d-2}$, as the difference between the two species scale vectors is given by a vector which is orthogonal to $\vec{\zeta}_{\rm t}$.}
	
Notice, however, that the same story does not apply immediately when the amount of supersymmetry preserved by our theory is reduced, since then the charge-to-mass and species vectors can `slide' (or jump) non-trivially depending on where we sit in moduli space, see Sections \ref{s:16supercharges} and \ref{s:8supercharges}. In any event, most of our efforts in the upcoming sections will be dedicated to show that, even in such cases, the pattern is still verified at any infinite distance boundary, and it does so in a way that can be easily understood from pictures similar to those shown in Figure \ref{fig:MT2radions} above.	
	
\subsection{Maximal supergravity in 9d}
\label{ss:9d}
	
Next, we will illustrate the above general scenarios in concrete examples, starting with the unique 9d $\mathcal{N}=2$ supergravity theory arising from compactifying M-theory on a two-dimensional torus. The $\zeta$-vectors for the towers of states in this particular set-up were already analyzed in \cite{Etheredge:2022opl}, whilst the species ones have been derived in the previous chapter. Here we will build upon these results and simply check if the pattern \eqref{eq:pattern} is verified, paying special attention to the way in which this happens.

Recall that the moduli space of M-theory on $\mathbf{T}^2$ is classically exact, it moreover presents a coset structure (c.f. \eqref{eq:9dmodspaceSSDC}) and is locally parametrized by the complex structure of the torus as well as the overall volume modulus. Per our discussion in Section \ref{ss:MthyT2SSDC}, we will forget in the following about the compact scalar field, since it seems to play no role whatsoever. We also use the basis \eqref{eq:canonicalnormalization} of canonically normalized fields $\{\hat U, \hat \tau \}$ to express any moduli-dependent quantity, including the relevant $\zeta$- and $\mathcal{Z}$-vectors.

As discussed in \cite{Etheredge:2022opl}, the relevant towers of states becoming light at the infinite distance limits of this moduli space are $\frac{1}{2}$-BPS particles. For this particular example, the convex hull determined by the scalar charge-to-mass vectors associated to all light towers is spanned by Kaluza-Klein modes with the following $\zeta$-vectors
\begin{equation}\label{eq:KKzetas9d}
	\vec{\zeta}_{\text{KK},\, 1} = \left( \frac{3}{\sqrt{14}},\frac{1}{\sqrt{2}} \right) \, , \qquad
	\vec{\zeta}_{\text{KK},\, 1'} = \left( \frac{3}{\sqrt{14}},-\frac{1}{\sqrt{2}}\right) \, ,
\end{equation}
as well as M2-branes wrapping the compactification manifold, with
\begin{equation}\label{eq:M2zeta9d}
	\vec{\zeta}_{\text{M2}} = \left( -\sqrt{\frac{8}{7}},0 \right) \, .
\end{equation}
We have adopted the notation $\vec{\zeta}=\left(\zeta^{\hat U}, \zeta^{\hat \tau} \right)$. Notice that they all satisfy the relation $|\vec{\zeta}|^2=8/7$, in accordance with eq. \eqref{eq:zeta&speciesveconemodulus} above for $d=9$ and $n=1$.
	
On the other hand, their associated species scale vectors, $\vec{\mathcal{Z}}$, were found to be
\begin{equation} \label{eq:SSvec9d}
	\begin{split} 
		\vec{\mathcal{Z}}_{\text{KK},\, 1} &= \left( \frac{3\sqrt{14}}{112},\frac{\sqrt{2}}{16} \right) \, , \qquad \vec{\mathcal{Z}}_{\text{KK},\, 1'} = \left( \frac{3\sqrt{14}}{112},-\frac{\sqrt{2}}{16}\right) \, ,\\
		\vec{\mathcal{Z}}_{\text{M2}} &= \left( -\frac{1}{2\sqrt{14}} , 0\right) \, ,
	\end{split}
\end{equation}
corresponding to the appropriate 10d Planck mass of the decompactified (dual) theories. 
	
Furthermore, as explained in Chapter \ref{ch:bounds}, a crucial ingredient when determining the set of all possible species scales is the concept of effective tower. Indeed, for intermediate directions between $\vec{\zeta}_{\text{KK},\, 1}$ and $\vec{\zeta}_{\text{KK},\, 1'}$, despite one KK tower being (in general) parametrically lighter than the other, one still needs to account for bound states thereof in order to properly compute the species scale in that asymptotic regime (see discussion around eq. \eqref{eq:conforeffectivetower}). Upon doing so, one arrives at the following species scale vector
\begin{equation}\label{eq:effectivespecies}
	\vec{\mathcal{Z}}_{{\rm KK},\, 2}=\frac{1}{9} \left( \vec{\zeta}_{{\rm KK},\, 1} +\vec{\zeta}_{{\rm KK},\, 1'} \right) = \left( \frac{\sqrt{14}}{21} , 0\right)\, ,
\end{equation}
to which we can associate an effective (averaged) mass scale and charge-to-mass vector as follows
\begin{equation}\label{eq:effectivemass}
	\vec{\zeta}_{{\rm KK},\,2}=\frac{1}{2} \left( \vec{\zeta}_{{\rm KK},\, 1} +\vec{\zeta}_{{\rm KK},\, 1'} \right) = \left( \frac{3}{\sqrt{14}} , 0\right)\, .
\end{equation}
The physical interpretation for \eqref{eq:effectivespecies} is clear --- it corresponds to the 11d Planck scale, whilst the charge-to-mass vector \eqref{eq:effectivemass} is a meaningful quantity only when one takes the decompactification limit in an isotropic way, namely for an asymptotic vector $\hat T = \partial_{\hat U}$. Still it may be useful to think in terms of `averaged' geometric quantities when computing the species scale vectors and checking the pattern \eqref{eq:pattern} explicitly, as we discuss later on in this section.
	
Apart from these, there is also another set of $\frac{1}{2}$-BPS states comprised by critical Type IIA strings arising from M2-branes wrapped on a non-trivial 1-cycle. Their oscillator modes were seen to lead to the following charge-to-mass vectors (c.f. eq. \eqref{eq:speciesscalevectorsstringsT2})
\begin{equation}\label{eq:9dstrings}
	\vec{\zeta}_{\text{osc}} = \left( -\frac{1}{2 \sqrt{14}},\frac{1}{2\sqrt{2}} \right) \, , \quad 
	\vec{\zeta}_{\text{osc'}} = \left( -\frac{1}{2 \sqrt{14}},- \frac{1}{2\sqrt{2}} \right) \, ,
\end{equation}
which coincide with those of their associated species scale and moreover satisfy $|\vec{\mathcal{Z}}_{\text{osc}}|^2=\frac{1}{d-2}=\frac{1}{7}$ (c.f. eq. \eqref{eq:zeta&speciesvecstring}).

\begin{figure}[htb]
	\begin{center}
		\includegraphics[scale=.55]{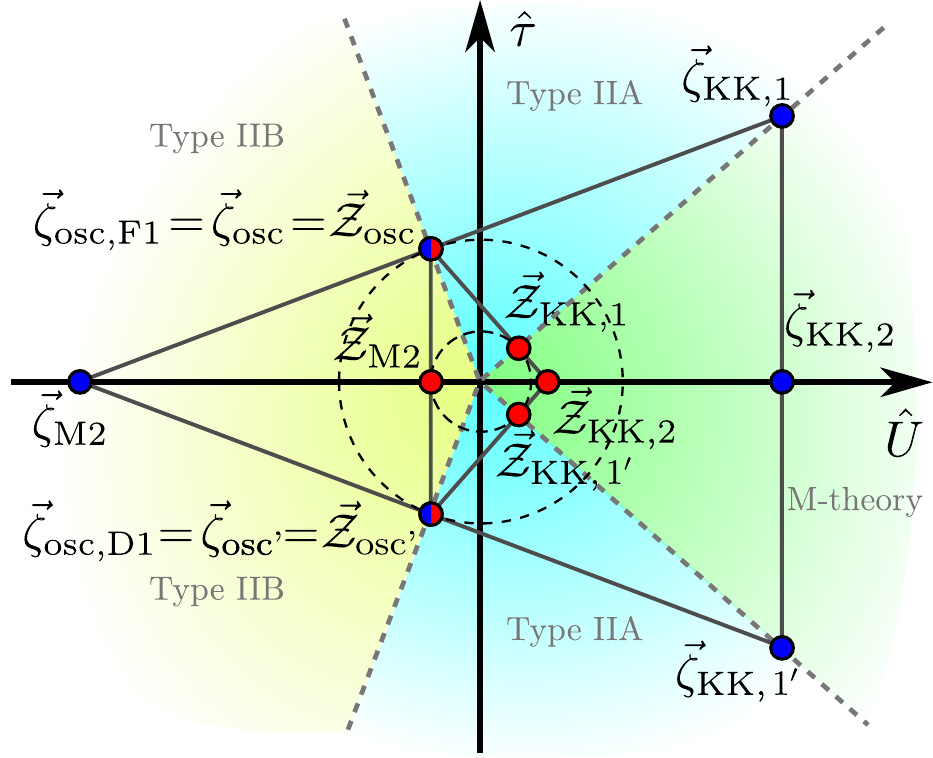}
		\caption{\small Convex hulls spanned by the species scale (red) and mass scales of the leading towers (blue) in nine-dimensional maximal supergravity. The 1-spheres of radii $\frac{1}{\sqrt{d-2}}=\frac{1}{\sqrt{7}}$ and $\frac{1}{\sqrt{(d-1)(d-2)}}=\frac{1}{\sqrt{56}}$ are plotted in dashed lines. We also depict the different duality frames of the theory using distinct shades. Notice that both Type IIA and Type IIB string theory have two different duality frames, whereas there is a single one for M-theory.} 
		\label{fig:T2SDC&SSDC}
	\end{center}
\end{figure}

In Figure \ref{fig:T2SDC&SSDC} we depict again the convex hulls associated to the towers of states along with their species scale vectors, which are constructed from the expressions \eqref{eq:KKzetas9d}-\eqref{eq:9dstrings}. Notice that there is a $\mathbb{Z}_2$-symmetry with respect to the $\hat \tau$-axis, which may be thought of as a discrete remnant of the U-duality group of the theory (more specifically it is given by its associated Weyl group, see footnote \ref{fnote.Weyl} below). Therefore, it is enough to focus just on the upper-half plane in order to check the condition \eqref{eq:pattern}.
	
First, notice that for those directions in which both $\vec{\zeta}_{\text{t}}$ and $\vec{\mathcal{Z}}_{\text{sp}}$ are aligned, namely when $\hat T$ is parallel to the vector $\vec{\zeta}_I$ associated to any leading tower, the condition $\vec{\zeta}_{\text{t}} \cdot \vec{\mathcal{Z}}_{\text{sp}}= \frac{1}{d-2} =\frac{1}{7}$ is satisfied. Moreover, this turns out to be sufficient for the pattern to hold also along intermediate directions. The reason behind is a duality between both convex hull diagrams. In fact, as one can see from Figure \ref{fig:T2SDC&SSDC}, the vertices from one correspond to edges of the other and viceversa, the latter being orthogonal to the former. Therefore, it follows that whenever we take $\vec{\zeta}_{\text{t}}$ (analogously $\vec{\mathcal{Z}}_{\text{sp}}$) to be given by any of the two vertices generating an edge of its corresponding diagram, its inner product with the dual $\vec{\mathcal{Z}}_{\text{sp}}$ (analogously $\vec{\zeta}_{\text{t}}$) orthogonal to such edge reduces to that of the previous `parallel' cases and thus satisfies the pattern \eqref{eq:pattern}.
	
\subsection{Maximal supergravity in 8d}
\label{ss:8d}
	
\begin{figure}
	\begin{center}
\subfigure[]{
				\includegraphics[width=0.45\textwidth]{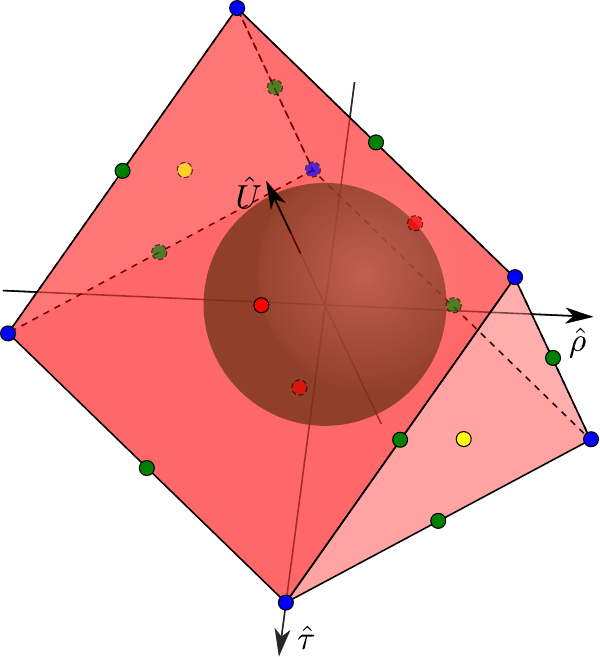}\label{sfig:mass8d}
			}
            \quad
			\subfigure[]{
				\includegraphics[width=0.45\textwidth]{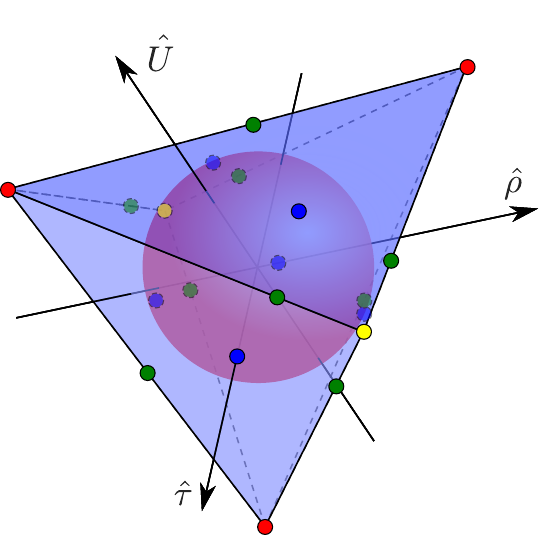}\label{sfig:species8d}
			}
			\caption{\small Convex hull conditions for the masses $\{\vec{\zeta}_I\}$ \textbf{(a)} and species scales $\{\vec{\mathcal{Z}}_J\}$ \textbf{(b)} of the leading towers in eight-dimensional maximal supergravity, containing the `extremal balls' of radii $\frac{1}{\sqrt{d-2}}=\frac{1}{\sqrt{6}}$ and $\frac{1}{\sqrt{(d-1)(d-2)}}=\frac{1}{\sqrt{42}}$, respectively. The string towers are depicted in red \fcolorbox{black}{red}{\rule{0pt}{6pt}\rule{6pt}{0pt}}, whilst KK towers associated to decompactification of one, two and three dimensions appear in blue \fcolorbox{black}{blue}{\rule{0pt}{6pt}\rule{6pt}{0pt}}, green \fcolorbox{black}{dark-green}{\rule{0pt}{6pt}\rule{6pt}{0pt}} and yellow \fcolorbox{black}{yellow}{\rule{0pt}{6pt}\rule{6pt}{0pt}}, respectively. Note that the string vectors coincide in the two diagrams.}\label{fig:convexhulls8d}
	\end{center}
\end{figure}

As our second example, we now take M-theory compactified on a $\mathbf{T}^3$, leading to 8d $\mathcal N=2$ supergravity, whose bosonic action was already discussed in Section \ref{ss:8dmaxsugra}. However, instead of choosing a parametrization which makes the U-duality group manifest, we take here the same approach as in Section \ref{ss:MthyT3SSDC} and consider the 9d theory from the previous example compactified on an additional circle, leading to eq. \eqref{eq:8dalternativeaction}. We will forget again about the axion fields, and moreover fix some convenient basis $\{ \hat U, \hat \tau, \hat \rho\}$ of canonically normalized saxions (see discussion around eq. \eqref{eq:8dcanonical}), which allows us to read off most of the scalar charge-to-mass vectors characterizing the infinite towers of states from the previous 9d example.\footnote{This procedure is explained in detail in Appendix \ref{ap:generalities}.} Therefore, for the KK towers one obtains
\begin{equation}
	\begin{split} 
		\vec{\zeta}_{\text{KK},\, 1} &= \left( \frac{1}{\sqrt{2}} , \frac{1}{\sqrt{42}}, \frac{3}{\sqrt{14}} \right) \, , \qquad \vec{\zeta}_{\text{KK},\, 1'} = \left( -\frac{1}{\sqrt{2}} , \frac{1}{\sqrt{42}}, \frac{3}{\sqrt{14}} \right) \, ,\\
		\vec{\zeta}_{\text{KK},\, 1''} &= \left( 0 , \sqrt{\frac{7}{6}}, 0 \right) \, ,
	\end{split}
\end{equation}
where the last $\zeta$-vector arises from the extra $\mathbf{S}^1$ and the notation is $\vec{\zeta}=\left(\zeta^{\hat \tau}, \zeta^{\hat \rho}, \zeta^{\hat U}\right)$. Analogously, one finds a triplet of towers comprised by M2-branes wrapping different 2-cycles within $\mathbf{T}^3$, with the following charge-to-mass vectors
\begin{equation} \label{eq:M2vectorspattern}
	\begin{split} 
		\vec{\zeta}_{\text{M},\, 1} &= \left( \frac{1}{\sqrt{2}} , -\frac{5}{\sqrt{42}}, -\frac{1}{\sqrt{14}} \right) \, , \qquad \vec{\zeta}_{\text{M},\, 1'} = \left( -\frac{1}{\sqrt{2}} , -\frac{5}{\sqrt{42}}, -\frac{1}{\sqrt{14}} \right) \, ,\\
		\vec{\zeta}_{\text{M},\, 1''} &= \left( 0, \frac{1}{\sqrt{42}}, -\sqrt{\frac{8}{7}} \right)  \, ,
	\end{split}
\end{equation}
where the last one is inherited from the 9d set-up, whilst the first two are new (c.f. Table \ref{tab:BPSstates}). 
Notice that these vectors already generate the convex hull associated to the light towers, see Figure \ref{sfig:mass8d}. However, there also exist additional towers of states corresponding to the oscillation modes of critical (Type IIA) strings, whose $\zeta$-vectors read as
\begin{equation} \label{eq:stringvectorspattern}
	\begin{split} 
		\vec{\zeta}_{\text{osc}} &= \left( \frac{1}{2\sqrt{2}} , \frac{1}{\sqrt{42}}, -\frac{1}{2 \sqrt{14}} \right) \, , \qquad \vec{\zeta}_{\text{osc}'} = \left( -\frac{1}{2\sqrt{2}} , \frac{1}{\sqrt{42}}, -\frac{1}{2 \sqrt{14}} \right) \, ,\\
		\vec{\zeta}_{\text{osc}''} &= \left( 0 , -\sqrt{\frac{2}{21}}, \frac{1}{\sqrt{14}} \right)  \, ,
	\end{split}
\end{equation}
and which happen to lie at the extremal ball, thus saturating the sharpened Distance Conjecture \cite{Etheredge:2022opl}. The first two are inherited from the 9d example above (c.f. \eqref{eq:9dstrings}), whilst the third one arises from the M2-brane of 11d supergravity wrapped along the additional circle.
	
One can analogously compute the species scale vectors within each asymptotic direction of the 8d moduli space. This exercise was already performed in Section \ref{ss:MthyT3SSDC}, so we refrain from repeating it here and refer the reader interested in the details to that section. The resulting $\mathcal{Z}$-vectors are displayed in eqs. \eqref{eq:stringvectors}-\eqref{eq:KK&M2vectorscombined}, leading to the convex hull diagram shown in Figure \ref{sfig:species8d}. 

\begin{figure}[htb]
	\begin{center}
		\includegraphics[scale=.7]{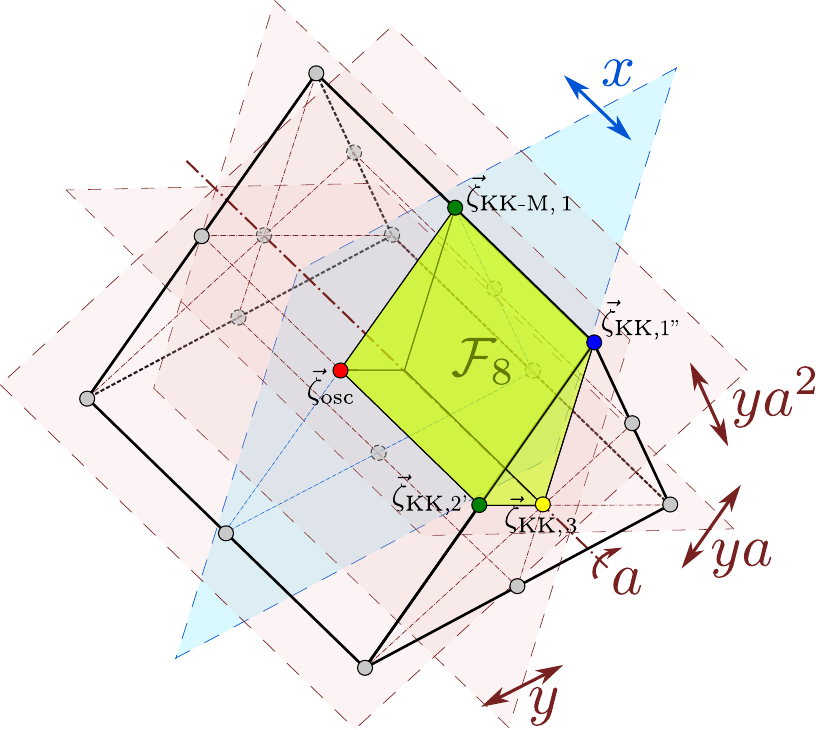}
		\caption{\small Sketch of the fundamental domain $\mathscr{F}_8$ of the $\mathsf{S_3}\times \mathsf{S_2}=\left \langle y,\,a: y^2=a^3=e,\, yay=a^{-1} \right\rangle\times\left\langle x: x^2=e\right\rangle$ symmetry group acting on the scalar charge-to-mass vectors associated to the relevant towers in 8d maximal supergravity. The figure also shows the towers spanning the fundamental domain as well as the individual actions of the symmetry group. A completely analogous fundamental domain for the species scale polytope from Figure \ref{sfig:species8d} can be built, since both $\{\vec{\zeta}_I\}$ and $\{\vec{\mathcal{Z}}_J\}$ present the same symmetries.} 
		\label{fig:sym}
	\end{center}
\end{figure}

In order to check whether the condition \eqref{eq:pattern} is satisfied or not, one can proceed as in the 9d example above and focus --- thanks to the U-duality group of the theory --- on a strictly smaller polyhedron. Indeed, since the symmetry group of the convex polytope is $\mathsf{S_2}\times \mathsf{S_3}$, it is enough for our purposes to take 1/12 of the full diagram, namely the one containing e.g., the set $\lbrace \vec{\zeta}_{\text{KK},\, 1''}, \vec{\zeta}_{\text{osc}}, \vec{\zeta}_{\text{KK-M},\,2}, \vec{\zeta}_{{\rm KK},\, 2'}, \vec{\zeta}_{{\rm KK},\, 3}\rbrace$. Figure \ref{fig:sym} depicts the aforementioned vertices and the fundamental domain they span, as well as the discrete symmetries associated to the diagram. One can now easily check that along these particular directions, the product $\vec{\zeta}_{\text{t}} \cdot \vec{\mathcal{Z}}_{\text{sp}}= \frac{1}{d-2} =\frac{1}{6}$ is verified, since the species scale and the charge-to-mass vectors are aligned. Furthermore, as it was also the case in our previous example, this is actually all we need to check in order to get convinced that the pattern holds along every other intermediate asymptotic direction as well. This follows again from the fact that the vertices spanning one convex hull are orthogonal to the faces of the other and viceversa, see Figure \ref{fig:convexhulls8d}.
	
\subsection{Maximal supergravity in $d<8$}
\label{ss:generaldim}
	
After the previous concrete examples, we will argue in what follows that the results discussed there hold more generally in the context of maximal supergravity. The strategy will be to isolate the key ingredients from the nine- and eight-dimensional set-ups and translate them into the more general case in $d$ spacetime dimensions. This is done in Section \ref{sss:sketch}, whilst the computational details are relegated to Section \ref{sss:generalcomputations}.
	
\subsubsection{A sketch of the proof}
\label{sss:sketch}
	
The argument proceeds in a recursive manner, relying essentially on the duality properties of the theory as well as the uniqueness of maximal supergravity for $d\leq9$. 
	
Let us start by noticing from the examples above that the charge-to-mass vectors associated to towers with density parameter $n$ lie always along a facet\footnote{Actually, they are located at the point of the facet closest to the origin.} of the convex hull polytope with \emph{dimension} equal to $n-1$ (see Figures \ref{fig:T2SDC&SSDC} and \ref{sfig:mass8d}), whilst those vectors controlling the species scale belong to a facet of \emph{codimension} $n$ (c.f. Figures \ref{fig:T2SDC&SSDC} and \ref{sfig:species8d}).\footnote{The vectors associated to string towers appear at facets of maximal (co-)dimension for the charge-to-mass (resp. species) diagram.} This holds in lower spacetime dimensions as well, since the length of the vectors is fully determined once $d$ and $n$ are specified (c.f. \eqref{eq:zeta&speciesveconemodulus}), and it is indeed a  clear manifestation of the duality between both convex hulls in the sense that the facets of one correspond to the vertices of the other, and viceversa. 
	
One also notices that the diagrams present some symmetry properties that reflect the U-duality group of the quantum theory (see Table \ref{tab:irreps} below). This, in turn, allows us to restrict ourselves to some \emph{fundamental domain}, i.e. a subset of the original convex hull containing all the relevant information for the diagram, whilst the remaining parts of the hull appear to be mere copies of the former, obtained upon acting with the different elements of the symmetry group. In fact, one may view such fundamental domain as the region whose boundaries precisely arise as fixed submanifolds under some element(s) of the symmetry group, which moreover coincides with the Weyl subgroup associated to the U-duality group (see Figure \ref{fig:sym}).\footnote{\label{fnote.Weyl}Consider some EFT with a $n$-dimensional moduli space $\mathcal{M}_{\rm mod}$ parametrized by the scalars $\{\varphi^i\}$, $i=1, \ldots, n$. The U-duality group $G$ of said theory transforms the scalars $\{\varphi^i\}$ in a way such that the different states of the EFT are mapped to one another. However, if we are interested only in non-compact scalars (thus ignoring compact axionic fields), some of the transformations of $G$ might affect only the compact ones, which we left fixed. These transformations are the elements of a \emph{maximal torus of $G$, $T_G\hookrightarrow G$}, which is the maximal Abelian, connected and compact subgroup of $G$. As in general $T_G$ is not a normal subgroup of $G$, in order to properly quotient $G$ by $T_G$, one introduces the normalizer $N_G (T_G)=\{g\in G:gT_G=T_Gg\}$, corresponding to the largest subgroup of $G$ such that $T_G$ is a normal subgroup. Then the \emph{Weyl group} of $G$ is defined as ${\rm W(G)}:=N_G(T_G)/T_G$, and it will correspond to the symmetries of the non-compact scalars (and thus of the different vectors under consideration). ${\rm W(G)}$ is then some finite (there are only so many ways of exchanging points) subgroup of $\mathsf{GL(\mathbb{R}^{k})}$, where $k\leq n$ is the number of unbounded moduli.}
	
Therefore, what we need first to know is how to select a fundamental domain $\mathscr{F}_d$, in practice. For this, we note that the towers of states with $n=1$ arrange themselves into a \emph{single} irreducible representation of the U-duality group for $d<9$, as shown in the second column of Table \ref{tab:irreps}. These include perturbative (i.e. KK, winding modes, etc.) as well as non-perturbative states (wrapped branes, KK-monopoles, etc.), and for us it will be enough to focus on just one of them, which we take to be of perturbative nature, namely a Kaluza-Klein vector. Hence, we work inductively, starting from M-theory compactified on $\mathbf{T}^k$ down to $d+1=11-k$ dimensions, where we assume the pattern \eqref{eq:pattern} to hold. Then, we dimensionally reduce on an extra circle, leading to M-theory on $\mathbf{T}^k \times \mathbf{S}^1 \cong \mathbf{T}^{k+1}$, and we consider the cone of asymptotic directions comprised by the large radius direction (of the additional $\mathbf{S}^1$) and the KK replica of the vectors determining some fundamental domain, $\mathscr{F}_{d+1}$, in the parent $(d+1)$-dimensional theory. 
Upon doing so, one can easily check (see Section \ref{sss:generalcomputations} below) that eq. \eqref{eq:pattern} is verified along any asymptotic trajectory within $\mathscr{F}_d$. Finally, since the pattern has already been shown to hold for $k=1,2,3$ (corresponding to maximal supergravity in ten, nine and eight dimensions, respectively), one concludes that it extends to all lower dimensional cases as well.
	
\begin{table}[h!!]\begin{center}
		\renewcommand{\arraystretch}{1.00}
		\begin{tabular}{|c|c|c|c|c|}
			\hline
			$d$ &  U-duality group  & Irrep. &  $\{\vec{\zeta}_I\}$ sym. group & Order\\
			\hline 
                \hline
			10A & 1  & $\mathbf{1}$ & 1 & 1 \\
			10B & $\mathsf{SL(2,\mathbb{Z})}$  & $\mathbf{2}$ & $\mathbb{Z}_2 \simeq \mathsf{S_2}$ & 2 \\
			9 & $\mathsf{SL(2,\mathbb{Z})}$   & $\mathbf{2} \oplus \mathbf{1}$ & $\mathbb{Z}_2 \simeq \mathsf{S_2}$ & 2 \\
			8 & $\mathsf{SL(2,\mathbb{Z})} \times \mathsf{SL(3,\mathbb{Z})}$  & $(\mathbf{2},\mathbf{3} )$ & $\mathsf{S_2}\times \mathsf{S_3}$ & 12 \\
			7 & $\mathsf{SL(5,\mathbb{Z})}$  & $\mathbf{10}$ & $\mathsf{S_5}$ & 120 \\
			6 & $\mathsf{SO(5, 5, \mathbb{Z})}$  & $\mathbf{16}$ & $\mathsf{\text{W}\, (Spin(5,5))}$ & $1\,920$ \\
			5 & $\mathsf{E_{6 (6)} (\mathbb{Z})}$  & $\mathbf{27}$ & $\mathsf{\text{W}\, (E_6)}$ & $51\,840$ \\
			4 & $\mathsf{E_{7 (7)} (\mathbb{Z})}$  & $\mathbf{56}$ & $\mathsf{\text{W}\, (E_7)}$ & $2\,903\,040$ \\
			3 & $\mathsf{E_{8 (8)} (\mathbb{Z})}$  & $\mathbf{248}$ & $\mathsf{\text{W}\, (E_8)}$ & $719\,953\,920$ \\
				\hline
		\end{tabular}
		\caption{\small U-duality representations of the particle multiplets in M-theory on $\mathbf{T}^{k}$ \cite{Obers:1998fb} for $10\geq d\geq 3$. Note that there are two possibilities for $d=10$, corresponding to ten-dimensional Type IIA and Type IIB supergravities. The second column shows the vector and charge representations for $n=1$ BPS towers, which for $d<9$ arrange into a single irrep. Additionally, the symmetry group acting on the $\zeta$ (equivalently $\mathcal{Z}$)-vectors is displayed, which corresponds to the Weyl subgroup of the associated U-duality group, as well as its finite order \cite{wilson2009finite}. The latter controls the number of copies of $\mathscr{F}_d$ that comprise the convex hull of $\zeta$- or $\mathcal{Z}$-vectors.}
		\label{tab:irreps}
	\end{center}
\end{table}
	
\subsubsection{Relevant computations}
\label{sss:generalcomputations}
The aim of this subsection is to provide some of the details that corroborate our previous claims regarding the analysis of the pattern \eqref{eq:pattern} in $d<8$ maximal supergravity. Let us assume that we have already fixed a fundamental domain $\mathscr{F}_d$, as outlined in Section \ref{sss:sketch}. Such polytope is thus generated by the reference $n=1$ tower, with charge-to-mass vector $\vec{\zeta}_{\rm KK, 1}$, together with the KK replica of those vectors determining the fundamental domain of the theory in one dimension higher (see Figure \ref{fig:sym}). In the following, we will denote the latter as $\lbrace \vec{\zeta}_{{\rm KK,}\,n+1} \rbrace$, with $n\in\{1,\ldots,10-d,\infty\}$. 
First, we notice that whenever we focus on a given direction determined by some $\vec{\zeta}$ within $\mathscr{F}_d$, the pattern is automatically satisfied, since both the species and charge-to-mass vectors are associated to one and the same tower and thus parallel to each other (c.f. \eqref{eq:patternKKn}). The non-trivial task is to show that eq. \eqref{eq:pattern} is still satisfied along intermediate directions as well, where the vectors $\lbrace \vec{\zeta}_{\text{t}}, \vec{\mathcal{Z}}_{\text{sp}}\rbrace$ are no longer aligned. 
To do so, we first prove the following claim:
\begin{theorem}\label{claim1}
	The leading tower of states within $\mathscr{F}_d$ always corresponds to $\vec{\zeta}_{\rm KK,\, 1}$. Additional towers $m_I$ can become light at the same rate along certain asymptotically geodesic trajectories, characterized by some normalized tangent vector $\hat{T}$.
\end{theorem}
This can be easily shown upon computing the inner product between $\vec{\zeta}_{\rm KK,\, 1}$ and any other charge-to-mass vector belonging to the set $\lbrace \vec{\zeta}_{{\rm KK,}\,n+1} \rbrace$. One finds 
\begin{equation}\label{eq:proofKK1dominant}
	\vec{\zeta}_{\rm KK,\, 1} \cdot \vec{\zeta}_{\text{KK},\, n+1} = \vec{\zeta}_{\rm KK, 1} \cdot \left[\frac{1}{n+1} \left( \vec{\zeta}_{\rm KK,\, 1} + n\, \vec{\zeta}_{{\rm KK,}\, n} \right) \right] = \frac{d+n-1}{(d-2) (n+1)} = |\vec{\zeta}_{{\rm KK},\, n+1}|^2\, ,
\end{equation}
where we have used eq. \eqref{eq:effectivezeta} in the second equality. The fact that it coincides with $|\vec{\zeta}_{{\rm KK},\, n+1}|^2$ implies, geometrically, that the segment of the hull determined by both vectors is indeed orthogonal to $\vec{\zeta}_{{\rm KK},\, n+1}$ itself (see e.g., Figure \ref{fig:MT2radions}). Now, given any normalized tangent vector $\hat{T}$, we can split it into parallel and perpendicular components with respect to the plane spanned by $\vec{\zeta}_{\rm KK,\, 1}$ and $\vec{\zeta}_{\text{KK,}\, n+1}$, such that $\hat{T}=\hat{T}^\parallel+\hat{T}^\perp$, where $\hat{T}^{\parallel}=a\,\vec{\zeta}_{\rm KK,\, 1}+b\,\vec{\zeta}_{\text{KK,}\, n+1}$, and with $a,\,b\geq 0$. Therefore, we have
\begin{align}
	\hat{T}\cdot(\vec{\zeta}_{\rm KK,\, 1} - \vec{\zeta}_{\text{KK},\, n+1})&=\hat{T}^\parallel\cdot(\vec{\zeta}_{\rm KK,\, 1} - \vec{\zeta}_{{\rm KK},\, n+1})\notag\\
	&=a\, \vec{\zeta}_{\rm KK,\, 1}\cdot(\vec{\zeta}_{\rm KK,\, 1} - \vec{\zeta}_{{\rm KK},\, n+1})+ b\,\vec{\zeta}_{{\rm KK},\, n+1}\cdot(\vec{\zeta}_{\rm KK,\, 1} - \vec{\zeta}_{{\rm KK},\, n+1})\notag\\
	&=a\,\underbrace{(|\vec{\zeta}_{{\rm KK},\, 1}|^2-|\vec{\zeta}_{{\rm KK},\, n+1}|^2)}_{>0}\geq 0,
\end{align}
so that the $\vec{\zeta}_{\rm KK,\, 1}$ tower always becomes light faster than $\vec{\zeta}_{\text{KK},\, n+1}$ except for $a=0$, namely when $\hat{T}^{\parallel} \propto \vec{\zeta}_{\text{KK},\, n+1}$, in which case they do so at the same rate. 
This ends our proof of Claim \ref{claim1} above.
On the other hand, the species scale strongly depends on the chosen asymptotic trajectory (see e.g., Figure \ref{fig:convexhulls8d}). Hence, in order to check the pattern \eqref{eq:pattern}, one needs to demonstrate the following statement:
\begin{theorem}\label{claim2}
	For any possible species scale vector spanning $\mathscr{F}_d$, that we collectively denote $\lbrace \vec{\mathcal{Z}}_{\rm{KK,}\,n+1} \rbrace$ with $n\in\{1,\ldots,10-d,\infty\}$, we find:
	\begin{subequations}
		\begin{equation}\label{eq:claim2a}
			\vec{\zeta}_{\rm{KK,}\, 1} \cdot \vec{\mathcal{Z}}_{\rm{KK},\, n+1} = \frac{1}{d-2}\, ,
		\end{equation}
		\begin{equation}\label{eq:claim2b}
		\vec{\zeta}_{\rm{KK},\, n'+1} \cdot \vec{\mathcal{Z}}_{\rm{KK},\, n+1}=\frac{1}{d-2}\, .
		\end{equation}
	\end{subequations}
In particular, the second equality holds provided the parent vectors satisfy the pattern in the higher $(d+1)$-dimensional theory. 
\end{theorem}
Note that the first part of the claim above trivially follows from eqs. \eqref{eq:eff-vector} and \eqref{eq:proofKK1dominant}. The second statement, however, requires a bit more work. Intuitively, it means that the condition \eqref{eq:pattern} is consistent (or preserved) under dimensional reduction. 
Thus, we take, without loss of generality, some vector $\vec{\mathcal{Z}}_{\text{KK},\, n+1}$ as the one dominating certain asymptotic region of moduli space within the fundamental domain, and we consider the inner product \eqref{eq:claim2b}. Here, $\vec{\zeta}_{\text{KK},\, n'+1}$ is taken to be any other charge-to-mass vector within $\mathscr{F}_d$ such that it verifies the pattern with respect to $\vec{\mathcal{Z}}_{\text{KK},\, n+1}$ in the parent $(d+1)$-dimensional theory. Recall that, upon dimensionally reducing some vectors $\vec{\zeta}^{\,(d+1)}_{\text{KK},\, n'}$ and $\vec{\mathcal{Z}}^{\,(d+1)}_{\text{KK},\, n}$ on a circle, one gets (c.f. Section \ref{s:consistencydimreduc})
\begin{align}
	\vec{\zeta}_{\text{KK},\, n'} = \left( \vec{\zeta}^{\,(d+1)}_{\text{KK},\, n'}\, ,\, \frac{1}{\sqrt{(d-1)(d-2)}}\right)\, , \quad  \vec{\mathcal{Z}}_{\text{KK},\, n+1} = \left( \vec{\mathcal{Z}}^{\,(d+1)}_{\text{KK},\, n}\, ,\, \frac{1}{\sqrt{(d-1)(d-2)}}\right)\, ,
\end{align}
where the first components of both vectors are directly inherited from the ones of the theory in $d+1$ dimensions, whilst the last entry corresponds to the $\mathbf{S}^1$ radion direction (see also Appendix \ref{ap:generalities}). Hence, requiring $\vec{\zeta}_{\text{KK},\, n'+1}$ to verify the pattern in the higher-dimensional theory translates into the following statement
\begin{equation}\label{eq:patternd+1}
	\vec{\zeta}^{\,(d+1)}_{\text{KK},\, n'} \cdot \vec{\mathcal{Z}}^{\,(d+1)}_{\text{KK},\, n} = \frac{1}{d-1}\, ,
\end{equation}
such that we finally obtain
\begin{align}
	\vec{\zeta}_{\text{KK},\, n'+1} \cdot \vec{\mathcal{Z}}_{\text{KK},\, n+1} &= \left[\frac{1}{n'+1} \left( \vec{\zeta}_{\text{KK},\, 1} + n'\, \vec{\zeta}_{\text{KK},\, n'} \right) \right] \cdot \vec{\mathcal{Z}}_{\text{KK},\, n+1} \notag\\
	& = \frac{1}{n'+1} \vec{\zeta}_{\text{KK},\, 1} \cdot \vec{\mathcal{Z}}_{\text{KK},\, n+1} + \frac{n'}{n'+1} \left[ \vec{\zeta}^{\,(d+1)}_{\text{KK},\, n'} \cdot \vec{\mathcal{Z}}^{\,(d+1)}_{\text{KK},\, n} + \frac{1}{(d-1)(d-2)}\right] \notag\\
	&=\frac{1}{d-2}\, ,
\end{align}
where in order to arrive at the last equality one needs to use eqs. \eqref{eq:claim2a} and \eqref{eq:patternd+1} above. This completes the proof of Claim \ref{claim2}, which ensures that both convex hull diagrams, namely that associated to the $\zeta$-vectors and the species one, are completely dual to each other (with respect to a sphere of radius $\frac{1}{\sqrt{d-2}}$), as also happened for the 9d and 8d cases. Therefore, according to our discussion in Section \ref{sss:sketch}, the immediate consequence of this is that the pattern \eqref{eq:pattern} holds in complete generality for flat space compactifications with maximal supergravity.

For completeness, let us mention that this property holds as well between vectors in- and outside the selected fundamental region (see e.g., Figures \ref{fig:T2SDC&SSDC} and \ref{fig:convexhulls8d}). Notice that this follows from the analysis restricted to $\mathscr{F}_d$ just performed, since any vector outside the fundamental domain can be reached from another one within the latter via the action of some element $g\in G$ of the finite symmetry group $G$ of the diagram. However, since $G$ is a subgroup of the U-duality group of the theory (c.f. Table \ref{tab:irreps}), and this itself is a subgroup of the coset which parameterizes the moduli space (see e.g., \cite{Cecotti:2015wqa}), the scalar product defined with respect to the bi-invariant metric $G_{i j}$ is thus automatically preserved.

\section{Examples in set-ups with 16 supercharges}
\label{s:16supercharges}
	
As we lower the level of supersymmetry, Kaluza-Klein replica are not necessarily BPS anymore, and the vectors generating the convex hull of the towers and the species scale can change upon exploring different regions of the moduli space. Satisfying the pattern in those cases becomes less trivial and provides strong evidence for it beyond maximal supergravity. In this section, we will discuss certain slices of the moduli space of theories preserving 16 supercharges. First, in Section \ref{ss:het s1}, we discuss Heterotic string theory on a circle, for which all asymptotic corners in the space of vacua are well-known \cite{Aharony:2007du}. In that case, it is still possible to define a flat metric\footnote{When referring to a `flat' frame in a certain moduli space we always ignore the compact (axionic) directions, since taking them into account usually introduces a non-vanishing curvature, thus obstructing the definition of a global flat chart.} which will allow us to draw the convex hull in a global fashion \cite{Etheredge:2023odp}, and discuss how it changes as we move in moduli space. For completeness, we also briefly discuss the case of M-theory on $K3$ in Section \ref{ss:Mthy7dpattern}.
	
\subsection{Heterotic string theory in 9d}
\label{ss:het s1}
	
\begin{figure}
\begin{center}
	\subfigure[]{
			\includegraphics[width=0.4\textwidth]{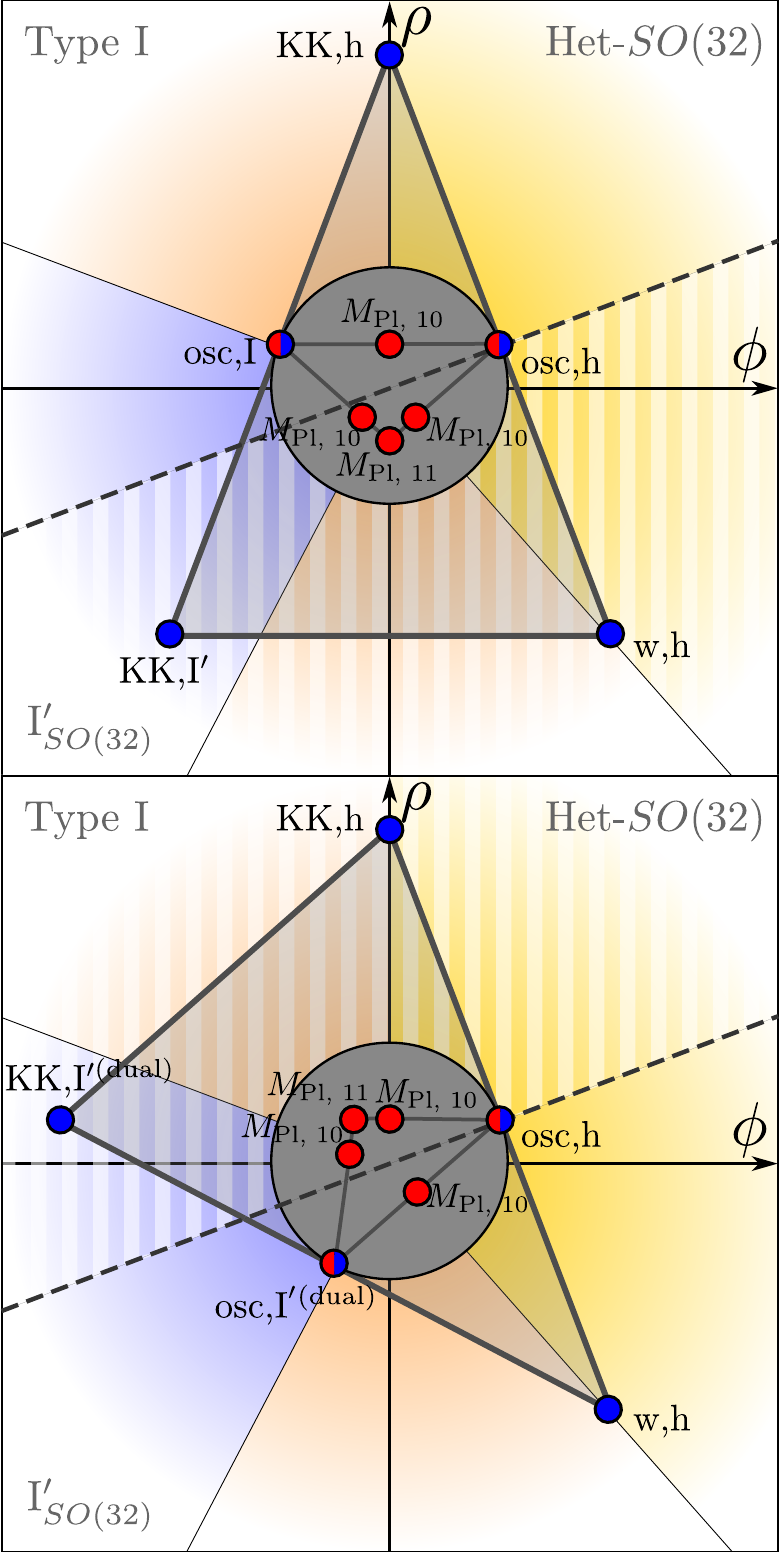}\label{sfig:so32}
		}
        \quad
	\subfigure[]{
			\includegraphics[width=0.4\textwidth]{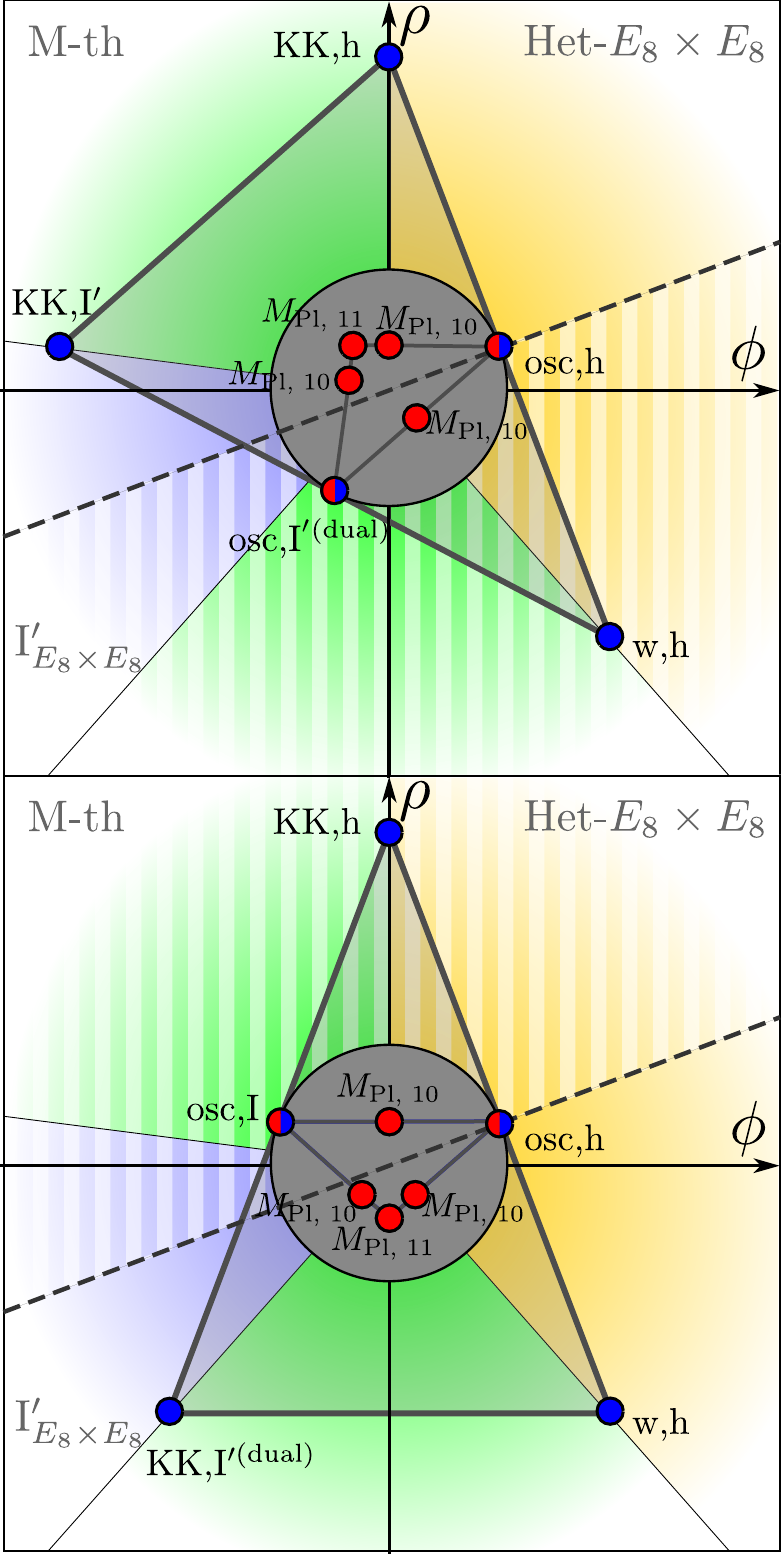}\label{sfig:e8e8}
		}
	\caption{\small Scalar charge-to-mass vectors for the towers (blue) and species scales (red) observed for the \textbf{(a)} $\mathsf{SO(32)}$ and \textbf{(b)} $\mathsf{E_8\times E_8}$ slices of the moduli space of the Heterotic string on $\mathbf{S}^1$, depending on whether the infinite distance limits (along the \emph{non}-dashed regions) are above or below the self-dual line (dashed), following the convention for the canonically normalized moduli as in \cite{Etheredge:2023odp}.}\label{fig:hets1}
\end{center}
\end{figure}

A typical example of a theory with 16 supercharges is that obtained by the compactification of the heterotic string on $\mathbf{S}^1$. This results in an 18-dimensional moduli space $\mathcal{M}_{\text{het}}=\mathbb{R}\times \mathsf{SO(17,1;\mathbb{Z})\backslash SO(17,1;\mathbb{R})/SO(17)}$, parametrizing the 10d dilaton $\phi$, radion $\rho$ and the 16 Wilson lines. We can then study two-dimensional $\{\phi,\rho\}$ slices of $\mathcal{M}_{\text{het}}$ with fixed Wilson line moduli. In particular, we will be interested in two concrete slices of the moduli space of rank 16 (for the gauge group), which can be obtained by compactifying the $\mathsf{SO(32)}$ and $\mathsf{E_8\times E_8}$ 10d heterotic string theories on a circle, with all Wilson lines turned off. We expect equivalent results for the disconnected components of the moduli space with lower rank \cite{Aharony:2007du, Etheredge:2023odp}. Depending on the values taken by the dilaton and radion v.e.v.s, the theory can be better presented in terms of a different dual description, resulting in a finite chain of duality frames, as shown in Figure \ref{fig:hets1} and described in more detail in \cite{Aharony:2007du, Etheredge:2023odp}. Both slices present a self-dual line at $\rho=\frac{1}{\sqrt{7}}\phi$ (the dashed line in Figure \ref{fig:hets1} below) splitting each diagram in two mirrored regions.
	
\begin{figure}[htb]
\begin{center}
	\includegraphics[scale=0.8]{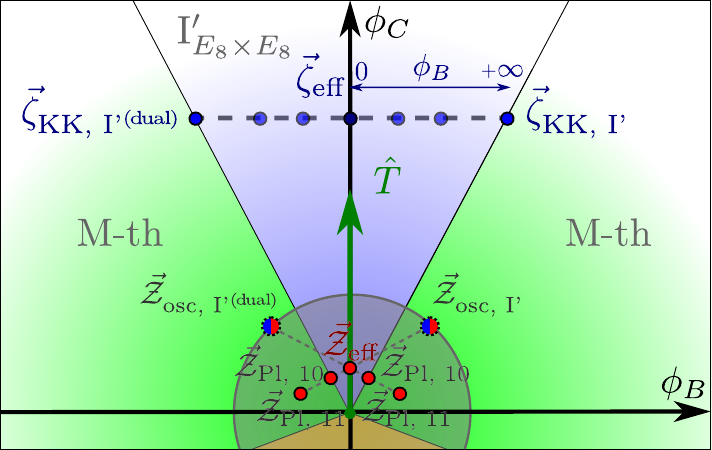}
	\caption{\small Details of the $\mathsf{E_8\times E_8}$ slice of $\mathcal{M}_{\rm het}$, parameterized in terms of $\{\phi_B,\phi_C \}$. When moving with $\hat{T}=(0,1)$, thus parallel to the self-dual line, the Type I$'$ KK tower (and its dual) has a scalar charge-to-mass vector $\vec{\zeta}_{\rm KK,\; I'}$ whose expression depends on the distance $\phi_B$ of the trajectory to the self-dual line (c.f. eq. \eqref{eq: sliding}), coalescing for $\phi_B\to 0$ to $(0,\frac{5}{2\sqrt{7}})$. The fixed $\vec{\mathcal{Z}}_{\rm eff}=\left(0,\frac{2}{5\sqrt{7}} \right)$ vector satisfying the pattern is also depicted. Additional $\mathcal{Z}$-vectors associated to the obstructed emergent string towers as well as the heavier Planck masses are also presented. The $\mathsf{SO(32)}$ slice has an analogous behavior, with $\vec{\zeta}_{\rm KK,\;I'}$ located on the other side of the self-dual line, see \cite{Etheredge:2023odp}.} 
\label{fig:sliding}
\end{center}
\end{figure}

The most interesting duality frame is that corresponding to Type I$'$ string theory, which is an orientifolded version of Type IIA on a circle, with two $O8^-$ planes at the endpoints of the interval and 16 D8-branes, whose location determines the gauge group (16 of then stacked on one orientifold for $\mathsf{SO(32)}$ and a symmetric pair of 8 D8-stacks for $\mathsf{E_8\times E_8}$), with the dilaton running between the $O8^-$ planes and the branes \cite{Polchinski:1995df}. As a result, 
the large radius limit of Type I$'$ leads to decompactification to a running solution of massive Type IIA in 10 dimensions (rather than a higher dimensional vacuum). This makes the scalar charge-to-mass vector of the Type I$'$ KK tower (which is non-BPS) to change non-trivially as we move in moduli space. The main result of \cite{Etheredge:2023odp} shows that warping effects make this vector to \emph{slide} in a perpendicular fashion as we move along a trajectory parallel to self-dual line and change the distance to the latter, see Figure \ref{fig:sliding}. As a function of the asymptotic direction, though, it is simply seen as a \emph{jumping} of the KK vector from one unwarped value to the other as we cross the self-dual line. (Notice that the jump occurs in opposite directions for the $\mathsf{SO(32)}$ or $\mathsf{E_8\times E_8}$ theories.) This implies that, for each duality frame, the location of the $\zeta$-vectors of the towers is the same as in the moduli space of 9d maximal supergravity --- i.e. with 32 supercharges, which becomes clear upon comparing Figure \ref{fig:hets1} with Figure \ref{fig:T2SDC&SSDC} of Section \ref{ss:9d}. The lower level of supersymmetry plays only an important role when determining how to `glue' the different patches altogether, which occurs in a very non-trivial way.
	
Hence, as long as we do not move parallel to the self-dual line in the Type I$'$ region, the relation \eqref{eq:pattern} is still satisfied, since the distribution of the towers and the species vectors is locally the same as in maximal supergravity. Each region will be characterized by a different realization of the species scale (either the 10d string scale or the 11d Planck scale), such that the convex hulls of the towers and species scale are dual to each other and the pattern is thus realized. The tower vectors were already described in \cite{Etheredge:2023odp}, so we are simply computing the species vectors as well here in order to represent everything together in Figure \ref{fig:hets1} above.
	
It remains to be seen, though, whether the pattern will also hold if moving parallel to the self-dual line in the Type I$'$ region. As explained, this limit decompactifies to a running solution in massive Type IIA with a non-trivial spatial dependence of the dilaton. In particular, this changes the exponential rate of the KK tower in comparison to the unwarped result \eqref{eq:zeta&speciesveconemodulus}, as computed in \cite{Etheredge:2023odp}.  For the $\mathsf{E_8\times E_8}$ slice\footnote{The $\mathsf{SO(32)}$ is analogous but with slightly more cumbersome expressions, see Section 3 in \cite{Etheredge:2023odp}.} one has
\begin{equation}\label{eq: sliding}
	\frac{m_{\rm KK,\, I'}}{M_{\rm Pl;\, 9}}\sim e^{-\frac{5}{2\sqrt{7}}\phi_C+\frac{3}{2}\phi_B}(1+3e^{2\phi_B})^{-1}\Longrightarrow \vec{\zeta}_{\rm KK,\, I'}=\left(\frac{1}{2}-\frac{2}{1+3e^{2\phi_B}},\frac{5}{2\sqrt{7}}\right)
\end{equation}
which is written in a basis of flat coordinates $\{\phi_B,\phi_C\}$.\footnote{This amounts to a clockwise $\frac{\pi }{2}+\arctan\left(\frac{1}{\sqrt{7}}\right)$ rotation from the $\{\phi,\rho\}$ coordinates shown in Figure \ref{fig:hets1}.} Each of these coordinates measures, respectively, the moduli space distance perpendicular and parallel to the self-dual line in the Type I$'$ frame. As already mentioned, this implies that the Type I$'$ KK modes move orthogonal to the self-dual line as a function of $\phi_B$, see Figure \ref{fig:sliding}.
At each side of the self-dual line (i.e. in each of the Type I$'$ frames) we seem to have a different tower of KK states, whose scalar charge-to-mass ratio coincides when moving exactly along the interface. We expect that these towers actually correspond to different sets of states that are mapped to each other upon performing the duality. If that is the case, they should both contribute to $\LSP$, yielding a lower value for the species cut-off (i.e. a larger value of the exponential rate) than what each tower alone would provide. The Type I$'$ string oscillator modes, though, are not expected to contribute since the string perturbative limit is obstructed. Computing this species scale from top-down string theory would constitute a project on its own, so we leave it for future work. Here, we will simply determine what should be the value for $\LSP$ along the self-dual line such that the pattern holds even for these decompactifications to running solutions. We hope that this can be useful to elucidate the fate of the pattern in these special cases.
	
Along the self-dual line, the scalar charge-to-mass vector of the KK towers is given by $\vec{\zeta}_{\rm eff}=\left(0,\frac{5}{2\sqrt{7}}\right)$, with an associated species vector $\vec{\mathcal{Z}}_{\rm eff}$ that should also point towards this direction. For decompactification limits, the species scale can be computed in terms of an \emph{effective tower} $m_{\text{eff},\, n}\sim n^{1/p_{\rm eff}}m_{\rm eff,\, 0}$ with  $p_{\rm eff}=\sum_i p_i$ and $m_{i,\, n}\sim n^{1/p_{i}}m_{i,\, 0}$, see Section \ref{ss:MultipleTowers} for details on this. We do not expect $p_{\rm eff}=1$ since this would correspond to having a single KK tower decompactifying one dimension, nor $p_{\rm eff}=2$ since it would rather indicate a double decompactification. In fact, for the pattern to hold, one can check that the required value for the density parameter is somewhat in between, namely $p_{\rm eff}=\frac{4}{3}$, which can be obtained upon identifying
\begin{equation}
	\frac{\Lambda_{\rm eff}}{M_{\text{Pl;}\, 9}}=\left(\frac{m_{\rm eff}}{M_{\text{Pl;}\, 9}}\right)^{\frac{p_{\rm eff}}{9-2+p_{\rm eff}}}=e^{-\frac{2}{5\sqrt{7}}\phi_C}\Longrightarrow \vec{\mathcal{Z}}_{\rm eff}=\left(0,\frac{2}{5\sqrt{7}} \right)\, .
\end{equation}
This value would imply $\vec{\zeta}_{\rm eff}\cdot\vec{\mathcal{Z}}_{\rm eff}=\vec{\zeta}_{\rm KK,\, I'}\cdot\vec{\mathcal{Z}}_{\rm eff}=\vec{\zeta}_{\rm KK,\, I'^{\,(dual)}}\cdot\vec{\mathcal{Z}}_{\rm eff}=\frac{1}{7}$, satisfying the pattern for any $\phi_B\geq 0$. Along the self-dual line, the Type I$'$ radion --- measured in 10d Planck units --- and string coupling scale as $R_{\rm I'}M_{\rm Pl; 10}=g_{\rm I'}^{-5/4}\sim e^{\frac{5\sqrt{7}}{16}\phi_C}$. This implies that the species cut-off should scale as $\frac{\LSP}{M_{\rm Pl;\, 10}}\sim (R_{\rm I'}M_{\rm Pl;\, 10})^{-\frac{32}{175}}\sim g_{\rm I'}^{\frac{8}{35}}$, although it is not possible for us to elucidate the separate dependence on the radion and the dilaton. It would be interesting to check, directly from string theory, whether this behaviour of the species scale is indeed realized and the structure of the KK towers (taking into account the large warping associated to decompactifying to a running solution) is such that effectively implies  $p_{\rm eff}=\frac{4}{3}$. Hence, whether the pattern is fulfilled in this particular asymptotic direction remains open and is left for future investigation. Similarly, it is easy to see that with these state of affairs the bound \eqref{eq:lowerboundspecies} introduced in Chapter \ref{ch:bounds} would be equally satisfied along every possible infinite distance direction, including those parallel to the self-dual line. 
	
	
	

\subsection{M-theory in 7d}\label{ss:Mthy7dpattern}
	
Let us now turn to our second example and consider M-theory compactified on a $K3$ surface, leading to a supersymmetric set-up in 7d with 16 supercharges as well. This setting exhibits many features that will be explained in more detail when discussing 4d theories arising from Calabi--Yau compactifications. Furthermore, our analysis here nicely complements the work performed in \cite{Lee:2019xtm}, where the emphasis was placed on the leading tower of states rather than the species scale. Here we will focus again on \emph{attractive} $K3$ two-folds, since in that case all relevant moduli dependence arises just from the K\"ahler sector.

To check the pattern in the present context, we need to know the explicit form of the moduli space metric, which is captured as usual by the kinetic terms in the scalar lagrangian. The full bosonic action was already discussed in Section \ref{sss:MtheoryonK3}, so we will refer oftentimes to the material presented there. Thus, in the attractive case, the relevant line element reads
\begin{equation}\label{eq:7dMthymetric}
	\begin{aligned}
		ds_{\rm VM}^2\, = \, \frac{9}{20} d\mathcal{V}_{K3}^2 + \mathsf{G}_{a b}\, d\tilde{t}^a d \tilde{t}^b\, ,
	\end{aligned}
\end{equation}
where $\mathcal{V}_{K3}$ denotes the overall internal volume and $\{ \tilde{t}^a\}$ are constrained K\"ahler moduli, see discussion around eq. \eqref{eq:7dMthyattractive}. The latter describe a (classically exact) subspace of the group coset \eqref{eq:cosetspace7d}, which admits a natural metric given by
\begin{equation}\label{eq:7dmodspacemetricpattern}
	\begin{aligned}
		\mathsf{G}_{a b} = \frac{t_a t_b}{\mathcal{V}_{K3}}- \eta_{a b} = \tilde{t}_a \tilde{t}_b -\eta_{a b}\, ,
	\end{aligned}
\end{equation}
where the indices are lowered with the intersection form $\eta_{a b}$. 
	
Regarding the infinite distance boundaries of such moduli space, there are several of them, according to which moduli are sent to infinity: the large volume point, the small `radius' limit, a \emph{unique} type of infinite distance degeneration at constant $\mathcal{V}_{K3}$ and combinations thereof. We discuss each of them in turn.
	
\subsubsection*{The large/small volume limits}
	
Let us start with the large volume singularity $\mathcal{V}_{K3} \to \infty$, which of course lies at infinite distance in the field space metric defined from eq. \eqref{eq:7dMthymetric} above. It corresponds to the full decompactification limit, where the $K3$ manifold grows large and we come back effectively to 11d supergravity. Thus, the infinite tower of asymptotically light states is given by the KK tower, whose mass is given by
\begin{equation}
	\frac{m_{\text{KK, K3}}}{M_{\text{Pl};\, 7}} = \mathcal{V}_{K3}^{-9/20} \quad \Longrightarrow \quad \vec{\zeta}_{\text{KK, K3}} = \left( \frac{9}{20} \frac{1}{\mathcal{V}_{K3}}, 0, \ldots, 0 \right)\, ,
\end{equation}
where we have used that the 7d and 11d Planck scales are related by $M_{\text{Pl};\, 7}^5= M_{\text{Pl};\, 11}^5 \mathcal{V}_{K3}$. The associated species scale corresponds to the 11d Planck mass, such that upon taking the inner product between $\vec{\zeta}_{\text{KK, K3}}$ and $\vec{\mathcal{Z}}_{\text{sp}} = \frac{4}{9} \vec{\zeta}_{\text{KK, K3}}$ (c.f. eq. \eqref{eq:eff-vector}) we find that $\vec{\zeta}_{\text{KK, K3}} \cdot \vec{\mathcal{Z}}_{\text{sp}}= \frac{1}{5}$, in agreement with \eqref{eq:pattern}.
	
The small `radius' limit, namely $\mathcal{V}_{K3} \to 0$, is of different physical nature. One can argue that it corresponds to an emergent string limit, where an asymptotically tensionless and weakly coupled Heterotic string emerges at infinite distance. Indeed, it is possible to construct an Heterotic-like string by wrapping the M5-brane on the whole $K3$ surface \cite{Cherkis:1997bx,Park_2009}, with a tension in 7d Planck units which reads as follows
\begin{equation}
	\frac{T_{\text{M5}}}{M_{\text{Pl};\, 7}^2} = \mathcal{V}_{K3}^{3/5} \quad \Longrightarrow \quad \vec{\zeta}_{\text{osc, M5}} = \left( -\frac{3}{10} \frac{1}{\mathcal{V}_{K3}}, 0, \ldots, 0 \right)\, .
\end{equation}
Moreover, there are additional $\frac{1}{2}$-BPS states arising from wrapped M2-branes on certain holomorphic curves within the $K3$, which correspond to perturbative winding modes of the dual Heterotic string on $\mathbf{T}^3$.\footnote{Note that since $H^2(K3, \mathbb{Z})$ defines a lattice of signature $(3,19)$ there are precisely 3 non-equivalent holomorphic curve classes with non-negative self-intersection, and thus non-contractible. These should correspond to the 3 winding modes sectors of the dual Heterotic string on $\mathbf{T}^3$.} Their mass operator can be deduced from the DBI action (c.f. eq. \eqref{eq:DBICSactionMpbranes}), and yields 
\begin{equation}\label{eq:M2mass&vectors}
	\frac{m_{\text{M2}}^{(a)}}{M_{\text{Pl};\, 7}} = t^a\, \mathcal{V}_{K3}^{-1/5} = \tilde{t}^a\, \mathcal{V}_{K3}^{3/10} \quad \Longrightarrow \quad \vec{\zeta}^{\,(a)}_{\text{M2}} = \left( -\frac{3}{10} \frac{1}{\mathcal{V}_{K3}}, 0, \ldots, -\frac{1}{\tilde{t}^a}, \ldots, 0 \right)\, ,
\end{equation}
where the non-zero entries correspond to the overall volume component and the one associated to the rescaled $\tilde{t}^a$ modulus (see discussion after eq. \eqref{eq:7dMthymetric}). It is therefore clear that upon contracting $\vec{\zeta}_{\text{t}}= \lbrace \vec{\zeta}_{\text{osc, NS5}}, \vec{\zeta}^{\,(a)}_{\text{M2}} \rbrace$ with $\vec{\mathcal{Z}}_{\text{sp}}=\vec{\zeta}_{\text{osc, NS5}}$, one obtains $\vec{\zeta}_{\text{t}} \cdot \vec{\mathcal{Z}}_{\text{sp}}= \frac{1}{5}$, thus fulfilling the pattern.
	
\subsubsection*{Infinite distance at fixed (overall) volume}
	
Let us consider now infinite distance limits with the overall volume kept fixed and constant. In fact, as demonstrated in \cite{Lee:2019xtm} (see also earlier related works in \cite{Lee:2018urn,Lee:2018spm}), for such a limit to exist it must be possible to select some $\omega_0 =\sum_a c^a \omega_a \in H^{1,1}(X_3,\mathbb{Z})$ (with $c^a \geq 0$) such that\footnote{The fact that the limit \eqref{eq:T2limit7d} lies at infinite distance with respect to the metric \eqref{eq:7dmodspacemetricpattern} follows from the asymptotic dependence of $\mathsf{G}_{a b}$:
\begin{equation}
	\notag \Delta= \int_1^{\infty} \dd \sigma \sqrt{\mathsf{G}_{a b} \frac{d \tilde{t}^a}{d \sigma} \frac{d \tilde{t}^b}{d \sigma}}\, \sim\, \int_1^{\infty} \dd \log \left(\tilde{t}^0 \right)\, \to\, \infty\, ,
\end{equation}
where we have used that $\mathsf{G}_{i j}=\eta_{0i} \eta_{0j} \left( \tilde{t}^0\right)^2 + \mathcal{O} (\sigma^0)$, $\mathsf{G}_{0 j}=\eta_{0j} \eta_{i 0} \tilde{t}^j \tilde{t}^0 - \eta_{0i} + \mathcal{O} (1/\sigma^2)$ and $\mathsf{G}_{0 0}= \eta_{0j} \eta_{i 0} \tilde{t}^i \tilde{t}^j$.}
\begin{equation}\label{eq:T2limit7d}
	J= t^0 \omega_0 + t^i \omega_i 
	\, , \qquad \text{with}\ \ t^0 = \sigma,\ t^i = \frac{a^i}{\sigma}, \quad \sigma \to \infty\, ,
\end{equation}
where $i=1, \ldots, 19,$ and the basis $\lbrace \omega_0, \omega_i\rbrace$ verifies that $\omega_0 \cdot \omega_0=0$ and $\sum_i a^i\, \omega_0 \cdot \omega_i= \mathcal{V}_{K3} + \mathcal{O}(1/\sigma^2)$. Geometrically, the very existence of such a limit enforces the attractive $K3$ to admit some elliptic fibration over a $\mathbb{P}^1$-base, with the genus-one fibre $\mathcal{C}_0$ being Poincaré dual to the K\"ahler cone generator $\omega_0$. Such holomorphic curve shrinks upon taking the limit \eqref{eq:T2limit7d}, whilst the base grows at the same rate so as to keep the overall $\mathcal{V}_{K3}$ fixed and finite.
	
\begin{figure}[htb]
	\begin{center}
		\includegraphics[scale=.35]{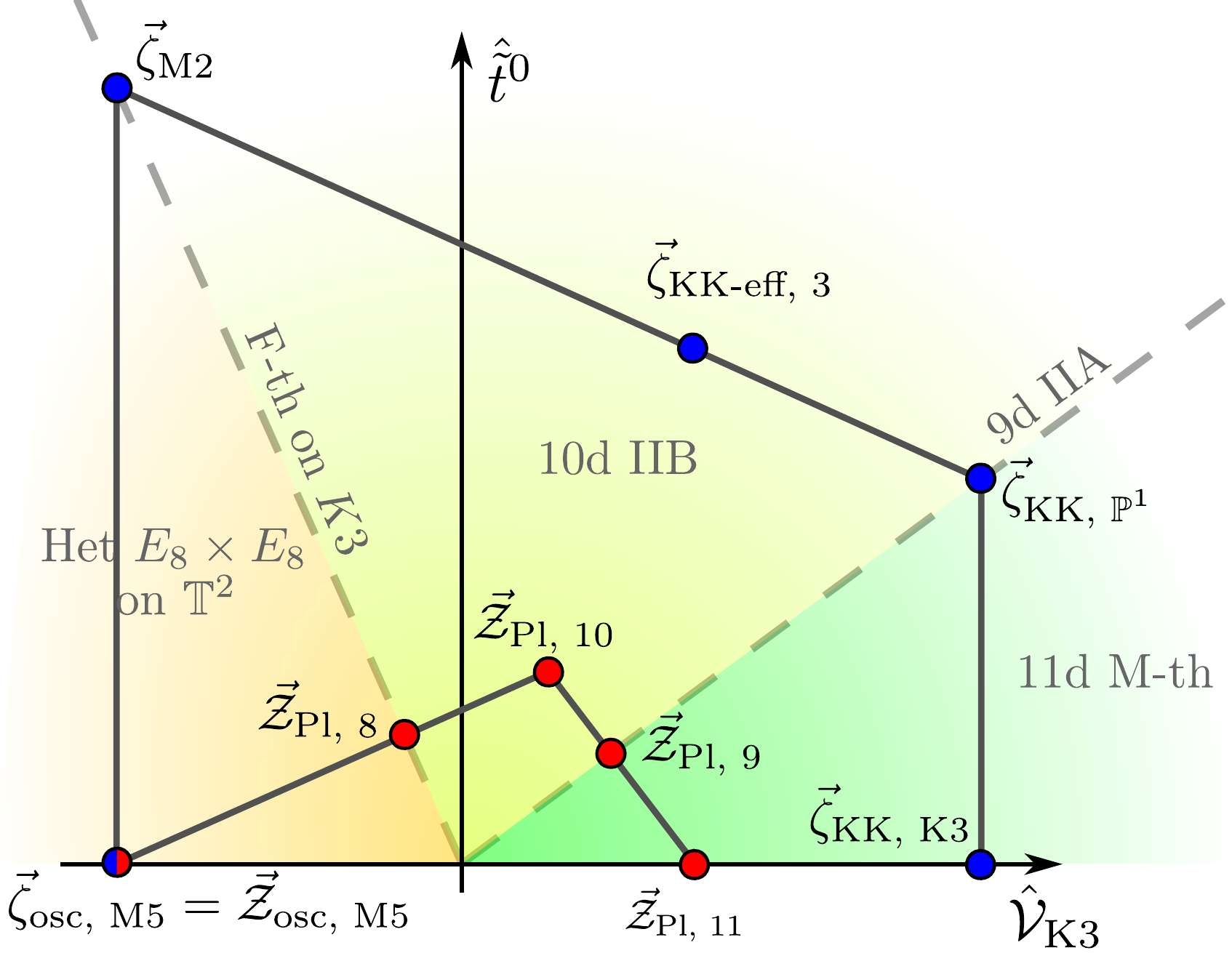}
		\caption{\small Convex hulls for the lightest towers (blue) and species scale (red) in M-theory compactified on an attractive $K3$ surface, using a flat frame $\lbrace \hat{\tilde{t}}^0, \hat{\mathcal{V}}_{K3}\rbrace$, in which the equations of the different vectors are $\vec{\zeta}_{\rm osc,\; M5}=\vec{\mathcal{Z}}_{\rm osc,\; M5}=\left(-\frac{1}{\sqrt{5}},0\right)$, $\vec{\zeta}_{\rm M2}=\left(-\frac{1}{\sqrt{5}},1\right)$, $\vec{\zeta}_{\rm KK-eff,\; 3}=\left(\frac{2}{3\sqrt{5}},\frac{2}{3}\right)$, $\vec{\zeta}_{{\rm KK},\; \mathbb{P}^1}=\left(\frac{3}{2\sqrt{5}},\frac{1}{2}\right)$, $\vec{\zeta}_{{\rm KK},\; K3}=\left(\frac{3}{2\sqrt{5}},0\right)$, $\vec{\mathcal{Z}}_{\rm Pl,\; 8}=\left(-\frac{1}{6\sqrt{5}},\frac{1}{6}\right)$, $\vec{\mathcal{Z}}_{\rm Pl,\; 10}=\left(\frac{1}{4\sqrt{5}},\frac{1}{4}\right)$, $\vec{\mathcal{Z}}_{\rm Pl,\; 9}=\left(\frac{3}{7\sqrt{5}},\frac{1}{7}\right)$ and $\vec{\mathcal{Z}}_{\rm Pl,\; 11}=\left(\frac{2}{3\sqrt{5}},0\right)$. It is easy to see that both polytopes are dual to each other (with respect to the 1-sphere of radius $\frac{1}{\sqrt{d-2}}=\frac{1}{\sqrt{5}}$), and thus the pattern is satisfied. The different limiting theories, which can be deduced by looking at the dominant species scale in each region of the moduli space, are also shown for completeness.} 
		\label{fig:CHMthy7d}
	\end{center}
\end{figure}

Given the behavior of the different 2-cycles along the limit \eqref{eq:T2limit7d}, there exist potentially two kinds of infinite towers of states. First, there are the supergravity KK modes associated to the $\mathbb{P}^1$-base, whose volume grows asymptotically. The mass scale of such tower behaves as follows
\begin{equation}\label{eq:massP1}
	\frac{m_{\text{KK},\, \mathbb{P}^1}}{M_{\text{Pl};\, 7}}=\frac{1}{\left(\tilde{t}^0\right)^{1/2}\, \mathcal{V}_{K3}^{9/20}} \quad \Longrightarrow \quad \vec{\zeta}_{\text{KK},\, \mathbb{P}^1} = \left(\frac{9}{20} \frac{1}{\mathcal{V}_{K3}}, \frac{1}{2 \tilde{t}^0}, 0, \ldots, 0 \right)\, ,
\end{equation}
so that it becomes (exponentially) light upon probing the $\tilde{t}^0 \to \infty$ limit. In addition, there is a second infinite set of states becoming light even faster, which arise from M2-branes wrapping the genus-one fibre. Their mass is controlled by the volume of the latter
\begin{equation}\label{eq:M2massFthy}
	\frac{m_{\text{M2}}}{M_{\text{Pl};\, 7}} = \mathcal{V}_{\mathcal{C}_0}\, \mathcal{V}_{K3}^{-1/5}=\frac{\mathcal{V}_{K3}^{3/10}}{\tilde{t}^0} \quad \Longrightarrow \quad \vec{\zeta}_{\text{M2}} = \left( -\frac{3}{10} \frac{1}{\mathcal{V}_{K3}}, \frac{1}{\tilde{t}^0}, 0, \ldots, 0 \right)\, ,
\end{equation}
and they can be seen to correspond to the dual KK replica implementing the duality between M-theory on $K3$ and F-theory on $K3\times \mathbf{S}^1$ \cite{Vafa:1996xn,Lee:2019xtm}. However, in order to correctly interpret what is the resolution of the singularity in QG, we need to study the behavior of the species scale. One can thus associate two such scales, one for each tower, as follows (c.f. eq. \eqref{eq:eff-vector})
\begin{align}\label{eq:species7d}
	\notag \frac{\Lambda_{\rm M2}}{M_{\text{Pl};\, 7}} &\simeq \left(m_{\text{M2}}\right)^{1/6} = \frac{\mathcal{V}_{K3}^{1/20}}{\left( \tilde{t}^0\right)^{1/6}}\quad \Longrightarrow \quad \vec{\mathcal{Z}}_{\text{Pl},\, 8} = \left( -\frac{1}{20} \frac{1}{\mathcal{V}_{K3}}, \frac{1}{6\, \tilde{t}^0}, 0,\ldots, 0 \right)\, ,\\
	\frac{\Lambda_{\text{Pl},\, 9}}{M_{\text{Pl};\, 7}} &\simeq \left(m_{\text{KK},\, \mathbb{P}^1}\right)^{2/7} = \frac{1}{\left(\tilde{t}^0\right)^{1/7}\, \mathcal{V}_{K3}^{9/70}} \quad \Longrightarrow \quad \vec{\mathcal{Z}}_{\text{Pl},\, 9} = \left(\frac{9}{70} \frac{1}{\mathcal{V}_{K3}}, \frac{1}{7 \tilde{t}^0}, 0, \ldots, 0 \right)\, , 
\end{align}
which coincide with the 8d Planck scale\footnote{This can be easily checked upon identifying $R_8 = \frac{(\tilde{t}^0)^{5/6}}{\mathcal{V}_{K3}^{1/4}}$, where $R_8$ denotes the radius (in 8d Planck units) of the F-theory circle, as well as the relation between the 8d and 7d Planck scales, namely $M_{\text{Pl};\, 7}^5= M_{\text{Pl};\, 8}^5 2\pi R_8$.} (in the F-theory frame) and the 9d Planck scale, respectively. We are not done yet though, since both sets of states can be combined together forming bound states, namely the wrapped M2-branes may have non-trivial momentum along the $\mathbb{P}^1$-base. Furthermore, such `mixed' states contribute to the computation of a third candidate for the species scale, whose $\mathcal{Z}$-vector reads (see eq. \eqref{eq:effectiveKKspeciesvector})
\begin{align}\label{eq:speciesKK10d}
	\vec{\mathcal{Z}}_{\text{Pl},\, 10} = \frac{1}{8} \left( \vec{\zeta}_{\text{M2}} + 2 \vec{\zeta}_{\text{KK},\, \mathbb{P}^1}\right) = \left(\frac{3}{40} \frac{1}{\mathcal{V}_{K3}}, \frac{1}{4 \tilde{t}^0}, 0, \ldots, 0 \right)\, , 
\end{align}
thus signalling towards decompactification to 10d Type IIB string theory. In Figure \ref{fig:CHMthy7d} all these vectors are plotted --- both for the mass and species scale, including those relevant in the large/small $K3$ volume limit, as previously discussed. 
	
With this, we are now ready to check what is the minimum $\LSP$ dominating the asymptotic physics along the limit \eqref{eq:T2limit7d}. Indeed, it is easy to see either from the formulae above or the diagram in Figure \ref{fig:CHMthy7d}, that this becomes the 10d Planck scale. Therefore, such limit may be interpreted as some `nested' decompactification, first from 7d M-theory to 8d F-theory (as remarked in \cite{Lee:2019xtm}) and then up to ten dimensions, effectively sending all supersymmetry breaking defects (i.e. D7-branes and $O7$-planes) to infinity and restoring maximal chiral supergravity in 10d. Hence, a quick computation reveals that the pattern $\vec{\zeta}_{\text{M2}} \cdot \vec{\mathcal{Z}}_{\text{Pl},\, 10}= \frac{1}{5}$ is also verified in this limit (to leading order in $1/\tilde{t}^0$). 
	
\subsubsection*{Intermediate limits}
	
To conclude, let us briefly comment on the possibility of superimposing any of the previous limits, thus sending both the overall $K3$ volume and the $\tilde{t}^0$ K\"ahler modulus to infinity at different rates, a priori. In fact, upon comparing the different species scale that can arise (and even compete) at distinct corners of the moduli space, one can indeed separate these asymptotic regions into different sectors, depending on which specific scale dominates (see Figure \ref{fig:CHMthy7d}). In any event, it is straightforward to verify that the pattern is respected in all such cases, due to the non-trivial gluing conditions between the different patches.

\section{Examples in 4d $\mathcal{N}=2$ EFTs}
\label{s:8supercharges}
	
We now turn to theories with 8 supercharges. In particular, we will focus on 4d $\mathcal{N}=2$ set-ups arising upon compactifying Type II string theory on Calabi--Yau three-folds. The singularity structure of the moduli space of these theories is very rich and has been thoroughly studied in the literature, providing for different types of infinite distance limits. In Sections \ref{ss:preliminary}-\ref{ss:commentsTypeIIB} we discuss the vector multiplet sector by studying different concrete examples as well as presenting general arguments in favour of satisfying the pattern. Section \ref{ss:hypers} analyzes the effect of (towers of) instanton corrections on singularities located classically at infinite distance, which are nevertheless excised and deflected within the true quantum hypermultiplet moduli space.
	
\subsection{The vector multiplet moduli space}
\label{ss:preliminary}
	
Recall from Section \ref{sss:4dN=2basics} that the moduli space of 4d $\mathcal{N}=2$ theories factorizes at the two-derivative level in two pieces: the vector multiplet and the hypermultiplet sectors. For concreteness, we focus on theories obtained upon compactifying Type IIA string theory on a Calabi--Yau three-fold $X_3$, although we will make some comments regarding the Type IIB counterpart later on in Section \ref{ss:commentsTypeIIB}.

Since we will only be interested in the computation of the relevant scalar charge-to-mass vectors as well as the corresponding species scale, we restrict ourselves to the scalar and gravitational sectors of the 4d action \eqref{eq:IIAaction4d}, effectively forgetting about the vector fields. Thus, in the low energy regime, the relevant piece of the action reads
\begin{equation}\label{eq:IIAscalaraction4d}
	\begin{aligned}
		\ S^{\text{4d}}_{\rm IIA}\, \supset\, & \frac{1}{2\kappa^2_4} \int \mathcal{R} \star 1 -2 G_{a\bar b}\, d z^a\wedge \star d\bar z^b + 2h_{pq}\, d q^p \wedge \star d q^q\, ,
	\end{aligned}
\end{equation}
where the fields $z^a=b^a+{\rm i}t^a$, $a=1,\ldots, h^{1,1}$, describe the (complexified) K\"ahler sector of the theory, whereas the scalars in the various hypermultiplets (including e.g., the complex structure moduli) are denoted by $q^p$. In the following, we will particularize to the vector multiplet moduli space, leaving the analysis of the hypermultiplet sector for Section \ref{ss:hypers}. 

The explicit expression for metric associated to the complex fields $\{ z^a \}$ is (c.f. eq. \eqref{eq:kahlersectormetric})
\beq\label{eq:kahlersectormetricpattern}
	G_{a\bar b}=\partial_a \partial_{\bar{b}} K_{\text{ks}}= -\partial_a \partial_{\bar{b}} \log \left( 8 \mathcal{V}_{X_3} \right)\, ,
\eeq
where $\mathcal{V}_{X_3}$ is the classical volume of the three-fold measured in string units. By Mirror Symmetry, this effective theory can be equivalently described as arising from compactifying Type IIB on the mirror Calabi--Yau, such that the role of K\"ahler and complex structure moduli get exchanged (see Section \ref{ss:dualitieswithlowersusy} for details). The different types of infinite distance limits in the vector multiplet sector can then be systematically classified using the theory of Mixed Hodge Structures \cite{Grimm:2018ohb, Grimm:2018cpv}. However, in the present work, we will analyze each of these limits using the language of Type IIA compactifications, since the microscopic interpretation of the corresponding asymptotic limit (either decompactification or emergent string limit \cite{Lee:2019wij}) becomes more apparent from this point of view.
	
\subsubsection*{Classification of infinite distance limits at large volume}
\label{sss:largevolume}
	
From the perspective of Type IIA string theory, we need to particularize to the large volume patch, where one can safely ignore both $\alpha'$ and worldsheet instanton contributions which further correct the form of the metric displayed in \eqref{eq:kahlersectormetricpattern}. Still, the structure of possible infinite distance singularities is very rich as we review in what follows. Thus, according to \cite{Corvilain:2018lgw,Grimm:2018cpv,Grimm:2018ohb}, we can parametrize infinite distance limits within the K\"ahler cone in terms of trajectories of the form
\beq \label{eq:singlefieldlim}
	\lbrace t^i\rbrace = t^1,\ldots , t^{n}\to \infty\, ,\qquad n\leq h^{1,1}(X_3)\, ,
\eeq
with $b^i$ approaching finite values. The several distinct types of infinite distance limits have been thoroughly studied and classified by different means in \cite{Corvilain:2018lgw,Lee:2019wij}, and can be divided into three classes shown in Table \ref{tab:intersN=2} below, depending on the behavior of the intersection numbers $\mathcal{K}_{abc}$ with the asymptotic direction taken. More details about the notation in terms of Roman numerals can be found in \cite{Grimm:2018ohb}, whilst that in terms of $J$-class A/B can be found in \cite{Lee:2019wij} (see also \cite{Lee:2019tst}). Geometrically, these three classes correspond to different fibration structures: the \emph{unique} limit in which the overall volume of $X_3$ blows up uniformly, thus corresponding to the large volume point; the ones in which the CY$_3$ possesses an elliptic fibration over some K\"ahler two-fold; and those in which the three-fold develops either some $K3$ or $\mathbf{T}^4$ fibration over a $\mathbb{P}^1$-base. We will consider in the upcoming subsections specific examples of each representative class of limit followed by a general analysis of each singularity type, providing all of them more evidence in favour of the pattern here described. 
	
In Table \ref{tab:limitsN=2}, we summarize the microscopic interpretation of the leading tower of states becoming light at each type of infinite distance limit, as well as the physical realization of the species scale for each case. Recall that, in this section, we consider infinite distance limits lying purely within the vector multiplet moduli space, while all hypermultiplet scalars (including the 4d dilaton) remain fixed. To achieve this, we will sometimes need to co-scale properly certain ten-dimensional variables \cite{Lee:2019wij}. For instance, if we want to keep the 4d dilaton, $\varphi_4 =\phi-\frac{1}{2} \log \mathcal{V}$, fixed and \emph{finite}, one needs to rescale accordingly the 10d dilaton $\phi$, which will bring us to the strong coupling regime of Type IIA as we will see below in more detail. For other limits involving also the hypermultiplets, see Section \ref{ss:hypers}.

\begin{table}[h!!]\begin{center}
	\renewcommand{\arraystretch}{0.80}
		\begin{tabular}{|c|c|c|}
		\hline 
		Type \cite{Corvilain:2018lgw} & Type \cite{Lee:2019wij} & Intersection numbers \\
		\hline 
		\hline 
		IV$_d$ & --- & $\text{rk}(\boldsymbol{\kappa}^{(n)}) = \text{rk}(\boldsymbol{\kappa}^{(n)}_a)= 1$ and $\text{rk}(\boldsymbol{\kappa}^{(n)}_{a b})=d$   \\
		III$_c$ & $J$-class A & $\text{rk}(\boldsymbol{\kappa}^{(n)}) = 0$, $\text{rk}(\boldsymbol{\kappa}^{(n)}_a)= 1$ and $\text{rk}(\boldsymbol{\kappa}^{(n)}_{a b})=c+2$ \\
		II$_b$ & $J$-class B & $\text{rk}(\boldsymbol{\kappa}^{(n)}) = \text{rk}(\boldsymbol{\kappa}^{(n)}_a)= 0$ and $\text{rk}(\boldsymbol{\kappa}^{(n)}_{a b})=b$ \\
		\hline
	\end{tabular}
\caption{\small Infinite distance limits in the large volume regime within the vector multiplet moduli space of Type IIA compactified on a CY$_3$. They can be classified in terms of the behavior of the triple intersection numbers $\mathcal{K}_{abc}$ via Mixed Hodge Theory \cite{Corvilain:2018lgw}, or using a purely geometrical analysis \cite{Lee:2019wij}. The following notation has been introduced: $\boldsymbol{\kappa}^{(n)}_{a b} = \sum_{i=1}^n \mathcal{K}_{i a b}$, $\boldsymbol{\kappa}^{(n)}_{a} = \sum_{i,j=1}^n \mathcal{K}_{i j a}$, $\boldsymbol{\kappa}^{(n)} = \sum_{i,j,k=1}^n \mathcal{K}_{i j k}$ and $\text{rk}(\cdot)$ denotes the rank of the corresponding matrix.}
\label{tab:intersN=2}
\end{center}
\end{table}
\begin{table}[h!!]\begin{center}
\renewcommand{\arraystretch}{1.00}
	\begin{tabular}{|c|c|c|c|c|}
		\hline
		Type \cite{Corvilain:2018lgw} & Type \cite{Lee:2019wij} &  Fibration structure  &  Dominant Tower & $\LSP$\\
		\hline 
		\hline 
		IV$_d$ & --- & Unspecified  & D0 & $M_{\text{Pl};\, 5}$\\
		III$_c$  & $J$-class A & Elliptic Fibration & D0 and D2 on $\mathbf{T}^2$ & $M_{\text{Pl};\, 6}$\\
		II$_a$ & $J$-class B & $K3$ or $\mathbf{T}^4$ Fibration & NS5 on $K3/\mathbf{T}^4$ & $\sqrt{T_{\rm NS5}}$\\
		\hline
	\end{tabular}
	\caption{\small Infinite distance limit classification according to refs. \cite{Corvilain:2018lgw} and \cite{Lee:2019wij}. We also show the kind of asymptotic fibration structure exhibited by the three-fold as well as the dominant tower(s) of states controlling the species scale for each case.}
	\label{tab:limitsN=2}
\end{center}
\end{table}


\subsection{Type IV limits: M-theory circle decompactification}
\label{ss:typeIVlimits}
	
\subsubsection{Example 1: the Quintic}
\label{sss:ExampleI}
	
As our first example, we consider a one-modulus case and we explore the large volume point, which is always present within the vector multiplet moduli space \cite{Corvilain:2018lgw}. For concreteness, we particularize to the quintic three-fold studied in \cite{Candelas:1987se,Candelas:1990rm}, which may be defined as a family of degree 5 hypersurfaces in $\mathbb{P}^4$. Such three-fold presents 101 complex parameters (appearing in the quintic polynomial) associated to complex structure deformations, as well as a single (complexified) K\"ahler structure modulus that we denote by $z=b+{\rm i}t$. Within the vector multiplet moduli space one finds three singularities: the large volume point at $z\to {\rm i}\infty$, the conifold locus located at $z \simeq 1.21\, {\rm i}$, and the Landau-Ginzburg orbifold point, which happens for $z = \frac{1}{2} \left( 1+ {\rm i} \cot{\pi/5}\right)$ \cite{Blumenhagen:2018nts}. 
	
Close to the large volume point, the K\"ahler potential behaves as \cite{Candelas:1990rm}
\beq\label{eq:KahlerpotLV}
	e^{-K_{\text{ks}}} = \frac{256 \pi^6 }{9375}t^3 + \mathcal{O}\left(t^0 \right)\, ,
\eeq
with $t= \text{Im}\, z$, such that the moduli space metric can be approximated by
\beq\label{eq:quinticmetric}
	G_{z\bar z}= \frac{3}{4} \frac{1}{(\text{Im}\, z)^2} + \mathcal{O}\left(1/t^5 \right)\, .
\eeq
Next, we need to compute the scalar charge-to-mass vector associated to the leading infinite tower of states, as well as the corresponding species scale. Regarding the former point, there is indeed a plethora of perturbative (e.g., KK modes) and non-perturbative states becoming light upon exploring the large volume singularity (see e.g., \cite{Font:2019cxq,Corvilain:2018lgw,Lee:2019wij}). The former can be easily seen to be subleading, whilst the latter arise as $\frac{1}{2}$-BPS bound states of D0- and D2-branes wrapping minimal 2-cycles of the CY$_3$, whose mass is controlled by the normalized $\mathcal{N}=2$ central charge\footnote{We do not consider here magnetically charged states corresponding to wrapped D4- or D6-particle states \cite{Ceresole:1995ca}, since they do not become massless in the limit of interest (see e.g., \cite{Font:2019cxq}).}
\beq\label{eq:centralcharge}
	\frac{m_{n_{2p}}}{M_{\text{Pl;}\, 4}} = \sqrt{8\pi } e^{K_{\text{ks}}/2} |Z_{\text{IIA}}| = \sqrt{\frac{\pi}{ \mathcal{V}_{X_3}}}\, |n_0+n_{2,a} z^a|\, ,
\eeq
where $n_0, n_{2,a}\in \mathbb{Z}$ correspond to D0- and D2-brane charges, respectively, and the subscript $a$ indicates the holomorphic 2-cycle wrapped by the 2-brane. For the quintic, given that $h^{1,1}=1$, the previous mass formula reduces to
\beq\label{eq:centralchargequintic}
	\frac{m_{n_{2p}}}{M_{\text{Pl;}\, 4}} \sim \frac{|n_0+n_{2} z|}{ t^{3/2}}\, .
\eeq
Any state with D2-brane charge (i.e. $n_2\neq 0$) will scale as $m_{\rm D2}\sim t^{-1/2}M_{\rm Pl;\, 4}$, while for $n_2=0$ we have instead $m_{\rm D0}\sim t^{-3/2}M_{\rm Pl;\, 4}$. This means, in particular, that the leading tower becomes that comprised by D0-branes alone, whilst there is another one (which is additive, in the sense of Section \ref{ss:MultipleTowers})  made out of bound states of D0- and D2-particles \cite{Corvilain:2018lgw}.\footnote{\label{fnote:stabilityBPS}In general, it is difficult to properly argue for the existence of an \emph{infinite} tower of states which become asympotically stable \cite{Grimm:2018ohb,Palti:2021ubp}. This is why in the original work \cite{Grimm:2018ohb}, the monodromy transformations characterizing the infinite distance singularities were exploited, since it allows to argue at least for the existence of the \emph{monodromy tower}, which may or may not be the leading one. In certain circumstances, however, we may instead use dualities to support the existence of the tower, as happens in the present case, where the D0 bound states correspond to the KK replicas of the 5d fields along the M-theory circle, see Section \ref{sss:IIA/Mthy}.}
	
Therefore, from eq. \eqref{eq:centralchargequintic}, we obtain
\begin{equation}\label{eq:D0zeta4d}
	\vec{\zeta}_{\text{D0}} = -\partial_t \log m_{\text{D0}} = \frac{3}{2t}\, \quad \Longrightarrow \quad |\vec{\zeta}_{\text{D0}}| = \sqrt{\frac{3}{2}}\, ,
\end{equation}
where  we have used the field space metric \eqref{eq:quinticmetric} to compute the norm of the charge-to-mass vector, namely $|\vec{\zeta}_{\text{D0}}|= 2G^{z\bar z} \partial_z \log m_{\text{D0}}\, \partial_{\bar z} \log m_{\text{D0}}$. Note that this precisely matches that of a KK decompactification of one extra dimension, c.f. \eqref{eq:zeta&speciesveconemodulus}. This is of course no coincidence since the D0-branes correspond to the KK tower of the M-theory circle, so that the large volume limit induces a circle decompactification to a 5d $\mathcal{N}=1$ theory described in terms of M-theory on the same three-fold $X_3$ (see Section \ref{sss:IIA/Mthy} below). 	
	
The species scale can then be computed as usual for a single KK tower (see e.g., \eqref{Mp}), leading to
\beq \label{eq:D0tower}
	\frac{\LSP}{M_{\text{Pl;}\, 4}}\, \simeq\, \left( \frac{m_{\text{D0}}}{M_{\text{Pl;}\, 4}} \right)^{1/3}\, \sim\, \frac{1}{\mathcal{V}_{X_3}^{1/6}}\, \sim\, \frac{1}{t^{1/2}}\, ,
\eeq
which goes to zero upon exploring the $t \to \infty$ limit, as expected. It moreover matches with the 5d Planck scale, as we show later explicitly. Hence, from eq. \eqref{eq:D0tower} one obtains
\begin{equation}\label{eq:D0Zeta4d}
	\vec{\mathcal{Z}}_{\text{sp}} = -\partial_t \log \LSP = \frac{1}{2t}\, ,
\end{equation}
such that upon contracting with \eqref{eq:D0zeta4d} using the moduli space metric \eqref{eq:quinticmetric} we find
\begin{equation}
	\vec{\zeta}_{\text{D0}} \cdot \vec{\mathcal{Z}}_{\text{sp}} = \frac{1}{2}\, ,
\end{equation}
thus fulfilling the pattern in the present $d=4$ set-up.
	
\subsubsection{General story}
\label{sss:IIA/Mthy}
	
The above large volume singularity is always present within the vector multiplet moduli space of any Type IIA CY$_3$ compactification, such that the results found for the quintic can be easily extended to the more general case, as we argue in the following. 
	
Recall from Section \ref{sss:4dN=2basics} that the relevant piece of the 4d lagrangian obtained from Type IIA compactified on a generic three-fold is
\begin{equation}\label{eq:IIAlagrangian4d}
	\mathcal{L}_{\text{IIA}}^{\text{4d}} \supset \dfrac{1}{2\kappa_{4}^2}\,  \sqrt{- g} \left[\mathcal{R} - G_{a b}(\tilde{t})\, \partial \tilde{t}^a \cdot \partial \tilde{t}^b - \frac{1}{6} \left( \partial \log \mathcal{V}_{X_3} \right)^2 - 2\left( \partial \varphi_4 \right)^2 \right]\, ,
\end{equation}
where we defined $G_{a b} = 2 G_{a \bar b}$ (c.f. \eqref{eq:kahlersectormetricpattern}) and we have split the K\"ahler coordinates into restricted ones, $\tilde{t}^a = t^a/\mathcal{V}_{X_3}^{1/3}$ --- which satisfy the constraint $\mathcal{K}_{abc}\tilde{t}^a \tilde{t}^b \tilde{t}^c = 6$ --- and the overall volume modulus $\mathcal{V}_{X_3}$. Now, since we take a limit here where $\mathcal{V}_{X_3} \to \infty$ with the 4d dilaton fixed and finite, the 10d dilaton needs to be co-scaled, such that we end up probing the strong $g_s$ regime, i.e. $\phi \to \infty$, which can be better described by M-theory. Comparing then the lagrangian \eqref{eq:IIAlagrangian4d} with the one obtained by dimensionally reducing M-theory on the same manifold times a circle of radius $R_5$ (in 5d Planck units), which reads \cite{Cadavid:1995bk}
\begin{equation}
	\mathcal{L}_{\text{M-th}}^{\text{4d}} \supset \dfrac{1}{2\kappa_{4}^2}\,  \sqrt{- g} \left[\mathcal{R} - G_{a b}(\tilde{t})\, \partial \tilde{t}^a \cdot \partial \tilde{t}^b - \frac{3}{2} \left( \partial \log R_5 \right)^2 - \frac{1}{2} \left( \partial \log \mathcal{V}_5 \right)^2 \right]\, ,
\end{equation}
we arrive at the following moduli identifications (taking also into account quantum corrections \cite{Gopakumar:1998ii,Gopakumar:1998jq,Lawrence:1997jr})
\beq
\label{eq:modulimatching}
	R_5^3 = \mathcal{V}_{X_3}\, , \qquad \mathcal{V}_5=e^{-2\varphi_4}\, ,
\eeq
where $\mathcal{V}_5$ denotes the volume of the three-fold measured in 11d Planck units.
	
With these identifications at hand it is now easy to see how the masses of the D0- and D2-particles in 4d Planck units are translated into 5d quantities
\begin{equation}\label{eq:massesD0D2}
	\begin{aligned}
		\frac{m_{\text{D0}}}{M_{\text{Pl;}\, 4}} & =\sqrt{8\pi } e^{K_{\text{ks}}/2} = \sqrt{\frac{\pi}{\mathcal{V}_{X_3}}} = \frac{m_{\text{KK},\, 5}}{M_{\text{Pl;}\, 4}}\, ,\\
		\frac{m_{\text{D2}}}{M_{\text{Pl;}\, 4}} & =\sqrt{8\pi } e^{K_{\text{ks}}/2} |t^a| = \frac{\sqrt{\pi}\, \tilde t^a}{\mathcal{V}_{X_3}^{1/6}} = \frac{m_{\text{M2}}}{M_{\text{Pl;}\, 4}}\, ,
	\end{aligned}
\end{equation}
where in the last expression we have considered a single D2-brane wrapping some 2-cycle once and for simplicity we have set the corresponding axion v.e.v. $b^a$ to zero. Proceeding similarly to what we did in the quintic example, and taking the limit $\mathcal{V}_{X_3} \to \infty$ (whilst keeping the $\tilde t^a$ fixed and non-degenerate) we obtain the following charge-to-mass and species vectors
\begin{equation}
	\left(\zeta_{\text{D0}}\right)_{\mathcal{V}_{X_3}} = - \partial_{\mathcal{V}_{X_3}} \log (m_{\text{D0}}) = \frac{1}{2 \mathcal{V}_{X_3}}\, , \qquad
	\left(\mathcal{Z}_{\text{sp}}\right)_{\mathcal{V}_{X_3}} = - \partial_{\mathcal{V}_{X_3}} \log(\LSP) = \frac{1}{6 \mathcal{V}_{X_3}}\, ,
\end{equation}
where the remaining components, namely those arising from log-derivatives with respect to the $\tilde t^a$ fields, are vanishing. Note that the species scale, as computed from \eqref{eq:D0tower}, coincides asymptotically with the 5d Planck mass, which can be related to the 4d one by $M_{\text{Pl};\, 5}^2 2\pi R_5 = M_{\text{Pl;}\, 4}^2$. Therefore, upon contracting them using the moduli space metric in \eqref{eq:IIAlagrangian4d}, i.e. $G_{\mathcal{V}_{X_3}\mathcal{V}_{X_3}}=\frac{1}{6\mathcal{V}_{X_3}^2}$, we find that $\vec{\zeta}_{\text{D0}} \cdot \vec{\mathcal{Z}}_{\text{sp}} = \frac{1}{2}$ is again fulfilled.	
	
Interestingly, there is an alternative very simple way to show the realization of the pattern in general for this type of limit. Indeed, the leading tower of states is made of D0-branes, so that we can write $\zeta_{\text{D0}}^a=\frac{G^{a b}}{2}\frac{\partial K_{\text{ks}}}{\partial t^{b}}$. Furthermore, since we decompactify a single extra dimension, the species scale vector is given by $\vec{\mathcal{Z}}_{\text{sp}}=\frac{1}{3} \vec{\zeta}_{\text{D0}}$ (c.f. eq. \eqref{eq:eff-vector}). Therefore, we may have alternatively computed the inner product as follows
\beq 
\label{noscale}
	\vec{\zeta}_{\text{D0}} \cdot \vec{\mathcal{Z}}_{\text{sp}} = \frac{1}{12} \frac{\partial K_{\text{ks}}}{\partial t^{a}} G^{a b} \frac{\partial K_{\text{ks}}}{\partial t^b} = \frac{1}{2}\, ,
\eeq
where in order to arrive at the right-hand side, one needs to use the no-scale property of the vector-multiplet metric \eqref{eq:kahlersectormetricpattern}, namely $K_a G^{a b} K_b=6$. 		
	
\subsection{Type III limits: Partial decompactification}
\label{ss:typeIIIlimits}
	
\subsubsection{Example 2: Type IIA on $\mathbb{P}^{1,1,1,6,9} [18]$}
\label{sss:ExampleII}
	
Let us now consider Type IIA string theory compactified on the three-fold $X_3=\mathbb{P}^{1,1,1,6,9}$ [18], which may be seen as a smooth elliptic fibration over a $\mathbb{P}^2$-base, with $h^{1,1}=2$ \cite{Candelas:1994hw}. We denote the (real-valued) K\"ahler moduli by $\lbrace t^1, t^2 \rbrace$, which at large volume control the $\mathcal{N}=2$ K\"ahler potential
\begin{equation}\label{eq:kahlerpotP11169}
	e^{-K_{\text{ks}}} = \frac{4}{3}\mathcal{K}_{abc}t^a t^b t^c = 12(t^1)^3 + 12(t^1)^2 t^2 + 4 t^1(t^2)^2 + \ldots\, ,
\end{equation}
with $\mathcal{K}_{abc}$ being the triple intersection numbers in an integral basis of $H^2(X_3)$ and the ellipsis denotes further perturbative and non-perturbative $\alpha'$-corrections. From this, we can can easily compute both the moduli space metric and its inverse. In particular, the latter reads as
\begin{equation}\label{eq:fullinversemetric}
	G^{-1}\, = \, \left(
	\begin{array}{cc}
		2(t^1)^2 +\frac{3(t^1)^4}{3t^1t^2+(t^2)^2}& - \frac{3(t^1)^2 \left(t^1+t^2\right)}{t^2}  \\
		- \frac{3(t^1)^2 \left(t^1+t^2\right)}{t^2} & 9(t^1)^2+ \frac{9(t^1)^3}{t^2} +3t^1t^2 +(t^2)^2 \\
	\end{array}
	\right) \, .
\end{equation}
On the other hand, the infinite distance boundaries present in this example were analyzed from the MHS point of view in \cite{Grimm:2018cpv}, and three types of infinite distance limits were found therein: \emph{(i)} $t^1 \to \infty$ with $t^2$ finite (a Type IV$_1$ singularity), \emph{(ii)} $t^2 \to \infty$ with $t^1$ finite (a Type III$_0$ singularity) and \emph{(iii)} $t^1, t^2 \to \infty$ (a Type IV$_2$ singularity). The asymptotic regime in the latter case can be divided into two subregions (i.e., growth sectors) depending on whether $t^1\gg t^2 $ or $t^2\gg t^1 $ as $t^1, t^2 \to \infty$. 	
	
\begin{figure}[t!]
	\begin{center}
	\subfigure[]{
			\includegraphics[width=0.4\textwidth]{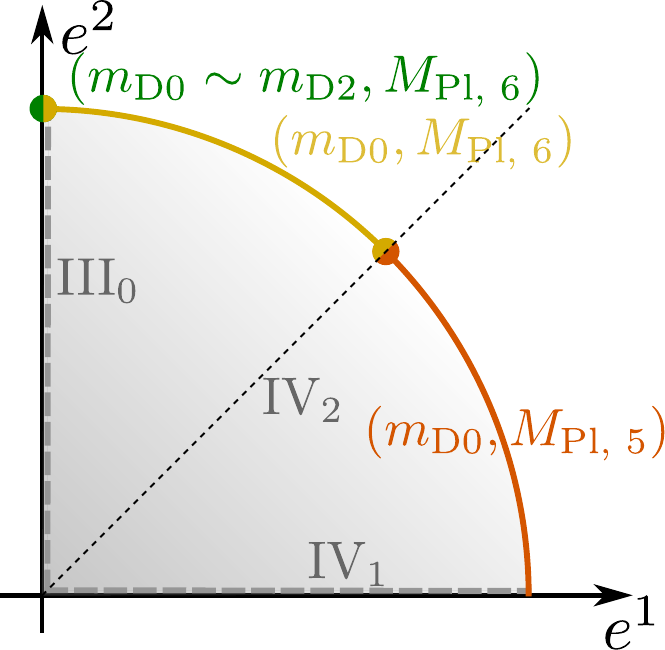}\label{fig:asympt_lim_IIAP11169-lim}
		}
        \quad
	\subfigure[]{
			\includegraphics[width=0.5\textwidth]{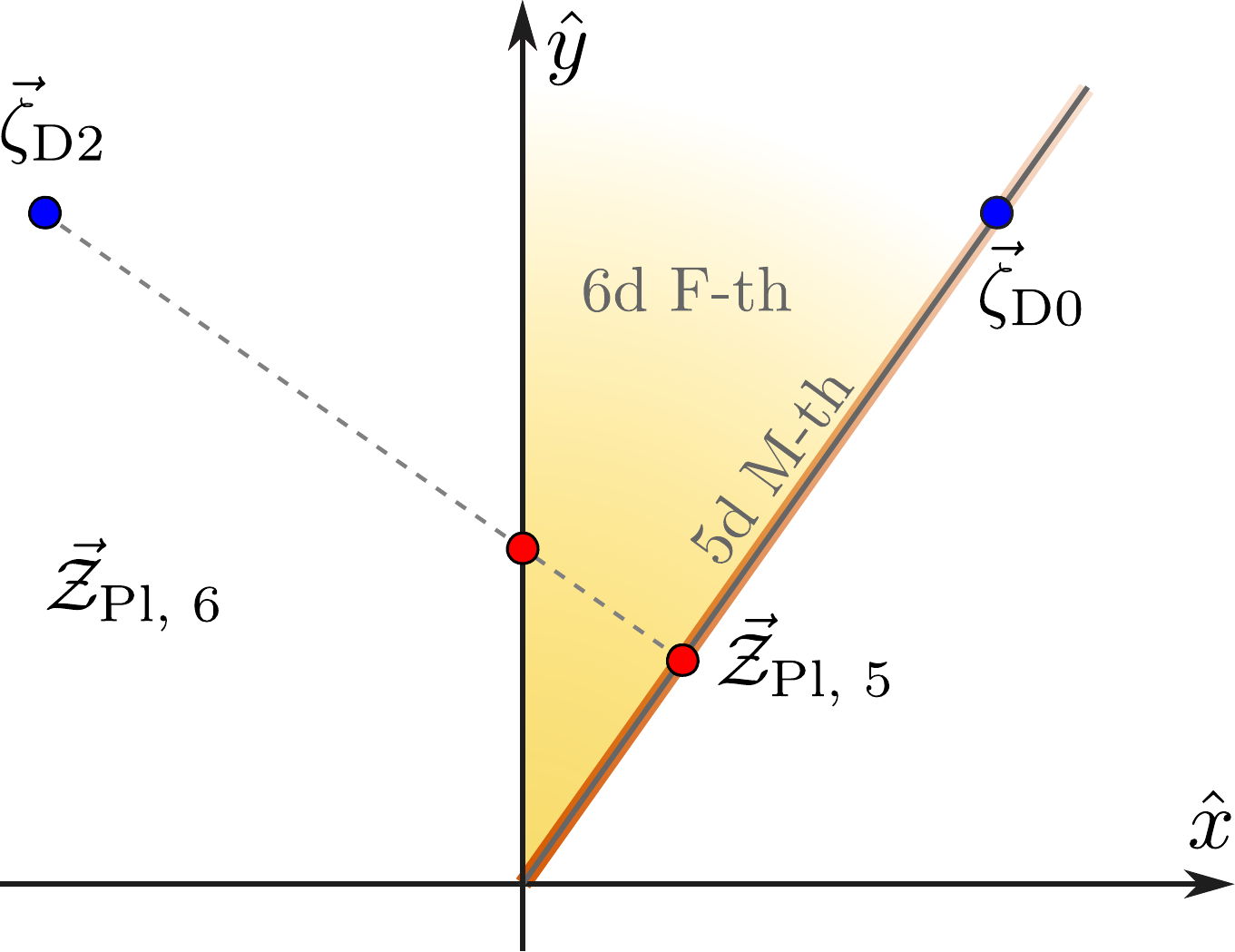}\label{fig:asympt_lim_IIAP11169-vec}
		}
		\caption{\small \textbf{(a)} Classification of infinite distance limits for Type IIA string theory on $\mathbb{P}^{1,1,1,6,9}[18]$ according to their singularity type \cite{Corvilain:2018lgw}, as well as their leading tower and species scales. \textbf{(b)} Relevant scalar charge-to-mass (blue) and species vectors (red) in the flat frame \eqref{eq:flatcoordsP11169}. In particular, one finds $\vec{\zeta}_{\rm D0}=\left(\frac{1}{\sqrt{2}},1\right)$, $\vec{\mathcal{Z}}_{\rm Pl,\; 6}=\left(0, \frac{1}{2}\right)$ and $\vec{\mathcal{Z}}_{\rm Pl,\; 5}=\left(\frac{1}{3\sqrt{2}}, \frac{1}{3}\right)$.}
		\label{fig:asympt lim  IIAP11169}
	\end{center}
\end{figure}

In what follows, we will study each of them in turn so as to show that the pattern
\begin{equation} \label{eq:patternP11169}
	\left.\vec{\zeta}_{\text{t}}\cdot\vec{\mathcal{Z}}_{\rm sp}\right|_{\mathbf{t}(\sigma)}=\left.\left(G^{a b}\partial_{a}\log m_{\text{tower}}\,\partial_{b}\log \Lambda_{\rm sp}\right)\right|_{\mathbf{t}(\sigma)} = \frac{1}{2}\, ,
\end{equation}
indeed holds for any trajectory $\mathbf{t}(\sigma)$ within each region, upon replacing $\LSP$ with the properly identified species scale in each growth sector. This is summarized in Figure \ref{fig:asympt_lim_IIAP11169-lim}, where the leading towers of states and species scales are explicitly indicated.
	
Notice that in this example, unlike the situation in simple toroidal compactifications where the $\zeta$-vectors remain fixed (c.f. Section \ref{s:maxsugra}), both the mass formulae and the metric $G_{a b}$ vary non-trivially across the moduli space. Indeed, for the latter one finds
\begin{equation}\label{eq:asymptmetricP11169}
	G|_{t^1\gg t^2}=
	\begin{pmatrix}
		\frac{3}{2(t^1)^2}& \frac{1}{2(t^1)^2}\\
		\frac{1}{2(t^1)^2}&\frac{1}{6(t^1)^2}
	\end{pmatrix}
	+\mathcal{O}\left(\frac{t^2}{(t^1)^3}\right)\,,\quad
	G|_{t^2\gg t^1}=
	\begin{pmatrix}
		\frac{1}{2(t^1)^2}& \frac{3}{2(t^2)^2}\\
		\frac{3}{2(t^2)^2}&\frac{1}{(t^2)^2}
	\end{pmatrix}+\mathcal{O}\left(\frac{t^1}{(t^2)^3}\right)\, ,
\end{equation}
which exhibits quite different behaviours depending on the infinite distance regime that we explore. Consider first those limits with $t^1\gg t^2 \gg 1$. As one can see, the coordinate $t^2$ becomes asymptotically irrelevant, thus not affecting the expression for the metric $G_{a b}$ nor the relevant masses or species cut-off (see below). This means, in particular, that within this growth sector the moduli space becomes effectively one-dimensional. On the other hand, for limits where $t^2\gg t^1\gg 1$ there are in fact subleading $t^1$--\,dependent terms that appear in the metric \eqref{eq:asymptmetricP11169}, as well as in the mass and species cut-off. More precisely, one can introduce a globally defined flat chart,\footnote{This is actually possible since the slice of $\mathcal{M}_{\rm VM}$ parametrized by $\{t^1,t^2\}$ is Riemann flat, as one may easily verify.} parametrized by the coordinates
\begin{equation}\label{eq:flatcoordsP11169}
 \begin{aligned}
     \hat{x} &= \frac{\sqrt{2}}{6} \left( \log \left( 3(t^1)^3 + 3(t^1)^2t^2 + t^1(t^2)^2 \right) - \log \left( \frac{1+\left( \frac{t^2}{3t^1 + t^2} \right)^{3/2}}{1-\left( \frac{t^2}{3t^1 + t^2} \right)^{3/2}} \right)^2\right)\, ,\\
     \hat{y} &= \frac{1}{3} \left( \log \left( 3(t^1)^3 + 3(t^1)^2t^2 + t^1(t^2)^2 \right) + \log \left( \frac{1+\left( \frac{t^2}{3t^1 + t^2} \right)^{3/2}}{1-\left( \frac{t^2}{3t^1 + t^2} \right)^{3/2}} \right)\right)\, ,
 \end{aligned}
\end{equation}
which maps the K\"ahler cone $\{t^1\geq 0,\,t^2\geq 0\}$ into the region $\{\hat{y}\geq \sqrt{2}\hat{x}\geq 0\}$. Using these coordinates one can readily check that the growth sector $t^1\gtrsim t^2 \gg 1$ is indeed mapped asymptotically to the one-dimensional `boundary' $\{\hat{y}=\sqrt{2}\hat{x}\}$, whilst the other sector covers up the remaining part of the cone, see Figure \ref{fig:asympt_lim_IIAP11169-vec}.

Incidentally, note that the fact that the moduli space --- when described using the flat chart \eqref{eq:flatcoordsP11169} --- ends precisely along the line determined by the vector $\vec{\mathcal{Z}}_{\text{Pl},\, 5}$, prevents the lower bound
\begin{equation}\label{eq:SSDC4d}
	|\vec{\mathcal{Z}}_{\text{sp}}|=\lambda_{\rm sp} \geq \frac{1}{\sqrt{6}}\, ,
\end{equation}
from being immediately violated. This provides further evidence for the latter, which was discussed and tested only in string theory set-ups with maximal supersymmetry, see Chapter \ref{ch:bounds} for details.

\subsubsection*{Growth sector $t^1\gg t^2$ with $t^1, t^2 \to \infty$}
	
This includes the particular case of sending $t^1 \to \infty$ with $t^2$ finite (i.e. a type IV$_1$ singularity), since it shares the same leading behaviour of the mass of the towers as well as the species scale. The three-fold volume behaves asymptotically as follows
\begin{equation}\label{eq:volP11169n=3}
	\mathcal{V}_{X_3} = \frac{3}{2} (t^1)^3 \left(1 + \mathcal{O}\left(t^2/t^1\right) \right)\, .
\end{equation}
%
%
%
Following the discussion of the previous section, this limit corresponds again to decompactifying to 5d M-theory, as expected from being a type IV singularity. 
Thus, it is clear that the pattern will hold along this set of limits due to the general argument given around eq. \eqref{noscale}, but let us show it explicitly here for illustrative purposes. By repeating the same kind of computations as in the previous example we find 
\begin{equation}\label{eq:massesD0D2P11169}
	\frac{m_{\text{D0}}}{M_{\text{Pl;}\, 4}} =\sqrt{8\pi } e^{K_{\text{ks}}/2} \sim \frac{1}{(t^1)^{3/2}}\, , \qquad \frac{m_{\text{D2}}}{M_{\text{Pl;}\, 4}} =\sqrt{8\pi } e^{K_{\text{ks}}/2} t^1 \sim \frac{1}{(t^1)^{1/2}}\, ,
\end{equation}
for the mass scale of the D0- and D2-particles, respectively. Strictly speaking, there are two possibilities for obtaining four-dimensional BPS states from wrapped D2-branes, since there exist two different non-trivial classes of holomorphic curves within $\mathbb{P}^{1,1,1,6,9}[18]$. The one corresponding to the mass scale computed in \eqref{eq:massesD0D2P11169} is associated to the `horizontal' class, namely the generic fibre of the elliptic fibration. For the other `vertical' class, since the supersymmetric cycle wrapped by the 2-brane is topologically equivalent to a $\mathbb{P}^{1}$-curve that is moreover contractible, there is only a finite number of associated Gopakumar-Vafa (GV) invariants which are non-zero (see e.g., \cite{Candelas:1994hw, Hosono:1993qy}). This means, in turn, that such D2-particles do not give rise to an infinite tower of states becoming massless along the $t^1 \to \infty$ limit, such that we can safely ignore them for our purposes here. 
	
Similarly, as discussed in the previous subsection, the species scale can be computed through D0-brane state counting, arriving at the following result
\beq\label{eq:speciesn=3} 
	\frac{\LSP}{M_{\text{Pl;}\, 4}}\, \simeq\, \left( \frac{m_{\text{D0}}}{M_{\text{Pl;}\, 4}} \right)^{1/3} \sim\, \frac{1}{(t^1)^{1/2}}\, ,
\eeq
which is nothing but the 5d Planck scale.
	
With this, we now have all the necessary information so as to check whether the condition \eqref{eq:patternP11169} is satisfied or not. Thus, let us first compute the charge-to-mass vectors of the leading tower of states, namely the D0-brane bound states, as well as the species vector obtained from eq. \eqref{eq:speciesn=3} above. The former is given by
\beq\label{eq:zetaD0n=3} 
\begin{aligned}
    \vec{\zeta}_{\text{D0}} &= \left( \frac{\left(3t^1+t^2\right)^2}{6 (t^1)^3 + 6 (t^1)^2 t^2 + 2t^1(t^2)^2}\, , \, \frac{3t^1+2t^2}{6 (t^1)^2 + 6 t^1 t^2 + 2(t^2)^2}\right)\\
    &=\left( \frac{3}{2t^1}, \frac{1}{2t^1}\right) + \mathcal{O}\left(t^2/(t^1)^2\right)\, ,
\end{aligned}
\eeq
where the notation is $\vec{\zeta}=\left( \zeta_{t^1}, \zeta_{t^2} \right)$. The latter, on the other hand, is simply proportional to the charge-to-mass vector associated to the D0-branes, namely $\vec{\mathcal{Z}}_{\text{sp}}=\frac{1}{3} \vec{\zeta}_{\text{D0}}$. Hence, upon contracting both vectors using the inverse moduli space metric \eqref{eq:fullinversemetric}, one finds that indeed $\vec{\zeta}_{\text{D0}} \cdot \vec{\mathcal{Z}}_{\text{sp}}= \frac{1}{d-2}=\frac{1}{2}$ is verified \emph{exactly}, namely even before performing the expansion in $t^2/t^1$.

\subsubsection*{Growth sector $t^2\gg t^1$ with $t^1, t^2 \to \infty$}
	
For the other growth sector, the situation turns out to be quite different. First, note that it includes the particular case of sending $t^2 \to \infty$ with $t^1$ finite (i.e. a Type III$_0$ singularity) and, as can be easily checked, the volume \eqref{eq:kahlerpotP11169} is dominated by the last term in the right-hand side
\begin{equation}\label{eq:volumen=2limit}
	\mathcal{V}_{X_3} = \frac{1}{2} t^1 (t^2)^2 \left(1 + \mathcal{O}\left(t^1/t^2\right) \right)\, ,
\end{equation}
which implies the following asymptotic dependence for the inverse metric components (to leading order in $t^2$)
\begin{equation}\label{eq:modspacemetricP11169n=2}
	G^{-1}\, = \, \left(
	\begin{array}{cc}
		2(t^1)^2 & -3 (t^1)^2  \\
		-3 (t^1)^2 & (t^2)^2 \\
	\end{array}
	\right) + \ldots\, .
\end{equation}
The QG resolution of the singularity requires from a double decompactification to 6d F-theory on the same elliptic three-fold $X_3$ \cite{Lee:2019wij,Castellano:2022bvr, Marchesano:2022axe}. This may be intuitively understood by looking again at the asymptotic behavior of the mass scales of the infinite towers of light states\footnote{For the D2-branes wrapping the elliptic fibre $k \in \mathbb{Z} \setminus \lbrace 0 \rbrace$ times one obtains \cite{Klemm:2012sx, Klemm:1996hh}
\beq \label{eq:GVinvariantsT2limit}
	n_{\textbf{k}} = \chi_E(X_3) = 2 \left ( h^{1,1} (X_3) -  h^{2,1} (X_3) \right)\, ,
\eeq
thus behaving like a KK spectrum associated to a circle reduction from 5d to 4d.}
\begin{equation}\label{eq:massesD0D2P11169n=2}
	\frac{m_{\text{D0}}}{M_{\text{Pl;}\, 4}}  \sim \frac{1}{\sqrt{t^1} t^2}\, , \qquad \frac{m_{\text{D2}}}{M_{\text{Pl;}\, 4}} \sim \frac{\sqrt{t^1}}{t^2}\, ,
\end{equation}
which present both the same dependence, contrary to the previous case (c.f. \eqref{eq:massesD0D2P11169}). Additionally, one can form $\frac{1}{2}$-BPS bound states of D0- and D2-particles upon turning on some (quantized) worldvolume flux $\mathcal{F}$ for the wrapped D2-brane \cite{Polchinski:1998rr}. As a consequence, and following the algorithmic computation of the species scale proposed in Chapter \ref{ch:SpeciesIntro}, one arrives at a cut-off of the form
\beq 
	\frac{\LSP}{M_{\text{Pl;}\, 4}} \simeq \left(N_{\text{D0}}\, N_{\text{D2}}\right)^{1/2} \sim \frac{1}{\sqrt{t^2}}\, ,
\eeq
where $N_{\text{D2p}}$ counts the number of D$2p$-brane states falling below the species scale. This moduli dependence of the species scale indeed matches with the 6d Planck scale (see discussion around \eqref{eq:6dPlanckscale} below). We can then compute the scalar charge-to-mass vectors for the two towers of states, which to leading order in $1/t^2$, read as
\begin{equation}\label{eq:zetavectorsD0D2n=2}
	\vec{\zeta}_{\text{D0}} = \left( \frac{1}{2 t^1}, \frac{1}{t^2} \right) + \mathcal{O}\left(t^1/(t^2)^3\right)\, , \qquad \vec{\zeta}_{\text{D2}} = \left( -\frac{1}{2 t^1}, \frac{1}{t^2} \right) + \mathcal{O}\left(t^1/(t^2)^3\right)\, .
\end{equation}
Analogously, one finds for the species vector
\begin{equation}\label{eq:speciesvectorIIA}
	\vec{\mathcal{Z}}_{\text{sp}} = \left( \frac{3}{4 t^2}, \frac{1}{2t^2} \right) + \mathcal{O}\left(t^1/(t^2)^3\right)\, ,
\end{equation}
such that upon taking the product with respect to the inverse metric \eqref{eq:modspacemetricP11169n=2}, the pattern \eqref{eq:pattern} still holds for \emph{both} towers. In this case, however, it turns out being crucial to take into account that $t^1/t^2 \to 0$ asymptotically along the limit of interest.

\subsubsection{General story}
\label{sss:IIA/Fthy}
	
The previous example contained two types of limits, one belonging to the category of Section \ref{ss:typeIVlimits} and a new one: A partial decompactification to 6d F-theory. Let us elaborate a bit more on this second case, which corresponds to the regime where $t^2$ grows faster than $t^1$. From \eqref{eq:modspacemetricP11169n=2}, one can check that the length of relevant vectors associated to the tower of bound states behave as follows
\begin{equation}
	|\vec{\zeta}_{\text{eff}}|= \left|\frac{1}{2} (\vec{\zeta}_{\text{D0}} + \vec{\zeta}_{\text{D2}}) \right|= 1 + \mathcal{O}\left( \left(t^1/t^2\right)^2\right)\, , \qquad
	|\vec{\mathcal{Z}}_{\text{sp}}| = \frac{1}{2} + \mathcal{O}\left( \left(t^1/t^2\right)^2\right)\, ,
\end{equation}
which indeed coincide with the typical values for Kaluza-Klein vectors associated to a two-dimensional compact manifold, matching with the microscopic interpretation of the singularity as a decompactification from 4d to 6d. Our aim here will be to show that this is generically the case whenever we explore a type $\mathbf{T}^2$ limit in the language of \cite{Lee:2019wij} (or a Type III singularity in the language of \cite{Grimm:2018ohb}). Subsequently, this will allow us to argue that the pattern $\eqref{eq:pattern}$ holds in general for such kind degenerations. 
	
Let us consider an infinite distance limit in which the curve associated to the fastest growing modulus has the intersection numbers of a Type III singularity (see second row in Tables \ref{tab:intersN=2} and \ref{tab:limitsN=2}). Geometrically, this corresponds to a limit in which the Calabi--Yau three-fold exhibits an elliptic fibration  over a K\"ahler surface $B_2$
\begin{equation}\label{eq:ellfibration}
			\begin{aligned}
				\pi: \qquad \mathbf{T}^2 \hookrightarrow &\;X_{3} \\
				&\;\; \downarrow \qquad , \\ &\ \ B_{2}
			\end{aligned}
\end{equation}
where the volume of the latter grows faster than the fiber (i.e.  belongs to the type $\mathbf{T}^2$ class).
After resolving any Kodaira-Néron type of singularity that may be present \cite{Weigand:2018rez}, we can then divide the K\"ahler moduli into two sets: those parametrizing fibral curves, $\lbrace t^a_f \rbrace$, and the ones inherited from the base, $\lbrace t^{\alpha}_b \rbrace$. These fields arise as the expansion coefficients of the K\"ahler 2-form $J$ over a basis $\lbrace \omega_A \rbrace = \lbrace \omega_a, \omega_{\alpha} \rbrace$ of $H^{1,1}(X_3, \mathbb{Z})$, as follows
\begin{equation}
		J= t^A \omega_A = t^a_f \omega_a + t^{\alpha}_b \omega_{\alpha} \, , 
\end{equation}
with $\alpha=1,\ldots, h^{1,1}(B_2)$, whilst the index $a$ runs from 1 to $n$, with $n-1$ being the sum of the ranks of the Mordell-Weil group and the non-Abelian gauge algebras realized at co-dimension one degenerations $\Delta \subset B_2$ \cite{wazir_2004}.
Therefore, the limit we want to study corresponds to
\begin{equation}\label{eq:n=2limit}
	t^a_f = \text{const.}\, , \qquad t^{\alpha}_b = \xi^{\alpha} \sigma\, , \qquad \text{with}\, \, \sigma\to \infty\, , 
\end{equation}
accompanied by a suitable co-scaling of the 10d dilaton --- so as to keep fixed all moduli in the hypermultiplet sector. Microscopically, the quantum gravity resolution of the singularity requires from a double decompactification to F-theory on the same elliptic three-fold $X_3$, as we review in the following.
	
On the one hand, at the level of the spectrum, one finds --- at least in the simplest instances --- only two infinite sets of asymptotically light states: those comprised by D0-branes and D2-branes wrapping the elliptic fibre class. Notice that, since the volume of the latter 2-cycle, which we denote by $\mathcal{V}_{\mathbf{T}^2}$, does not diverge in the limit \eqref{eq:n=2limit}, the central charges associated to both towers of states are controlled by the same quantity, namely the (square root of the) overall three-fold volume:
\begin{equation}\label{eq:massesD0D2Fthylimit}
	\frac{m_{\text{D0}}}{M_{\text{Pl;}\, 4}}  = \sqrt{\frac{\pi}{\mathcal{V}_{X_3}}}\, , \qquad \frac{m_{\text{D2}}}{M_{\text{Pl;}\, 4}} = \sqrt{\frac{\pi}{\mathcal{V}_{X_3}}}\, \mathcal{V}_{\mathbf{T}^2}\, .
\end{equation}
and indeed they furnish the Kaluza-Klein replica along the torus of the 6d F-theory massless fields. 
	
From this set of asymptotically light towers, one can easily compute the species scale dominating the infinite distance limit. In fact, upon using Type IIA/F-theory duality \cite{Lee:2019wij}, we conclude that the species scale should be controlled parametrically by the six-dimensional Planck mass, namely\footnote{The second equality in \eqref{eq:6dPlanckmass} follows from the moduli identifications $R_5 = \mathcal{V}_{X_3}^{1/3}$ (c.f. \eqref{eq:modulimatching}) as well as $R_6^{-4/3}=\frac{\mathcal{V}_{\mathbf{T}^2}}{\mathcal{V}_{X_3}^{1/3}}$ \cite{Corvilain:2018lgw}.}
\beq \label{eq:6dPlanckmass}
	M_{\text{Pl; 6}}^2 \simeq\frac{M_{\text{Pl;}\, 4}^2}{R_6^{2/3}R_5} = \frac{M_{\text{Pl;}\, 4}^2 \mathcal{V}_{\mathbf{T}^2}^{1/2}}{\mathcal{V}_{X_3}^{1/2}} \sim \frac{1}{\sigma^{1/2}}\, ,
\eeq
with $R_6$ and $R_5$ being the corresponding radii of the 6d-to-5d and 5d-to-4d circle compactifications (measured in the 6d and 5d Planck units respectively). Indeed, one can check that
\beq\label{eq:6dPlanckscale}
	\left(\frac{M_{\text{Pl; 6}}}{M_{\text{Pl;}\, 4}}\right)^2 \simeq \left( \frac{m_{\text{D0}}\, m_{\text{D2}}}{M_{\text{Pl;}\, 4}^2}\right)^{\frac{1}{2}} \sim \left(\frac{\LSP}{M_{\text{Pl;}\, 4}}\right)^2\, ,
\eeq
in agreement with the usual species counting computation. 
	
On the other hand, for the K\"ahler potential one finds the following leading asymptotic behavior \cite{Lee:2019wij,Cota:2022maf}
\begin{equation}\label{eq:kahlerpotn=2}
	K_{\text{ks}}= - \log \left(\frac{1}{2} \left( c_a t^a_f\right)\eta_{\alpha \beta} t_b^{\alpha} t_b^{\beta} + \mathcal{O} (\sigma)\right)\, ,
\end{equation}
where $c_a$ are some positive coefficients\footnote{\label{fnote:ellipticclass}The coefficients $c_a$ in eq. \eqref{eq:kahlerpotn=2} determine the generic elliptic fibre class $[\mathcal{C}_{\mathbf{T}^2}]$. Hence, in terms of a basis $\lbrace \mathcal{C}_f^a \rbrace$ of generators of the relative Mori cone $\mathsf{Mori}(X_3/B_2)$, the former becomes $\mathcal{C}_{\mathbf{T}^2}= \sum_a c_a \mathcal{C}_f^a$, where the notation follows that of \cite{Cota:2022maf}.} determined by the particular fibration structure of the three-fold and $\eta_{\alpha \beta}$ denote the intersection numbers of the two-fold base \cite{Corvilain:2018lgw}. Notice, in particular, that the basis $\lbrace \omega_A\rbrace = \lbrace \omega_a, \omega_{\alpha} \rbrace$ verifies that $\omega_a \wedge \omega_b \wedge \omega_c = \mathcal{K}_{a b c}=0$. Therefore, upon inserting the leading order expansion \eqref{eq:kahlerpotn=2} into the definition of the vector multiplet metric $G_{A B}$, one finds
%
\begin{align}\label{eq:metricn=2}
	G_{\alpha \beta} &= \frac{1}{2} \frac{\partial^2 K_{\text{ks}}}{\partial t^{\alpha}_b \partial t^{\beta}_b} = G^{(\rm lead.)}_{\alpha \beta} + \mathcal{O}\left(1/\sigma^3 \right)\, , \quad G_{\alpha b} = \frac{1}{2} \frac{\partial^2 K_{\text{ks}}}{\partial t^{\alpha}_b \partial t^{b}_f} = G^{(\rm lead.)}_{\alpha b} + \mathcal{O}\left(1/\sigma^2 \right)\, ,\notag\\
	G_{a b} &= \frac{1}{2} \frac{\partial^2 K_{\text{ks}}}{\partial t^a_f \partial t^b_f} = G^{(\rm lead.)}_{a b} + \mathcal{O}\left(1/\sigma \right)\, ,
\end{align}
%
with the following explicit expression for the leading-order matrices in eq. \eqref{eq:metricn=2} above
\begin{align}\label{eq:constmatricesn=2lim}
	\notag G^{(\text{lead.})}_{\alpha \beta} &=\frac{2 \left(\mathcal{K}_{\alpha \gamma a} t_b^{\gamma} t^a_f\right) \left( \mathcal{K}_{\beta \delta b} t_b^{\delta} t^b_f\right)}{\left( \mathcal{K}_{a \gamma \delta} t^a_f t_b^{\gamma} t_b^{\delta} \right)^2} - \frac{\mathcal{K}_{\alpha \beta a} t^a_f}{\mathcal{K}_{a \gamma \delta} t^a_f t_b^{\gamma} t_b^{\delta}}\, , \quad G^{(\text{lead.})}_{a b} =\frac{\left(\mathcal{K}_{a \alpha \beta} t_b^{\alpha} t_b^{\beta}\right) \left( \mathcal{K}_{b \gamma \delta} t_b^{\gamma} t_b^{\delta}\right)}{2\left( \mathcal{K}_{a \gamma \delta} t^a_f t_b^{\gamma} t_b^{\delta} \right)^2}\, ,\\
    G^{(\text{lead.})}_{\alpha b} &=\frac{\left(\mathcal{K}_{\alpha \beta a} t_b^{\beta} t^a_f\right) \left( \mathcal{K}_{b \gamma \delta} t_b^{\gamma} t_b^{\delta}\right)}{\left( \mathcal{K}_{a \gamma \delta} t^a_f t_b^{\gamma} t_b^{\delta} \right)^2} - \frac{\mathcal{K}_{\alpha b \gamma} t_b^{\gamma}}{\mathcal{K}_{a \gamma \delta} t^a_f t_b^{\gamma} t_b^{\delta}}\, ,
\end{align}
where $\mathcal{K}_{a \alpha \beta} = c_a \eta_{\alpha \beta}$. It is easy to see that these matrices have full rank except for $G^{(\rm lead.)}_{a b}$, which has rank one.\footnote{Actually, the (sub-)matrix $G^{(\text{lead.})}_{\alpha b}$, despite having full rank in general, can be identically zero in special circumstances, given that there are two terms with opposite sign in eq. \eqref{eq:constmatricesn=2lim}. This is the case when e.g., the fibration is non-degenerate, as happens in the two-moduli example discussed in Section \ref{sss:ExampleII}.}
	
Armed with all this, one can readily check upon using the no-scale property of the metric $G^{(\text{lead.})}_{\alpha \beta}$, namely the identity
\begin{equation}\label{eq:noscale}
	\frac{\partial K_{\text{ks}}}{\partial t^{\alpha}_b} G_{(\text{lead.})}^{\alpha \beta} \frac{\partial K_{\text{ks}}}{\partial t^{\beta}_b} = 4\, ,
\end{equation}
that the product
\begin{equation}
	\vec{\zeta}_{\text{D0, D2}}\cdot\vec{\mathcal{Z}}_{\rm sp}=\left(G^{A B}\partial_{A}\log m_{\text{tower}}\,\partial_{B}\log \Lambda_{\rm sp}\right) \stackrel{~\eqref{eq:n=2limit}~}{=} \frac{1}{2}\, , \qquad A, B= \lbrace a, \beta \rbrace\, ,
\end{equation}
is indeed satisfied for any trajectory of the form specified in \eqref{eq:n=2limit} above. To see this, it is important to realize that any term involving derivatives with respect to the fibral moduli, $\{t^a_f\}$, provides ultimately a contribution to the scalar product $\vec{\zeta}_{\text{t}}\cdot\vec{\mathcal{Z}}_{\rm sp}$ which is of $\mathcal{O}\left(1/\sigma\right)$ or higher, such that it goes away upon taking the infinite distance limit. For this same reason, the result also applies to more general cases in which the fiber volume is also sent to infinity, but at a slower rate than that of the base.
	
\subsection{Type II limits: Emergent string limits}
\label{ss:typeIIlimits}
	
\subsubsection{Example 3: Type IIA on $\mathbb{P}^{1,1,2,2,6} [12]$}
\label{sss:ExampleIII}
	
\begin{figure}[htb]
	\begin{center}
	\subfigure[]{
			\includegraphics[width=0.4\textwidth]{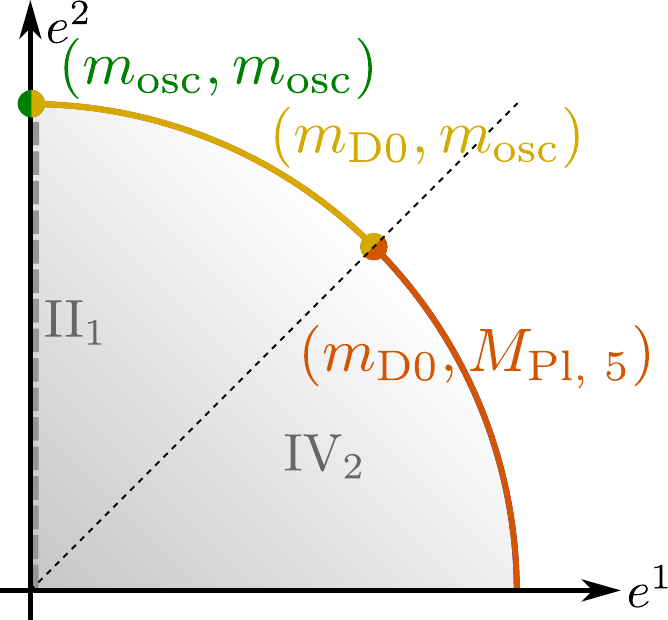}\label{fig:asympt_lim_IIAP1126-lim}
		}
        \quad
	\subfigure[]{
			\includegraphics[width=0.45\textwidth]{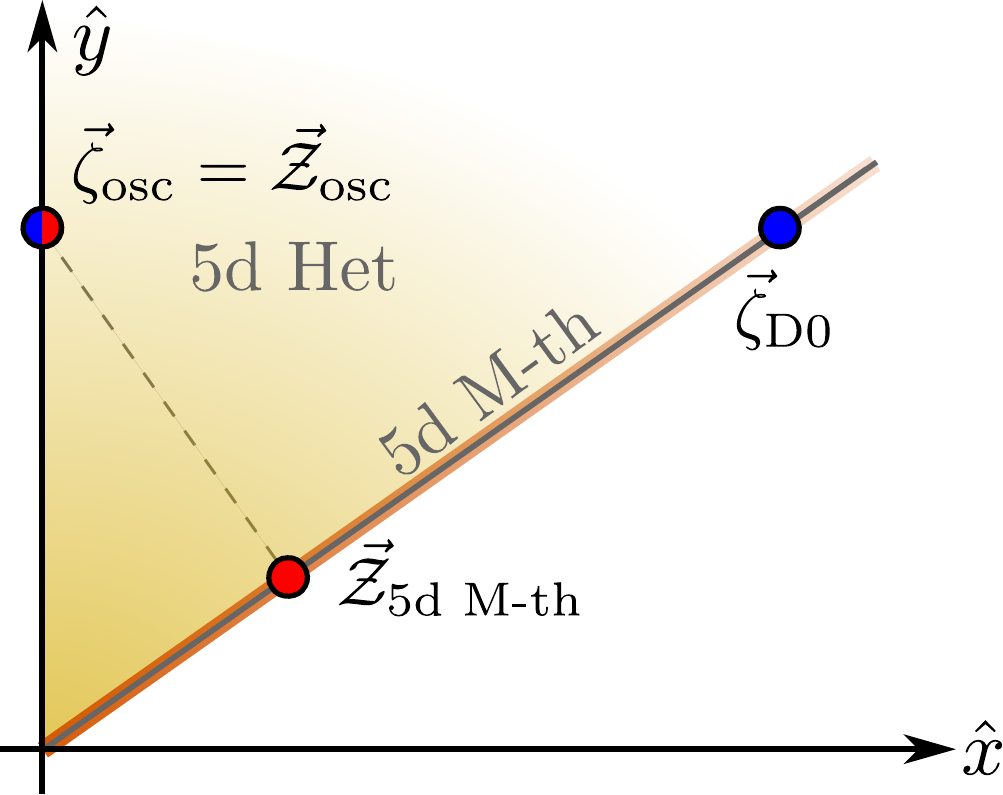}\label{fig:asympt_lim_IIAP1126-vec}
		}
		\caption{\small \textbf{(a)} Classification of infinite distance limits for Type IIA string theory on $\mathbb{P}^{1,1,2,2,6}[12]$ according to their singularity type \cite{Corvilain:2018lgw}, as well as their leading tower and species scales. \textbf{(b)} Relevant scalar charge-to-mass (blue) and species vectors (red) in the flat frame \eqref{e:mod 11226}. In particular, one finds $\vec{\zeta}_{\rm osc}=\vec{\mathcal{Z}}_{\rm osc}=\left(0,\frac{1}{\sqrt{2}}\right)$, $\vec{\zeta}_{\rm D0}=\left(1,\frac{1}{\sqrt{2}}\right)$ and $\vec{\mathcal{Z}}_{\rm Pl,\; 5}=\left(\frac{1}{3}, \frac{1}{3\sqrt{2}}\right)$.}
		\label{fig:asympt lim  IIAP1126}
	\end{center}
\end{figure}

As our final example, we consider Type IIA string theory compactified on the three-fold $X_3=\mathbb{P}^{1,1,2,2,6}$ [12]. Topologically, such two-parameter CY$_3$ can be seen as a $K3$ fibration over a $\mathbb{P}^1$-base, whose K\"ahler moduli $\lbrace t^1, t^2 \rbrace$ appear in the K\"ahler potential as follows\footnote{Here $t^2$ measures the classical volume of the $\mathbb{P}^1$-base, whilst $t^1$ parameterizes the volume of a second $\mathbb{P}^1$ corresponding to a rational curve (of non-negative self-intersection) inside the $K3$-fibre \cite{Candelas:1993dm}.}
\begin{equation}\label{eq:kahlerpotP11226}
	e^{-K_{\text{ks}}} = \frac{16}{3}(t^1)^3 + 8(t^1)^2 t^2 + \ldots\, ,
\end{equation}
where the ellipsis denotes further $\alpha'$ as well as worldsheet instanton corrections, which are both suppressed in the large volume patch. The explicit (inverse) moduli space metric that derives from the K\"ahler potential above is given by 
\begin{equation}\label{eq:fullinversemetricP11226}
	G^{-1}\, = \, \left(
	\begin{array}{cc}
		(t^1)^2 & - \frac{2}{3} (t^1)^2  \\
		- \frac{2}{3} (t^1)^2 & \frac{4}{3}(t^1)^2+\frac{8}{3} t^1 t^2 +2 (t^2)^2 \\
	\end{array}
	\right) \, .
\end{equation}
Using the nomenclature of MHS, we have the following infinite distance limits (see e.g., \cite{kerr2019polarized, Bastian:2021eom}): \emph{(i)} $t^1 \to \infty$ with $t^2$ finite (a Type IV$_2$ singularity), \emph{(ii)} $t^2 \to \infty$ with $t^1$ finite (a Type II$_1$ singularity), and \emph{(iii)} $t^1,t^2 \to \infty$ (a Type IV$_2$ singularity). The type IV singularities will again correspond to M-theory circle decompactifications, so the analysis of Section \ref{sss:IIA/Mthy} carries over. In fact, as it was the case in the example from Section \ref{sss:ExampleII}, all $t^2$--\,dependence disappears when taking the limit $t^1\gg t^2 \gg 1$, such that the moduli space becomes effectively one-dimensional. In addition, one may define global flat coordinates for the slice parametrized by $\lbrace t^1, t^2\rbrace$, which read
\begin{equation}\label{e:mod 11226}
 \begin{aligned}
     \hat{x} &= \frac{1}{3}\log \left(2 (t^1)^3\right)\, , \qquad \hat{y} =\frac{1}{\sqrt{2}}\log \left( \frac{2t^1+3t^2}{2^{2/3}}\right)\, ,
 \end{aligned}
\end{equation}
with the restriction that $\hat{y} \geq \frac{1}{\sqrt{2}}\hat{x}\geq 0$. These can then be used to depict the different relevant $\zeta$- and $\mathcal{Z}$-vectors in the present set-up, as shown in Figure \ref{fig:asympt lim  IIAP1126}. Let us stress that the regime $t^1\gtrsim t^2 \gg 1$ is mapped again to the boundary $\{\hat{y}=\frac{1}{\sqrt{2}}\hat{x}\}$, ensuring that the bound \eqref{eq:SSDC4d} is non-trivially satisfied. 
	
On the other hand, things get more interesting upon probing the limit $t^2 \to \infty$ (either with $t^1$ fixed or growing at a smaller rate), since the QG resolution of the corresponding Type II singularity is of a different kind than the ones discussed so far. The purely geometric analysis of \cite{Lee:2019wij} shows that it  corresponds to an emergent string limit, in which a critical Heterotic string arising from a NS5-brane wrapping the $K3$-fibre \cite{Harvey:1995rn} becomes asymptotically tensionless at the infinite distance boundary. It is thus clear that the pattern \eqref{eq:pattern} is also satisfied in this case, since the species scale is set by the string scale, whose exponential rate must be given by $\frac{1}{\sqrt{d-2}}$ --- if corresponding to a fundamental string propagating in $d$ dimensions. Nevertheless, let us check this explicitly by computing the relevant vectors within this regime. We first calculate the leading contribution to the three-fold volume in \eqref{eq:kahlerpotP11226}, which asymptotically scales as follows 
\begin{equation}\label{eq:volTypeIIlimit}
	\mathcal{V}_{X_3} = 2 (t^1)^2\, t^2 \left(1 + \mathcal{O}\left(t^1/t^2\right) \right)\, .
\end{equation}
Next we need to determine both the charge-to-mass vectors associated to the leading tower of states and the appropriate species scale. There is indeed a plethora of potentially light towers, both of perturbative and non-perturbative nature. First of all, one finds a critical string arising from a NS5-brane wrapped on the $K3$ surface, whose tension behaves as
\begin{equation}
	T_{\text{NS5}} = \frac{2\pi}{ \ell_s^2 g_s^2} \mathcal{V}_{K3}\, ,
\end{equation}
with $\ell_s = 2\pi \sqrt{\alpha'}$ being the fundamental string length and $\mathcal{V}_{K3}= (t^1)^2$ denotes the volume of the $K3$-fibre. Notice that along the $t^2$--\,limit that we consider here, the volume of the fibre either remains constant or grows at a smaller rate. 
Hence, upon properly co-scaling the 10d dilaton so as to keep its 4d counterpart fixed and finite (see discussion at the end of Section \ref{ss:preliminary}) one arrives at 
\begin{equation}\label{eq:NS5vector}
	\frac{T_{\text{NS5}}}{M_{\text{Pl;}\, 4}^2} = \frac{\mathcal{V}_{K3}}{ 2 \mathcal{V}_{X_3}} \sim \frac{1}{t^2} \quad \Longrightarrow \quad \vec{\zeta}_{\text{osc, NS5}} = \vec{\mathcal{Z}}_{\text{osc, NS5}} = \left( \frac{1}{3t^2}, \frac{1}{2 t^2} \right) + \mathcal{O}\left( t^1/(t^2)^2\right)\, .
\end{equation}
%
%
%
Apart from these, there are also additional infinite towers of states which become asymptotically massless in the limit of interest. These can be seen to correspond to Kaluza-Klein modes associated to the diverging $\mathbb{P}^1$-base, with characteristic mass
\begin{equation}\label{eq:KKP1scale}
	\frac{m^2_{\text{KK}, \, \mathbb{P}^1}}{M_{\text{Pl;}\, 4}^2} = \frac{e^{2\varphi_4}}{4 \pi \mathcal{V}_{\mathbb{P}^1}} \sim \frac{1}{t^2}\, ,
\end{equation}
as well as non-perturbative states arising from D0- and D2-branes wrapping the rational curve within the $K3$-fibre, 
whose masses scale as follows
\begin{equation}\label{eq:D0D2emergenthet}
	\frac{m_{\text{D0}}}{M_{\text{Pl;}\, 4}} = \frac{\sqrt{\pi}}{\mathcal{V}_{X_3}^{1/2}}\sim \frac{1}{t^1 (t^2)^{1/2}}\, , \qquad \frac{m_{\text{D2}}}{M_{\text{Pl;}\, 4}} = \frac{\sqrt{\pi} t^1}{\mathcal{V}_{X_3}^{1/2}}\sim \frac{1}{(t^2)^{1/2}}\, .
\end{equation}
%
%
%
The latter infinite set of D2-branes are mapped through Type IIA/Heterotic duality (c.f. Section \ref{ss:dualitieswithlowersusy}) to winding modes of the dual Heterotic string on $\widehat{K3} \times \mathbf{T}^2$ \cite{Harvey:1995fq, Kawai:1996te}. Note that all these towers decay at the same rate than the emergent string along the limit $t^2\to \infty$ with $t^1$ fixed.
	
From the above mass formulae one may readily compute the associated charge-to-mass vectors upon taking derivatives with respect to the non-compact K\"ahler fields,\footnote{Strictly speaking, the vector $\vec{\zeta}_{\text{KK},\, \mathbb{P}^1}$ presents an additional non-trivial component due to the dependence of the KK scale on the 4d dilaton in \eqref{eq:KKP1scale}. Such component, despite not contributing to the inner product \eqref{eq:patternemergentheterotic} below, must be taken into account when computing the length of the charge-to-mass vector, which then matches eq. \eqref{eq:zeta&speciesveconemodulus} for $d=4$ and $n=2$.} yielding
\begin{equation}\label{eq:zetavectorsn=1}
\begin{aligned}
	\vec{\zeta}_{\text{KK},\, \mathbb{P}^1} &= \left( 0, \frac{1}{2 t^2} \right)\, , \qquad \vec{\zeta}_{\text{D0}} = \left( \frac{1}{t^1}, \frac{1}{2 t^2} \right) + \mathcal{O}\left( t^1/(t^2)^2\right)\, ,\\
    \vec{\zeta}_{\text{D2}} &= \left( \frac{1}{3t^2}, \frac{1}{2 t^2} \right) + \mathcal{O}\left( t^1/(t^2)^2\right)\, .
\end{aligned}
\end{equation}
Therefore, taking into account that the species counting is dominated by the excitation modes of the dual Heterotic string, such that $\vec{\mathcal{Z}}_{\text{sp}}=\vec{\mathcal{Z}}_{\text{osc, NS5}}$, one can explicitly check that
\begin{equation}\label{eq:patternemergentheterotic}
	\vec{\zeta}_{\text{t}} \cdot \vec{\mathcal{Z}}_{\text{osc, NS5}} = \frac{1}{2}\, ,
\end{equation}
where $\text{t}\in \lbrace \text{KK, D0, D2, NS5} \rbrace $ includes all the light leading towers, and we have made use of the inverse metric shown in eq. \eqref{eq:fullinversemetricP11226}. In fact, the above inner product holds exactly (i.e. even before taking the infinite distance limit) for all charge-to-mass vectors except for $\vec{\zeta}_{\text{KK},\, \mathbb{P}^1}$, in which case \eqref{eq:pattern} is satisfied at leading order in $t^1/t^2$.
	
To summarize, consider some limit of the form $\mathbf{t}(\sigma) = \left( t^1, t^2 \right) = \left(\sigma^{e^1}, \sigma^{e^2}\right)$, with the vector $\mathbf{e}$ belonging to the first quadrant of $\mathbf{S}^1$. If  $e^2 > e^1$, we get an emergent string limit and the analysis presented above readily applies. If $e^1 \geq e^2$, we rather decompactify to 5d M-theory and the general argument of Section \ref{sss:ExampleI} carries over so that the pattern equally holds. In Figure \ref{fig:asympt_lim_IIAP1126-lim} these limits, as well as the leading towers and species, are depicted, while in \ref{fig:asympt_lim_IIAP1126-vec} the associated scalar charge-to-mass vectors (which are constant in flat coordinates) are represented. Hence, the pattern
\begin{equation}\label{eq:patternP11226}
	\left.\vec{\zeta}_{\text{t}}\cdot\vec{\mathcal{Z}}_{\rm sp}\right|_{\mathbf{t}(\sigma)}=\left.\left(G^{a b}\partial_{a}\log m_{\text{tower}}\,\partial_{b}\log \Lambda_{\rm sp}\right)\right|_{\mathbf{t}(\sigma)} = \frac{1}{2}\, ,
\end{equation}
is verified for any such asymptotic trajectory $\mathbf{t}(\sigma)$. 
	
Finally, let us remark here that some of the towers of particles arising in the present set-up, such as the one associated to $m_{\text{KK}, \, \mathbb{P}^1}$, suffer from the \emph{sliding} phenomenon first described within the Heterotic string theory context in ref. \cite{Etheredge:2023odp} (c.f. Section \ref{ss:het s1}). Moreover, note that the charge-to-mass and species vectors arrangement in Figure \ref{fig:asympt_lim_IIAP1126-vec} corresponds to a rotated version of that shown in Figure \ref{sfig:KKstring}, thus making manifest that they both share the same physical origin.

\subsubsection*{The Seiberg-Witten point and worldsheet instantons}

It is interesting at this point to extend the previous large volume analysis to other infinite distance degenerations which crucially require from both perturbative and non-perturbative $\alpha'$-corrections so as to be properly defined. The simplest such instance happens actually in the present $\mathbb{P}^{1,1,2,2,6} [12]$ example, when sitting close to the Seiberg-Witten singularity \cite{Seiberg:1994rs}. This can be reached upon considering the intersection locus between the conifold discriminant and the $t^2 \to \infty$ divisor, see e.g., \cite{Kachru:1995fv, Curio:2000sc} for details. In what follows, it will be convenient to introduce some local patch which vanish by construction at Seiberg-Witten point, as follows
\beq \label{eq:normalcoords}
	z^1=1-1728\, e^{b^1+\i t^1}\, , \qquad z^2= \frac{4\, e^{b^2+\i t^2}\, 1728^2 \left(e^{b^1+\i t^1} \right)^2}{\left( 1-1728\, e^{b^1+\i t^1}\right)^2}\, .
\eeq
Indeed, the singularity located at $z^1=z^2=0$ lies at infinite distance in the full K\"ahler metric (i.e. taking into account worldsheet instantons), and therefore the relation \eqref{eq:pattern} should hold as well. This limit is moreover closely related to the one previously discussed, with the crucial difference that now the fibre volume reaches its minimum possible value --- which is of stringy size \cite{Aspinwall:1993xz, Greene:1996tx,Greene:2000ci}, whilst that of the $\mathbb{P}^1$-base remains divergent. Hence, in order to check the pattern we need to know both the asymptotic form of the moduli space metric as well as the period vector close to the Seiberg-Witten point. These had been already computed and thoroughly studied in the literature, so we will only need to adapt the relevant results for our purposes here. 

Let us consider first the periods expanded around the point of interest. After solving the Picard-Fuchs system of differential equations \cite{Cox:2000vi,Hori:2003ic}, one finds the following convenient basis of solutions (in the coordinate frame \eqref{eq:normalcoords})
\begin{equation} \label{eq:periodvectorPicardFuchs}
    \varpi_{\rm sw} = \frac{1}{\pi} \begin{pmatrix}
    1 + \frac{5}{36} z^1 \\
    z^1 \\
    -\sqrt{z^1} \\
   \frac{\i}{\pi} \left(\log(z^2) - 6 \log2 + 7\right) \sqrt{z^1}\\
   \frac{\i}{2\pi} \left(5+\log \left(z^2 (z^1)^2 \right)\right) \left( 1 + \frac{5}{36} z^1\right)\\
   \frac{\i}{2\pi} \left(1 + \log \left(z^2 (z^1)^2 \right)\right) z^1
\end{pmatrix}\, ,
\end{equation}
where we only display the leading order terms in an expansion around $z^1=z^2=0$. Next, one needs to perform some analytic continuation translating the basis of integral periods around the large volume point to the Seiberg-Witten locus. This can be done upon multiplying \eqref{eq:periodvectorPicardFuchs} by the matrix \cite{Lee:2019wij}
\begin{equation} \label{eq:analyticContmatrix}
    \mathsf{M} = \begin{pmatrix}
       \frac{1}{2\mathcal{X}} & \frac{\mathcal{X}}{2} & \frac12 & 0 & 0 & 0 \\
    \frac{\i}{2\mathcal{X}} & -\frac{\i \mathcal{X}}{2} & 0 & 0 & 0 & 0 \\
    \i \left( \xi_1 + \frac{\xi_2}{\mathcal{X}^2}\right) & \i \left( \xi_3 + \xi_4 \mathcal{X}^2\right) & 0 & 1 & -\frac{1}{2\mathcal{X}} & -\frac{\mathcal{X}}{2} \\
    \frac{1}{\mathcal{X}} & \mathcal{X} & 0 & 0 & 0 & 0 \\
    -2\xi_1 + \frac{2\xi_2} {\mathcal{X}^2} & 2 \xi_3-2\xi_4 \mathcal{X}^2 & 0 & 0 & \frac{\i}{\mathcal{X}} & -\i \mathcal{X} \\
    0 & 0 & 0 & 2 & 0 & 0
\end{pmatrix}\, ,
\end{equation}
where $\mathcal{X}=\frac{\Gamma \left( \frac34 \right)^4}{\sqrt{3} \pi^2}$ and $\{\xi_i \}$ are numerical constants given by
\beq
	\xi_1 = -4.16688\, , \quad \xi_2 =0.130737\, , \quad \xi_3 = -0.99488\, , \quad \xi_4 = 7.51362\, ,
\eeq
which yields the following (leading-order) period vector 
\begin{equation} \label{eq:periodvectorSW}
    \Pi_{\rm sw} = \begin{pmatrix}
    X^0 \\
           X^1 \\
           X^2\\
           \mathcal{F}_1 \\
           \mathcal{F}_2\\
           \mathcal{F}_0
\end{pmatrix} := \frac{1}{\pi} \begin{pmatrix}
    \frac{1}{2\mathcal{X}}\\
    \frac{\i}{2\mathcal{X}} \\
    -\frac{\i}{4 \pi \mathcal{X}} \log \left(z^2 (z^1)^2 \right) \\
   -\frac{1}{2 \pi \mathcal{X}} \log \left(z^2 (z^1)^2 \right)\\
   \frac{1}{\mathcal{X}}\\
   \frac{2\i}{\pi} \sqrt{z^1} \log z^2
\end{pmatrix}\, .
\end{equation}
The above quantity determines the quantum-corrected volumes of even-dimensional supersymmetric cycles, and thus the corresponding BPS masses of D-brane particles (and strings), when seen from the 4d perspective. In particular, we will be interested in the volumes associated to the objects already described in eqs. \eqref{eq:NS5vector}-\eqref{eq:D0D2emergenthet}, whose masses are captured by the 4d $\mathcal{N}=2$ central charge as follows
\beq \label{eq:masstower}
	m_{\rm tow} = \left| Z(\mathbf{q}_{\text{t}})\right| = \sqrt{8\pi} e^{\frac{K_{\rm ks}}{2}}\,\mathbf{\Pi}_{\rm sw} \cdot \mathbf{q}_{\text{t}}\, ,
\eeq
where $\mathbf{q}_{\text{t}}$ is the relevant vector of charges and the dot in \eqref{eq:masstower} denotes the usual Cartesian product.

With this, we are now ready to check whether the relation \eqref{eq:pattern} still survives at the Seiberg-Witten point. First, we determine the asymptotic expression for the K\"ahler metric and, subsequently, we compute the masses of the relevant towers of states as well as the species cut-off. Regarding the former, the K\"ahler potential for the moduli fields in the coordinate system \eqref{eq:normalcoords} is computed to be
\beq \label{eq:Kahlerpot}
	K_{\rm ks} = - \log \left( \frac{4 y^1 + 2y^2}{\pi^2 \mathcal{X}^2} - \frac{16}{\pi^2} e^{-2 \pi y^1} \left( \frac14 y^2 - \frac{1}{2\pi}\right) + \ldots\right)\, ,
\eeq
where we have defined the complex coordinates $x^i + \i y^i := \frac{1}{2 \pi \i} \log (z^i)$, in terms of which the infinite distance point is located at $y^i \to \infty$. Notice that \eqref{eq:Kahlerpot} consists of a leading polynomial contribution and a second \emph{exponentially suppressed} piece that is nonetheless necessary so as to have a well-defined metric \cite{Bastian:2021eom, Bastian:2021hpc}.\footnote{Indeed, if we ignore the instanton-like terms in \eqref{eq:Kahlerpot} we obtain the following metric for the real fields $\{ y^i\}$
\begin{equation}
    G = \frac{1}{2\left( 2 y^1 + y^2\right)^2} \begin{pmatrix}
		4 \quad  2\\2 \quad  1
	\end{pmatrix}\, ,
\end{equation}
which is clearly degenerate \cite{Bastian:2021eom}. Hence, the inclusion of such exponentially suppressed contributions is crucial for computing the product \eqref{eq:pattern}.} Using the real coordinate system $\{ x^i, y^i\}$ and the fact that the leading term within $K_{\rm ks}$ preserves the shift symmetry of the axions, the line element associated to the previous K\"ahler potential can be computed to be
\begin{align} \label{eq:lineelement}
	ds^2 &= \frac{\partial^2 K_{\rm ks}}{\partial z^i \partial \bar{z}^j} dz^i d\bar{z}^j = \frac12 G_{i j} \left( dy^i dy^j + db^idb^j\right)\, ,
\end{align}
with
\begin{align} \label{eq:realmetricSWpoint}
	G_{i j} &= \frac12 \frac{\partial^2 K_{\rm ks}}{\partial y^i \partial y^j} \notag\\
    &= \begin{pmatrix}
\frac{2 e^{2 \pi y^1} \pi^2 (e^{2 \pi y^1} + 2 (-2 + \pi y^2) (2 + \pi (2 y^1 + y^2)) \mathcal{X}^2)}{( e^{2 \pi y^1} \pi (2 y^1 + y^2) - 2 (-2 + \pi y^2) \mathcal{X}^2)^2} & \frac{ e^{2 \pi y^1} \pi^2 (e^{2 \pi y^1} - 2 (3 + 2 \pi y^1) \mathcal{X}^2)}{( e^{2 \pi y^1} \pi (2 y^1 + y^2) - 2 (-2 + \pi y^2) \mathcal{X}^2)^2} \\
\frac{ e^{2 \pi y^1} \pi^2 (e^{2 \pi y^1} - 2 (3 + 2 \pi y^1) \mathcal{X}^2)}{( e^{2 \pi y^1} \pi (2 y^1 + y^2) - 2 (-2 + \pi y^2) \mathcal{X}^2)^2} & \frac{\pi^2 (e^{2 \pi y^1} - 2 \mathcal{X}^2)^2}{2 ( e^{2 \pi y^1} \pi (2 y^1 + y^2) - 2 (-2 + \pi y^2) \mathcal{X}^2)^2}
\end{pmatrix}\, ,
\end{align}
which indeed satisfies $\det G_{ij} \neq 0$.

Next, we turn to the energy scales of the relevant towers of states becoming massless (in 4d Planck units) at the Seiberg-Witten singularity. These correspond to the ones already discussed in eqs. \eqref{eq:NS5vector}-\eqref{eq:D0D2emergenthet}, whose moduli dependence now read as (c.f. eq. \eqref{eq:masstower})
\begin{equation}\label{eq:BPSmassesSW}
	\begin{aligned}
	\frac{T_{\text{NS5}}}{\Mpf^2} &= \frac{\mathcal{V}_{K3}}{ 2 \mathcal{V}} = 4|X^0|^2\,  e^{K_{\rm ks}}\, \left|\frac{\mathcal{F}_2-X^0}{X^0}\right| = \frac{1}{4 y^1 + 2y^2 - 16 \mathcal{X}^2 e^{-2 \pi y^1} \left( \frac14 y^2 - \frac{1}{2\pi}\right)}\, ,\\
    \frac{m_{\text{D0}}}{\Mpf} &= \frac{\sqrt{\pi}}{\mathcal{V}^{1/2}} = \sqrt{8 \pi}\, |X^0|\,  e^{K_{\rm ks}/2} = \frac{\pi^{1/2}}{\left( 2 y^1 + y^2 - 8 \mathcal{X}^2 e^{-2 \pi y^1} \left( \frac14 y^2 - \frac{1}{2\pi}\right)\right)^{1/2}}\, ,\\
    \frac{m_{\text{D2}}}{\Mpf} &= \frac{\sqrt{\pi}\, \mathcal{V}_{\mathbb{P}_f^1}}{\mathcal{V}^{1/2}} = \sqrt{8 \pi}\, |X^0|\,  e^{K_{\rm ks}/2}\,  \left|\frac{X^1}{X^0}\right| = \frac{\pi^{1/2}}{\left( 2 y^1 + y^2 - 8 \mathcal{X}^2 e^{-2 \pi y^1} \left( \frac14 y^2 - \frac{1}{2\pi}\right)\right)^{1/2}}\, ,\\
    \frac{m_{\text{KK}, \, \mathbb{P}^1}}{\Mpf} &= \frac{e^{\varphi_4}}{\sqrt{4 \pi} \left(\mathcal{V}_{\mathbb{P}_b^1} \right)^{1/2}} = \frac{e^{\varphi_4}}{\sqrt{4 \pi}}\, \left|\frac{X^0}{X^2}\right|^{1/2} =\frac{e^{\varphi_4}}{\sqrt{4 \pi}}\, \frac{1}{\left( 2 y^1 + y^2 - 8 \mathcal{X}^2 e^{-2 \pi y^1} \left( \frac14 y^2 - \frac{1}{2\pi}\right)\right)^{1/2}}\, .
	\end{aligned}
\end{equation}
Notice that, similarly to what happened in the limit discussed around eq. \eqref{eq:volTypeIIlimit}, all four mass scales end up being determined by the same quantity, namely the overall volume modulus. Therefore, the same considerations apply, such that upon computing the product \eqref{eq:patternP11226} one obtains
\begin{equation}\label{eq:NS5norm}
  \begin{aligned}
	 \vec{\zeta}_{\text{t}}\cdot\vec{\mathcal{Z}}_{\rm sp} &= \frac{e^{4\pi y^1} \pi y^2 -4e^{2\pi y^1} \mathcal{X}^2-4 \left( -2+\pi y^2\right) \mathcal{X}^4}{2e^{4\pi y^1} \pi y^2 -4e^{2\pi y^1} \left(2+\pi \left(2y^1 + y^2\right)\right) \mathcal{X}^2}\\
     &= \frac{1}{2} + \mathcal{X}^2\, \frac{2y^1+y^2}{y^2} e^{-2\pi y^1} + \mathcal{O}\left( e^{-4\pi y^1}\right)\; 
  \end{aligned}
\end{equation}
in agreement with eq. \eqref{eq:pattern} for $d=4$.
	
\subsubsection{General story}
\label{sss:IIA/heterotic}	
	
	
Here we want to generalize our previous discussion so as to systematically check the pattern for any Type II singularity (in the MHS nomenclature) located within the large volume patch. The following analysis builds on the intuition gained from the example above and it parallels that from Section \ref{sss:IIA/Fthy}.   
	
First, recall that this class of limits can be equivalently described in a purely geometrical way as exhibiting some kind of asymptotic surface fibration \cite{doran2016mirror}, where the fibre is isomorphic to either a $K3$ or a $\mathbf{T}^4$ two-fold (see Table \ref{tab:limitsN=2}). Therefore, let us assume that the CY$_3$ admits such a fibration structure over a $\mathbb{P}^1$-base 
\begin{equation}\label{eq:K3/T4fibration}
			\begin{aligned}
				\rho: \qquad K3/\mathbf{T}^4 \hookrightarrow &\;X_{3} \\
				&\;\; \downarrow \qquad . \\ &\ \ \mathbb{P}^1
			\end{aligned}
\end{equation}
For simplicity, we will require the fibration \eqref{eq:K3/T4fibration} to not present any degenerations.\footnote{One could also relax this assumption and allow both for finite and infinite distance singularities. In those cases, the analysis becomes slightly more complicated, so we restrict ourselves to the non-degenerate set-up, since this is enough to prove our point here.} Therefore we can separate the K\"ahler moduli as follows
\begin{equation}\label{eq:Kahlermoduliemergentstring}
	J= t^A \omega_A = t^0 \omega_0 + t_f^{\alpha} \omega_{\alpha} \, . 
\end{equation}
where $t^0$ measures the volume of the $\mathbb{P}^1$-base, and $\lbrace t^{\alpha}_f\rbrace$, with $\alpha= 1, \ldots, h^{1,1}-1$, are instead associated to the $K3/\mathbf{T}^4$ fibre.  Notice that the generator $\omega_0$ satisfies $\omega_0^2=0$, thus implying that the triple intersection numbers verify $\mathcal{K}_{00 A}=0$.

Hence, we now consider the particular infinite distance limit described by 
\begin{equation}\label{eq:n=1limit}
	t_f^{\alpha}  = \text{const.}\, , \qquad t^0= \sigma\, , \qquad \text{with}\, \, \sigma\to \infty\, , 
\end{equation}
which indeed belongs to the $K3/\mathbf{T}^4$ class. Microscopically, \eqref{eq:n=1limit} is believed to correspond to an emergent Heterotic (or Type II) string limit, where the critical string arises from compactifying a NS5-brane on the generic $K3$ (respectively $\mathbf{T}^4$) fibre.\footnote{This is difficult to prove in general, since one would need to study the excitation spectrum associated to the world-volume theory of the wrapped NS5-brane and match it (at all mass levels) with that of the fundamental dual string, which is of course a very non-trivial task.} Mirroring our discussion in Section \ref{sss:IIA/Fthy}, we both look at the relevant light spectrum and moduli space metric. Regarding the former, one finds a $\frac{1}{2}$-BPS string obtained by wrapping the NS5-brane on the generic fibre (that is assumed to be \emph{fundamental}), D0-branes as well as D2-branes wrapped on 2-cycles within the fibre class, and a (double) KK tower associated to the base of the fibration. Their mass/tension read as
\begin{align}
	\notag \frac{T_{\text{NS5}}}{M_{\text{Pl;}\, 4}^2} &= \frac{\mathcal{V}_{\text{fib}}}{ 2 \mathcal{V}_{X_3}}\, , \qquad \frac{m_{\text{D0}}}{M_{\text{Pl;}\, 4}} = \frac{\sqrt{\pi}}{\mathcal{V}_{X_3}^{1/2}}\, ,\\
	\frac{m_{\text{D2}}^{(\alpha)}}{M_{\text{Pl;}\, 4}} &= \frac{\sqrt{\pi} t_f^{\alpha}}{\mathcal{V}_{X_3}^{1/2}}\, , \qquad \frac{m_{\text{KK}, \, \mathbb{P}^1}}{M_{\text{Pl;}\, 4}} = \frac{e^{\varphi_4}}{\sqrt{4 \pi \mathcal{V}_{\mathbb{P}^1}}}\, ,
\end{align}
with $\mathcal{V}_{X_3}$ the overall three-fold volume, $\mathcal{V}_{\text{fib}}= \frac{1}{2} \int_{X_3} J\wedge J \wedge \omega_0$ that of the fibre and $\mathcal{V}_{\mathbb{P}^1} = t^0$ controls the volume of the $\mathbb{P}^1$-base. 
	
On the other hand, the K\"ahler potential presents the following leading asymptotic behavior \cite{Lee:2019wij}
\begin{equation}\label{eq:kahlerpotn=1}
	K_{\text{ks}}= - \log \left(t^0\, \eta_{\alpha \beta} t_f^{\alpha} t_f^{\beta} + \mathcal{O} \left((t^0)^0\right)\right)\, ,
\end{equation}
where $\eta_{\alpha \beta} = \mathcal{K}_{0 \alpha \beta}$ denotes the intersection form associated to the $K3/\mathbf{T}^4$-fibre. From this, one can compute the moduli space metric, which can be expanded as a power series in $1/t^0$ as follows
%
\begin{align}\label{eq:metricn=1}
	G_{\alpha \beta} &= \frac{1}{2} \frac{\partial^2 K_{\text{ks}}}{\partial t_f^{\alpha} \partial t_f^{\beta}} = G^{(\rm lead.)}_{\alpha \beta} + \mathcal{O}\left(1/t^0\right)\, , \quad G_{0 \alpha} = \frac{1}{2} \frac{\partial^2 K_{\text{ks}}}{\partial t^0 \partial t_f^{\alpha}} = \frac{1}{(t^0)^2} G^{(\rm lead.)}_{0 \alpha} + \mathcal{O}\left(1/(t^0)^3\right)\, ,\notag\\
	G_{0 0} &= \frac{1}{2} \frac{\partial^2 K_{\text{ks}}}{\partial t^0 \partial t^0} = \frac{1}{2(t^0)^2} + \mathcal{O}\left(1/(t^0)^3\right)\, .
\end{align}
%
The above leading-order matrices can be explicitly computed in terms of the K\"ahler moduli \eqref{eq:Kahlermoduliemergentstring}, yielding 
\begin{align}\label{eq:constmatricesn=1lim}
	\notag G^{(\text{lead.})}_{\alpha \beta} &=\frac{2 \left(\eta_{\alpha \gamma} t_f^{\gamma}\right) \left(\eta_{\beta \delta} t_f^{\delta}\right)}{\left( \eta_{\gamma \delta} t^{\gamma}_f t^{\delta}_f \right)^2} - \frac{\eta_{\alpha \beta}}{\eta_{\gamma \delta} t^{\gamma}_f t^{\delta}_f}\, ,\\
    G^{(\text{lead.})}_{0 \alpha} &=\frac{\mathcal{K}_{\alpha \beta \gamma} t^{\beta}_f t^{\gamma}_f}{2 \eta_{\gamma \delta} t^{\gamma}_f t^{\delta}_f} - \frac{1}{3}\frac{\left( \eta_{\alpha \beta} t_f^{\beta}\right) \left( \mathcal{K}_{\delta \gamma \lambda} t^{\delta}_f t^{\gamma}_f t^{\lambda}_f\right)}{\left( \eta_{\gamma \delta} t^{\gamma}_f t^{\delta}_f \right)^2}\, ,
\end{align}
which have moreover full rank. For the (sub-)matrix $G^{(\text{lead.})}_{\alpha \beta}$ this follows from assuming the generic $K3/\mathbf{T}^4$-fibre to be non-degenerate, thus ensuring that the intersection form $\eta_{\alpha \beta}$ in eq. \eqref{eq:constmatricesn=1lim} is non-degenerate as well.
	
With this, we can finally prove that the pattern \eqref{eq:pattern} holds for the present Type II degenerations. Indeed, using the fact that (to leading order in $1/t^0$)
\begin{equation}\label{eq:noscalen=1}
	\frac{\partial K_{\text{ks}}}{\partial t^0} G^{0 0} \frac{\partial K_{\text{ks}}}{\partial t^0} = 2\, ,
\end{equation}
which can be regarded as a no-scale property, the condition
\begin{equation}\label{eq:patternn=1limits}
	\vec{\zeta}_{\text{t}} \cdot\vec{\mathcal{Z}}_{\text{osc, NS5}}=\left(G^{A B}\partial_{A}\log m_{\text{tower}}\,\partial_{B}\log \Lambda_{\rm sp}\right) \stackrel{~\eqref{eq:n=1limit}~}{=} \frac{1}{2}\, , \qquad A, B= \lbrace a, \beta \rbrace\, ,
\end{equation}
is indeed satisfied for all $\text{t}= \lbrace \text{KK, D0, D2, NS5} \rbrace$. We would like to stress that eq. \eqref{eq:patternn=1limits} holds to leading order in $1/t^0$, since any term involving derivatives with respect to the fibral moduli $\{t_f^{\alpha}\}$ contributes at an order $\mathcal{O}\left(1/t^0\right)$ or higher. Once again, this is the reason why the result also applies to more general limits in which the fiber volume is also sent to infinity, but at a slower rate than that of (any curve within) the base.
	
\subsection{Comments about the complex structure moduli space of Type IIB}
\label{ss:commentsTypeIIB}
	
Let us briefly mention here how the previous analysis extends to the vector multiplet moduli space of Type IIB string theory compactified on the (mirror) three-fold $Y_3$. In principle, via Mirror Symmetry, a similar story should also hold for the complex structure moduli space of Type IIB on $Y_3$, where the charge-to-mass and species vectors must behave in the same fashion as in the Type IIA counterpart. In practice, however, the microscopic physics is oftentimes lurked, preventing us from performing a clean geometrical analysis as in the previous sections. The reason for this is two-fold: First, it is difficult to argue for the existence of infinite towers of BPS bound states, since not every BPS charge may be actually populated due to the possible presence of walls of marginal stability (see footnote \ref{fnote:stabilityBPS}).\footnote{Notice that the results of ref. \cite{Palti:2021ubp} suggest that a tower of electric BPS states would always exist as long as we also have the corresponding BPS extremal black hole solution for large charges.} Therefore, it is usually not at all clear which is the lightest tower, whose $\zeta$-vector we would need to compute. Relatedly, the fact that we cannot determine all towers of states becoming light for each limit means that the species scale can be hard to calculate, in general.
	
Our aim here will be to comment on how some of these difficulties can be sidestepped, using both techniques from the Mixed Hodge Structure  literature (see e.g., \cite{Grimm:2018ohb, Grimm:2018cpv}) as well as building on our previous Type IIA analysis. Thus, regarding the leading tower of states, we will assume that there is \emph{at least} one tower given by D3-branes wrapping the fastest shrinking 3-cycle. This can be motivated from the examples of Sections \ref{sss:ExampleII} and \ref{sss:ExampleIII}, where there was always some D0 or D2-brane tower becoming light at the fastest rate (even in the emergent string limits, c.f. \eqref{eq:D0D2emergenthet}). These states are all mapped through Mirror Symmetry to certain D3-branes wrapping special Lagrangian 3-cycles. From this, one can deduce at least one co-leading scalar charge-to-mass vector $\vec{\zeta}_{\rm D3}$, whose components read
\begin{equation}
	\left(\zeta_{\text{D3}}\right)_i = - \frac{1}{2} \frac{\partial K_{\text{cs}}}{\partial \text{Im}\, z^i}\, ,
\end{equation}
where $\lbrace z^i\rbrace$ denote the complex structure moduli and $K_{\text{cs}}$ is the associated K\"ahler potential (c.f. eq. \eqref{eq:CSmetric}).
	
To compute the species scale, on the other hand, one needs to know not only how many towers there are but also their microscopic degeneracy. Here, we will avoid having to deal with these subtleties by looking instead at certain moduli dependent functions that correct the 4d $\mathcal{N}=2$ two-derivative lagrangian, which according to our discussion in Chapter \ref{ch:Higherdimops} should capture at least the asymptotic behavior of the species scale within $\cM_{\rm VM}$. Following the original works \cite{vandeHeisteeg:2022btw,vandeHeisteeg:2023ubh}, we take the topological genus-one partition function $\mathcal{F}_1$, whose exact expression is (c.f. eq. \eqref{eq:F1IIA})
\beq
	\mathcal{F}_1 =   \frac {1}{2}\left( 3+h^{2,1}-\frac {\chi_{E} (Y_3)}{12}\right)K_{\text{cs}} + \frac {1}{2}\log \det G_{i \bar j} + \log|f|^2\, ,
\label{eq:F1}
\eeq
to give a proxy for the number of species in the vector multiplet sector. Here, $h^{2,1}$ is the (complex) dimension of the complex structure moduli space, $\chi_{E}$ denotes the Euler characteristic of the three-fold $Y_3$, $G_{i \bar j}$ is the moduli space metric derived from the K\"ahler potential $K_{\text{cs}}$ (c.f. eq. \eqref{eq:CSmetric}) and $f(z^i)$ is an holomorphic anomaly which can be generically fixed upon comparing with the known asymptotic behavior of $\mathcal{F}_1$ \cite{Cecotti:1992vy,Bershadsky:1993ta}.
	
For concreteness, we particularize in what follows to the large complex structure (LCS) regime, where a plethora of infinite distance degenerations may occur. Let us note in passing that the argument works equally well for any other such singularity, not necessarily belonging to the LCS patch. We will thus need the leading order behavior of $\mathcal{F}_1$, which is given by \cite{Bershadsky:1993ta,Bershadsky:1993cx} 
\beq
	\mathcal{F}_1 = \frac{1}{24} \int_{X_3} J\wedge c_2(TX_3) + \ldots = \frac{1}{24} c_{2,\, i}\, \text{Im}\, z^i + \ldots\, ,
\label{eq:F1LCS}
\eeq
where $X_3$ is the mirror three-fold with associated K\"ahler 2-form $J$, $c_2(TX_3)$ denotes its second Chern class and the ellipsis indicates further contributions which are subleading when $\text{Im}\, z^i \gg 1$. From this, one obtains
\begin{equation}
	\left(\mathcal{Z}_{\text{sp}}\right)_i = - \partial_i \log \LSP = \frac{1}{2} \partial_i \log \mathcal{F}_1 = \frac{c_{2,\, i}}{2 \int J\wedge c_2} + \mathcal{O} \left( \frac{\log \text{Im}\, z^i}{\text{Im}\, z^i} \right)\, ,
\end{equation}
where we have used that $\LSP= M_{\text{Pl;}\, 4}\, N^{-1/2}$, with $N= \mathcal{F}_1$. Therefore, what we want to show here is that the product
\begin{equation}\label{eq:IIBproduct}
	\vec{\zeta}_{\text{t}} \cdot \vec{\mathcal{Z}}_{\text{sp}} =-\frac{1}{4} K_i\, G^{i j}\, \frac{c_{2,\, j}}{\int J\wedge c_2} = -\frac{1}{2} K_i\, K^{i j}\, \frac{c_{2,\, j}}{\int J\wedge c_2}\, ,
\end{equation}
gives $\frac{1}{2}$ regardless of the kind of limit that we explore. Note that in the previous expression we have substituted the metric element along the saxionic directions $G_{i j}$ in favour of $ K_{i j} = \partial_i \partial_j K_{\rm cs}$. 
	
In a nutshell, this follows from the homogeneous dependence of the quantities $e^{K_{\rm cs}}$ and $\int J\wedge c_2$ with respect to the complex structure moduli $z^i$. Indeed, for Type II, III and IV degenerations in the complex structure moduli space, Mixed Hodge Theory tells us that the K\"ahler potential behaves to leading order as follows (see e.g., \cite{Grimm:2018cpv})
\begin{equation}
	K_{\text{cs}} \to K_{\text{cs}} - \omega \log \sigma\, , \qquad \text{as}\ \ \text{Im}\, z^a \to \sigma\, \text{Im}\, z^a\, \quad \text{with }\sigma\rightarrow \infty\, ,
\end{equation}
with $\omega=1,2,3$ respectively, and where the set $\lbrace z^a \rbrace \subseteq \lbrace z^i \rbrace$ denotes those moduli which are sent to infinity upon approaching the corresponding infinite distance boundary. From the above relation one can prove a number of useful identities. In particular, one finds
\begin{align}
	\text{Im}\, z^a \partial_a K_{cs} &= -\omega + \ldots\, , \qquad \text{Im}\, z^a\, \text{Im}\, z^b\, \partial_a \partial_b K_{cs} = \omega + \ldots\, ,
\end{align}
which can then be used to show that
\begin{subequations}
	\begin{align}
		K_a K^{a b} K_b &= \omega+\dots \, ,\\
		K^{ab} K_b &= -\text{Im}\, z^a+\dots \, ,
	\end{align}
\end{subequations}
where the corrections in all previous equations vanish asymptotically. Note that the first relation is nothing but the familiar no-scale condition of the metric $G_{i j}$, whilst upon plugging the second one into eq. \eqref{eq:IIBproduct} we obtain
\begin{equation}
	\vec{\zeta}_{\text{t}} \cdot \vec{\mathcal{Z}}_{\text{sp}} = \frac{\text{Im}\, z^a\, c_{2,\, a}}{2\int J\wedge c_2} = \frac{1}{2}\, ,
\end{equation}
where one needs to use that $\text{Im}\, z^a\, \partial_a \log \left( \int J\wedge c_2\right) =1$ in order to arrive at the final result. This follows again from the asymptotic homogeneity of the integrated second Chern class, and proves that the pattern holds in the Type IIB set-up as well.

\subsection{The hypermultiplet moduli space}
\label{ss:hypers}

Up to now we have restricted ourselves to a purely classical analysis, where quantum effects can be safely neglected. The purpose of this subsection is to study the fate of the pattern \eqref{eq:pattern} within heavily quantum corrected moduli spaces, thus providing strong evidence for its robustness. We will still restrict ourselves to 4d $\mathcal{N}=2$ set-ups, now focusing on the hypermultiplet sector, which locally decouples from its vector multiplet counterpart, c.f. eq. \eqref{productmoduli}. 

In Type IIA CY$_3$ compactifications, the hypermultiplet moduli space describes a quaternionic-K\"ahler space parametrized by $4(h^{2,1} (X_3)+1)$ real scalars. The exact field content as well as its higher-dimensional origin was already explained in Section \ref{sss:4dN=2basics}. Here we will focus mostly on the non-compact directions, which are parametrized by the complex structure moduli $\{ z^i\}$ and the 4d dilaton $\varphi_4$. Classically, the sigma-model metric for this restricted set of fields is very simple and reads (c.f. eq. \eqref{eq:classicalhypermetric} for the full line element)
\begin{align}\label{eq:classicalhypermetricnoaxions}
	ds^2_{\rm HM} = 2\left( d \varphi_4\right)^2 + G_{i \bar j} d z^i d z^{\bar j} + (\text{axions})\, ,
\end{align}
where we recall that $G_{i \bar j}$ defines the Weil-Petersson metric on the space of complex structures \cite{Candelas:1990pi}
\begin{equation}\label{eq:CSmetricpattern}
	G_{i \bar j} = \partial_{z^i}\partial_{\bar z^j} K_{\text{cs}} \, ,\qquad \text{with}\quad K_{\text{cs}}=-\log \left({\rm i} \int_{X_3}\Omega_3 \wedge \bar \Omega_3 \right)\, .
\end{equation}
Quantum-mechanically, however, the above line element receives both perturbative and non-perturbative corrections, the latter due to e.g., Euclidean D2-brane instantons wrapping special Lagrangian (sLag) 3-cycles \cite{Becker:1995kb}. These are, in general, very difficult to obtain (see Appendix \ref{ap:hypermetric} for details), but in principle they could strongly modify any classical statement based on geodesic trajectories and asymptotic moduli behaviour. 
Our discussion here will closely follow the analysis presented in \cite{Marchesano:2019ifh,Baume:2019sry}, where the effect of the aforementioned instanton corrections on certain classical infinite distance singularities was studied with some detail. 

\subsubsection{Classical infinite distance points}
\label{sss:classivalvsquantum}
	
In the following, we will focus on trajectories within $\mathcal{M}_{\rm HM}$ which lie entirely along the non-compact directions, namely we set the axion v.e.v.s to zero value. This moreover allows one to compute the relevant perturbative and non-perturbative corrections to the classical metric components \cite{Marchesano:2019ifh,Baume:2019sry} (see Appendix \ref{ss:exactmetric}). In particular, we will be interested in studying the realization of the pattern \eqref{eq:pattern} along a certain family of trajectories, which we parametrize as follows\footnote{Recall that since we focus now on trajectories lying entirely in the hypermultiplet moduli space, the overall volume $\mathcal{V}_{X_3}$ of the CY$_3$ is assumed to be fixed. Hence, the 10d and 4d dilaton agree up to this constant (although large) volume factor.}
\begin{equation}\label{eq:generictraj}
	\text{Im}\, z^i \sim \sigma^{e^1}\, , \qquad e^{-\phi}\sim \sigma^{e^2}\, , \qquad \sigma \to \infty\, ,
\end{equation}
with $e^1, e^2 \geq 0$. Note that such paths correspond to geodesic trajectories with respect to the classical hypermultiplet metric \eqref{eq:classicalhypermetricnoaxions}. We now consider different scenarios depending on the precise values of $\mathbf{e}=(e^1, e^2)$.
	
\subsubsection*{Weak string coupling point}
	
For the case in which we take $\mathbf{e}=(0, e^2)$, the only contribution to the classical moduli space distance $\Delta_{\rm HM}$ arises from the 4d dilaton piece. The lightest tower of states correspond to the oscillation modes of the fundamental string, whose mass behaves asymptotically as follows (we set $e^2=1$ without loss of generality)
\begin{equation}\label{eq:fundstringmass}
	\left(\frac{m_s}{M_{\text{Pl;}\, 4}} \right)^2= \frac{ e^{2\varphi_4}}{4 \pi} \sim \frac{1}{\sigma^2}\, ,
\end{equation}
thus leading to a charge-to-mass vector with vanishing components except for that associated to the 4d dilaton field, namely 
\begin{equation}
	\vec{\zeta}_{\text{osc}} = \left(\zeta_{\varphi_4}\, ,\, \ldots\right)= \left(-1, 0\, , \ldots, 0\right)\, .
\end{equation}
Notice that, since the volume of the three-fold is kept fixed, the associated KK-scale also behaves like \eqref{eq:fundstringmass} asymptotically, i.e. $m_{\rm KK,\, 6} = m_s/\mathcal{V}_{X_3}^{1/6} \sim \sigma^{-1}$. Its charge-to-mass vector, $\vec{\zeta}_{\text{KK},\, 6}$, may be easily obtained as well
\begin{equation}\label{eq:KKCYzetavector}
	\vec{\zeta}_{\text{KK},\, 6} = \left(\zeta_{\varphi_4},\zeta_{\mathcal{V}_{X_3}}\, ,\, \ldots\right)= \left(-1,\, \frac{1}{6\mathcal{V}_{X_3}}\, ,0\, , \ldots, 0\right)\, ,
\end{equation}
where the extra non-trivial component corresponds to the overall volume direction.\footnote{Note that upon computing the norm of the vector \eqref{eq:KKCYzetavector} using the metrics in eqs. \eqref{eq:classicalhypermetricnoaxions} and \eqref{eq:IIAlagrangian4d} one gets $|\vec{\zeta}_{\text{KK}}|=\sqrt{\frac{2}{3}}$, in agreement with \eqref{eq:zeta&speciesveconemodulus} for $d=4$ and $n=6$.} 

On the other hand, the species cut-off coincides with the string scale, such that upon taking the inner product between the previous vectors and $\vec{\mathcal{Z}}_{\text{sp}}=\vec{\zeta}_{\text{osc}}$, one gets $\vec{\zeta}_{\text{t}} \cdot \vec{\mathcal{Z}}_{\text{osc}} = \frac{1}{2}$, in agreement with \eqref{eq:pattern}. To show this, one needs to use that $G_{\varphi_4 \varphi_4}=2$ as well as the factorization of the vector multiplet and hypermultiplet metrics, c.f. eq. \eqref{productmoduli}.
	
For completeness, let us here mention that even though the scaling of the 10d dilaton in \eqref{eq:generictraj} has been chosen so as to probe the weak coupling behavior of the fundamental Type IIA string, one could in principle consider trajectories with $e^2 \leq 0$, thus exploring the strong coupling regime. It turns out, however, that both kind of limits are related by $\mathsf{SL(2, \mathbb{Z})}$ duality (see Appendix \ref{ss:SL2Z} for details), such that everything said so far trivially extends to this dual scenario as well. In particular, for the S-dual limit the dominant critical string becoming light corresponds to a D4-brane wrapping the reference sLag 3-cycle of the CY manifold, which has $\mathbf{T}^3$ topology \cite{Strominger:1996it}, 
and is mapped via Mirror Symmetry precisely to a D1-string in the Type IIB dual picture \cite{Alvarez-Garcia:2021pxo}.
	
\subsubsection*{Large complex structure point}
	
Let us now turn to the other possibility, namely we consider the case $\mathbf{e}=(e^1, 0)$ in \eqref{eq:generictraj}, thus exploring the LCS point at fixed dilaton v.e.v. 
Note that the string scale is now fixed in Planck units, such that it can no longer provide for the leading tower of states. Moreover, even though the overall three-fold volume is kept constant, the fact that we take a large complex structure limit means that the compact manifold behaves in a highly anisotropic way. This can be confirmed by looking at the volume of supersymmetric 3-cycles $\Gamma=n^I A_I + n_J B^J$, where $\{A_I, B^J \}$ define an integral symplectic basis of $H_3(X_3)$, such that
\begin{equation}\label{eq:symplecticpairing}
	A_I \cdot B^J = - B^J \cdot A_I= \delta^J_I\, .
\end{equation}
The aforementioned volumes can be computed in string units as follows \cite{Lee:2019wij}
\begin{equation}\label{eq:slagvolumes}
	\mathcal{V}_{\Gamma} = \int_{\Gamma} \dd^3 y\, \sqrt{g} = \left(\frac{8 \mathcal{V}_{X_3}}{{\rm i} \int\Omega_3 \wedge \bar \Omega_3 } \right)^{1/2} \text{Im}\, \int_{\Gamma} e^{-{\rm i} \theta}\Omega_3 \, ,
\end{equation}
where $\theta$ determines the appropriate calibration 3-form. For the limit of interest, such volumes are controlled by the period vector $\Pi(z^i) = \left(Z^0, Z^i, \mathcal{F}_j, \mathcal{F}_0\right)^{T}$, as well as the would-be K\"ahler potential \eqref{eq:CSmetricpattern}, thus leading to the following schematic behavior
\begin{equation}
	\mathcal{V}_{\Gamma} \sim \left \lbrace (z^i)^{-3/2}, (z^i)^{-1/2}, (z^i)^{1/2}, (z^i)^{3/2} \right\rbrace\, .
\end{equation}
Therefore, it becomes clear that the relevant set of asymptotically light states are linked to the fastest shrinking/growing 3-cycles, namely the one associated to the reference period (i.e. $A_0$) and its symplectic dual ($B^0$), respectively. These determine the KK scale, which behaves as follows (we henceforth set $e^1=1$ for concreteness)
\begin{equation}\label{eq:KKSYZ}
	\left(\frac{m_{\text{KK},\, B^0}}{M_{\text{Pl;}\, 4}} \right)^2= \frac{1}{\mathcal{V}_{B^0}^{2/3}} \left(\frac{m_s}{M_{\text{Pl;}\, 4}} \right)^2 \sim \frac{1}{\sigma}\, ,
\end{equation}
and the tension of the dual Type IIA string arising from a D4-brane wrapping the reference $A_0$-cycle (see discussion after \eqref{eq:KKCYzetavector})
\begin{equation}\label{eq:D4SYZ}
	\left(\frac{T_{\text{D4}}}{M_{\text{Pl;}\, 4}^2} \right)= \frac{\mathcal{V}_{A_0}}{g_s} \left(\frac{m_s}{M_{\text{Pl;}\, 4}} \right)^2 \sim \frac{1}{\sigma^{3/2}}\, .
\end{equation}
Notice that, since the KK tower \eqref{eq:KKSYZ} is parametrically heavier than the mass scale of the emergent dual Type IIA string, the limit thus explored is \emph{pathological}, as defined in \cite{Lee:2019wij} (see also \cite{Baume:2019sry, Alvarez-Garcia:2021pxo}), in the sense that upon approaching the singularity it seems that one can in principle retrieve a fundamental string in less than ten spacetime dimensions. Despite this abnormal behavior, the pattern \eqref{eq:pattern} seems to be nevertheless satisfied, as one can readily confirm:
\begin{align}\label{eq:patternviolation}
	\vec{\zeta}_{\rm{D4}} \cdot \vec{\mathcal{Z}}_{\text{D4}} &= \frac{\partial \log m_{\rm{D4}}}{\partial \text{Im}\, z^i}\, G^{ij}\, \frac{\partial \log \LSP}{\partial \text{Im}\, z^j} + \frac{\partial \log m_{\rm{D4}}}{\partial \varphi_4}\, G^{\varphi_4 \varphi_4}\, \frac{\partial \log \LSP}{\partial \varphi_4}\notag\\
	&=\frac{3}{8}+\frac{1}{8}=\frac{1}{2}\, ,
\end{align}
where we have defined $G_{i j} =2 G_{i \bar j}$ and we made use of the no-scale property of $K_{\text{cs}}$ close to the LCS point, which reads $K_i K^{i \bar j} K_{\bar j} = 3$. Similarly, for the scalar product between the lightest KK tower and the species scale one finds
\begin{align}\label{eq:patternviolationII}
	\vec{\zeta}_{\text{KK},\, B^0} \cdot \vec{\mathcal{Z}}_{\text{D4}} &= \frac{\partial \log m_{\text{KK},\, B^0}}{\partial \text{Im}\, z^i}\, G^{ij}\, \frac{\partial \log \LSP}{\partial \text{Im}\, z^j} + \frac{\partial \log m_{\text{KK},\, B^0}}{\partial \varphi_4}\, G^{\varphi_4 \varphi_4} \frac{\partial \log \LSP}{\partial \varphi_4}\notag\\
	&=\frac{1}{4}+\frac{1}{4}=\frac{1}{2}\, .
\end{align}
At this point, one would be tempted to conclude that the pattern \eqref{eq:pattern} also seems to hold for the hypermultiplet sector in $\mathcal{N}=2$ theories. However, as already mentioned, such moduli space receives strong quantum corrections, such that it is not clear at all whether the conclusions drawn from the present classical analysis will survive after taking into account perturbative and non-perturbative $g_s$--\,corrections. In the following, we will argue (building on earlier works in the topic \cite{Marchesano:2019ifh, Baume:2019sry, Alvarez-Garcia:2021pxo}), that the effect of including such quantum corrections is to correct the pathological behavior exhibited in  eqs. \eqref{eq:KKSYZ} and \eqref{eq:D4SYZ}, while ensuring that the pattern is still fulfilled.

\subsubsection{Non-perturbative corrections}
\label{sss:instantons}
	
As explained in \cite{Marchesano:2019ifh, Baume:2019sry}, the reason why the previous classical analysis is incomplete hinges on the presence of large quantum corrections which had been ignored so far. Such quantum effects arise from Euclidean D2- as well as NS5-brane instantons, and when taken into account, they may strongly modify the tree-level hypermultiplet metric displayed in eq. \eqref{eq:classicalhypermetric}. In fact, the classical LCS singularity above gets heavily corrected and is traded at the quantum level for another infinite distance degeneration, now at weak 4d string coupling. However, a careful analysis of these matters becomes rather intricate, requiring moreover from the introduction of several new tools. Therefore, in order to not complicate unnecessarily the main discussion in this section, we summarize here the upshot and the main intuition behind it, leaving the details for Appendix \ref{ap:hypermetric} (see in particular the discussion in Section \ref{ss:detailshyper}). 
	
The argument goes as follows. One can indeed exploit the $\mathsf{SL(2,\mathbb{Z})}$ symmetry that the hypermultiplet moduli space enjoys (even at the quantum level) to translate any limit of the form \eqref{eq:generictraj} into a \emph{dual} one at weak string coupling and fixed complex structure moduli. Hence, it suffices to know how the weak coupling point is affected by the aforementioned quantum corrections. Fortunately, we do not expect neither perturbative nor non-perturbative effects to play any important role at weak coupling, since those should be suppressed along the limit $g_4=g_s \mathcal{V}_{X_3}^{-1/2} \to 0$. This can be confirmed upon looking at how the exact moduli space metric deviates from the tree-level one. Indeed, there are additional terms which at leading order behave as follows \cite{Gunther:1998sc,Becker:1995kb} (see Appendices \ref{ss:exactmetric} and \ref{ss:detailshyper} for details) 
\begin{equation}
	\delta  \dd s_{\rm HM}^2 =  \delta  \dd s_{\rm HM}^2\rvert_{\text{1-loop}} + \delta  \dd s_{\rm HM}^2\rvert_{\text{D-inst}}\, \sim\, g_4^2\, +\, \sum_{\gamma} \Omega_{\gamma}\, e^{-S_{m,\, k_I}} \to 0\, ,
\end{equation}
where the sum runs over all (towers of) D2-brane instantons with action denoted by $S_{m,\, k_I} \sim \frac{1}{g_4}$ (c.f. eq. \eqref{eq:D2instantonaction}). Hence, it is enough to use the classical approximation \eqref{eq:classicalhypermetric} for all practical purposes here, such that we conclude that the calculations performed after \eqref{eq:fundstringmass} remain valid, and the pattern is still verified for all trajectories of the form \eqref{eq:generictraj}.		
	
\subsubsection{Intertwining the vector and hypermultiplet sectors}
\label{sss:mixedlimits}
	
Finally, let us briefly consider the possibility of taking limits which imply moving both in the vector and hypermultiplet moduli spaces. As a representative example, we analyze in what follows the large volume limit at fixed 10d string dilaton, corresponding to decompactification from 4d to 10d Type IIA supergravity. In terms of the appropriate 4d variables, we send $\mathcal{V}_{X_3}\to\infty$ and, consequently, $\varphi_4=\phi-\frac{1}{2}\log\mathcal{V}_{X_3} \to -\infty$. This means, in particular, that the string mass becomes light in 4d Planck units
\begin{equation}
	m_s = (4\pi)^{-1/2}\, M_{\rm Pl;\, 4}\, e^{\varphi_4} =  (4\pi)^{-1/2}\, M_{\rm Pl;\, 4}\, e^{\phi}\, \mathcal{V}_{X_3}^{-1/2}\to 0\;.
\end{equation}
Furthermore, for such a decompactification limit, the overall KK tower becomes asymptotically massless at a faster rate,
\begin{equation}
	m_{\rm KK,\, 6} = m_s\, \mathcal{V}_{X_3}^{-1/6}= (4\pi)^{-1/2}\, M_{\rm Pl;\, 4}\, e^{\varphi_4} \,\mathcal{V}_{X_3}^{-1/6} = (4\pi)^{-1/2}\, M_{\rm Pl;\, 4}\, e^{\phi}\, \mathcal{V}_{X_3}^{-2/3}\to 0\,,
\end{equation}
so that it corresponds to the leading tower, since the D0/D2-brane towers are slightly/much heavier than $m_{s}$ for the limit at hand. Regarding $\LSP$, we note that the 10-dimensional Planck mass scales asymptotically like the string scale,
\begin{equation}
	M_{\rm Pl;\, 10} = (4\pi)^{1/8}\, m_s\, e^{-\phi/4} = (4\pi)^{-3/8}\, M_{\rm Pl;\, 4}\, e^{\frac{3}{4}\varphi_4}\, \mathcal{V}_{X_3}^{-1/8} = (4\pi)^{-3/8}\, M_{\rm Pl;\, 4}\, e^{\frac{3}{4}\phi}\, \mathcal{V}_{X_3}^{-1/2}\, ,
\label{Mpl104d}
\end{equation}
so that we conclude that the species scale is set by the string scale. Therefore, sticking to the $\{\varphi_4,\mathcal{V}_{X_3}\}$ basis, one obtains
\begin{equation}
	\vec{\zeta}_{\rm t}=\vec{\zeta}_{\rm KK,\, 6}=\left(-1,\frac{1}{6\mathcal{V}_{X_3}}, 0\, , \ldots, 0\right),\qquad \vec{\mathcal{Z}}_{\rm sp}=\vec{\mathcal{Z}}_{\rm osc}= \left(-1, 0\, , \ldots, 0\right)\, ,
\end{equation}
for the charge-to-mass and species vectors, such that upon using the relevant metric components it can be readily checked that indeed 
\begin{equation}
	\vec{\zeta}_{\rm t}\cdot \vec{\mathcal{Z}}_{\rm sp}=\frac{1}{d-2}=\frac{1}{2}\, ,
\end{equation}
in agreement with \eqref{eq:pattern}.\footnote{This particular limit is analogous to the large volume limit of a toroidal decompactification. It is then also verified that $\vec{\zeta}_{\rm KK,\, 6}\cdot \vec{\mathcal{Z}}_{\rm Pl,\, 10}=\frac12$ with $\vec{\mathcal{Z}}_{\rm Pl,\, 10}= \left(-\frac{3}{4},\frac{1}{8\mathcal{V}_{X_3}}, 0\, , \ldots, 0\right)$, as derived from \eqref{Mpl104d}.}
	
In general, one can take several combinations of limits involving moduli from both sectors of the 4d $\mathcal{N}=2$ moduli space, resulting in different microscopic interpretations of the singularities. Some of them will be subjected to strong quantum corrections, as previously discussed, but nonetheless we expect the pattern to be satisfied in all such cases, as they will simply correspond to combinations of the building blocks already discussed.

\section{On the quest for a bottom-up rationale}
\label{s:bottomup}
		
In the previous sections we have provided significant evidence for the asymptotic constraint \eqref{eq:pattern} in string theory compactifications. This pattern provides a very sharp relation between the growth of the density of states and the rate at which they become light at infinite distance: The more dense the tower is, the slower the mass goes to zero. A natural question to ask at this point would be whether this pattern is a lamppost effect of the string theory landscape or rather some general feature of quantum gravity. To answer this, we need to provide some bottom-up explanation for the latter, independently of string theory. While we do not have yet such argument (e.g., based on black hole physics), we are able to identify and motivate some sufficient conditions that allow the pattern to hold in a general way.

\subsubsection*{Three sufficient conditions}

To start with, let us note that the Distance Conjecture \cite{Ooguri:2006in} already ensures that the mass of the leading tower --- and consequently the species scale --- decreases \emph{exponentially} with the moduli space distance $\Delta_{\phi}$ when approaching some infinite distance boundary in field space, c.f. Section \ref{s:SDC} for details. This can be further motivated from a bottom-up perspective by the Emergence Proposal which, as explained in Chapter \ref{ch:Emergence}, states that all the IR dynamics in quantum gravity emerges from integrating out the dual massive degrees of freedom. From this, it follows automatically that the product $\vec{\zeta}_{\rm t}\cdot \vec{\mathcal{Z}}_{\rm sp}$ must approach some constant asymptotically, but not necessarily the same `universal' one for all infinite distance limits. Such constant seems to be a priori model-dependent, and it is somewhat surprising that we always obtain the same value in string theory. To argue for this, we propose three \emph{sufficient} conditions which together ensure that \eqref{eq:pattern} is fulfilled along any asymptotic direction. 	
	
\begin{center}
	\textbf{Condition 1}: \textit{The exponential rates $\{\lambda_I\}$ of the different towers $\{m_I\}$ are continuous over the asymptotic regions where they are defined. Furthermore, $\vec{\zeta}_{\rm t}\cdot\vec{\mathcal{Z}}_{\rm sp}$ must be well defined along any asymptotic direction}. 
\end{center}

This means that the exponential rate $\lambda_{\rm t}=\hat{T}\cdot\vec{\zeta}_{\rm t}$ of the leading tower is purely determined by the asymptotic direction $\hat{T}$, regardless of the particular geodesic we follow towards it. This does not require $\vec{\zeta}_{\rm t}$ to remain constant along parallel trajectories, being allowed to change or \emph{slide} in the components perpendicular to $\hat{T}$.\footnote{Note that, given two parallel trajectories reaching infinity in moduli space, if the tower becomes asymptotically light with different exponential rates for each of them, then $m_{\rm t}$ would take \emph{parametrically} distinct values between points separated by some finite distance. To avoid this, $\lambda_{\rm t}$ should remain constant along parallel trajectories.} It implies, though, that the change in $\vec{\zeta}_{\rm t}$ has to be seen as a discrete \emph{jump} in terms of the asymptotic direction. This can occur either because: \emph{(i)} the microscopic interpretation of the leading tower changes as a different tower starts dominating, in which case the decay rate for both towers automatically coincide in the transition region and $\lambda_{\rm t}$ is continuous, or \emph{(ii)} because a complicated moduli dependence of the mass makes $\vec{\zeta}_{\rm t}$ to \emph{jump} when crossing some \emph{sliding loci}  (see Section \ref{ss:het s1} for a detailed example in Heterotic string theory). In this latter scenario, we further need to require that $\vec{\zeta}_{\rm t}\cdot\vec{\mathcal{Z}}_{\rm sp}$ remains well-defined, otherwise the product will depend on the specific trajectory that is chosen. Hence, a consequence of \emph{Condition 1} is that we can divide the set of infinite distance limits into regions over which the vectors $\vec{\zeta}_{\rm t}$ and $\vec{\mathcal{Z}}_{\rm sp}$ take some definite expressions, such that their product is indeed constant. 

\begin{center}
	\textbf{Condition 2}: \textit{For every infinite distance limit along which several towers decay at the same rate, there must exist bound states involving all of them, such that the species scale is given by the associated multiplicative species}. 
\end{center}

Consider several towers $\{m_1,\ldots,m_k\}$ becoming light at the same rate along some trajectory (or interface) with unit tangent vector $\hat{T}$, so that $\lambda_{\rm t}=\hat{T}\cdot\vec{\zeta}_{1}=\ldots=\hat{T}\cdot\vec{\zeta}_{k}$. These towers a priori span a lattice of states labeled by the quantum numbers $(n_1,\ldots,n_k)$, with a spectrum of the form $m_{i,\, n_i} = n_i^{1/p_i}\, m_i$ for each individual tower. Now, as explained in Chapter \ref{ch:SpeciesIntro}, if there exists a (sub-)lattice which is populated by particle states, then the total number of species is \emph{multiplicative}, namely it behaves as $N\sim \prod_{i=1}^k N_i$ when $N_i \gg1$. Therefore, its associated species scale can be effectively computed as follows
\begin{equation}\label{eq:multspecies}
	\Lambda_{\rm sp}=M_{\text{Pl;}\, d}\left(\frac{M_{{\rm Pl;}\,d}}{m_{\rm eff}}\right)^{-\frac{p_{\rm eff}}{d-2+p_{\rm eff}}}\, ,
\end{equation}	
in terms of an effective mass and density parameters
\begin{equation}\label{e:eff tow}
	m_{\rm eff}=(m_1^{p_1} \ldots m_k^{p_k})^{1/p_{\rm eff}},\qquad p_{\rm eff}=\sum_{i=1}^kp_i\; .
\end{equation}
In that case, the resulting species vector is given by
\begin{align}\label{e: Z mult}
	\vec{\mathcal{Z}}_{\rm sp}&=\frac{1}{d-2+p_{\rm eff}}\sum_{i=1}^k(d-2+p_i)\vec{\mathcal{Z}}_I=\notag\\
	&=\frac{p_{\rm eff}}{d-2+p_{\rm eff}}\vec{\zeta}_{\rm eff}\perp{\rm Hull}(\{\vec{\zeta}_1, \ldots ,\vec{\zeta}_k\})\; ,
\end{align}
which is moreover orthogonal to the hull spanned by the $\zeta$-vectors and dominates over the individual species scales (c.f. Chapter \ref{ch:bounds}). This implies that $\vec{\zeta}_1\cdot \vec{\mathcal{Z}}_{\rm sp}=\ldots= \vec{\zeta}_k\cdot \vec{\mathcal{Z}}_{\rm sp}$, such that the product \eqref{eq:pattern} takes the same value in every different adjacent region (as well as in the interface). For \emph{additive} species, though, we do not obtain any additional species vector, and thus $\vec{\zeta}_{\rm t}\cdot \vec{\mathcal{Z}}_{\rm sp}$ would generically change upon crossing the interfaces, see Figure \ref{fig:add mult} below. This is why \emph{Condition 2} requires the existence of the (sub-)lattice of bound states yielding a multiplicative number of species, which can be further motivated by Swampland considerations such as the Completeness Hypothesis \cite{Banks:2010zn,Polchinski:2003bq}, that applies e.g., when the quantum numbers $\{n_k \}$ correspond to gauge charges under some massless $\mathsf{U(1)}$ gauge fields.\footnote{Notice that the scenario of additive species would also result in independent towers of states becoming light at the same rate, naively implying different massless gravitons asymptotically, which goes against general Swampland expectations \cite{Bedroya:2023tch,Kim:2019ths}.}

\begin{figure}[htb]
	\begin{center}
        \subfigure[]{
		\includegraphics[width=0.4\textwidth]{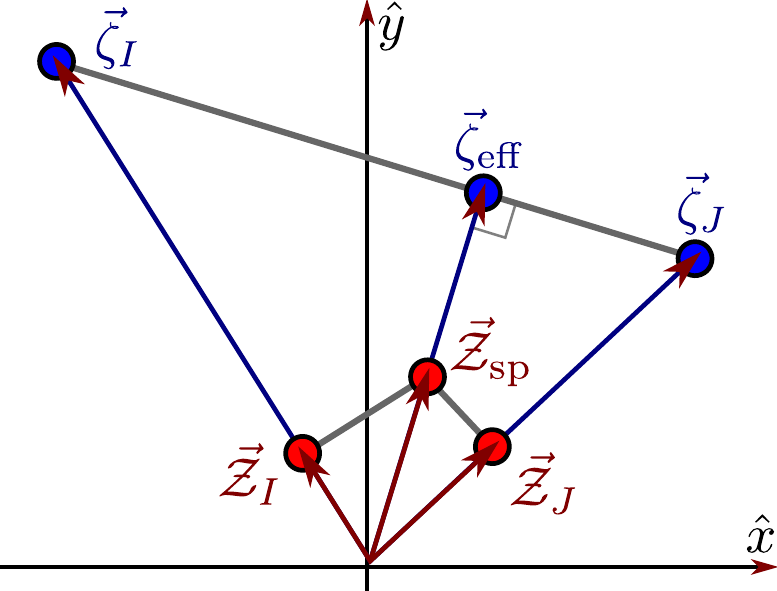}\label{fig:mult}
        }
        \quad
	\subfigure[]{
		\includegraphics[width=0.3\textwidth]{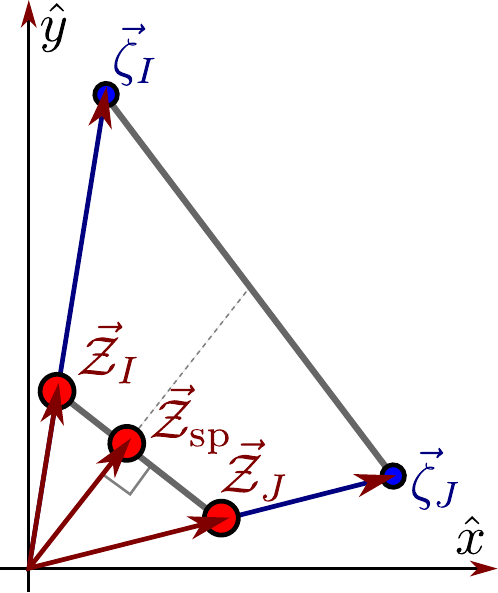}\label{fig:add}
        }
	\caption{\small Sketch of the possible behaviour of the species scale $\Lambda_{\rm sp}$ along limits for which two (or more) leading towers become light at the same rate. Whenever there exists an effective tower $\vec{\zeta}_{\rm eff}$ of bound states, the associated \emph{multiplicative species} \textbf{(a)} dominates over the individual $\vec{\mathcal{Z}}_I$ and $\vec{\mathcal{Z}}_J$, and is perpendicular to the facet spanned by the individual towers, ${\rm Hull}(\{\vec{\zeta}_I,\vec{\zeta}_J\})$. On the contrary, if the effective towers are absent, the resulting \emph{additive species} \textbf{(b)} $\vec{\mathcal{Z}}_{\rm sp}$ is associated to the sum of the states given by each tower alone, being moreover perpendicular to ${\rm Hull}(\{\vec{\mathcal{Z}}_I,\vec{\mathcal{Z}}_J\})$ and only providing for the actual cut-off when both individual species fall at the same rate. In this case, $\vec{\mathcal{Z}}_{\rm sp}$ is not expected in general to be orthogonal to ${\rm Hull}(\{\vec{\zeta}_I,\vec{\zeta}_J\})$.}
		\label{fig:add mult}
	\end{center}
\end{figure}

\begin{center}
	\textbf{Condition 3}: \textit{For every connected component of the space of infinite distance limits, there exists at least one direction associated to an emergent string limit or the homogeneous decompactification of an internal cycle to a higher dimensional vacuum}. 
\end{center}

With the previous two conditions, we have divided the moduli space into different regions and shown that  $\vec{\zeta}_{\rm t}\cdot\vec{\mathcal{Z}}_{\rm sp}$  remains constant across those. The only thing missing is to set this constant to  $\frac{1}{d-2}$, which occurs if there exists \emph{at least one} asymptotic direction resulting in a string perturbative limit or a decompactification to a higher dimensional vacuum. This resembles but it is a weaker condition than the Emergent String Conjecture \cite{Lee:2019wij}, as we explain in the following.

\subsubsection*{Relation to Emergent String Conjecture}

To conclude, we want to comment on the relation between the pattern \eqref{eq:pattern} and the Emergent String Conjecture (see Section \ref{s:SDC}), since they are clearly linked and one might wonder to what extent the former follows from the latter, and viceversa. In a nutshell, this conjecture holds that every infinite distance limit should either correspond to a decompactification to higher dimensions or to a perturbative limit where a critical string becomes weakly coupled and tensionless. Therefore, other potential descriptions where the lightest object corresponds to a higher-dimensional $p$-brane (with $p \geq 2$) would be thus forbidden, which has been argued to be related to the consistency of the conjecture under dimensional reduction \cite{Alvarez-Garcia:2021pxo}.
	
On the other hand, in the present section  we have identified some sufficient conditions that allow the pattern to hold universally in moduli space, so that we can compare them directly with the Emergent String Conjecture. \emph{Condition 1} does not follow from the latter, since it is actually a requirement on the asymptotic structure of the towers and how the $\zeta$- and $\mathcal{Z}$-vectors are allowed to change as we move within moduli space. \emph{Condition 3} clearly follows from it, even though it is a weaker statement. The interesting connection, however, is associated to \emph{Condition 2}, which is the most important feature underlying the pattern. A priori, it is not obvious whether the Emergent String Conjecture implies such condition, or why the latter requirement is stronger, as we explain in the following. Consider for instance some decompactification limit in which we have several Kaluza-Klein towers so that several directions open up asymptotically. If all these towers are truly Kaluza-Klein towers from the perspective of the \emph{same} duality frame, then it is guaranteed that we will populate the lattice of KK quantum numbers and thus satisfy \emph{Condition 2}. This is because for very large momenta, one can use the WKB approximation to compute the eigenvalues of the laplacian of the internal space, and they are such that the number of modes with mass smaller than or equal to some large energy $\Lambda_{\rm UV}$ scales roughly as follows
\begin{equation}
	N \sim \left( \frac{\Lambda_{\rm UV}}{m_{\rm KK}}\right)^n\, ,
\end{equation}
with $n$ being equal to the total number of decompactifying dimensions. Note that this has precisely the structure of the \emph{effective tower} (c.f. eq. \eqref{e:eff tow}) in the multiplicative species scenario, such that \emph{Condition 2} holds. However, the Emergent String Conjecture does not require a priori that the individual limits associated to each tower can be interpreted as decompactification points from the perspective of the same dual frame. For instance, in the case in which we take a limit along which a KK tower decays at the same rate than a tower of winding modes, even if both towers signal a decompactification limit towards some dual frame, they do not do so within the same duality description and therefore the total number of species is actually additive. We denote this as a case of \emph{non-compatible} decompactification limits. Hence, if we only had these two separate towers, we would not get a lattice of bound states thereof such that \emph{Condition 2} --- and consequently the pattern --- would not hold. However, in practice, whenever this scenario occurs in string theory, we always get additionally a tower of string oscillator modes precisely along the direction where the KK and winding modes decay at the same rate, so that we realize an emergent string limit (rather than decompactifying two extra dimensions), ensuring that \eqref{eq:pattern} is satisfied. This seems to be always the case even in more complicated top-down constructions, where we are not simply considering circle decompactifications and we do not have winding modes of a perturbative string but rather towers of particles coming from wrapped branes. Nevertheless, even in those cases, the rich network of string dualities always allow us to identify some critical string becoming tensionless along the interface between the different  \emph{non-compatible} decompactification limits. We want to remark that this is indeed crucial for the pattern to hold, and from a bottom-up perspective, it imposes a non-trivial constraint on how the different infinite distance limits glue together within the moduli space.
	
Therefore, if we interpret the Emergent String Conjecture as the milder claim that the leading tower must be either a Kaluza-Klein one --- in some dual frame --- or an emergent string, then it does not immediately imply \emph{Condition 2} and is strictly weaker than the pattern. For instance, the above scenario of \emph{non-compatible} decompactification limits would still be consistent with this mild version of the conjecture even if we did not have the string becoming tensionless at the interface. However, if we interpret the Emergent String Conjecture as the claim that there must be either a single dual frame where all the leading towers can be seen as KK towers or we get an emergent string providing for the leading one, then it automatically implies \emph{Condition 2}. In that case, the pattern would essentially follow from it, barring some subtleties related to the sliding of the $\zeta$-vectors, c.f. \emph{Condition 1} above. This latter possibility would be very interesting, as the pattern would then open new avenues to try to provide a bottom-up explanation for the conjecture itself, which so far has only been motivated by string theory examples.\footnote{See \cite{Basile:2023blg,Basile:2024dqq,Bedroya:2024ubj} however for recent efforts in trying to address this point.} 
In fact, using recent results \cite{Bedroya:2024uva} which argue that asymptotically the mass scale of the lightest tower $m_{\rm t}$ can be detected as well by (neutral) black holes which undergo some phase transition, one may rewrite \eqref{eq:pattern} equivalently as follows
\beq\label{eq:patternblackholes}
	\frac{\vec\nabla \Lambda_{\rm BH}}{\Lambda_{\rm BH}} \cdot\frac{\vec\nabla \LSP}{\LSP}= \frac{1}{d-2}\, ,
\eeq
where $\ell_{\rm BH} = \Lambda_{\rm BH}^{-1}$ defines the size of the corresponding singular black hole and is such that $\ell_{\rm BH} \geq \ell_{\rm sp}$. In realistic examples taken from the quantum gravity landscape, this transition typically coincides with the one described by Gregory and Laflamme \cite{Gregory:1993vy,Gregory:1994bj} (for decompactification limits) or rather with the Horowitz-Polchinski solution \cite{Horowitz:1996nw,Horowitz:1997jc} (in the emergent string case). Hence, a bottom-up argument linking the variation over the moduli space of these two energy scales or equivalently their entropies, namely
\beq\label{eq:patternentropies}
	\vec\nabla \log S_{\rm BH} \cdot \vec \nabla S_{\rm BH,\, min}= d-2\, ,
\eeq
may serve as a good starting point for addressing this important question in the future.

\section{Summary}

In this chapter we have pointed out an interesting relation that it is satisfied in all known examples of infinite distance limits in the moduli space of string theory compactifications, regardless of the level of supersymmetry or the topology/geometry of the internal space. This pattern moreover provides a sharp connection between the asymptotic value of the variation rate (in moduli space) of the species cut-off and the mass of the leading tower of states, given by \eqref{eq:pattern}. We checked that it holds for multi-field geodesic trajectories where several moduli are taken to infinity at the same time, even if the species scale is not only determined by the leading tower of states but captures information of other subleading ones. 
	
At the very least, it can be regarded as a common thread underlying all known string theory examples that have been explored so far, and makes manifest the very constrained structure behind the vast casuistics of different types of infinite distance degenerations, as well as how they can fit together in a given moduli space. We suspect, though, that the universality of a relation like \eqref{eq:pattern} is rooted in a deeper underlying quantum gravity principle, rather than being just a lamppost effect of all known string constructions. Hence, an important step forward in our understanding of the pattern would be to search for a purely bottom-up rationale that could explain the latter independently of string theory. Promising avenues along this direction include phrasing the problem in terms of black holes or holographic entropy bounds (c.f. eqs. \eqref{eq:patternblackholes} and \eqref{eq:patternentropies}), since the pattern seems to relate two different special scales in black hole physics \cite{Bedroya:2024uva}. Alternatively, one could also think of the number of species as a measure of the density of states in any theory coupled to Einstein gravity, so that the less dense the tower is, the faster it can become light according to \eqref{eq:patternN}. Another interesting direction would be to use S-matrix bootstrap techniques, since the species cut-off can be understood as the scale at which the semi-classical Einstein gravity description breaks down and higher-derivative terms start dominating over the tree-level Einstein term, see Chapter \ref{ch:Higherdimops}. It would be fantastic if one could derive a precise link between $\LSP$ and e.g., the scale of the first massive spin-2 field of some Kaluza-Klein tower.
	
In a similar vein, one could argue that in fact finding the physics behind the pattern would presumably have profound consequences for the Swampland program, since it implies a refined formulation of the Distance conjecture that constrains the nature of the towers and imposes a sharp bound on how fast it becomes light. Indeed, if the pattern holds then it automatically implies a lower bound on the exponential rate of the tower given by $\lambda_{\text{t}} \geq \frac{1}{\sqrt{d-2}}$, which supports the idea put forward in \cite{Etheredge:2022opl} and it is closely related to the Emergent String Conjecture \cite{Lee:2019wij}. Relatedly, the relation \eqref{eq:pattern} also constrains the exponential decay rate of the species cut-off, whose convex hull condition (c.f. also Chapter \ref{ch:bounds}) becomes dual --- in the polytope sense --- to the one imposed on the towers. Furthermore, it provides a clear recipe to determine the species scale upon knowledge of the leading tower of states along different directions. It would be equally interesting to explore how it could be extended to the interior of the moduli space, where the notion of a leading tower of states is no longer well-defined.\footnote{See \cite{Rudelius:2023spc,Bedroya:2024uva} for recent attempts along this direction.}
	
As a byproduct, we identified in Section \ref{s:bottomup} three sufficient conditions that the towers of states and the asymptotic geometry in moduli space must satisfy to allow for the pattern to hold. Interestingly, the most important condition resembles a sort of (sub-)Lattice WGC where the role of the gauge charges is played by the levels of the towers. This condition also follows from a strong interpretation of the Emergent String Conjecture. Hence, many ideas in the Swampland program get interconnected and can be re-derived from this simple equation relating the variation of the species scale and the leading tower of states becoming light asymptotically.





\part{\scshape  \textcolor{MyDarkRed}{Conclusions}} 
\label{part:Conclussions}
\chapter{Final Comments} \label{ch:Conclusions}\thispagestyle{simple}

The main theme of this thesis has been the investigation of the regime of validity associated with any effective field theory weakly coupled to Einstein gravity. This is encapsulated in the concept of the \emph{quantum gravity cut-off} $\LQG$, defined as the energy scale beyond which quantum-gravitational effects can no longer be neglected, thereby invalidating the original effective field theory description. The significance of this scale is twofold: on one hand, it is intimately linked with physical phenomena associated with the underlying UV completion of gravity, such as extra dimensions or fundamental vibrating strings. On the other hand, from a modern perspective, it is also crucial for studying how different effective descriptions of gravitational interactions at low energies deal with non-trivial infra-red constraints, such as the non-existence of exact global symmetries in the theory. The rigorous study of these issues constitutes the central quest of the Swampland program \cite{Vafa:2005ui}, for which a detailed understanding of the cut-off scale $\LQG$ could thus have a significant impact.


After reviewing and introducing in Part \ref{partI} the main theoretical tools employed in this thesis, we turned in Part \ref{part:QGscale} to the core ideas discussed in this work. Based on our experience from other non-renormalizable field theories, we concluded that the most natural candidate for $\LQG$ should be the Planck scale $\Mpd$. This is precisely the energy scale associated with the gravitational coupling constant itself, namely Newton's constant $G_N$, and it signals the point where strong back-reaction effects are to be expected, giving rise to extreme phenomena in gravity such as the formation of black holes. However, as originally discussed in \cite{Han:2004wt, Dvali:2007hz, Dvali:2007wp}, this naive picture seems to fail in the presence of a large number of light degrees of freedom $N$. Indeed, using different theoretical arguments invoking black hole physics, non-perturbative (i.e. holographic) considerations as well as perturbation theory of the graviton state, one can argue for a \emph{species scale} $\LSP$ instead as the relevant energy cut-off in semi-classical gravity. Crucially, this scale is sensitive to the aforementioned number of degrees of freedom, is bounded from above by $\Mpd$ — coinciding with the latter when $ N= \mathcal{O}(1)$ — and can be parametrically lowered (when measured in Planck units) in the presence of a large number of species. In Chapter \ref{ch:SpeciesIntro}, we also investigated the behavior of $\LSP$ close to infinite distance and weak coupling regions within the EFT, where this number $N$ grows exponentially, as per the Distance \cite{Ooguri:2006in} and Weak Gravity conjectures \cite{Heidenreich:2015nta, Heidenreich:2016aqi, Montero:2016tif, Andriolo:2018lvp}. There, it was found precise agreement with our expectations arising from Kaluza-Klein theories of gravity with extra dimensions and string perturbation theory, where $\LSP$ is given either by the higher-dimensional Planck mass or the fundamental string scale, respectively. Furthermore, a completely general bottom-up algorithm was provided to compute the species scale in the presence of several towers of states becoming light.

Subsequently, in Part \ref{part:StringTheoryTests}, we tested the general picture advocated in Chapter \ref{ch:SpeciesIntro} using various concrete string theory constructions. More specifically, in Chapter \ref{ch:Higherdimops} we studied the behavior of certain supersymmetric operators arising in diverse string theory compactifications, which involve curvature invariants with mass dimension greater than two. This allowed us to elucidate the energy scale suppressing these operators with respect to the tree-level kinematics of gravity, i.e. the Einstein-Hilbert term. Interestingly, it was found that the corresponding cut-off turns out being the aforementioned species scale, whose physical interpretation changes depending on the asymptotic corner of the theory that we probe. However, several relevant observations were made. First, it was shown, upon studying further BPS terms in the supergravity action, that the EFT expansion in terms of a unique clear-cut scale $\LSP$ only arises close to infinite distance boundaries, where the classical dimension of the operators provides a good approximation and the quantum corrections (e.g., anomalous dimensions) are rendered parametrically small. Second, we noticed that along decompactification limits, the scale suppressing the higher-curvature operators was sometimes given by the Kaluza-Klein scale instead of the quantum gravity cut-off. This occurs when the finite threshold contributions to the operator under study dominate over the ‘bare’ suppression by $\LSP$, which happens precisely when the latter is irrelevant (in the Wilsonian sense) already in the higher-dimensional theory. Crucially, for energies well above the Kaluza-Klein scale, the non-local threshold effects induced by the tower get washed away (which usually requires a non-trivial resummation procedure), and the only surviving suppression corresponds to that controlled by $\LSP$, as expected.

In Chapter \ref{ch:Emergence}, we investigated the precise role of the species scale within an intriguing conjecture in quantum gravity usually referred to as the \emph{Emergence Proposal} \cite{Harlow:2015lma, Grimm:2018ohb, Heidenreich:2018kpg, Palti:2019pca}. This conjecture holds that the kinematics of the low energy degrees of freedom entering the EFT (including the graviton itself), emerge upon integrating out the dual infinite number of heavy modes in the UV complete theory. Our aim was to elucidate whether this proposal is respected — in its weakest versions — by the numerous string theory constructions available in the literature. Interestingly, we found that in this regard it seems crucial to identify the physical cut-off of the EFT with the species scale, thereby ensuring that the proposal is verified at leading order, regardless of the nature of the infinite distance limit, the number of non-compact dimensions and the amount of supersymmetry preserved. 


Finally, in Part \ref{part:pattern}, we focused on the investigation of universal constraints exhibited by $\LSP$ close to the infinite distance boundaries in field space. As a result, we were able to motivate a lower bound on the exponential decay rate $\lambda_{\rm sp}$ of the species cut-off within these regimes, which presents various interesting features singling it out from other potential candidates, such as its preservation under dimensional reduction, and the explicit non-trivial verification in maximally supersymmetric theories. Furthermore, inspired by certain hidden symmetries exhibited by the convex hull diagrams constructed from the decay rates of the towers and species cut-offs within particular string theory examples, we proposed and analyzed in detail in Chapter \ref{ch:pattern} a certain asymptotic equality relating the variation of both quantities over field space. This relation, which we dubbed \emph{the pattern}, has strong implications for the asymptotic behavior of the decay rates as well as for the possible duality phases that can be glued together within a given quantum gravity theory. Accordingly, we thoroughly investigated using a range of string theory constructions in various dimensions and with different amounts of supersymmetry, how this pattern is non-trivially satisfied at each possible infinite distance boundary of moduli space, paying special attention to the global structures emerging at infinity. This also allowed us to extract a minimal set of requirements which, when imposed as bottom-up constraints, directly imply the fulfillment of the pattern. This is significant for various reasons. First, it would be intriguing to understand the physics behind this constraint, potentially using black hole or unitarity arguments. Second, given the intimate connection with the Emergent String Conjecture \cite{Lee:2019wij}, it opens a pathway to argue for the latter from a bottom-up perspective, something that has been lacking since its conception.

Overall, the work presented in this thesis provides new insights into both the nature and importance of the quantum gravity scale, which is linked to the holographic principle and can thus be motivated and understood regardless of explicit UV completions of gravity, such as string theory. While it ultimately agrees with the behavior expected from a top-down perspective, the fact that it can already be detected in the low energy realm suggests that it may be intimately related to non-trivial infra-red constraints that gravitational EFTs must feature. Therefore, a sharp and complete understanding of this scale, as well as any possible universal constraint exhibited by the latter, provides a useful and fruitful approach to understanding quantum gravity from a low energy point of view, as well as to deduce possible phenomenological consequences that are not apparent from the field theory perspective. We thus believe that the present work can serve to motivate further investigations on this exciting topic, uncovering new connections and consequences that may teach us valuable lessons about Nature.


\chapter*{Comentarios Finales}
\selectlanguage{spanish}\thispagestyle{simple}

El tema principal de esta tesis ha consistido en la investigación del régimen de validez asociado a cualquier teoría de campos efectiva acoplada débilmente a la gravedad de Einstein. Este último quedaría encapsulado en el concepto de la \emph{escala de gravedad cuántica} $\LQG$, definida como la escala de energía más allá de la cual los efectos cuántico-gravitacionales se vuelven significativos y no pueden ignorarse de forma sistemática, invalidando así la descripción original en términos de una teoría de campos efectiva. La importancia de esta escala es a su vez doble. Por un lado, resulta estar íntimamente ligada a fenómenos físicos asociados a la compleción en el ultravioleta de la gravedad, que incluiría elementos nuevos como la existencia de dimensiones extra o de objetos extendidos en la teoría. Por otro lado, desde una perspectiva quizá más moderna, también resulta ser crucial para estudiar cómo diferentes descripciones efectivas de las interacciones gravitacionales implementan las restricciones infrarrojas que gravedad cuántica implicaría a bajas energías, como la no existencia de simetrías globales exactas en la teoría. El estudio riguroso de estos temas constituye la búsqueda central del programa de la Ciénaga \cite{Vafa:2005ui}, para el cual una comprensión detallada de la escala de gravedad cuántica $\LQG$ podría tener un impacto considerable.

En consecuencia, después de introducir en la Parte \ref{partI} las principales herramientas teóricas empleadas en esta tesis, procedemos en la Parte \ref{part:QGscale} a abordar las ideas centrales discutidas en este trabajo. Así, basándonos en nuestra experiencia con otras teorías de campos no renormalizables, obtenemos nuestro primer candidato para $\LQG$, que identificamos con la masa de Planck $\Mpd$. Esta cantidad se corresponde precisamente con la escala de energías asociada a la constante de acoplamiento gravitacional, es decir, la constante de Newton $G_N$, y marca el punto de inflexión donde los efectos gravitacionales se vuelven dominantes, dando lugar a fenómenos extremos como la formación de agujeros negros. Sin embargo, como se discutió originalmente en \cite{Han:2004wt, Dvali:2007hz, Dvali:2007wp}, este dibujo plausible parece fallar cuando nuestra teoría bajo estudio presenta un gran número de grados de libertad $N$. De hecho, utilizando diferentes argumentos teóricos que involucran la física de agujeros negros, consideraciones no perturbativas así como teoría de perturbaciones del cuanto de la gravitación, puede argumentarse en favor de la denominada \emph{escala de especies} $\LSP$ como la escala de energía relevante en gravedad semi-clásica. Crucialmente, esta escala resulta ser sensible al mencionado número de grados de libertad, presenta una cota superior dada por $\Mpd$ --- coincidiendo con esta última cuando $N= \mathcal{O}(1)$ --- y además puede disminuir paramétricamente respecto a la escala de Planck en presencia de un gran número de especies. Asimismo, en el Capítulo \ref{ch:SpeciesIntro} investigamos el comportamiento exhibido por $\LSP$ cerca de puntos a distancias infinita así como regiones de acoplamiento débil dentro de la teoría efectiva, donde este número $N$ crece exponencialmente, según las conjeturas de Distancia \cite{Ooguri:2006in} y Gravedad Débil \cite{Heidenreich:2015nta, Heidenreich:2016aqi, Montero:2016tif, Andriolo:2018lvp}. Allí, se recupera una concordancia precisa con nuestras expectativas derivadas de las teorías de dimensiones extra (o de Kaluza-Klein) así como con la teoría perturbativa de supercuerdas, donde $\LSP$ es dado por la masa de Planck de mayor dimensión o la escala de la cuerda fundamental, respectivamente. Además, se propuso un algoritmo completamente general para el cálculo de la escala de especies en presencia de varias torres de estados que se vuelven ligeras.

Posteriormente, en la Parte \ref{part:StringTheoryTests}, tratamos de corroborar las ideas centrales defendidas en el Capítulo \ref{ch:SpeciesIntro} utilizando varias construcciones concretas dentro de teoría de cuerdas. Específicamente, en el Capítulo \ref{ch:Higherdimops}, estudiamos el comportamiento de ciertos operadores supersimétricos que aparecen en diversas compactificaciones de teoría de cuerdas, los cuales involucran invariantes de la curvatura cuya dimensión clásica es mayor que dos. Esto nos permitió dilucidar cuál es la escala de energía que suprime dichos operadores con respecto al término cinético de la gravedad. Curiosamente, se encontró que la energía así hallada se corresponde con la antes mencionada escala de especies, cuya interpretación física cambia dependiendo de la frontera asintótica de la teoría que investiguemos. Sin embargo, como consecuencia de este estudio se hicieron varias observaciones de carácter relevante. En primer lugar, se mostró al estudiar ciertos términos adicionales en la acción de supergravedad, que la expansión efectiva de la teoría en términos de una única escala $\LSP$ solo surge cerca de los límites a distancia infinita, donde la dimensión clásica de los operadores proporciona una buena aproximación y las correcciones cuánticas quedan paramétricamente suprimidas. En segundo lugar, notamos que a lo largo de los límites de descompactificación, la escala que suprime estos operadores a veces es dada por la escala de Kaluza-Klein en lugar de la propia de gravedad cuántica. Esto ocurre precisamente cuando las contribuciones cuánticas finitas inducidas por la torre ligera dominan sobre la supresión ‘neta’ dada por $\LSP$. Es crucial para la consistencia de la teoría, no obstante, que para energías muy superiores que la escala de Kaluza-Klein, dichos efectos cuánticos se desvanezcan, dejando pues como única supresión la correspondiente a $\LSP$.

En el Capítulo \ref{ch:Emergence}, investigamos el papel que la escala de especies juega dentro de la conocida como la \emph{Propuesta de Emergencia} \cite{Harlow:2015lma, Grimm:2018ohb, Heidenreich:2018kpg, Palti:2019pca}. Dicha conjetura sostiene que la cinemática de los grados de libertad a baja energía que ingresan en la teoría efectiva (incluyendo el propio cuanto de la gravitación), emerge al integrar un número infinito de modos pesados en la teoría completa. Nuestro objetivo fue dilucidar si esta propuesta es respetada --- en sus versiones más débiles --- por la gran cantidad de construcciones de teoría de cuerdas disponibles en la literatura. Curiosamente, encontramos que, en este sentido, parece crucial identificar la escala de gravedad cuántica con la escala de especies, asegurando así que la propuesta se verifique en todos los ejemplos estudiados en este trabajo, independientemente de la naturaleza del límite bajo consideración, el número de dimensiones no compactas así como la cantidad de supersimetría preservada.

Finalmente, en la Parte \ref{part:pattern}, nos enfocamos en la investigación de aquellas restricciones universales exhibidas por $\LSP$ cerca de los límites a distancia infinita en el espacio de módulos de la teoría. Como resultado, pudimos motivar un límite inferior sobre la tasa de decaimiento exponencial $\lambda_{\rm sp}$ que la escala de especies satisface dentro de estos regímenes. Además, inspirados por ciertas simetrías exhibidas por los diagramas convexos construidos a partir de las tasas de decaimiento tanto de las torres como de la escala de especies (usando ejemplos particulares en teoría de cuerdas), propusimos y estudiamos en el Capítulo \ref{ch:pattern} una sorprendente igualdad asintótica que relaciona la variación de ambas cantidades a lo largo del espacio de módulos. Esta relación, que denotamos como \emph{el patrón}, tendría fuertes implicaciones sobre el comportamiento asintótico de las tasas de decaimiento antes mencionadas, así como para las posibles fases de dualidad que puedan aparecer dentro de una teoría de gravedad cuántica dada. En consecuencia, utilizando una gran variedad de construcciones derivadas de teoría de cuerdas en diversas dimensiones y con diferente cantidad de supersimetría, investigamos a fondo cómo este patrón se verifica de manera no trivial en cada posible límite a distancia infinita del espacio de módulos, prestando especial atención a las complejas estructuras globales que emergen en el infinito. Esto también nos permitió extraer un conjunto mínimo de requisitos que, una vez impuestos como restricciones en la teoría, implican automáticamente el cumplimiento del patrón. Esto último resulta ser significativo por varias razones. Primero, sería interesante entender la física detrás de esta restricción, potencialmente utilizando argumentos de agujeros negros o de unitariedad. Segundo, dada la conexión íntima con la Conjetura de la Cuerda Emergente \cite{Lee:2019wij}, el patrón abriría una nueva vía para argumentar en favor de esta última desde una perspectiva infrarroja.

En general, el trabajo presentado en esta tesis proporciona nuevas ideas sobre la naturaleza e importancia de la escala de gravedad cuántica, la cual estaría vinculada al principio holográfico y, por tanto, puede ser motivada y entendida independientemente de la teoría subyacente de gravedad cuántica. Asimismo, si bien en última instancia esta escala de especies concuerda con el comportamiento esperado desde una perspectiva ultravioleta, el hecho de que pueda detectarse ya en el ámbito de bajas energías sugiere que puede estar íntimamente relacionada con restricciones infrarrojas no triviales que las teorías efectivas gravitacionales deben presentar. Por lo tanto, una comprensión precisa y completa de dicha escala, así como cualquier posible ligadura que esta haya de satisfacer, proporcionaría un enfoque útil para entender la gravedad cuántica desde un punto de vista efectivo, así como para deducir posibles consecuencias fenomenológicas que puedan derivar de forma no trivial de esta última.  

\selectlanguage{british}
\part{\scshape  \textcolor{MyDarkRed}{Appendices}} \label{part:appendices}
\appendix  
\renewcommand{\thesection}{\Alph{chapter}.\arabic{section}}
\chapter{Conventions}
\label{ap:conventions}

In this appendix we summarize the conventions used throughout the main text.

\subsubsection*{Metric signature}

In general, when referring to $d$-dimensional Minkowski space, we denote the global flat coordinates as $\{x^{\mu}\}$, with $\mu=0,1,\ldots,d-1$. Moreover, our convention for the metric signature is the mostly plus one, namely
\beq\label{eq:Minksignature}
 \eta_{\mu \nu} = \text{diag} \left(-1, +1, \ldots, +1 \right)\, .
\eeq
Similarly, for those cases where in addition to a $d$-dimensional flat background we also have some internal compact space (of real dimension $n$) $\mathcal{X}_n$, we denote by $\{y^m\}$, $m=1, 2, \ldots n$, any local set of coordinates for the latter.

\subsubsection*{Differential forms}

On the other hand, when writing down local lagrangians describing the dynamics associated to tensor-like fields (of any rank), we adopt differential form notation. In particular, $p$-forms living in some $d$-dimensional manifold $\mathcal{M}$ may be expanded as follows
\beq\label{eq:pforms}
 C_p= \frac{1}{p!} C_{\mu_1 \ldots \mu_p} dx^{\mu_1} \wedge \ldots \wedge dx^{\mu_p}\, ,
\eeq
where the subindex indicates the rank of the anti-symmetric tensor and $\wedge$ denotes the exterior product within the algebra $\Omega(\mathcal{M}) = \bigoplus_{p=0}^{d} \Omega_p(\mathcal{M})$. For instance, taking the product between a $p$-form $C_p$ and a $q$-form $A_q$, yields the following $(p+q)$-form
\beq\label{eq:extproduct}
 C_p \wedge A_q= \frac{1}{p! q!} C_{\mu_1 \ldots \mu_p} A_{\nu_1 \ldots \nu_q} dx^{\mu_1} \wedge \ldots \wedge dx^{\mu_p} \wedge dx^{\nu_1} \wedge \ldots \wedge dx^{\nu_q}\, .
\eeq
These may be interpreted as generalized Abelian gauge fields subject to the redundancy condition
\beq\label{eq:gaugepformtransf}
 C_p \to C_p + d \omega_{p-1}\, ,
\eeq
where $d= \partial_{\mu} dx^{\mu}$ denotes the exterior derivative --- which acts as a map from $\Omega_p(\mathcal{M})$ to $\Omega_{p+1}(\mathcal{M})$ --- whereas $\omega_{p-1} \in \Omega_{p-1}(\mathcal{M})$. Additionally, one may define field strengths $F_{p+1}$ for the above gauge fields $C_p$ through the exterior derivative
\beq\label{eq:fieldstrength}
 F_{p+1} := dC_p= \frac{1}{p!} \partial_{\mu_0}C_{\mu_1 \ldots \mu_p} dx^{\mu_0} \wedge dx^{\mu_1} \wedge \ldots \wedge dx^{\mu_p}\, ,
\eeq
which are left invariant under the transformation \eqref{eq:gaugepformtransf}.\footnote{More generally, the field strength is invariant under any shift of the form $C_p \to C_p + \xi_{p}$, with $\xi_{p} \in \text{Ker}\, (\dd)$. These transformations can be classified in terms of the cohomology group $H^p (\mathcal{M})$ associated to the exterior derivative, such that the ones displayed in eq. \eqref{eq:gaugepformtransf} correspond to the trivial class, whereas in general there might be additional non-trivial classes which provide for \emph{large} gauge transformations, i.e. those which are not continuously connected to the identity.} In terms of those, the usual kinetic terms for the $p$-form gauge fields would read as
\beq\label{eq:kinetictermforms}
 -\frac12 \int F_{p+1} \wedge \star F_{p+1} = -\frac12 \int \frac{1}{p!} F_{\mu_1 \ldots \mu_{p+1}} F^{\mu_1 \ldots \mu_{p+1}}\, \star 1 = -\frac12 \int \dd^dx\, \sqrt{-g}\, \frac{1}{p!} F_{\mu_1 \ldots \mu_{p+1}} F^{\mu_1 \ldots \mu_{p+1}}\, ,
\eeq
where $\star 1 = \sqrt{-g}\, dx^1 \wedge dx^{2} \wedge \ldots \wedge dx^{d}$ denotes the volume form on $\mathcal{M}$.

\subsubsection*{Curvature tensors}

In all our discussions from the main text regarding the dynamics of the spacetime metric $g_{\mu \nu} (x)$, the conventions for the different curvature invariants are the following.

First, we define the Levi-Civita connection $\Gamma^{\sigma}_{\mu \nu}$ as usual
\beq\label{eq:LeviCivita}
 \Gamma^{\sigma}_{\mu\nu}=\begin{Bmatrix}
\sigma\\ \mu \nu
\end{Bmatrix}=\frac{1}{2}\, g^{\sigma \lambda} \left( \partial_{\mu}g_{\nu\lambda}+\partial_{\nu}g_{\mu\lambda} - \partial_{\lambda}g_{\mu\nu} \right)\, ,
\eeq
where $g^{\mu \nu}$ is the inverse metric. From this one can readily compute the Riemann tensor
\beq\label{eq:Riemanntensor}
\mathcal{R}^{\sigma}_{\ \lambda\mu\nu}=\partial_{\mu}\Gamma^{\sigma}_{\nu\lambda}-\partial _{\nu}\Gamma^{\sigma}_{\mu\lambda}+\Gamma^{\eta}_{\mu\lambda}\Gamma^{\sigma}_{\mu\eta}-\Gamma^{\eta}_{\mu\lambda}\Gamma^{\sigma}_{\nu\eta}\, ,
\eeq
together with the associated Ricci tensor and curvature scalar
\beq\label{eq:Ricci&curvaturescalar}
 \mathcal{R}_{\mu \nu} = \mathcal{R}^{\sigma}_{\ \mu \sigma \nu}\, , \qquad \mathcal{R}= g^{\mu \nu}\mathcal{R}_{\mu \nu}\, .
\eeq
These quantities enter both in the action functional for Einstein gravity
\beq
\begin{aligned}\label{eq:conventionalEH}
    S_{\rm EH} \left[ g_{\mu \nu} (x) \right]= \frac{1}{2 \kappa_d^2} \int \mathcal{R} \star 1\ +\ S_{\rm matter} \left[ \phi (x),\, \Psi (x),\, \ldots \right]\, ,
\end{aligned}
\eeq
where $\kappa_d^2 = 8\pi G_N$ and $G_N$ denotes Newton's gravitational constant, as well as in the corresponding classical equations of motion
\beq
\begin{aligned}
    \mathcal{R}_{\mu \nu}- \frac12 g_{\mu \nu} \mathcal{R} = \kappa_d^2 T_{\mu \nu}\, ,
\end{aligned}
\eeq
where $T_{\mu \nu}$ is the energy-momentum tensor of the matter fields, see Section \ref{ss:basics} for details.

\subsubsection*{Spinors and Clifford algebra}

The Clifford algebra in $d$ spacetime dimensions is generated by the gamma matrices $\gamma^{\mu}$, which satisfy the following anti-commutation rules 
\beq\label{eq:Cliffordalgebra}
\begin{aligned}
    \lbrace \gamma^{\mu}, \gamma^{\nu}\rbrace= 2g^{\mu \nu}\, .
\end{aligned}
\eeq
Using a locally flat frame such that $\gamma^{a}= e^a_{\mu} \gamma^{\mu}$, the above relation can be conveniently rewritten as
\beq
\begin{aligned}
    \lbrace \gamma^{a}, \gamma^{b}\rbrace= 2 \eta^{a b}\, , \qquad a, b = 0, 1, \ldots, d-1\, ,
\end{aligned}
\eeq
where we have introduced a vielbein $e^a = e^a_{\mu}(x) dx^{\mu}$ which locally diagonalizes the spacetime metric, namely it satisfies
\beq
\begin{aligned}\label{eq:apvielbeindef}
    g^{\mu \nu} e^a_{\mu} e^b_{\nu} = \eta^{a b}\, .
\end{aligned}
\eeq
Note that $\{ e^a_{\mu}\}$ is defined up to local Lorentz rotations of the form
\beq
\begin{aligned}
    e^a_{\mu} \to \Lambda^a_{\ b} e^b_{\mu}\, , \qquad \Lambda^a_{\ b} \in \mathsf{SO(1, d-1)}\, ,
\end{aligned}
\eeq
which obviously leave the condition \eqref{eq:apvielbeindef} unchanged.

The spinor fields $\psi(x)$ hence arise as (Grassmann-valued) representations of the above algebra. These crucially depend on the number of spacetime dimensions as well as the metric signature (i.e. whether it is of Lorentzian or Riemannian type). For instance, in eleven-dimensional Minkowski space, the irreducible representation of the Clifford algebra \eqref{eq:Cliffordalgebra} involves Majorana spinors, which are real-valued such that $\psi^* = \psi$ (c.f. Section \ref{ss:Mthy11d}). On the other hand, in even-dimensional spacetimes, it is possible to introduce an additional gamma matrix 
\beq
\begin{aligned}\label{eq:gammad+1}
    \gamma^{d+1}=\i^{\frac{d}{2}-1} \prod_{k=0}^{d-1}\, ,
\end{aligned}
\eeq
which squares to the identity operator, commutes with any other gamma matrix and moreover projects onto states of definite chirality. Hence, one can define Weyl spinors as follows
\beq
\begin{aligned}
    \psi_{\pm} = \left(1 \pm \gamma^{d+1}\right) \psi\, ,
\end{aligned}
\eeq
thus satisfying $\gamma^{d+1}\, \psi_{\pm} = \pm\, \psi_{\pm}$. Let us mention that both Majorana and Weyl conditions are compatible in dimensions $d=2$ (mod 8).

\subsubsection*{Units}

In this work we employ natural units, namely we set $\hbar=c=1$ from the start --- unless stated otherwise. This convenient choice leaves us with only one inequivalent physical magnitude (or quantity), for instance energy, denoted here by $[E]$. To measure those, we will oftentimes switch between two different sets of units that are customarily used when considering gravitational effective field theories arising as low energy limits of string theory.

The first one is the most natural choice in string theory, where we measure every quantity in terms of the string length $\ell_s=2 \pi \sqrt{\alpha'}$. Accordingly, by studying physical processes involving gravitons in the external legs one finds that the strength of gravitational interactions in $d$ spacetime dimensions --- i.e. the gravitational coupling $\kappa_d$ --- reads as
\beq
\begin{aligned}
    2\kappa_d^2= e^{2\varphi_d}\, \frac{\ell_s^{d-2}}{2\pi}\, ,
\end{aligned}
\eeq
where $\varphi_d= \phi - \frac12 \log \mathcal{V}_{10-d}$ denotes the (vacuum expectation value of the) $d$-dimensional dilaton, which depends both on the 10d one and the volume of the internal space (measured in string units as well).

Alternatively, the conventional choice of units in gravity involves sitting in the Einstein frame, where the two-derivative lagrangian has the form displayed in \eqref{eq:conventionalEH}. Moreover, it is common practice to associate an energy scale to the gravitational interactions themselves, which is usually referred to as the (reduced) Planck mass
\beq
   \Mpd = \kappa_d^{-\frac{1}{d-2}} \, .
\eeq
In terms of length-scales, one analogously defines the so-called Planck length as follows
\beq
   \ell_d = \frac{(4\pi)^{\frac{1}{d-2}}}{\Mpd}\, ,
\eeq
which can be easily related to the string scale previously introduced by the following mathematical relation
\beq
   \ell_d^{d-2} = \ell_s^{d-2} e^{2\varphi_d}\, .
\eeq

\chapter{Relevant Automorphic Forms}
\label{ap:Massform}

This appendix serves as a mathematical compendium for the relevant set of automorphic functions that appear at several instances in the thesis. We particularize to the discrete groups $\mathsf{SL(2, \mathbb{Z})}$ and $\mathsf{SL(3, \mathbb{Z})}$, since they capture the different U-duality symmetries arising in $d=10,\, 9$ and 8 maximal supergravity theories, which are thoroughly discussed in Parts \ref{part:StringTheoryTests} and \ref{part:pattern}. A similar analysis can be done for the (bigger) U-duality groups that appear upon reducing the number of non-compact spacetime dimensions, see e.g., \cite{Green:2010kv} for details. 

\section{Mathematical preliminaries}\label{ss:mathdefs}

An automorphic function $\varphi$ of a given Lie group $G$ is defined as a map from a space $\mathcal{M}$ to $\mathbb{R}$ (or more generally $\mathbb{C}$), where $\mathcal{M}$ admits some natural $G$-group action. Such automorphic function, $\varphi: \mathcal{M} \to \mathbb{R}$, is moreover left invariant under the corresponding group action, namely
\beq\label{eq:automorphicfn}
 \varphi (g \cdot p) = \varphi (p)\, , \qquad \forall p \in \mathcal{M},\quad \text{and}\quad \forall g \in G\, .
\eeq
This means, in particular, that the function $\varphi$ can be unambiguously defined on the quotient space $\mathcal{M}/G$.

In general, for a given pair $(\mathcal{M}, G)$ there can be more than one non-trivial automorphic form, and in certain cases the set $\lbrace \varphi\rbrace$ may even be infinite --- a simple example being the pair $(\mathbb{H}, \mathsf{SL(2, \mathbb{Z})})$, with $\mathbb{H}$ the upper-half plane. In addition, whenever $\mathcal{M}$ has a non-trivial boundary $\partial \mathcal{M}$, it is convenient to classify the set of automorphic functions depending on their behaviour at $\partial \mathcal{M}$. This includes the case of hyperbolic spaces, where despite their non-compactness, one can define some boundary after a process of `one-point' compactification. In such instances, the boundary $\partial \mathcal{M}$ lies at infinite distance (in the natural bi-invariant metric), see below. 

\section{$\mathsf{SL(2, \mathbb{Z})}$ Maas waveforms} \label{s:SL2Waveforms}

In this section we particularize to the case in which the group $G$ is isomorphic to $\mathsf{SL(2, \mathbb{Z})}$. We will restrict ourselves to the set of automorphic functions of $\mathsf{SL(2, \mathbb{Z})}$ which are moreover real analytic, since they appear as (generalized) `Wilson coefficients' in the EFT expansion of some gravitational effective field theories (see Chapter \ref{ch:Higherdimops}). In fact, there exists a very convenient and economic way to generate such analytic functions as eigenfunctions of some appropriate elliptic operator. Now, since we want these functions to be automorphic forms as well, we can simply take the hyperbolic Laplace operator, which is both elliptic and $\mathsf{SL(2, \mathbb{Z})}$-invariant (in a precise sense that we specify below). This operator reads
\beq\label{eq:SL2Laplacian}
\Delta_2 = \tau_2^2 \left( \frac{\partial^2}{\partial \tau_1^2} + \frac{\partial^2}{\partial \tau_2^2}\right)\, ,
\eeq
where as usual we define $\tau=\tau_1 + \text{i} \tau_2$. Note that this is nothing but the laplacian operator associated to the hyperbolic metric \eqref{eq:IIB10dSL2}. Therefore, the eigenfunctions of this operator --- which are moreover modular invariant --- are called singular Maas forms \cite{DHoker:2022dxx}. Here we will be interested, for reasons that will become clear later on, in a subgroup of such set of functions, those denoted simply as Maas forms, which have the additional property of growing polynomially (instead of exponentially) with $\tau_2$, as $\tau_2 \to \infty$. An example of Maas form that plays a key role in the discussion from the main text are the so-called non-holomorphic Eisenstein series \cite{DHoker:2022dxx}
\beq\label{eq:nonholoEisenstein}
\frac{\Gamma(\ell)}{2\pi^{\ell}}\, E_{\ell}^{sl_2}(\tau, \bar \tau) \equiv \pi^{-\ell}\, \Gamma(\ell)\, \frac{1}{2} \sum_{(m, n) \in \mathbb{Z}^2 \setminus \lbrace (0,0) \rbrace} \frac{\tau_2^\ell}{\left| m+n\tau\right|^{2\ell}}\, ,
\eeq
which converge absolutely if $\text{Re}\, \ell >1$. It can be shown (upon using that the fractional linear transformation in eq. \eqref{eq:SdualitytransIIBdilaton} conmutes with the operator $\Delta_2$), that indeed $E_{\ell}^{sl_2}(\tau)$ are both automorphic and eigenfunctions of the hyperbolic laplacian, with eigenvalue given by $\ell(\ell-1)$. The polynomial growth of the Eisenstein series can be also easily understood, since upon taking the limit $\tau_2 \to \infty$, the infinite series is clearly dominated by the terms with $n=0$, which grows as $\tau_2^\ell$. More precisely, the functions $E_{\ell}^{sl_2}(\tau)$ have an alternative Fourier expansion in $\tau_1$, which can be obtained upon Poisson resumming\footnote{\label{fnote:Poissonresummation}The Poisson resummation identity reads as follows \begin{equation}
    \notag \sum_{n \in \mathbb{Z}} F(x+na) = \frac{1}{a} \sum_{k \in \mathbb{Z}} \tilde{F} \left(\frac{2\pi k}{a} \right) e^{2\pi i k x/a}\, ,
\end{equation}
with $\tilde{F} (\omega)=\int_{-\infty}^{\infty} \dd x\, F(x) e^{-i \omega x}$ the Fourier transform of $F(x)$.} on the integer $n$, yielding
\begin{align}\label{eq:nonpertexpansion}
	\notag E_{\ell}^{sl_2} =\, & \bigg[ 2\zeta(2\ell) \tau_2^{\ell} + 2\pi^{1/2}\frac{\Gamma(\ell-1/2)}{\Gamma(\ell)} \zeta(2\ell-1) \tau_2^{1-\ell}\\
 &+ \frac{8 \pi^\ell \tau_2^{1/2}}{\Gamma(\ell)} \sum_{m=1}^{\infty} m^{\ell-1/2} \sigma_{1-2\ell} (m)\, \cos(2\pi m \tau_1)\, K_{\ell-1/2} (2\pi m \tau_2)\bigg]\, ,
\end{align}
where $\sigma_{1-2\ell} (m) = \sum_{d|m} d^\ell$ runs over all divisors $d$ of $m$, and $K_\ell(y)$ is the modified Bessel function of second kind, which is defined as follows
\begin{equation}
    K_\ell(y)=\frac{1}{2} \int_0^{\infty} \dd x\, x^{\ell-1} \exp \left[ -\frac{y}{2} \left( x + \frac{1}{x}\right)\right]\, ,
    \end{equation}
and decays asymptotically as $ K_\ell(y) \sim \sqrt{\frac{\pi}{2y}} e^{-y}$ for $y \to \infty$.

Let us finally mention that the modular form $\left(2\pi^{\ell}\right)^{-1} \Gamma(\ell) E_{\ell}^{sl_2}(\tau)$, when seen as a function also of the variable $\ell$, has a meromorphic continuation to all $\ell\in \mathbb{C}$, which is thus analytic everywhere except for simple poles at $\ell=0,1$. Moreover, if the divergence for $\ell=1$ is `extracted', namely upon selecting the constant term (with respect to $\ell$) in the Laurent series for $E_{\ell}^{sl_2}$ at $\ell=1$, one obtains the following function \cite{DHoker:2022dxx}
\beq
2\pi\left(\gamma_{\text{e}} - \log 2\right) - \pi \log \left( \tau_2\,|\eta(\tau)|^4 \right)\, ,
\eeq
where $\gamma_{\text{e}}$ is the Euler-Mascheroni constant and $\eta(\tau)$ denotes the Dedekind eta function, which may be defined as
\beq \label{eq:Dedekind}
\eta(\tau) = q^{\frac{1}{24}} \prod_{k=1}^{\infty} \left( 1-q^k\right)\, , \qquad q=e^{2\pi \i \tau}\, .
\eeq
To conclude, let us note that even though the function $\hat{E}_{1}^{sl_2}(\tau) = - \pi\log \left( \tau_2\,|\eta(\tau)|^4 \right)$ arises as some sort of analytic extension of $E_{1}^{sl_2}(\tau)$, it is actually not a Maas form, since $\Delta_2 \hat{E}_{1}^{sl_2}(\tau)$ is not proportional to $\hat{E}_{1}^{sl_2}(\tau)$ itself but it rather gives a constant value. This can be easily checked upon noting that $\partial \bar \partial \hat{E}_{1}^{sl_2}(\tau)= \frac{\pi}{4 \tau^2_2}$, as well as $\Delta_2=4 \tau_2^2 \partial \bar \partial$, where we have defined $\partial = \partial/\partial \tau$ and $\bar \partial = \partial/\partial \bar \tau$. In any event, what remains true is that the large modulus behaviour of $\hat{E}_{1}^{sl_2}(\tau)$ matches with that expected for $E_{\ell=1}^{sl_2}(\tau)$, since upon using the Fourier series expansion for $\eta(\tau)$
\beq
\eta(\tau) = q^{\frac{1}{24}} \left( 1-q-q^2+q^5 + \mathcal{O}(q^7) \right)\, ,
\eeq
one finds the following relevant asymptotic expression 
\beq \label{eq:asymptotic behavior}
-\pi \text{log} \left(\tau_2\,|\eta(\tau)|^4\right)\, \sim\, -\pi \text{log} \left(\tau_2\,e^{-\frac{\pi \tau_2}{3}}\right)\, \sim\, \frac{\pi^2}{3} \tau_2 - \pi \text{log} (\tau_2)\, ,
\eeq
whose first term precisely is $2 \zeta(2) \tau_2$.

\section{$\mathsf{SL(3, \mathbb{Z})}$ Maas waveforms}

We consider now the case where $G=\mathsf{SL(3, \mathbb{Z})}$. The motivation comes from the fact that it captures the U-dualities arising in maximal supergravity in eight dimensions. Therefore, following the same strategy as before, let us first introduce the appropriate elliptic $\mathsf{SL(3, \mathbb{Z})}$-invariant operator, namely the Laplace operator on the coset space $\mathsf{SL(3, \mathbb{R})}/\mathsf{SO(3)}$
\beq \label{eq:laplacianSL3}
\Delta_3 = 4\tau_2^2 \partial_{\tau} \partial_{\bar \tau} + \frac{1}{\nu \tau_2} \left| \partial_b -\tau \partial_c \right|^2 + 3 \partial_{\nu} \left( \nu^2 \partial_{\nu}\right)\, ,
\eeq
where the local parametrization in \eqref{eq:laplacianSL3} has been chosen to make contact with Type IIB string theory compactified on $\mathbf{T}^2$ (see Section \ref{ss:MthyT3}). Note that the previous coordinates can be compactly grouped into the following $3\times3$ matrix (see e.g., \cite{Kiritsis:1997em})
\beq
 \mathcal{B}= \nu^{1/3} \begin{pmatrix}
		\frac{1}{\tau_2} \quad  \frac{\tau_1}{\tau_2} \quad \frac{c+\tau_1 b}{\tau_2}\\ \frac{\tau_1}{\tau_2} \quad  \frac{|\tau|^2}{\tau_2} \quad \frac{\tau_1 c+|\tau|^2 b}{\tau_2}\\ \frac{c+\tau_1 b}{\tau_2} \quad  \frac{\tau_1 c+|\tau|^2 b}{\tau_2} \quad \frac{1}{\nu} + \frac{|c+\tau b|^2}{\tau_2}
	\end{pmatrix}\, ,
\eeq
which moreover satisfies $\mathcal{B}=\mathcal{B}^{\text{T}}$ as well as $\det \mathcal{B}=1$. The usefulness of the matrix $\mathcal{B}$ rests on the fact that it transforms in the adjoint representation of $\mathsf{SL(3, \mathbb{Z})}$, namely upon performing some transformation $\mathcal{A} \in \mathsf{SL(3, \mathbb{Z})}$, one finds that
\beq \label{eq:Btransf}
  \mathcal{B} \rightarrow \mathcal{A}^{\text{T}}\, \mathcal{B}\, \mathcal{A}\, .
\eeq
With this, we are now ready to define the Eisenstein $\mathsf{SL(3, \mathbb{Z})}$ series of order $\ell$:
\beq\label{eq:SL3Eisenstein}
E_{\ell}^{sl_3} = \sum_{\mathbf{n}\, \in\, \mathbb{Z}^3 \setminus \lbrace \vec{0} \rbrace} \left( \sum_{i, j =1}^3 n_i\, \mathcal{B}^{ij}\, n_j\right)^{-\ell} = \sum_{\mathbf{n}\, \in\, \mathbb{Z}^3 \setminus \lbrace \vec{0} \rbrace} \nu^{-\ell/3} \left[ \frac{\left| n_1 + n_2 \tau + n_3 \left( c+\tau b\right)\right|^2}{\tau_2} + \frac{n_3^2}{\nu}\right]^{-\ell}\, ,
\eeq
where $\mathcal{B}^{ij}$ denote the components of the inverse matrix $\mathcal{B}^{-1}$. Note that the above expression is manifestly $\mathsf{SL(3, \mathbb{Z})}$-invariant, since the vector $\textbf{n} = \left( n_1, n_2, n_3\right)$ transforms as $\textbf{n} \rightarrow \mathcal{A}^{\text{T}}\, \textbf{n}$ under the duality group. In addition, as it was also the case for the non-holomorphic Eisenstein series defined in eq. \eqref{eq:nonholoEisenstein} above, the functions $E_{\ell}^{sl_3}$ are eigenvectors of the laplacian $\Delta_3$, satisfying
\beq 
  \Delta_3 E_{\ell}^{sl_3} = \frac{2\ell (2\ell-3)}{3} E_{\ell}^{sl_3}\, .
\eeq
Let us also mention that the series $E_{\ell}^{sl_3}$, when viewed as a function of $\ell$, are absolutely convergent for $\ell>3/2$, whilst $E_{3/2}^{sl_3}$ is logarithmically divergent. This is reminiscent of the situation for the $\mathsf{SL(2, \mathbb{Z})}$ Eisenstein series $E_{\ell}^{sl_2}$, which had a simple pole for $\ell=1$. Therefore, proceeding analogously as in that case, one may define
\beq \label{eq:regularisation}
  \hat{E}_{3/2}^{sl_3} \equiv \lim_{\ell\to 3/2} \left( E_{\ell}^{sl_3} - \frac{2\pi}{\ell-3/2} - 4\pi(\gamma_{\text{e}}-1)\right)\, ,
\eeq
where again $\gamma_{\text{e}}$ denotes the Euler-Mascheroni constant. Such newly defined function is no longer singular and remains invariant under $\mathsf{SL(3, \mathbb{Z})}$ transformations, with the price of not being a zero-mode of the laplacian \eqref{eq:laplacianSL3} anymore.

\subsubsection*{Fourier-like expansions}

In what follows, our aim will be to rewrite the $\mathsf{SL(3, \mathbb{Z})}$ Eisenstein series in a way which makes manifest the perturbative and non-perturbative origin of the different terms that appear in the expansion, similarly to what we did for the $\mathsf{SL(2, \mathbb{Z})}$ case. We closely follow Appendix A of \cite{Kiritsis:1997em}. First, let us introduce the following integral representation
\begin{align}\label{eq:integralrep}
	\notag E_{\ell}^{sl_3} &= \frac{\pi^\ell}{\Gamma(\ell)} \int_0^{\infty} \frac{\dd x}{x^{1+\ell}} \sum_{\mathbf{n}\, \in\, \mathbb{Z}^3 \setminus \lbrace \vec{0} \rbrace} \exp \left[-\frac{\pi}{x} \left( \sum_{i, j =1}^3 n_i\, \mathcal{B}^{ij}\, n_j\right) \right]\\
 &= \nu^{-\ell/3} \frac{\pi^\ell}{\Gamma(\ell)} \int_0^{\infty} \frac{\dd x}{x^{1+\ell}} \sum_{\mathbf{n}\, \in\, \mathbb{Z}^3 \setminus \lbrace \vec{0} \rbrace} \exp \left[-\frac{\pi}{x} \left( \frac{\left| n_1 + n_2 \tau + n_3 \left( c+ \tau b\right) \right|^2}{\tau_2} + \frac{n_3^2}{\nu}\right) \right]\, ,
\end{align}
which can be shown to coincide with the defining series \eqref{eq:SL3Eisenstein} after performing the change of variables $y=x^{-1}$ and using the definition of the $\Gamma$-function, namely 
\begin{align}
	\Gamma(z) = \int_0^{\infty} \dd y\, y^{z-1} e^{-y}\, .
\end{align}
After carefully separating the sum in the integers $n_i$ and performing a series of Poisson resummations (see footnote \ref{fnote:Poissonresummation}), one arrives at a Fourier expansion of the form \cite{Kiritsis:1997em,Basu:2007ru,Basu:2007ck}
\begin{align}\label{eq:instexpSL3}
	\notag E_{\ell}^{sl_3} &= 2\nu^{-\ell/3} \tau_2^\ell \zeta(2\ell) + 2 \sqrt{\pi} T_2 \left( \tau_2 \nu^{1/3}\right)^{3/2-\ell} \frac{\Gamma(\ell-1/2)}{\Gamma(\ell)} \zeta(2\ell-1) + 2\pi \nu^{2\ell/3-1} \frac{\zeta(2\ell-2)}{\ell-1}\\
 &+  2 \frac{\pi^\ell \sqrt{\tau_2}}{\Gamma(\ell) \nu^{\ell/3}} \sum_{m,n \neq 0} \left| \frac{m}{n}\right|^{\ell-1/2} e^{2\pi \text{i} m n\tau_1}\, K_{\ell-1/2} (2\pi |m n| \tau_2)\, +\, \sum_{m, n \in \mathbb{Z} \setminus \lbrace (0,0) \rbrace} \mathcal{I}^\ell_{m, n}\, ,
\end{align}
where we have defined $T_2 \equiv \text{Im}\, T$, with $T= b+ \text{i} \left( \nu \tau_2\right)^{-1/2}$, and 
\begin{align}
	\mathcal{I}^\ell_{m, n} = 2\frac{\pi^\ell \nu^{\ell/6-1/2}}{\Gamma(\ell) \tau_2^{\ell/2-1/2}} \sum_{k \neq 0} \left| \frac{m+n\tau}{k}\right|^{\ell-1} e^{2\pi \text{i} k \left[n(c+\tau_1 b)- (m+n\tau_1)b \right]}\, K_{\ell-1} \left(2\pi |k|\frac{\left| m+n\tau \right|}{\sqrt{\nu \tau_2}}\right)\, .
\end{align}
Notice that upon using the expansion for the $\mathsf{SL(2, \mathbb{Z})}$ series in eq. \eqref{eq:nonpertexpansion}, one can group the terms which depend on $\nu^{-\ell/3}$ into the following expression
\begin{align}\label{eq:SL3&SL2}
	 E_{\ell}^{sl_3} &= \nu^{-\ell/3} E_{\ell}^{sl_2}(\tau) + 2\pi \nu^{2\ell/3-1} \frac{\zeta(2\ell-2)}{\ell-1}\, +\, \sum_{m, n \in \mathbb{Z} \setminus \lbrace (0,0) \rbrace} \mathcal{I}^\ell_{m, n}\, .
\end{align}
From a string theory perspective, each of these terms in the expansion can be given a physical interpretation in terms of instanton corrections, see Chapter \ref{ch:Higherdimops} for details. 

Furthermore, there exists another set of coordinates on $\mathsf{SL(3, \mathbb{R})}/\mathsf{SO(3)}$ apart from those employed in eq. \eqref{eq:laplacianSL3}, in which $\{\nu, \tau\}$ are exchanged with $\{\varphi_8, T\}$, where $e^{-2\varphi_8}=\tau_2^{3/2} \nu^{-1/2}$. From the Type IIB point of view, they correspond to the complexified K\"ahler modulus of $\mathbf{T}^2$ as well as the eight-dimensional dilaton (see Section \ref{ss:MthyT3}). Using such parametrization, one can expand $E_{\ell}^{sl_3}$ around `weak coupling' as follows 
\begin{align}\label{eq:instexpSL3-2}
	\notag E_{\ell}^{sl_3} &= 2 \zeta(2\ell) e^{-\frac{4\ell}{3} \varphi_8} + \pi^{1/2} \frac{\Gamma(\ell-1/2)}{\Gamma(\ell)} e^{-\left(\frac{2\ell}{3}-1 \right)\varphi_8} E_{\ell-1/2}^{sl_2} (T)\\
    \notag&+  \frac{2\pi^\ell}{\Gamma(\ell)} T_2^{\ell/2-1/4} e^{-\left(\frac{\ell}{3}-\frac{1}{2} \right)\varphi_8} \sum_{m,n \neq 0} \left| \frac{m}{n}\right|^{\ell-1/2} e^{2\pi \text{i} m n\tau_1}\, K_{\ell-1/2} (2\pi |m n| \tau_2)\\
    &+\frac{2\pi^\ell}{\Gamma(\ell)} T_2^{1/2} e^{\left(\frac{2\ell}{3}-1 \right)\varphi_8} \sum_{k \neq 0} \left| \frac{m+n\tau}{k}\right|^{\ell-1} e^{2\pi \text{i} k \left[n(c+\tau_1 b)- (m+n\tau_1)b \right]}\, K_{\ell-1} \left(2\pi |k| \left| m+n\tau \right| T_2\right)\, ,
\end{align}
where one should view $\tau_2$ as a function of $\lbrace\varphi_8, T_2\rbrace$ in the previous expression.

\subsubsection*{The $E_{3/2}^{sl_3}$ series}

To close this section, let us briefly discuss the particular case of the $\mathsf{SL(3, \mathbb{Z})}$ Eisenstein series of order-$3/2$, since it plays a crucial role in our analysis in Section \ref{ss:MthyT3}. In fact, as already mentioned, $E_{\ell}^{sl_3}$, when seen as a function of the variable $\ell$, has a simple pole at $\ell=3/2$.\footnote{This is easy to see from eq. \eqref{eq:instexpSL3} above, since the functions $\zeta(1+x)$ as well as $\Gamma(x)$ present simple poles at $x=0$. Indeed, one obtains the following expansions around the pole:
\beq
\notag \zeta(1+\epsilon)= \frac{1}{\epsilon} + \gamma_{\text{e}} + \mathcal{O}(\epsilon)\, , \qquad \Gamma(\epsilon)=\frac{1}{\epsilon} - \gamma_{\text{e}} + \mathcal{O}(\epsilon)\, .
\eeq} Regularizing in a way that preserves automorphicity (see eq. \eqref{eq:regularisation}), one finds for the series expansion the following expression
\begin{align}\label{eq:Eisenstein3/2}
	\notag \hat{E}_{3/2}^{sl_3} &= 2\zeta(3) \frac{\tau_2^{3/2}}{\nu^{1/2}} + \frac{2 \pi^2}{3} T_2 + \frac{4\pi}{3} \log \nu\\
 &+  4\pi \sqrt{\frac{\tau_2}{\nu}} \sum_{m,n \neq 0} \left| \frac{m}{n}\right| e^{2\pi \text{i} m n\tau_1}\, K_{1} (2\pi |m n| \tau_2)\, +\, \sum_{m, n \in \mathbb{Z} \setminus \lbrace (0,0) \rbrace} \mathcal{I}^{3/2}_{m, n}\, ,
\end{align}
which in the limit \eqref{eq:instexpSL3-2} becomes \cite{Green:2010wi}
\begin{align}\label{eq:Eisenstein3/2-2}
	\hat{E}_{3/2}^{sl_3} &= 2\zeta(3) e^{-2\varphi_8} + 2 \hat{E}_{1}^{sl_2} (T) + \frac{4 \pi}{3} \varphi_8 + \mathcal{O} \left( \exp(-(T_2 e^{2\varphi_8})^{-1/2}),\exp(-(T_2^{-1} e^{2\varphi_8})^{-1/2})\right)\, .
\end{align}

\chapter{A Heat Kernel Primer}
\label{ap:heatkernel}

In this appendix we review the basics of the heat kernel formalism, which provides a useful technique to perform covariant computations in field theory and gravity. As a proof of concept, we apply these ideas so as to determine the contribution to the Einstein-Hilbert term induced by a tower of massive particles. See also Section \ref{ss:Emergencegraviton} for a complementary discussion.

\section{The heat kernel expansion}
\label{s:intro}

The heat kernel or `inverse mass' expansion (see e.g., \cite{Schubert:2001he,Vassilevich:2003xt} for reviews on the topic) allows one to compute in a manifestly gauge and diffeomorphism invariant fashion the corrections to the Wilsonian/quantum effective action induced by integrating out a particle at one loop in perturbation theory. The basic idea hinges on the following mathematical identity (due to Schwinger):
\beq\label{eq:Schwingerintegral}
\int_{\varepsilon \to 0^+}^{\infty} \frac{\dd \tau}{\tau}\, e^{-\tau A} = -\log A + \text{const.}\, .
\eeq
This relation allows us to define in a convenient way the logarithm of the determinant of any (trace class) operator $D$ as a Gaussian integral 
\beq\label{eq:logdetop}
\log \det D = \text{tr} \log D = - \int_{0}^{\infty} \frac{\dd \tau}{\tau}\, \text{tr}\, e^{-\tau D} = \sum_n \log \lambda_n\, ,
\eeq
where we have assumed the spectrum of eigenvalues $\{ \lambda_n \}$ of the operator $D$ to be discrete for simplicity, and the integral above should be interpreted in the regularized sense \eqref{eq:Schwingerintegral}. Moreover, one can promote the definition \eqref{eq:logdetop} to incorporate as well those cases in which $D$ is some differential operator with a continuous spectrum, such that the trace above includes a priori both discrete sums as well as spacetime integrals. For instance, one may consider differential operators which arise as deformations of the laplacian (therefore appearing in the kinetic terms of the theory) of the form
\beq\label{eq:laplaciandeformation}
D = -\nabla_{\mu} \nabla^{\mu} + X\, ,
\eeq
where $\nabla_{\mu}$ denotes the corresponding gauge and covariant derivative, whilst $X$ captures the deformation (e.g., mass terms, spin-orbit couplings, etc.).

Furthermore, it is customary to define the \emph{heat kernel} associated to a given operator $D$ as follows
\beq\label{eq:heatkerneldef}
K(\tau; x, y) = e^{-\tau D}\, ,
\eeq
where the name originates from the fact that $K(\tau; x, y)$ indeed solves the heat equation
\beq\label{eq:heateq}
\frac{\partial K(\tau; x, y)}{\partial \tau} + D\, K(\tau; x, y) = 0\, ,
\eeq
with the boundary condition $K(\tau; 0,0)= \delta (x-y)$. In simple cases, such as when $\nabla_{\mu} = \partial_{\mu}$ and $X=0$, one can explicitly solve the equation \eqref{eq:heateq} to find
\beq\label{eq:heatkernelfree}
K_{\rm free}(\tau; x, y) = \frac{1}{(4\pi\tau)^{d/2}} e^{-\frac{(x-y)^2}{4\tau}}\, ,
\eeq
where $d$ is the spacetime dimension. More generally, though, it is hard to obtain an analytic solution to the heat kernel equation. However, it is actually possible to perform a small proper time expansion which gives a Taylor series of the form
\beq\label{eq:heatkernelgeneric}
K(\tau; x, y) = K_{\rm free}(\tau; x, y) \sum_{n=0}^{\infty} a_n(x, y) \tau^n\, ,
\eeq
whose coefficients $a_n(x, y)$ are known as the \emph{Seeley-deWitt coefficients} and characterize in a universal way the operator whose determinant we are interested in computing. Inserting this back into the one-loop determinant arising from the path integral one obtains the following formal expression
\beq\label{eq:heatkernel}
S[g, A] = \int_{0}^{\infty} \frac{\dd \tau}{\tau}\, e^{-\tau m^2} \int \frac{\dd^dx\, \sqrt{-g}}{(4\pi \tau)^{\frac{d}{2}}}\, \sum_{n=0}^{\infty} a_n(x) \tau^n\, .
\eeq
Note that the parameters $a_n(x) = a_n(x, y=x)$ are always regular, although they can of course lead ultimately to some UV divergence upon integration over Scwhinger proper time, $\tau\in \mathbb{R}_+$. Such divergences are ultra-violet in nature and always come from the lower part of the integration domain --- namely when $\tau \to 0$, which corresponds to small loops in the target space. For example, in $d=4$ the terms with $n=0,1,2$ all become UV divergent and thus subject to the renormalization procedure. On the other hand, from the point of view of the Emergence proposal --- and following our discussion in Chapter \ref{ch:Emergence}, in this thesis we are interested in imposing the species/quantum gravity scale as the UV cut-off, and indeed the wordline/heat kernel formalism here described allows us to do so in a manifestly gauge invariant way. In practice, we just need to restrict the integration domain to $\tau \in [\varepsilon, \infty)$, where by dimensional analysis one identifies $\varepsilon= \LSP^{-2}$. For instance, in the case of a minimally coupled complex scalar $\phi(x)$ one finds the following first Seeley-deWitt coefficients \cite{Bastianelli:2008cu}
\begin{align} \label{eq:dewittcoeffscalar}
	 a_0 &= 1\, , \qquad a_1 = \frac{1}{6} \mathcal{R}\, , \qquad a_2=-\frac{1}{12} F_{\mu \nu} F^{\mu \nu}\, ,
\end{align}
whilst for a (minimally coupled) Weyl fermion $\Psi(x)$ one rather obtains \cite{Bastianelli:2008cu}
\begin{align} \label{eq:dewittcoefffermion}
	a_0 = -2\, , \qquad a_1 = \frac{1}{6} \mathcal{R}\, , \qquad  a_2=-\frac{1}{3} F_{\mu \nu} F^{\mu \nu}\, .
\end{align}
These can be seen to contribute to the renormalization of the vacuum energy, the Planck mass and (in case we add an extra $\mathsf{U(1)}$ gauge field) the electric charge, respectively. 

\section{A one-loop correction to the Einstein-Hilbert action}
\label{s:oneloopEH}

As a simple application of this formalism, let us compute the one-loop contribution to the kinetic term of the graviton due to either a spin-0 or spin-$\frac{1}{2}$ field, c.f. eqs. \eqref{eq:dewittcoeffscalar} and \eqref{eq:dewittcoefffermion} above. In both cases we get
\beq\label{eq:heatkernelEH}
S^{\text{1-loop}}_{\text{EH}} = \int \dd^dx\, \sqrt{-g}\, \mathcal{R} \int_{\varepsilon}^{\infty} \dd \tau\, e^{-\tau m^2}  \frac{1}{6(4\pi \tau)^{\frac{d}{2}}}\, ,
\eeq
such that upon integrating over Schwinger proper time leads to
\beq \label{eq:integratedoneloop}
\frac{1}{6(4\pi)^{\frac{d}{2}}}\, \varepsilon^{\frac{d-2}{2}}\, E_{\frac{d}{2}} (m^2 \varepsilon)\, ,
\eeq
which can be seen to decrease as $m^2\varepsilon$ increases, namely when the field mass gets close to the UV cut-off. Moreover, this can be expanded for any spacetime dimension whenever $m^2 \varepsilon \ll 1$ by using the asymptotic properties of the exponential integral function, namely
\beq\label{eq:Expfnasymptoticsodd}
\int_{\varepsilon}^{\infty} \frac{\dd \tau}{\tau^{\frac{d}{2}}}\, e^{-\tau m^2} \sim m^{d-2} \left[  \Gamma (-d/2+1) - \sum_{n=0}^{\infty} (-)^n \frac{\left( m^2 \varepsilon \right)^{-\frac{d}{2}+n+1}}{n! \left( -\frac{d}{2}+n+1\right)}\right]\, ,
\eeq
as $m^2 \varepsilon \to 0^+$, where we assumed $d \notin 2 \mathbb{Z}_{\geq0}$. For $d \in 2 \mathbb{Z}_{\geq0}$, one obtains an additional logarithmic term
\beq\label{eq:Expfnasymptoticseven}
\int_{\varepsilon}^{\infty} \frac{\dd \tau}{\tau^{\frac{d}{2}}}\, e^{-\tau m^2} \sim m^{d-2} \left[ c_{\frac{d}{2}-1}+ \frac{(-1)^{\frac{d}{2}}}{\left(\frac{d}{2}-1 \right)!} \log (m^2 \varepsilon) - \sum_{n=0}^{\infty} (-)^n \frac{\left( m^2 \varepsilon \right)^{-\frac{d}{2}+n+1}}{n! \left( -\frac{d}{2}+n+1\right)}\right]\, ,
\eeq
where $c_{\frac{d}{2}-1}$ denotes some $d$-dependent numerical coefficient of value
\beq
c_{\frac{d}{2}-1}= \frac{(-1)^{\frac{d}{2}-1}}{\left(\frac{d}{2}-1 \right)!} \left(\gamma-\sum_{n=1}^{d/2-1} \frac{1}{n} \right)\, ,
\eeq
and $\gamma \approx 0.577$ is the Euler-Mascheroni constant. Therefore, upon identifying $\LSP=\varepsilon^{-2}$ and substituting either asymptotic expression at leading order in \eqref{eq:heatkernel}, one finds
\beq \label{eq:leading1loopEH}
S^{\text{1-loop}}_{\text{EH}} = \frac{\LSP^{d-2}}{3(d-2)(4\pi)^{\frac{d}{2}}}\int \dd^dx\, \sqrt{-g}\, \mathcal{R}\, .
\eeq
Hence, assuming that we have $N$ of these fields, we recover the result advocated in eq. \eqref{eq:finalquantumEH} from the main text.

\subsubsection*{Example: Kaluza-Klein theory on $\mathbf{S}^1$}

For illustrative purposes, we analyze here an explicit example in which one can perform the summation of the one-loop contribution to the graviton kinetic term due to an \emph{infinite} tower of states. For simplicity, and since this already provides a couple of nice insights, we consider the simplest Kaluza-Klein scenario of a $D=d+1$ dimensional theory compactified on a circle. We will concentrate on the contribution to the Einstein-Hilbert term associated to Kaluza-Klein replicas of a $D$-dimensional complex scalar $\phi(s, z)$.

In fact, it is easy to see that for a spectrum of this sort, when performing the one-loop integral \eqref{eq:heatkernelEH} and after expanding the result in powers of $m^2\varepsilon$ (c.f. eqs. \eqref{eq:Expfnasymptoticsodd} and \eqref{eq:Expfnasymptoticseven}), each term contributes after summing over all the modes roughly the same way, i.e. precisely as $N \LSP^{d-2}$, such that they can all be in principle resummed. This latter observation was actually to be expected based on the following heuristic argument. The intuition comes from the fact that our KK states in the $d$-dimensional theory are nothing but the momentum excitations modes of the massless $D$-dimensional scalar field, i.e. before compactifying on the $\mathbf{S}^1$. Hence, from the higher dimensional perspective, one expects these massless fields to provide for a one-loop correction of the form 
\beq\label{eq:masslessheatkernelEH}
S^{\text{1-loop}}_{\text{EH}} = \int \dd^{D}x\, \sqrt{-\hat g}\, \hat{\mathcal{R}} \int_{\varepsilon}^{\infty} \dd \tau\,  \frac{1}{6(4\pi \tau)^{\frac{D}{2}}}= \frac{\varepsilon^{-\frac{D-2}{2}}}{3 (D-2)(4\pi)^{\frac{D}{2}}} \int \dd^{D}x\, \sqrt{-\hat g}\, \hat{\mathcal{R}}\, ,
\eeq
thus proportional to $\LSP^{d-1} \simeq M_{\text{Pl};\, d+1}^{d-1}$. However, one should also take into account the extra $\sqrt{-\hat g}$ factor in \eqref{eq:masslessheatkernelEH}, which when integrated over the $(d+1)$-dimensional spacetime $\mathcal{M}^{d+1} \cong \mathbb{R}^{1,\, d-1} \times \mathbf{S}^1_{\mathsf{R}}$ behaves roughly as $2\pi \mathsf{R} \LSP^{d-1} \simeq (\LSP/m_{\rm KK}) \LSP^{d-2} \simeq N \LSP^{d-2}$, where we substituted $m_{\rm KK}=1/\mathsf{R}$. 

Being slightly more careful, the Schwinger integral in $\mathcal{M}^{d+1} \cong \mathbb{R}^{1,\, d-1} \times \mathbf{S}^1_{\mathsf{R}}$ reads as follows
\beq\label{eq:Schwingercircle}
\begin{aligned}
\mathcal{A}=\frac{1}{6(4\pi)^{\frac{d}{2}}} \int_{\varepsilon}^{\infty} \frac{\dd \tau}{\tau^{\frac{d}{2}}}\, \frac{1}{2\pi \mathsf{R}} \sum_{n=-\infty}^{\infty} e^{-\tau \frac{n^2}{\mathsf{\mathsf{R}}^2}}= \frac{1}{3\times 2^{d+2} \pi^{\frac{d+1}{2}}} \int_{0}^{\varepsilon^{-1}} \dd \hat{\tau}\, \hat{\tau}^{\frac{d-3}{2}}\, \sum_{\omega=-\infty}^{\infty} e^{- \hat{\tau} \left( \pi \mathsf{R} \omega\right)^2}\, ,
\end{aligned}
\eeq
where we have performed a Poisson resummation (c.f. footnote \ref{fnote:Poissonresummation})
\begin{align}
    \sum_{n\in\mathbb{Z}} e^{-(2\pi n)^2 a/2}=\frac{1}{\sqrt{2\pi a}} \sum_{\omega\in\mathbb{Z}} e^{-\omega^2/2a}\, ,
\end{align}
and we defined a new variable $\hat{\tau} = \tau^{-1}$. It is clear from the above expression that the UV divergent part is now associated to having no winding of the particle worldline along the circle, namely when $\omega=0$. Separating both pieces we find
\beq\label{eq:UVconvS1}
\begin{aligned}
\mathcal{A}_{\omega\neq0}\, &\sim\, \frac{1}{3\times 2^{d+1} \pi^{\frac{d+1}{2}}}\, \sum_{\omega \geq 1} \left(\pi \mathsf{R} \omega \right)^{-d+1}\, \Gamma \left( \frac{d-1}{2}\right) \\
&= \frac{1}{3\times 2^{d+1} \pi^{\frac{3d-1}{2}} \mathsf{R}^{d-1}}\, \Gamma \left( \frac{d-1}{2}\right)\, \zeta(d-1)\, ,
\end{aligned}
\eeq
for the UV finite part (we only keep the leading order term in $\varepsilon$), as well as 
\beq\label{eq:UVdivS1}
\begin{aligned}
\mathcal{A}_{\omega=0}\, &=\, \frac{1}{3\times 2^{d+2} \pi^{\frac{d+1}{2}}} \int_{0}^{\varepsilon^{-1}} \dd \hat{\tau}\, \hat{\tau}^{\frac{d-3}{2}} = \frac{\LSP^{d-1}}{3\times 2^{d+1} \pi^{\frac{d+1}{2}} (d-1)}\, ,
\end{aligned}
\eeq
for the divergent piece, which agrees with our previous estimation \eqref{eq:masslessheatkernelEH}.

Notice that the dominant contribution comes from the $\omega=0$ term. However, there is a second finite threshold correction to the $d$-dimensional EH term scaling like $m_{\rm KK}^{d-2}$.\footnote{Actually, one can also arrive at \eqref{eq:UVconvS1} upon imposing $\zeta$-function regularization directly to eq. \eqref{eq:Schwingercircle}, see e.g., \cite{Alvarez:2022hjn}.}

\chapter{Loop Calculations}
\label{ap:Loops}
		
In this appendix we provide detailed calculations of the one-loop Feynman graphs presented in Section \ref{s:selfenergybosons} from the main text. In particular, we focus on those diagrams that contribute to the wave-function renormalization of a scalar, a 1-form gauge field and a Weyl spinor coming from loops of massive scalar and fermions particles. To do so, we compute the amputated one-loop diagram corresponding to each of these processes, shown in Figures \ref{fig:scalarpropagator}-\ref{fig:kineticfermionsbas}. As an important remark, even though we discuss our set-up for the different relevant cases in Lorentzian spacetimes, when performing any loop calculation we will analytically continue the relevant integrals so as to work with Euclidean signature instead, which simplifies the analysis considerably.
		
\section{Self-energy of a modulus}
\label{ap:Loopsscalar}
		
Let us begin by considering a real modulus $\phi$, coupled to massive (real) scalars $\{\sigma^{(n)}\}$ or Dirac fermions $\{\psi^{(n)}\}$ through their mass terms as follows
\begin{align}
			S_{\mathrm{kin,} \phi}\, &=\, -\dfrac{1}{2} \int   d\phi \wedge \star d\phi \, , \label{eq:Skinphi} \\
			S_{\sigma^{(n)}} \, &=\,  - \dfrac{1}{2} \int   \left( d\sigma^{(n)} \wedge \star d\sigma^{(n)}\,  +\,  m_n(\phi)^2 \sigma^{(n)} \sigma^{(n)} \right) \star 1\, , \label{eq:Ssigman}\\
			S_{\psi^{(n)}} \, &= \,  \int  \left( \i \overline{\psi^{(n)}}\,  \slashed{D} \, \psi^{(n)} \, -\,  m_n(\phi) \, \overline{\psi^{(n)}}\psi^{(n)} \right) \star 1\, . \label{eq:Spsin}
\end{align}
We keep in mind that the label $n \in \mathbb{N}$ will eventually denote the step in the tower in which either the scalars $\{\sigma^{(n)}\}$ or the fermions $\{\psi^{(n)}\}$ are organized, with their masses $m_n(\phi)$ increasing accordingly, but for now the computation is meant to be quite general. In the context of Emergence, we are interested in the computation of the wave-function renormalization of the scalar field $\phi$ in $d$ spacetime dimensions due to scalar and fermionic loops. The idea is thus to extract the momentum-dependent part of the exact propagator of the massless modulus $\phi$ at $\mathcal{O}(\hbar)$ in the Wilsonian effective action after integrating out the heavy fields, which takes the form
\beq\label{eq:exactpropscalar}
		D(p^2)=\frac{1}{p^2-\Pi(p^2)}\, ,
\eeq
after deforming the contour of integration and analytically extending the results to Euclidean signature, i.e. $\Bar{g}_{\mu \nu}= \delta_{\mu \nu}$ (see e.g., \cite{Nair:2005iw}). Here, $\Pi(p^2)$ corresponds to the (amputated) one-loop Feynman diagram displayed in Figure \ref{fig:scalarpropagator}.
		
\subsubsection*{Scalar loop}
		
Let us begin by considering the contribution due to a loop of scalars $\{\sigma^{(n)}\}$, which is shown in Figure \ref{fig:scalarloopscalar} and  reads (taking into account an overall $1/2$ symmetry factor of the diagram)
\beq
		\Pi_n(p^2) \ = \frac{\lambda_n^2}{2} \int \frac {\text{d}^dq}{(2\pi)^d} \frac {1}{(q^2+m_n^2)} \frac {1}{\left((q-p)^2+m_n^2\right)}\, ,
\label{eq:selfenergyscalar(ap)}
\eeq
with the coupling $\lambda_n=2m_n(\partial_\phi m_n)$ coming from the trilinear vertex arising after expanding the mass term in eq. \eqref{eq:Ssigman} around the modulus v.e.v. at linear order. Since we are interested in the correction to the propagator, we need to extract the term proportional to $p^2$, so that we take a derivative with respect to $p^2$ and evaluate the result at $p=0$ to obtain\footnote{Notice that naively one would also obtain a term proportional to $1/|p|$ after taking the derivative with respect to $p^2$, but this would correct the linear term in the momentum expansion, which can be seen to be absent when the detailed computation is performed (as required by Lorentz invariance).} 
\beq
		\frac {\partial \Pi_n(p^2)}{\partial p^2} \bigg\rvert_{p=0}  = - \frac{\lambda_n^2}{2} \int \frac {\text{d}^dq}{(2\pi)^d} \frac {1}{(q^2+m_n^2)^3}\, .
\label{eq:sigmaa_ap}
\eeq
From this expression we expect the integral to be divergent for $d\geq 6$ and convergent otherwise. However, since we will always keep in mind the idea of introducing the UV cut-off associated to QG, namely the species scale, we perform the momentum integral up to a maximum scale $\Lambda$, which yields the following general expression
\beq
		\begin{split}
			\frac{\partial \Pi_n(p^2)}{\partial p^2} \bigg\rvert_{p=0}  \, =  \, - \lambda_n^2  \ \frac{ \pi^{d/2}  }{8\,  (2 \pi)^d \, \Gamma(d/2) } \ \frac{\Lambda^d}{m_n^6} \ & \left[ - \frac{(d-6)m_n^4+(d-4)m_n^2\Lambda^2}{(\Lambda^2+m_n^2)^2} \right.  \\
			& \quad \left.  + \left(d+ \frac{8}{d}-6\right) \ _2{\cal F}_1\left( 1,\frac{d}{2};\frac{d+2}{2}; -\frac{\Lambda^2}{m_n^2}\right) \right]  \, ,
		\end{split}
\label{eq:scalarloopscalarexact}
\eeq
with $_2{\cal F}_1(a,b;c;d)$ the ordinary (or Gaussian) hypergeometric function. Given the kind of towers that we are dealing with (c.f. Section \ref{s:speciesscale}), the two relevant asymptotic limits for this expression are \emph{(i)} $\Lambda \gg m_n$ (for most states of KK-like towers) and \emph{(ii)} $\Lambda \simeq m_n$ (for most states of stringy towers). In order to study each of these limits in turn, we will also distinguish between $d>6$, $d=6$ and $d<6$, given that the divergence of the corresponding expressions in the large $\Lambda$ limit is different for these three cases.
		
Let us begin by considering the limit, $\Lambda \gg m_n$, which dominates the contributions coming from KK-like towers. In this case, the integral diverges polynomially  with $\Lambda$ for $d>6$ as
\beq
		\frac {\partial \Pi_n^{(d>6)}(p^2)}{\partial p^2} \bigg\rvert_{p=0}\, = \, - \lambda_n^2 \ \frac{\pi^{d/2}}{(2 \pi)^d\ \Gamma\left( d/2 \right) \ (d-6)}   \ \Lambda^{d-6} \, + \, \mathcal{O}\left(\Lambda^{d-8}\, m_n^2\right) + \mathrm{const.} \ ,
\label{eq:scalarloopscalarsd>6}
\eeq
such that the leading term goes like $\Lambda^{d-6}$. For $d<6$ one can expand eq. \eqref{eq:scalarloopscalarexact} to obtain
\beq
		\frac {\partial \Pi_n^{(d<6)}(p^2)}{\partial p^2} \bigg\rvert_{p=0}\, = \, - \lambda_n^2 \ \frac{\pi^{\frac{d+2}{2}}}{16 \ (2 \pi)^d  \ \Gamma\left( d/2 \right)} \frac{(d-2)(d-4)}{\sin\left( d \pi/2\right)} \ \frac{1}{m_n^{6-d}} + \, \mathcal{O}\left(\frac{1}{\Lambda^{6-d}}\right)  \ .
\label{eq:scalarloopscalarsd<6}
\eeq
		\begin{table}[t]\begin{center}
				\renewcommand{\arraystretch}{2.00}
				\begin{tabular}{|c||c|c|c|c|c|}
					\hline
					$d$ & 2 & 3 & 4 & 5 & 6 \\
					\hline 
					$\dfrac {\partial \Pi_n}{\partial p^2} \bigg\rvert_{p=0}$ &
					$-\dfrac{1}{16 \pi}\dfrac{\lambda_n^2}{m_n^4}$ & 
					$ -\dfrac{1}{64 \pi  }\dfrac{\lambda_n^2}{m_n^3}$ &
					$ -\dfrac{1}{64 \pi^2}\dfrac{\lambda_n^2}{m_n^2}$ & 
					$-\dfrac{9}{128 \pi^2}\dfrac{\lambda_n^2}{m_n}$ &
					$ -\dfrac{\lambda_n^2 }{256 \pi ^3} \log \left(\dfrac{\Lambda ^2}{m_n^2}\right) $ \\
					\hline 
					\hline
					$d$ &  7 & 8 & 9 & 10 & 11\\
					\hline 
					$\dfrac {\partial \Pi_n}{\partial p^2} \bigg\rvert_{p=0}$ &
					$  -\dfrac{\lambda_n^2  \, \Lambda }{240 \pi^4}  $ &
					$-\dfrac{\lambda_n^2  \, \Lambda^2  }{3072 \pi^4} $ &
					$ -\dfrac{\lambda_n^2  \, \Lambda^3  }{10080 \pi^5}$ &
					$ -\dfrac{\lambda_n^2  \, \Lambda^4   }{98304 \pi^5}$ &
					$  -\dfrac{\lambda_n^2  \, \Lambda^5 }{302400 \pi^6} $  \\
					\hline
				\end{tabular}
				\caption{Leading contribution to the wave-function renormalization of a modulus field due to a loop of massive scalars, as given by eq. \eqref{eq:scalarloopscalarexact}, in the limit $\Lambda\gg m_n$ for different number of spacetime dimensions $2 \leq d \leq 11$.}
				\label{tab:scalarloopscalarLambda>>m}\end{center}
		\end{table}  
Note that the leading piece here is actually the constant term in $\Lambda$, which was irrelevant in \eqref{eq:scalarloopscalarsd>6} but provides instead the leading correction for $d<6$. Additionally, the piece proportional to $(d-2)(d-4)\sin^{-1}\left( d \pi/2\right)$ must be defined as a limit, thus taking a value equal to $\{ 4/\pi,\, 3,\, 1\}$ for $d=2,\, 4$, $d=1,\, 5$, and $d=3$, respectively. Let us also remark that for $d<6$ the loop integral is convergent such that, at the QFT level, no UV cut-off (nor UV regulator whatsoever) is actually necessary. Finally, for the marginal case $d=6$, we get the expected leading logarithmic correction
\beq
		\frac {\partial \Pi_n^{(d=6)}(p^2)}{\partial p^2} \bigg\rvert_{p=0} \, =\, -  \frac{\lambda_n^2 }{256 \pi^3 }\log \left( \frac{\Lambda^2}{m_n^2} \right) \ + \ \mathcal{O}\left( \Lambda^0 \right)  \, .
\label{eq:scalarloopscalard=6}
\eeq
A summary of the relevant leading term for a different number of spacetime dimensions can be found in Table \ref{tab:scalarloopscalarLambda>>m}.
		
Consider now the alternative limiting case, namely $\Lambda \simeq m_n$, which gives an upper bound for the states whose contribution to the loop must be included. Notice that this is the dominant term for towers of string oscillator modes. In this case, we can expand eq. \eqref{eq:scalarloopscalarexact} for any $d$, yielding the following expression
\beq
		\begin{split}
			\frac {\partial \Pi_n(p^2)}{\partial p^2} \bigg\rvert_{p=0} \, = \, -  \frac{\lambda_n^2 \  \pi^{d/2}}{32 \ (2 \pi)^d\ \Gamma(d/2) }   &\left\{ 10-2d+(d-2)(d-4) \left[\psi\left( \frac{d+2}{4}\right)-  \psi\left( \frac{d}{4} \right)  \right] \right\} \Lambda^{d-6}   \\
			& \quad + \mathcal{O}(\Lambda-m_n)   \, ,
		\end{split}
\label{eq:scalarloopscalarLambda=m}
\eeq
where $\psi(z)$ represents the digamma function.\footnote{The digamma function, $\psi(z)$, is defined as the logarithmic derivative of the familiar gamma function $\Gamma(z)$ with respect to its argument, namely
\beq
			\notag \psi(z) = \frac{d}{dz} \log \left( \Gamma(z)\right) = \frac{\Gamma'(z)}{\Gamma(z)}\, .
\eeq
} Notice that since in this limit $\Lambda \simeq m_n$, we recover the same leading asymptotic dependence with $\Lambda$ and $m_n$ as in eqs. \eqref{eq:scalarloopscalarsd>6}-\eqref{eq:scalarloopscalard=6}. The precise form of the leading term for different number of spacetime dimensions is summarized in Table \ref{tab:scalarloopscalarLambda=m}.
		\begin{table}[t]\begin{center}
				\renewcommand{\arraystretch}{2.00}
				\begin{tabular}{|c||c|c|c|c|c|}
					\hline
					$d$ & 2 & 3 & 4 & 5 & 6 \\
					\hline 
					$\dfrac {\partial \Pi_n}{\partial p^2} \bigg\rvert_{p=0}$ &
					$-\dfrac{3}{64 \pi}\dfrac{\lambda_n^2}{m_n^4}$ & 
					$ -\dfrac{1}{128 \pi}\dfrac{\lambda_n^2}{m_n^3}$ &
					$ -\dfrac{1}{256 \pi^2}\dfrac{\lambda_n^2}{m_n^2}$ & 
					$-\dfrac{(3\pi-8)}{768 \pi^3}\dfrac{\lambda_n^2}{m_n}$ &
					$ -\dfrac{8\log(2)-5}{2048 \pi ^3} \lambda_n^2  $ \\
					\hline 
					\hline
					$d$ &  7 & 8 & 9 & 10 & 11\\
					\hline 
					$\dfrac {\partial \Pi_n}{\partial p^2} \bigg\rvert_{p=0}$ &
					
					$\begin{aligned}[t] -\tfrac{(16-5\pi) }{2560 \pi^4} \times  & \\  \lambda_n^2  \, \Lambda & \end{aligned}$ &
					$\begin{aligned}[t]-\tfrac{(17-24\log(2)) }{24576 \pi^4} \times & \\ \lambda_n^2  \, \Lambda^2 & \end{aligned}$  &
					$ \begin{aligned}[t]-\tfrac{(105\pi-328)}{322560 \pi^5} \times & \\ \lambda_n^2  \, \Lambda^3 & \end{aligned}$ &
					$ \begin{aligned}[t]-\tfrac{(16\log(2)-11)  }{131072 \pi^5} \times & \\ \lambda_n^2  \, \Lambda^4 & \end{aligned}$ &
					$ \begin{aligned}[t] -\tfrac{(992-315\pi) }{9676800 \pi^6} \times & \\ \lambda_n^2  \, \Lambda^5 & \end{aligned}$  \\
					\hline
				\end{tabular}
				\caption{Leading contribution to the wave-function renormalization of a modulus field due to a loop of massive scalars, given by eq. \eqref{eq:scalarloopscalarexact}, in the limit $\Lambda \simeq m_n$ for different number of spacetime dimensions $2 \leq d \leq 11$.}
				\label{tab:scalarloopscalarLambda=m}\end{center}
		\end{table}  

Let us remark that the leading asymptotic dependence with the corresponding energy scale (i.e. with the UV cut-off or the mass of the particle running in the loop) is the same for the two limiting cases, $\Lambda \gg m_n$ and $\Lambda\simeq m_n$, with only numerical prefactors differing between the two expressions. Thus, since these limits actually bound the contribution of a given particle to the loop, we can safely use any of the above asymptotic relations in order to calculate the field dependent contribution of the towers to the relevant kinetic terms.

\subsubsection*{Fermionic loop}
		
We now consider the contribution to the scalar metric from a loop of fermions, with a coupling induced by the mass term as specified in the action \eqref{eq:Spsin}. The calculation is similar to the scalar loop above, and the corresponding Feynman diagram, displayed in Figure \ref{fig:scalarloopfermion}, gives the following correction
\begin{equation}\label{eq:selfenergyfermion(ap)}
			\begin{aligned}
				\Pi_n(p^2) \ &= -\mu_n^2  \int \frac {\text{d}^dq}{(2\pi)^d}\ \text{tr} \left (\frac {1}{\i \slashed{q}+m_n}\ \frac {1}{\i (\slashed{q}-\slashed{p})+m_n} \right)\\
				&= -\mu_n^2  \int \frac {\text{d}^dq}{(2\pi)^d} \text{tr} \left ( \frac{(-\i \slashed{q} + m_n)(-\i (\slashed{q}-\slashed{p})+m_n)}{(q^2+m_n^2)((q-p)^2+m_n^2)} \right)\ . 
			\end{aligned}
\end{equation}
Here, the relevant coupling constant is $\mu_n = \partial_\phi m_n(\phi)$, and notice that there is an extra minus sign with respect to \eqref{eq:selfenergyscalar(ap)} due to the fact that the particle is of fermionic nature. By recalling that the dimensionality of the Dirac matrices in $d$ spacetime dimensions is $\fdim$ (where $\lfloor x \rfloor$ denotes the largest integer less than or equal to $x$), and using the following identities
\begin{equation}
			\label{eq:scalarloopfermionstraces}
			\text{tr} \left(\gamma^{\mu} \right)=\ 0\,  , \qquad
			\text{tr} \left(\gamma^{\mu} \gamma^{\nu}\right)=\ \fdim \delta^{\mu \nu}\ ,
\end{equation}
we can explicitly perform the trace in \eqref{eq:selfenergyfermion(ap)}, which leads to
\begin{equation}
			\text{tr}\left\{ (-\i \slashed{q} + m_n)(-\i(\slashed{q}-\slashed{p})+m_n) \right\} \, =\, -\fdim (q^2 - p\cdot q -m_n^2)\, .
\end{equation}
Thus, upon extracting the part that is linear in $p^2$ we arrive at
\begin{equation}\label{eq:fermionloopddim}
			\frac{\partial \Pi_n(p^2)}{\partial p^2} \bigg\rvert_{p=0} \, = \,   -\mu_n^2\, \fdim  \int \frac {\text{d}^dq}{(2\pi)^d} \frac{1}{(q^2+m_n^2)^2} \ + \ 2 m_n^2 \, \mu_n^2\ \fdim \int \frac {\text{d}^dq}{(2\pi)^d} \frac{1}{(q^2+m_n^2)^3} \, ,
\end{equation}
where we have used the fact that some terms quadratic in $q$ cancel identically between themselves and that those linear in $q$ vanish after integration along the angular directions. Notice that the second piece is exactly the same as the contribution from $\fdim$ real scalars (recall that $\lambda_n= 2 m_n (\partial_\phi m_n)=2 m_n \mu_n$), but with \emph{opposite} sign. Thus, we can use all the results from our previous computations in order to evaluate the exact contribution. (Note that in the case in which the number of fermionic degrees of freedom equals the bosonic ones --- as e.g., in supersymmetric set-ups --- there is an exact cancellation between these two pieces.) The first term in \eqref{eq:fermionloopddim}, however, has a different (although similar) structure, and it is expected to be divergent for $d\geq 4$. Its precise form after imposing a UV cut-off $\Lambda$ for the momentum integral is therefore
\begin{equation}
			\frac{\partial \Pi_n(p^2)}{\partial p^2} \bigg\rvert_{p=0}  \, =  \, - \mu_n^2  \ \frac{ \fdim \pi^{d/2}  }{  (2 \pi)^d \, \Gamma(d/2) } \ \frac{\Lambda^d}{m_n^4} \  \left[  \frac{m_n^2}{\Lambda^2+m_n^2}  + \left(\frac{2}{d}-1\right) \ _2{\cal F}_1\left( 1,\frac{d}{2};\frac{d+2}{2}; -\frac{\Lambda^2}{m_n^2}\right) \right]  \, .
\label{eq:scalarloopfermionexact}
\end{equation}
		\begin{table}[t]\begin{center}
				\renewcommand{\arraystretch}{2.00}
				\begin{tabular}{|c||c|c|c|c|c|}
					\hline
					$d$ & 2 & 3 & 4 & 5 & 6 \\
					\hline 
					$\dfrac {\partial \Pi_n}{\partial p^2} \bigg\rvert_{p=0}$ &
					$-\dfrac{1}{2 \pi}\dfrac{\mu_n^2}{m_n^2}$ & 
					$ -\dfrac{1}{4 \pi  }\dfrac{\mu_n^2}{m_n}$ &
					$ -\dfrac{\mu_n^2 }{4 \pi^2 } \log \left(\dfrac{\Lambda ^2}{m_n^2}\right)$ & 
					$ -\dfrac{\mu_n^2  \, \Lambda }{3 \pi^2}$ &
					$ -\dfrac{\mu_n^2  \, \Lambda^2 }{16 \pi^3} $ \\
					\hline 
					\hline
					$d$ &  7 & 8 & 9 & 10 & 11\\
					\hline 
					$\dfrac {\partial \Pi_n}{\partial p^2} \bigg\rvert_{p=0}$ &
					$  -\dfrac{\mu_n^2  \, \Lambda^3 }{45 \pi^4}  $ &
					$-\dfrac{\mu_n^2  \, \Lambda^4  }{192 \pi^4} $ &
					$ -\dfrac{\mu_n^2  \, \Lambda^5  }{525 \pi^5}$ &
					$ -\dfrac{\mu_n^2  \, \Lambda^6   }{2304 \pi^5}$ &
					$  -\dfrac{\mu_n^2  \, \Lambda^7 }{6615 \pi^6} $  \\
					\hline
				\end{tabular}
				\caption{Leading contribution to the wave-function renormalization of a modulus field due to a loop of massive fermions, as given by eq. \eqref{eq:scalarloopfermionexact}, in the limit $\Lambda\gg m_n$ for different number of spacetime dimensions $2 \leq d \leq 11$.}
				\label{tab:scalarloopfermionLambda>>m}\end{center}
		\end{table}  
Now, in the limit $\Lambda \gg m_n$, which as we said is particularly relevant for most states in a KK-like tower, the leading contribution to the propagator in $d>4$ takes the form
\begin{equation}
			\frac {\partial \Pi_n^{(d>4)}(p^2)}{\partial p^2} \bigg\rvert_{p=0}\, = \, -\mu_n^2\,  \frac{2^{\lfloor \frac{d+2}{2} \rfloor} \pi^{d/2}}{(2 \pi)^d\ \Gamma\left( d/2 \right) \ (d-4)}   \ \Lambda^{d-4} \, + \, \mathcal{O}\left(\Lambda^{d-6}\, m_n^2\right) + \mathrm{const.} \ ,
			\label{eq:scalarloopfermionsd>4}
\end{equation}
which is very similar to the scalar contribution \eqref{eq:scalarloopscalarsd>6} but with a different power of the cut-off. Similarly, for $d<4$ the dominant term (which corresponds to the `const.' piece in the previous expansion) reads\footnote{Notice that in the context of the Swampland program one typically studies EFTs in $d\geq 4$, but we also include here the results in lower dimensions for completeness.}
\begin{equation}
			\frac {\partial \Pi_n^{(d<4)}(p^2)}{\partial p^2} \bigg\rvert_{p=0}\, = \, - \mu_n^2 \ \frac{2^{\lfloor\frac{d-2}{2} \rfloor} \ \pi^{\frac{d+2}{2}} }{(2 \pi)^d  \ \Gamma\left( d/2 \right)} \ \frac{(2-d)}{\sin\left( d \pi/2\right)} \ \frac{1}{m_n^{4-d}} + \, \mathcal{O}\left(\frac{1}{\Lambda^{4-d}}\right)  \ ,
\label{eq:scalarloopfermionsd<4}
\end{equation}
where once again for $d=2$ the quotient $(2-d) \sin^{-1}\left( d \pi/2\right)$ is defined as a limit and takes a value of $2/ \pi$. For the marginal case, we recover instead the expected logarithmic divergence
\beq
		\frac {\partial \Pi_n^{(d=4)}(p^2)}{\partial p^2} \bigg\rvert_{p=0} \, =\, -  \frac{\mu_n^2 }{4 \pi^2 }\log \left( \frac{\Lambda^2}{m_n^2} \right) \ + \ \mathcal{O}\left( \Lambda^0 \right)  \, .
\label{eq:scalarloopfermionsd=6}
\eeq
The precise leading contributions for all relevant values of $d$ are summarized in Table \ref{tab:scalarloopfermionLambda>>m}.
		
Taking now the other relevant limit, namely $\Lambda \simeq m_n$, we can similarly expand eq. \eqref{eq:scalarloopfermionexact} to obtain the following expression
\begin{equation}
			\frac {\partial \Pi_n(p^2)}{\partial p^2} \bigg\rvert_{p=0} \, = \, - \mu_n^2 \ \frac{2^{\lfloor\frac{d}{2}-2 \rfloor}  \pi^{d/2}}{ (2 \pi)^d\ \Gamma(d/2) }   \left\{ 2+(d-2)\left[\psi\left( \frac{d}{4}\right)-  \psi\left( \frac{d+2}{4} \right)  \right] \right\} \Lambda^{d-4} + \, \mathcal{O}(\Lambda-m_n)   \, .
\label{eq:scalarloopfermionLambda=m}
\end{equation}
As in the scalar case, since we have $\Lambda\simeq m_n$, the asymptotic dependence with the relevant scale is the same as the one in the $\Lambda \gg m_n$ limit, and only numerical prefactors change. The leading terms in \eqref{eq:scalarloopfermionLambda=m} for $2\leq d \leq 11$ are outlined in Table \ref{tab:scalarloopfermionLambda=m}.
		\begin{table}[t]\begin{center}
				\renewcommand{\arraystretch}{2.00}
				\begin{tabular}{|c||c|c|c|c|c|}
					\hline
					$d$ & 2 & 3 & 4 & 5 & 6 \\
					\hline 
					$\dfrac {\partial \Pi_n}{\partial p^2} \bigg\rvert_{p=0}$ &
					$-\dfrac{1}{4 \pi}\dfrac{\mu_n^2}{m_n^2}$ & 
					$ -\dfrac{(\pi -2)}{8 \pi^2  }\dfrac{\mu_n^2}{m_n}$ &
					$ -\frac{(2\log(2)-1)}{8 \pi^2 } \mu_n^2 $ & 
					$ \begin{aligned}[t]-\tfrac{(10-3\pi)}{24 \pi^3}\times &\\ \mu_n^2  \, \Lambda & \end{aligned}$ &
					$ \begin{aligned}[t]-\tfrac{(3-4\log(2))}{32 \pi^3}\times &\\ \mu_n^2  \, \Lambda^2 & \end{aligned}$ \\
					\hline 
					\hline
					$d$ &  7 & 8 & 9 & 10 & 11\\
					\hline 
					$\dfrac {\partial \Pi_n}{\partial p^2} \bigg\rvert_{p=0}$ &
					$ \begin{aligned}[t] -\tfrac{ (15\pi-46)}{360 \pi^4}\times &\\  \mu_n^2  \, \Lambda^3 & \end{aligned}$ &
					$\begin{aligned}[t]-\tfrac{(3\log(2)-2)}{96 \pi^4}\times &\\ \mu_n^2  \, \Lambda^4  & \end{aligned}$ &
					$ \begin{aligned}[t]-\tfrac{(334-105\pi) }{12600 \pi^5}\times &\\ \mu_n^2  \, \Lambda^5 & \end{aligned}$ &
					$ \begin{aligned}[t]-\tfrac{ (17-24\log(2)) }{4608 \pi^5}\times &\\ \mu_n^2  \, \Lambda^6 & \end{aligned}$ &
					$ \begin{aligned}[t] -\tfrac{(315\pi-982) }{264600 \pi^6}\times &\\  \mu_n^2  \, \Lambda^7 & \end{aligned}$  \\
					\hline
				\end{tabular}
				\caption{Leading contribution to the wave-function renormalization of a modulus field due to a loop of massive fermions, as given by eq. \eqref{eq:scalarloopfermionexact}, in the limit $\Lambda\simeq m_n$ for different number of spacetime dimensions $2 \leq d \leq 11$.}
				\label{tab:scalarloopfermionLambda=m}\end{center}
		\end{table}  

\section{Self-energy of a gauge 1-form}
\label{ap:Loops1-form}
		
We consider now a 1-form, $A_\mu$, with field strength $F_{\mu \nu }\, =\,2\  \partial_{[\mu} A_{\nu]}$, coupled to massive (complex) scalars $\{\chi^{(n)}\}$ or fermions $\{\psi^{(n)}\}$ through the following action
\begin{align}
			S_{\mathrm{kin,} A_1}\, &= \, -\dfrac{1}{4\, g^2} \int \dd^d x \sqrt{-g} \   F_{\mu\nu} F^{\mu \nu}\, , \label{eq:SkinA1} \\
			S_{\sigma^{(n)}} \, &=\,  - \dfrac{1}{2} \int \dd^d x \sqrt{-g} \ \left(  D_\mu \chi^{(n)} \overline{D^\mu \chi^{(n)}}\,  +\,  m_n^2 \ \chi^{(n)} \overline{\chi^{(n)}} \right) \, , \label{eq:SchinA1}\\
			S_{\psi^{(n)}} \, &= \,  \int \dd^d x \sqrt{-g} \   \left( \i \overline{\psi^{(n)}}\,  \slashed{D} \, \psi^{(n)} \, -\,  m_n \ \overline{\psi^{(n)}}\psi^{(n)} \right) \, . \label{eq:SpsinA1}
\end{align}
Here, the overline denotes complex conjugation for the scalars as well as Dirac conjugation for the fermions, whilst $D_\mu$ represents the appropriate covariant derivative of the fields minimally coupled to $A_1$, defined as 
\begin{equation}
			D_\mu \chi^{(n)} \, = \, \left(\partial_\mu -\i q_n A_\mu \right) \chi^{(n)} \, , \qquad D_\mu \psi^{(n)} \, = \, \left(\partial_\mu -\i q_n A_\mu \right) \psi^{(n)}\, .
\end{equation}
We will be concerned in what follows with the corrections to the propagator of $A_1$ induced by quantum loops from integrating out heavy scalar and fermion fields. As a remark, we will not elaborate on the subtleties associated to gauge invariant regularization, which are made manifest specially when imposing a UV cut-off $\Lambda$. 
Let us just mention that gauge invariance in the presence of a hard cut-off can be ensured rigorously (see Appendix \ref{ap:heatkernel} for details), but we will take a pragmatic approach here by focusing only on the dependence of the required amplitudes with $\Lambda$, instead of watching carefully that the correct tensorial structure is maintained even at the quantum level --- which is of course related to the preservation of gauge invariance. To do so, we use the Lorenz gauge (i.e. $\partial_\mu A^\mu=0$), since it can also be easily generalized to arbitrary $p$-form gauge fields. The propagator then takes the form (on a flat background with Euclidean metric $\Bar{g}_{\mu \nu}= \delta_{\mu \nu}$)\footnote{\label{fn:FeynmantHooftgauge}Strictly speaking, in order to fix the tensorial structure of the propagator as in eq. \eqref{eq:A1propagatorapp}, one has to impose additionally the Feynman-`t Hooft gauge, which is an instance of the more general $R_{\xi}$-gauges, with $\xi$ fixed to be equal to 1.} 
\begin{equation}
\label{eq:A1propagatorapp}
			D^{\mu \nu} (p^2) \, = \, \left( \dfrac{p^2}{g^2} \delta ^{\mu \nu} - \Pi^{\mu \nu}(p^2) \right)^{-1}\, ,
\end{equation}
where $\Pi^{\mu \nu}$ is zero at tree level, and gives the amputated Feynman diagram from the loops shown in Figure \ref{fig:1-formpropagator}. By using again our gauge choice, we can extract the tensorial dependence as follows 
%
\begin{equation}
			\label{eq:A1loopamplitudeapp}
			\Pi^{\mu \nu} (p^2) \, = \, \Pi(p^2) \delta ^{\mu \nu} \, .
\end{equation}
We are thus interested in extracting the piece proportional to $p^2$ within $\Pi(p^2)$, as arising from the aforementioned loop corrections.
		
\subsubsection*{Scalar loop}
		
We begin by considering the coupling of the 1-form to a complex scalar, $\chi^{(n)}$, with mass $m_n$ and charge $q_n$, as given by the action \eqref{eq:SchinA1}. The relevant one-loop Feynamn diagram is shown in Figure \ref{fig:1-formloopscalar}, and it reads
\beq
		\Pi^{\mu \nu}_n(p) \, =\,   g^2 \, q_n^2 \int \frac {\text{d}^dq}{(2\pi)^d} \frac {(2q-p)^{\mu} (2q-p)^{\nu}}{(q^2+m_n^2)\left( (q-p)^2+m_n^2\right)} \, .
\label{eq:A1scalar}
\eeq
From all the terms in the numerator, we only need to keep track of the ones $\propto q^\mu q^\nu$. The reason being that the ones proportional to $p^\mu p^\nu$ amount essentially to a change of gauge, which as we argued is not important for our purposes here, whilst the ones linear in $q^\mu$ instead turn out to either cancel identically or produce also linear terms in $q^\mu$ after taking the derivative with respect to $p^2$ and setting $p$ to zero, which then also cancel after the angular integration. Moreover, we can explicitly use Lorentz invariance to replace
\begin{equation}
\label{eq:qmuqnuaverage}
			q^\mu q^\nu \ \longrightarrow \ \dfrac{q^2}{d} \, \delta^{\mu \nu}\, ,
\end{equation}
under the integral sign in \eqref{eq:A1scalar}. Notice that this gives rise at the end of the day to the tensor structure announced in \eqref{eq:A1loopamplitudeapp}. Thus, the precise form of the relevant piece of the amputated Feynman diagram yields
\begin{equation}\label{eq:1-formscalarloopprop}
			\frac{\partial \Pi^{\mu \nu}_n(p^2)}{\partial p^2} \bigg\rvert_{p=0} \, = \, -g^2\,  q_n^2\,   \frac{4}{d} \, \delta^{\mu\nu} \, \int \dfrac{d^d q}{(2\pi)^d} \dfrac{q^2}{(q^2+m_n^2)^3} \, .
\end{equation}
As happened with the modulus case, we expect this integral to behave differently depending on the number of spacetime dimensions. In particular, it seems to diverge for $d\geq4$, but we will introduce a cut-off for any $d$ since at the end of the day we are interested in integrating up to a physical UV scale beyond which our EFT weakly coupled to Einstein gravity stops being valid. The exact expression gives therefore
\beq
		\begin{split}
			\frac{\partial \Pi_n(p^2)}{\partial p^2} \bigg\rvert_{p=0}   =  \, - g^2 \, q_n^2  \ \frac{ \pi^{d/2}  }{d\,  (2 \pi)^d \, \Gamma(d/2) } \ \frac{\Lambda^{d+2}}{m_n^6} \ & \left[ - \frac{(d-4)m_n^4+(d-2)m_n^2\Lambda^2}{(\Lambda^2+m_n^2)^2} \right.  \\
			& \quad \left.  + \dfrac{d(d-2)}{d+2} \ _2{\cal F}_1\left( 1,\frac{d+2}{2};\frac{d+4}{2}; -\frac{\Lambda^2}{m_n^2}\right) \right]  \, ,
		\end{split}
\label{eq:1-formloopscalarexact}
\eeq
where $\Pi_n(p^2)$ captures the part of the diagram after extracting the tensorial piece (c.f. eq. \eqref{eq:A1loopamplitudeapp}).
		
In analogy with the massless scalar case, the two relevant asymptotic limits that we take for this expression are \emph{(i)} $\Lambda \gg m_n$ (for most states of KK-like towers) and \emph{(ii)} $\Lambda \simeq m_n$ (for most states of stringy towers). In the first case, the integral diverges polynomially with $\Lambda$ for $d>4$ as 
\begin{equation}
			\frac {\partial \Pi_n^{(d>4)}(p^2)}{\partial p^2} \bigg\rvert_{p=0} = \, -g^2 \, q_n^2\  \frac{8\ \pi^{d/2}}{(2 \pi)^d\ \Gamma\left( d/2 \right) \ d\, (d-4)}    \ \Lambda^{d-4} \, + \, \mathcal{O}\left(\Lambda^{d-6}\, m_n^2\right) + \mathrm{const.} \ ,
\label{eq:1formloopfermionsd>4}
\end{equation}
whereas in lower dimensions it is convergent and the leading contribution is given by
\begin{equation}
			\frac {\partial \Pi_n^{(d<4)}(p^2)}{\partial p^2} \bigg\rvert_{p=0} = \, - g^2 \, q_n^2 \ \frac{\pi^{\frac{d+2}{2}} }{2 \, (2 \pi)^d  \ \Gamma\left( d/2 \right)} \ \frac{(2-d)}{\sin\left( d \pi/2\right)} \ \frac{1}{m_n^{4-d}} + \, \mathcal{O}\left(\frac{1}{\Lambda^{4-d}}\right)  \ ,
\label{eq:1formloopfermionsd<4}
\end{equation}
with the quotient $(2-d)/\sin\left( d \pi/2\right)$ defined as a limit with value $2/ \pi$ for $d=2$. Finally, for the marginal case, we get the expected logarithmic behaviour familiar from (scalar) QED
\beq
		\frac {\partial \Pi_n^{(d=4)}(p^2)}{\partial p^2} \bigg\rvert_{p=0} \, =\, -  \frac{g^2 \, q_n^2 }{16 \pi^2 }\log \left( \frac{\Lambda^2}{m_n^2} \right) \ + \ \mathcal{O}\left( \Lambda^0 \right)  \, .
\label{eq:1formloopscalard=4}
\eeq
The exact leading contributions for different values of $d$ are summarized in Table \ref{tab:1-formloopscalarLambda>>m}.
		\begin{table}[t]\begin{center}
				\renewcommand{\arraystretch}{2.00}
				\begin{tabular}{|c||c|c|c|c|c|}
					\hline
					$d$ & 2 & 3 & 4 & 5 & 6 \\
					\hline 
					$\dfrac {\partial \Pi_n}{\partial p^2} \bigg\rvert_{p=0}$ &
					$-\dfrac{1}{4 \pi}\dfrac{g^2 \, q_n^2}{m_n^2}$ & 
					$ -\dfrac{1}{8 \pi  }\dfrac{g^2 \, q_n^2}{m_n}$ &
					$ -\dfrac{g^2 \, q_n^2 }{16 \pi^2 } \log \left(\dfrac{\Lambda ^2}{m_n^2}\right)$ & 
					$ -\dfrac{g^2 \, q_n^2  \, \Lambda }{15 \pi^2}$ &
					$ -\dfrac{g^2 \, q_n^2  \, \Lambda^2 }{192 \pi^3} $ \\
					\hline 
					\hline
					$d$ &  7 & 8 & 9 & 10 & 11\\
					\hline 
					$\dfrac {\partial \Pi_n}{\partial p^2} \bigg\rvert_{p=0}$ &
					$  -\dfrac{g^2 \, q_n^2  \, \Lambda^3 }{630 \pi^4}  $ &
					$-\dfrac{g^2 \, q_n^2  \, \Lambda^4  }{6144 \pi^4} $ &
					$ -\dfrac{g^2 \, q_n^2  \, \Lambda^5  }{18900 \pi^5}$ &
					$ -\dfrac{g^2 \, q_n^2  \, \Lambda^6   }{184320 \pi^5}$ &
					$  -\dfrac{g^2 \, q_n^2  \, \Lambda^7 }{582120 \pi^6} $  \\
					\hline
				\end{tabular}
				\caption{Leading contribution to the wave-function renormalization of a gauge 1-form due to a loop of massive charged complex scalars, as given by eq. \eqref{eq:1-formloopscalarexact}, in the limit $\Lambda\gg m_n$, for different number of spacetime dimensions $2 \leq d \leq 11$. }
				\label{tab:1-formloopscalarLambda>>m}\end{center}
		\end{table}  
In the other relevant limit, namely when $\Lambda \simeq m_n$, the expansion of eq. \eqref{eq:1-formloopscalarexact} produces instead 
\begin{equation}
			\begin{split}
				\frac {\partial \Pi_n(p^2)}{\partial p^2} \bigg\rvert_{p=0}  =  - g^2 q_n^2 \ \frac{ \pi^{d/2}}{ 4 d\, (2 \pi)^d\ \Gamma(d/2) }  & \left\{ 2(d-3) +d(d-2)\left[\psi\left( \frac{d+2}{4}\right)-  \psi\left( \frac{d+4}{4} \right)  \right] \right\} \Lambda^{d-4} \\
				& \quad +  \mathcal{O}(\Lambda-m_n)   \, .
			\end{split}
\label{eq:1-formloopscalarLambda=m}
\end{equation}
The precise values for this expression in different number of spacetime dimensions are summarized in Table \ref{tab:1-formloopscalarLambda=m}.
		\begin{table}[t]\begin{center}
				\renewcommand{\arraystretch}{2.00}
				\begin{tabular}{|c||c|c|c|c|c|}
					\hline
					$d$ & 2 & 3 & 4 & 5 & 6 \\
					\hline 
					$\dfrac {\partial \Pi_n}{\partial p^2} \bigg\rvert_{p=0}$ &
					$-\dfrac{1}{16 \pi}\dfrac{g^2 \, q_n^2}{m_n^2}$ & 
					$ -\dfrac{(3\pi -8)}{48 \pi^2  }\dfrac{g^2 \, q_n^2}{m_n}$ &
					$ \begin{aligned}[t]-\tfrac{(8\log(2)-5)}{128 \pi^2 } \times &\\g^2 \, q_n^2 &\end{aligned}$  & 
					$ \begin{aligned}[t]-\tfrac{(16-5\pi)}{160 \pi^3}\times &\\ g^2 \, q_n^2  \, \Lambda & \end{aligned}$ &
					$ \begin{aligned}[t]-\tfrac{( 17-24\log(2))}{1536 \pi^3}\times &\\ g^2 \, q_n^2  \, \Lambda^2 & \end{aligned}$ \\
					\hline 
					\hline
					$d$ &  7 & 8 & 9 & 10 & 11\\
					\hline 
					$\dfrac {\partial \Pi_n}{\partial p^2} \bigg\rvert_{p=0}$ &
					$ \begin{aligned}[t] -\tfrac{ (105\pi-328)}{20160 \pi^4}\times &\\  g^2 \, q_n^2  \, \Lambda^3 & \end{aligned}$ &
					$\begin{aligned}[t]-\tfrac{(16\log(2)-11)}{8192 \pi^4}\times &\\ g^2 \, q_n^2  \, \Lambda^4  & \end{aligned}$ &
					$ \begin{aligned}[t]-\tfrac{(992-315\pi ) }{604800 \pi^5}\times &\\ g^2 \, q_n^2  \, \Lambda^5 & \end{aligned}$ &
					$ \begin{aligned}[t]-\tfrac{ (167-240\log(2)) }{1474560 \pi^5}\times &\\ g^2 \, q_n^2  \, \Lambda^6 & \end{aligned}$ &
					$ \begin{aligned}[t] -\tfrac{(385\pi-1208) }{10348800 \pi^6}\times &\\  g^2 \, q_n^2  \, \Lambda^7 & \end{aligned}$  \\
					\hline
				\end{tabular}
				\caption{Leading contribution to the wave-function renormalization of a gauge 1-form due to a loop of massive charged complex scalars, given by eq. \eqref{eq:1-formloopscalarexact}, in the limit $\Lambda \simeq m_n$, for different number of spacetime dimensions $2 \leq d \leq 11$.}
				\label{tab:1-formloopscalarLambda=m}\end{center}
		\end{table}  

\subsubsection*{Fermionic loop}
		
Let us consider now the effect of the coupling of the 1-form to a spin-$\frac{1}{2}$ fermion $\psi^{(n)}$, with mass $m_n$ and charge $q_n$ (c.f. action \eqref{eq:SpsinA1}. The corresponding one-loop Feynamn diagram is displayed in Figure \ref{fig:1-formloopfermion}, and it takes the form
\begin{equation}\label{eq:A1selfenergyfermion}
			\begin{aligned}
				\Pi^{\mu\nu}_n(p^2) \ &= - (\i g)^2 \, q_n^2  \int \frac {\text{d}^dq}{(2\pi)^d}\ \text{tr} \left (\frac {1}{\i \slashed{q}+m_n}\ \gamma^\mu \ \frac {1}{\i (\slashed{q}-\slashed{p})+m_n} \ \gamma^\nu  \right) \\
				&= g^2\ q_n^2  \int \frac {\text{d}^dq}{(2\pi)^d} \text{tr} \left ( \frac{(-\i \slashed{q} + m_n)\, \gamma^\mu \, (-\i(\slashed{q}-\slashed{p})+m_n)\, \gamma^\nu }{(q^2+m_n^2)((q-p)^2+m_n^2)} \right)\ .  
			\end{aligned}
\end{equation}
In order to perform the traces of the numerator we make use of the relations \eqref{eq:scalarloopfermionstraces}, as well as
\begin{equation}
			\text{tr} \left(\gamma^{\mu} \gamma^{\nu} \gamma^{\rho} \gamma^{\sigma}\right)= \fdim \left( \delta^{\mu \nu}\delta^{\rho \sigma}-\delta^{\mu \rho}\delta^{\nu \sigma}+\delta^{\mu \sigma}\delta^{\rho \nu} \right)\ ,
\label{eq:traceidentityfour}
\end{equation}
to obtain
\begin{equation}
\label{eq:traces1-formloop}
			\begin{split}
				\text{tr}& \Big \lbrace (-\i \slashed{q} + m_n)\, \gamma^\mu \, (-\i(\slashed{q}-\slashed{p})+m_n)\, \gamma^\nu \Big \rbrace \\
				& =\, \fdim \, \left( \delta^{\mu \nu} \, (q^2 - q\cdot p + m_n^2)- 2\,  q^\mu q^\nu+p^\mu q^\nu+p^\nu q^\mu  \right) \, . 
			\end{split}
\end{equation}
By differentiating with respect to $p^2$ so as to select the piece that contributes to the propagator, and after taking into account the fact that  linear and cubic  terms in $q^\mu$ yield zero upon performing the angular integration over $q$ --- together with identical cancellations in the term proportional to $(q\cdot p)$ and \eqref{eq:qmuqnuaverage}, we get
\begin{equation}\label{eq:1formfermionloopcompleteexpression}
\begin{aligned}
	\frac{\partial \Pi^{\mu \nu}_n(p^2)}{\partial p^2} \bigg\rvert_{p=0}  &= \, -\fdim\, g^2\,  q_n^2 \, \delta^{\mu\nu}  \int \dfrac{d^d q}{(2\pi)^d} \dfrac{1}{(q^2+m_n^2)^2}\\
    &+ \,  \fdim\,  g^2\,  q_n^2\,   \frac{2}{d} \, \delta^{\mu\nu}  \int \dfrac{d^d q}{(2\pi)^d} \dfrac{q^2}{(q^2+m_n^2)^3}\, .    
\end{aligned}
\end{equation}
Similarly to the modulus case, the second piece has the same form as the scalar contribution but with an opposite sign. In fact, by taking into account that we are now considering a complex scalar with two real degrees of freedom, it can be seen that in the presence of an equal number of fermionic and bosonic degrees of freedom with identical mass and charge, the cancellation between the scalar contribution and this second term from the fermions would be exact. Therefore, the precise expression for this correction (along with its asymptotic expansions) can be easily obtained from eqs. \eqref{eq:1-formloopscalarexact}-\eqref{eq:1-formloopscalarLambda=m} by simply multiplying by a factor of $-\fdim/2$.
		
Let us now focus on the first term. Notice that, after extracting the tensorial structure, it gives exactly the same contribution as \eqref{eq:fermionloopddim} upon substituting $\mu_n^2 \,  \to \,  g^2 \, q_n^2$. Hence, we can use, \emph{mutatis mutandis}, the corresponding formulae from the modulus section, that we summarize here for completeness. By introducing a UV cut-off $\Lambda$ and upon performing the integral, the first term in \eqref{eq:1formfermionloopcompleteexpression} reads as
\begin{equation}
			\frac{\partial \Pi_n(p^2)}{\partial p^2} \bigg\rvert_{p=0}  \, =  \, - g^2\, q_n^2  \ \frac{ \fdim \pi^{d/2}  }{  (2 \pi)^d \, \Gamma(d/2) } \ \frac{\Lambda^d}{m_n^4} \  \left[  \frac{m_n^2}{\Lambda^2+m_n^2}  + \left(\frac{2}{d}-1\right) \ _2{\cal F}_1\left( 1,\frac{d}{2};\frac{d+2}{2}; -\frac{\Lambda^2}{m_n^2}\right) \right]  \, .
\label{eq:1-formloopfermionexact}
\end{equation}
		\begin{table}[t]\begin{center}
				\renewcommand{\arraystretch}{2.00}
				\begin{tabular}{|c||c|c|c|c|c|}
					\hline
					$d$ & 2 & 3 & 4 & 5 & 6 \\
					\hline 
					$\dfrac {\partial \Pi_n}{\partial p^2} \bigg\rvert_{p=0}$ &
					$-\dfrac{1}{2 \pi}\dfrac{g^2\, q_n^2}{m_n^2}$ & 
					$ -\dfrac{1}{4 \pi  }\dfrac{g^2\, q_n^2}{m_n}$ &
					$ -\dfrac{g^2\, q_n^2 }{4 \pi } \log \left(\dfrac{\Lambda ^2}{m_n^2}\right)$ & 
					$ -\dfrac{g^2\, q_n^2  \, \Lambda }{3 \pi^2}$ &
					$ -\dfrac{g^2\, q_n^2  \, \Lambda^2 }{16 \pi^3} $ \\
					\hline 
					\hline
					$d$ &  7 & 8 & 9 & 10 & 11\\
					\hline 
					$\dfrac {\partial \Pi_n}{\partial p^2} \bigg\rvert_{p=0}$ &
					$  -\dfrac{g^2\, q_n^2  \, \Lambda^3 }{45 \pi^4}  $ &
					$-\dfrac{g^2\, q_n^2  \, \Lambda^4  }{192 \pi^4} $ &
					$ -\dfrac{g^2\, q_n^2  \, \Lambda^5  }{525 \pi^5}$ &
					$ -\dfrac{g^2\, q_n^2  \, \Lambda^6   }{2304 \pi^5}$ &
					$  -\dfrac{g^2\, q_n^2  \, \Lambda^7 }{6615 \pi^6} $  \\
					\hline
				\end{tabular}
				\caption{Leading contribution to the wave-function renormalization of a gauge 1-form due to a loop of massive charged fermions, as given by eq. \eqref{eq:1-formloopfermionexact}, in the limit $\Lambda\gg m_n$ for different number of spacetime dimensions $2 \leq d \leq 11$.}
				\label{tab:1-formloopfermionLambda>>m}\end{center}
		\end{table}  
In the limit $\Lambda \gg m_n$, the leading piece from the fermionic loop to the propagator when $d>4$ is
\begin{equation}
			\frac {\partial \Pi_n^{(d>4)}(p^2)}{\partial p^2} \bigg\rvert_{p=0}\, = \, -g^2\, q_n^2\,  \frac{2^{\lfloor \frac{d+2}{2} \rfloor} \pi^{d/2}}{(2 \pi)^d\ \Gamma\left( d/2 \right) \ (d-4)}   \ \Lambda^{d-4} \, + \, \mathcal{O}\left(\Lambda^{d-6}\, m_n^2\right) + \mathrm{const.}\, .
\label{eq:1-formloopfermionsd>4}
\end{equation}
Similarly, for $d<4$ the dominant contribution takes the form
\begin{equation}
			\frac {\partial \Pi_n^{(d<4)}(p^2)}{\partial p^2} \bigg\rvert_{p=0}\, = \, - g^2\, q_n^2 \ \frac{2^{\lfloor\frac{d-2}{2} \rfloor} \ \pi^{\frac{d+2}{2}} }{(2 \pi)^d  \ \Gamma\left( d/2 \right)} \ \frac{(2-d)}{\sin\left( d \pi/2\right)} \ \frac{1}{m_n^{4-d}} + \, \mathcal{O}\left(\frac{1}{\Lambda^{4-d}}\right)\, ,
\label{eq:1-formloopfermionsd<4}
\end{equation}
with the quotient $(2-d)/\sin\left( d \pi/2\right)$  defined as a limit with value $2/ \pi$ for $d=2$. For the marginal case,  the expected logarithmic divergence is obtained
\beq
		\frac {\partial \Pi_n^{(d=4)}(p^2)}{\partial p^2} \bigg\rvert_{p=0} \, =\, -  \frac{g^2\, q_n^2 }{4 \pi^2 }\log \left( \frac{\Lambda^2}{m_n^2} \right) \ + \ \mathcal{O}\left( \Lambda^0 \right)\, .
\label{eq:1-formloopfermionsd=}
\eeq
These results are summarized in Table \ref{tab:1-formloopfermionLambda>>m}.
		
Taking instead the limit $\Lambda \simeq m_n$ in eq. \eqref{eq:1-formloopfermionexact}, one arrives at
\begin{equation}
			\frac {\partial \Pi_n(p^2)}{\partial p^2} \bigg\rvert_{p=0} \, = \, - g^2\, q_n^2 \ \frac{2^{\lfloor\frac{d}{2}-2 \rfloor}  \pi^{d/2}}{ (2 \pi)^d\ \Gamma(d/2) }   \left\{ 2+(d-2)\left[\psi\left( \frac{d}{4}\right)-  \psi\left( \frac{d+2}{4} \right)  \right] \right\} \Lambda^{d-4} + \, \mathcal{O}(\Lambda-m_n)\, .
\label{eq:1-formloopfermionLambda=m}
\end{equation}
As happened with the scalar modulus before, since we have $\Lambda\simeq m_n$, the asymptotic dependence with the relevant scale coincides with the $\Lambda \gg m_n$ limit, and only the numerical prefactors change. The relevant leading terms for $2\leq d \leq 11$ are shown in Table \ref{tab:1-formloopfermionLambda=m}.
		\begin{table}[t]\begin{center}
				\renewcommand{\arraystretch}{2.00}
				\begin{tabular}{|c||c|c|c|c|c|}
					\hline
					$d$ & 2 & 3 & 4 & 5 & 6 \\
					\hline 
					$\dfrac {\partial \Pi_n}{\partial p^2} \bigg\rvert_{p=0}$ &
					$-\dfrac{1}{4 \pi}\dfrac{g^2\, q_n^2}{m_n^2}$ & 
					$ -\dfrac{(\pi -2)}{8 \pi^2  }\dfrac{g^2\, q_n^2}{m_n}$ &
					$ -\frac{(2\log(2)-1)}{8 \pi^2 } g^2\, q_n^2 $ & 
					$ \begin{aligned}[t]-\tfrac{(10-3\pi)}{24 \pi^3}\times &\\ g^2\, q_n^2  \, \Lambda & \end{aligned}$ &
					$ \begin{aligned}[t]-\tfrac{(3- 4\log(2))}{32 \pi^3}\times &\\ g^2\, q_n^2  \, \Lambda^2 & \end{aligned}$ \\
					\hline 
					\hline
					$d$ &  7 & 8 & 9 & 10 & 11\\
					\hline 
					$\dfrac {\partial \Pi_n}{\partial p^2} \bigg\rvert_{p=0}$ &
					$ \begin{aligned}[t] -\tfrac{ (15\pi- 46)}{360 \pi^4}\times &\\  g^2\, q_n^2  \, \Lambda^3 & \end{aligned}$ &
					$\begin{aligned}[t]-\tfrac{(3\log(2)-2)}{96 \pi^4}\times &\\ g^2\, q_n^2  \, \Lambda^4  & \end{aligned}$ &
					$ \begin{aligned}[t]-\tfrac{(334-105\pi) }{12600 \pi^5}\times &\\ g^2\, q_n^2  \, \Lambda^5 & \end{aligned}$ &
					$ \begin{aligned}[t]-\tfrac{ (17-24\log(2)) }{4608 \pi^5}\times &\\ g^2\, q_n^2  \, \Lambda^6 & \end{aligned}$ &
					$ \begin{aligned}[t] -\tfrac{(315\pi -982) }{264600 \pi^6}\times &\\  g^2\, q_n^2  \, \Lambda^7 & \end{aligned}$  \\
					\hline
				\end{tabular}
				\caption{Leading contribution to the wave-function renormalization of a gauge 1-form due to a loop of massive charged fermions, as given by eq. \eqref{eq:1-formloopfermionexact}, in the limit $\Lambda\simeq m_n$ for different number of spacetime dimensions $2 \leq d \leq 11$.}
				\label{tab:1-formloopfermionLambda=m}\end{center}
		\end{table} 

\section{Self-energy of a Weyl fermion}
\label{ap:LoopsWeylfermion}
	
To close up this appendix, we will consider a \emph{chiral} (i.e. we restrict to even-dimensional spacetimes) spin-$\frac{1}{2}$ field, $\chi$, coupled to massive (complex) scalars $\{\phi^{(n)}\}$ and Dirac fermions, $\{\Psi^{(n)}\}$, through the following Yukawa-like interactions
\begin{equation}\label{eq:fermion&bosonYukawas(ap)}
			\mathcal{Y}_n\ \overline{\phi^{(n)}} \left(\psi^{(n)}\chi \right)\, ,  
\end{equation}
where $\mathcal{Y}_n$ denotes the coupling constant and $n \in \mathbb{Z} \setminus \lbrace0\rbrace$ labels the massive fields. We also use $\psi^{(n)}$ to denote the Weyl fermion of the same chirality as $\chi$, which pairs up with its charge conjugate (say the one labeled by $-n$) so as to form the aforementioned massive Dirac spin-$\frac{1}{2}$ field, i.e. $\Psi^{(n)}=\left(\psi^{(n)}, \overline{\psi^{(-n)}} \right)^{\text{T}}$. 
		
In the following, and for simplicity, we will use Dirac fermions all along so as to perform the relevant loop integrals. Therefore, in order to take into account that the massless field $\chi$ is chiral we define a new Dirac fermion $\mathcal{X}$, which reduces to $\chi$ upon using the familiar chirality projector $P_{-}=\frac{1}{2}(1-\gamma^{d+1})$, i.e. $\chi = P_{-}\ \mathcal{X}$, c.f. eq. \eqref{eq:gammad+1}. With this in mind, it is easy to see that the interaction \eqref{eq:fermion&bosonYukawas(ap)} above can be written in terms of $\{\Psi^{(n)},\, \mathcal{X}\}$ as follows
\begin{equation}\label{eq:interactionsfermion2(ap)}
			\mathcal{Y}_n\ \phi^{(n)} \left( \overline{\Psi^{(n)}} P_{-} \mathcal{X} \right) + \text{h.c.}\, .  
\end{equation}
The idea then is to extract again the momentum-dependent part of the exact propagator associated to the massless fermion $\chi$ at $\mathcal{O}(\hbar)$ in the effective action, which after analytically extending to Euclidean signature reads formally as
\beq\label{eq:Euclexactpropagator(ap)}
		S(\slashed{p})= \frac{1}{\i\slashed{p}}\, P_{-} + \frac{1}{\i\slashed{p}}\, P_{-}\, \left(\i \Sigma(\slashed{p})\right)\, \frac{1}{\i\slashed{p}}\, P_{-} + \ldots\, ,
\eeq
where the fermion self-energy $\i \Sigma(\slashed{p})$ corresponds in this case to the (amputated) one-loop Feynman graph displayed in Figure \ref{fig:kineticfermionsbas}. (Notice that this is nothing but the fermionic analogue of $\Pi(p^2)$ in \eqref{eq:exactpropscalar}.) 
		
\subsubsection*{Loop computation}
		
We will concentrate on the first diagram\footnote{The analysis involving massive vectors as in Figure \ref{fig:fermionloopvector} should give us analogous results.} in Figure \ref{fig:kineticfermionsbas} involving Dirac fermions $\{\Psi^{(n)}\}$ with masses $\{m_n^{{\text{f}}}\}$ as well as complex bosonic scalars $\{\phi^{(n)}\}$ with mass given by $\{m_n^{{\text{b}}}\}$, which reads
\beq
		\i \Sigma_n(\slashed{p}) \ = |\mathcal{Y}_n|^2 \int \frac{\text{d}^dq}{(2\pi)^d} \frac{P_{-} \left( -\i \slashed{q} + m_n^{\text{f}}\right) P_{+}}{q^2+(m_n^{\text{f}})^2} \frac {1}{(q-p)^2+(m_n^{{\text{b}}})^2}\, ,
\label{eq:selfenergychi(ap)}
\eeq
where the projection operators $P_{\pm}$ arise from the Feynman rules associated to the interaction \eqref{eq:interactionsfermion2(ap)}. There are several interesting things to notice before moving on with the loop computation. First, and due to the anti-commutation properties between $\gamma^{d+1}$ and the $\gamma^{\mu}$ (namely $\lbrace \gamma^{\mu}, \gamma^{d+1} \rbrace=0$), the operators $P_{\pm}$ project out the term proportional to $m_n^{{\text{f}}}$ in the numerator of eq. \eqref{eq:selfenergychi(ap)} above whilst keeping the one $\propto \slashed{q}$. This ultimately translates into the fact that the self-energy provides no net contribution at $\mathcal{O}(\hbar)$ for the mass of the chiral field $\chi$.\footnote{This is actually ensured to be true at all orders in perturbation theory due to the chirality of the fermionic field $\chi(x)$.} We also notice that the self-energy includes the projector $P_{+}$, as it should since it is associated to the \emph{chiral} massless fermion, $\chi$.
		
Thus, in order to extract the wave-function renormalization one needs to focus on the piece in the self-energy linear in $p$. Therefore, one can mimic the discussion in the preceding sections by taking derivatives with respect to $p^{\mu}$, and then evaluating the resulting expression at $p=0$. Upon doing so one finds
\beq
		\begin{aligned}\label{eq:Euclwavefunctionfermion(ap)}
			\frac{\partial \Sigma_n(\slashed{p})}{\partial p^{\mu}} \bigg\rvert_{p=0} = \frac{-2 |\mathcal{Y}_n|^2 \delta_{\mu \nu} \gamma^{\nu}\ P_{+}}{d} \int \frac{\text{d}^dq}{(2\pi)^d} \frac{q^2}{\left[ q^2 + (m_n^{{\text{f}}})^2 \right] \left[ q^2 + (m_n^{{\text{b}}})^2 \right]^2}\, .
		\end{aligned}
\eeq
Notice that this has the correct sign so as to renormalize the wave-function of the massless fermion appropriately in eq. \eqref{eq:Euclexactpropagator(ap)}.

Now, in order to study the kind of corrections induced by the above diagram, we will first specialize to the easier case in which both towers present identical mass gaps, namely when $m_n^{{\text{b}}} = m_n^{{\text{f}}} = m_n$. One is thus lead to perform the following integral in momentum space (after introducing a UV cut-off $\Lambda$), which we already encountered in Section \ref{ap:Loops1-form} before (c.f. eq. \eqref{eq:1-formscalarloopprop})
\beq\label{eq:momentumintegralfermionsamemasses}
		\begin{aligned}
			\frac{\partial \Sigma_n(\slashed{p})}{\partial p^{\mu}} \bigg\rvert_{p=0} = \frac{-2 |\mathcal{Y}_n|^2 \delta_{\mu \nu} \gamma^{\nu}\ P_{+}}{d} \int_{|q| \leq \Lambda} \frac{\text{d}^dq}{(2\pi)^d} \frac{q^2}{\left(q^2 + m_n^2 \right)^3} \, .
		\end{aligned}
\eeq
Of course, this is not a coincidence, since one place in which this kind of diagrams naturally appears is in supersymmetric gauge theories, see discussion in Section \ref{sss:emergencefermion} in the main text. The behaviour of such integral depends, among various things, on the ratio $\Lambda/m_n$ as well as the spacetime dimension, $d$. For concreteness, let us show in here the explicit results for the case in which $\Lambda/m_n \gg 1$. For $d>4$, the integral diverges polynomially as 
\begin{equation}
			\frac{\partial \Sigma_n^{(d>4)}(\slashed{p})}{\partial p^{\mu}} \bigg\rvert_{p=0} = \, -|\mathcal{Y}_n|^2 \delta_{\mu \nu} \gamma^{\nu}\ P_{+}\  \frac{4\ \pi^{d/2}}{(2 \pi)^d\ \Gamma\left( d/2 \right) \ d\, (d-4)}    \ \Lambda^{d-4} \, + \, \mathcal{O}\left(\Lambda^{d-6}\right) + \mathrm{const.}\, ,
\label{eq:fermionloopd>4}
\end{equation}
whereas in lower dimensions it is convergent and the leading contribution is given by
\begin{equation}
			\frac{\partial \Sigma_n^{(d<4)}(\slashed{p})}{\partial p^{\mu}} \bigg\rvert_{p=0} = \, - |\mathcal{Y}_n|^2 \delta_{\mu \nu} \gamma^{\nu}\ P_{+} \ \frac{\pi^{\frac{d+2}{2}} }{4 \, (2 \pi)^d  \ \Gamma\left( d/2 \right)} \ \frac{(2-d)}{\sin\left( d \pi/2\right)} \ \frac{1}{m_n^{4-d}} + \, \mathcal{O}\left(\frac{1}{\Lambda^{4-d}}\right)\, ,
\label{eq:fermionloopd<4}
\end{equation}
with the quotient $(2-d)/\sin\left( d \pi/2\right)$ defined as a limit with value $2/ \pi$ for $d=2$. Finally, for the marginal case, we get the usual logarithmic behaviour
\beq
		\frac{\partial \Sigma_n^{(d=4)}(\slashed{p})}{\partial p^{\mu}} \bigg\rvert_{p=0} \, =\, -  \frac{|\mathcal{Y}_n|^2 \delta_{\mu \nu} \gamma^{\nu}\ P_{+}}{32 \pi^2 }\, \log \left( \frac{\Lambda^2}{m_n^2} \right) \ + \ \mathcal{O}\left( \Lambda^0 \right)\, .
\label{eq:fermionloopd=4}
\eeq

Let us come back to the more general expression, i.e. eq. \eqref{eq:Euclwavefunctionfermion(ap)}, in which we take the states running in the loop to have different masses. Performing the momentum integral we arrive at the analogue of \eqref{eq:momentumintegralfermionsamemasses}, namely
\begin{equation}
			\begin{aligned}
				\frac{\partial \Sigma_n(\slashed{p})}{\partial p^{\mu}} \bigg\rvert_{p=0} &= \, -|\mathcal{Y}_n|^2 \delta_{\mu \nu} \gamma^{\nu}\ P_{+}\  \frac{2\ \pi^{d/2}}{(2 \pi)^d\ \Gamma\left( d/2 \right)\ d} \ \frac{\Lambda^{d+2}}{\left[ (m_n^{{\text{f}}})^3 -m_n^{{\text{f}}} (m_n^{{\text{b}}})^2 \right]^2} \left[  \frac{(m_n^{{\text{f}}})^2\left[(m_n^{{\text{f}}})^2 - (m_n^{{\text{b}}})^2 \right]}{(m_n^{{\text{b}}})^2\ (\Lambda^2+(m_n^{{\text{b}}})^2)} \right.  \\
				& \quad \left.  +\, \dfrac{2}{d+2} \ _2{\cal F}_1\left( 1,\frac{d+2}{2};\frac{d+4}{2}; -\frac{\Lambda^2}{(m_n^{{\text{f}}})^2}\right) \right.  \\
				& \quad \left.  +\, \dfrac{(m_n^{{\text{f}}})^2\left[(m_n^{{\text{b}}})^2(d-2) - (m_n^{{\text{f}}})^2 d\right]}{(m_n^{{\text{b}}})^4\ (d+2)} \ _2{\cal F}_1\left( 1,\frac{d+2}{2};\frac{d+4}{2}; -\frac{\Lambda^2}{(m_n^{{\text{b}}})^2}\right) \right] \ ,
			\end{aligned}
\label{eq:momentumintegralfermiondiffmasses(ap)}
\end{equation}
which of course reduces to the previous expressions whenever the masses are taken to be equal.

\chapter{Generalities on charge-to-mass and species vectors}
\label{ap:generalities}

In this appendix we present a derivation of the formulae associated to the computation of the relevant scalar charge-to-mass and species vectors that are extensively used in Part \ref{part:pattern} of the thesis. Section \ref{s:compactificationNmfd} focuses on general compactifications of a $D$-dimensional gravitational theory on some Ricci-flat closed manifold of real dimension $n\in \mathbb{N}$. In Section \ref{ss:nestedcompactifications} we generalize the analysis to the case in which the compact space is a product of the form $\mathcal{X}'_n=\mathcal{X}_{n_1}\times \ldots \times\mathcal{X}_{n_N}$, with $n_i$ denoting the dimensionality of the corresponding submanifold. In both cases we take the opportunity to revisit the universal pattern presented in Chapter \ref{ch:pattern}, checking it explicitly.
	
\section{Compactification on an $n$-dimensional cycle}\label{s:compactificationNmfd}
	
Let us start by studying the kind of charge-to-mass vectors that typically appear in string-motivated EFTs. In order to be as general as possible, we consider a $D$-dimensional theory compactified down to $d=D-n$ spacetime dimensions. We denote $\mathcal{V}_n$ the overall volume modulus associated to the internal compact manifold, $\mathcal{X}_n$, measured in $D$-dimensional Planck units. Suppose that we focus on a sector of the theory described by the following simple action \cite{Etheredge:2022opl}
\begin{equation}\label{eq:higherDdim}
	S_{D} \supseteq \int \dd^{D}x\, \sqrt{-g_D}\,  \left[  \frac{1}{2\kappa_{D}^2}\mathcal{R}_{D} - \frac{1}{2}\left(\partial \hat \phi \right)^2 \right]\, ,
\end{equation}
where $\hat{\phi}$ is some generic canonically normalized modulus. Note that one may also think of $\hat \phi$ as parametrizing some fixed (asymptotically) geodesic trajectory in a multi-moduli set-up. Upon compactification on the $n$-fold $\mathcal{X}_n$, one arrives at
\begin{equation}\label{eq:ddim}
	S_{d} \supseteq \int \dd^{d}x\, \sqrt{-g_d}\,  \left[ \frac{1}{2\kappa_{d}^2} \left(\mathcal{R}_{d} - \frac{d+n-2}{n (d-2)} \left(\partial \log \mathcal{V}_n \right)^2 \right)- \frac{1}{2} \left(\partial \hat \phi \right)^2 \right]\, ,
\end{equation}
where we have retained only the scalar-tensor sector of the lower dimensional theory, ignoring possible extra fields arising in the dimensional reduction process.\footnote{To obtain \eqref{eq:ddim} in such form one needs to perform a Weyl rescaling of the $d$-dimensional metric as follows $g_{\mu \nu} \to g_{\mu \nu} \mathcal{V}_n^{-\frac{2}{d-2}}$.} One can then define a canonically normalized volume modulus
\begin{equation}\label{eq:canonicalvolume}
	\hat \rho = \frac{1}{\kappa_d}\sqrt{\frac{d+n-2}{n(d-2)}} \log \mathcal{V}_n\, ,
\end{equation}
which indeed controls the overall Kaluza-Klein scale associated to the compact internal space
\begin{equation}\label{eq:KKscale}
	m_{\text{KK},\, n} \sim  M_{\text{Pl};\, d}\, e^{-\kappa_d \sqrt{\frac{d+n-2}{n (d-2)}} \hat \rho}\, .
\end{equation}
As customary, this tower of states becomes exponentially light when taking the decompactification limit $\hat \rho \to \infty$. In terms of scalar charge-to-mass vectors one would then write
\begin{equation}\label{eq:kkcharge2mass}
	\vec{\zeta}_{\text{KK},\, n} = \left( 0 , \sqrt{\frac{d+n-2}{n (d-2)}} \right)\, ,
\end{equation}
where the first (last) entry corresponds to the normalized modulus $\hat \phi$ ($\hat \rho$).
	
Let us also assume that the scalar $\hat \phi (x)$ is non-compact, and that the higher dimensional theory satisfies the Distance Conjecture \cite{Ooguri:2006in}. Therefore, there should exist an infinite tower of particles with mass behaving asymptotically as follows
\begin{equation}\label{eq:SDCDdim}
	m_{\text{tow}} \sim  M_{\text{Pl};\, D}\, e^{-\kappa_D \lambda_D \hat \phi}\, ,
\end{equation}
where $\lambda_D$ is nothing but the $D$-dimensional scalar charge-to-mass ratio along the positive $\hat \phi$-direction. If such tower of particles is inherited by the lower-dimensional theory, they would present a mass which in Planck units depends on both $\hat \phi$ and the volume modulus $\hat \rho$ through the relation
\begin{equation}\label{eq:SDCddim}
	m_{\text{tow}} \sim  M_{\text{Pl};\, d}\, \exp\left\{-\kappa_d \lambda_D \hat \phi - \kappa_d \sqrt{\frac{n}{(d+n-2)(d-2)}} \hat \rho\right\}\, ,
\end{equation}
where the second term in the exponent arises just from the ratio $M_{\text{Pl};\, D}/M_{\text{Pl};\, d}$. Again, in terms of scalar charge-to-mass vectors one obtains
\begin{equation}\label{eq:chargetomasstower}
	\vec{\zeta}_{\text{t}} = \left( \lambda_D , \sqrt{\frac{n}{(d+n-2)(d-2)}} \right)\, .
\end{equation}
Note that if $\hat{\phi}$ denotes the $D$-dimensional dilaton in some string theory, then $\lambda_D=\frac{1}{\sqrt{D-2}}=\frac{1}{\sqrt{d+n-2}}$ \cite{Etheredge:2022opl, vandeHeisteeg:2023ubh}, whilst if it corresponds to a volume modulus from a higher compactification (i.e. from $D'=D+n'$ to $D$ spacetime dimensions), then $\lambda_D=\sqrt{\frac{D+n'-2}{n'(D-2)}}=\sqrt{\frac{d+n+n'-2}{n'(d+n-2)}}$. Remarkably, this also encompasses the case in which one of the moduli corresponds to some dilatonic field, since upon taking the limit $n'\to \infty$ the first entry of the scalar charge-to-mass vector becomes $\frac{1}{\sqrt{D-2}}$.
	
For the species scale, on the other hand, we will distinguish between two possibilities, as predicted by the Emergent String Conjecture \cite{Lee:2019wij}. First of all, if the limit corresponds to an emergent critical string, the QG cut-off will be given by the string scale since the set of light states will be dominated by an exponentially large number of string excitation modes. Because of this, one has
\begin{equation} \label{eq:stringmassdependence}
	\LSP \sim m_{\rm string}\sim M_{\text{Pl};\, D}\, \exp\left\{-\kappa_D\frac{1}{D-2} \hat \phi\right\}\, .
\end{equation}
Hence $\vec{\mathcal{Z}}_{\rm osc}=\vec{\zeta}_{\rm osc}$, so that in this limit
\begin{equation}\label{eq:generalstringlimit}
	\vec{\zeta}_{\text{t}}\cdot\vec{\mathcal{Z}}_{\rm sp}=|\vec{\zeta}_{\rm osc}|^2=\frac{1}{d-2}\, ,
\end{equation}
thus fulfilling \eqref{eq:pattern}. Notice that \eqref{eq:generalstringlimit} above is also verified when $\vec{\zeta}_{\text{t}}=\vec{\zeta}_{\text{KK},\, n}$ (see Figure \ref{sfig:KKstring}), since for an emergent string limit the KK tower falls at the same rate as the string mass \cite{Lee:2019wij}. Otherwise, one could retrieve a critical string in $d<10$.
	
The second possibility would correspond to explore some decompactification limit, namely when the tower from \eqref{eq:SDCDdim} is of Kaluza-Klein nature (in some duality frame). In such a case, one would have three different species vectors: those which are parallel to the original $\zeta$-vectors and a new one arising as an effective combination thereof. For the former, one can write 
\begin{equation}\label{eq:Zvectorgeneral}
	\vec{\mathcal{Z}}_{{\rm KK},\, n'}=\frac{n'}{d+n'-2}\vec{\zeta}_{{\rm KK},\, n'}\, , \qquad \vec{\mathcal{Z}}_{{\rm KK},\, n}=\frac{n}{d+n-2}\vec{\zeta}_{{\rm KK},\, n}\, ,
\end{equation}
where $\vec{\zeta}_{{\rm KK},\, n}$ is given by \eqref{eq:kkcharge2mass} above and with
\begin{equation}
	\vec{\zeta}_{{\rm KK},\, n'} = \left( \sqrt{\frac{d+n+n'-2}{n' (d+n-2)}} , \sqrt{\frac{n}{(d+n-2)(d-2)}} \right)\, ,
\end{equation}
thus satisfying $|\vec{\zeta}_{{\rm KK},\, n'}|^2=\frac{d+n'-2}{n' (d-2)}$. Therefore, whenever we explore an asymptotic direction parallel to one of these two, the species scale will be parametrically controlled by the Planck scale of the $(d+n')$-dimensional (resp. $(d+n)$) theory. As an example, upon taking the limit $\hat\phi, \hat\rho \to \infty$ along the $\vec{\zeta}_{{\rm KK},\, n'}$\,-direction one finds
\begin{align} \label{eq:kkspeciesvector}
	\LSP \sim M_{{\rm Pl};\, d+n'}\sim M_{\text{Pl};\, d}\, \left(\frac{m_{{\rm KK},\, n'}}{M_{\text{Pl};\, d}}\right)^{\frac{n'}{d+n'-2}}\, ,
\end{align}
with $m_{{\rm KK},\, n'}$ denoting the mass scale of the corresponding KK-like tower. For intermediate directions, however, the dominant species vector is that obtained by combining the previous ones as follows
\begin{equation}\label{eq:effectiveKKspeciesvector}
	\vec{\mathcal{Z}}_{{\rm KK},\, n+n'} = \frac{1}{d+n+n'-2}\left( n'\, \vec{\zeta}_{{\rm KK},\, n'} + n\, \vec{\zeta}_{{\rm KK},\, n}  \right)\, ,
\end{equation}
which is indeed controlled by the Planck scale of the $(d+n+n')$-dimensional parent theory, see Figure \ref{sfig:twoKK}. With this we can now check if the pattern \eqref{eq:pattern} is satisfied. Once again, for the directions determined by any of the three species vectors one easily verifies that $\vec{\zeta}_{\text{t}} \cdot \vec{\mathcal{Z}}_{\text{sp}}=\frac{1}{d-2}$. In particular, when probing the $\vec{\mathcal{Z}}_{{\rm KK},\, n+n'}\,$-direction what one effectively does is decompactifying both cycles at the same rate, such that the total KK mass yields a charge-to-mass vector of the form
\begin{equation} \label{eq:effectivezeta}
	\vec{\zeta}_{{\rm KK},\, n+n'} = \frac{1}{n+n'}\left( n'\, \vec{\zeta}_{{\rm KK},\, n'} + n\, \vec{\zeta}_{{\rm KK},\, n}  \right)\, ,
\end{equation}
which happens to lie at the point closest to the origin within the polytope generated by $\vec{\zeta}_{{\rm KK},\, n'}$ and $\vec{\zeta}_{{\rm KK},\, n}$, see Figure \ref{sfig:twoKK}.
	
For intermediate cases, given that the species scale is determined by $\vec{\mathcal{Z}}_{{\rm KK},\, n+n'}$ together with the fact that $\vec{\zeta}_{{\rm KK},\, n+n'}$ is orthogonal to the line joining the two $\zeta$-vectors, one finds that \eqref{eq:pattern} still holds for any asymptotically light tower.
	
\section{Generalization to `nested' compactifications}
\label{ss:nestedcompactifications}	
	
The previous analysis can be easily generalized to the case in which our $D$-dimensional theory is compactified down to $d=D-n$ on an $n$-dimensional manifold given by the Cartesian product $\mathcal{X}_n=\mathcal{X}_{n_1}\times \ldots \times\mathcal{X}_{n_N}$, with $n=\sum_{i=1}^N n_i$. This can be alternatively seen as a step-by-step (or `nested') compactification
\begin{equation}\label{eq:compactchain}
	\notag D=d+\sum_{i=1}^N n_i\to d+\sum_{i=2}^N n_i\to \ldots \to d+n_N\to d\, ,
\end{equation}
where the order of the compactification chain is unimportant and only amounts to a certain rotation of the associated scalar charge-to-mass vectors, hence not affecting neither their length nor the angles subtended between them. With this in mind, one finds that the KK tower obtained from the decompactifying any $\mathcal{X}_{n_i} \subset \mathcal{X}_n$ is given by
\begin{equation}\label{eq: gen eq}
	\zeta^j_{{\rm KK},\, n_i}=\left\{
	\begin{array}{ll}
		0& \qquad \text{if }j<i\\
		\sqrt{\frac{d+\sum_{l=i}^Nn_l-2}{n_i(d+\sum_{l=i+1}^Nn_l-2)}}& \qquad \text{if }i=j\\
		\sqrt{\frac{n_j}{(d+\sum_{l=j}^Nn_l-2)(d+\sum_{l=j+1}^Nn_l-2)}}& \qquad \text{if }j>i
	\end{array}
	\right.
\end{equation} 
Notice that this also encompasses the case in which one of the moduli corresponds to some $D$-dimensional dilaton, upon setting $n_0\to \infty$, so that the zero-th entry of the scalar charge-to-mass vector becomes $\frac{1}{\sqrt{D-2}}$.
	
On the other hand, given a subset $\{\vec{\zeta}_{{\rm KK},\, m_j}\}_{j=1}^M\subseteq \{\vec{\zeta}_{{\rm KK},\, n_i}\}_{i=1}^N$, one can show that
\begin{equation}\label{eq: mult decompact}
	\vec{\zeta}_{{\rm KK},\, \sum_j m_j}=\frac{1}{\sum_{j=1}^M m_j}\sum_{j=1}^M m_j\, \vec{\zeta}_{{\rm KK},\, m_j}\, ,
\end{equation}
corresponds to the `effective' KK tower associated to the joint decompactification of $\mathcal{X}_{m_1}\times \ldots \times \mathcal{X}_{m_M}$, where the volume of each of the cycles grows at the same rate. Incidentally, it can be seen to coincide with the point of the polytope spanned by $\{\vec{\zeta}_{{\rm KK},\, m_j}\}_{j=1}^M$ located closest to the origin.
	
Taking infinite distance limits, the easiest possibility would be an emergent string limit, for which $\vec{\zeta}_{\text{t}}\cdot\vec{\mathcal{Z}}_{\rm sp}=|\vec{\zeta}_{\rm osc}|^2=\frac{1}{d-2}$ is trivially fulfilled. The other option would correspond to explore a decompactification limit from $d$ to $d+\sum_{j=1}^M m_j$ dimensions, with $\{m_j\}_{j=1}^M\subseteq\{n_i\}_{i=1}^N$, where we allow the possibility of a dilaton-like direction by taking $m_0\to\infty$. In this case the species scale will be parametrically given by the Planck scale of the ($d+\sum_{j=1}^M m_j$)-dimensional theory,\footnote{If $m_0\to\infty$ then the species scale is again given by the fundamental string scale.} so that
\begin{align} \label{eq:kkspeciesvectorgeneral}
	\LSP&\sim M_{{\rm pl},\, d+\sum_{j=1}^M m_j}\sim M_{\text{Pl};\, d}\exp\left\{-\kappa_d\frac{\sum_{j=1}^M m_j}{(d+\sum_{j=1}^M m_j-2)(d-2)}\hat{\rho}\right\}\notag\\
	&\sim M_{\text{Pl};\, d}\left(\frac{m_{{\rm KK},\, \sum_{j=1}^M m_j}}{M_{{\rm Pl},d}}\right)^{\frac{\sum_{j=1}^M m_j}{d+\sum_{j=1}^M m_j-2}},
\end{align}
where $\hat{\rho}$ is the normalized modulus denoting the volume being decompactified. As a result, we find
\begin{equation}
	\vec{\mathcal{Z}}_{\rm sp}=\frac{\sum_{j=1}^M m_j}{d+\sum_{j=1}^M m_j-2}\vec{\zeta}_{{\rm KK},\, \sum_{j=1}^M m_j}=\frac{1}{d+\sum_{j=1}^M m_j-2}\sum_{j=1}^M m_j\, \vec{\zeta}_{{\rm KK},\, m_j}\, ,
\end{equation}
where \eqref{eq: mult decompact} is used. Now, for the leading tower, we have two possibilities. First of all, we might be moving in the joint compactification direction, so $\vec{\zeta}_{\text{t}}=\vec{\zeta}_{{\rm KK},\, \sum_{j=1}^M m_j}$, and thus
\begin{equation}
	\vec{\zeta}_{\text{t}}\cdot\vec{\mathcal{Z}}_{\rm sp}=\frac{\sum_{j=1}^M m_j}{d+\sum_{j=1}^M m_j-2} |\vec{\zeta}_{{\rm KK},\, \sum_{j=1}^M m_j}|^2=\frac{1}{d-2}\, .
\end{equation}
The other possibility is that we move in some other direction, where while still decompactifying  $\mathcal{X}_{m_1}\times \ldots \times \mathcal{X}_{m_M}$, not all cycles do so at the same rate. Then we will have a leading tower $\vec{\zeta}_{\text{t}}=\vec{\zeta}_{{\rm KK},{m_{i_0}}}\in \{\vec{\zeta}_{{\rm KK},m_j}\}_{j=1}^M$, so that
\begin{align}
	\vec{\zeta}_{\text{t}}\cdot\vec{\mathcal{Z}}_{\rm sp}&= \frac{1}{d+\sum_{j=1}^M m_j-2}\sum_{j=1}^M m_j\, \vec{\zeta}_{{\rm KK},\, {m_{i_0}}}\cdot\vec{\zeta}_{{\rm KK},\, m_j}\notag\\
	&=\frac{1}{d+\sum_{j=1}^M m_j-2}\left[m_{i_0}|\, \vec{\zeta}_{{\rm KK},\,{m_{i_0}}}|^2+\sum_{j\neq i_0}m_j\, \vec{\zeta}_{{\rm KK},\,{m_{i_0}}}\cdot\vec{\zeta}_{{\rm KK},\, m_j}\right]\notag\\
	&=\frac{1}{d+\sum_{j=1}^M m_j-2}\frac{d+\sum_{j=1}^Mm_{j}-2}{d-2}=\frac{1}{d-2}\, ,
\end{align}
where for the last sum in the second line we have used \eqref{eq: gen eq}. The generalization of this, for which several (but not all) of the cycles decompactify the fastest at the same pace is straightforward, as $\vec{\zeta}_{\text{t}}$ will be a convex combination of KK vectors (actually determined by the closest point to the origin of the polytope generated by the latter). Indeed, this follows from the fact that all possible $\vec{\zeta}_{\rm t}$ are located in the polytope spanned by the $\vec{\zeta}_{\text{KK},\, m_j}$ vectors corresponding to dimensions being decompactified, to which $\vec{\mathcal{Z}}_{\rm sp}$ is perpendicular, by construction.

\chapter{Details on the Hypermultiplet Metric}
\label{ap:hypermetric}
	
The material presented in this appendix is complementary to the discussion in Section \ref{ss:hypers}, where the fate of the pattern presented in Chapter \ref{ch:pattern} within certain heavily quantum-corrected moduli spaces was analyzed. Here we provide more details regarding the relevant non-perturbative corrections, as well as their contribution to the exact hypermultiplet metric. Section \ref{ss:exactmetric} briefly summarizes the procedure employed in \cite{Alexandrov:2014sya} to obtain the aforementioned line element, upon using the twistorial formulation of quaternionic-K\"ahler spaces. Section \ref{ss:SL2Z} reviews the duality properties of the hypermultiplet moduli space arising from Type II compactifications on CY three-folds \cite{Bohm:1999uk, Robles-Llana:2007bbv}, both at the classical and quantum levels. Finally, in Section \ref{ss:detailshyper} we use these considerations to argue how the pattern survives at the quantum level in a non-trivial way.
	
\section{The moduli space metric}
\label{ss:exactmetric}
	
The exact hypermultiplet metric for Type IIA string theory compactified on a CY$_3$ has been recently computed exactly to all orders in $g_s$ incorporating the contributions of \emph{mutually local} D2-brane instantons in \cite{Alexandrov:2014sya}. The strategy followed in that work was to exploit the twistorial description of quaternionic-K\"ahler manifolds (see e.g., \cite{Alexandrov:2008ds,Alexandrov:2010qdt}), combined with certain symmetries which are also expected to be preserved at the quantum level. In the following we will briefly review such computation in order to explicitly show the very non-trivial metric one arrives at, which is strongly corrected both at the perturbative and non-perturbative level, thus putting naively in danger any conclusion drawn from the tree-level metric displayed in \eqref{eq:classicalhypermetric}.
	
The crucial ingredient to obtain the hypermultiplet metric is the so-called \emph{contact potential} $\chi^{\rm IIA}$, which is a real-valued function defined over a twistor space $\mathcal{Z}$ constructed as a $\mathbb{P}^1$-bundle over the moduli space $\mathcal{M}_{\rm HM}$. It moreover has a connection given by the $\mathsf{SU(2)}$ part, $\vec{p}=\left( p^+, p^-, p^3\right)$, of the Levi-Civita connection on $\mathcal{M}_{\rm HM}$, which in turn determines the holomorphic contact structure associated to $\mathcal{Z}$ (see e.g., the review \cite{Alexandrov:2011va}). Therefore, one may define a holomorphic 1-form as follows
\begin{align}\label{eq:holomorphic1form}
	\mathcal{X}=-4 \text{i} \chi^{\text{IIA}} Dt\, ,
\end{align}
where $t$ is a complex coordinate on $\mathbb{P}^1$ and $Dt= d t +p^+-\text{i} p^3t+p^- t^2$. Now, in order to obtain the metric on $\mathcal{M}_{\rm HM}$ one first computes the contact potential $\chi^{\rm IIA}$ including all D-instanton corrections, which reads \cite{Alexandrov:2009zh}
\begin{align}\label{eq:chiIIAtwistor}
	\chi^{\rm IIA} =\, \frac{\mathcal{R}^2}{2} e^{-K_{\text{cs}}} + \frac{\chi_{E}(X_3)}{96\pi} -\frac{\text{i} \mathcal{R}}{16 \pi^2} \sum_{\gamma} \Omega(\gamma) \left( Z_{\gamma} \mathcal{J}_{\gamma}^{(1, +)} + \bar{Z}_{\gamma} \mathcal{J}_{\gamma}^{(1, -)}\right)\, ,
\end{align}
where $\mathcal{R}=e^{-\phi} \mathcal{V}_{A_0}/2$ is the mirror dual of the ten-dimensional IIB dilaton, $K_{\text{cs}}$ is the complex structure K\"ahler potential, and $Z_{\gamma}(z)=q_I z^I-p^I \mathcal{F}_I$ denotes the central charge function of a D2-instanton with integral charges $\gamma=\left( q_I, p^I\right)$. Their degeneracy is captured by the Donaldson-Thomas invariants $\Omega(\gamma)$, which count (in a BPS indexed way) the relevant instantons within the class $[\gamma] \in H_3(X_3, \mathbb{Z})$ \cite{Alexandrov:2013yva}.\footnote{The Donaldson-Thomas invariants $\Omega(\gamma)$ can be related, upon using Mirror Symmetry, to the genus-0 Gopakumar-Vafa invariants in the Type IIB dual description \cite{Gopakumar:1998ii, Gopakumar:1998jq}, see discussion after eq. \eqref{eq:alpha'correctedKahlerpot}.} We have also defined the twistorial integrals \cite{Alexandrov:2014sya}
\begin{align}\label{eq:Jintegral}
	\mathcal{J}_{\gamma}^{(1, \pm)}= \pm \int_{\ell_{\gamma}} \frac{\dd t}{t^{1 \pm 1}}\, \log \left( 1-\sigma_{\gamma} e^{-2\pi \text{i} \Theta_{\gamma}(t)}\right)\, ,
\end{align}
where $\ell_{\gamma}$ is a BPS ray on $\mathbb{P}^1$, $\sigma_{\gamma}$ is a sign function that we will take to be $+1$ in the following, and $\Theta_{\gamma}(t)$ are functions defined over the twistor space $\mathcal{Z}$ which, in the case of mutually local instantons, are given by
\begin{align}\label{eq:Thetafn}
	\Theta_{\gamma}(t) = q_I \xi ^I -p_I \tilde{\xi}_I + \mathcal{R} \left( t^{-1} Z_{\gamma}-t \bar{Z}_{\gamma}\right)\, .
\end{align}
As a next step, one needs to determine the $\mathsf{SU(2)}$ connection $\vec{p}$ as functions on the base $\mathcal{M}_{\text{HM}}$ and the complex coordinate $t \in \mathbb{P}^1$, from which one extracts the triplet of quaternionic 2-forms $\vec{\omega}$ as follows 
\begin{align}
	\vec{\omega}=-2 \left(d \vec{p} + \frac{1}{2} \vec{p} \times \vec{p} \right)\, .
\end{align}
The advantage of knowing $\vec{\omega}$ is that these are defined by the almost complex structures $\vec{\mathcal{I}}$ characterizing the quaternionic-K\"ahler manifold $\mathcal{M}_{\text{HM}}$ as well as by its metric. Therefore, upon specifying e.g., $\mathcal{I}^3$ by providing a basis of holomorphic 1-forms on $\mathcal{M}_{\text{HM}}$, one may retrieve the metric via the relation $g(X,Y)=\omega^3 (X, \mathcal{I}^3 Y)$, for all $X,Y \in T\mathcal{M}_{\text{HM}}$. Once all this has been done, one arrives at the quantum-corrected line element (we henceforth set all magnetic charges $p^I=0$, which can be achieved via some symplectic rotation) \cite{Alexandrov:2014sya}:
\begin{align}\label{eq:quantumhypermetric}
	\nonumber d s^2_{\text{HM}} &= \frac{1}{2\left(\chi^{\text{IIA}}\right)^2} \left( 1-\frac{\chi^{\text{IIA}}}{\mathcal{R}^2 U}\right)(d \chi^{\text{IIA}})^2 + \frac{1}{2 \left(\chi^{\text{IIA}}\right)^2\left( 1-\frac{\chi^{\text{IIA}}}{\mathcal{R}^2 U}\right)} \left( d \varrho - \tilde{\xi}_J d\xi^J+\xi^J d\tilde{\xi}_J + \mathcal{H} \right)^2\\
	\nonumber  &+\frac{\mathcal{R}^2}{2 \left(\chi^{\text{IIA}}\right)^2} \left| z^I \mathcal{Y}_I\right|^2 + \frac{1}{2 \chi^{\text{IIA}} U} \left| \mathcal{Y}_I M^{IJ} \bar{v}_J - \frac{\text{i} \mathcal{R}}{2\pi} \sum_{\gamma} \Omega_{\gamma} \mathcal{W}_{\gamma} d Z_{\gamma}\right|^2\\
	\nonumber &-\frac{1}{2 \chi^{\text{IIA}}} M^{IJ} \left( \mathcal{Y}_I + \frac{\text{i} \mathcal{R}}{2\pi} \sum_{\gamma} \Omega_{\gamma}\, q_I \mathcal{J}_{\gamma}^{(2, +)} \left( d Z_{\gamma}-U^{-1} Z_{\gamma}\, \partial e^{-K_{\text{cs}}}\right)\right)\\
	\nonumber &\times \left( \bar{\mathcal{Y}}_J - \frac{\text{i} \mathcal{R}}{2\pi} \sum_{\gamma'} \Omega_{\gamma'}\, q'_J \mathcal{J}_{\gamma'}^{(2, -)} \left( d \bar{Z}_{\gamma'}-U^{-1} \bar{Z}_{\gamma'}\, \bar{\partial} e^{-K_{\text{cs}}}\right)\right)\\
	\nonumber &+\frac{\mathcal{R}^2\, e^{-K_{\text{cs}}}}{2 \chi^{\text{IIA}}} \Bigg( G_{i \bar j} d z^i d z^{\bar j} - \frac{1}{\left( 2\pi U\right)^2} \left|  \sum_{\gamma} \Omega_{\gamma} \mathcal{W}_{\gamma} Z_{\gamma} \right|^2 \left| \partial K_{\text{cs}}\right|^2\\
	&+\frac{e^{K_{\text{cs}}}}{2\pi} \sum_{\gamma} \Omega_{\gamma} \mathcal{J}_{\gamma}^{(2)}\left| d Z_{\gamma}-U^{-1} Z_{\gamma}\, \partial e^{-K_{\text{cs}}}\right|^2\Bigg)\, ,
\end{align}
where $\mathcal{Y}_I$ is a (1,0)-form adapted to $\mathcal{I}^3$ which reads
\begin{align}\label{eq:holomorphic1formJ3}
	\mathcal{Y}_I= d\tilde{\xi}_I -\mathcal{F}_{IK} d\xi^K - \frac{1}{8\pi^2} \sum_{\gamma} \Omega_{\gamma} q_I d \mathcal{J}_{\gamma}^{(1)}\, ,
\end{align}
whilst $U$ denotes some real function that is defined as follows\footnote{Note that the quantity $U$ defined in \eqref{eq:Udef} can be intuitively thought of as an instanton corrected version of the complex structure K\"ahler potential.}
\begin{align}\label{eq:Udef}
	U= e^{-K_{\text{cs}}} - \frac{1}{2\pi} \sum_{\gamma} \Omega_{\gamma} \left| Z_{\gamma}\right|^2\mathcal{J}_{\gamma}^{(2)} + v_I M^{IJ} \bar{v}_J\, ,
\end{align}
with the matrix $M^{IJ}$ being the inverse of $M_{IJ}=-2\text{Im}\, \mathcal{F}_{IJ} - \sum_{\gamma} \Omega_{\gamma} \mathcal{J}_{\gamma}^{(2)} q_I q_J$, and the vector $v_I$ is given by
\begin{align}
	v_I=\frac{1}{4\pi} \sum_{\gamma} \Omega_{\gamma} q_I \left( Z_{\gamma} \mathcal{J}_{\gamma}^{(2, +)} + \bar{Z}_{\gamma} \mathcal{J}_{\gamma}^{(2, -)}\right)\, .
\end{align}
We have also introduced the quantities $\mathcal{W}_{\gamma}=\bar{Z}_{\gamma} \mathcal{J}_{\gamma}^{(2)}- \mathcal{J}_{\gamma}^{(2, +)}v_I M^{IJ} q_J$ and $\mathcal{H}$, the latter being a 1-form generalizing the K\"ahler connection on the complex structure moduli space (see \cite{Alexandrov:2014sya} for details); as well as the following twistorial integrals (c.f. \eqref{eq:Jintegral}) 
\begin{align}\label{eq:JintegralII}
	\mathcal{J}_{\gamma}^{(2, \pm)}&= \pm \int_{\ell_{\gamma}} \frac{\dd t}{t^{1 \pm 1}}\, \frac{1}{\sigma_{\gamma} e^{-2\pi \text{i} \Theta_{\gamma}(t)}-1}\, , \qquad \mathcal{J}_{\gamma}^{(2)}= \int_{\ell_{\gamma}} \frac{\dd t}{t}\, \frac{1}{\sigma_{\gamma} e^{-2\pi \text{i} \Theta_{\gamma}(t)}-1}\, , \notag\\
	\mathcal{J}_{\gamma}^{(1)}&= \int_{\ell_{\gamma}} \frac{\dd t}{t}\, \log\left(1-\sigma_{\gamma} e^{-2\pi \text{i} \Theta_{\gamma}(t)} \right)\, ,
\end{align}
which may be rewritten in terms of Bessel functions, thus capturing the exponentially suppressed behavior --- at large central charge --- associated to D-instanton effects.
	
Several comments are in order. First, notice how cumbersome the quantum-corrected metric becomes when compared with its classical analogue in \eqref{eq:classicalhypermetric}. 
Particularly interesting are the corrections to the metric components associated to the non-compact scalars, namely the 4d dilaton and the complex structure moduli. Regarding the former, it is the contact potential $\chi^{\rm IIA}$ which may be taken to parametrize the quantum hypermultiplet moduli space.\footnote{In fact, the real function $\chi^{\rm IIA}$ can be physically identified with the quantum-exact four-dimensional dilaton $\varphi_4$ \cite{deWit:2006gn}, and it plays a role similar to a would-be K\"ahler potential \cite{deWit:1999fp,deWit:2001brd}.} As for the latter, we clearly see that the classical piece $G_{i \bar j} d z^i d z^{\bar j}$ receives strong instanton corrections which can even overcome the tree-level contribution \cite{Marchesano:2019ifh}. Moreover, there also appear cross-terms of the form $(d \chi^{\text{IIA}} d z^i + \text{c.c.})$, which arise from the 1-form $d \mathcal{J}_{\gamma}^{(1)}$ inside $\mathcal{Y}_I$ in \eqref{eq:holomorphic1formJ3} above. Hence, a direct evaluation of the pattern discussed in Chapter \ref{ch:pattern} at infinite distance points within $\mathcal{M}_{\text{HM}}$ in principle requires from the use of the full lime element \eqref{eq:quantumhypermetric}, which can become rather involved depending on the limit of interest. Therefore, it is highly non-trivial for the inner product $\vec{\zeta}_{\text{t}} \cdot \vec{\mathcal{Z}}_{\text{sp}}$ to verify \eqref{eq:pattern} at any infinite distance boundary, even if it does so already at the classical level.
	
\subsection{The contact potential $\chi^{\rm IIA}$}
\label{sss:chiIIA}
	
Before moving on, let us have a closer look at the contact potential to get a grasp on its physical meaning. This will also provide us with some useful formulae that will be used several times in the following.
	
Therefore, we start from the twistorial expression for $\chi^{\rm IIA}$, as shown in eq. \eqref{eq:chiIIAtwistor}, which may be written as follows \cite{Alexandrov:2008gh}
\begin{equation}\label{eq:fullcontactpotential}
	\chi^{\rm IIA}= \chi^{\rm IIA}_{\rm class} + \chi^{\rm IIA}_{\rm quant}\, .
\end{equation}
The first term corresponds to the classical piece
\begin{equation}\label{eq:classicalcontactpot}
	\chi^{\rm IIA}_{\rm class} = \frac{\mathcal{R}^2}{2} e^{-K_{\text{cs}}}\, ,
\end{equation}
such that $\chi^{\rm IIA}_{\rm class}$ matches with $e^{-2\varphi_4}$, as one can easily check upon using eqs. \eqref{eq:CSmetric} and \eqref{eq:slagvolumes}. On the other hand, for the quantum corrected piece, $\chi^{\rm IIA}_{\rm quant}$ and in the particular case of mutually local instantons arising from D2-branes\footnote{This set of instantons is mapped by Mirror Symmetry to D($-1$) and D1-instantons wrapping holomorphic 0- and 2-cycles within the CY three-fold, respectively \cite{Robles-Llana:2007bbv}.} wrapping sLag representatives of the 3-cycle classes $[A_I]$, one finds \cite{Robles-Llana:2007bbv,Alexandrov:2014sya,Cortes:2021vdm}
\begin{equation}\label{eq:quantumchi}
	\chi^{\rm IIA}_{\rm quant} = \frac{\chi_{E}(X_3)}{96\pi} + \frac{\mathcal{R}}{2\pi^2} \sum_{\gamma} \Omega (\gamma) \sum_{m=1}^{\infty} \frac{|k_I z^I|}{m} \cos \left( 2\pi m k_I \zeta^I\right) K_1 \left( 4\pi m \mathcal{R}|k_I z^I|\right)\, ,
\end{equation}
where the term proportional to the Euler characteristic of the three-fold, $\chi_{E}(X_3)=2( h^{1,1}(X_3)-h^{2,1}(X_3))$, comes from a one-loop $g_s$-correction, whilst the second piece arises from the non-perturbative D2-brane instantons. To actually see how \eqref{eq:quantumchi} arises from eq. \eqref{eq:chiIIAtwistor} above, one needs to substitute the definition of the quantities $\mathcal{J}_{\gamma}^{(1, \pm)}$ (c.f. eq. \eqref{eq:Jintegral}), then expand the logarithm around $\Theta_{\gamma}=0$ and finally rewrite the integrals in terms of the modified Bessel function upon using the following identity
\begin{align}\label{eq:Besselintegral}
	\int_{0}^{\infty} \frac{\dd y}{y}\, \left(ay + \frac{b}{y}\right) e^{-\left(ay+b/y\right)/2} = 4 \sqrt{ab} K_1 \left( \sqrt{ab}\right)\, .
\end{align}

Notice that the contribution to \eqref{eq:quantumchi} associated to the D2-instantons is controlled by their BPS central charge, which coincides (up to order one factors) with the corresponding 4d action
\begin{equation}\label{eq:D2instantonaction}
	S_{m,\, k_I} = 4\pi m \mathcal{R} |k_I z^I| + 2\pi {\rm i} m k_I \zeta^I\, ,
\end{equation}
where $k_I=(k_0, \mathbf{k})$ denote the (quantized) instanton charges. The axionic v.e.v.s $\zeta^I$ measure the oscillatory part of the corrections, whereas the non-compact scalars $(z^I, \mathcal{R})$ determine their `size' through the modified Bessel function $K_1(y)$.
	
\section{$\mathsf{SL(2,\mathbb{Z})}$ duality}
\label{ss:SL2Z}
	
Here we provide some details regarding the $\mathsf{SL(2,\mathbb{Z})}$ invariance that the Type IIA hypermultiplet metric inherits from its dual Type IIB compactification via Mirror Symmetry. This will moreover highlight the effect that the D2-brane instanton corrections have on certain (classical) infinite distance singularities $\mathcal{M}_{\rm HM}^{\rm IIA}$ in the large complex structure (LCS) limit studied in Section \ref{ss:hypers} (see also Section \ref{ss:detailshyper} below).
	
\subsection{The classical metric}
	
Let us first exhibit the duality of the theory at the classical level. The tree-level metric was shown in \eqref{eq:classicalhypermetric} above, and we repeat it here for the comfort of the reader:
\begin{align}\label{eq:classicalhypermetricII}
	h_{p q}\, d q^p d q^q &= \left( d \varphi_4\right)^2 + G_{i \bar j} d z^i d z^{\bar j} + \frac{e^{4\varphi_4}}{4} \left( d \varrho - \left( \tilde{\xi}_J d\xi^J-\xi^J d\tilde{\xi}_J \right)\right)^2 \notag\\
	& -\frac{e^{2\varphi_4}}{2} \left( \text{Im}\, \mathcal{U}\right)^{-1\ IJ} \left( d\tilde{\xi}_I -\mathcal{U}_{IK} d\xi^K\right) \left( d\tilde{\xi}_J -\bar{\mathcal{U}}_{JL} d\xi^L\right)\, ,
\end{align}
where the different fields describing the hypermultiplet sector of Type IIA on the three-fold $X_3$ were discussed around \eqref{eq:CSmoduli}. In order to uncover the $\mathsf{SL(2,\mathbb{Z})}$ invariance of the action at tree-level, it is useful to switch to the Type IIB mirror description, where the symmetry is manifest, and then map the duality transformations back to the original Type IIA set-up. Regarding the first step,  we will simply state here the relevant identifications, whilst referring the reader interested in the details to the original references \cite{Candelas:1990rm, Aspinwall:1993nu}. These read 
\begin{align}\label{eq:mirrormap}
	\xi^0 &= \tau_1\, , \quad \xi^i = c^i-\tau_1 b^i\, , \quad z^i=b^i+\text{i}t^i\, , \quad \mathcal{R}=\frac{\tau_2}{2}\, , \notag\\
	\tilde{\xi}_0 &= c^0-\frac{1}{2} \rho_j b^j + \frac{1}{2} \kappa_{ijk} c^i b^j b^k - \frac{1}{6} \tau_1\, \kappa_{ijk} b^i b^j b^k\, , \quad \tilde{\xi}_i=\rho_i - \kappa_{ijk} c^j b^k + \frac{1}{2} \tau_1\, \kappa_{ijk} b^j b^k\, , \notag\\
	\varrho &= 2b^0 + \tau_1 c^0 + \rho_j \left( c^j - \tau_1 b^j\right)\, ,
\end{align}
where $\tau=\tau_1 + \text{i} \tau_2= C_0 + \text{i}\, e^{-\phi_{\rm{IIB}}}$ is the Type IIB axio-dilaton, $\vartheta^i \equiv b^i+ \text{i} t^i$ denote the (complexified) K\"ahler moduli of the mirror three-fold $Y_3$, $\{ c^i, \rho_i\}$ arise as period integrals of the RR and 2-form and 4-form fields $\{C_2, C_4\}$ over integral bases of $H_2(Y_3)$ and $H_4(Y_3)$, respectively; and finally $\{b^0, c^0 \}$ are scalar fields dual to the four-dimensional components of the 2-forms $C_2$ and $B_2$. We stress that the complex structure moduli $\{z^i\}$ appearing in the mirror map above should be taken as the `flat' (inhomogeneous)  coordinates associated to the expansion of the prepotential around the LCS point \cite{Hori:2003ic}. Therefore, upon applying such map to the line element displayed in \eqref{eq:classicalhypermetricII} one obtains \cite{Ferrara:1989ik}
\begin{align}\label{classicalhypermetric}
	\nonumber h_{pq} d q^p d q^q &= (d \varphi_4)^2 + G_{i \bar j} d \vartheta^i d \bar \vartheta^j + \frac{1}{24} e^{2\varphi_4}\cK (d C_0)^2\\
	\nonumber  &+\frac{1}{6}e^{2\varphi_4}\cK G_{i \bar j} \left(d c^i-C_0 d b^i\right) \left(d c^j-C_0 d b^j\right)\\
	&+\frac{3}{8\cK}e^{2\varphi_4}G^{i \bar j}\left(d \rho_i - \kappa_{ikl} c^k d b^l\right)\left(d \rho_j - \kappa_{jmn} c^m d b^n\right)\\
	\nonumber &+\frac{3}{2\cK}e^{2\varphi_4}\left(d c^0-\frac{1}{2}(\rho_id b^i -b^i d \rho_i)\right)^2\\
	\nonumber&+\frac{1}{2}e^{4\varphi_4}\left(d b^0+C_0d c^0 +c^i d \rho_i+\frac{1}{2}C_0(\rho_id b^i- b^i d \rho_i)-\frac{1}{4} \kappa_{ijk}c^i c^jd  b^k\right)^2\, .
\end{align}
Now, as already mentioned, the 4d theory inherits from the 10d supergravity a continuous $\mathsf{SL(2,\mathbb{R})}$ symmetry which is broken down to a discrete $\mathsf{SL(2,\mathbb{Z})}$ subgroup by non-perturbative effects (see Section \ref{s:dualities} for details). The action of any such element $\mathcal{A} \in \mathsf{SL(2,\mathbb{Z})}$ on the Type IIB coordinates reads as \cite{Gunther:1998sc, Bohm:1999uk}
\begin{align}\label{eq:SdualitytransIIB}
	\tau \rightarrow \frac{a\tau + b}{c\tau+d}\,,\qquad
	t^i \rightarrow |c\tau+d| t^i \,,\qquad
	\begin{pmatrix}
		c^i\\b^i
	\end{pmatrix}
	\rightarrow
	\begin{pmatrix}
		a \quad  b\\c \quad  d
	\end{pmatrix}
	\begin{pmatrix}
		c^i\\b^i
	\end{pmatrix}\, , 
\end{align}
where we have only displayed the transformations that are most relevant for our purposes here.\footnote{\label{fnote:SL2Zcontactpotential}Notice that under \eqref{eq:SdualitytransIIB}, the 4d dilaton transforms non-trivially, namely $e^{-2\varphi_4} \rightarrow \frac{e^{-2\varphi_4}}{|c\tau+d|}$.} One can then easily check that these are already enough so as to prove the invariance of the first two rows in \eqref{classicalhypermetric} under $\mathsf{SL(2,\mathbb{Z})}$.
	
Finally, it is now straightforward to translate the S-duality transformations \eqref{eq:SdualitytransIIB} into a set of analogous ones in the Type IIA mirror dual compactification upon using the mirror map \eqref{eq:mirrormap}. This leads to
\begin{align}\label{eq:SdualitytransIIA}
	&\Xi \rightarrow \frac{a\, \Xi + b}{c\, \Xi+d}\,,\qquad
	\text{Im}\, z^i \rightarrow |c\, \Xi+d|\, \text{Im}\, z^i \,,\qquad \notag\\
	&\begin{pmatrix}
		\xi^i + \xi^0\, \text{Re}\, z^i\\ \text{Re}\, z^i
	\end{pmatrix}
	\rightarrow
	\begin{pmatrix}
		a \quad  b\\c \quad  d
	\end{pmatrix}
	\begin{pmatrix}
		\xi^i + \xi^0\, \text{Re}\, z^i\\ \text{Re}\, z^i
	\end{pmatrix}\, , 
\end{align}
where we have defined the complex field $\Xi= \xi^0 + 2 \text{i} \mathcal{R}$. Note that this is again sufficient to show the invariance of the metric components in \eqref{eq:classicalhypermetricII} associated to the 4d dilaton, the complex structure and the $\xi^I$ coordinates.
	
\subsection{Quantum corrections}
	
One can go beyond the previous tree-level analysis and study $\mathsf{SL(2,\mathbb{Z})}$ duality once quantum corrections have been taken into account. Following the discussion of Section \ref{ss:exactmetric}, we will only consider the effect of `electric' D2-brane instantons, i.e. those wrapping the $A_I$\,-cycles introduced in \eqref{eq:symplecticpairing}.  
	
Recall that the quantum hypermultiplet metric can be effectively encoded into the contact (or tensor) potential, $\chi^{\rm IIA}$, which reads (see Section \ref{sss:chiIIA})
\begin{align}\label{eq:apchiIIA}
	\chi^{\rm IIA} &=\, \frac{\mathcal{R}^2}{2} \frac{{\rm i} \int\Omega \wedge \bar \Omega }{|Z^0|^2} + \frac{\chi_{E}(X_3)}{96\pi} \notag\\
	&+ \frac{\mathcal{R}}{2\pi^2} \sum_{\gamma} \Omega (\gamma) \sum_{m=1}^{\infty} \frac{|k_I z^I|}{m} \cos \left( 2\pi m k_I \xi^I\right) K_1 \left( 4\pi m \mathcal{R}|k_I z^I|\right) \, ,
	\end{align}
where the first, second and third terms correspond to the classical, one-loop and D2-instanton contributions, respectively. Now, instead of trying to show how the exact hypermultiplet metric \eqref{eq:quantumhypermetric} still respects $\mathsf{SL(2,\mathbb{Z})}$ duality, we will concentrate on rewriting the above expression in a way which manifestly reflects the symmetry. This will allow us to relate certain non-perturbative corrections to classically-derived terms, thus providing more evidence in favour of our argumentation in Section \ref{ss:detailshyper} below. 
	
Let us start by extracting a common $\sqrt{\mathcal{R}}$ factor from each of the three terms in \eqref{eq:apchiIIA}, yielding
\begin{align}\label{eq:modinvariantchi}
	\frac{\chi^{\rm IIA}}{\sqrt{\mathcal{R}}} &=\, \frac{\mathcal{R}^{3/2}}{2} \frac{{\rm i} \int\Omega \wedge \bar \Omega }{|Z^0|^2} + \frac{\chi_{E}(X_3)}{96\pi} \mathcal{R}^{-1/2} \notag\\
	&+ \frac{\mathcal{R}^{1/2}}{2\pi^2} \sum_{\gamma} \Omega (\gamma) \sum_{m=1}^{\infty} \frac{|k_I z^I|}{m} \cos \left( 2\pi m k_I \xi^I\right) K_1 \left( 4\pi m \mathcal{R}|k_I z^I|\right) \, .
\end{align}
Therefore, given that the contact potential transforms under $\mathsf{SL(2, \mathbb{Z})}$ precisely the same way as $\sqrt{\mathcal{R}}$ does (see footnote \ref{fnote:SL2Zcontactpotential}), we can now concentrate on finding a modular invariant expression for the right-hand side of \eqref{eq:modinvariantchi}. To do so, we first expand the classical term around the LCS, as follows
\begin{equation}\label{eq:chiclassical}
	\begin{aligned}
		\frac{\mathcal{R}^{3/2}}{2} \frac{{\rm i} \int\Omega \wedge \bar \Omega }{|Z^0|^2} =&\,  4 \mathcal{R}^{3/2} \bigg[ \frac{1}{3!} \kappa_{ijk} v^i v^j v^k + \frac{\zeta(3) \chi_{E}(X_3)}{4(2 \pi)^3}\\
		&+ \frac{1}{2(2 \pi)^3} \sum_{\textbf{k}>0} n_{\textbf{k}}\, \text{Re}\, \left \lbrace \text{Li}_3 \left( e^{2\pi \text{i}k_i z^i} \right) + 2\pi k_i v^i \text{Li}_2 \left( e^{2\pi \text{i}k_i z^i} \right)\right \rbrace \bigg ]\, ,
	\end{aligned}
\end{equation}
where $\zeta(x)$ denotes the Riemann zeta function, $\text{Li}_k (x)= \sum_{j=1}^{\infty} \frac{x^j}{j^k}$ is the polylogarithm function and we have defined $v^i \equiv \text{Im}\, z^i$ in the above expression. The physical interpretation of each term is clear: the first piece corresponds to the classical volume term of the mirror dual Type IIB compactification on $Y_3$, whilst the second and third ones arise as perturbative and non-perturbative $\alpha'$-corrections that modify the former away from the large volume point. The integers $n_{\textbf{k}}$ denote the genus-zero Gopakumar-Vafa invariants that `count' the multiplicity of holomorphic 2-cycles in a given class $[k_i \gamma^i] \in H^+_2(Y_3, \mathbb{Z})$.
	
Next, we divide the instanton piece in \eqref{eq:modinvariantchi} into two different terms, namely we separate the contributions associated to D2-branes wrapped on the SYZ cycle from those wrapping the remaining $A_I$\,-cycles. The reason for doing so will become clear in the following. This leads to
\begin{align}\label{eq:chiD2}
	\frac{\chi^{\rm IIA}_{\text{D2}}}{\sqrt{\mathcal{R}}} =\, &\frac{\mathcal{R}^{1/2} \chi_{E}(X_3)}{8\pi^2} \sum_{k_0, m \neq 0} \left| \frac{k_0}{m}\right| e^{ 2\pi \i m k_0 \xi^0} K_1 \left( 4\pi \mathcal{R}|m k_0|\right)\notag\\
	&+ \frac{\mathcal{R}^{1/2}}{4\pi^2} \sum_{\textbf{k}>0} n_{\textbf{k}} \sum_{m\neq0, k_0 \in \mathbb{Z}} \frac{|k_I z^I|}{|m|} e^{ 2\pi \i m k_I \xi^I} K_1 \left( 4\pi m \mathcal{R}|k_I z^I|\right)\, ,
\end{align}
where we have substituted the Donaldson-Thomas invariants $\Omega(\gamma)$ by $\chi_{E}(X_3)/2$ and $n_{\textbf{k}}$ for $\gamma = \left(k_0 \neq 0, \textbf{k}=0 \right)$ and $\gamma = \left(k_0 \in \mathbb{Z}, \textbf{k} > 0 \right)$, respectively. 
	
With this, we are finally ready to rewrite \eqref{eq:modinvariantchi} in a manifestly modular invariant way. Notice that the first term in eq. \eqref{eq:chiclassical} is left unchanged under the set of transformations in \eqref{eq:SdualitytransIIA}, reflecting the fact that the tree-level hypermultiplet metric at LCS/Large Volume is modular invariant. Consider now the terms which are proportional to the Euler characteristic of the three-fold, $\chi_{E}(X_3)$. They read
\begin{align}\label{eq:chiEuler}
	\frac{\chi^{\rm IIA}_{\chi_{E}}}{\sqrt{\mathcal{R}}} =\, &\frac{\chi_{E}(X_3)}{2(2\pi)^3} \bigg[ 2\mathcal{R}^{3/2} \zeta(3) + \frac{\pi^2}{6} \mathcal{R}^{-1/2} + 4\pi \mathcal{R}^{1/2} \sum_{k_0 \neq 0, m > 0} \left| \frac{k_0}{m}\right| e^{ 2\pi \i m k_0 \xi^0} K_1 \left( 4\pi \mathcal{R}|m k_0|\right) \bigg]\, .
\end{align}
which from eq. \eqref{eq:nonpertexpansion} we recognize to be
\begin{align}\label{eq:chiEulerII}
	\frac{\chi^{\rm IIA}_{\chi_{E}}}{\sqrt{\mathcal{R}}}\, =\, \frac{\chi_{E}(X_3)}{2(2\pi)^3} \sum_{m, n \in \mathbb{Z}\setminus \lbrace (0,0) \rbrace} \frac{\mathcal{R}^{3/2}}{|m\, \Xi +n|^3}\, ,
\end{align}
and is indeed modular invariant.
	
Finally, we group together those terms containing sums over Gopakumar-Vafa invariants, such that, after performing a Poisson resummation over the unconstrained integer $k_0$ (c.f. footnote \ref{fnote:Poissonresummation}), one finds \cite{Robles-Llana:2007bbv}
\begin{align}\label{eq:chiGV}
	\frac{\chi^{\rm IIA}_{\text{GV}}}{\sqrt{\mathcal{R}}}\, =\, \frac{1}{(2\pi)^3} \sum_{\textbf{k}\neq 0} n_{\textbf{k}} \sum_{m, n \in \mathbb{Z}\setminus \lbrace (0,0) \rbrace} \frac{\mathcal{R}^{3/2}}{|m\, \Xi +n|^3} \left( 1+2\pi |m\, \Xi +n| k_i v^{i}\right) e^{-S_{m, n}}\, ,
\end{align}
where $S_{m, n}= 2 \pi k_i \left( |m\, \Xi +n| v^i + \i m \left( \xi^i+\xi^0 \text{Re}\, z^i \right) -in \text{Re}\, z^i\right)$. This last term can be seen to be the mirror dual of the quantum corrections arising from Euclidean Type IIB $(p,q)$-strings, and it tells us that the exponentially suppressed terms within the complex structure K\"ahler potential --- close to the LCS point --- are related by $\mathsf{SL(2, \mathbb{Z})}$ duality to certain D2-brane instanton contributions. In fact, it is precisely this relation the one that plays a key role when evaluating the pattern \eqref{eq:pattern} after taking into account stringy quantum corrections, see Section \ref{ss:detailshyper} below.
	
\section{The evaluation of the pattern within $\mathcal{M}_{\rm HM}$}
\label{ss:detailshyper}
	
In Section \ref{ss:hypers} from the main text, we were interested in evaluating the relation \eqref{eq:pattern} for certain trajectories lying entirely within the hypermultiplet moduli space $\mathcal{M}_{\rm HM}$. Such infinite distance paths were of the form
\begin{equation}\label{eq:generictrajII}
	\text{Im}\, z^i \sim \sigma^{e^1}\, , \qquad e^{-\varphi_4}\sim \sigma^{e^2}\, , \qquad \sigma \to \infty\, ,
\end{equation}
with $e^1, e^2 \geq 0$, thus including both the weak coupling and large complex structure points. Classically, i.e. without taking into account D-instanton corrections, both kind of limits were shown to fulfill the pattern. Quantum-mechanically, however, one expects large instanton contributions to modify the computation, at least in some cases. The purpose of this subsection is to put all the machinery previously described into work in order to prove that eq. \eqref{eq:pattern} still holds even after taking into account all relevant quantum effects, as advertised in Section \ref{sss:instantons}. We analyze each of these limits in turn.	
	
\subsubsection*{Weak coupling point}
	
In this case, since the singularity that is being approached is at weak string coupling, we do not expect neither perturbative nor non-perturbative effects to become important, and indeed the classical analysis from Section \ref{sss:classivalvsquantum} should be reliable. This can be readily confirmed upon looking at the behavior of the sum in eq. \eqref{eq:quantumchi}, since for $\mathcal{R} \to \infty$ and $z^I$ finite one finds
\begin{equation}\label{eq:asympotic Bessel infinity}
	K_1 \left( 4\pi m \mathcal{R}|k_I z^I|\right) \sim \sqrt{\frac{1}{8m \mathcal{R}|k_I z^I|}}\, e^{-4\pi m \mathcal{R}|k_I z^I|} 
	\, ,
\end{equation}
such that $\chi^{\rm IIA}_{\rm quant}= \text{const.}\, + \mathcal{O}\left( e^{-\mathcal{R}|k_I z^I|}\right) \ll \chi^{\rm IIA}_{\rm class}$ asymptotically. Similarly, the moduli space metric deviates from the tree-level one by additional terms which at leading order behave as follows (c.f. \eqref{eq:quantumhypermetric}) 
\begin{equation}
	\delta  d s_{\rm HM}^2 =  \delta  d s_{\rm HM}^2\rvert_{\text{1-loop}} + \delta  d s_{\rm HM}^2\rvert_{\text{D-inst}}\, \sim\, \frac{\chi_{E}(X_3)}{\chi^{\rm IIA}}\, +\, \sum_{\gamma} \Omega_{\gamma}\, e^{-S_{m,\, k_I}}\, ,
\end{equation}
and thus it is enough to use the classical approximation \eqref{eq:classicalhypermetricII}. Therefore, we conclude that the calculations performed after \eqref{eq:fundstringmass} remain valid, and the pattern is still verified.
	
Let us also say a few words about the S-dual limit, since it will play a crucial role in what follows. As we mention in the main text, the weak coupling singularity here discussed translates into a physically equivalent one at both strong coupling and LCS, namely $\left( \mathcal{R}' \sim \sigma^{-1} , \text{Im}\, z^{i\, '} \sim \sigma \right)$. Notice that $\mathcal{R}'\, \text{Im}\, z^{j\, '} \to \text{const.}$ , which means, in practice, that the tree-level piece of $\chi^{\rm IIA}$ still dominates over the quantum corrections, i.e. the D2-brane instanton contributions decouple.\footnote{This is not completely true, since the instanton sum can still lead to additional \emph{finite distance} degenerations, which are the S-dual versions of the conifold loci \cite{Candelas:1989js}.} Hence, one can again safely use the classical metric \eqref{eq:classicalhypermetricII} to compute the inner products between the relevant charge-to-mass and species vectors. These are associated to the D4-string, with tension
\begin{equation}\label{eq:D4SYZSdual}
	\left(\frac{T_{\text{D4}}}{\Mpf^2} \right)= \frac{2 \mathcal{R}'}{\left(\chi^{\rm IIA}\right)'} = \frac{1}{\chi^{\rm IIA}} \sim \frac{1}{\sigma^{2}}\, ,
\end{equation}
and the KK scale
\begin{equation}\label{eq:KKSYZSdual}
	\left(\frac{m_{\text{KK},\, B^0}}{\Mpf} \right)^2 \sim \frac{1}{\text{Im}\, z^{i\, '} \left(\chi^{\rm IIA}\right)'} \sim  \frac{1}{\text{Im}\, z^{i}\, \chi^{\rm IIA}} \sim \frac{1}{\sigma^2}\, ,
\end{equation}
where in order to arrive at the second equalities we have used the S-duality transformation rules (see eq. \eqref{eq:SdualitytransIIA}). 
	
\subsubsection*{Large complex structure point}
	
A slightly different story holds for the second kind of limit, namely that corresponding classically to large complex structure at fixed 4d dilaton
\begin{equation}\label{eq:LCSfixeddilaton}
	z^j ={\rm i} \xi^j \sigma\, , \qquad \varphi_4= \text{const.}\, , \qquad \sigma \to \infty\, .
	\end{equation}
This limit is indeed the mirror dual to the one explored in \cite{Marchesano:2019ifh,Baume:2019sry}. In terms of the relevant coordinates controlling the behavior of the contact potential, such trajectories are of the form $(z^j (\sigma), \mathcal{R} (\sigma)) \sim \left( {\rm i} \sigma, \sigma^{-3/2}\right)$, which means that for small enough instanton charges $k_I$, the correction term controlled by the Bessel function in \eqref{eq:quantumchi} will behave as 
\begin{equation}\label{eq:strongorrections}
	K_1 \left( 4\pi m \mathcal{R}|k_I z^I|\right) \sim\frac{1}{4\pi m \mathcal{R}|k_I z^I|}\, .
\end{equation}
More precisely, the charges must be such that
\begin{equation}\label{eq:instantoncondition}
	4\pi m \mathcal{R}|k_0 + k_i z^i| \ll 1\, ,
\end{equation}
for the associated D2-instantons to contribute significantly to the tensor potential $\chi^{\rm IIA}_{\text{quant}}$. As already noted in \cite{Marchesano:2019ifh}, this parallels the behavior of the exponentially light towers of D3-brane bound states appearing in the mirror dual vector multiplet moduli space \cite{Grimm:2018ohb}.
	
To see what is the upshot of including such quantum corrections into the hypermultiplet metric along the limit specified by \eqref{eq:LCSfixeddilaton}, one can follow the same strategy as in \cite{Baume:2019sry} and exploit the $\mathsf{SL(2, \mathbb{Z})}$ duality of the theory. This allows us to translate the aforementioned limit into a simpler one where we can readily identify the relevant asymptotic physics. Indeed, after performing the duality we end up exploring the following `classical' limit
\begin{equation}\label{eq:smallCSsmalldilaton}
	\text{Im}\, z^{j\, '} \sim \sigma^{-1/2}\, , \qquad \mathcal{R}' = \frac{e^{-\phi\, '} \mathcal{V}_{A_0}'}{2} \sim \sigma^{3/2}\, , \qquad e^{2\varphi_4'} \sim \sigma^{-3/2}\, ,
\end{equation}
where one should think of $z^{i\, '}= \frac{1}{2\pi {\rm i}} \log x^i$ as flat complex structure variables defined close to the LCS point ($x^i \to 0$), see below. Notice that this is nothing but the mirror dual of the F1 limit studied in \cite{Baume:2019sry}. There, the relevant quantum corrections to the classical Type IIB hypermultiplet metric are induced by $\alpha'$ and worldsheet instantons, whilst D-brane contributions decouple. Importantly, here such `corrections' are already captured by the \emph{exact} complex structure metric \eqref{eq:CSmetric}, thus simplifying the analysis enormously.
	
Therefore, recall that away from the LCS point, the periods of the holomorphic $(3,0)$-form $\Omega$ receive corrections from their flat values, namely \cite{B_hm_2000, Hori:2003ic} (see also Section \ref{ss:dualitieswithlowersusy} for details)
\begin{equation}\label{eq:analyticCSvariables}
	z^{j\, '} = \frac{1}{2\pi {\rm i}} \log x^j + \mathcal{O}(x^i)\, ,
\end{equation}
such that upon increasing $x^i$ towards one, the logarithmic approximation for $z^{i\, '}$ stops being valid and the polynomial corrections clearly dominate. Hence, instead of reaching a point where $\text{Im}\, z^{i\, '} \to 0$ asymptotically, what happens is that the complex structure variables generically approach some constant $\mathcal{O}(1)$ value (see e.g., \cite{Blumenhagen:2018nts,Joshi:2019nzi,Alvarez-Garcia:2021mzv,Cota:2023uir}). 
This does not prevent, however, the $\mathcal{R}$ coordinate from keep flowing towards weak coupling, such that a more accurate parametrization of the asymptotic trajectory would be the following
\begin{equation}\label{eq:smallCSsmalldilatoncorrected}
	\text{Im}\, z^{j\, '} = \text{const.}\, , \qquad \mathcal{R}' \sim \sigma^{3/2}\, , \qquad e^{2\varphi_4'} \sim \sigma^{-3}\, .
\end{equation}
Notice that this belongs to the family of geodesics in \eqref{eq:generictrajII} with $\mathbf{e}=(0, 3/2)$. Hence, our previous analysis for the weak coupling singularity around \eqref{eq:asympotic Bessel infinity} also applies here and we conclude that the pattern still holds.
	
From the original perspective, though, a direct evaluation of the scalar product \eqref{eq:pattern} seems to be rather involved, since the metric receives strong corrections (c.f. eq. \eqref{eq:quantumhypermetric}) that deviate from the simple block diagonal form displayed in \eqref{eq:classicalhypermetricII} above. However, let us stress again that we do not need to do this, as we already know what is the S-dual limit of \eqref{eq:smallCSsmalldilatoncorrected}: It corresponds to an infinite distance trajectory of the form $\left( \mathcal{R} \sim \sigma^{-3/2} , \text{Im}\, z^{i} \sim \sigma^{3/2} \right)$, thus located at strong coupling and LCS (see discussion around \eqref{eq:D4SYZSdual}). Incidentally, this nicely explains why the pattern was still verified along the classically obstructed limit \eqref{eq:LCSfixeddilaton}, since the products in eqs. \eqref{eq:patternviolation} and \eqref{eq:patternviolationII} are formally identical to the ones that need to be computed along the present quantum corrected trajectory.
\newpage

\bibliographystyle{JHEP}
\bibliography{alberbib}
\inputencoding{utf8}

\newpage
\thispagestyle{empty}
\phantom{lala}
\end{document}